\documentclass[11pt]{article}

\usepackage[top=2cm,bottom=2cm,left=1in,right=1in]{geometry}
\usepackage{graphicx}
\usepackage[export]{adjustbox}   % trim/clip panels of the merged tikzDevice figures
\usepackage{amsfonts, amssymb, amsmath, amsthm, booktabs, dsfont, float, hyperref, mathrsfs, euscript, enumitem, bm, bbm, pifont, multirow, tabularx, multicol, tikz}
\usepackage{accents}   % \accentset for \rhopad, \underaccent for \ubar; load after amsmath
\usepackage[onehalfspacing]{setspace}
\usepackage[labelfont=bf,textfont=it]{caption}
\usepackage{subcaption}
\definecolor{ptonorange}{RGB}{232,119,34}
\definecolor{figorange}{RGB}{213,94,0}
\definecolor{NavyBlue}{cmyk}{0.94,0.54,0,0}
\hypersetup{pdfborder = {0 0 0},colorlinks=true,allcolors=figorange,
            citecolor=NavyBlue}
\colorlet{DarkRed}{red!70!black}
\colorlet{DarkGreen}{green!70!black}
\definecolor{tvblue}{HTML}{1F4E79}
\usepackage{natbib}
\makeatletter
\renewcommand\NAT@nmfmt[1]{{\color{black}\NAT@up#1}}
\makeatother
\usepackage[title]{appendix}
\usepackage{placeins}
\usepackage{threeparttable}
\usepackage[normalem]{ulem}
\usepackage{algorithm}
\usepackage{algpseudocode}

\usepackage{pgfplots}
\pgfplotsset{compat=1.18}
\usepgfplotslibrary{fillbetween}
\usetikzlibrary{
arrows.meta,
automata,
backgrounds,
bending,
calc,
decorations.pathreplacing,
shapes,
shapes.geometric,
shapes.misc,
shapes.multipart,
positioning}

\everydisplay\expandafter{\the\everydisplay
  \abovedisplayskip=12.66pt plus 3.165pt minus 7.385pt
  \belowdisplayskip=12.66pt plus 3.165pt minus 7.385pt
  \abovedisplayshortskip=0pt plus 3.165pt
  \belowdisplayshortskip=6.858pt plus 3.693pt minus 3.165pt}
\allowdisplaybreaks                 % Allow multiple-page

\theoremstyle{plain}
\newtheorem{thm}{Theorem}
\newtheorem{prop}{Proposition}

\newtheorem{assumptionLS}{Assumption}
\renewcommand{\theassumptionLS}{LS}

\numberwithin{equation}{section}

\makeatletter
\newcommand{\@endmark}[1]{%
  \unskip\nobreak\hfil\penalty50\hskip1em\hbox{}\nobreak\hfil#1%
  \parfillskip=\z@\finalhyphendemerits=\z@\relax}
\newcommand{\exmarkclose}{\@endmark{\ding{169}}}   % end of example (as in the SA)
\newcommand{\remarkclose}{\@endmark{\ding{168}}}   % end of remark  (as in the SA)
\makeatother

\theoremstyle{definition}
\newtheorem{example_tmp}{Example}[section]
\newenvironment{example}
	{ \begin{example_tmp} 	}
	{\exmarkclose\end{example_tmp}}

\newenvironment{examplecont}[1]
	{\par\addvspace{\medskipamount}\noindent\textbf{#1, continued.}\enspace\ignorespaces}
	{\exmarkclose\par\addvspace{\medskipamount}}

\theoremstyle{definition}
\newtheorem{remark_tmp}{Remark}
\newenvironment{remark}
	{ \begin{remark_tmp} 	}
	{\remarkclose\end{remark_tmp}}
\theoremstyle{definition}
\allowdisplaybreaks[1]

\makeatletter
\DeclareMathOperator{\E}{\mathbb{E}} 					 % Expected Value
\DeclareMathOperator{\V}{\mathbb{V}} 					 % Variance
\DeclareMathOperator{\R}{\mathbb{R}} 					 % Real numbers
\DeclareMathOperator{\N}{\mathbb{N}} 					 % Natural numbers
\DeclareMathOperator*{\argmax}{arg\,max} 				 % Argmax of a set
\DeclareMathOperator*{\argsup}{arg\,sup} 				 % Argmin of a set
\DeclareMathOperator*{\arginf}{arg\,inf} 				 % Argmin of a set
\DeclareMathOperator*{\esssup}{ess\,sup}

\newcommand{\Pbase}{P_{*}}                           % Baseline distribution (edit here to change its symbol globally)
\newcommand{\calA}{\mathcal{A}}

\newcommand{\calD}{\mathcal{D}}
\newcommand{\calE}{\mathcal{E}}
\newcommand{\calF}{\mathcal{F}}

\newcommand{\calP}{\mathcal{P}}

\newcommand{\calV}{\mathcal{V}}

\newcommand{\calY}{\mathcal{Y}}
\newcommand{\calZ}{\mathcal{Z}}

\newcommand{\indic}{\mathds{1}}

\newcommand{\op}{o_{\mathbb{P}}}

\newcommand{\de}{\mathrm{d}}

\newcommand{\rint}{\mathrm{ri}}

\newcommand{\supp}{\mathrm{supp}}

\newcommand{\TV}{\mathsf{TV}}

\newcommand{\rhopad}{\accentset{\rightharpoonup}{\rho}}

\newcommand{\normal}{\mathsf{N}}

\NewDocumentCommand{\vertiii}{sO{}m}{%
  \IfBooleanTF{#1}{% automatic scaling, use with care
    \left|\opnormkern\left|\opnormkern\left|
    #3
    \right|\opnormkern\right|\opnormkern\right|
  }{
    \mathopen{#2|\opnormkern #2|\opnormkern #2|}
    #3
    \mathclose{#2|\opnormkern #2|\opnormkern #2|}
  }%
}
\newcommand{\opnormkern}{\mkern-1.5mu\relax}% adjust for the font

\newcommand{\cprob}{\overset{\mathbb{P}}{\longrightarrow}}

\newcommand{\independent}{\mathrel{\perp\!\!\!\perp}}

\DeclareFontFamily{U}{mathx}{\hyphenchar\font45}
\DeclareFontShape{U}{mathx}{m}{n}{
      <5> <6> <7> <8> <9> <10>
      <10.95> <12> <14.4> <17.28> <20.74> <24.88>
      mathx10
      }{}
\DeclareSymbolFont{mathx}{U}{mathx}{m}{n}
\DeclareFontSubstitution{U}{mathx}{m}{n}
\DeclareMathSymbol{\bigtimes}{1}{mathx}{"91}

\newcommand{\dimm}{\mathsf{d}_m}
\newcommand{\dimt}{\mathsf{d}_\theta}
\newcommand{\dimz}{\mathsf{d}_z}
\newcommand{\dime}{\mathsf{d}_{\mathtt{e}}}
\newcommand{\dimk}{\mathsf{d}_{\mathtt{k}}}

\newcommand{\distas}[1]{\mathbin{\overset{#1}{\kern\z@\sim}}}  % distributed as \distas[]
\newcommand{\Kmin}{\underline{K}_{\delta}}
\newcommand{\Kmax}{\overline{K}_{\delta}}
\newcommand{\Tmin}{\underline{\Theta}_{\delta}}
\newcommand{\Tmax}{\overline{\Theta}_{\delta}}
\newcommand{\Lmin}{\underline{\Lambda}_{\delta}}
\newcommand{\Lmax}{\overline{\Lambda}_{\delta}}
\newcommand{\Xmin}{\underline{\Xi}_{\delta}}
\newcommand{\Xmax}{\overline{\Xi}_{\delta}}
\newcommand{\Tminhat}{\widehat{\underline{\Theta}}_{\delta,n}}
\newcommand{\Tmaxhat}{\widehat{\overline{\Theta}}_{\delta,n}}

\makeatother

\newcommand{\legTVbounds}{$\widehat{\underline{\tau}}(\delta),\,\widehat{\overline{\tau}}(\delta)$}
\newcommand{\legTVboundsX}{$\widehat{\underline{\tau}}_{X}(\delta),\,\widehat{\overline{\tau}}_{X}(\delta)$}
\newcommand{\legLeebound}{$\widehat{\underline{\tau}}_{\mathsf{Lee}},\,\widehat{\overline{\tau}}_{\mathsf{Lee}}$}
\newcommand{\legLeeboundX}{$\widehat{\underline{\tau}}_{\mathsf{Lee},X},\,\widehat{\overline{\tau}}_{\mathsf{Lee},X}$}
\newcommand{\blo}{\underline{\mathsf{b}}}
\newcommand{\bup}{\overline{\mathsf{b}}}
\newcommand{\blohat}{\widehat{\underline{\mathsf{b}}}_{n}}

\renewenvironment{abstract}{\small\quotation}{\endquotation}

\makeatletter
\let\lnk@makefntext\@makefntext
\long\def\@makefntext#1{\lnk@fntarget\lnk@makefntext{#1}}
\def\lnk@fntarget{%
  \ifx\Hy@footnote@currentHref\@empty\else
    \ifx\Hy@footnote@currentHref\lnk@lastfn\else
      \global\let\lnk@lastfn\Hy@footnote@currentHref
      \Hy@raisedlink{\hyper@anchorstart{\Hy@footnote@currentHref}\hyper@anchorend}%
    \fi
  \fi}
\let\lnk@lastfn\@empty
\let\lnk@deferred@thm@head\deferred@thm@head
\def\deferred@thm@head#1{%
  \ifx\@empty\dth@counter
    \lnk@deferred@thm@head{#1}%
  \else
    \lnk@deferred@thm@head{%
      \Hy@raisedlink{\hyper@anchorstart{\Hy@dth@currentHref}\hyper@anchorend}#1}%
    \let\dth@counter\@empty
  \fi}
\makeatother

\definecolor{bgcolor}{RGB}{250,249,245}   
\newcommand{\bgcolor}{bgcolor}            
\pagecolor{bgcolor}

\begin{document}
\hypersetup{pageanchor=false}

\title{\vspace{-.25in}\textbf{Sensitivity Analysis in Population Shares}\thanks{I am
thankful to my committee, Matias Cattaneo, Kevin Dano, David Lee, and 
Ulrich M{\"u}ller, for their guidance, patience, and encouragement throughout this
project. I also thank Thomas
Bearpark, Francesco Bilotta, Giampaolo Bonomi, Andrea Cerrato, Federico Crippa, Carlo
Cusumano, 
Pietro Dedola, Nicolas Hommel, Bo Honor\'e, Michal Koles\'ar, Elena Manresa, Matt Masten, 
Geert Mesters, Mikkel Plagborg-M{\o}ller, Chiara Motta, Frank Schorfheide, and Bartolomeo
Stellato
for comments that made the paper better, and Claude for
giving me back a lot of hours. All remaining mistakes are obviously mine. All methods are
implemented in the companion \texttt{R} and \texttt{Python} packages
\texttt{tvbounds} (\url{https://github.com/filippopalomba/tvbounds}).
}\bigskip}
\author{Filippo Palomba\addtocounter{footnote}{2}\thanks{Department of Economics.
\href{mailto:fpalomba@princeton.edu}{fpalomba@princeton.edu}.}\\[.6em]
Princeton University}
\date{\today\\[.6em]}
\maketitle

{\setlength{\leftmargini}{1.2em}
\begin{abstract}\onehalfspacing
    \noindent Empirical work often relies on assumptions on the distribution of
    random variables, such as potential outcomes or latent variables, to identify
    some parameter of interest. We introduce a unified framework for assessing the
    sensitivity of a target estimand to relaxing these distributional assumptions for
    a given share of the population. We derive bounds on the values the estimand can
    take as this share varies between 0 and 1, and compute the bounds using tractable
    finite-dimensional programs. We propose plug-in estimators for the bounds and
    characterize their large sample properties. We also show that our bounds recover
    exact nonparametric sharp bounds at no additional computational cost, and
    introduce summary measures that separate misspecification uncertainty from
    sampling uncertainty and rank studies by their robustness. We apply our framework
    to assess the robustness of (\textit{i}) average treatment effects in randomized
    experiments with attrition, (\textit{ii}) counterfactual predictions in structural
    models, and (\textit{iii}) formula instruments. We provide \texttt{R} and
    \texttt{Python} packages implementing all our methods.
\end{abstract}}

\textit{Keywords}: sensitivity analysis, robustness,
partial identification, convex duality.

\thispagestyle{empty}
\clearpage
\hypersetup{pageanchor=true}
\onehalfspacing 
\pagestyle{plain}
\addtocounter{page}{-1}
\section{Introduction}
    \label{sec:intro}

A long tradition has stressed the need for sensitivity analyses to assess the
fragility of a study's conclusions \citep{leamer1983LetsTakeCon,
rosenbaum1983AssessingSensitivityUnobserved, leamer1985SensitivityAnalysesWould,
tamer2015SensitivityAnalysisSome}. Consider a policymaker deciding which of several
structural models or natural experiments should inform policy, 
a manager allocating a finite budget across pilot studies, or 
a drug regulator weighing the evidence from competing trials. 
These alternative empirical analyses report point estimates and confidence intervals 
for a common target parameter, but all rest on untestable, sometimes not even falsifiable,
distributional assumptions: that attrition in a randomized experiment is random, that
unobserved preference shocks in an indirect utility follow an extreme value
distribution, or that the exogenous shocks in a formula instrument are uniformly
distributed. When the studies disagree, the decision maker needs to know how much of
the disagreement reflects misspecification rather than sampling uncertainty, and two
questions arise. Can we quantify how sensitive each conclusion is to
distributional assumptions, so that studies can be ranked? And can we do so in units
interpretable enough to communicate the ranking to a broad audience of stakeholders?

This paper answers both questions with a framework that assesses the sensitivity of
a target estimand to violations of distributional assumptions and measures the extent
of a violation by the share of the population that does not satisfy such assumptions.
The framework gauges, for instance, how much an average treatment effect would
change if nonresponse were nonrandom for $60\%$ of the experimental treated units, how
much a
counterfactual welfare prediction from a structural model would change if $30\%$ of
the agents drew shocks that were not extreme value distributed, or how much an
estimated elasticity would change if $40\%$ of the time the exogenous shocks were
assigned by a mechanism other than the one the researcher postulated.

Formally, the object of sensitivity analysis is an
expectation $\E_{\Pbase{}}[g(Z;\theta)]$, where $g$ is a known function, $Z$ is a random
vector,
$\theta$ is a structural parameter
disciplined by moment equalities or inequalities, and $\Pbase{}$  
is the \textit{baseline distribution}, the model the econometrician 
would use absent any concern for misspecification.  
In this scenario, our goal is to obtain the smallest and largest values
the estimand can take
over a nonparametric neighborhood of candidate distributions that 
allow up to a fraction $\delta$ of the population to violate the 
distributional assumption imposed by $\Pbase{}$. We refer to these two
values as the \textit{sensitivity bounds} at $\delta$, to the neighborhood as the
\textit{robustness set}, and to $\delta$ as the \textit{sensitivity parameter}.
In a randomized experiment with attrition,
$\delta$ bounds the share of treated respondents for whom attrition is not
random. In a discrete choice model, it bounds the share of agents whose
taste shocks do not follow the postulated distribution. With formula 
instruments, it bounds the probability that the shocks were
assigned by a different mechanism. 

We show that the sensitivity bounds are the optimal values of two symmetric
infinite-dimensional programs, which in general admit no closed form solution. Using
tools from convex analysis \citep*{rockafellar1970ConvexAnalysis,
rockafellar2009VariationalAnalysis, kuhn2025DistributionallyRobustOptimization}, we
recast them as finite-dimensional convex programs, so that the entire path of bounds
can be traced as $\delta$ ranges between 0 and 1. At $\delta=0$, the bounds
reduce to the value of the estimand under the baseline specification $\Pbase{}$. At
$\delta=1$, we prove that they coincide with the sharp nonparametric bounds, namely the
tightest bounds on the estimand implied by the model's remaining restrictions once the
distributional assumption is dropped entirely.
As a consequence, our procedure can also be used to
compute nonparametric sharp bounds directly from a tractable program whose dimensionality
grows only with the number of moment conditions imposed.
Put differently, $\delta$ trades off informativeness against
credibility, in the spirit of \textit{the law of decreasing credibility} of
\citet{manski2003PartialIdentificationProbability}. At $\delta=0$, the estimand is
point identified, but only under the full distributional assumption. At
$\delta=1$, the bounds rest on the model's remaining restrictions alone, but they are at
their widest.

Regarding estimation and inference, we propose simple plug-in estimators for 
the bounds. We allow for both a known baseline, as when $\Pbase{}$ is
the distribution of taste shocks in a discrete choice model, and an
estimated one, as when $\Pbase{}$ is the outcome distribution of the treated
respondents in a
randomized experiment. We establish the large sample properties
of the estimators by showing their consistency, deriving their limiting distribution,
and proposing a bootstrap procedure to construct valid confidence intervals for the
bounds.

We also propose summary measures that separate misspecification uncertainty
from sampling uncertainty and make studies comparable. Together with the bounds, they
answer the two questions raised in our opening paragraphs. Finally, we
apply the framework to
the online tutoring experiment of \citet{carlana2025ApartConnectedOnline}, in which
attrition threatens identification, to the marital college premium application of
\citet{christensen2023CounterfactualSensitivityRobustness}, 
and to the formula instruments of
\citet{borusyak2023NonrandomExposureExogenous}, who estimate the employment effects
of market access growth from Chinese high-speed rail construction.

\textbf{Related literature.}
Some recent work studies the sensitivity of counterfactuals in structural models.
In \citet{christensen2023CounterfactualSensitivityRobustness}, the values a
counterfactual takes are bounded as the distribution of latent variables ranges over
a $\phi$-divergence neighborhood of a parametric specification while the structural
model's moment restrictions are maintained.
\citet{gu2024WassersteinRobustCounterfactuals} bound scalar counterfactuals in
partially identified structural models over Wasserstein neighborhoods of the latent
distribution using random set theory, and
\citet*{lieber2026CounterfactualsDiscreteChoice} bound counterfactuals in
nonparametric discrete choice models when the latent density lies in a pointwise
band around a known reference density, using tools from graph theory and mechanism
design. This paper differs from these works mainly in that our sensitivity parameter
$\delta$ admits a population share interpretation and our framework 
applies beyond the structural models these papers consider.

Our framework is also related to a large body of work that develops
sensitivity exercises within specific identification problems: contaminated and
corrupted sampling \citep*{horowitz1995IdentificationRobustnessContaminated};
missing data and attrition \citep*{kline2013SensitivityMissingData,
ober-reynolds2026RobustnessMissingData}; selection on unobservables and
unmeasured confounding \citep*{rosenbaum1983AssessingSensitivityUnobserved,
imbens2003SensitivityExogeneityAssumptions, altonji2005SelectionObservedUnobserved,
masten2018IdentificationTreatmentEffects, bonvini2022SensitivityAnalysisProportion,
masten2024AssessingSensitivityUnconfoundedness, masten2025GeneralApproachRelaxing};
the exclusion and exogeneity of instruments
\citep*{masten2021SalvagingFalsifiedInstrumental, diegert2026AssessingSensitivityIV};
the distribution of the exogenous shocks that generate formula
instruments \citep*{sheng2026WhatMagicFormula};
random assignment and rank invariance
\citep*{masten2020InferenceBreakdownFrontiers}; and parallel trends in
difference-in-differences designs \citep*{rambachan2023MoreCredibleApproach}.
Relative to these papers, we provide a single framework that nests many of the 
exercises described above, can include moment restrictions and
structural parameters, provides a general estimation and inference theory,
and allows the sensitivity parameter to maintain an analogous
population share interpretation for a wide family of settings. 
In this sense, our sensitivity parameter is a \textit{common currency} 
for misspecification.

Another strand of literature focused on local sensitivity
analysis, which measures how quantities of interest respond to infinitesimal
perturbations of the model or of its moment conditions
\citep*{kitamura2013RobustnessInfinitesimalNeighborhoods,
andrews2017MeasuringSensitivityParameter, mukhin2018SensitivityRegularEstimators,
andrews2020InformativenessDescriptiveStatistics,
armstrong2021SensitivityAnalysisUsing, bonhomme2022MinimizingSensitivityModel},
whereas we allow for global misspecification, in that $\delta$ can be large and does
not shrink as the sample size grows. We show how the two 
approaches can be linked via the
shadow price of robustness, the derivative of the
sensitivity bounds with respect to $\delta$, which is a local sensitivity measure at
small budgets.

Finally, on the methodological side, we contribute to a line of work in economics that
uses
convex duality to turn optimization problems over infinite-dimensional sets of
distributions into tractable ones \citep{hansen2001RobustControlModel,
ekeland2010OptimalTransportationFalsifiability, schennach2014EntropicLatentVariable}.
Closest
to our work, \citet{christensen2023CounterfactualSensitivityRobustness} use duality to
recast the optimization of a counterfactual over a $\phi$-divergence neighborhood,
subject to moment restrictions, as a finite-dimensional convex program. Their duality
results are stated for
divergences generated by strictly convex and differentiable functions, a class that
excludes the distance on which our neighborhoods are built.

\begin{figure}[h]
\caption{Sensitivity analysis in population shares: a visual roadmap.}
\centering
\resizebox{.9\textwidth}{!}{% Roadmap figure: sensitivity analysis as a function of the sensitivity parameter.
% Static adaptation of the deck's "This Paper" picture
% (presentation/26-9-9-princeton/main-presentation.tex, \thispaperpic);
% each style below is the neutral variant of the deck's \tpset call.
% Included via \resizebox{.8\textwidth}{!}{\input{figures/roadmap_sensitivity.tex}}.
\providecolor{figorange}{RGB}{213,94,0}
\providecolor{tvblue}{HTML}{1F4E79}
\tikzset{
  zdot/.style={figorange},
  ztick/.style={thick},
  ztext/.style={},
  oline/.style={black!60, dotted, thick},
  otick/.style={thick},
  otext/.style={},
  nptext/.style={},
  carrow/.style={thick, black!70},
  ctext/.style={},
  cbox/.style={draw=tvblue, thick, rounded corners=3pt, fill=tvblue!8},
  bline/.style={figorange, very thick},
  bandfill/.style={figorange, opacity=0.2},
  bandline/.style={figorange, thick, dashed},
  bdline/.style={black!75, dashed, thick},
  bdtext/.style={},
  bddot/.style={black!75},
  cbdline/.style={black!75, dashed, thick},
  cbdtext/.style={},
  cbddot/.style={black!75},
  brline/.style={thick, black!75},
  brtext/.style={},
  mbrline/.style={thick, black!75},
  mbrtext/.style={},
}%
\begin{tikzpicture}[x=7.4cm, y=2.4cm, font=\scriptsize,
  >={Stealth[length=2mm,width=1.6mm]},
  declare function={
    hfun(\x) = 0.3*\x + 0.7*(2*\x - \x*\x);
    ubnd(\x) = 0.5 + 0.6*hfun(\x);
    lbnd(\x) = 0.5 - 0.75*hfun(\x);
    ciw(\x)  = 0.07 + 0.03*\x;
  }]
  \def\tauzero{0.5}
  \def\deltab{0.4918}
  \def\deltac{0.3910}

  \draw[->, thick] (-0.015,0) -- (1.12,0) node[right, align=left] {\textit{sensitivity}\\\textit{parameter}};
  \draw[->, thick] (0,-0.35) -- (0,1.5);
  \node[rotate=90, anchor=south] at (-0.03,0.55) {\textit{quantity of interest}};

  \fill[zdot] (0,\tauzero) circle (2.2pt);
  \draw[ztick] (0,0.025) -- (0,-0.025);
  \node[ztext, anchor=north east, inner sep=1.5pt] at (0,-0.02) {0};
  \node[ztext, anchor=north, align=center] at (0,-0.45) {no concern for\\misspecification};

  \draw[->, carrow] (0.2,-0.46) -- (0.8,-0.46);
  \node[ctext, anchor=north] (concern) at (0.5,-0.49) {increasing concern for misspecification};

  \draw[->, carrow] ($(concern.south)+(0,-0.01)$) -- ++(0,-0.16);
  \node[cbox, anchor=north, align=center, inner xsep=6pt, inner ysep=3.5pt] at ($(concern.south)+(0,-0.18)$)
    {\textbf{our currency:} increasing share of population\\for which the distributional assumption fails};

  \draw[oline] (1,-0.4) -- (1,1.32);
  \draw[otick] (1,0.025) -- (1,-0.025);
  \node[otext, anchor=north west, inner sep=1.5pt] at (1,-0.02) {1};
  \node[nptext, anchor=north, align=center] at (1,-0.45) {nonparametric\\sharp bounds};

  \draw[bdline] (\deltab,0) -- (\deltab,1.2);
  \node[bdtext, anchor=south west, align=center, inner sep=2pt] at (\deltab,1.2) {breakdown \\ budget};
  \draw[cbdline] (\deltac,0) -- (\deltac,1.2);
  \node[cbdtext, anchor=south east, align=center, inner sep=2pt] at (\deltac,1.2) {certified\\breakdown budget};

  \fill[bandfill]
    plot[smooth, samples=60, domain=0:1] (\x,{ubnd(\x)+ciw(\x)})
    -- plot[smooth, samples=60, domain=1:0] (\x,{ubnd(\x)}) -- cycle;
  \fill[bandfill]
    plot[smooth, samples=60, domain=0:1] (\x,{lbnd(\x)})
    -- plot[smooth, samples=60, domain=1:0] (\x,{lbnd(\x)-ciw(\x)}) -- cycle;
  \draw[bandline, smooth, samples=60, domain=0:1] plot (\x,{ubnd(\x)+ciw(\x)});
  \draw[bandline, smooth, samples=60, domain=0:1] plot (\x,{lbnd(\x)-ciw(\x)});

  \draw[bline, smooth, samples=60, domain=0:1] plot (\x,{ubnd(\x)});
  \draw[bline, smooth, samples=60, domain=0:1] plot (\x,{lbnd(\x)});

  \fill[bddot] (\deltab,0) circle (1.6pt);
  \fill[cbddot] (\deltac,0) circle (1.6pt);

  \draw[decorate, decoration={brace, mirror, amplitude=3pt}, brline]
    (\deltac,-0.15) -- (\deltab,-0.15);
  \node[brtext, anchor=north] at ({(\deltac+\deltab)/2},-0.19) {sampling uncertainty};
  % above the axis: the lower bound crosses y=0 at delta_b, so this strip is
  % clear all the way to delta=1. Unmirrored brace -> nub up, label above.
  \draw[decorate, decoration={brace, amplitude=3pt}, mbrline]
    (\deltab,0.10) -- (1,0.10);
  \node[mbrtext, anchor=south] at ({(\deltab+1)/2},0.145) {misspecification uncertainty};
\end{tikzpicture}}
\label{fig:roadmap}
\end{figure}
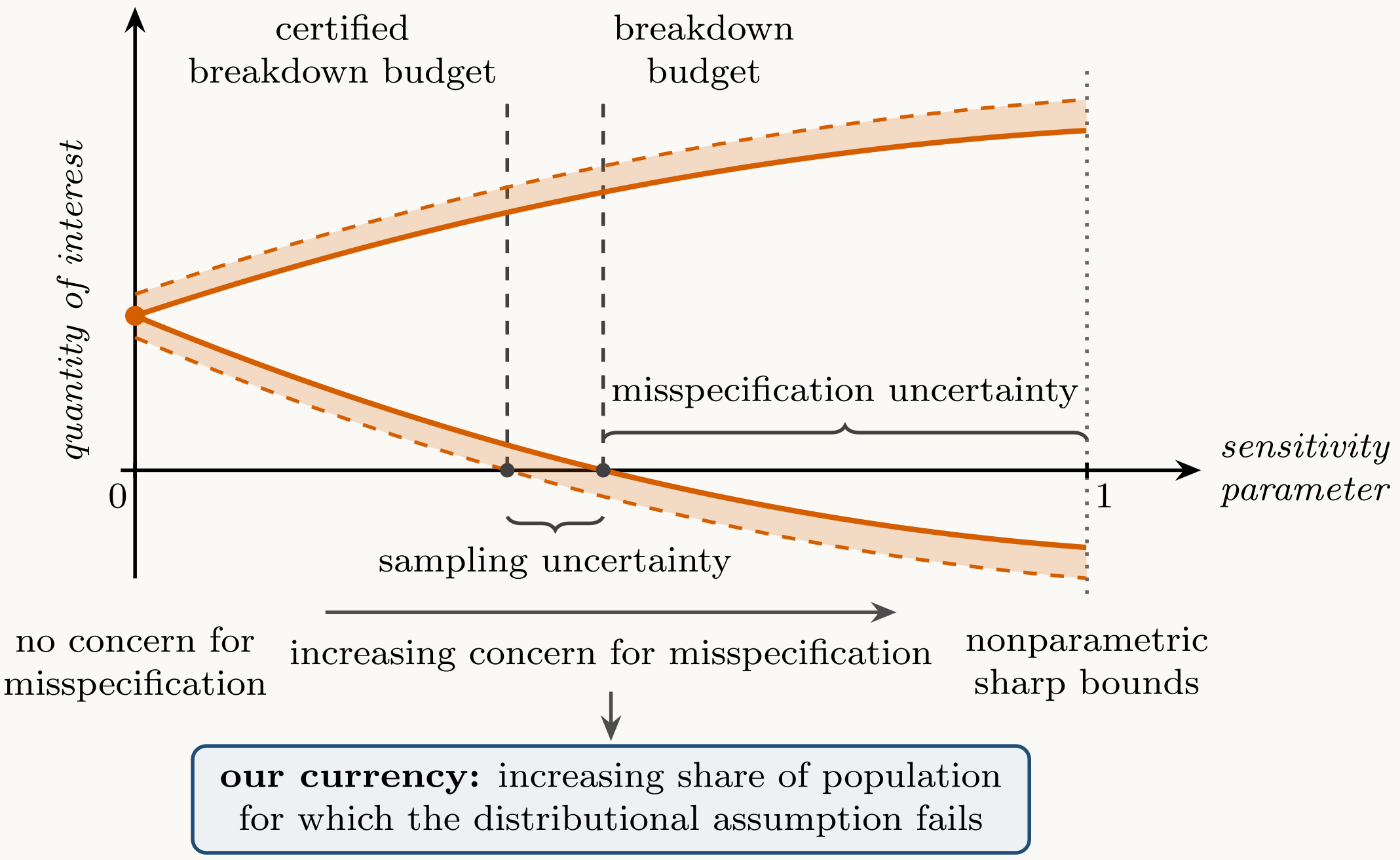

\textbf{Paper organization.} Figure~\ref{fig:roadmap} collects the main output of a
prototypical sensitivity exercise conducted with our framework and organizes the rest
of the paper. Section~\ref{sec:motivating examples} presents three leading
applications that motivate our sensitivity exercise. Section~\ref{sec:setup} formally
defines the framework nesting these applications and the optimization problems behind
the thick lines, and explains why moving rightward on the horizontal axis raises the
share of the population for which the distributional assumption imposed by
$\Pbase{}$ is allowed to fail. Section~\ref{sec:theory} shows how to compute,
estimate, and conduct inference on the sensitivity bounds, and thereby delivers the
thick lines and the dashed bands around them. Section~\ref{sec:usage} proves that at
$\delta=1$ the bounds coincide with the exact nonparametric sharp bounds and discusses
how to use them in practice by means of summary measures, such as the breakdown and
certified breakdown budgets of Figure~\ref{fig:roadmap}. The breakdown budget is the
smallest share of the population for which the sensitivity bounds contain zero, whereas
the certified breakdown budget, which also accounts for sampling uncertainty, is the
smallest share for which the dashed bands do. Section~\ref{sec:applications} deploys
the framework in the three empirical
applications introduced in Section~\ref{sec:motivating examples}, and
Section~\ref{sec:conclusion} concludes. The supplementary materials collect all proofs,
establish the existence of solutions to the dual and outer problems,
relax the simplifying conditions of Section~\ref{sec:theory}, allow for moment
inequalities and nuisance parameters, and work out further applications.

\textbf{Notation.} For a set $\calA$, $\Delta(\calA)$ denotes the set of
probability distributions on $\calA$ and, for $P,Q\in\Delta(\calA)$,
$\TV(P\,\|\,Q):=\sup_{A\subseteq\calA}|P(A)-Q(A)|\in[0,1]$ denotes the total variation
distance between them. We write $P\ll Q$ when $P$ is absolutely continuous with respect
to $Q$, $\de P/\de Q$ for the corresponding density, $\supp(P)$ for the support of $P$,
$Pf:=\E_P[f(Z)]$ for a random vector $Z$ distributed according to $P$ and a
$P$-integrable function $f$, and $P\text{-}\esssup_{z}f(z)$ for the smallest
$c\in[-\infty,+\infty]$ such that $f\leq c$ holds $P$-almost surely. We let
$[n]:=\{1,\ldots,n\}$ for $n\in\N$, $\N_0:=\N\cup\{0\}$, $\R_+:=[0,\infty)$, $0_d$ be the
$d$-dimensional zero vector, $\indic\{\cdot\}$ the indicator function, $x^\top$ the
transpose of $x$, and $\|\cdot\|$ the Euclidean norm. For a convex set $C\subseteq\R^d$,
$\mathrm{int}(C)$ and $\rint(C)$ denote its interior and relative interior. Finally,
$\rightsquigarrow$ denotes weak convergence, $\op(1)$ a term converging to zero in
probability, and $\normal(0,\Sigma)$ the Gaussian distribution with covariance
matrix $\Sigma$.

\section[Motivating Examples]{Motivating Examples}
    \label{sec:motivating examples}

In this section, we illustrate how three common economic applications rest on
untestable distributional assumptions and are thus amenable to sensitivity analysis:
a randomized controlled trial (RCT, henceforth) with attrition, counterfactuals in
structural models following \citet{christensen2023CounterfactualSensitivityRobustness},
and the recentered instrumental variables estimator of
\citet{borusyak2023NonrandomExposureExogenous}. Many economically relevant scenarios
beyond these three fit the framework (see
Section~SA6
of the supplementary materials).

\begin{example}[RCTs with attrition]\label{ex:rct}
Consider a randomized experiment that grants half of the applicants
access to a free job training course and then, two years later, 
runs a follow-up survey to record each applicant's weekly wage.\footnote{A 
famous program of this kind is Job Corps, which offers vocational training 
to disadvantaged youth aged 16--24 and remains active today with more than 120 
campuses across the United States and Puerto Rico, roughly 25,000 students 
enrolled at a time, and a budget of about \$1.8 billion for fiscal year 
2026 \citep{congress2026ConsolidatedAppropriationsAct}.}
Wages, however, are observed only for the applicants who follow up.
Worse, trainees stay in touch with the agency while the control-group applicants are the
hardest
to track down, so respondents may end up being selected on
particular dimensions, confounding the effects of the program. Practitioners 
typically assume attrition is ignorable or, equivalently, that nonresponse 
is as good as random, an assumption known as ``missing completely at random'' 
(\textsf{MCAR}, henceforth) that we define precisely below.\footnote{
Perhaps surprisingly, this practice is the norm rather than the exception. In a
systematic review of survey-based articles published in the top-five economics
journals, \citet*{dutz2026SelectionSurveysUsing} find that nonresponse rates
average $50\%$ when researchers collect their own data and $19\%$ when using pre-existing
household surveys. Despite that, nearly half of
the articles do not discuss nonresponse bias at all, while the remaining
articles either assert that nonresponse is ignorable or correct for it by
reweighting on observables.
Linking their own survey to administrative records on the
full population, they document substantial nonresponse bias that survives
adjustment for observables, even in a survey run by a national statistical
agency.
There is further empirical evidence in economics that agents, when given
the choice of whether to report, make that decision on the basis of
observable or unobservable characteristics
\citep*{mu2026WhatsMissingEnvironmental,bollinger2019TroubleTailsWhat,
crippa2025CorporateOmissionsCorrecting}.}

Formally, consider a population described by $(Y(0),Y(1),S(0),S(1),D)\sim P_0$, where
$D\in\{0,1\}$ is
a binary treatment, $Y(d)\in\calY$, $d\in\{0,1\}$, are potential outcomes taking values in
a closed set $\calY\subseteq\R$,
$S(d)\in\{0,1\}$, $d\in\{0,1\}$, are potential response indicators, and $P_0$ satisfies
random assignment of $D$,
that is $D\independent (S(0),S(1),Y(0),Y(1))$.
Observed and potential quantities are linked by $Y=DY(1)+(1-D)Y(0)$ and
$S=DS(1)+(1-D)S(0)$. The econometrician observes $(YS, S,D)$ for every
applicant, as the outcome $Y$ is observed only for respondents, that is, when $S=1$.
The pair of potential response indicators partitions the population into
four groups:
\begin{equation*}
G:=\begin{cases}
\mathsf{AO} & \text{if } (S(0),S(1))=(1,1),\\
\mathsf{C} & \text{if } (S(0),S(1))=(0,1),\\
\mathsf{NO} & \text{if } (S(0),S(1))=(0,0),\\
\mathsf{DF} & \text{if } (S(0),S(1))=(1,0),
\end{cases}
\end{equation*}
namely the \textit{always-observed}, who respond regardless of the arm they are
assigned to; the \textit{compliers}, who respond only if treated; the
\textit{never-observed}, who drop out of the sample regardless of the arm;
and the \textit{defiers}, who respond only if untreated.\footnote{This
taxonomy deliberately mirrors the imperfect compliance taxonomy of
\citet*{angrist1996IdentificationCausalEffects}, with response to the
follow-up survey playing the role of treatment take-up.}
As a parameter of interest, we consider the average treatment effect (ATE, henceforth) for
the always-observed,\footnote{Many more
causal parameters are identified under \textsf{MCAR}. We focus on this particular one
because it is often the object of interest of the attrition literature and because it
can be nonparametrically partially identified without bounded $\calY$ when \textsf{MCAR}
does
not hold. We show how to conduct sensitivity analysis when the ATE
over the whole population is the target estimand in Section~SA6.4.3
of the supplementary materials.}
which is defined as
\begin{equation*}
\tau_0:=\E_{P_0}[Y(1)-Y(0)\mid G=\mathsf{AO}].
\end{equation*}
For simplicity, we assume throughout that treatment weakly encourages response,
i.e., $S(1)\geq S(0)$ almost surely (\textsf{Mono}), an assumption known as
``monotonicity,'' which in this notation reads $P_0[G=\mathsf{DF}]=0$. In
Remark~SA-15 of the supplementary
materials we explain how to conduct the same sensitivity exercise without \textsf{Mono}.
Absent defiers, the respondents in the control arm are exactly the
always-observed, hence 
\begin{equation}\label{eq: alt control AO mean}
   \mu^{\mathsf{AO}}(0):=\E_{P_0}[Y(0)\mid G=\mathsf{AO}]=\E_{P_0}[Y\mid D=0,S=1]
\end{equation}
is nonparametrically identified, and we can focus on the treated arm.

In the treated arm, the data identify the conditional distribution of $Y$
given $D=1$ and $S=1$ under $P_0$, which we denote by $P_{\mathsf{T}}$ and
whose mean we denote by $\mu_{\mathsf{T}}:=\E_{P_{\mathsf{T}}}[Y]$, but not who
among the treated respondents is a complier and who an always-observed. Let
$P_{\mathsf{C}}$ and $P_{\mathsf{AO}}$ denote the conditional distributions of
$Y(1)$ given $G=\mathsf{C}$ and given $G=\mathsf{AO}$ under $P_0$, respectively, and let
\begin{equation}\label{eq: alt share of compliers}
\pi:=P_0[G=\mathsf{C}\mid D=1,S=1]=
\frac{P_0[S=1\mid D=1]-P_0[S=1\mid D=0]}{P_0[S=1\mid D=1]}
\end{equation}
be the complier share among treated respondents, where the second equality
follows from random assignment and \textsf{Mono}, so that $\pi$ is
identified by comparing the response rates of the two arms. We assume
that $P_0[S=1\mid D=0]>0$, that treatment strictly raises the response rate, and that
$\E_{P_0}[|Y(d)|]<\infty$, $d\in\{0,1\}$, so that $\pi\in(0,1)$ and the means above are
finite. Since $S(1)=1$ if and only if $G\in\{\mathsf{AO},\mathsf{C}\}$,
random assignment yields the mixture decomposition
\begin{equation}\label{eq: alt mixture}
P_{\mathsf{T}}=\pi P_{\mathsf{C}}+(1-\pi)P_{\mathsf{AO}},
\end{equation}
in which neither component is identified.
The decomposition \eqref{eq: alt mixture} is key for two reasons. First,
\textsf{MCAR}, the assumption that response is independent of
potential outcomes, $G\independent(Y(0),Y(1))$, implies that the
distribution of $Y(1)$ is the same for compliers and always-observed, so that
in the treated arm \textsf{MCAR} amounts to 
\begin{equation}\label{eq: alt MCAR}
P_{\mathsf{C}}=P_{\mathsf{AO}}=P_{\mathsf{T}},
\end{equation}
which identifies the unobserved complier distribution with the observed
distribution of treated respondents.
Second, \eqref{eq: alt mixture} ties the always-observed to the compliers:
for any candidate complier distribution $Q\in\Delta(\calY)$, the
mixture identity forces the always-observed distribution to be
$(P_{\mathsf{T}}-\pi Q)/(1-\pi)$, whose mean is
$(\mu_{\mathsf{T}}-\pi\E_Q[Y])/(1-\pi)$. Subtracting \eqref{eq: alt control AO mean}, we
obtain the map
\begin{equation}\label{eq: alt tau map}
\tau(Q):=\frac{\mu_{\mathsf{T}}-\pi\,\E_Q[Y]}{1-\pi}-\mu^{\mathsf{AO}}(0),\qquad Q\in\Delta(\calY),
\end{equation}
in which $\E_Q[Y]$ is the only quantity not identified from the data.
At the true complier distribution the map recovers the parameter of
interest, $\tau(P_{\mathsf{C}})=\tau_0$. Under
\textsf{MCAR}, \eqref{eq: alt MCAR} gives $\tau_0=\tau(P_{\mathsf{T}})$, so that
$\tau_0$ is point identified by the difference in mean outcomes among respondents
\citep{rubin1976InferenceMissingData},
\begin{equation*}
\tau_{\mathsf{MCAR}}:=\tau(P_{\mathsf{T}})
=\E_{P_0}[Y\mid D=1,S=1]-\E_{P_0}[Y\mid D=0,S=1].
\end{equation*}
However, \textsf{MCAR} is untestable and hard to
justify by economic theory. 
Hence, a stakeholder
might be interested in quantifying how far $\tau_0$ can lie from
$\tau_{\mathsf{MCAR}}$ when \textsf{MCAR} fails.

The way we model relaxations of \textsf{MCAR} is by requiring that
the outcome distributions of compliers and always-observed be close rather
than equal. Put differently, we consider candidate pairs
$(Q_{\mathsf{C}},Q_{\mathsf{AO}})$ of distributions on $\calY$ that satisfy
$P_{\mathsf{T}}=\pi Q_{\mathsf{C}}+(1-\pi)Q_{\mathsf{AO}}$ and
$\TV(Q_{\mathsf{C}}\,\|\,Q_{\mathsf{AO}})\leq\delta$. As we show
in Section~\ref{sec:setup}, with this formulation, $\delta$ 
is the share of treated respondents or, equivalently, the share of compliers
who may not satisfy \textsf{MCAR}.

The pair $(Q_{\mathsf{C}},Q_{\mathsf{AO}})$ is fully
described by its first component. Indeed, 
the mixture constraint \eqref{eq: mixture constraint for TV} gives that
$Q_{\mathsf{AO}}=(P_{\mathsf{T}}-\pi Q_{\mathsf{C}})/(1-\pi)$ and that
$\TV(Q_{\mathsf{C}}\,\|\,P_{\mathsf{T}})\leq (1-\pi)\delta$.\footnote{The equivalence
follows from the
identity
$\TV(Q_{\mathsf{C}}\,\|\,P_{\mathsf{T}})=(1-\pi)\TV(Q_{\mathsf{C}}\,\|\,Q_{\mathsf{AO}})$,
which total variation inherits from being a norm-induced metric and
which $\phi$-divergences do not enjoy in general.}
This reformulation replaces the unidentified reference
distribution $Q_{\mathsf{AO}}$ by the identified $P_{\mathsf{T}}$, and lets us define
the set of candidate complier distributions as
\begin{align}\label{eq: alt robustness set}
\calP_{\textsf{RCT}}(\delta) :=
\big\{Q\in\Delta(\calY):\ &\TV(Q\,\|\,P_{\mathsf{T}}) \leq (1-\pi)\delta,\nonumber\\
&P_{\mathsf{T}}=\pi Q+(1-\pi)Q_{\mathsf{AO}}\ \text{for some }Q_{\mathsf{AO}}\big\}.
\end{align}
The sensitivity bounds are then the extreme values that the causal estimand in 
\eqref{eq: alt tau map} can take over the robustness set defined in
\eqref{eq: alt robustness set}. Formally, for $\delta\in[0,1]$
\begin{equation*}
\underline{\tau}(\delta):=\inf_{Q\in\calP_{\textsf{RCT}}(\delta)}\tau(Q)\qquad\text{and}\qquad
\overline{\tau}(\delta):=\sup_{Q\in\calP_{\textsf{RCT}}(\delta)}\tau(Q).
\end{equation*}
\end{example}

\begin{example}[Counterfactuals in structural models]\label{ex:structural}
Counterfactual predictions from structural models depend on the distribution specified for
latent variables such as taste shocks, unobserved heterogeneity, or productivity shocks,
which is typically not nonparametrically identifiable and is therefore chosen for
tractability. A stakeholder may want to know how sensitive the predictions are to this
untestable choice.
As a running example, we revisit one of the leading empirical applications of
\citet{christensen2023CounterfactualSensitivityRobustness} (\textsf{CC23}, henceforth): 
the marital college premium in the marriage-matching model that
\citet*{chiappori2017PartnerChoiceInvestment} estimate on American
Community Survey data.

Consider a frictionless one-to-one matching model with transferable utility. Men
and women each belong to one of $J\in\N$ education groups, indexed by
$[J]:=\{1,\ldots,J\}$, which we term 
``types.'' We let $j=0$ denote the outside option of remaining unmatched.
A type-$b\in[J]$ woman receives utility $\nu_{0b}$ if unmatched
and $t_{jb}+\nu_{jb}$ if matched with a type-$j\in[J]$ man, where $t_{jb}$ is
the systematic component of match utility and $\nu_{jb}$ captures
idiosyncratic marital preferences. For simplicity, we assume
that the distribution of females' preference shocks does not depend on their
type, so we drop the $b$ subscript and consider a single random vector $U =
(\nu_0,\ldots,\nu_J)^\top$.\footnote{See
Section~SA6.2, and
Remark~SA-16 in particular, in the
supplementary materials for a discussion of this common-shock representation
and its alternatives.}
We let $\widetilde{P}$
denote the distribution of latent preference shocks $(\nu_0,\ldots,\nu_J)$
postulated by the econometrician, which we assume to be a Gumbel (type I extreme value)
distribution.\footnote{Another common choice
is to assume that $\widetilde{P}$ is a standard Gaussian distribution, which
gives a probit instead of a logit discrete choice model.}

We fix two female types $b,b'\in[J]$, whose marital college
premium is the object of interest, and let $\varrho\in\R^{2J}$ collect their
observed marriage probabilities with each male type.
The model pins down the structural parameter
$\vartheta:=(t_{jb},t_{jb'},j\in[J])$, ranging over a parameter space
$\Theta\subseteq\R^{2J}$, via a set of moment equalities
\begin{align*}
    \E_{\widetilde{P}}[s(U;\vartheta)] = (\varrho, 0_{2(J+1)}),\qquad 
    s(U;\vartheta) =(s_1(U;\vartheta)^\top,s_2(U;\vartheta)^\top)^\top,
\end{align*}
where $s(U;\vartheta)$
is the moment function, which we now describe in detail. First,
\[s_1(U; \vartheta)=
\begin{bmatrix}
\Big(\indic\Big\{j=\argmax\limits_{j'\in\{0, \ldots, J\}}\Big(t_{j'
b}+\nu_{j'}\Big)\Big\}\Big)_{j=1}^J \\
\Big(\indic\Big\{j=\argmax\limits_{j'\in\{0, \ldots, J\}}\Big(t_{j'
b'}+\nu_{j'}\Big)\Big\}\Big)_{j=1}^J
\end{bmatrix}.\]
Ties are broken toward the lowest indexed alternative.
The moment conditions $\E_{\widetilde{P}}[s_1(U;\vartheta)]=\varrho$ thus
equate the observed marriage probabilities $\varrho$ for female types $b$ and $b'$
to the model implied choice probabilities under $\widetilde{P}$.
Note that because the moment function must be a known
function of $(U,\vartheta)$ alone, the observed marriage probabilities
$\varrho$ enter as the target of the moment conditions rather than as an argument of
$s_1$.

Second, one may impose some normalizations on the female shocks through
\[s_2(U;\vartheta) = (\nu_j, \nu_j^2-\pi^2/6)_{j=0}^J.\]
Therefore, the moment conditions $\E_{\widetilde{P}}[s_2(U;\vartheta)]=0_{2(J+1)}$
impose that each shock must have mean zero and variance $\pi^2/6$
under $\widetilde{P}$, the same variance as a standard Gumbel distribution.

Consider the marital college premium (i.e., the difference in indirect
utilities) for female types $b$ and $b'$,
\[
k(\vartheta;\widetilde{P}) := \E_{\widetilde{P}}[f(U;\vartheta)],
\quad 
f(U;\vartheta)
=
\max_{j\in\{0,\dots,J\}}\{t_{jb}+\nu_{j}\} -
\max_{j\in\{0,\dots,J\}}\{t_{jb'}+\nu_{j}\},
\]
where $t_{0b}=t_{0b'}=0$. A stakeholder
interested in the output of the model above may ask what
happens to the premium if not all women draw their shocks
from a Gumbel distribution.
To answer this question, we let $P$ range over the robustness set
\begin{align}\label{eq: robustness set structural}
\calP_{\mathsf{STR}}(\vartheta;\delta) := \big\{P\in\Delta(\R^{J+1}):\,
&\E_{P}[s(U;\vartheta)]=(\varrho,0_{2(J+1)}),\nonumber\\
&P = (1-\delta)\widetilde{P} + \delta R,\ R\in\Delta(\R^{J+1})
\big\},
\end{align}
which collects all the populations of women that (\textit{i}) reproduce
the observed marriage probabilities and satisfy the imposed normalizations and
(\textit{ii})
consist of a share $\delta\in[0,1]$ of
women with arbitrary shocks and a share $1-\delta$ of women with Gumbel shocks. 
The sensitivity bounds on the premium are then
\begin{equation*}
\underline{\mathsf{k}}(\delta):=\inf_{\vartheta\in\Theta}\
\inf_{P\in\calP_{\mathsf{STR}}(\vartheta;\delta)}\E_{P}[f(U;\vartheta)],
\qquad
\overline{\mathsf{k}}(\delta):=\sup_{\vartheta\in\Theta}\
\sup_{P\in\calP_{\mathsf{STR}}(\vartheta;\delta)}\E_{P}[f(U;\vartheta)],
\end{equation*}
where the inner infimum and supremum are set to $+\infty$ and $-\infty$, 
respectively, whenever $\calP_{\mathsf{STR}}(\vartheta;\delta)$ is empty. 
Therefore, the values of $\vartheta$ at which no distribution within $\delta$ of the
Gumbel 
baseline matches the observed marriage probabilities are infeasible.
\end{example}

\begin{example}[Recentered IV]\label{ex:riv}

Many causal questions in economics involve treatments or instruments that
combine multiple sources of variation according to a known formula. Examples
include shift-share (or Bartik) instruments, treatments capturing spillovers
in social or transportation networks, or simulated instruments
for policy eligibility
\citep*[see][for a review]{borusyak2023DesignBasedIdentificationFormula}.
\citet{borusyak2023NonrandomExposureExogenous} (\textsf{BH23}, henceforth)
point out that exogeneity of the shocks entering the formula does not, by
itself, deliver a consistent instrumental variables estimator, because units
are nonrandomly exposed to the shocks. The
solution \textsf{BH23} propose is to recenter the instrument, subtracting
from it the \textit{expected instrument}, hat is, the average instrument 
across counterfactual shocks. Computing the expected
instrument requires the researcher to specify the
distribution of the shocks, or ``shock
assignment process.'' In a randomized experiment the assignment process is given by
the randomization protocol, but in natural experiments the assignment
process is rarely known or justified by economic theory. A stakeholder
may accordingly ask how much the recentered IV estimate would change under a
different assignment process.

Formally, let $i\in[n], n\in\N$, index the units of the sample and consider the causal or
structural model
\begin{equation*}
y_i=\beta x_i+\varepsilon_i,
\end{equation*}
where $y_i\in\R$ is a demeaned outcome, $x_i\in\R$ is a demeaned endogenous
regressor, $\varepsilon_i\in\R$ is an unobserved residual, and $\beta\in\R$
is the parameter of interest. Let $v=(v_1,\ldots,v_K)\in\calV\subseteq\R^K$,
$K\in\N$, be a vector of exogenous shocks, let $w$ collect predetermined
covariates of unrestricted dimension, and let
\begin{equation*}
z_i:=f_i(v;w)
\end{equation*}
be the candidate instrument of unit $i\in[n]$, where
$f_i(\cdot\,;w):\calV\to\R$ are known, bounded, nonstochastic
functions.\footnote{Boundedness holds in the empirical application
of Section~\ref{subsec:app-fiv}, where $\calV$ is a finite set.}
\textsf{BH23} assume that the shocks are exogenous, that is,
\begin{align}\label{eq: exogeneity of shocks}
    v\independent(\varepsilon_1,\ldots,\varepsilon_n)\mid w,
\end{align}
and show that this alone does not make $z_i$ orthogonal to $\varepsilon_i$,
because units are nonrandomly exposed to the shocks through the formula.

Letting $\widetilde{P}\in\Delta(\calV)$ denote the
assignment process postulated by the econometrician,
the expected instruments are $\mu_i(\widetilde{P}):=\E_{\widetilde{P}}[f_i(v;w)]$
and the recentered instruments are
$\widetilde{z}_i:=z_i-\mu_i(\widetilde{P})$, $i\in[n]$.\footnote{We assume for
simplicity that the distribution of $v\mid w$ does not depend on $w$, as is
the case for the application we consider in Section~\ref{subsec:app-fiv}.}
The recentered IV estimator of \textsf{BH23} is then
\begin{equation}\label{eq:riv}
\widehat{\beta}_*
:=\frac{\sum_{i=1}^{n}\widetilde{z}_i\,y_i}{\sum_{i=1}^{n}\widetilde{z}_i\,x_i}.
\end{equation}
The setup also covers the case in which the treatment itself is generated
by the formula, so that $z_i=x_i$ as in the empirical application of \textsf{BH23} that we
revisit in Section~\ref{subsec:app-fiv}.

Our sensitivity exercise maintains exogeneity of the shocks in
\eqref{eq: exogeneity of shocks} 
and relaxes the assumption that their assignment process is known and equal to
$\widetilde{P}$. For this exercise, we hold the
realized sample (outcomes, treatments, covariates, and the realized
shocks $\check{v}$) fixed, so that the recentered estimate is a function
of the assignment process alone. Because
\eqref{eq:riv} depends on $\widetilde{P}$ only through the
expected instruments, replacing the postulated assignment process by a
candidate one $P\in\Delta(\calV)$ produces the estimate
\begin{equation}\label{eq:riv-candidate}
\widehat{\beta}(P):=\frac{G_y(P)}{G_x(P)},
\end{equation}
where
\begin{alignat*}{2}
G_y(P)&:=g_y(\check{v})-\E_P[g_y(v)],\qquad&
g_y(\cdot)&:=\sum_{i=1}^{n}y_if_i(\cdot\,;w),\\
G_x(P)&:=g_x(\check{v})-\E_P[g_x(v)],\qquad&
g_x(\cdot)&:=\sum_{i=1}^{n}x_if_i(\cdot\,;w).
\end{alignat*}
By construction $\widehat{\beta}(\widetilde{P})=\widehat{\beta}_*$, so the
exercise is centered at the reported estimate. For any
$\delta\in[0,1]$, we let the candidate assignment process $P$ range over
\begin{equation}\label{eq:robustness-set-fiv}
\calP_{\mathsf{FI}}(\delta):=
\big\{P\in\Delta(\calV):\ P=(1-\delta)\widetilde{P}+\delta R\
\text{ for some }R\in\Delta(\calV)\big\}.
\end{equation}
Thus, up to a share $\delta$ of the shock draws may be assigned by a mechanism
other than the one the researcher postulated. Throughout the exercise we maintain that the
recentered first stage is uniformly bounded away from zero over the largest
robustness set, that is,
\begin{equation}\label{eq:fiv-uniform-first-stage}
\inf_{P\in\calP_{\mathsf{FI}}(1)}|G_x(P)|>0.
\end{equation}
As $P$ ranges over $\calP_{\mathsf{FI}}(1)=\Delta(\calV)$, the
expectation $\E_P[g_x(v)]$ ranges over an interval with endpoints
$\inf_{v'\in\calV}g_x(v')$ and $\sup_{v'\in\calV}g_x(v')$, so that
condition \eqref{eq:fiv-uniform-first-stage} holds if and only if
$g_x(\check{v})$ lies strictly above this supremum or strictly below this
infimum. In particular, it requires $\check{v}\notin\calV$.
Because the robustness sets are nested, condition \eqref{eq:fiv-uniform-first-stage}
guarantees that the estimator is well defined under every candidate assignment process and
at every budget.\footnote{In Section~SA6.3 of the
supplementary materials we relax this assumption.} Since the analysis is conditional on
the realized
sample, the left-hand side of \eqref{eq:fiv-uniform-first-stage} is directly computable,
and
in the empirical application of Section~\ref{subsec:app-fiv} we verify that it holds.
Finally, we define the sensitivity bounds as
\begin{equation*}
\underline{\beta}(\delta)
:=\inf_{P\in\calP_{\mathsf{FI}}(\delta)}\widehat{\beta}(P),
\qquad
\overline{\beta}(\delta)
:=\sup_{P\in\calP_{\mathsf{FI}}(\delta)}\widehat{\beta}(P).
\end{equation*}
At $\delta=0$ the robustness set reduces to $\{\widetilde{P}\}$ and both
bounds collapse to $\widehat{\beta}_*$.
\end{example}

\section{Setup}\label{sec:setup}

This section formally defines our framework for sensitivity analysis in terms of
population
shares not satisfying the distributional assumption, and shows how the three leading
applications of Section~\ref{sec:motivating examples} fit into it despite their apparent
differences: the first has no structural parameter but an estimated baseline, the second
has
structural parameters but a known baseline distribution, and the third has a known
baseline
and no structural parameter. In the supplementary materials we present our fully fledged
framework, which also allows for moment inequalities and external nuisance parameters.

\textbf{Estimand.} Let $\Theta\subseteq\R^{\dimt},\dimt\in\N_0$ 
be the parameter space, $\calZ\subseteq\R^{\dimz}, \dimz\in\N$ a closed set 
collecting the possible realizations of a random vector $Z$.
The quantity of interest is a scalar expectation over a distribution $P\in\Delta(\calZ)$
of a known function $g:\calZ\times\Theta\to\R$
\begin{equation}\label{eq: estimand}
\E_P[g(Z;\theta)]=\int_{\calZ}g(z;\theta)\,\de P(z).
\end{equation}
Leading examples are a counterfactual in a structural model, where $\theta$ plays the role
of 
a structural parameter, a causal quantity such as an average treatment effect,
or an estimated elasticity.

\textbf{Moment restrictions.} Let $m:\calZ\times\Theta\to\R^{\dimm}$,
$\dimm\in\N_0$ be a known moment function and 
$\rhopad_0\in\R^{\dimm}$ a vector of target moments. We require that
\begin{align}\label{eq: moments}
    \E_P[m(Z;\theta)]=\rhopad_0.
\end{align}

In a structural model these are the conditions that identify the structural
parameter $\theta$, for instance by matching model implied choice
probabilities with observed ones, while in a randomized experiment they can carry
moments of pre-treatment covariates added to tighten the bounds.
We allow for the components of $\rhopad_0$ to be either
known or estimated. Section~\ref{subsec:estimation} makes the
distinction explicit. When no moment function is present, we set $\dimm=0$
and $\R^0=\{0\}$, interpret \eqref{eq: moments} as vacuous, and suppress $m$
and $\rhopad_0$ from the notation.

\textbf{Baseline distribution.} The econometrician postulates that
$Z$ is distributed according to a \textit{baseline distribution}
$\Pbase{}$, the distribution she would use absent any concern
for misspecification. We let $\Pbase{}\in\calP_\theta(\calZ)$, where
$\calP_\theta(\calZ)$ denotes the set of distributions in $\Delta(\calZ)$ under
which $|g(\cdot\,;\theta)|$ and $\|m(\cdot\,;\theta)\|$ are integrable at
$\theta\in\Theta$.
For brevity, we omit the dependence on $\theta$ and write
$\calP(\calZ)\equiv\calP_\theta(\calZ)$. We
reserve $P_0$ for the true distribution of $Z$. The baseline may be a parametric
specification for a 
distribution that is not nonparametrically identifiable (e.g., a Gumbel distribution 
for idiosyncratic taste shocks) or an identified feature of $P_0$ (e.g., the
outcome distribution of observed respondents), in which case it must be
estimated. In either case the identifying assumption is the equality
between the law of $Z$ and $\Pbase{}$, and the sensitivity exercise relaxes
it. Both cases are covered by our framework. 

\textbf{Distributional assumptions and robustness sets.} Different assumptions are
represented by
different distributions in $\calP(\calZ)$. To control the degree of relaxation of
the distributional assumption embedded in $\Pbase{}$,
or equivalently the degree of misspecification, we introduce a \textit{sensitivity 
parameter} $\delta\in[0,1]$ to bound the distance between
the baseline distribution $\Pbase{}$ and any candidate alternative distribution 
$P\in\calP(\calZ)$. In this paper, the distance is measured in terms of
the share of the population in the alternative distribution $P$ 
for which the distributional assumption embedded in $\Pbase{}$ may fail. We call
the set of distributions that satisfy the moment restrictions in 
\eqref{eq: moments} and are within a distance $\delta$ of the baseline
distribution a \textit{robustness set}. We consider two types of robustness sets, each
suited to a different setting
among those presented in Section~\ref{sec:motivating examples}. 

\textit{First}, consider settings
where the statistical model naturally yields a mixture constraint
\begin{align}\label{eq: mixture constraint for TV}
    \Pbase{} = \kappa P + (1-\kappa)R,\qquad\text{for some}\ R\in\calP(\calZ),
\end{align}
for some known $\kappa\in(0,1)$. In words, the population described by
$\Pbase{}$ consists of two sub-populations, or groups: a $\kappa$-share of
its units belongs to the group whose distribution is $P$ and the remaining
$(1-\kappa)$-share belongs to a second group whose distribution is $R$.
This is the case of Example~\ref{ex:rct}, where \eqref{eq: alt mixture}
shows that the observed treated units are a mixture of two groups, the
compliers and the always-observed units, with $P$ playing the role of the
complier distribution, $R$ that of the always-observed distribution, and
$\kappa=\pi$ the complier share in \eqref{eq: alt share of compliers}. 
In this case, we define the \textit{total variation robustness set} as
\begin{align}\label{eq: TV rob set}
    \calP_{\TV}(\theta;\delta) := \big\{ P\in\calP(\calZ): &\TV(P\,\|\,R) \leq \delta,\ 
    \E_P[m(Z;\theta)]=\rhopad_0,\nonumber\\
    &\Pbase = \kappa P + (1-\kappa)R\ \text{for some}\ R\in\calP(\calZ) \big\}.
\end{align}
The reason \eqref{eq: TV rob set} is appealing is that $\delta$ is the
share of the population for which the distributional assumption in $\Pbase{}$ may fail.
The interpretation of $\delta$ just given deserves some elaboration.
The constraint $\TV(P\,\|\,R)=\delta$ only tells
us that the distributions $P$ and $R$ of the two groups share a fraction $1-\delta$ of
their
probability mass, in the sense that they admit the mixture
representations $P=(1-\delta)C+\delta Q_P$ and
$R=(1-\delta)C+\delta Q_R$ for a common distribution
$C\in\calP(\calZ)$ and some group-specific distributions 
$Q_P,Q_R\in\calP(\calZ)$.
Then, substituting the two representations above into 
the mixture constraint \eqref{eq: mixture constraint for TV}, we get
\[
\Pbase{}=(1-\delta)\,C+\delta\,\bigl(\kappa Q_P+(1-\kappa)Q_R\bigr).
\]
Put differently, for a $(1-\delta)$-share of the population, drawn from the common
distribution $C$, the distribution of $Z$ does not depend on the group a unit belongs to,
and only the remaining $\delta$-share, drawn from the group-specific
$\kappa Q_P+(1-\kappa)Q_R$, is allowed to differ across groups. The distributional
assumption embedded in the baseline, namely that the two groups are drawn from the same
distribution, is thus recovered at $\delta=0$ and is gradually relaxed as $\delta$ grows.

\textit{Second}, when the statistical model does not directly yield a
mixture constraint, we consider a \textit{contamination robustness set} in the spirit
of \citet{huber1964RobustEstimationLocation}, which is defined as
\begin{align}\label{eq: cont rob set}
    \calP_{\mathsf{C}}(\theta;\delta) := \big\{ P\in\calP(\calZ):\ 
    &P = (1-\delta) \Pbase{} + \delta R\ \text{for some}\ R\in\calP(\calZ),\nonumber\\ 
    &\E_P[m(Z;\theta)]=\rhopad_0 \big\}.
\end{align}
In words, a distribution $P$ is a valid candidate distribution if it is
a mixture of the baseline distribution $\Pbase{}$ and an arbitrary distribution $R$,
with $\delta$ being the share of the population that is drawn from $R$. In this case,
the interpretation of $\delta$ is identical to the one given above for the 
total variation robustness set.

\textbf{Sensitivity bounds.} The \textit{sensitivity bounds} are 
the extreme values the estimand \eqref{eq: estimand} can take over the 
robustness sets \eqref{eq: TV rob set} or \eqref{eq: cont rob set}, 
depending on the application:
\begin{equation}\label{eq:bounds}
\underline{\mathsf{b}}_\lozenge(\delta):=\inf_{\theta\in\Theta}\
\inf_{P\in\calP_\lozenge(\theta;\delta)}\E_P[g(Z;\theta)]
\quad\text{and}\quad
\overline{\mathsf{b}}_\lozenge(\delta):=\sup_{\theta\in\Theta}\
\sup_{P\in\calP_\lozenge(\theta;\delta)}\E_P[g(Z;\theta)],
\end{equation}
for $\lozenge\in\{\TV,\mathsf{C}\}$, with the convention that the inner infimum
and supremum are $+\infty$ and $-\infty$, respectively, when the robustness set
is empty at $\theta$. For each fixed
$\theta$, we refer to the optimization over $P$ as the \textit{inner problem} and to
the optimization over $\theta$ as the \textit{outer problem}. Section~\ref{sec:theory}
shows how to solve the inner problems in
\eqref{eq:bounds} exactly despite their infinite-dimensional nature.

The sensitivity exercise we propose in this work reports
$[\underline{\mathsf{b}}_\lozenge(\delta),\overline{\mathsf{b}}_\lozenge(\delta)]$ 
for $\delta \in[0,1]$. At $\delta=0$ the interval collapses to the baseline 
value of the estimand $\E_{\Pbase{}}[g(Z;\theta)]$.\footnote{This statement
presumes that there exists a
$\theta_*\in\Theta$ for which the baseline satisfies the moment conditions in
\eqref{eq: moments}, so that at $\delta=0$ the baseline is the only
distribution in the robustness set at the only feasible parameter value, a common
assumption
in the majority of the studies involving (generalized) method of moments estimators
\citep[see, e.g., Section 2.2.3 in][]{newey1994Chapter36Large}.} As $\delta$ grows, the
sensitivity bounds
widen monotonically because the robustness sets are nested. The smallest $\delta$ at
which the interval contains a given value, zero say, thus answers the question
``how large does the share of the population for which the baseline assumption fails
need to be to overturn the conclusion?'' Section~\ref{sec:usage} formalizes this
answer through summary measures.   

We now map the three leading applications into this framework.

\begin{examplecont}{Example~\ref{ex:rct}}
We can map the randomized experiment with attrition into the framework as follows. 
First, there is no structural parameter, so $\dimm=\dimt=0$ and $\Theta=\{0\}$. 
The random vector of the framework is the treated potential outcome,
that is $Z=Y(1)$ and $\calZ=\calY$. The target estimand is $g(y;\theta)=y$.
The baseline is the outcome distribution of the treated
respondents, $\Pbase{}=P_{\mathsf{T}}$, which is identified by the data and
must therefore be estimated, whereas the distribution being perturbed is the
complier distribution $P_{\mathsf{C}}$. In this notation, \textsf{MCAR} in
the form \eqref{eq: alt MCAR} is exactly the identifying equality between
the distribution of $Z$, here $P_{\mathsf{C}}$, and the baseline $P_{\mathsf{T}}$. 
The mixture decomposition \eqref{eq: alt mixture} is the mixture constraint
\eqref{eq: mixture constraint for TV} with $\kappa=\pi$, the complier share
in \eqref{eq: alt share of compliers}, with $P$ playing the role of the
complier distribution and $R$ that of the always-observed distribution.
The relevant robustness set is thus the total variation one in
\eqref{eq: TV rob set}, and in fact
$\calP_{\TV}(\theta;\delta)=\calP_{\mathsf{RCT}}(\delta)$ for every
$\delta\in[0,1]$. In line with the
interpretation given above, $\delta$ is the share of the
treated respondents for whom \textsf{MCAR} may fail.\footnote{Equivalently, $\delta$
can be interpreted as the share of compliers or always-observed units for 
whom \textsf{MCAR} may fail.} With these choices, the bounds in \eqref{eq:bounds} are
bounds on the
complier mean,
\begin{equation*}
\underline{\mathsf{b}}_{\TV}(\delta)
=\inf_{P\in\calP_{\mathsf{RCT}}(\delta)}\E_{P}[Y]
\qquad\text{and}\qquad
\overline{\mathsf{b}}_{\TV}(\delta)
=\sup_{P\in\calP_{\mathsf{RCT}}(\delta)}\E_{P}[Y].
\end{equation*}
The complier mean is the only quantity in \eqref{eq: alt tau map} that the
data do not identify, and $\tau(\cdot)$ is affine and decreasing in it, so
the sensitivity bounds of Example~\ref{ex:rct} are
\begin{equation*}
\underline{\tau}(\delta)
=\frac{\mu_{\mathsf{T}}-\pi\,\overline{\mathsf{b}}_{\TV}(\delta)}{1-\pi}
-\mu^{\mathsf{AO}}(0)
\qquad\text{and}\qquad
\overline{\tau}(\delta)
=\frac{\mu_{\mathsf{T}}-\pi\,\underline{\mathsf{b}}_{\TV}(\delta)}{1-\pi}
-\mu^{\mathsf{AO}}(0),
\end{equation*}
and both reduce to the \textsf{MCAR} estimand
$\tau_{\mathsf{MCAR}}$ at $\delta=0$. Two features of this application are
worth anticipating.
First, the mixing weight $\pi$ is estimated alongside the baseline. Second, when outcomes
are continuously distributed,
$\underline{\tau}(\delta)$ and $\overline{\tau}(\delta)$
admit a closed form, given in Theorem~SA-34 of the
supplementary materials, and
$[\underline{\tau}(1),\overline{\tau}(1)]$ coincides with the sharp
nonparametric bounds for $\tau_0$ of Proposition 1a in
\citet{lee2009TrainingWagesSample}. When pre-treatment covariates are
added to tighten the bounds, 
there is no closed form for $\delta\in(0,1)$, whereas at $\delta=1$
the sharp nonparametric bounds of Proposition 1b in
\citet{lee2009TrainingWagesSample} and Lemma 3 in
\citet{semenova2025GeneralizedLeeBounds} are recovered
(see Section~SA6.1.1 of the supplementary
materials).
\end{examplecont}

\begin{examplecont}{Example~\ref{ex:structural}}
In this case, the random vector $Z$ is the vector of preference shocks, that is $Z=U$ and
so
$\calZ=\R^{J+1}$. The structural parameter collects the systematic
components of match utility, $\theta=\vartheta$ and $\Theta\subseteq\R^{2J}$,
the estimand is the marital college premium, $g=f$, and the moment function
is $m=s$, so that $\dimm=4J+2$ and $\rhopad_0=(\varrho,0_{2(J+1)})$. The
first $2J$ components of $\rhopad_0$ are the observed marriage
probabilities, which are estimated from the data, whereas the remaining
$2(J+1)$ components are the normalizations. The baseline
is the Gumbel specification of the shocks, $\Pbase{}=\widetilde{P}$, which
is known rather than estimated. Unlike Example~\ref{ex:rct}, the baseline is
not itself a mixture of two groups, so the relevant robustness set is the contamination
one: \eqref{eq: robustness set structural} is exactly \eqref{eq: cont rob set}
with the choices above, that is,
$\calP_{\mathsf{STR}}(\vartheta;\delta)=\calP_{\mathsf{C}}(\vartheta;\delta)$
for every $\vartheta\in\Theta$ and $\delta\in[0,1]$. Its two groups are the
women whose shocks are Gumbel, a share $1-\delta$ of the population, and the women whose
shocks are drawn from an arbitrary distribution $R$, a share $\delta$, so that $\delta$
is the share of women for whom the Gumbel assumption may fail. Finally,
the sensitivity bounds are
$\underline{\mathsf{b}}_{\mathsf{C}}(\delta)=\underline{\mathsf{k}}(\delta)$
and $\overline{\mathsf{b}}_{\mathsf{C}}(\delta)=\overline{\mathsf{k}}(\delta)$
for every $\delta\in[0,1]$. In contrast with Example~\ref{ex:rct}, these
bounds admit no closed form, and the dual problems of
Section~\ref{sec:theory} are the route through which we compute them in
Section~\ref{subsec:app-structural}.
\end{examplecont}

\begin{examplecont}{Example~\ref{ex:riv}}
Contrary to Examples~\ref{ex:rct} and~\ref{ex:structural}, the target of
this sensitivity exercise is the ratio $\widehat{\beta}(P)$ of two
expectations, which is not of the form \eqref{eq: estimand}. Its numerator
and denominator, however, are both affine in the candidate assignment
process $P$, and this is what we leverage to cast the problem in the desired
form. Indeed, for $b\in\R$ we have $\widehat{\beta}(P)=b$ if
and only if $G_y(P)-b\,G_x(P)=0$, and
\begin{equation*}
G_y(P)-b\,G_x(P)=g_b(\check{v})-\E_P[g_b(v)],
\qquad
g_b(\cdot):=g_y(\cdot)-b\,g_x(\cdot),
\end{equation*}
in view of \eqref{eq:riv-candidate}. In words, the recentered estimator
attains the value $b$ under the candidate assignment process $P$ exactly
when $\E_P[g_b(v)]$ equals the realized value $g_b(\check{v})$. Once $b$
is fixed, $g_b$ is a known function on $\calV$, and the exercise fits the
framework upon setting $Z=v$, $\calZ=\calV$, $g(v;\theta)=g_b(v)$,
$\Theta=\{0\}$, $\dimm=\dimt=0$, and
$\Pbase{}=\widetilde{P}$, the assignment process of the exogenous shocks 
postulated by the econometrician. With these choices, the robustness set
\eqref{eq:robustness-set-fiv} is the contamination set
\eqref{eq: cont rob set}, that is
$\calP_{\mathsf{FI}}(\delta)=\calP_{\mathsf{C}}(\theta;\delta)$ for every
$\delta\in[0,1]$, where we use that $g_b$ is bounded, so that
every $P\in\Delta(\calV)$ belongs to $\calP(\calZ)$. The interpretation of
$\delta$ is that the exogenous
shocks are drawn a fraction $(1-\delta)$ of the time from the postulated assignment
process
$\widetilde{P}$ and a fraction $\delta$ of the time from an arbitrary assignment process,
so that
$\delta$ is the share of the population
for which the assumption that the shocks are assigned 
by $\widetilde{P}$ may fail. Write
$\underline{\mathsf{b}}_{\mathsf{C}}(\delta;g_b)$ and
$\overline{\mathsf{b}}_{\mathsf{C}}(\delta;g_b)$ for the bounds in
\eqref{eq:bounds} obtained with $g=g_b$, making the dependence on $b$
explicit. Because $\calP_{\mathsf{FI}}(\delta)$ is convex and
$P\mapsto\E_P[g_b(v)]$ is linear, the values of $\E_P[g_b(v)]$ generated
by the robustness set form an interval whose endpoints are these two
bounds. It follows that every value $b$ taken by $\widehat{\beta}(P)$ over
$\calP_{\mathsf{FI}}(\delta)$ belongs to the set
\begin{equation*}
\mathsf{B}(\delta):=\Big\{b\in\R:\ \underline{\mathsf{b}}_{\mathsf{C}}(\delta;g_b)
\leq g_b(\check{v})\leq\overline{\mathsf{b}}_{\mathsf{C}}(\delta;g_b)\Big\}.
\end{equation*}
The sensitivity bounds of
Example~\ref{ex:riv} are therefore the endpoints of $\mathsf{B}(\delta)$,
\begin{equation*}
\underline{\beta}(\delta)=\inf\mathsf{B}(\delta)
\qquad\text{and}\qquad
\overline{\beta}(\delta)=\sup\mathsf{B}(\delta).
\end{equation*}
Computing them thus amounts to solving the two inner problems in
\eqref{eq:bounds} for each $b$. 
\end{examplecont}

\section{Implementation}
    \label{sec:theory}

The inner problems in \eqref{eq:bounds} optimize over an infinite-dimensional
space of distributions and, outside special cases such as the RCT with attrition of
Example~\ref{ex:rct}, admit no closed form. In this section we show that they can
nonetheless be solved exactly through a convex program in $\dimm+2$ variables which,
solved over a grid of values for $\delta$, traces the thick orange lines of
Figure~\ref{fig:roadmap}.
Section~\ref{subsec:duality} derives the finite-dimensional dual of the inner problem for
each robustness set of Section~\ref{sec:setup} and states conditions under which its value
coincides with that of the primal in \eqref{eq:bounds}.
Sections~\ref{subsec:estimation}--\ref{subsec:large-sample} then turn to finite samples:
the former introduces plug-in estimators of the bounds, and the latter derives their limit
distribution and proposes a resampling procedure to form valid confidence intervals.

\subsection{Duality}\label{subsec:duality}
Fix $\theta\in\Theta$ and $\lozenge\in\{\TV,\mathsf{C}\}$, and write
\begin{equation}\label{eq: primal inner problem}
\overline{V}_{\mathsf{P}}^{\lozenge}(\theta;\delta):=
\sup_{P\in\calP_\lozenge(\theta;\delta)}\E_P[g(Z;\theta)]
\end{equation}
for the value of the inner problem in the upper bound of \eqref{eq:bounds}.
The lower bound is recovered by
applying every result of this section to $-g$ in place of $g$, so we focus
on the upper bound throughout.

The estimand and the moment restrictions are linear in $P$, and both
robustness sets are convex, because the mixture representations in
\eqref{eq: TV rob set} and \eqref{eq: cont rob set} are preserved under
convex combinations and the total variation distance is jointly convex in its
two arguments. The inner problem is therefore a convex program over probability
distributions, and we construct its dual with elementary tools from convex analysis,
following the logic of the duality results for distributionally robust optimization in
\citet*{kuhn2025DistributionallyRobustOptimization}.\footnote{For 
further details, we refer the interested reader to
Section~SA2 of the supplementary materials. 
For a textbook treatment of the relevant convex analysis tools, see
\citet[Sections 6, 12 and 31]{rockafellar1970ConvexAnalysis} and
\citet{rockafellar1974ConjugateDualityOptimization}.}  
The appeal of the dual is that it looks at the problem from the point of view of its
constraints rather than that of the candidate distribution, and thereby reduces its
dimensionality from that of the space of distributions to
$\dimm+2$.

For $\theta\in\Theta$ and $\delta\in(0,1]$, the optimization problems dual to
\eqref{eq: primal inner problem} with 
robustness sets \eqref{eq: TV rob set} and \eqref{eq: cont rob set} are
respectively
\begin{align}\label{eq:dual-tv}
\overline{V}_{\mathsf{D}}^{\TV}(\theta;\delta):=
\inf_{\substack{(\zeta,\eta,\lambda)\in\\ \R\times\R_+\times\R^{\dimm}}}
\Big\{&\zeta+\eta(1-\kappa)\delta+\lambda^{\top}\rhopad_0 \nonumber\\
&+\E_{\Pbase{}}\big[\mathsf{r}_{\TV}\big(g(Z;\theta)-\lambda^{\top}m(Z;\theta)-\zeta,
\ \eta\big)\big]\Big\}
\end{align}
and
\begin{align}\label{eq:dual-c}
\overline{V}_{\mathsf{D}}^{\mathsf{C}}(\theta;\delta):=
\inf_{\substack{(\zeta,\eta,\lambda)\in\\ \R\times\R_+\times\R^{\dimm}}}
&\Big\{\zeta+\eta\delta+\lambda^{\top}\rhopad_0
+\E_{\Pbase{}}\big[\mathsf{r}_{\mathsf{C}}\big(g(Z;\theta)-\lambda^{\top}m(Z;\theta)-\zeta,
\ \eta\big)\big]\Big\}\nonumber\\
\text{s.to}\quad
&\ \zeta+\tfrac{\eta}{2}\ \geq\
\sup_{z\in\calZ}\big\{g(z;\theta)-\lambda^{\top}m(z;\theta)\big\},
\end{align}
where, for $x\in\R$ and $\eta\geq0$,
\begin{equation}\label{eq:integrands}
\begin{aligned}
\mathsf{r}_{\TV}(x,\eta)&:=
\max\Big\{-\frac{\eta}{2},\ x,
\ \frac{x}{\kappa}-\frac{(1-\kappa)\,\eta}{2\kappa}\Big\},\\
\mathsf{r}_{\mathsf{C}}(x,\eta)&:=
\begin{cases}
\max\Big\{(1-\delta)\,x-\frac{\delta\,\eta}{2},\ x\Big\}, &
x\leq\eta/2,\\[4pt]
+\infty, & x>\eta/2.
\end{cases}
\end{aligned}
\end{equation}
Some comments on the two duals are in order. First, 
both programs have $\dimm+2$ variables, one per restriction of the primal problem 
other than the mixture restriction: $\zeta \in\R$ is the multiplier
associated with the constraint that $P$ is a probability distribution; $\eta\in\R_+$
to the budget constraint, which limits the share of the population whose
distribution may depart from the baseline; $\lambda \in\R^{\dimm}$ 
to the moment equalities. Second, 
the three expressions inside the maximum defining $\mathsf{r}_{\TV}$ are the
values of the associated Lagrangian when a unit of the baseline population is removed
from the group described by $P$, kept at its baseline weight, or assigned
the largest weight compatible with the mixture restriction; the two
expressions in $\mathsf{r}_{\mathsf{C}}$ are its values when the unit is
scaled down to the share $1-\delta$ or kept at its baseline weight.
Third, \eqref{eq:dual-c} carries one additional constraint for each point of $\calZ$,
because candidate distributions in $\calP_{\mathsf{C}}(\theta;\delta)$ may place mass
outside
the support of $\Pbase{}$, and the constraints make such relocations unprofitable. No such
constraint appears in \eqref{eq:dual-tv}, because every $P\in\calP_{\TV}(\theta;\delta)$
is
absolutely continuous with respect to $\Pbase{}$, the mixture restriction forcing
$\kappa P\leq\Pbase{}$. Fourth, $\mathsf{r}_{\mathsf{C}}$ embeds an implicit constraint
of its own. Since $\mathsf{r}_{\mathsf{C}}(x,\eta)=+\infty$ whenever $x>\eta/2$, the
objective of \eqref{eq:dual-c} is finite only if
$\zeta+\eta/2\geq g(z;\theta)-\lambda^{\top}m(z;\theta)$ for $\Pbase{}$-almost every
$z\in\calZ$. The pointwise constraint in \eqref{eq:dual-c} imposes the same inequality
at every $z\in\calZ$ and therefore implies the implicit one.
Proposition~\ref{prop:ignore-constraint} below gives a formal condition under which
the converse also holds, so that the pointwise constraint is redundant and can be
deleted.

The next theorem shows that solving \eqref{eq:dual-tv} and \eqref{eq:dual-c} yields
the same value as \eqref{eq: primal inner problem}. Toward this goal,
define the sets of attainable moments
\begin{align*}
\Gamma_{\TV}(\theta;\delta)&:=\big\{\E_P[m(Z;\theta)]:\ P\in\calP(\calZ),\
\TV(P\,\|\,R)\leq\delta,\\
&\qquad\qquad\ \Pbase{}=\kappa P+(1-\kappa)R\ \text{for some }R\in\calP(\calZ)\big\},
\nonumber\\
\Gamma_{\mathsf{C}}(\theta;\delta)&:=\big\{\E_P[m(Z;\theta)]:\ P\in\calP(\calZ),\
P=(1-\delta)\Pbase{}+\delta R\ \text{for some }R\in\calP(\calZ)\big\},\nonumber
\end{align*}
which collect the moment vectors generated by the distributions satisfying
every restriction defining the corresponding robustness set other than the
moment equalities in \eqref{eq: moments}.
\begin{thm}[Strong duality]\label{thm:duality}
Let $\theta\in\Theta$ and $\delta\in(0,1]$.
\begin{enumerate}[label=(\roman*),ref=(\roman*),leftmargin=2em,topsep=2pt,itemsep=2pt]
\item\label{thm:duality-tv} If $\rhopad_0\in\rint(\Gamma_{\TV}(\theta;\delta))$,
then
$\overline{V}_{\mathsf{P}}^{\TV}(\theta;\delta)=\overline{V}_{\mathsf{D}}^{\TV}(\theta;\delta)$.
\item\label{thm:duality-c} If $\sup_{z\in\calZ}|g(z;\theta)|<\infty$ and
$\rhopad_0\in\rint(\Gamma_{\mathsf{C}}(\theta;\delta))$, then
$\overline{V}_{\mathsf{P}}^{\mathsf{C}}(\theta;\delta)=\overline{V}_{\mathsf{D}}^{\mathsf{C}}(\theta;\delta)$.
\end{enumerate}
\end{thm}
The condition $\rhopad_0\in\rint(\Gamma_\lozenge(\theta;\delta))$ is a
Slater condition. In words, it asks that the moment restrictions remain
attainable, within the same budget $\delta$, after a small perturbation of the target
$\rhopad_0$
in every direction along which the attainable moments vary. The relative
interior formulation accommodates models whose moment components satisfy
exact affine relations. When no moment
restrictions are present, the condition holds under the conventions of
Section~\ref{sec:setup}. The boundedness of $g$ in part~\ref{thm:duality-c} is the price
of letting the candidate distributions place mass everywhere in $\calZ$, even where
$\Pbase{}$ does not. 

The pointwise constraint, potentially an infinite collection of inequalities, is what
makes
\eqref{eq:dual-c} more burdensome to compute than \eqref{eq:dual-tv}, and we now give a
condition under which it can be dropped without affecting the value of the program. For
$\theta\in\Theta$ and $\delta\in(0,1]$ define
\begin{equation*}
\overline{V}^{\mathsf{C},\ll}_{\mathsf{P}}(\theta;\delta):=
\sup_{P\in\calP^{\ll}_{\mathsf{C}}(\theta;\delta)}\E_P[g(Z;\theta)],\qquad
\calP^{\ll}_{\mathsf{C}}(\theta;\delta):=
\calP_{\mathsf{C}}(\theta;\delta)\cap\{P:\,P\ll\Pbase{}\},
\end{equation*}
for the value of the inner problem over the contamination robustness set restricted
to distributions dominated by the baseline, write
$\overline{V}^{\mathsf{C},\ll}_{\mathsf{D}}(\theta;\delta)$ for
the infimum of the objective of \eqref{eq:dual-c} over all of
$\R\times\R_+\times\R^{\dimm}$, that is, for the value of \eqref{eq:dual-c} with the
pointwise constraint deleted, and let
\begin{equation*}
\Gamma^{\ll}_{\mathsf{C}}(\theta;\delta):=\big\{\E_P[m(Z;\theta)]:\
P\in\calP(\calZ),\ P\ll\Pbase{},\
P=(1-\delta)\Pbase{}+\delta R,\ R\in\calP(\calZ)\big\}
\end{equation*}
be the attainable moment set of the restricted problem, a subset of
$\Gamma_{\mathsf{C}}(\theta;\delta)$. The inequalities
$\overline{V}^{\mathsf{C},\ll}_{\mathsf{P}}(\theta;\delta)\leq \overline{V}^{\mathsf{C}}_{\mathsf{P}}(\theta;\delta)$
and
$\overline{V}^{\mathsf{C},\ll}_{\mathsf{D}}(\theta;\delta)\leq \overline{V}^{\mathsf{C}}_{\mathsf{D}}(\theta;\delta)$
hold in general, the first because
$\calP^{\ll}_{\mathsf{C}}(\theta;\delta)\subseteq\calP_{\mathsf{C}}(\theta;\delta)$ and
the second
because deleting a constraint enlarges the feasible set.
\begin{prop}[Deleting the pointwise constraint]\label{prop:ignore-constraint}
Let $\theta\in\Theta, \delta\in(0,1]$,
$\rhopad_0\in\rint(\Gamma^{\ll}_{\mathsf{C}}(\theta;\delta))$, 
$\sup_{z\in\calZ}|g(z;\theta)|<\infty$, 
and, for every $\lambda\in\R^{\dimm}$,
\begin{equation}\label{eq:esssup-condition}
\sup_{z\in\calZ}\{g(z;\theta)-\lambda^{\top}m(z;\theta)\}
=\Pbase{}\text{-}\esssup_{z\in\calZ}\{g(z;\theta)-\lambda^{\top}m(z;\theta)\},
\end{equation}
then
$\overline{V}^{\mathsf{C}}_{\mathsf{P}}(\theta;\delta)
=\overline{V}^{\mathsf{C},\ll}_{\mathsf{P}}(\theta;\delta)
=\overline{V}^{\mathsf{C},\ll}_{\mathsf{D}}(\theta;\delta)$.
\end{prop}
Proposition~\ref{prop:ignore-constraint} contains two results. The first equality states
that, under its regularity conditions, removing the distributions $P\not\ll\Pbase{}$ from
$\calP_{\mathsf{C}}(\theta;\delta)$ does not change the value of
\eqref{eq: primal inner problem}. The second states that this value is delivered by the
dual \eqref{eq:dual-c} without the pointwise constraint. Condition
\eqref{eq:esssup-condition} holds, for example, whenever $\supp(\Pbase{})=\calZ$ and
$z\mapsto g(z;\theta)-\lambda^{\top}m(z;\theta)$ is lower semicontinuous for every
$\lambda$, as in the formula instrument setting of Example~\ref{ex:riv}, and also for the
matching indicators of Example~\ref{ex:structural}, which are not continuous.

\subsection{Estimation}\label{subsec:estimation}

Having recast the inner problems as tractable convex programs, we now turn to estimating
the bounds, that is, the thick orange lines in Figure~\ref{fig:roadmap}. Toward this goal,
we introduce some notation. Partition the moment function as
$m=(m_{\mathtt{e}}^{\top},m_{\mathtt{k}}^{\top})^{\top}$, where the target
of the moment function $m_{\mathtt{e}}:\calZ\times\Theta\to\R^{\dime},\dime\in\N_0$ 
is estimated and that of
$m_{\mathtt{k}}:\calZ\times\Theta\to\R^{\dimk},\dimk\in\N_0$ is known and 
normalized to zero. Throughout this section,
$\rhopad_0:=(\rho_0^{\top},{0}_{\dimk}^{\top})^{\top}$
where $\rho_0\in\R^{\dime}$ denotes the moment target to be estimated.
Accordingly, partition the multipliers as
$\lambda=(\lambda_{\mathtt{e}}^{\top},\lambda_{\mathtt{k}}^{\top})^{\top}$ with
$\lambda_{\mathtt{e}}\in\R^{\dime}$ and $\lambda_{\mathtt{k}}\in\R^{\dimk}$.
Throughout, we let $\rho\in\R^{\dime}$, $P\in\Delta(\calZ)$, and
$\xi:=(\zeta,\eta,\lambda)$
collect the dual variables. To treat the two robustness sets at once, let
$c_{\TV}:=1-\kappa$ and $c_{\mathsf{C}}:=1$ be the factors that multiply
the term $\eta\delta$ in \eqref{eq:dual-tv} and \eqref{eq:dual-c},
respectively. For $\lozenge\in\{\TV,\mathsf{C}\}$, we write the objective of
the corresponding dual problem as
\begin{equation*}
\overline{\mathsf{J}}^{\lozenge}(\xi,\theta;\rho,P):=
\zeta+c_\lozenge\,\eta\delta
+\lambda_{\mathtt{e}}^{\top}\rho+P\overline{\ell}^{\lozenge}_{\xi,\theta},
\quad
\overline{\ell}^{\lozenge}_{\xi,\theta}(z):=
\mathsf{r}_\lozenge\big(g(z;\theta)
-\lambda^{\top}m(z;\theta)-\zeta,\ \eta\big),
\end{equation*}
with $\mathsf{r}_\lozenge$ as in \eqref{eq:integrands}, and its feasible set as
$\overline{\calF}^{\TV}(\theta):=\R\times\R_+\times\R^{\dimm}$ and
\begin{align*}
\overline{\calF}^{\mathsf{C}}(\theta):=
\Big\{\xi\in\R\times\R_+\times\R^{\dimm}:\
\zeta+\tfrac{\eta}{2}\geq\sup_{z\in\calZ}
\big\{g(z;\theta)-\lambda^{\top}m(z;\theta)\big\}\Big\},
\end{align*}
so that $\overline{\calF}^{\mathsf{C}}(\theta)$ is the set cut out by the
pointwise constraint of \eqref{eq:dual-c}, on which the integrand
$\mathsf{r}_{\mathsf{C}}$ is finite, whereas we recall that \eqref{eq:dual-tv} has no such
constraint. Symmetrically, let $\underline{\mathsf{J}}^{\lozenge}$,
$\underline{\ell}^{\lozenge}_{\xi,\theta}$, and $\underline{\calF}^{\lozenge}(\theta)$ be
defined as $\overline{\mathsf{J}}^{\lozenge}$, $\overline{\ell}^{\lozenge}_{\xi,\theta}$,
and $\overline{\calF}^{\lozenge}(\theta)$ with $-g$ in place of $g$, so that, in
particular,
$\underline{\calF}^{\TV}(\theta)=\overline{\calF}^{\TV}(\theta)$. Further, let
$\calP_\lozenge(\theta;\delta,\rho,P)$ denote the robustness set
\eqref{eq: TV rob set} or \eqref{eq: cont rob set} with $(\rhopad_0,\Pbase{})$
replaced by $((\rho^{\top},{0}_{\dimk}^{\top})^{\top},P)$, so that
$\calP_\lozenge(\theta;\delta)$ equals
$\calP_\lozenge(\theta;\delta,\rho_0,\Pbase{})$.
In line with the convention following \eqref{eq:bounds}, we define the criterion functions
\begin{equation}\label{eq:criterion}
\Kmax^{\lozenge}(\theta;\rho,P):=
\begin{cases}
\displaystyle\inf_{\xi\in\overline{\calF}^{\lozenge}(\theta)}
\overline{\mathsf{J}}^{\lozenge}(\xi,\theta;\rho,P)
&\text{if}\ \calP_\lozenge(\theta;\delta,\rho,P)\neq\emptyset,\\[2pt]
-\infty&\text{otherwise},
\end{cases}
\end{equation}
and
\begin{equation}\label{eq:criterion-min}
\Kmin^{\lozenge}(\theta;\rho,P):=
\begin{cases}
\displaystyle-\inf_{\xi\in\underline{\calF}^{\lozenge}(\theta)}
\underline{\mathsf{J}}^{\lozenge}(\xi,\theta;\rho,P)
&\text{if}\ \calP_\lozenge(\theta;\delta,\rho,P)\neq\emptyset,\\[2pt]
+\infty&\text{otherwise}.
\end{cases}
\end{equation}
Finally, let
\begin{equation*}
\mathsf{D}_\lozenge(\theta;\rho,P)
:=\inf_{Q\in\calP_\lozenge(\theta;\delta,\rho,P)}\TV(Q\,\|\,P)
\ \text{and}\
\Theta^{\lozenge}_\delta(\rho,P):=\big\{\theta\in\Theta:\
\mathsf{D}_\lozenge(\theta;\rho,P)<c_\lozenge\,\delta\big\}
\end{equation*}
be the smallest distance from $P$ to the robustness set and the set of strictly
feasible parameter values.
The sensitivity bounds can thus be written as
\begin{equation*}
\underline{\mathsf{b}}_\lozenge(\delta,\rho,P):=
\inf_{\theta\in\Theta^{\lozenge}_\delta(\rho,P)}\Kmin^{\lozenge}(\theta;\rho,P)
\quad\text{and}\quad
\overline{\mathsf{b}}_\lozenge(\delta,\rho,P):=
\sup_{\theta\in\Theta^{\lozenge}_\delta(\rho,P)}\Kmax^{\lozenge}(\theta;\rho,P),
\end{equation*}
for $\lozenge\in\{\TV,\mathsf{C}\}$. At the true inputs $(\rho_0,\Pbase{})$,
these coincide with the sensitivity bounds $\underline{\mathsf{b}}_\lozenge(\delta)$
and $\overline{\mathsf{b}}_\lozenge(\delta)$ in \eqref{eq:bounds}.
The plug-in estimators replace the two inputs by
$\widehat\rho_n$, an estimator of $\rho_0$, and by the empirical measure $\widehat P_{n}$
obtained from an i.i.d.\ sample from $\Pbase{}$, which we recall is an identified feature
of $P_0$, the true distribution of $Z$. Formally, the plug-in estimators are
$\widehat{\underline{\mathsf{b}}}_{\lozenge,n}(\delta):=\underline{\mathsf{b}}_\lozenge(\delta,\widehat\rho_n,\widehat P_{n})$
and
$\widehat{\overline{\mathsf{b}}}_{\lozenge,n}(\delta):=\overline{\mathsf{b}}_\lozenge(\delta,\widehat\rho_n,\widehat P_{n})$.
When the baseline is instead known but the moment target is estimated, as in
Example~\ref{ex:structural}, the estimators are
simply $\underline{\mathsf{b}}_\lozenge(\delta,\widehat\rho_n,\Pbase{})$ and
$\overline{\mathsf{b}}_\lozenge(\delta,\widehat\rho_n,\Pbase{})$.

\begin{remark}[Over-identified models]\label{rem:overid}
When the model is over-identified, i.e., when there are more moments ($\dimm$) than
parameters ($\dimt$), there might not exist a $\theta\in\Theta$ for which
$\E_{\Pbase{}}[m(Z;\theta)]$ equals the estimated target
$(\widehat\rho_n^{\top},0_{\dimk}^{\top})^{\top}$.
We refer to Section 3.3 in \cite{christensen2023CounterfactualSensitivityRobustness}
for two alternative solutions to handle this case.
\end{remark}

\subsection{Large Sample Properties}\label{subsec:large-sample}

Our large sample theory rests on two key properties of \eqref{eq:criterion},
which hold for both robustness sets: (\textit{i})
the objective $\overline{\mathsf{J}}^{\lozenge}$ is affine in $(\rho,P)$; (\textit{ii})
the feasible set $\overline{\calF}^{\lozenge}(\theta)$ does not depend on
$(\rho,P)$. As a consequence,
the map $(\rho,P)\mapsto\Kmax^{\lozenge}(\theta;\rho,P)$ is concave with
supergradient $(\lambda_{\mathtt{e}},\overline{\ell}^{\lozenge}_{\xi,\theta})$
at any dual minimizer, and derivatives of the bounds can be read off the dual solutions
by an envelope argument. The same two properties hold for
\eqref{eq:criterion-min}, so that $(\rho,P)\mapsto\Kmin^{\lozenge}(\theta;\rho,P)$ is
convex with subgradient
$-(\lambda_{\mathtt{e}},\underline{\ell}^{\lozenge}_{\xi,\theta})$.

The theory of this section provides large sample guarantees when both the moment target
$\rho_0$ and the baseline $\Pbase{}$ are estimated from the data.
Remark~\ref{rem:known-baseline}
below explains how it simplifies when $\Pbase{}$ is known, as in
Example~\ref{ex:structural}.
In terms of Figure~\ref{fig:roadmap}, we now turn to the dashed orange lines denoting the
confidence intervals of the bounds.

Before stating the conditions needed by the large sample theory, we introduce some
useful quantities. For $\lozenge\in\{\TV,\mathsf{C}\}$, let
\[\Tmax^{\lozenge}(\rho,P):= \argsup_{\theta\in\Theta}\Kmax^{\lozenge}(\theta\,;\rho,P)
\quad \text{and}\quad
\Xmax^{\lozenge}(\theta;\rho,P):= \arginf_{\xi\in\overline{\calF}^{\lozenge}(\theta)}
\overline{\mathsf{J}}^{\lozenge}(\xi,\theta;\rho,P)\]
be the set of outer optimizers of the dual problem and
the set of optimal dual multipliers at $(\theta,\rho,P)$, respectively, and 
\[
\Lmax^{\lozenge}(\theta;\rho,P):=\bigl\{\lambda_{\mathtt{e}}\in\R^{\dime}: 
(\zeta,\eta,\lambda) \in
\Xmax^{\lozenge}(\theta;\rho,P) 
\text{ with } \lambda=(\lambda_{\mathtt{e}}, \lambda_{\mathtt{k}})
 \text{ for some } (\zeta,\eta,\lambda_{\mathtt{k}})
\bigr\}
\]
be the projection of $\Xmax^{\lozenge}(\theta;\rho,P)$ onto
the space of multipliers $\lambda_{\mathtt{e}}$. The quantities
$\Tmin^{\lozenge}$, $\Xmin^{\lozenge}$, and $\Lmin^{\lozenge}$ are defined in the
same way from the program \eqref{eq:criterion-min}. The following 
assumption lists all the regularity conditions needed for the large sample theory.
\begin{assumptionLS}\label{assump:large-sample}
We observe an i.i.d.\ sample $\{Z_i\}_{i=1}^{n}$ from $\Pbase{}$. Furthermore, the 
following conditions hold:
\begin{enumerate}[label=(\roman*),ref=\theassumptionLS(\roman*),leftmargin=2.6em,topsep=2pt,itemsep=3pt]
\item\label{assump:ls-primitive} (Compactness.) $\calZ$ and $\Theta$ are nonempty compact 
sets.
\item\label{assump:ls-reg} (Continuity.) The functions
$g(z;\theta)$ and $m(z;\theta)$ are
jointly continuous on $\calZ\times\Theta$ and there exists a constant $L<\infty$ such that
for all $\theta,\theta'\in\Theta$
\[
\sup_{z\in\calZ}|g(z;\theta)-g(z;\theta')|\leq L\|\theta-\theta'\| 
\quad\text{and}\quad
\sup_{z\in\calZ}\|m(z;\theta)-m(z;\theta')\|\leq L\|\theta-\theta'\|.\] 
\item\label{assump:ls-feas} (Feasibility.)
$\Theta^{\lozenge}_\delta(\rho_0,\Pbase{})\neq\emptyset$.
\item\label{assump:ls-slater} (Slater condition.) For each
$\theta\in\Tmin^{\lozenge}(\rho_0,\Pbase{})\cup\Tmax^{\lozenge}(\rho_0,\Pbase{})$,
$\rhopad_0\in\mathrm{int}\bigl(\Gamma_\lozenge(\theta;\delta)\bigr)$.
\item\label{assump:ls-lin} (Asymptotic linearity.) We have
$\widehat\rho_n=\rho_0+\widehat P_{n}\psi+\op(n^{-1/2})$, with influence function
$\psi:\calZ\to\R^{\dime}$ satisfying $\Pbase{}\psi=0$ and
$\Pbase{}\|\psi\|^2<\infty$.
\item\label{assump:ls-lhc} (Lower hemicontinuity.) The
correspondences $\theta\mapsto\Xmin^{\lozenge}(\theta;\rho_0,\Pbase{})$ and
$\theta\mapsto\Xmax^{\lozenge}(\theta;\rho_0,\Pbase{})$ are lower hemicontinuous at each
$\theta\in\Tmin^{\lozenge}(\rho_0,\Pbase{})$ and each
$\theta\in\Tmax^{\lozenge}(\rho_0,\Pbase{})$, respectively.
\end{enumerate}
\end{assumptionLS}

Assumption~\ref{assump:ls-primitive} greatly simplifies the integrability conditions
and, with Assumptions~\ref{assump:ls-reg}
and~\ref{assump:ls-feas}, guarantees the existence of
solutions to the outer problem.\footnote{
We stress that some of the motivating examples do not satisfy compactness of $\calZ$ in 
Assumption~\ref{assump:ls-primitive}.
In Example~\ref{ex:rct} the outcome space $\calY$ is a closed,
possibly unbounded, subset of $\R$, 
whereas in Example~\ref{ex:structural}, $\calZ=\R^{J+1}$.
For Example~\ref{ex:rct},
Proposition~SA-35 of the supplementary
materials exploits the closed form of the bounds to replace compactness by a finite
second moment of the outcome and a density positive at the trimming quantiles, and it
also accounts for the estimation of the complier share.
For Example~\ref{ex:structural}, relaxing compactness of $\calZ$ would require
assuming a coercivity condition on the moment function, which we leave for future
work. In the same example, the matching indicators in $s_1$ also
violate the continuity requirement of Assumption~\ref{assump:ls-reg}, but only on the
set of ties, which is null under the Gumbel baseline, so that this requirement can be
relaxed as well.}
In Assumption~\ref{assump:ls-reg}, joint continuity is used, together with compactness,
to control the supremum in the pointwise constraint of
$\overline{\calF}^{\mathsf{C}}(\theta)$ and for the existence of solutions to the outer
problem,
whereas Lipschitz continuity is needed to invoke an appropriate functional central limit
theorem and is vacuous when there is no structural parameter, as in Examples~\ref{ex:rct}
and \ref{ex:riv}. Assumption~\ref{assump:ls-feas} ensures 
that the criterion functions are not identically $\pm\infty$.
Assumption~\ref{assump:ls-slater} is the Slater-type condition discussed after
Theorem~\ref{thm:duality}, stated with the interior rather than the relative interior,
which
rules out exact affine relations among the moment components. It also guarantees that the
dual minimizer sets
$\Xmin^{\lozenge}(\theta;\rho_0,\Pbase{})$ and $\Xmax^{\lozenge}(\theta;\rho_0,\Pbase{})$
are nonempty, compact, and convex (Lemma~SA-24 of
the supplementary materials), so that the multiplier sets
$\Lmin^{\lozenge},\Lmax^{\lozenge}$ are well defined.
Assumption~\ref{assump:ls-lin} requires the estimator of $\rho_0$ to be asymptotically
linear,
so that the joint law of $\sqrt{n}(\widehat\rho_n-\rho_0)$ and of the empirical process
$\mathbb{G}_n:=\sqrt{n}(\widehat P_{n}-\Pbase{})$ is amenable to the delta method used
below. Finally, Assumption~\ref{assump:ls-lhc} is used to establish the directional
differentiability of the value function.
Under Assumption~\ref{assump:large-sample} 
the plug-in estimators for the bounds are consistent.
\begin{thm}[Consistency]\label{thm:cons-EB}
Let Assumption~\ref{assump:large-sample} hold for
$\lozenge\in\{\TV,\mathsf{C}\}$. Then
\[
\underline{\mathsf{b}}_{\lozenge}(\delta,\widehat\rho_n,\widehat P_{n})
\cprob
\underline{\mathsf{b}}_{\lozenge}(\delta,\rho_0,\Pbase{})
\qquad\text{and}\qquad
\overline{\mathsf{b}}_{\lozenge}(\delta,\widehat\rho_n,\widehat P_{n})
\cprob
\overline{\mathsf{b}}_{\lozenge}(\delta,\rho_0,\Pbase{}).
\]
\end{thm}
To establish a distributional approximation for the bounds, we rely on a delta method for
maps that are only Hadamard directionally differentiable, that is, whose derivatives are
continuous but not necessarily linear.\footnote{For a formal definition of Hadamard
differentiability and Hadamard directional differentiability, see Section 2.2
in \cite{fang2019InferenceDirectionallyDifferentiable}.} The reason is that the bounds
are values of optimization problems. The maps
$(\rho,P)\mapsto(\underline{\mathsf{b}},\overline{\mathsf{b}})(\delta,\rho,P)$ are
Hadamard
differentiable when these programs have a single optimizer, but generally only
directionally
so when they have several, as
\citet[Section 6.1]{christensen2023CounterfactualSensitivityRobustness}
already noted for the case of known $\Pbase{}$. We therefore use the delta method for
Hadamard directionally differentiable functionals
\citep[see, e.g.,][]{shapiro1991AsymptoticAnalysisStochastic,
dumbgen1993NondifferentiableFunctionsBootstrap}. The next result first derives the
directional
derivative of the bounds at $(\rho_0,\Pbase{})$ and then establishes their distributional
approximation.

\begin{thm}[Directional differentiability and distributional
approximation]\label{thm:asy-EB}
Let Assumption~\ref{assump:large-sample} hold for
$\lozenge\in\{\TV,\mathsf{C}\}$. Then the maps
$(\rho,P)\mapsto(\underline{\mathsf{b}}_{\lozenge},\overline{\mathsf{b}}_{\lozenge})(\delta,\rho,P)$ 
are Hadamard directionally differentiable at $(\rho_0,\Pbase{})$, with
\begin{equation*}\label{eq:HD-EB}
\begin{aligned}
d\underline{\mathsf{b}}_{\lozenge}(\delta,\rho_0,\Pbase{})[h,\gamma]
&=\min_{\theta^*\in\Tmin^{\lozenge}(\rho_0,\Pbase{})}\
\max_{\xi^*\in\Xmin^{\lozenge}(\theta^*;\rho_0,\Pbase{})}
\big\{-(\lambda^*_{\mathtt{e}})^{\top}h-\gamma\,\underline{\ell}^{\lozenge}_{\xi^*,\theta^*}\big\},\\[4pt]
d\overline{\mathsf{b}}_{\lozenge}(\delta,\rho_0,\Pbase{})[h,\gamma]
&=\max_{\theta^*\in\Tmax^{\lozenge}(\rho_0,\Pbase{})}\
\min_{\xi^*\in\Xmax^{\lozenge}(\theta^*;\rho_0,\Pbase{})}
\big\{(\lambda^*_{\mathtt{e}})^{\top}h+\gamma\,\overline{\ell}^{\lozenge}_{\xi^*,\theta^*}\big\},
\end{aligned}
\end{equation*}
and so
\begin{equation*}
\sqrt{n}
\begin{pmatrix}
\underline{\mathsf{b}}_{\lozenge}(\delta,\widehat\rho_n,\widehat P_{n})
-\underline{\mathsf{b}}_{\lozenge}(\delta,\rho_0,\Pbase{})\\[4pt]
\overline{\mathsf{b}}_{\lozenge}(\delta,\widehat\rho_n,\widehat P_{n})
-\overline{\mathsf{b}}_{\lozenge}(\delta,\rho_0,\Pbase{})
\end{pmatrix}
\rightsquigarrow
\begin{pmatrix}
d\underline{\mathsf{b}}_{\lozenge}(\delta,\rho_0,\Pbase{})[Z_\psi,\mathbb{G}]\\[4pt]
d\overline{\mathsf{b}}_{\lozenge}(\delta,\rho_0,\Pbase{})[Z_\psi,\mathbb{G}]
\end{pmatrix},
\end{equation*}
where $\mathbb{G}$ is a $\Pbase{}$-Brownian bridge and
$Z_\psi=\mathbb{G}\psi$, the pair $(Z_\psi,\mathbb{G})$ being jointly
Gaussian with $\E[\mathbb{G}f\,Z_\psi]=\Pbase{}(f\psi)$.
\end{thm}

The novelty of Theorem~\ref{thm:asy-EB} lies in the fact that 
the limiting distribution carries the term
$\mathbb{G}\overline{\ell}^{\lozenge}_{\xi^*,\theta^*}$, which is also correlated with
$Z_\psi$ as both are
functionals of the same sample. When the sets of outer maximizers and dual
minimizers are singletons, the directional derivatives become 
linear and the limit is Gaussian with variance
\begin{equation*}
\sigma^2_*
=\V_{\Pbase{}}\big((\lambda^*_{\mathtt{e}})^{\top}\psi
+\overline{\ell}^{\lozenge}_{\xi^*,\theta^*}(Z)\big),
\end{equation*}
where the two terms in the variance capture the uncertainty due to 
the estimation of the moment target $\rho_0$ and of the baseline $\Pbase{}$, respectively,
and their covariance reflects that both are
estimated from the same sample.

Outside this singleton case the derivatives in Theorem~\ref{thm:asy-EB} are, in general,
nonlinear, and the standard bootstrap is invalid
\citep[Theorem 3.1]{fang2019InferenceDirectionallyDifferentiable}.
Indeed, the standard bootstrap may be interpreted as
relying on a (numerical) derivative that consistently estimates the directional
derivative only when the latter is linear
\citep[see Remark 3.5 in][]{fang2019InferenceDirectionallyDifferentiable}.
Accordingly, in the same spirit as
\cite{christensen2023CounterfactualSensitivityRobustness},
we specialize the resampling procedure proposed in 
\citet{fang2019InferenceDirectionallyDifferentiable} to our case. 
Concretely, we estimate the directional derivative as
\begin{align*}
\widehat{d\overline{\mathsf{b}}}_{\lozenge,n}(\delta)[h,\gamma]
&:=\sup_{\theta\in\Tmaxhat^{\lozenge}}\
\inf_{\xi\in\Xmax^{\lozenge}(\theta;\widehat\rho_{n},\widehat P_{n})}
\big\{\lambda_{\mathtt{e}}^{\top}h+\gamma\,\overline{\ell}^{\lozenge}_{\xi,\theta}\big\},\\
\Tmaxhat^{\lozenge}&:=\Big\{\theta\in\Theta^{\lozenge}_{\delta}(
\widehat\rho_{n},\widehat P_{n}):
\Kmax^{\lozenge}(\theta;\widehat\rho_{n},\widehat P_{n})\geq
\widehat{\overline{\mathsf{b}}}_{\lozenge,n}(\delta)-\widehat\nu_{n}
\sqrt{\tfrac{\log n}{n}}\Big\}
\end{align*}
where $\widehat\nu_{n}\geq0$ is a tuning sequence that is bounded and
bounded away from zero in probability. The lower-bound counterparts
$\Tminhat^{\lozenge}$ and $\widehat{d\underline{\mathsf{b}}}_{\lozenge,n}(\delta)$ are
defined
analogously, with $-g$ in place of $g$. Following
\citet[Section 6.2]{christensen2023CounterfactualSensitivityRobustness}, a convenient
data-driven choice of $\widehat\nu_n$ is the smallest diagonal entry of
the covariance matrix of $m(Z;\widehat\theta_n)$ under $\widehat P_{n}$, with
$\widehat\theta_n$ an approximate maximizer of the estimated outer
problem, which requires no simulation. 
Algorithm~\ref{alg:boot} details all the steps of the resampling procedure, whose
validity is proven in the supplementary
materials under a strengthening of Assumption~\ref{assump:large-sample}.
In particular, we show that the bootstrap statistics
$\widehat T_{n}^{*,b,\mathrm{low}}$ and $\widehat T_{n}^{*,b,\mathrm{up}}$ of
Algorithm~\ref{alg:boot} are consistent for their population counterparts in
Theorem~\ref{thm:asy-EB}, and the confidence sets they deliver have asymptotic
coverage at least $1-\alpha$ whenever the distribution functions of those
limits are continuous and strictly increasing at the quantiles used. 
\begin{algorithm}[t]
\caption{Confidence sets for the sensitivity bounds}
\label{alg:boot}
\begin{algorithmic}[1]
\State Compute $\widehat\rho_{n}$, $\widehat P_{n}$, the bounds
$\widehat{\underline{\mathsf{b}}}_{\lozenge,n}(\delta),\widehat{\overline{\mathsf{b}}}_{\lozenge,n}(\delta)$,
the near-argmax sets $\Tminhat^{\lozenge},\Tmaxhat^{\lozenge}$, and the dual
minimizer sets $\Xmax^{\lozenge}(\theta;\widehat\rho_{n},\widehat P_{n})$
and
$\Xmin^{\lozenge}(\theta;\widehat\rho_{n},\widehat P_{n})$.
\For{$b=1,\ldots,B$}
\State Draw an i.i.d.\ sample $\{Z_{i}^{*,b}\}_{i=1}^n$ from
$\widehat P_{n}$ and form
$\widehat P_{n}^{*,b}$ and $\widehat\rho_{n}^{*,b}$.
\State Set
$\widehat T_{n}^{*,b,\mathrm{up}}
:=\widehat{d\overline{\mathsf{b}}}_{\lozenge,n}(\delta)\big[\sqrt{n}(\widehat\rho_{n}^{*,b}-\widehat\rho_{n}),\,
\sqrt{n}(\widehat P_{n}^{*,b}-\widehat P_{n})\big]$
and the analog
$\widehat T_{n}^{*,b,\mathrm{low}}$.
\EndFor
\State For $u\in(0,1)$, let
$\widehat{\mathfrak{c}}_{u}^{\mathrm{low}}$ and $\widehat{\mathfrak{c}}_{u}^{\mathrm{up}}$
be the
empirical $u$-quantiles of $\{\widehat T_{n}^{*,b,\mathrm{low}}\}_{b=1}^{B}$
and $\{\widehat T_{n}^{*,b,\mathrm{up}}\}_{b=1}^{B}$, respectively. Valid 
lower and upper $100(1-\alpha)\%$ confidence sets are
\[
[\widehat{\underline{\mathsf{b}}}_{\lozenge,n}(\delta)
-\widehat{\mathfrak{c}}_{1-\alpha}^{\mathrm{low}}/\sqrt{n},+\infty)
\quad\text{and}\quad
(-\infty,\widehat{\overline{\mathsf{b}}}_{\lozenge,n}(\delta)
-\widehat{\mathfrak{c}}_{\alpha}^{\mathrm{up}}/\sqrt{n}].\]
\end{algorithmic}
\end{algorithm}

\begin{remark}[Known baseline]\label{rem:known-baseline}
When the baseline is known, the same argument with $P$ held fixed at
$\Pbase{}$ yields consistency, while the directional derivative becomes
\begin{equation*}
d\overline{\mathsf{b}}_{\lozenge}(\delta,\rho_0,\Pbase{})[h]
=\max_{\theta^*\in\Tmax^{\lozenge}(\rho_0,\Pbase{})}\
\min_{\lambda^*\in\Lmax^{\lozenge}(\theta^*;\rho_0,\Pbase{})}(\lambda^*)^{\top}h,
\end{equation*}
which is identical to that of
\citet[Theorem 6.2]{christensen2023CounterfactualSensitivityRobustness} even
though total variation sits outside the class of divergences considered there.
In this case, the Lipschitz requirement of
Assumption~\ref{assump:ls-reg} is not needed, and the lower
hemicontinuity of Assumption~\ref{assump:ls-lhc} may be imposed on the
multiplier projections $\Lmin^{\lozenge},\Lmax^{\lozenge}$ alone, since the
derivative then sees the dual optimizer only through them. Moreover, since
expectations against $\Pbase{}$ are computed exactly, the data enter only through
$\widehat\rho_n$. 
Accordingly, Assumption~\ref{assump:ls-lin} may be weakened to the requirement that
$\sqrt{n}(\widehat\rho_n-\rho_0)\rightsquigarrow\normal(0,\Sigma)$ for a
finite $\dime\times\dime$ matrix $\Sigma$, with $\widehat\rho_n$ computed from
whatever sample identifies $\rho_0$, which need not be a sample of $Z$. In 
Example~\ref{ex:structural}, $Z$ collects
latent preference shocks that are never observed while $\rho_0$ is estimated
from data on realized matches. Then,
Step 3 of Algorithm~\ref{alg:boot} draws only a
bootstrap replicate of $\widehat\rho_n$ and
the resulting procedure coincides with the algorithm of
\citet[][Section 6.2]{christensen2023CounterfactualSensitivityRobustness}
(see Remark~SA-11 of the supplementary materials for formal
statements).
\end{remark}

\section{How to Use the Sensitivity Bounds?}
    \label{sec:usage}

This section explains how the output of our sensitivity framework is meant to be
used. In terms of Figure~\ref{fig:roadmap}, Section~\ref{subsec:sharp} studies the 
right endpoint of the horizontal
axis, where at $\delta=1$ the bounds coincide with the exact nonparametric
sharp bounds, and Section~\ref{subsec:summary} introduces some
reportable summary measures.

\subsection{Exact Nonparametric Sharp Bounds}
\label{subsec:sharp}
Besides admitting the population share interpretation, measuring the relaxation of
the distributional assumption through shares has an important byproduct: the
sensitivity parameter never needs to exceed one. At $\delta=1$, the total variation
constraint in \eqref{eq: TV rob set} holds for every pair of groups, because the total
variation distance never exceeds one, and the mixture representation in
\eqref{eq: cont rob set} reads $P=R$, so that it places no restriction on $P$. What
remains binding at $\delta=1$ are the restrictions that the model itself imposes,
namely that $Z$ takes values in $\calZ$, that the moment restrictions
\eqref{eq: moments} hold at $\theta$, and, for the total variation robustness set, the
mixture constraint \eqref{eq: mixture constraint for TV}. Therefore, the sensitivity
bounds at $\delta=1$ coincide with the nonparametric sharp bounds on the
estimand \eqref{eq: estimand}. The next theorem formalizes this intuition. 
Toward this goal, we define
\begin{equation*}
\calE(\theta):=\big\{P\in\calP(\calZ):\ \E_P[m(Z;\theta)]=\rhopad_0\big\},
\end{equation*}
the set of distributions compatible with the moment restrictions at $\theta$, 
\[
\calD_{\TV}(\Pbase{}):=\big\{P\in\calP(\calZ):\ \kappa P\leq\Pbase{}\big\}
\qquad\text{and}\qquad
\calD_{\mathsf{C}}(\Pbase{}):=\calP(\calZ).
\]

\begin{thm}[Nonparametric sharp bounds]\label{thm:sharp}
Let $\lozenge\in\{\TV,\mathsf{C}\}$. Then, for $\theta\in\Theta$,
\[\calP_\lozenge(\theta;1)=\calD_\lozenge(\Pbase{})\cap\calE(\theta).\]
Consequently, the sensitivity bounds \eqref{eq:bounds} at $\delta=1$ satisfy
\begin{equation*}
\underline{\mathsf{b}}_{\lozenge}(1)
=\inf_{\theta\in\Theta}
\inf_{P\in\calD_\lozenge(\Pbase{})\cap\calE(\theta)}\E_P[g(Z;\theta)],\quad
\overline{\mathsf{b}}_{\lozenge}(1)
=\sup_{\theta\in\Theta}
\sup_{P\in\calD_\lozenge(\Pbase{})\cap\calE(\theta)}\E_P[g(Z;\theta)].
\end{equation*}
\end{thm}

Theorem~\ref{thm:sharp} follows directly from the definitions of the two
robustness sets. 
In words, $\underline{\mathsf{b}}_{\mathsf{C}}(1)$ and
$\overline{\mathsf{b}}_{\mathsf{C}}(1)$ are the infimum and the supremum of
$\E_P[g(Z;\theta)]$ over all $\theta\in\Theta$ and all $P\in\calE(\theta)$, that is,
over every distribution on $\calZ$ that satisfies the moment restrictions at some
parameter value, whereas $\underline{\mathsf{b}}_{\TV}(1)$ and
$\overline{\mathsf{b}}_{\TV}(1)$ further require the baseline to admit the mixture
representation \eqref{eq: mixture constraint for TV}. In both cases the bounds
are sharp, meaning that no distribution
compatible with the maintained restrictions generates a value of the estimand outside
$[\underline{\mathsf{b}}_\lozenge(1),\overline{\mathsf{b}}_\lozenge(1)]$, so that 
neither bound can be tightened without imposing restrictions beyond those maintained.

Theorem~\ref{thm:sharp} goes beyond sensitivity analysis as it 
makes the framework a device for computing
nonparametric sharp bounds. In some problems those bounds have known closed-form
expressions. For instance, in Example~\ref{ex:rct} 
the interval $[\underline{\mathsf{b}}_\TV(1),\overline{\mathsf{b}}_\TV(1)]$
delivers the trimming bounds of
\citet{lee2009TrainingWagesSample} when outcomes are continuously
distributed (see Theorem~SA-34 of the supplementary
materials). 
Similarly, in the missing data problem of
Section~SA6.4.3 of the supplementary materials,
where a share $p$ of the population is observed and its outcome law is the baseline,
the contamination bounds
$[\underline{\mathsf{b}}_\mathsf{C}(1-p),\overline{\mathsf{b}}_\mathsf{C}(1-p)]$
at a budget equal to the missing share
coincide with the sharp nonparametric bounds of 
\citet{Manski_1994, manski2003PartialIdentificationProbability}.
In other problems, such as the marital college premium of
Example~\ref{ex:structural}, no closed form is available, and the dual problems of
Section~\ref{sec:theory} return the sharp nonparametric bounds numerically.

\subsection{Summary Measures}\label{subsec:summary}

Throughout this subsection we let $\lozenge\in\{\TV,\mathsf{C}\}$ and suppress it from the
notation, so that $\blo(\delta)$ and $\bup(\delta)$ denote the bounds in
\eqref{eq:bounds}. We measure the robustness of a conclusion against a
\textit{reference value} that the analyst wants to rule out, such as zero for the sign of
the estimand or a policy relevant threshold. For simplicity, we set it to zero and take
the effect to be positive at the baseline, $\blo(0)>0$, the opposite case being
symmetric. The estimated counterparts we report in Section~\ref{sec:applications} are
plug-in estimators of the four summary measures defined below.
Section~SA5 of the supplementary materials studies the four
measures and establishes the large sample properties of the estimated certified breakdown
budget. Its formal results are stated for
the total variation robustness set without structural parameters, which is the case of
the attrition application of Section~\ref{subsec:app-rct}, but the properties below
hold for the contamination robustness set by the same arguments.

\textbf{Breakdown and certified breakdown.} The \textit{breakdown budget} is the smallest
budget at which the sensitivity bounds include the reference value, that is,
\begin{equation}\label{eq:breakdown}
\delta_{\mathsf{b}}\equiv\delta_{\mathsf{b}}(0)
:=\inf\big\{\delta\in[0,1]:\ \blo(\delta)\leq0\big\},
\end{equation}
with the convention that the infimum over an empty set equals $1$. The breakdown point
goes back to the robust statistics literature
\citep{hampel1968ContributionsTheoryRobust,donoho1983NotionBreakdownPoint} and has been
adopted in econometrics to quantify how much a maintained assumption can be relaxed before
a conclusion is overturned
\citep[see, among many,][]{horowitz1995IdentificationRobustnessContaminated,
kline2013SensitivityMissingData, masten2020InferenceBreakdownFrontiers,
ober-reynolds2026RobustnessMissingData}. By the population share interpretation of
Section~\ref{sec:setup}, no departure from the identifying assumption involving less than
a $\delta_{\mathsf{b}}$ share of the population can overturn the conclusion.

The breakdown budget captures solely misspecification uncertainty. To account for
sampling uncertainty as well, for $\alpha\in(0,1)$ and a sample size $n$ held fixed, we
define the \textit{certified breakdown budget}
\begin{equation}\label{eq:certified}
{\delta}_{\mathsf{cb}}(\alpha;n)
:=\inf\Big\{\delta\in[0,1]:\
\blo(\delta)
-\frac{\mathfrak{c}_{1-\alpha/2}\,{\sigma}(\delta)}{\sqrt{n}}
\leq0\Big\},
\end{equation}
where $\sigma(\delta)$ is the asymptotic standard deviation of the plug-in estimator
$\blohat(\delta)$ and $\mathfrak{c}_{1-\alpha/2}$ is the $(1-\alpha/2)$-th quantile of the
standard normal distribution.\footnote{This definition presumes that the problems
defining $\blo(\delta)$ have a unique optimizer, so that the limit of
$\sqrt{n}(\blohat(\delta)-\blo(\delta))$ is Gaussian by Theorem~\ref{thm:asy-EB}; see
Remark~SA-13 of the supplementary materials for
the attrition application and the general case.}
The critical value is two-sided because the exercise does not commit in advance to the
sign of $\blo(\delta)$. We say that a study \textit{certifies} robustness at a budget
$\delta$ and level $\alpha$ when its conclusion survives both misspecification of size
$\delta$ and sampling uncertainty at level $1-\alpha$. Similar measures appear in
\citet{cinelli2020MakingSenseSensitivity} for unobserved confounding and in
\citet{rambachan2023MoreCredibleApproach} for violations of parallel trends.

By construction, $\delta_{\mathsf{cb}}(\alpha;n)\leq\delta_{\mathsf{b}}$. Moreover,
$\delta_{\mathsf{cb}}(\alpha;n)$ is nondecreasing in $n$ and converges to
$\delta_{\mathsf{b}}$ as $n\to\infty$, and, under the conditions of
Proposition~SA-32 of the supplementary materials, its sample
counterpart is consistent for $\delta_{\mathsf{b}}$. Sampling uncertainty thus vanishes
but robustness does not, because $\delta_{\mathsf{b}}$ is a feature of the design alone.
Reporting the pair $(\widehat\delta_{\mathsf{b}},\widehat{\delta}_{\mathsf{cb}})$
therefore separates fragility by design (a small breakdown budget) from fragility due to
imprecision (a large gap between the two). The two budgets can also rank studies
differently: a decision maker weighing whether to fund a second round of data collection
should prefer the study with the larger breakdown budget, whereas one who has to act on
the evidence in hand should prefer the study with the larger certified budget.

\textbf{The shadow price of robustness.} We hold the structural parameter fixed and
suppress it from the notation. Under appropriate regularity conditions, the bound paths
are finite, monotone, and concave ($\bup$) and convex ($\blo$), respectively, on the range
of budgets at which the robustness set is nonempty, so that each path is differentiable at
all but at most countably many budgets. At every such budget we define the
\textit{shadow price of robustness} as
\begin{equation*}
\overline{\eta}(\delta):=\bup'(\delta)
\qquad\text{and}\qquad
\underline{\eta}(\delta):=-\blo'(\delta).
\end{equation*}
Both shadow prices are nonnegative and nonincreasing in the budget, so that the first unit
of misspecification is the most damaging. For the total variation robustness set, the
budget enters \eqref{eq:dual-tv} only through the additive term $\eta(1-\kappa)\delta$, so
that $\eta$ prices the total variation constraint as a Lagrange multiplier prices a
resource constraint, and $\overline{\eta}(\delta)=(1-\kappa)\,\eta^{*}$ wherever the
optimal multiplier $\eta^{*}$ is unique (Proposition~SA-31 of
the supplementary materials).

Unlike the breakdown budget, which concerns the sign of an effect, the shadow price also
speaks to its magnitude. Suppose that a decision maker is unwilling to commit to a single
budget and instead holds beliefs $\mu$ on $[0,1]$, with distribution function $F_\mu$.
Then, provided that $\blo$ is finite and continuous on $[0,1]$, she could evaluate the
design by integrating the worst case sensitivity bound $\blo(\delta)$ over her beliefs,
that is
\begin{equation*}
\E_\mu\big[\blo(\delta)\big]
=\blo(0)-\int_0^{1}\underline{\eta}(s)\,\big(1-F_\mu(s)\big)\,\de s,
\end{equation*}
so that the price of the $s$-th unit of misspecification is weighted by the probability
she assigns to a violation at least that large.

\textbf{The certification frontier.} For a budget $\delta<\delta_{\mathsf{b}}$, the
\textit{certification frontier}
\begin{equation}\label{eq:frontier}
n^{\star}(\delta;\alpha)
:=\frac{\mathfrak{c}_{1-\alpha/2}^{2}\,\sigma^{2}(\delta)}
{\blo(\delta)^{2}}
\end{equation}
is the sample size a study would need in order to certify robustness at budget $\delta$
and level $\alpha$. Under the conditions of Proposition~SA-33 of
the supplementary materials, its semi-elasticity with respect to $\delta$ behaves as
$2/(\delta_{\mathsf{b}}-\delta)$ as $\delta\to\delta_{\mathsf{b}}$, so that each
additional percentage point of certified robustness becomes increasingly expensive in
observations. More data, however, cannot make a study more robust: as $n\to\infty$ the
certified budget converges to $\delta_{\mathsf{b}}$ and no further, because
misspecification is of fixed size while sampling uncertainty shrinks at rate $n^{-1/2}$.
Put differently, the frontier prices certification, not robustness itself.

\section{Empirical Applications}
    \label{sec:applications}

We present three empirical applications. The first
revisits the online tutoring program of \citet{carlana2025ApartConnectedOnline}, a
randomized experiment in which attrition is moderate but strongly differential
across arms. The second revisits the marital college premium application of
\citet{christensen2023CounterfactualSensitivityRobustness} and focuses on
interpretation and computational cost. The third turns to formula
instruments: it assesses the sensitivity of the estimated elasticity of
\citet{borusyak2023NonrandomExposureExogenous} to the
postulated assignment process of the shocks. The supplementary materials
contain two further exercises on attrition. Section~SA7.1
revisits the transport subsidy experiment of \citet*{christensen2026DemandMobility},
in which attrition is severe but nearly balanced across arms, and
Section~SA7.2 runs the same sensitivity exercise on
every eligible trial in RCT-Bench, the public collection of randomized clinical
trials assembled by \citet*{shao2026HowShouldCovariates}, which currently comprises $125$
trials, asking how often attrition
threatens the conclusions of medical experiments in practice.

\subsection{Attrition in Randomized Experiments}\label{subsec:app-rct}
We revisit the Tutoring Online Program experiment of
\citet{carlana2025ApartConnectedOnline}, maintaining the setup of Example~\ref{ex:rct}
throughout. 

\textbf{Setup.} \citet{carlana2025ApartConnectedOnline}
evaluate the Tutoring Online Program (TOP), which offered free individual
online tutoring by university student volunteers to disadvantaged Italian
middle schoolers. The experiment ran in two waves, with individual-level randomization in
each: TOP
2020, during the COVID-19 school closures ($530$ treated, $529$
control), and TOP 2022, during regular in-person schooling ($607$
treated, $530$ control). The outcome is a standardized mathematics
performance index that is observed only for test takers (i.e., respondents).
Table~\ref{tab: attrition rates} shows that attrition varies
across waves and arms: the response rates of the treated and control arms are
$87.9\%$ and $46.3\%$ in 2020 and $92.3\%$ and $72.3\%$ in 2022, implying estimated
complier shares $\widehat{\pi}$ of $0.47$ and $0.22$ in the 2020 and
2022 samples, respectively. Our sensitivity exercise asks how much of the
experimental conclusion survives departures from ignorable
attrition. In addition, our framework allows us to compare and rank the 
two waves in terms of how robust their conclusions are.

\begin{table}[h]
\centering
\caption{Attrition and sample size by wave in the Tutoring Online Program.}
\label{tab: attrition rates}
\begin{tabular}{lccccc}
\toprule\toprule
 & \multicolumn{2}{c}{\textit{Treatment}} & \multicolumn{2}{c}{\textit{Control}} & \\
\cmidrule(lr){2-3}\cmidrule(lr){4-5}
Wave & $n$ & Attrition & $n$ & Attrition & $\widehat{\pi}$ \\
\midrule
TOP 2020 & 530 & 12.1\% & 529 & 53.7\% & 0.47 \\
\addlinespace[4pt]
TOP 2022 & 607 & 7.7\% & 530 & 27.7\% & 0.22 \\
\bottomrule\bottomrule
\end{tabular}

\end{table}

\begin{figure}[t]
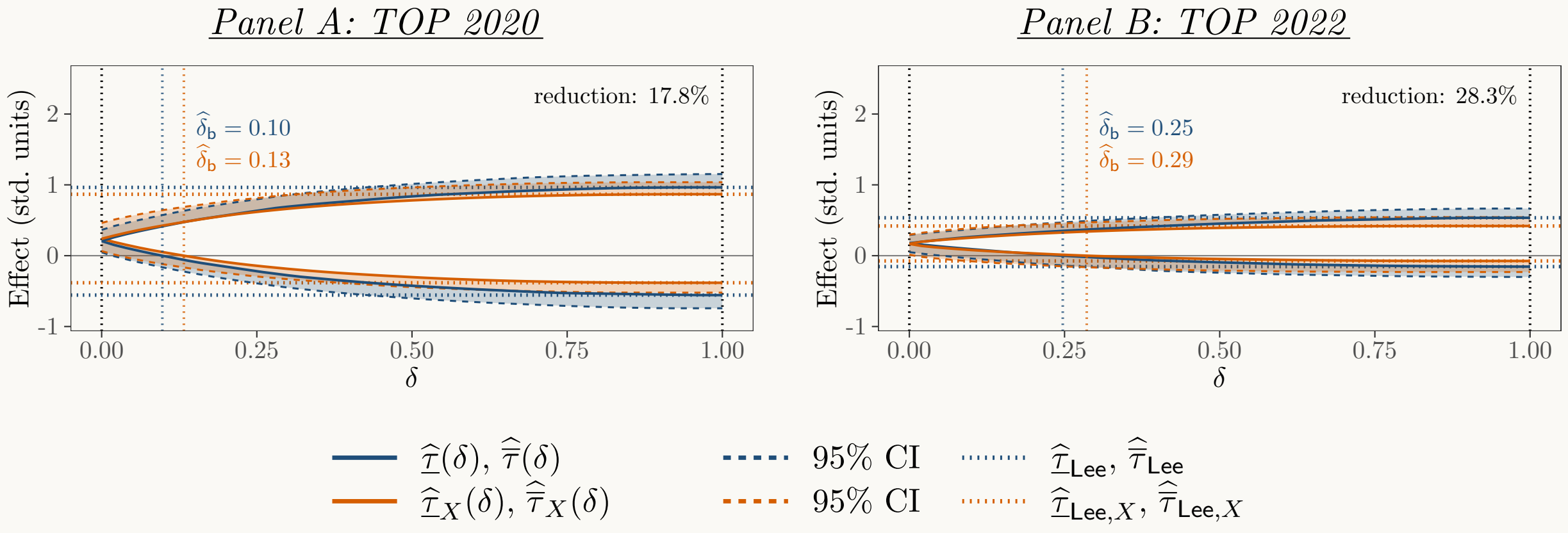

\caption{$\TV$ sensitivity bounds in the Tutoring Online Program.}
\centering
\begin{minipage}[t]{0.495\textwidth}
\centering
\uline{\textit{Panel A: TOP 2020}}\par\smallskip
\resizebox{\textwidth}{!}{%
  \adjustbox{trim=-11.66pt 72.05pt 0pt 0pt,clip}{%
    \input{figures/TV_bounds_performance_main_2020.tex}}}
\end{minipage}\hfill
\begin{minipage}[t]{0.495\textwidth}
\centering
\uline{\textit{Panel B: TOP 2022}}\par\smallskip
\resizebox{\textwidth}{!}{%
  \adjustbox{trim=-11.66pt 72.05pt 0pt 0pt,clip}{%
    \input{figures/TV_bounds_performance_main.tex}}}
\end{minipage}
\par\medskip
\resizebox{0.59\textwidth}{!}{% =====================================================================
% Shared legend for the two-panel RCT-with-attrition figure in main.tex.
% Extracted verbatim from the legend block of
% TV_bounds_performance_main.tex (tikzDevice output), so that the two
% panels of the merged figure can be shown without their own legends.
% Coordinates are the original picture coordinates; the bounding box is
% restricted to the legend block. Re-extracted 2026-08-22 after the legend
% moved from three rows of two to two rows of three (no covariates / with
% covariates); regenerate this file whenever the legend block changes.
% =====================================================================
\providecommand{\bgcolor}{white}% background inherited from the document
\begin{tikzpicture}[x=1pt,y=1pt]
\path[use as bounding box] (62.00, 8.50) rectangle (555.57, 62.91);
\begin{scope}
\path[clip] (  0.00,  0.00) rectangle (578.16,433.62);
\colorlet{fillColor}{\bgcolor}

\path[fill=fillColor] ( 62.00,  8.50) rectangle (555.57, 62.91);
\end{scope}
\begin{scope}
\path[clip] (  0.00,  0.00) rectangle (578.16,433.62);
\colorlet{fillColor}{\bgcolor}

\path[fill=fillColor] ( 70.50, 39.95) rectangle (113.18, 54.41);
\definecolor{drawColor}{RGB}{31,78,121}

\path[draw=drawColor,line width= 2.3pt,line join=round] ( 74.77, 47.18) -- (108.91, 47.18);
\end{scope}
\begin{scope}
\path[clip] (  0.00,  0.00) rectangle (578.16,433.62);
\colorlet{fillColor}{\bgcolor}

\path[fill=fillColor] (276.54, 39.95) rectangle (319.22, 54.41);
\definecolor{drawColor}{RGB}{31,78,121}

\path[draw=drawColor,line width= 2.3pt,dash pattern=on 4pt off 4pt ,line join=round] (280.80, 47.18) -- (314.95, 47.18);
\end{scope}
\begin{scope}
\path[clip] (  0.00,  0.00) rectangle (578.16,433.62);
\colorlet{fillColor}{\bgcolor}

\path[fill=fillColor] (402.47, 39.95) rectangle (445.14, 54.41);
\definecolor{drawColor}{RGB}{31,78,121}

\path[draw=drawColor,line width= 2.3pt,dash pattern=on 1pt off 3pt ,line join=round] (406.73, 47.18) -- (440.88, 47.18);
\end{scope}
\begin{scope}
\path[clip] (  0.00,  0.00) rectangle (578.16,433.62);
\colorlet{fillColor}{\bgcolor}

\path[fill=fillColor] ( 70.50, 17.00) rectangle (113.18, 31.45);
\definecolor{drawColor}{RGB}{213,94,0}

\path[draw=drawColor,line width= 2.3pt,line join=round] ( 74.77, 24.23) -- (108.91, 24.23);
\end{scope}
\begin{scope}
\path[clip] (  0.00,  0.00) rectangle (578.16,433.62);
\colorlet{fillColor}{\bgcolor}

\path[fill=fillColor] (276.54, 17.00) rectangle (319.22, 31.45);
\definecolor{drawColor}{RGB}{213,94,0}

\path[draw=drawColor,line width= 2.3pt,dash pattern=on 4pt off 4pt ,line join=round] (280.80, 24.23) -- (314.95, 24.23);
\end{scope}
\begin{scope}
\path[clip] (  0.00,  0.00) rectangle (578.16,433.62);
\colorlet{fillColor}{\bgcolor}

\path[fill=fillColor] (402.47, 17.00) rectangle (445.14, 31.45);
\definecolor{drawColor}{RGB}{213,94,0}

\path[draw=drawColor,line width= 2.3pt,dash pattern=on 1pt off 3pt ,line join=round] (406.73, 24.23) -- (440.88, 24.23);
\end{scope}
\begin{scope}
\path[clip] (  0.00,  0.00) rectangle (578.16,433.62);
\definecolor{drawColor}{RGB}{0,0,0}

\node[text=drawColor,anchor=base west,inner sep=0pt, outer sep=0pt, scale=  1.60] at (121.68, 41.67) {\legTVbounds};
\end{scope}
\begin{scope}
\path[clip] (  0.00,  0.00) rectangle (578.16,433.62);
\definecolor{drawColor}{RGB}{0,0,0}

\node[text=drawColor,anchor=base west,inner sep=0pt, outer sep=0pt, scale=  1.60] at (327.72, 41.67) {95\% CI};
\end{scope}
\begin{scope}
\path[clip] (  0.00,  0.00) rectangle (578.16,433.62);
\definecolor{drawColor}{RGB}{0,0,0}

\node[text=drawColor,anchor=base west,inner sep=0pt, outer sep=0pt, scale=  1.60] at (453.64, 41.67) {\legLeebound};
\end{scope}
\begin{scope}
\path[clip] (  0.00,  0.00) rectangle (578.16,433.62);
\definecolor{drawColor}{RGB}{0,0,0}

\node[text=drawColor,anchor=base west,inner sep=0pt, outer sep=0pt, scale=  1.60] at (121.68, 18.72) {\legTVboundsX};
\end{scope}
\begin{scope}
\path[clip] (  0.00,  0.00) rectangle (578.16,433.62);
\definecolor{drawColor}{RGB}{0,0,0}

\node[text=drawColor,anchor=base west,inner sep=0pt, outer sep=0pt, scale=  1.60] at (327.72, 18.72) {95\% CI};
\end{scope}
\begin{scope}
\path[clip] (  0.00,  0.00) rectangle (578.16,433.62);
\definecolor{drawColor}{RGB}{0,0,0}

\node[text=drawColor,anchor=base west,inner sep=0pt, outer sep=0pt, scale=  1.60] at (453.64, 18.72) {\legLeeboundX};
\end{scope}
\end{tikzpicture}}
\par
\begin{center}
\parbox[1]{\textwidth}{\footnotesize
\textit{Notes:}
Computed from the replication data of
\citet{carlana2025ApartConnectedOnline}. Each panel plots
the plug-in estimators of $\underline{\tau}(\delta)$ and $\overline{\tau}(\delta)$
described in
Section~\ref{subsec:estimation} over $\delta\in\{0,0.02,\ldots,1\}$, without and with
covariates, together with the bounds of \citet{lee2009TrainingWagesSample} and the outer
$95\%$
confidence bounds from $1{,}000$ iterations of the bootstrap of Algorithm~\ref{alg:boot}.
The
vertical dotted lines mark the estimated breakdown budgets $\widehat{\delta}_{\mathsf{b}}$
without and with covariates, and the annotation in each panel reports the reduction in the
width of the bounds at $\delta=1$ achieved by the covariates. The covariate
cells are the full interaction of indicators for grades six and seven, a
median split of the baseline INVALSI math score, mother's high school
completion, mother's white collar occupation, and single parent household.}
\end{center}
\label{fig:attrition}
\end{figure}

\textbf{Results.} Figure~\ref{fig:attrition} reports the bounds for TOP 2020 (Panel A)
and TOP 2022 (Panel B). In TOP 2020, the bounds start at $\delta=0$
from the difference in means among respondents, which corresponds to an estimated
$0.207$ standard deviations, and at $\delta=1$ reach the Lee bounds, $[-0.556,0.964]$.
The sign of the effect survives at most $10\%$ of the compliers departing from the
\textsf{MCAR} assumption. Once sampling uncertainty is taken into account, the sign
is certified at the $5\%$ level only up to $\widehat{\delta}_{\mathsf{cb}}=0.017$. 
We do not report the certified breakdown budgets in
Figure~\ref{fig:attrition} to avoid cluttering the panels. Conditioning on a handful
of pre-treatment covariates tightens the bounds at every budget, shrinking their
width at $\delta=1$ by $18\%$, and increases the breakdown budget to $0.13$ and the
certified breakdown budget to $0.028$.\footnote{We explain in detail how to add covariates
in Section~SA6.1.1 of the supplementary materials.}
In TOP 2022, the bounds start from a difference in means of $0.176$ standard deviations
and
reach the Lee bounds, $[-0.154,0.535]$. The sign of the effect survives at most $25\%$ of
the
compliers departing from the \textsf{MCAR} assumption, and it is certified at the $5\%$
level up
to $\widehat{\delta}_{\mathsf{cb}}=0.051$. Covariates tighten the bounds at every budget
above
$\delta=0.1$, shrinking their width at $\delta=1$ by $28\%$, and increase the breakdown
budget
to $0.29$. The sensitivity exercise thus ranks the two waves: TOP 2020 is the more fragile
one,
even though its difference in means among respondents is larger. In hindsight, this is not
surprising, as TOP 2020 ran during the school closures and is also the wave with the
larger
response differential across arms. Neither the point estimates nor the Lee bounds, which
do
not sign the effect in either wave, can make this distinction.

Table~\ref{tab:summary} reports, for
each specification, some of the summary measures of Section~\ref{subsec:summary}:
the breakdown budget $\widehat\delta_{\mathsf{b}}$, the certified breakdown budget
$\widehat\delta_{\mathsf{cb}}$ at the $5\%$ level, the shadow price of
robustness at the breakdown budget
$\widehat{\underline{\eta}}(\widehat{\delta}_{\mathsf{b}})$, and the certification
frontier evaluated five percentage points above the certified budget, together with its
implied cost (see the table notes). Three comments are in order.
First, in both waves the certified budget is far below the breakdown budget, so that most
of the apparent fragility is due to sampling uncertainty, not to misspecification
uncertainty. 
Second, certifying five additional percentage points of
robustness requires roughly $8{,}000$ additional students ($\$80$k per
point) in TOP 2020, but only about $1{,}200$ ($\$12$k per point) in
TOP 2022, whose smaller complier share makes the bounds widen more slowly. The
frontier thus answers the second round budgeting question raised in
Section~\ref{subsec:summary} and tells a decision maker funding a hypothetical second
round
where an additional dollar buys the most certified robustness.
Third, the shadow price $\widehat{\underline{\eta}}(\widehat{\delta}_{\mathsf{b}})$ is
minus the slope of the estimated lower bound in Figure~\ref{fig:attrition} at the budget
where it crosses zero. At the breakdown budget, each additional percentage point of
compliers departing from \textsf{MCAR} lowers the worst case by $0.017$ standard
deviations in TOP 2020, against $0.005$ in TOP 2022. As shadow prices are nonincreasing
in the budget, the lower bound of TOP 2020 falls at least this fast at every smaller
budget, while that of TOP 2022 never falls faster than its initial shadow price of
$1.00$. 
This is why TOP 2020 breaks down first despite its larger difference in
means.

\begin{table}[h]
\centering
\caption{Summary measures in the Tutoring Online Program.}
\label{tab:summary}
{\small
\begin{tabular}{lccccccc}
\toprule\toprule
 & $\widehat{\delta}_{\mathsf{b}}$ & $\widehat{\delta}_{\mathsf{cb}}$ &
 $\widehat{\underline{\eta}}(\widehat{\delta}_{\mathsf{b}})$ & $n$ &
 $n^{\star}$ & $\Delta n$ & \$/pp \\
\midrule
TOP 2020 & $0.098$ & $0.017$ & $1.74$ & 1,059 & 9,048 & $+$7,989 &
 $80$k \\[3pt]
TOP 2022 & $0.247$ & $0.051$ & $0.49$ & 1,137 & 2,326 & $+$1,189 &
 $12$k \\
\bottomrule\bottomrule
\end{tabular}}
\par
\begin{center}
\parbox[1]{\textwidth}{\footnotesize
\textit{Notes:} $\widehat{\delta}_{\mathsf{b}}$ is the plug-in breakdown budget
\eqref{eq:breakdown} and $\widehat{\delta}_{\mathsf{cb}}$ the certified
breakdown budget \eqref{eq:certified} at $\alpha=0.05$, both without
covariates; $\widehat{\underline{\eta}}(\widehat{\delta}_{\mathsf{b}})$
is the estimated shadow price of robustness at $\widehat{\delta}_{\mathsf{b}}$; $n$ is the
realized sample size and $n^{\star}$ the
certification frontier \eqref{eq:frontier} evaluated at
$\widehat{\delta}_{\mathsf{cb}}+0.05$, rounded up, with
$\Delta n=n^{\star}-n$; \$/pp is the implied cost per percentage point of
certified robustness at a marginal cost of $\$50$ per participant enrolled
and surveyed \citep{puri2019ChallengesRealWorldImpact}.}
\end{center}
\end{table}

Finally, we use the shadow price of robustness to offer a
different comparison of the two waves. Consider a decision maker who has no
information on the extent of the departure from \textsf{MCAR} and holds uniform
beliefs $\mu$ on $[0,1]$, so that $F_\mu(s)=s$ and the evaluation of
Section~\ref{subsec:summary} reads
$\E_\mu[\underline{\tau}(\delta)]=\underline{\tau}(0)-\int_0^{1}\underline{\eta}(s)(1-s)\,\de s$.
We compute the shadow prices as the slopes of the plug-in lower bound, without
covariates, between consecutive budgets of the grid of Figure~\ref{fig:attrition}.
In TOP 2020 the shadow price starts
at $2.74$, so that each of the first two percentage points of compliers departing
from \textsf{MCAR} lowers the worst case by about $0.027$ standard deviations, and
the weighted integral equals $0.556$, which gives
$\E_\mu[\widehat{\underline{\tau}}(\delta)]=0.207-0.556=-0.349$. In TOP 2022 the shadow
price
starts at $1.00$ and the integral equals $0.237$, which gives
$\E_\mu[\widehat{\underline{\tau}}(\delta)]=0.176-0.237=-0.061$. The decision maker thus
ranks
the waves in the opposite order to their differences in means because the larger point
estimate of TOP 2020 is more than offset by a shadow price that starts almost three
times as large.

\subsection{Counterfactuals in Structural Models}
\label{subsec:app-structural}

We revisit one of the leading applications of \textsf{CC23}: 
the sensitivity of the marital college premium in the matching model
of \citet{chiappori2017PartnerChoiceInvestment} to the Gumbel specification
of the taste shocks. 

\textbf{Setup.} For illustrative purposes, we use the data for a single
cohort instead of the 28 separate cohorts of \textsf{CC23}, namely cohort 14, composed
of white men born in 1953 and white wives one year younger. There are $J=5$ education
groups (high school dropouts, high school graduates, some college, college graduates,
and postgraduates), and we measure the premium between college graduate ($b$) and
some college ($b'$) women. We set $\Theta=[-12,12]^{2J}$ 
and plug in the estimated matching frequencies for $\varrho$. 
At the Gumbel baseline the premium is $-0.221$. 
We obtain confidence bounds from the version of
Algorithm~\ref{alg:boot} where $\Pbase{}$ is known. Expectations over $\Pbase{}$ are
approximated by $5{,}000$ scrambled Halton draws, while the outer optimization problem is
solved
with ten random restarts per bound and warm starts along the $\delta$-grid.
We compare two sensitivity exercises using the same data, moments, and simulation draws.
The first reports the bounds of \textsf{CC23}, which let the candidate distributions range
over the robustness set
\begin{align}\label{eq: robustness set cc23}
\calP_{\mathsf{CC23}}(\vartheta;\delta) := \Big\{P\in\calP(\R^{J+1}):\,
&P\ll\Pbase{},\ \E_{P}[s(U;\vartheta)]=(\varrho,0_{2(J+1)}),\nonumber\\
&\E_{\Pbase{}}\Big[\phi_{\mathsf{CC23}}\Big(\tfrac{\de P}{\de \Pbase{}}\Big)\Big]
\leq\delta\Big\},
\end{align}
where 
\begin{equation*}\label{eq: cc23 entropy}
\phi_{\mathsf{CC23}}(x):=
\begin{cases}
x\log x-x+1, & 0\leq x\leq e,\\[4pt]
\dfrac{1}{2e}(x-e)^2+(x-e)+1, & x>e.
\end{cases}
\end{equation*}
In words, the divergence used to measure distances between a candidate $P$ and the 
baseline $\Pbase{}$ coincides with the Kullback--Leibler divergence for $x\leq e$ and
continues
with a $\chi^2$-type quadratic tail for $x>e$. 
\textsf{CC23} adopt this hybrid because one of
their regularity conditions fails for pure Kullback--Leibler at the Gumbel
baseline. The second exercise delivers the bounds
$\underline{\mathsf{k}}(\delta)$ and $\overline{\mathsf{k}}(\delta)$ of
Example~\ref{ex:structural} over the robustness set
$\calP_{\mathsf{STR}}(\vartheta;\delta)$ in \eqref{eq: robustness set structural},
in which $P=(1-\delta)\Pbase{}+\delta R$, so that $\delta$ is the
share of women drawing non-Gumbel shocks.

\begin{figure}[htbp]
\caption{Sensitivity of the marital college premium on a common axis.}
\centering
\resizebox{.8\textwidth}{!}{\input{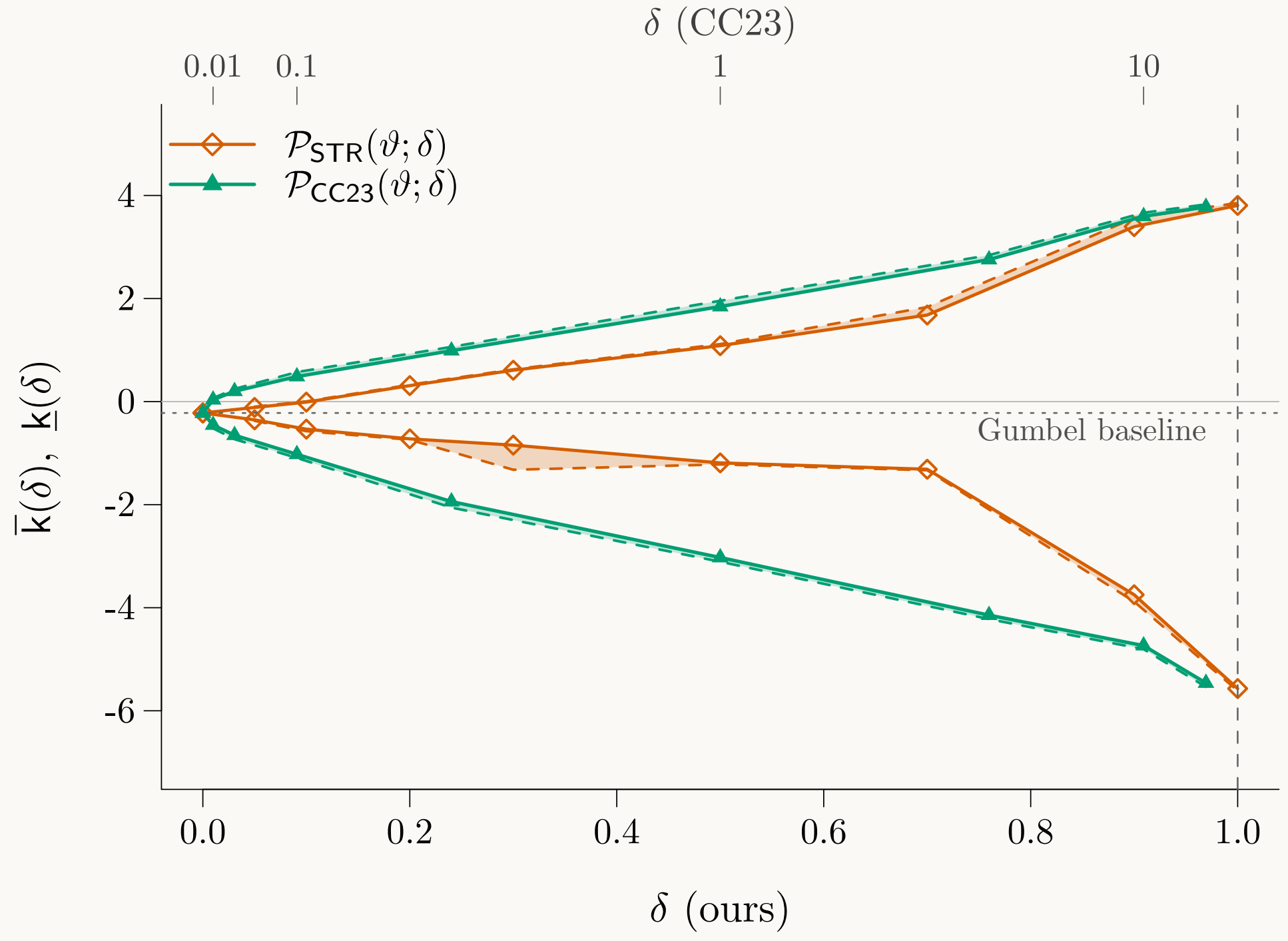}}
\par
\begin{center}
\parbox[1]{\textwidth}{\footnotesize
\textit{Notes:}
Computed from the cohort-14 marriage probabilities of white couples in the
data of \citet{chiappori2017PartnerChoiceInvestment}, as prepared by the
replication package of \textsf{CC23}. The
figure plots the bounds $\underline{\mathsf{k}}(\delta)$ and
$\overline{\mathsf{k}}(\delta)$ of Example~\ref{ex:structural} over the
robustness set $\calP_{\mathsf{STR}}(\vartheta;\delta)$ of
\eqref{eq: robustness set structural} (solid orange lines) and the bounds of
\textsf{CC23} over their robustness set
$\calP_{\mathsf{CC23}}(\vartheta;\delta)$ of \eqref{eq: robustness set cc23} 
(solid green lines). Our bounds are plotted against the bottom $x$-axis, the budget
$\delta$ of
\eqref{eq: robustness set structural}. The \textsf{CC23} bounds are drawn at the
horizontal
position $\delta/(1+\delta)$, which maps their budget $\delta\in[0,\infty)$ onto $[0,1)$,
and
the top $x$-axis reports the original values of $\delta$. Dashed lines are $95\%$
confidence bounds from $1{,}000$ draws of Algorithm~\ref{alg:boot}.}
\end{center}
\label{fig:gumbel}
\end{figure}

\begin{figure}[htbp]
\caption{Computational cost of the sensitivity bounds.}
\centering
\resizebox{.94\textwidth}{!}{\input{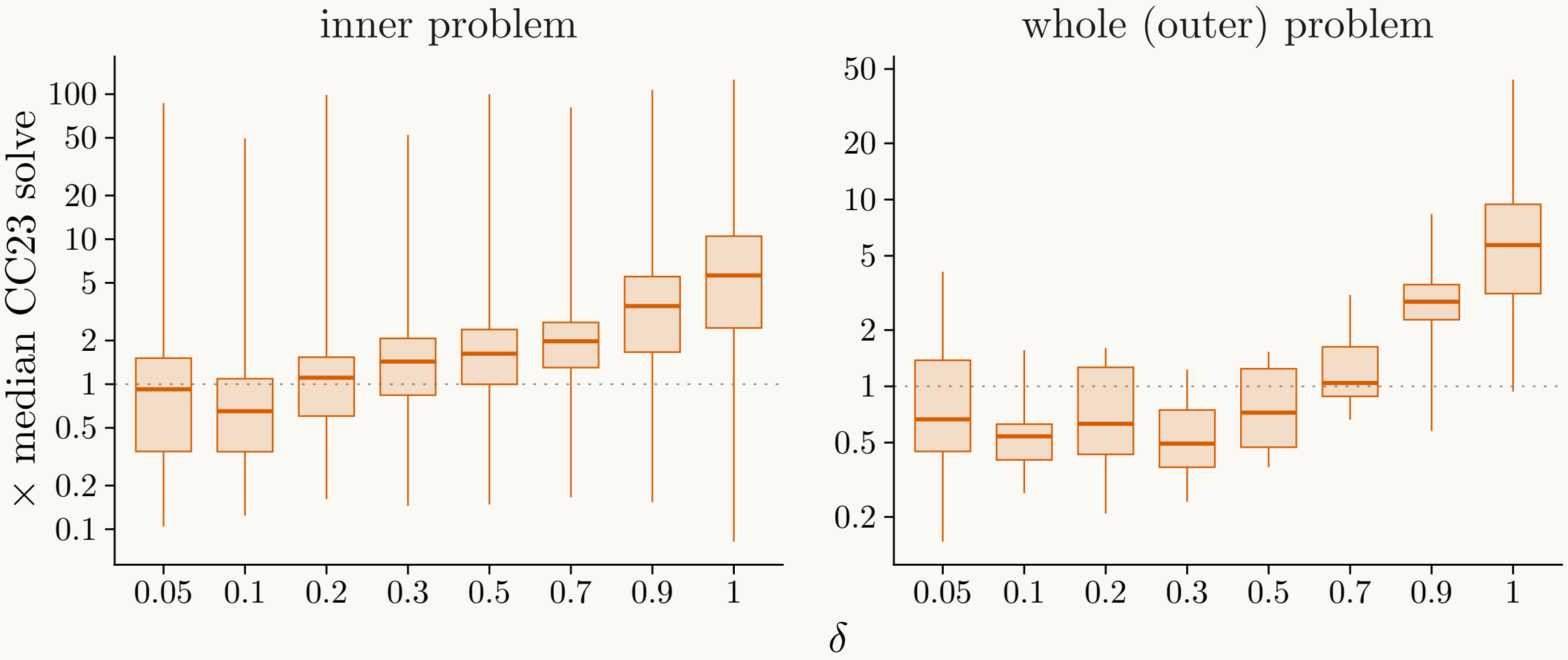}}
\par
\begin{center}
\parbox[1]{\textwidth}{\footnotesize
\textit{Notes:}
Computed from the replication data of
\textsf{CC23}, sharing the
simulation draws of Figure~\ref{fig:gumbel}. The panels report solve
times for the bounds over the robustness set
$\calP_{\mathsf{STR}}(\vartheta;\delta)$ of
\eqref{eq: robustness set structural}, for the inner problem at a given
$\vartheta$ (left panel) and for the whole problem (right panel), in multiples
of the median solve time of the corresponding \textsf{CC23} problem (dotted horizontal
line) using $\calP_{\mathsf{CC23}}(\vartheta;\delta)$ of \eqref{eq: robustness set cc23}, 
over the budget grid $\delta\in\{0.05,0.1,0.2,0.3,0.5,0.7,0.9,1\}$, displayed at
equally spaced positions. All programs are solved with Knitro 15.1 in Julia
1.12 on an Apple M5 laptop.}
\end{center}
\label{fig:timing-cost}
\end{figure}

\textbf{Results.} Figure~\ref{fig:gumbel} plots the two sets of bounds on a common axis.
Three comments are in order. 
First, the sign of the premium is fragile. The upper bound 
$\overline{\mathsf{k}}(\delta)$ is still
negative at $\delta=0.1$, where it equals $-0.010$, but it is already positive at
$\delta=0.2$, where it equals $0.310$. Hence, a share of women between a tenth
and a fifth drawing non-Gumbel shocks suffices to overturn the sign of the
baseline premium. Second, by Theorem~\ref{thm:sharp}, $\underline{\mathsf{k}}(1)$ and
$\overline{\mathsf{k}}(1)$ are the sharp nonparametric bounds on the premium. As expected,
the \textsf{CC23}
bounds approach the same envelope only in the large budget limit, without
attaining it at any finite budget, coming very close at a budget of roughly
$100$, the largest value \textsf{CC23} use in their application. 
Since our bounds attain the envelope,
they can also be used as a diagnostic for how well a divergence-based approximation
does at any given budget. Third, along the whole path sampling uncertainty
is an order of magnitude smaller than misspecification uncertainty,
suggesting that ambiguity, not noise, drives the width of the bounds.
Figure~\ref{fig:timing-cost} compares solve
times. A \textsf{CC23} inner solve takes on the order of a hundredth of a second at
$5{,}000$ draws, and the outer problem for one bound a few
seconds (median between $0.4$ and $5.3$ seconds). Our programs solve in comparable time:
up to $\delta=0.7$, the median inner solve 
takes less than twice as long as that of
\textsf{CC23}, while the median whole problem takes roughly the same time. Both
ratios rise to between five and six at $\delta=1$. 
Computing one bound over the
eight-point budget grid takes about $3$ minutes against $2.5$ for \textsf{CC23}. 

\subsection[Formula Instruments]{Formula Instruments}
\label{subsec:app-fiv}
We revisit one of the leading applications of \textsf{BH23}:
the effect of market access growth on the growth of urban employment across
 Chinese prefectures between 2007 and 2016.

\textbf{Setup.} \textsf{BH23} exploit the construction of the country's high-speed rail
network. The market access of prefecture $i\in[n]$ in year $t\in\{2007,2016\}$ is
\[
\mathsf{MA}_{it}\equiv \mathsf{MA}_{i}(v_t)=\sum_{j\in[n]}
\frac{\mathsf{Pop}_{j}}{\mathsf{Time}(v_t,\mathsf{Loc}_i,\mathsf{Loc}_j)},
\]
where $\mathsf{Pop}_{j}$ is the population of prefecture $j$ in the 2000 census and
$\mathsf{Time}(\cdot)$ is the predicted travel time between the locations $\mathsf{Loc}_i$
and $\mathsf{Loc}_j$ given the status $v_t$ of the railway lines in year $t$. The
endogenous regressor is $x_i=\log\mathsf{MA}_{i,2016}-\log\mathsf{MA}_{i,2007}$, the
outcome $y_i$ is the growth of urban employment over the same period, and the sample
consists of the $n=275$ prefectures for which the employment outcome is recorded. Because
market access grew most in the economic center of the country, a comparison with the
periphery may be confounded by unobserved determinants of employment growth.

\textsf{BH23} therefore use the timing of construction, rather than the location of the
network, as the source of exogenous variation. The shocks $v\in\{0,1\}^K$ are the 2016
completion indicators of the $K=149$ lines that were in service by 2016 (83) or completed
or under construction by 2019 (66), and $\calV=\{v^{(1)},\ldots,v^{(S)}\}$ collects the
$S=1{,}999$ counterfactual networks obtained by permuting completion status within groups
of lines that connect the same number of prefectures. The formula of
Example~\ref{ex:riv} is $f_i(v';w)=\log\mathsf{MA}_i(v')-\log\mathsf{MA}_{i,2007}$, where
$w=\big((\mathsf{Loc}_j,\mathsf{Pop}_j)_{j\in[n]},v_{2007}\big)$ is predetermined, so that
$z_i=f_i(\check{v};w)=x_i$ at the realized shocks $\check{v}$. Following \textsf{BH23},
$\widetilde{P}$ is the uniform distribution on $\calV$, so that
$\mu_i(\widetilde{P})=S^{-1}\sum_{s=1}^{S}f_i(v^{(s)};w)$. We consider both
specifications of \textsf{BH23}, the first controlling for a constant only and the second
adding distance to Beijing, latitude, and longitude. Recentering lowers the estimated
elasticity from $0.232$ to $\widehat\beta_*=0.084$ without controls and from $0.133$ to
$\widehat\beta_*=0.056$ with them. Condition
\eqref{eq:fiv-uniform-first-stage} holds in both specifications. The recentered
first stage at the postulated assignment process is $G_x(\widetilde{P})=9.92$
without controls and $9.24$ with them, whereas
$\sup_{v'\in\calV}g_x(v')-\E_{\widetilde{P}}[g_x(v)]$ equals $8.34$ and $5.66$,
respectively. 

\begin{figure}[htbp]
\caption{Sensitivity bounds for the employment effect of market access
growth.}
\centering
\resizebox{\textwidth}{!}{\input{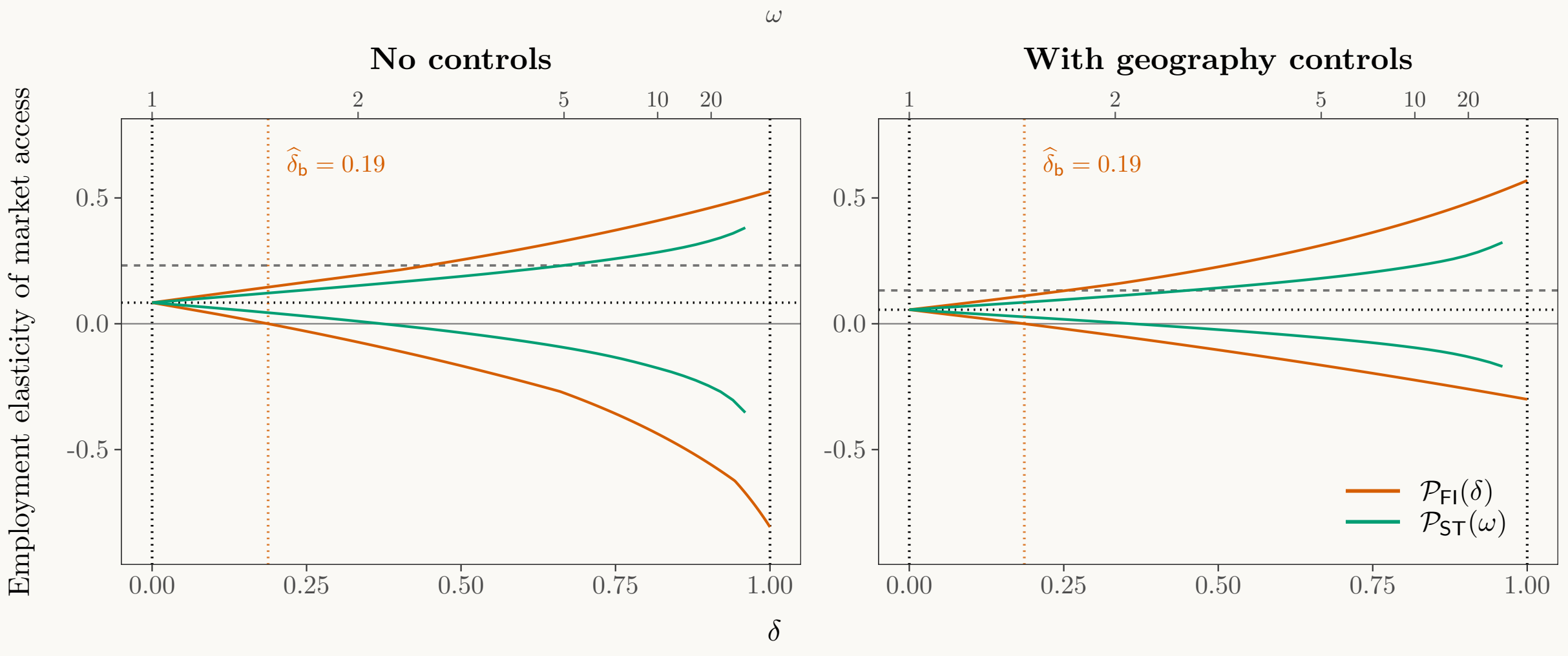}}
\par
\begin{center}
\parbox[1]{\textwidth}{\footnotesize
\textit{Notes:}
Computed from the replication data of \textsf{BH23}. 
Each panel plots the bounds
$\underline{\beta}(\delta)$ and $\overline{\beta}(\delta)$ of
Example~\ref{ex:riv} over the robustness set $\calP_{\mathsf{FI}}(\delta)$ of
\eqref{eq:robustness-set-fiv} (solid orange lines), and the
\textsf{ST26} bounds over the set \eqref{eq:st-set}
(solid green lines), the latter drawn against the rescaling
$(\omega-1)/(\omega+1)$ of their budget, whose original values are
reported on the top horizontal axis. The dotted horizontal line is the
recentered estimate $\widehat\beta_*$ of \eqref{eq:riv} and the dashed
horizontal line the unadjusted OLS estimate. The
orange vertical dotted line marks the estimated breakdown budget
$\widehat{\delta}_{\mathsf{b}}$ for the sign of the effect, defined for
$\widehat\beta_*$ as in \eqref{eq:breakdown}. The bounds carry no sampling
uncertainty (see Remark~SA-18 in the
supplementary materials).}
\end{center}
\label{fig:fiv}
\end{figure}

Figure~\ref{fig:fiv} reports two types of bounds. The first are our sensitivity bounds
$\underline{\beta}(\delta)$ and $\overline{\beta}(\delta)$ over the robustness set
$\calP_{\mathsf{FI}}(\delta)$ of \eqref{eq:robustness-set-fiv}
(Proposition~SA-37 in the supplementary
materials characterizes them and their computation). The second are the bounds of
\citet{sheng2026WhatMagicFormula} (\textsf{ST26}, henceforth), who ask the same question
for this very application and vary the assignment process over what they call ``the joint
sensitivity set''
\begin{equation}\label{eq:st-set}
\calP_{\mathsf{ST}}(\omega)
:=\Big\{P\in\Delta(\calV):\ P\ll\widetilde{P},\ 
\omega^{-1}\leq\de P/\de\widetilde{P}\leq\omega,\
\text{$\widetilde{P}$-a.s.}\Big\},
\end{equation}
for $\omega\geq 1$, which rescales the postulated probability of each configuration by a
factor of at most $\omega$ in either direction. Like ours, and unlike their marginal
sensitivity set, this set constrains the law of the configurations.

\textbf{Results.} Three features of this exercise are worth commenting on. 
First, the sign of the estimate is fragile:
the sensitivity bounds contain zero from a budget of 
$\widehat{\delta}_{\mathsf{b}}=0.19$ onwards, both without and with controls. In other
words, a mechanism other than the postulated one governing roughly a fifth of
the shock draws suffices for the estimated employment effect to lose its sign.
Second, the sensitivity bounds include the
unadjusted least squares estimate at
$\delta=0.45$ without controls, and already at $\delta=0.25$ with
them. Put differently, the gap between the OLS and the recentered IV estimates, 
which is what the recentering is designed to remove, can be reproduced by assignment
processes that are not far from the postulated one. Third, the
\textsf{ST26} bounds are nested in ours even though they need not be in general.
In this application, the nesting reflects the difference in the least (best) favorable
assignment
processes attaining each type of bound: the distributions attaining ours concentrate mass
on a
few configurations, which the multiplicative constraint in \eqref{eq:st-set} rules out by
capping
the reweighting of every configuration at the factor $\omega$.

\section{Conclusion}
    \label{sec:conclusion}

This paper developed a framework for global sensitivity analysis in which the
sensitivity parameter $\delta$ is the share of the population allowed to violate a
distributional assumption. The sensitivity bounds are the values of 
finite-dimensional programs, and at
$\delta=1$ they coincide with the exact nonparametric sharp bounds. We proposed plug-in
estimators of the bounds and
established their consistency, limit distribution, and the validity of a bootstrap
procedure. We also
proposed summary measures  that separate misspecification from
sampling uncertainty and make studies comparable. Applications to attrition in randomized
experiments, counterfactuals in structural models, and formula instruments illustrate
the questions these tools answer: which of two experiments to trust, how much of a
study's fragility is due to sampling uncertainty, and what a second round would have to
spend to certify more robustness.

\singlespacing\footnotesize
\begin{spacing}{0.8}
\setlength{\bibsep}{4pt plus .3ex}
\bibliographystyle{aer}
\bibliography{references}
\end{spacing}

\end{document}

\title{\vspace{-0.5cm}\textbf{Supplemental Appendix for \\
``Sensitivity Analysis in Population Shares''}}
\makeatletter\let\Title\@title\makeatother

\author{Filippo Palomba \\ \normalsize Department of Economics, Princeton
University}

\date{}

\pagenumbering{roman}
\pagestyle{empty}
\maketitle
\thispagestyle{empty}
\vspace{-0.5cm}
\noindent
\begin{minipage}[t]{0.47\textwidth}
\setlength\epigraphwidth{\linewidth}
\epigraph{\raggedright\textit{``Statistical inferences are based only in part upon the
observations. An equally important base is formed by prior assumptions about the
underlying situation.''}}{Peter J. Huber, \textit{Robust Statistics}}
\end{minipage}\hfill
\begin{minipage}[t]{0.47\textwidth}
\setlength\epigraphwidth{\linewidth}
\epigraph{\raggedright\textit{``A fragile inference is not worth taking seriously.
All scientific disciplines routinely subject their inferences to studies of fragility.
Why should economics be
different?''}}{Edward E. Leamer, \textit{Sensitivity Analyses Would Help}}
\end{minipage}

\begin{abstract}
\noindent 
This supplement collects all proofs, additional results, and technical
details on convex
duality, feasibility, estimation, inference, summary measures, and
applications of sensitivity analysis under total variation and contamination
neighborhoods.
Section~\ref{sec:introduction} fixes notation and sets up the statistical
framework used
throughout. Section~\ref{sec:convex-duality}
develops the core duality theory and introduces the total variation and
contamination neighborhoods.
Section~\ref{sec:feasibility-inner} establishes existence of solutions.
Section~\ref{sec:estimation-inference} develops the large sample theory.
Section~\ref{sec: summary measures} introduces the summary measures for sensitivity
analysis and derives their large sample properties.
Section~\ref{sec: applications of tv sensitivity} works through a
sequence of relevant applications.
Section~\ref{sec: empirical applications details} collects two additional
empirical applications on attrition.
Section~\ref{sec: proofs} contains all the proofs.
\end{abstract}

\tocloftpagestyle{empty}
\tableofcontents

\singlespacing
\setlength\abovedisplayskip{3pt}
\setlength\belowdisplayskip{3pt}
\setlength\abovedisplayshortskip{2pt}
\setlength\belowdisplayshortskip{2pt}
\setlength{\textfloatsep}{8pt}
\setlength{\intextsep}{8pt}
\setlength{\floatsep}{6pt}
\setlength\parindent{0pt}           % No indent in the whole document
\setlength{\parskip}{3pt}           % Space between paragraphs

\clearpage
\pagestyle{plain}
\pagenumbering{arabic}
\section{Introduction}\label{sec:introduction}
This section introduces the notation and sets up the statistical framework we
rely on throughout this supplement.

\subsection{Notation}

In this supplement, we let $n$ denote the sample size. 

\textbf{Sets and spaces.} We write $\R$ for the real line,
$\overline{\R}=\R\cup\{-\infty,+\infty\}$ for the extended reals, $\R_+$ and
$\R_{++}$ for
the nonnegative and positive reals, and $\N,\Z$ for the naturals and integers.
For $J\in\N$, we write $[J]:=\{1,\ldots,J\}$.
For a
closed set $\calZ$ of a Euclidean space, we write $\mathscr{F}$ for its Borel
$\sigma$-algebra, $\Delta(\calZ)$ for the probability measures on $(\calZ,\mathscr{F})$,
and $\calM_{+}(\calZ)$ for the cone of finite nonnegative measures on
$(\calZ,\mathscr{F})$. We write $C(\calZ;\R^k)$ for the space of continuous
functions from $\calZ$ to $\R^k$, and $C(\calZ):=C(\calZ;\R)$. For $1\leq p<\infty$,
we denote by $L_p(\mu)$ the space of
measurable $f$ with $\int_{\calZ}|f|^p\,\de\mu<\infty$. A subset $K$ of a real vector
space is a cone if $\lambda x\in K$ whenever $x\in K$ and $\lambda>0$,
so that $K$ is a union of half lines emanating from the origin.

\textbf{Linear algebra.} $0_k$ and $1_k$ are the $k$-dimensional vectors of zeros
and ones, $e_j$ is the $j$-th standard basis vector of a Euclidean space
whose dimension is clear from the context, and $^\top$ denotes transposition.
We write $|\cdot|$ or $\|\cdot\|_2$ for
the Euclidean norm and $\|\cdot\|_\infty$ for the sup-norm, and set
$[x]_+=\max\{x,0\}$ and $[x]_-=\max\{-x,0\}$.
For a point $x\in\R^k$ and a radius $r>0$, we write
\[
\calB_r(x):=\{u\in\R^k:\|u-x\|_2<r\}
\qquad\text{and}\qquad
\overline{\calB}_r(x):=\{u\in\R^k:\|u-x\|_2\leq r\}
\]
for the open and the closed ball of radius $r$ centered at $x$, respectively. A
neighborhood of $x$ is any set that contains $\calB_r(x)$ for some $r>0$, so that a
property holds on a neighborhood of $x$ if and only if it holds on $\calB_r(x)$ for
some $r>0$. We use the same notation on any space endowed with a norm or a
seminorm other than the Euclidean one, in which case we name the norm or
seminorm that replaces $\|\cdot\|_2$ in the two displayed sets.
For a point $a$ and a nonempty set $B$ of a Euclidean space,
$d(a,B):=\inf_{b\in B}\|a-b\|_2$ is the distance from $a$ to $B$, and for two nonempty
compact sets $A,B$ of a Euclidean space
\[
d_{\mathrm H}(A,B):=\max\Bigl\{\sup_{a\in A}d(a,B),\ \sup_{b\in B}d(b,A)\Bigr\}
\]
is their Hausdorff distance.

\textbf{Probability and measure.} Expectation and variance under a
distribution $P$ are $\E_P[\cdot]$ and $\V_P[\cdot]$, respectively, and
$\independent$ denotes independence. Without a subscript, $\PP$ and $\E$ denote
probability and expectation with respect to the joint distribution of the random variables
they involve. For two measures, $P\ll Q$ means $P$ is
absolutely
continuous with respect to $Q$, with Radon--Nikodym derivative $\de P/\de Q$,
while
$P\perp Q$ means they are mutually singular, i.e.\ there is a measurable
$\calD\in\mathscr{F}$ with $P(\calD)=0$ and $Q(\calZ\setminus \calD)=0$.
The total variation distance between $P,Q\in\Delta(\calZ)$ is
$\TV(P\,\|\,Q):=\sup_{\calA\in\mathscr{F}}|P(\calA)-Q(\calA)|$, which lies in $[0,1]$.
A point
$z\in\calZ$ is an atom of $Q\in\Delta(\calZ)$ if $Q(\{z\})>0$, and the support
$\supp(\mu)$ of a Borel measure $\mu$ on $\calZ$ is the smallest closed set of full
$\mu$-measure. For an event or condition $A$, $\indic\{A\}$ equals $1$ if $A$
holds and $0$ otherwise. We write $X\overset{d}{=}Y$ when $X$ and $Y$
have the same distribution, and $X\mid A\overset{d}{=}Y\mid B$ when the conditional
distribution of $X$ given $A$ equals that of $Y$ given $B$. For a distribution $P$ on a
closed set $\calY\subseteq\R$, $F_P(y):=P(\calY\cap(-\infty,y])$, $y\in\R$, is its
distribution function and $F_P^{-1}(u):=\inf\{y\in\calY:F_P(y)\geq u\}$, $u\in(0,1)$, is
its quantile function. We write $\Be(p)$ for the Bernoulli distribution with success
probability $p\in[0,1]$ and $\mathsf{Unif}([a,b])$ for the uniform distribution on an
interval $[a,b]\subset\R$ with $a<b$. Finally, $\delta^{\mathtt{Dirac}}_x$ denotes
the Dirac measure at a point
$x$.

\textbf{Convex analysis.} For a set $C\subseteq\R^k$ we write
$\mathrm{co}(C)$ for its convex hull, the set of finite convex
combinations of points of $C$, and $\operatorname{aff}(C)$ for its affine hull.
The
interior, closure, and relative interior of $C$ are
\[
\interior(C):=\{x:\exists\,\varepsilon>0,\
\calB_\varepsilon(x)\subseteq C\},\qquad
\cl(C):=\bigcap\{F\supseteq C:\ F\ \text{closed}\},
\]
\[
\rint(C):=\{x\in\operatorname{aff}(C):\exists\,\varepsilon>0,\
\calB_\varepsilon(x)\cap\operatorname{aff}(C)\subseteq C\}.
\]
For $f:\R^k\to\overline{\R}$, the effective domain and the subdifferential are
\[
\dom(f):=\{z\in\R^k:f(z)<\infty\},\qquad
\partial f(z):=\{v\in\R^k: f(z')\geq f(z)+v^\top(z'-z)\ \forall z'\in\R^k\},
\]
and $f$ is proper if $\dom(f)\neq\emptyset$ and $f(z)>-\infty$ for all
$z\in\R^k$. For a function $f$ defined on an interval of $\R$, we write
$\partial_+f(x)$ and $\partial_-f(x)$ for its right and left derivatives at $x$, that is,
the limits of $\{f(x+h)-f(x)\}/h$ as $h\downarrow0$ and as $h\uparrow0$, respectively,
whenever they exist. Finally, $\cl f$ is the closure (lower semicontinuous
hull) of $f$, and $f$ is closed if $\cl f=f$, which for a proper convex function
is
equivalent to lower semicontinuity, and $P\text{-}\esssup$ denotes the essential
supremum with respect to the distribution $P$.

\textbf{Asymptotic statements.} We write $x_n\to x$ for limits, including
one-sided ones, and use $x_n\downarrow x$ or $x_n\uparrow x$ only when the direction of
approach matters for the argument. For positive sequences
$\{a_n\},\{b_n\}$ we write $a_n=O(b_n)$ if $a_n\leq Mb_n$ for some $M$ and all
large $n$, and $a_n=o(b_n)$ if $a_n/b_n\to 0$. For sequences of random variables we
write $A_n=\op(B_n)$ and $A_n=\Op(B_n)$ for the usual stochastic-order analogs. We
denote convergence in probability by $\cprob$ and convergence in distribution by
$\rightsquigarrow$, and write $\normal(a,B)$ for a (possibly multivariate) Gaussian with
mean $a$ and variance--covariance $B$.

\textbf{Empirical process theory.} For a probability measure $P$ on
$\calZ$ and a $P$-integrable $f:\calZ\to\R$ we write $Pf:=\int_{\calZ}f\,\de P$, so
that $\E_{P}[f]=Pf$. For a class $\calH$ of functions and two measures $P,P'$ we
write $\|P-P'\|_{\calH}:=\sup_{f\in\calH}|Pf-P'f|$, and $\ell^{\infty}(\calH)$ denotes the
space of bounded real functions on $\calH$ equipped with the supremum norm. Given
observations $Z_1,\ldots,Z_n$ drawn i.i.d.\ from $P$, we write
$\widehat P_n:=n^{-1}\sum_{i=1}^n\delta^{\mathtt{Dirac}}_{Z_i}$ for their empirical
measure and $\mathbb G_{n}:=\sqrt n(\widehat P_{n}-P)$ for the empirical process, viewed
as the random map $f\mapsto\mathbb G_{n}f=\sqrt n(\widehat P_{n}f-Pf)$. Following
Section~2.1 in \citet{vandervaart1996WeakConvergenceEmpirical}, a class
$\calH\subseteq L_{2}(P)$ of measurable functions with $\sup_{f\in\calH}|f(z)-Pf|<\infty$
for every $z\in\calZ$ is called $P$-Donsker if $\mathbb G_{n}\rightsquigarrow\mathbb G$ in
$\ell^{\infty}(\calH)$, where $\rightsquigarrow$ is weak convergence in the sense of
Hoffmann-J{\o}rgensen and the limit $\mathbb G$ is a mean zero Gaussian process with
covariance $\E[\mathbb G f\,\mathbb G f']=P(ff')-Pf\,Pf'$, whose sample paths lie almost
surely in $UC(\calH,e_{P})$, the space of bounded functions on $\calH$ uniformly
continuous for the intrinsic semimetric $e_{P}(f,f')^{2}:=\V_{P}(f-f')$. The class $\calH$
is $P$-Glivenko--Cantelli if $\|\widehat P_{n}-P\|_{\calH}\to0$ in outer probability.

\subsection{Setup}\label{subsec: setup}
Let $\Theta\subseteq\R^{\dimt}$,
$\dimt\in\N$, be the parameter space and $\calZ\subseteq\R^{\dimz}$,
$\dimz\in\N$, a closed
set collecting the possible realizations of a random vector $Z\in\calZ$.
We equip $\calZ$ with its Borel $\sigma$-algebra
$\mathscr{F}$. The econometrician postulates that $Z$ is distributed according to a
baseline distribution $\Pbase{}\in\calP(\calZ)$. The class
$\calP(\calZ)\subseteq\Delta(\calZ)$, formally defined below in \eqref{eq: admissible P}
once the estimand and moment function have been introduced, collects the Borel
probability distributions under which the absolute value of the estimand's
integrand and the norm of the moment function are integrable. We impose this
restriction so that the expectations, moment restrictions, optimization
problems, and associated quantities introduced below are well defined.

The baseline
$\Pbase{}$ is the model one would use absent any concern for misspecification, and
the analysis
asks how a conclusion drawn from $\Pbase{}$ moves as the distribution is allowed to
depart from
it while still satisfying the model restriction.

\begin{rem}[Interpretation of the baseline]\label{rem: baseline pseudo true}
We denote the baseline distribution by $\Pbase$ rather than by $P_0$, which we reserve
for the distribution that generates the data. The starred notation
emphasizes that the baseline plays a
role analogous to that of a pseudo-true value in maximum likelihood theory:
it is the model the econometrician posits, and it need not coincide with the
truth. In the RCTs with attrition example of
Section~\ref{subsec: rct attrition}, for instance, the joint distribution
of
potential outcomes, response indicators, and treatment assignment is $P_0$, and the
data are drawn from the distribution of observables that $P_0$ induces. The baseline is
$\PT$, the identified
distribution of the outcome among treated respondents, and the point
identifying assumption that observations are missing completely at random is
the equality $P_{\mathsf{C}}=\PT$ between the unidentified outcome distribution
for the compliers and the baseline. The sensitivity exercise is accordingly
conducted around $\PT$, which incorporates the econometrician's point
identifying assumption, and not around $P_0$.
In the counterfactuals in structural models example of
Section~\ref{subsec: structural models} the baseline
is instead the distribution that the econometrician imposes on the
latent idiosyncratic utility shocks.
\end{rem}

\textbf{Estimand.} The quantity of interest is a scalar \textit{estimand},
obtained by
integrating an $\mathscr{F}$-measurable \textit{integrand}
$g:\calZ\times\Theta\to\R$
against the distribution of the data,
\[
\E_P[g(Z;\theta)]=\int_{\calZ} g(z;\theta)\,\de P(z).
\]
A leading case is a counterfactual or policy relevant quantity reported
by the analyst,
with $\theta$ indexing the structural parameter on which $g$ depends. 

\textbf{Model restrictions.} The model restricts the parameter through moment
conditions. Let $m:\calZ\times\Theta\to\R^{\dimm}$ be an $\mathscr{F}$-measurable
\textit{moment function}. We partition the moment function into four blocks,
\phantomsection\label{remark: M box structure}
\[
m(\cdot\,;\theta)
=\bigl(m_1(\cdot\,;\theta)^\top,m_2(\cdot\,;\theta)^\top,
m_3(\cdot\,;\theta)^\top,m_4(\cdot\,;\theta)^\top\bigr)^\top,
\]
where $m_i:\calZ\times\Theta\to\R^{\mathsf d_i}$,
$\mathsf d_i\in\{0,1,\ldots\}$, and
$\dimm=\sum_{i=1}^4\mathsf d_i$. The first and third blocks contain moment
inequalities, whereas the second and fourth contain moment equalities. We
collect the right-hand sides of the first two blocks in
$\rho=(\rho_1^\top,\rho_2^\top)^\top\in\R^{\dime}$, where
$\rho_1\in\R^{\mathsf d_1}$ and $\rho_2\in\R^{\mathsf d_2}$ are unknown vectors
of target moments, and
$\dime=\mathsf d_1+\mathsf d_2$. We also normalize the known right-hand sides of
the last two blocks to zero. The moment constraint set is thus
\[
\calM(\rho)
=
\Big(\bigtimes_{j=1}^{\mathsf d_1}(-\infty,\rho_{1,j}]\Big)
\times\{\rho_2\}
\times(-\infty,0]^{\mathsf d_3}
\times\{0\}^{\mathsf d_4}
\subseteq\R^{\dimm}.
\]
Thus,
\[
\E_P[m(Z;\theta)]\in\calM(\rho)
\quad\Longleftrightarrow\quad
\begin{pmatrix}
\E_P[m_1(Z;\theta)]\leq\rho_1\\
\E_P[m_2(Z;\theta)]=\rho_2\\
\E_P[m_3(Z;\theta)]\leq0\\
\E_P[m_4(Z;\theta)]=0
\end{pmatrix},
\]
where vector inequalities are interpreted componentwise. We also define
$\rhopad :=(\rho_1^\top,\rho_2^\top,
0_{\mathsf{d}_3}^\top,0_{\mathsf{d}_4}^\top)^\top \in\R^{\dimm}$.
We write $\rho_0\in\R^{\dime}$ for the population value of the target vector
and $\rhopad_0:=(\rho_0^\top, 0_{\mathsf{d}_3}^\top,0_{\mathsf{d}_4}^\top)^\top$.
Let $\lambda\in\R^{\dimm}$ denote the multiplier attached to the moment
restriction in the dual programs of Section~\ref{sec:convex-duality}. Partitioning it
conformably as
$\lambda=(\lambda_1^\top,\lambda_2^\top,
\lambda_3^\top,\lambda_4^\top)^\top$, define
$\lambda_{12}:=(\lambda_1^\top,\lambda_2^\top)^\top$ and
\[
\Rlam
:=\R_+^{\mathsf d_1}\times\R^{\mathsf d_2}
\times\R_+^{\mathsf d_3}\times\R^{\mathsf d_4}.
\]
For a set $C\subseteq\R^k$, we write $\iota_C$ for its convex indicator, which equals $0$
on $C$ and $+\infty$ off $C$, and $\iota^*_C(y):=\sup_{x\in C}y^\top x$ for its support
function, which is the convex conjugate of $\iota_C$ in the sense of
Definition~\ref{defi: convex conjugate}. The support function of the moment constraint set
is
\[
\iota^*_{\calM(\rho)}(\lambda)
=
\begin{cases}
\lambda_{12}^\top\rho,&\lambda\in\Rlam,\\
+\infty,&\lambda\notin\Rlam.
\end{cases}
\]
For the $\theta$ under consideration, we write
$\calM_{+,m}(\calZ)\subseteq\calM_+(\calZ)$ for the subcone of the measures $\nu$ with
finite moment $\int_{\calZ}\|m(z;\theta)\|\,\de\nu(z)<\infty$ and we omit the dependence
on $\theta$ for brevity.
When no moment function is present, all four block dimensions are zero. We
then use the conventions $m\equiv 0$,
$\calM(\rho)=\R^0$, and $\Rlam=\R^0$, so that the moment restriction is
vacuous and its support function contribution is zero. In this case we also suppress
$m,\rho,\lambda$, and $\lambda_{12}$ from the notation, interpret their
inner products as zero, and set $\calM_{+,m}(\calZ)=\calM_+(\calZ)$.

For each $\theta\in\Theta$, and suppressing the dependence on $\theta$ in the
notation, we formally define the class of admissible distributions as
\begin{align}\label{eq: admissible P}
    \calP(\calZ)\equiv \calP_\theta(\calZ)
:=
\left\{
P\in\Delta(\calZ):
\E_P[|g(Z;\theta)|]<\infty
\ \text{and}\
\E_P[\|m(Z;\theta)\|]<\infty
\right\}.
\end{align}
The first integrability requirement makes the estimand finite, while the
second makes the moment vector finite. Thus, every expectation entering the
moment restrictions and sensitivity bounds is well defined. Throughout
this supplement, unless explicitly stated otherwise, we maintain without
further mention the measurability and integrability conditions needed for all displayed
objects, optimization problems, and invoked results, and omit their routine verification
from the proofs. We say that a distribution $P{\in\calP(\calZ)}$ and a parameter $\theta$
are compatible with the model when $\E_P[m(Z;\theta)]\in\calM(\rho)$.

\textbf{Robustness set.} Sensitivity to deviations from $\Pbase{}$ is gauged by means
of a $\phi$-divergence. 
Let $\phi:\R\to\overline{\R}$ be an entropy function (see Definition~\ref{defi:
entropy function}) 
and $D_\phi(\cdot\,\|\,\Pbase{})$ the associated divergence from the baseline (see
Definition~\ref{defi: generalized phi divergence}). For a sensitivity parameter
$\delta\in\R_+$, also referred to as the divergence budget, the \textit{robustness
set} collects
the distributions that are both compatible with the model and within $\delta$ of
the
baseline in the divergence induced by $\phi$:
\begin{equation}\label{eq: robustness set definition}
    {\calP_\phi(\theta;{\rho},\Pbase{},\delta):=\left\{P\in{\calP(\calZ)}:\
    \E_P[m(Z;\theta)]\in\calM(\rho),\
    D_\phi(P\,\|\,\Pbase{})\leq\delta\right\}.
    }
\end{equation}
For an entropy function $\psi$ and a distribution $P\in\Delta(\calZ)$, the moment vectors
attainable within a budget $\delta$ of $P$ form the set
\begin{equation}\label{eq: attainable moment set}
\Gamma_\psi(\theta;P,\delta):=\bigl\{\E_Q[m(Z;\theta)]:\
Q\in\calP(\calZ),\ D_\psi(Q\,\|\,P)\leq\delta\bigr\}\subseteq\R^{\dimm},
\end{equation}
and $\Gamma^{\ll}_\psi(\theta;P,\delta)$ denotes its subset obtained by
restricting to the $Q\ll P$. We drop $\psi$ when $\psi=\phi$, $P$ when $P=\Pbase{}$, and
$\theta$ when it is held fixed, so that, for instance, $\Gamma(\delta)$ collects the
moment vectors attainable within divergence $\delta$ of the baseline. When no moment
function is present, we set all these sets equal to $\{0\}$.
We do not require the baseline distribution itself to satisfy the moment
restrictions at any parameter value. Under total variation, that is, for
$\phi=\phi_{\TV}$, where $\phi_{\TV}(s)=\frac12|s-1|$ for $s\geq0$ is the
total variation entropy of Example~\ref{ex: finite TV 1}, the smallest divergence
needed to reconcile the baseline with the
model is $\inf_{\theta\in\Theta}\mathsf{D}(\theta;\rho,\Pbase{})$, where $\mathsf{D}$ is
the minimum-divergence function of \eqref{eq:mindiv-primal} with
$\mathbb{U}=[0,+\infty)$ (see Remark~\ref{rem: mindiv specification}). When the baseline
satisfies the moment restrictions at some parameter value, this quantity is zero, and
the usual sensitivity analysis centered at a model-compatible baseline is obtained as a
special case.

\textbf{Sensitivity bounds.} Letting both $P$ and the compatible $\theta$ vary,
the
range of the estimand over the robustness set is delimited by its largest and
smallest
attainable values, namely the lower and upper sensitivity bounds, formally
defined as
\begin{align}
    \inf_{\theta\in\Theta}\ \inf_{P\in{\calP_\phi(\theta;{\rho},\Pbase{},\delta)}}
    \E_P[g(Z;\theta)]
    \quad\text{and}\quad
    \sup_{\theta\in\Theta}\ \sup_{P\in{\calP_\phi(\theta;{\rho},\Pbase{},\delta)}}
    \E_P[g(Z;\theta)],
    \label{eq: general optimization problem}
\end{align}
with the understanding that when the robustness set is empty for a given
$\theta\in\Theta$, 
the corresponding infimum or supremum is taken to be $+\infty$ or $-\infty$,
respectively, so that, should the set be empty for every $\theta\in\Theta$, the upper
bound is $-\infty$ and the lower $+\infty$.
For each fixed $\theta\in\Theta$, the inner optimization over
{$P\in\calP_\phi(\theta;{\rho},\Pbase{},\delta)$} in \eqref{eq: general optimization
problem} is the
\textit{inner problem}, whose optimal value is the most extreme value the
estimand attains
over the robustness set at that $\theta$, while the outer optimization over
$\theta\in\Theta$
is the \textit{outer problem}. We use these names for the two programs in
\eqref{eq: general optimization problem} throughout. Furthermore, the lower bound can be
obtained from the upper
bound by replacing the integrand $g$ with $-g$, since
$\inf_{\theta,P}\E_P[g(Z;\theta)]=-\sup_{\theta,P}\E_P[-g(Z;\theta)]$. We
accordingly
develop the theory for the upper bound and recover the lower bound by this
substitution.

\textbf{Addition of ``external'' nuisance parameters.} In some applications, the integrand
and the moment function carry an additional nuisance parameter
$\varsigma\in\calW$, with $\calW$ a metric space, not necessarily
finite-dimensional, so that the primitives of the problem become
$g(z;\theta,\varsigma)$ and $m(z;\theta,\varsigma)$. Leading examples of such
a parameter are a law of motion for state variables, a transition or a
conditional choice probability in dynamic discrete choice models, or a mixing
proportion such as the share of compliers $\pi$ in the RCTs-with-attrition
application of Section~\ref{subsec: rct attrition}. Unlike $\theta$, the
nuisance parameter is not optimized over. Rather, it has a true value
$\varsigma_0\in\calW$,
which is consistently estimable from the data in a preliminary step,
separately from the rest of the model, hence the terminology ``external'' nuisance
parameter. The model restrictions are accordingly required to hold at the true value,
{$\E_P[m(Z;\theta,\varsigma_0)]\in\calM(\rho)$}, the estimand becomes
$\E_P[g(Z;\theta,\varsigma_0)]$, and the optimization in \eqref{eq: general
optimization problem} continues to run over $(\theta,P)$ only, with the
nuisance held fixed at $\varsigma_0$.
{The definition of $\calP(\calZ)$ is then read with
$g(\cdot;\theta,\varsigma_0)$ and $m(\cdot;\theta,\varsigma_0)$ in place of
$g(\cdot;\theta)$ and $m(\cdot;\theta)$, with $\varsigma_0$ held fixed.}
In estimation, $\varsigma_0$ is
replaced with a consistent first-step estimator $\widehat{\varsigma}$.
With the nuisance parameter held fixed, the results of
Sections~\ref{sec:convex-duality} and~\ref{sec:feasibility-inner} are
unaffected and it would suffice to read $(\theta,\varsigma_0)$ in place of $\theta$ 
in all of the assumptions and results. The large sample theory of
Section~\ref{sec:estimation-inference} is a different matter: there, the
first-step estimator $\widehat{\varsigma}$ carries sampling uncertainty of
its own, which
propagates to the estimated bounds, as we discuss in
Remark~\ref{rem:nuisance-EB}.

\subsection{Correspondence with the Paper}
\label{subsec: correspondence with the paper}

The paper considers two robustness sets, the total variation robustness set
$\calP_{\TV}(\theta;\delta)$ and the contamination robustness
set $\calP_{\mathsf{C}}(\theta;\delta)$, defined in
\textup{(3.4)} and \textup{(3.5)} of the main text. This
supplement
instead works, in some of its parts, with a generic entropy function $\phi$. 
We now outline how the
general results developed here cover both robustness sets of the
main text, and
Table~\ref{tab: main sa correspondence} then shows where to locate in this
supplement the results reported in the paper.

\textbf{Both sets are constrained total variation neighborhoods.} Each of the two
sets is defined through a mixture restriction. In the total variation robustness
set, the baseline is a mixture of the candidate distribution and of the
distribution of a second group, $\Pbase{}=\kappa P+(1-\kappa)R$ with
$\kappa\in(0,1)$, and the two groups are within total variation distance $\delta$
of each other. In the contamination robustness set, the candidate distribution is
a mixture of the baseline and of an unrestricted contaminating distribution,
$P=(1-\delta)\Pbase{}+\delta R$.
Lemma~\ref{lemma: contamination as constrained tv} shows that each of these sets
is a total variation neighborhood of the baseline intersected with a one-sided
dominance constraint. The first one collects the distributions within total
variation distance $(1-\kappa)\delta$ of the baseline that satisfy
$\kappa P\leq\Pbase{}$, whereas the second one collects the distributions within
total variation distance $\delta$ of the baseline that satisfy
$P\geq(1-\delta)\Pbase{}$, in which case the total variation constraint is
redundant. The formal mappings are in \eqref{eq: paper tv set mapping} and
\eqref{eq: paper contamination set mapping}.

\textbf{The mixture restrictions are absorbed into the entropy.} By
Lemma~\ref{lemma: mixture density bounds}, each dominance constraint is a
pointwise bound on the density of the candidate distribution with respect to the
baseline. Such
a bound is the convex indicator of an interval evaluated at the density, and it can
therefore be added to the total variation entropy.
Proposition~\ref{prop: induced entropy} shows that the resulting functions,
\[
\overline{\phi}_{\TV}:=\phi_\TV+\iota_{[0,1/\kappa]}
\qquad\text{and}\qquad
\underline{\phi}_{\TV}:=\phi_\TV+\iota_{[1-\delta,\infty)},
\]
are again entropy functions in the sense of
Definition~\ref{defi: entropy function}, and that the two robustness sets of the
main text are the divergence neighborhoods they induce, that is,
\[
\calP_{\TV}(\theta;\delta)
=\calP_{\overline{\phi}_{\TV}}\bigl(\theta;\rho_0,\Pbase{},(1-\kappa)\delta\bigr)
\qquad\text{and}\qquad
\calP_{\mathsf{C}}(\theta;\delta)
=\calP_{\underline{\phi}_{\TV}}(\theta;\rho_0,\Pbase{},\delta).
\]

\textbf{Duality.} Because both sets are robustness sets of the form
\eqref{eq: robustness set definition}, the supremum and the infimum of the
estimand over them are dualized by the general theory of
Section~\ref{sec:convex-duality}, with no argument specific to either set. Weak
duality is Theorem~\ref{theorem: weak duality}, strong duality is
Theorem~\ref{thm: strong duality}, and
Theorem~\ref{theorem: strong duality with mixture constraint} specializes them to
the two induced entropies, its first part covering the total variation robustness
set and its second part the contamination one. The two duals differ in one respect. 
With the total
variation robustness set, the inner dual is an unconstrained convex program. In the
contamination robustness set the contaminating distribution may place mass
outside the support of the baseline at the finite cost
$\underline{\phi}_{\TV}^\infty(1)=1/2$, and the dual problem has an additional
constraint. Section~\ref{subsec: ignoring dual constraint} provides 
conditions under which this constraint can be deleted without changing
the value of the dual (Theorem~\ref{thm: ignoring dual constraint} and
Proposition~\ref{cor: ac unconstrained dual}).

\textbf{Existence.} Section~\ref{sec:feasibility-inner} establishes existence of
dual solutions and of solutions to the outer problem for generic entropies, under
Assumption~\ref{assump: dual primitives} when the recession is finite and under
Assumption~\ref{assump: dual primitives superlinear} when it is not. The total
variation robustness set falls under case (b) of
Assumption~\ref{assump: dps-entropy} with $\bar{x}=1/\kappa$, because
$\overline{\phi}_{\TV}$ is finite at both endpoints of its domain. Therefore
Theorem~\ref{thm: existence of dual solutions},
Lemma~\ref{lemma: outer value regularity}, and
Theorem~\ref{thm: existence full problem} apply to it, as recorded in
Remark~\ref{remark: feasibility mixture}. The contamination robustness set is not
covered by these statements, because $\underline{\phi}_{\TV}(0)=+\infty$ violates
Assumption~\ref{assump: dp-entropy}. Existence of dual solutions for this set is
instead established in Lemma~\ref{lemma: family duality-iii}, which shows that
the set of dual minimizers is nonempty, compact, and convex for every member of a
family of robustness sets containing both sets of the main text.
Its proof is direct
and relies only on a weak duality inequality with explicit slack and on a
constraint qualification, under the compact support and continuity conditions
maintained in the large sample theory.

\textbf{Inference.} Section~\ref{sec:estimation-inference} does not treat the two
robustness sets separately. It works with the family of
Section~\ref{subsec:ls-setup}, indexed by an interval $\mathbb{U}$ of admissible
values for the density of the candidate distribution with respect to the baseline,
of which the total variation robustness set is the member $\mathbb{U}=[0,1/\kappa]$
and the contamination robustness set is the member
$\mathbb{U}=[1-\delta,+\infty)$. By Lemma~\ref{lemma: family entropy}, every member
of the family is a divergence neighborhood whose dual integrand is a maximum of
finitely many affine functions, and by Lemma~\ref{lemma: family duality} strong
duality and existence of dual solutions hold for every member under a single
constraint qualification. The large sample arguments use only these common
features, together with the fact that the pointwise dual constraint, present for
the contamination robustness set and absent for the total variation one, depends
neither on the estimated moments nor on the baseline. Hence consistency
(Theorem~\ref{thm:cons-EB}), the asymptotic distribution
(Theorem~\ref{thm:HD-EB}), and the validity of the bootstrap
(Theorem~\ref{thm:boot-EB}) hold for both robustness sets at once. The first two
results are obtained under Assumption~\ref{assump: est-EB}, and the third one
additionally requires Assumption~\ref{assump: M-joint-EB}. The case of a known
baseline distribution is discussed in Remark~\ref{rem:known-baseline}.

\begin{table}[t]
\centering
\caption{Correspondence between the formal results in the paper and this supplement.}
\label{tab: main sa correspondence}
\begin{tabular}{l l >{\raggedright\arraybackslash}m{6.4cm}}
\toprule\toprule
Paper & Supplement & Supporting results \\
\midrule
Theorem~1
& Theorem~\ref{theorem: strong duality with mixture constraint}
& Theorems~\ref{theorem: weak duality} and~\ref{thm: strong duality},
  Lemma~\ref{lemma: mixture density bounds},
  Proposition~\ref{prop: induced entropy},
  Example~\ref{ex: tv truncated duals},
  Lemmas~\ref{lemma: contamination as constrained tv} and~\ref{lemma: family entropy} \\
\addlinespace
Proposition~1
& Theorem~\ref{thm: ignoring dual constraint} and
  Proposition~\ref{cor: ac unconstrained dual}
& Proposition~\ref{prop: induced entropy},
  Corollary~\ref{lemma: implicit regularization of entropy},
  Example~\ref{ex: tv truncated duals},
  Lemmas~\ref{lemma: contamination as constrained tv} and~\ref{lemma: family entropy} \\
\addlinespace
Theorem~2
& Theorem~\ref{thm:cons-EB}
& Lemmas~\ref{lem:donsker-EB} and~\ref{lem:aux-EB} \\
\addlinespace
Theorem~3
& Theorem~\ref{thm:HD-EB}
& Lemmas~\ref{lem:donsker-EB}, \ref{lem:aux-EB}
  and~\ref{lemma: family duality} \\
\addlinespace
Theorem~4
& Theorem~\ref{thm: endpoint sharp bounds}
& Lemmas~\ref{lemma: divergence lebesgue decomposition}
  and~\ref{lemma: mixture density bounds}, Proposition~\ref{prop: induced entropy} \\
\bottomrule\bottomrule
\end{tabular}
\end{table}

\clearpage
\section{Convex Duality}\label{sec:convex-duality}
This section introduces some key ideas from convex analysis on which the rest of this
supplement relies. Section~\ref{subsec:primer-convex} first defines the
concepts of convex conjugate, recession function, perspective function, and entropy
function, and then proceeds to discuss different definitions of $\phi$-divergences that
have been used in the
literature. Section~\ref{subsec:convex-dual} leverages this framework to recast
the
infinite-dimensional inner problem in~\eqref{eq: general optimization problem}
as a
tractable finite-dimensional program. Section~\ref{subsec:strong-duality}
identifies the
conditions under which strong duality holds. Section~\ref{subsec:mixture-neighborhoods}
introduces an additional mixture constraint to the
problem in \eqref{eq: general optimization problem}. Both robustness sets
of the main text are of this form, and
the formulation recurs in all applications of
Section~\ref{sec: applications of tv sensitivity}.
Section~\ref{subsec: ignoring dual constraint} provides primitive conditions under which
the pointwise constraint appearing in the duals derived
here can be ignored, that is, deleted without affecting the value of the
dual problem. Finally, Section~\ref{subsec: contamination tv} closes
the section by specializing to total variation and recasting the contamination
neighborhoods as constrained total variation neighborhoods.

\subsection{A Primer in Convex Analysis}\label{subsec:primer-convex}
We start by defining some fundamental concepts from convex analysis.

\begin{defi}[Convex conjugate] \label{defi: convex conjugate}
    The convex conjugate of a function $f:\R^k\to\overline{\R}$, $k\in\N$, is the
    function
    $f^*:\R^k\to\overline{\R}$
    \[f^*(y) = \sup_{z\in\R^k}\{y^\top z - f(z)\}.\]
    The function $f^{**}$, termed the convex biconjugate of $f$, is the convex
    conjugate
    of $f^*$, that is
    \[f^{**}(z) = \sup_{y\in\R^k}\{y^\top z - f^*(y)\}.\]
\end{defi}
The convex conjugate is a way of representing the epigraph of a function via
supporting
hyperplanes. For each slope $y$, it records the maximal gap between the
linear function $z\mapsto y^\top z$ and $f$.

\begin{defi}[Recession function,
{\citet[][Theorem~8.5]{rockafellar1970ConvexAnalysis}}]
\label{defi: recession function}
    Let $f:\R^k\to\overline{\R}$, $k\in\N$, be proper, convex, and lower
    semicontinuous. The recession function of $f$ is the function
    $f^\infty:\R^k\to\overline{\R}$
    defined as
    \[f^\infty(z) = \lim_{\alpha\to\infty}\frac{1}{\alpha}\{f(z_0+\alpha z) -
    f(z_0)\},\]
    where $z_0\in\R^k$ is any point with $f(z_0)<\infty$ (under the stated
    conditions on $f$, the limit exists in $\overline{\R}$ and does not
    depend on the choice of $z_0$).
\end{defi}

The recession function $f^\infty$ captures the asymptotic behavior (as
$\alpha\to\infty$)
of the function $f$ along the direction $z$.
It is a positively homogeneous function, i.e., $f^\infty(\alpha z) = \alpha
f^\infty(z)$
for all $\alpha>0$ and $z\in\R^k$.
The recession function is also convex and lower semicontinuous.

\begin{defi}[Perspective function] \label{defi: perspective function}
    Let $f:\R^k\to\overline{\R}$, $k\in\N$, be proper, convex, and lower
    semicontinuous. The perspective of $f$ is the function
    $f^\pi:\R^k\times\R\to\overline{\R}$
    defined as
    \[f^\pi(z,t) =
    \begin{cases}
    tf(z/t),\quad &t>0,\\
    f^\infty(z), \quad &t=0,\\
    +\infty, \quad &t<0.
    \end{cases}\] 
\end{defi}

The perspective function $f^\pi$ is useful to define the concept of generalized
$\phi$-divergence we rely on throughout this supplement (see Definition
\ref{defi: generalized
phi divergence}).
It is itself convex and lower semicontinuous (see
Lemma~\ref{lemma: entropy function proper cvx lsc} for the perspectives of an entropy
function and of its conjugate).

Before introducing the generalized $\phi$-divergence, we need to define the
concept of
entropy function.
\begin{defi}[Entropy function]\label{defi: entropy function}
An entropy function $\phi:\R\to\overline{\R}$ is a nonnegative lower
semicontinuous convex function
with $\phi(1)=0$ and $\phi(s) = +\infty$ for all $s<0$.
\end{defi}
Table \ref{tab: phi conjugate recession} below collects the entropy functions $\phi$,
their convex conjugates $\phi^*$, and their recession constants $\phi^\infty(1)$ for
popular $\phi$-divergences. 
\begin{table}[htbp]
\centering
\caption{Entropy functions $\phi$, their convex conjugates $\phi^\ast$, and
recession constants $\phi^\infty(1)$.}
\label{tab: phi conjugate recession}
\begin{tabular}{lccc}
\toprule\toprule
\textit{Divergence} & $\phi(t)$ & $\phi^*(t)$ & $\phi^\infty(1)$ \\
\midrule

Total variation
& $\tfrac12 |t-1|$
& $\begin{cases} -\tfrac12, & t<-\tfrac12\\ t, & -\tfrac12\le t\le\tfrac12\\ +\infty, & t>\tfrac12\end{cases}$
& $\tfrac12$ \\[0.6em]

Kullback--Leibler
& $t\log t - t + 1$
& $e^{t}-1$
& $\infty$ \\[0.6em]

Likelihood
& $-\log t + t - 1$
& $\begin{cases} -\log(1-t), & t<1\\ +\infty, & t\ge1\end{cases}$
& $1$ \\[0.6em]

Pearson $\chi^2$
& $(t-1)^2$
& $\begin{cases} -1, & t<-2\\ t+\tfrac{t^2}{4}, & t\ge-2\end{cases}$
& $\infty$ \\[0.6em]

Neyman $\chi^2$
& $\dfrac{(t-1)^2}{t}$
& $\begin{cases} 2-2\sqrt{1-t}, & t\le1\\ +\infty, & t>1\end{cases}$
& $1$ \\[0.6em]

Squared Hellinger
& $(1-\sqrt{t})^2$
& $\begin{cases} \dfrac{t}{1-t}, & t<1\\ +\infty, & t\ge1\end{cases}$
& $1$ \\[0.6em]

Jensen--Shannon
& $t\log\!\left(\dfrac{2t}{t+1}\right) + \log\!\left(\dfrac{2}{t+1}\right)$
& $\begin{cases} -\log(2-e^{t}), & t<\log2\\ +\infty, & t\ge\log2\end{cases}$
& $\log 2$ \\
\bottomrule\bottomrule
\end{tabular}

\end{table}

Now we have all the ingredients to define the family of generalized
$\phi$-divergences
\citep*[][Definition~2.5]{csiszar1963InformationstheoretischeUngleichungUnd,
morimoto1963MarkovProcessesTheorem,
ali1966GeneralClassCoefficients,
kuhn2025DistributionallyRobustOptimization}.
\begin{defi}[Generalized $\phi$-divergence] \label{defi: generalized phi
divergence}
    The generalized $\phi$-divergence of $P\in\calM_+(\calZ)$ with respect to
    $\Pbase{}\in\Delta(\calZ)$ is given by
    \[D_\phi(P\,\|\,\Pbase{}) = \int_\calZ \phi^\pi\left(\frac{\de P}{\de
    \varrho}(z),\frac{\de
    \Pbase{}}{\de \varrho}(z)\right)\de \varrho(z),\]
    where $\phi$ is an entropy function and $\varrho\in\calM_+(\calZ)$ is any
    measure dominating both $P$ and $\Pbase{}$, that is, $P,\Pbase{}\ll\varrho$.
\end{defi}
\begin{rem}\label{remark: unique representation of generalized phi divergence}
    The divergence does not depend on the choice of the dominating
    measure $\varrho$, see Proposition~2.6 in
    \cite*{kuhn2025DistributionallyRobustOptimization} or Theorem~7.26 in
    \citet{polyanskiy2025InformationTheoryCoding}.
\end{rem}

The next lemma expresses the divergence of a measure through its
Lebesgue decomposition with respect to the baseline. We use it repeatedly below.
\begin{lemma}[Lebesgue decomposition of the divergence]
\label{lemma: divergence lebesgue decomposition}
    Let $\phi$ be an entropy function in the sense of Definition~\ref{defi:
    entropy function}, $\Pbase{}\in\Delta(\calZ)$, and $\nu\in\calM_+(\calZ)$ with
    Lebesgue decomposition $\nu=w\Pbase{}+\nu^\perp$, where $\nu^\perp\perp \Pbase{}$
    and $w:=\de\nu_{\mathtt{ac}}/\de \Pbase{}$ is the Radon--Nikodym derivative of
    the absolutely continuous part of $\nu$ with respect to $\Pbase{}$. Then
    \[D_\phi(\nu\,\|\,\Pbase{}) = \int_\calZ\phi(w(z))\,\de \Pbase{}(z) +
    \phi^\infty(1)\,\nu^\perp(\calZ),\]
    where the product $\phi^\infty(1)\,\nu^\perp(\calZ)$ evaluates to $0$ if
    $\nu^\perp(\calZ)=0$.
\end{lemma}
\begin{flushright}
[\hyperref[proof: divergence lebesgue decomposition]{Proof}]
\end{flushright}

Definition~\ref{defi: generalized phi divergence} above is very general and
allows for
entropy functions that do not grow superlinearly (e.g., the total variation
entropy). We say that $\phi$ grows superlinearly if
$\lim_{s\to\infty}\phi(s)/s=+\infty$, or equivalently $\phi^\infty(1)=+\infty$. When
the entropy function grows superlinearly, the divergence
$D_\phi(P\,\|\,\Pbase{})$
is finite only if $P\ll \Pbase{}$.
Indeed, let $P=w\Pbase{}+P^\perp$ be the Lebesgue decomposition of
$P\in\calM_+(\calZ)$ with respect to $\Pbase{}$. By
Lemma~\ref{lemma: divergence lebesgue decomposition},
\[
D_\phi(P\,\|\,\Pbase{})
=\int_\calZ\phi(w(z))\,\de\Pbase{}(z)+\phi^\infty(1)\,P^\perp(\calZ),
\]
where the integral is nonnegative because $\phi\geq0$, and the second term equals
$+\infty$ as soon as $P^\perp(\calZ)>0$, because $\phi^\infty(1)=+\infty$. Hence a finite
divergence forces $P^\perp=0$, that is $P\ll\Pbase{}$, in which case
$w=\de P/\de\Pbase{}$ and the second term vanishes.
This also yields the following more common
definition for $\phi$-divergences \citep*{hansen2001RobustControlModel,
maccheroni2006AmbiguityAversionRobustness}:
\begin{align}
    D_\phi(P\,\|\,\Pbase{}) = 
    \begin{cases}
    \displaystyle \int \phi\!\left(\frac{\de P}{\de \Pbase{}}(z)\right)\,\de \Pbase{}(z),
    & P\ll
    \Pbase{},\\[0.6em]
    +\infty, & \text{otherwise.}
    \end{cases}
\label{eq: less general phi divergence definition}
\end{align}
Most importantly for our purposes, the definition in
\eqref{eq: less general phi divergence definition} assigns a finite value
only when $P\ll\Pbase{}$. Therefore, any neighborhood of finite radius under this
definition does not include distributions that are not absolutely continuous
with respect to the baseline, which would be possible under divergences with
finite recession, such as total variation or squared Hellinger distance.

\begin{ex}[Absolute continuity and total variation]\label{ex: finite TV 1}
The total variation entropy is $\phi_{\TV}(s)=\frac{1}{2}|s-1|$ for $s\geq 0$
and
$\phi_{\TV}(s)=+\infty$ for $s<0$. Its recession function is
$\phi_{\TV}^{\infty}(s) =
s/2$ for $s\geq 0$ and $\phi_{\TV}^{\infty}(s)=+\infty$ for $s<0$.
Let $Q,R\in\Delta(\calZ)$ be mutually singular, let $a\in[0,1]$, and define
$P_a=(1-a)Q+aR$. Taking $\varrho=Q+R$ as the dominating measure, mutual
singularity gives
\[
\frac{\de Q}{\de\varrho}=1,\qquad
\frac{\de P_a}{\de\varrho}=1-a
\quad Q\text{-almost surely},
\]
whereas
\[
\frac{\de Q}{\de\varrho}=0,\qquad
\frac{\de P_a}{\de\varrho}=a
\quad R\text{-almost surely}.
\]
Consequently,
\[
D_{\phi_{\TV}}(P_a\,\|\,Q)
=\phi_{\TV}(1-a)+\phi_{\TV}^{\infty}(a)
=\frac{a}{2}+\frac{a}{2}=a.
\]
Thus, for every $a>0$, the divergence is finite although $P_a\not\ll Q$.
Taking $Q=\delta^{\mathtt{Dirac}}_0$, $R=\delta^{\mathtt{Dirac}}_1$, and
$a=1$ gives an example with discrete variables, whereas taking
$Q=\mathsf{Unif}([0,1])$, $R=\delta^{\mathtt{Dirac}}_2$, and
$a\in(0,1)$ gives an example with continuous variables.
\end{ex}

We also record a property of total variation that is used repeatedly below.
\begin{lemma}[Total variation as a $\phi$-divergence]\label{lem:tv_posneg}
Let $P,Q\in\Delta(\calZ)$. If $\varrho\in\calM_+(\calZ)$
dominates $P$ and $Q$, with $p=\de P/\de\varrho$ and
$q=\de Q/\de\varrho$, then
\begin{equation*}
\TV(P\,\|\,Q)=\frac{1}{2}\int|p-q|\,\de\varrho=D_{\phi_{\TV}}(P\,\|\,Q),
\end{equation*}
and the supremum defining $\TV(P\,\|\,Q)$ is attained at $\{p\geq q\}$.
\end{lemma}
\begin{flushright}
[\hyperref[proof:lem:tv_posneg]{Proof}]
\end{flushright}

The next series of lemmas establishes some useful properties of the convex
conjugate, the recession %
function, and perspective
function of an entropy function $\phi$.
\begin{lemma}[Properties of entropy functions]\label{lemma: entropy function
proper cvx lsc}
    Let $\phi$ be an entropy function in the sense of Definition \ref{defi:
    entropy
    function}. Then $\phi^*$, $\phi^\pi$, and $(\phi^*)^\pi$ are proper, convex, and lower
    semicontinuous.
\end{lemma}
\begin{flushright}
[\hyperref[proof: entropy function proper cvx lsc]{Proof}]
\end{flushright}

\begin{lemma}[A lower bound for the perspective of the conjugate
function]\label{lemma: perspective larger than argument}
    Let $\phi$ be an entropy function in the sense of Definition \ref{defi:
    entropy
    function}. Then, for all $z\in\R$, $t\in\R$, and $\check{x}\in\dom(\phi)$,
    \[(\phi^*)^\pi(z,t)\geq z \check{x} - t\phi(\check{x}).\]
    Thus, for $\check{x}=1$, it follows that
    \[(\phi^*)^\pi(z,t)\geq z.\]
\end{lemma}
\begin{flushright}
[\hyperref[proof: perspective larger than argument]{Proof}]
\end{flushright}

\begin{lemma}[{\citet[][Lemma~4.11]{kuhn2025DistributionallyRobustOptimization}}]
    If $\phi$ is an entropy function in the sense of Definition \ref{defi:
    entropy
    function}, $t \in \R$, $\beta \in \R_{+}$, and $\eta \in \R_{++}$, then
    \begin{align*}
    \sup _{\alpha \in \R} \{t \alpha-\eta \phi^\pi(\alpha, \beta)\}=    
    \begin{cases}
    \beta \eta \phi^*(t / \eta) & \text { if } \beta>0, \\
    \eta \iota_{\cl(\dom(\phi^*))}(t / \eta) & \text { if } \beta=0.
    \end{cases}
    \end{align*}
    \label{lemma: lemma 4.11 kuhn}
\end{lemma}

\begin{lemma}[{\citet[][Lemma~4.12]{kuhn2025DistributionallyRobustOptimization}}]
    If $\phi$ is an entropy function in the sense of Definition \ref{defi:
    entropy
    function}, then
    \[
    \cl(\dom(\phi^*))=(-\infty,\phi^\infty(1)],
    \]
    where $(-\infty,+\infty]$ is understood as $\R$.
    \label{lemma: lemma 4.12 kuhn}
\end{lemma}

The next {corollary} shows that if the domain of the conjugate of an entropy function is
bounded
from above, then the recession function of $\phi$ must be such that
$\phi^\infty(1)<\infty$. Furthermore, the threshold beyond which the
conjugate $\phi^*$ is infinite is precisely $\phi^\infty(1)$. The equivalence plays
a central role in Section~\ref{subsec: ignoring dual constraint}, where it
identifies the implicit constraint that the conjugate embeds in the objective function of
the dual of the problem in \eqref{eq: general optimization problem}.

\begin{cor}[Implicit regularization of the entropy]\label{lemma: implicit
regularization of entropy}
Let $\phi$ be an entropy function in the sense of Definition
\ref{defi:
entropy function}. Then, the following are equivalent:
\begin{enumerate}[label=(\arabic*)]
    \item $\phi^{\infty}(1)<\infty$;
    \item $\sup \dom(\phi^*) < +\infty$;
    \item there exists $M\in\R$ such that $\phi^*(t)=+\infty$ for all $t>M$.
\end{enumerate}
Moreover, if any of the above holds, then
$\sup\dom(\phi^*)=\phi^{\infty}(1)$, so that in (3) we can pick
$M=\phi^{\infty}(1)$.
\end{cor}
\begin{flushright}
[\hyperref[proof: implicit regularization of entropy]{Proof}]
\end{flushright}

For any function $f:\R^k\to\overline\R$, the definition of the convex
conjugate implies the biconjugate inequality $f^{**}\leq f$.
This inequality will suffice for weak duality. Strong duality instead relies
on the following optimization result.
A convex set is polyhedral if it is the intersection of finitely many closed
half spaces. A proper convex function is polyhedral if its epigraph
$\operatorname{epi}f:=\{(x,t)\in\R^{k+1}:t\geq f(x)\}$ is a
polyhedral convex set. In particular, the indicator $\iota_C$ of a
polyhedral convex set $C$ is a polyhedral convex function.

\begin{lemma}[Fenchel's duality theorem,
{\citet[][Theorem~31.1]{rockafellar1970ConvexAnalysis}}]
\label{lemma: fenchel duality polyhedral}
Let $f,q:\R^k\to\overline\R$ be proper convex functions. If
\[
\rint(\dom(f))\cap\rint(\dom(q))\neq\emptyset,
\]
then
\[
\inf_{x\in\R^k}\{f(x)+q(x)\}
=
\sup_{x^*\in\R^k}\{-f^*(x^*)-q^*(-x^*)\}.
\]
If $q$ is polyhedral, the same conclusion holds under the weaker condition
\[
\rint(\dom(f))\cap\dom(q)\neq\emptyset.
\]
Whenever the common value is finite, the supremum is attained.
\end{lemma}

\subsection{Weak Duality}\label{subsec:convex-dual}
Throughout this and the next two subsections we suppress the dependence on
$\theta$ and on the
fixed baseline $\Pbase{}$ to reduce notational clutter, abbreviating
{$\calP_\phi(\theta;{\rho},\Pbase{},\delta)$
as $\calP_\phi({\rho},\delta)$.} Accordingly, every result below should be
interpreted as holding for a fixed
$\theta\in\Theta$ and the given baseline $\Pbase{}$.
The inner problem over {$\calP_\phi({\rho},\delta)$} in \eqref{eq: general
optimization problem}
is an infinite-dimensional optimization problem that we show can be recast as a
finite-dimensional one.
The estimand
\[\E_{P}[g(Z)] = \int_\calZ g(z)\de P(z)\]
is linear in the underlying measure $P\in\calM_+(\calZ)$, and the form of
the resulting dual depends
only on the entropy function $\phi$.
Heuristically, the next (weak) duality result presented in Theorem~\ref{theorem:
weak duality} is obtained by first
recasting the inner problem as a linear program over the space of signed
measures, then
obtaining the conjugate and biconjugate of the primal value function, and
concluding {with the biconjugate inequality stated
above.}

\begin{thm}[Weak duality]\label{theorem: weak duality}
    Let $\phi$ be an entropy function in the sense of
    Definition~\ref{defi: entropy function}, and let $g$ be integrable with
    respect to every
    $\nu\in\calM_{+,m}(\calZ)$ satisfying
    $D_\phi(\nu\,\|\,\Pbase{})<\infty$. Then
    \begin{align*}
        \sup_{P\in\calP_\phi({\rho},\delta)}\E_P[g(Z)]
        \leq
        \begin{cases}
        \inf\limits_{\substack{(\zeta,\eta,\lambda)\in\\
        \R\times\R_+\times\Rlam}}
        \lambda_{12}^\top\rho+\zeta+\eta\delta
        +\E_{\Pbase{}}\!\left[
        (\phi^*)^\pi\big(g(Z)-\lambda^\top m(Z)-\zeta,\eta\big)
        \right],\\
        \text{s.t.}\quad
        \zeta+\eta\phi^\infty(1)
        \geq
        \sup\limits_{z\in\calZ}
        \{g(z)-\lambda^\top m(z)\}.
        \end{cases}
    \end{align*}
    The product $\eta\phi^\infty(1)$ is understood to equal $+\infty$ when
    $\eta=0$ and $\phi^\infty(1)=+\infty$.
\end{thm}
\begin{flushright}
[\hyperref[proof: weak duality]{Proof}]
\end{flushright}

Two special cases are worth noting: (i) if $\phi^\infty(1)=+\infty$,
the pointwise constraint
is vacuous and the dual is an unconstrained problem; (ii) if instead
attention is restricted to
distributions that are absolutely continuous with respect to the baseline $\Pbase{}$,
the constraint is already implied by the perspective
 and can be dropped from the dual problem. For a formal explanation, see Remark
 \ref{rem: only abs cont distributions} below. A detailed explanation of when the
pointwise constraint can be deleted without affecting the value of the dual
is given in Theorem~\ref{thm: ignoring dual constraint} of
Section~\ref{subsec: ignoring dual constraint}.

\begin{rem}[Only absolutely continuous
distributions]\label{rem: only abs cont distributions}
Consider only distributions that are absolutely continuous with respect to the
baseline $\Pbase{}$. In this case, the robustness set becomes
\[
{\calP^{\ll}_\phi(\rho,\delta)}
:=
\left\{
P\in{\calP(\calZ)}:
{P\ll\Pbase{}},\
{\E_P[m(Z)]\in\calM(\rho)},\
D_\phi(P\,\|\,\Pbase{})\leq\delta
\right\}.
\]
For every such $P$ the singular part of $P$ with respect to $\Pbase{}$
vanishes, so that
Lemma~\ref{lemma: divergence lebesgue decomposition} gives
$D_\phi(P\,\|\,\Pbase{})=\int\phi(\de P/\de\Pbase{})\,\de\Pbase{}$. It is the singular
part that generates the pointwise constraint in the dual problem. In the proof of
Theorem~\ref{theorem: weak duality}, the constraint stems from the second integral
in~\eqref{eq: average + supremum}, which runs over the set where the density of
$\Pbase{}$ with respect to the dominating measure $\varrho$ vanishes. For measures that
are absolutely continuous with respect to $\Pbase{}$ we may take $\varrho=\Pbase{}$, so
that this set is empty and the integral disappears.
\end{rem}

\subsection{Strong Duality}\label{subsec:strong-duality}
For a divergence budget $b\in\R_+$, recall from \eqref{eq: attainable moment set} the set
$\Gamma(b)\equiv\Gamma_\psi(\theta;\Pbase{},b)$ of moments attainable when $m(z)$ is
integrated against distributions within divergence $b$ of $\Pbase{}$. It is convex,
because the set of distributions with $D_\phi(P\,\|\,\Pbase{})\leq b$ is convex, in view
of the convexity of $D_\phi(\cdot\,\|\,\Pbase{})$, and $\Gamma(b)$ is its image under
the linear map $P\mapsto\E_P[m(Z)]$.
It is nonempty because $\E_{\Pbase{}}[m(Z)]\in\Gamma(b)$ as
$D_\phi(\Pbase{}\,\|\,\Pbase{})=0\leq b$. The value function underlying the inner problem
is
\begin{align}
    h(a,b,x) = \inf_{\nu\in\calM_{+,m}(\calZ)} \left\{-\int g(z)\de \nu(z): \int
    m(z)\de
    \nu(z) = x,\: \int \de \nu(z) = a,\: D_\phi(\nu\,\|\,\Pbase{})\leq b\right\},
    \label{aux eq: function domain}
\end{align}
which is jointly convex in $(a,b,x)$ by the argument used in the proof of
Theorem~\ref{theorem: weak duality} to establish the convexity of the value function.
{Let
\[
K_\rho:=\{1\}\times\{\delta\}\times\calM(\rho).
\]
With this notation, the primal value function equals $-\inf_{y\in K_\rho}h(y)$. Because
$K_\rho$ is polyhedral, so is its indicator $\iota_{K_\rho}$. Therefore,
once we show that  $K_\rho\cap\rint(\dom(h))\neq\emptyset$ and $h$ is proper,
we can apply Lemma~\ref{lemma: fenchel duality polyhedral} to $h+\iota_{K_\rho}$. The next
lemma shows that both requirements hold.}

\begin{lemma}[Requirements for Fenchel duality]\label{lemma: fenchel requirements}
    Let $\delta>0$, let $\phi$ be an entropy function in the sense of
    Definition~\ref{defi: entropy function} that is continuous at $1$, and
    suppose $\int_{\calZ}|g(z)|\,\de\nu(z)<\infty$ for every
    $\nu\in\calM_{+}(\calZ)$ with $D_\phi(\nu\,\|\,\Pbase{})<\infty$. Then, the
    value function $h$ in \eqref{aux eq: function domain} is proper and
    \[
    (1,\delta,\om)\in\rint(\dom(h))
    \qquad\text{for every }\om\in\rint(\Gamma(\delta)).
    \]
\end{lemma}
\begin{flushright}
[\hyperref[proof: fenchel requirements]{Proof}]
\end{flushright}

We can now state the main result of this section.

\begin{thm}[Strong duality]\label{thm: strong duality}
    Let $\delta>0$, let $\phi$ be an entropy function in the sense of
    Definition~\ref{defi: entropy function} that is continuous at $1$, and
    let $\int_{\calZ}|g(z)|\,\de\nu(z)<\infty$ for every
    $\nu\in\calM_{+}(\calZ)$ satisfying
    $D_\phi(\nu\,\|\,\Pbase{})<\infty$, and suppose
    \[
    \calM(\rho)\cap\rint(\Gamma(\delta))\neq\emptyset.
    \]
    Then
    \begin{align*}
        \sup_{P\in\calP_\phi({\rho},\delta)}\E_P[g(Z)]
        =
        \begin{cases}
        \inf\limits_{\substack{(\zeta,\eta,\lambda)\in\\
        \R\times\R_+\times\Rlam}}
        \lambda_{12}^\top\rho+\zeta+\eta\delta
        +\E_{\Pbase{}}\Big[(\phi^*)^\pi
        \big(g(Z)-\lambda^\top m(Z)-\zeta,\eta\big)\Big],\\
        \text{s.t.}\quad
        \zeta+\eta\phi^\infty(1)
        \geq
        \sup\limits_{z\in\calZ}
        \{g(z)-\lambda^\top m(z)\}.
        \end{cases}
    \end{align*}
    The product $\eta\phi^\infty(1)$ is understood to equal $+\infty$ when
    $\eta=0$ and $\phi^\infty(1)=+\infty$.
\end{thm}
\begin{flushright}
[\hyperref[proof: strong duality]{Proof}]
\end{flushright}

\begin{rem}[Meaning of the compatibility condition]
The set $\Gamma(\delta)$ collects all moment vectors that distributions within
the divergence budget $\delta$ from the baseline $\Pbase{}$ can attain. 
Thus, the condition
\[
\calM(\rho)\cap\rint(\Gamma(\delta))\neq\emptyset
\]
requires at least one of the moment vectors compatible with the model constraints to lie
in the
relative interior of the attainable moment set. Put differently, small perturbations,
in every direction along which the attainable moments vary, of some vector in
$\calM(\rho)$ must remain attainable for the same divergence budget. Note that the
interior is relative
to the affine hull of $\Gamma(\delta)$, so the condition accommodates equality
restrictions and lower-dimensional moment sets. This condition is useful
because it is precisely the qualification under which
Lemma~\ref{lemma: fenchel duality polyhedral} applies, so that strong
duality holds without requiring
$\calM(\rho)$ to be compact or full dimensional. When no moment restrictions
are present, the condition is automatically satisfied under the convention introduced in
Section~\ref{subsec: setup}.
\end{rem}

We close this part by making the dual in
Theorem~\ref{thm: strong duality} explicit for the total variation
distance. Throughout this illustrative example, we
implicitly maintain the regularity and compatibility conditions of that
theorem.

\begin{ex}[Total variation I]\label{ex: tv conjugate perspective}
Recall that in Example~\ref{ex: finite TV 1} we
recorded that the recession function of the total variation entropy $\phi_{\TV}$
is 
$\phi_\TV^\infty(s)=s/2$ if $s\geq0$ and $+\infty$ otherwise, 
whereas the convex conjugate of $\phi_{\TV}$ is
\[
\phi_\TV^*(y)=\sup_{s\geq0}\Big\{ys-\tfrac12|s-1|\Big\}=
\begin{cases}
-\tfrac12, & y<-\tfrac12,\\
y, & -\tfrac12\leq y\leq\tfrac12,\\
+\infty, & y>\tfrac12,
\end{cases}
\]
so that $\dom(\phi_\TV^*)=(-\infty,\tfrac12]$, in agreement with
Lemma~\ref{lemma: lemma 4.12 kuhn}. Taking the
perspective of $\phi_\TV^*$ in the sense of Definition~\ref{defi: perspective
function}
gives
\[
(\phi_\TV^*)^\pi(z,t)=
\begin{cases}
-\tfrac{t}{2}, & t>0,\ z<-\tfrac{t}{2},\\
z, & t>0,\ -\tfrac{t}{2}\leq z\leq\tfrac{t}{2},\\
+\infty, & t>0,\ z>\tfrac{t}{2},\\
0, & t=0,\ z\leq0,\\
+\infty, & t=0,\ z>0,\\
+\infty, & t<0.
\end{cases}
\]
Because $\phi_\TV^\infty(1)=\tfrac12<\infty$, the pointwise constraint of
Theorem~\ref{thm: strong duality} is not vacuous and reads
$\zeta+\tfrac{\eta}{2}\geq\sup_{z\in\calZ}\{g(z)-\lambda^\top m(z)\}$.
Theorem~\ref{thm: ignoring dual constraint} below provides conditions
under which it can
be deleted without affecting the value of the dual. Under this constraint
the perspective integrand collapses to $\max\{g(Z)-\lambda^\top m(Z)-\zeta,\,-\eta/2\}$. 
Recalling from Section~\ref{subsec: setup} that
$\iota^*_{\calM(\rho)}(\lambda)=\lambda_{12}^\top\rho$ for $\lambda\in\Rlam$, the dual of
Theorem~\ref{thm: strong duality} becomes
\[
\sup_{P\in\calP_{\phi_\TV}({\rho},\delta)}\E_P[g(Z)]=
\begin{cases}
\inf\limits_{\substack{(\zeta, \eta, \lambda)\in\\\R\times\R_+\times\Rlam}}
\lambda_{12}^\top\rho+\zeta+\eta\delta
+\E_{\Pbase{}}\big[\max\{g(Z)-\lambda^\top
m(Z)-\zeta,\,-\tfrac{\eta}{2}\}\big],\\[4pt]
\text{s.t.}\quad \zeta+\tfrac{\eta}{2}\geq\sup_{z\in\calZ}\{g(z)-\lambda^\top
m(z)\}.
\end{cases}
\]
\end{ex}

\subsection{Duality Theory Under Mixture Constraints}
\label{subsec:mixture-neighborhoods}
We now modify the geometry of
{$\calP_\phi({\rho},\delta)$} by adding a new constraint in the form of a mixture
restriction. We do so in two complementary ways, each dictated by a distinct and
economically relevant class of applications. In particular, for some fixed
$\kappa\in(0,1)$, we define
\begin{equation}\label{eq: P mix baseline mixture}
{\calP^{\ll}_{\mathsf{mix}}(\rho,\delta)}
:=
\Big\{
P \in {\calP(\calZ)} \ : \
D_\phi(P\,\|\,\Pbase{})\leq \delta,\
{\E_P[m(Z)]\in \calM(\rho)},\
\exists\, R\in \Delta(\calZ) :
\Pbase{}=\kappa P+(1-\kappa)R
\Big\}
\end{equation}
and
\begin{equation}\label{eq: P mix baseline component}
{{\calP}_{\mathsf{mix}}(\rho,\delta)}
:=
\left\{
P \in {\calP(\calZ)}
\ : \
D_\phi(P\,\|\,\Pbase{})\leq \delta,\;
{\E_P[m(Z)]\in \calM(\rho)},\;
\exists\, R\in\Delta(\calZ) : P=\kappa \Pbase{}+(1-\kappa)R
\right\}.
\end{equation}
Our notation suppresses the dependence of the robustness sets on
$\kappa$. This is 
a deliberate choice because we interpret $\kappa$ as a fixed nuisance parameter which
will 
be treated as known in the large sample theory of Section~\ref{sec:estimation-inference}.
In
some special cases we also quantify the uncertainty in estimating $\kappa$ (see
Section~\ref{subsec: rct attrition} and Proposition \ref{prop: attrition inference}).

In the first robustness set, the baseline distribution must be a mixture
of the candidate distribution $P$ and some other distribution $R$.
This is the geometry of randomized
experiments with attrition (Section~\ref{subsec: rct attrition}), contaminated
data
(Section~\ref{subsec: contaminated data}), and departures from randomization
(Section~\ref{subsec: randomization}). In the second
robustness set, the roles are reversed
and the baseline is instead a component of the mixture.
This is the geometry of missing data (Section~\ref{subsec: missing data}). As
Proposition~\ref{prop: induced entropy} below establishes,
each mixture restriction amounts to replacing $\phi$ by a modified
entropy, after which both duals are obtained by the dual construction underlying
Theorem~\ref{theorem: weak duality} together with the strong duality of
Theorem~\ref{thm: strong duality}, applied to that modified entropy.

We stress that the two restrictions above, besides altering the geometry
of the robustness set, also change the interpretation of the
sensitivity parameter $\delta$, a point we return to in
Section~\ref{subsec: contamination tv} and, in particular, in
Remark~\ref{rem: different interpretations tv}.

Before deriving the duals under the two mixture restrictions,
we present two auxiliary results that give a different interpretation to the
mixture constraints and
help clarify their implications. The first,
Lemma~\ref{lemma: mixture density bounds}, shows that although each mixture restriction is
phrased through the existence of a mixing distribution $R$, it is in fact equivalent to a
pointwise bound on
the Radon--Nikodym derivative of the candidate distribution with respect to the
baseline distribution.
The two cases are mirror images of one another: when the baseline is itself the
mixture, the Radon--Nikodym derivative is
bounded above by $1/\kappa$, whereas when the baseline is one of the mixing
components, the
density of the absolutely continuous part is bounded below by $\kappa$. We
record both
characterizations together, as parts~(i) and~(ii) below. 

\begin{lemma}[Mixture restrictions as Radon--Nikodym bounds]\label{lemma:
mixture density bounds}
Fix $\kappa\in(0,1)$ and let $P\in\Delta(\calZ)$.

\begin{enumerate}[label={\textbf{(\roman*)}},ref=\thelemma(\roman*),leftmargin=*,itemsep=10pt,topsep=6pt]
\item\label{lemma: mixture density bounds-i} \uline{\textbf{Baseline as the
mixture.}} The following are equivalent:
\begin{enumerate}[label=\textup{(\arabic*)}]
    \item there exists $R\in\Delta(\calZ)$ such that
    \[
    \Pbase{}=\kappa P+(1-\kappa)R;
    \]
    \item $\kappa P \leq \Pbase{}$ as measures, that is,
    \[
    \kappa P(A)\leq \Pbase{}(A)\qquad \forall\, A\in\mathscr{F};
    \]
    \item $P\ll \Pbase{}$ and
    \[
    \frac{\de P}{\de \Pbase{}}\leq \frac1\kappa
    \qquad \Pbase{}\text{-a.s.}
    \]
\end{enumerate}

\item\label{lemma: mixture density bounds-ii} \uline{\textbf{Baseline as a
mixing component.}} The following are equivalent:
\begin{enumerate}[label=\textup{(\arabic*)}]
    \item there exists $R\in\Delta(\calZ)$ such that
    \[
    P=\kappa \Pbase{}+(1-\kappa)R;
    \]
    \item $P\geq \kappa \Pbase{}$ as measures, that is,
    \[
    P(A)\geq \kappa \Pbase{}(A)
    \qquad
    \forall A\in \mathscr{F};
    \]
    \item the density $u$ of the absolutely continuous part of $P$ in its
    Lebesgue decomposition $P=u\Pbase{}+P^\perp$ with respect to $\Pbase{}$,
    $P^\perp\perp\Pbase{}$, satisfies
    \[
    u\geq \kappa
    \qquad
    \Pbase{}\text{-a.s.}
    \]
\end{enumerate}
In particular, if $P\ll \Pbase{}$, then this restriction is equivalent to
\[
\frac{\de P}{\de \Pbase{}}\geq \kappa
\qquad
\Pbase{}\text{-a.s.}
\]
\end{enumerate}
\end{lemma}
\begin{flushright}
[\hyperref[proof: mixture density bounds]{Proof}]
\end{flushright}

The second auxiliary result, Proposition~\ref{prop: induced entropy}, 
builds on the characterization presented in Lemma \ref{lemma: mixture density
bounds} 
to show that the mixture restrictions can be absorbed into the entropy function,
inducing a new modified entropy function. This is because each mixture
restriction is the convex indicator of an interval evaluated at the density, which can be
added to the entropy function. The result allows us to treat the
robustness sets
{$\calP^{\ll}_{\mathsf{mix}}(\rho,\delta)$ and
$\calP_{\mathsf{mix}}({\rho},\delta)$} as standard 
robustness sets of the form defined in \eqref{eq: robustness set definition}
with the original entropy function replaced by a modified one.

\begin{prop}[Induced entropy functions]\label{prop: induced entropy}
Fix $\kappa\in(0,1)$ and $\delta\in\R_+$, and let $\phi$ be an entropy
function in the sense of Definition~\ref{defi: entropy function}.
\begin{enumerate}[label=\textbf{(\roman*)},ref=\thethm(\roman*),leftmargin=*,itemsep=10pt,topsep=6pt]
\item\label{prop: induced entropy-i} \uline{\textbf{Baseline as the mixture.}}
The function
$\overline{\phi}:=\phi+\iota_{[0,1/\kappa]}$ is an entropy function in the sense
of
Definition~\ref{defi: entropy function}, with recession
$\overline{\phi}^\infty(1)=+\infty$. Moreover
\[
{\calP^{\ll}_{\mathsf{mix}}(\rho,\delta)
=\calP_{\overline{\phi}}({\rho},\delta)},
\]
where $\calP_{\overline{\phi}}(\rho,\delta)$ is the robustness set
\eqref{eq: robustness set definition} for the entropy $\overline{\phi}$.

\item\label{prop: induced entropy-ii} \uline{\textbf{Baseline as a mixing
component.}} The
function $\underline{\phi}:=\phi+\iota_{[\kappa,\infty)}$ is an entropy function
in the sense
of Definition~\ref{defi: entropy function}, with recession
$\underline{\phi}^\infty(1)=\phi^\infty(1)$. Moreover
\[
{\calP_{\mathsf{mix}}({\rho},\delta)
=\calP_{\underline{\phi}}({\rho},\delta)},
\]
where $\calP_{\underline{\phi}}(\rho,\delta)$ is the robustness set
\eqref{eq: robustness set definition} for the entropy $\underline{\phi}$.
\end{enumerate}
\end{prop}
\begin{flushright}
[\hyperref[proof: induced entropy]{Proof}]
\end{flushright}

We now present the main result of this subsection, which provides the duals of the
inner problem in
\eqref{eq: general optimization problem} when the robustness set is defined
via either \eqref{eq: P mix baseline mixture} or \eqref{eq: P mix baseline
component}. The proof of the result
follows the steps of that of Theorem~\ref{thm: strong duality}, with the entropy
function replaced by the one identified in Proposition~\ref{prop: induced entropy},
which arises from the mixture restriction. The two geometries differ, however, in the
assumptions needed to guarantee strong duality.
When the baseline distribution is itself the mixture,
the restriction is equivalent to the bound on the Radon--Nikodym derivative
$\de P/\de \Pbase{}\leq 1/\kappa$ (see Lemma~\ref{lemma: mixture density bounds-i}).
As
such, the candidate distributions are forced to be absolutely continuous with
respect to
$\Pbase{}$, the modified entropy $\overline{\phi}$ satisfies
$\overline{\phi}^\infty(1)=+\infty$, and the pointwise
dual constraint in Theorem~\ref{thm: strong duality} is vacuous, as its
left-hand side equals $+\infty$ for every $\eta\geq0$ under the
convention on the product $\eta\,\overline{\phi}^\infty(1)$.
When instead the baseline is a component of the mixture, the restriction only
bounds from below the density $u$ of the absolutely continuous part of $P$,
$u\geq\kappa$ $\Pbase{}$-almost surely (see
Lemma~\ref{lemma: mixture density bounds-ii}), and leaves the singular part of $P$
unrestricted. The resulting modified entropy
$\underline{\phi}$ inherits the
recession of $\phi$, $\underline{\phi}^\infty(1)=\phi^\infty(1)$ (equal to
$1/2$ for the total variation entropy). When this recession is finite, the
constraint in Theorem~\ref{thm: strong duality} is not vacuous because
singular mass has finite cost $\phi^\infty(1)$.

\begin{thm}[Strong duality under mixture constraints]
\label{theorem: strong duality with mixture constraint}
Fix $\kappa\in(0,1)$ and $\delta>0$, let $\phi$ be an entropy function in
the sense of Definition~\ref{defi: entropy function} that is continuous at $1$, and, for
any entropy function $\psi$, let $\Gamma_\psi(\delta)$ be the
attainable moment set of \eqref{eq: attainable moment set}.

\begin{enumerate}[label={\textbf{(\roman*)}},ref=\thethm(\roman*),leftmargin=*,itemsep=10pt,topsep=6pt]
\item\label{theorem: strong duality with mixture constraint-i}
\uline{\textbf{Baseline as the mixture.}}
Let $\overline{\phi}:=\phi+\iota_{[0,1/\kappa]}$ be the induced entropy of
Proposition~\ref{prop: induced entropy-i}, and suppose
\[
\calM(\rho)\cap\rint(\Gamma_{\overline{\phi}}(\delta))
\neq\emptyset.
\]
Then
\[
\sup_{P\in\calP^{\ll}_{\mathsf{mix}}(\rho,\delta)}
\E_P[g(Z)]
=
\inf_{\substack{(\zeta,\eta,\lambda)\in\\
\R\times\R_+\times\Rlam}}
\lambda_{12}^\top\rho+\zeta+\eta\delta
+\E_{\Pbase{}}\!\left[
(\overline{\phi}^*)^\pi
\big(g(Z)-\lambda^\top m(Z)-\zeta,\eta\big)
\right],
\]
where, by definition of $\overline{\phi}$,
\[
\overline{\phi}^*(y)
=
\sup_{0\leq s\leq1/\kappa}\{ys-\phi(s)\}.
\]

\item\label{theorem: strong duality with mixture constraint-ii}
\uline{\textbf{Baseline as a mixing component.}}
Let $\underline{\phi}:=\phi+\iota_{[\kappa,\infty)}$ be the induced entropy
of Proposition~\ref{prop: induced entropy-ii}, and suppose
$\int_{\calZ}|g(z)|\,\de\nu(z)<\infty$ for every
$\nu\in\calM_+(\calZ)$ with
$D_{\underline{\phi}}(\nu\,\|\,\Pbase{})<\infty$ and
\[
\calM(\rho)\cap\rint(\Gamma_{\underline{\phi}}(\delta))
\neq\emptyset.
\]
Then
\[
\sup_{P\in\calP_{\mathsf{mix}}({\rho},\delta)}
\E_P[g(Z)]
=
\begin{cases}
\inf\limits_{\substack{(\zeta,\eta,\lambda)\in\\
\R\times\R_+\times\Rlam}}
\lambda_{12}^\top\rho+\zeta+\eta\delta
+\E_{\Pbase{}}\!\left[
(\underline{\phi}^*)^\pi
\big(g(Z)-\lambda^\top m(Z)-\zeta,\eta\big)
\right],\\
\text{s.t.}\quad
\zeta+\eta\phi^\infty(1)
\geq
\sup\limits_{z\in\calZ}
\{g(z)-\lambda^\top m(z)\},
\end{cases}
\]
where, by definition of $\underline{\phi}$,
\[
\underline{\phi}^*(y)
=
\sup_{s\geq\kappa}\{ys-\phi(s)\}.
\]
The product $\eta\phi^\infty(1)$ is understood to equal $+\infty$ when
$\eta=0$ and $\phi^\infty(1)=+\infty$.
\end{enumerate}
\end{thm}
\begin{flushright}
[\hyperref[proof: strong duality with mixture constraint]{Proof}]
\end{flushright}

Part (i) requires no integrability condition beyond the maintained
$\Pbase{}\in\calP(\calZ)$, whereas part (ii) does. The asymmetry between the
integrability conditions of the two parts
reflects the geometry of the two neighborhoods. In part (i), the domain
of $\overline{\phi}$ is the bounded interval $[0,1/\kappa]$, so that
$\overline{\phi}^\infty(1)=+\infty$. Lemma~\ref{lemma: divergence lebesgue decomposition}
then implies that every $\nu\in\calM_+(\calZ)$ with
$D_{\overline{\phi}}(\nu\,\|\,\Pbase{})<\infty$ has no singular part with respect to
$\Pbase{}$ and a density $\de\nu/\de\Pbase{}\leq1/\kappa$ $\Pbase{}$-almost surely. Hence
$\int_{\calZ}|g(z)|\,\de\nu(z)\leq\kappa^{-1}\E_{\Pbase{}}[|g(Z)|]<\infty$ for every such
$\nu$. In part (ii), the domain of $\underline{\phi}$ is $[\kappa,\infty)$, so the same
lemma bounds the density only from below, and $\nu$ may carry a singular part when
$\phi^\infty(1)<\infty$. Thus no such implication holds.

We conclude this subsection by making the induced entropies
$\underline{\phi}$ and $\overline{\phi}$ explicit for the total variation
entropy.

\begin{ex}[Total variation II]\label{ex: tv truncated duals}
For the total variation entropy $\phi_\TV$ of Example~\ref{ex: finite TV 1},
both modified entropies admit closed form conjugates.
For the upper truncation
$\overline{\phi}_{\TV}=\phi_\TV+\iota_{[0,1/\kappa]}$ of Theorem~\ref{theorem:
strong duality with mixture constraint-i},
\begin{align*}
\overline{\phi}_{\TV}^*(y)=\sup_{0\leq u\leq
1/\kappa}\Big\{yu-\tfrac12|u-1|\Big\}&=
\begin{cases}-\tfrac12,& y<-\tfrac12,\\ y,& -\tfrac12\leq y\leq\tfrac12,\\
\tfrac{y}{\kappa}+\tfrac{\kappa-1}{2\kappa},& y>\tfrac12,\end{cases}\\
(\overline{\phi}_{\TV}^*)^\pi(z,t)&=\begin{cases}\max\Big\{-\tfrac{t}{2},\ z,\
\tfrac{z}{\kappa}-\tfrac{(1-\kappa)\,t}{2\kappa}\Big\},& t\geq0,\\
+\infty,& t<0.\end{cases}
\end{align*}
At $t=0$ the first branch equals $\max\{z,0\}/\kappa$, the recession
function of $\overline{\phi}_{\TV}^*$, and for $t\geq0$ it coincides with the total
variation integrand $\mathsf{r}_{\TV}(z,t)$ in \textup{(4.4)} of the main
text.
For the lower truncation
$\underline{\phi}_{\TV}=\phi_\TV+\iota_{[\kappa,\infty)}$ of
Theorem~\ref{theorem: strong duality with mixture constraint-ii}, since
$\phi_\TV^\infty(1)=\tfrac12$
the pointwise constraint reads
$\zeta+\eta/2\geq\sup_{z\in\calZ}\{g(z)-\lambda^\top m(z)\}$, and
\begin{align*}
\underline{\phi}_{\TV}^*(y)=\sup_{u\geq\kappa}\Big\{yu-\tfrac12|u-1|\Big\}&=
\begin{cases}\kappa y+\tfrac{\kappa-1}{2},& y<-\tfrac12,\\ y,& -\tfrac12\leq
y\leq\tfrac12,\\ +\infty,& y>\tfrac12,\end{cases}\\
(\underline{\phi}_{\TV}^*)^\pi(z,t)&=\begin{cases}\kappa
z+\tfrac{\kappa-1}{2}t,&
t>0,\ z<-t/2,\\ z,& t>0,\ -t/2\leq z\leq t/2,\\ +\infty,& t>0,\ z>t/2,\\ \kappa
z,&
t=0,\ z\leq0,\\ +\infty,& t=0,\ z>0,\\ +\infty,& t<0.\end{cases}
\end{align*}
\end{ex}

\subsection{Ignoring the Pointwise Dual Constraint}
\label{subsec: ignoring dual constraint}
The general dual of Theorems~\ref{theorem: weak duality}
and~\ref{thm: strong duality} carries the pointwise constraint
\begin{equation}\label{eq: pointwise dual constraint}
    \zeta + \eta\,\phi^\infty(1) \;\geq\;
    \sup_{z\in\calZ}\{g(z)-\lambda^\top m(z)\},
\end{equation}
which prices the singular mass that a candidate distribution may place at any
point of $\calZ$. Relocating mass toward a point $z$ gains
$g(z)-\lambda^\top m(z)$ per unit in the Lagrangian, and the multipliers on
the normalization constraint, $\zeta$, and divergence constraint, $\eta$,
must render such relocations unprofitable everywhere on $\calZ$.
As in Theorem~\ref{theorem: weak duality}, here and in the rest of this
subsection the product $\eta\,\phi^\infty(1)$ evaluates to $+\infty$ when $\eta=0$ and
$\phi^\infty(1)=+\infty$.
When the entropy function grows superlinearly, as in
\cite*{christensen2023CounterfactualSensitivityRobustness}, singular mass is
infinitely expensive because $\phi^\infty(1)=+\infty$ and the constraint becomes vacuous.
For entropies with linear
growth, such as $\phi_\TV$, the recession function is such that $\phi^\infty(1)<\infty$,
and thus the constraint is not vacuous. 

Handling \eqref{eq: pointwise dual constraint} in practice can be demanding
from a computational perspective, since it adds one feasibility constraint per
support point.
It is therefore natural to ask when \eqref{eq: pointwise dual constraint} can be ignored, 
that is, deleted from the dual without changing its value. Beyond the
superlinear case, the two results of this subsection provide sufficient conditions for
doing so. Under regularity conditions on the primitives of the problem,
Theorem~\ref{thm: ignoring dual constraint} shows that restricting the
robustness set to distributions that are absolutely continuous with
respect to $\Pbase{}$ leaves the value of the inner primal problem unchanged. In turn,
Proposition~\ref{cor: ac unconstrained dual} shows that this
new restricted inner primal problem admits a dual problem that is identical to the
dual we obtained in Theorem~\ref{theorem: weak duality} but without
\eqref{eq: pointwise dual constraint}.

Before stating the results of this subsection, we introduce two
quantities. The first quantity is the restriction of the robustness set to absolutely
continuous distributions, $\calP^{\ll}_\phi(\rho,\delta)$, defined in
Remark~\ref{rem: only abs cont distributions} as the set of
$P\in\calP_\phi(\rho,\delta)$ with $P\ll\Pbase{}$. The second is the
moment set generated by absolutely continuous distributions,
$\Gamma^{\ll}(b)\subseteq\Gamma(b)$ of \eqref{eq: attainable moment set},
the analog of $\Gamma(b)$ that considers only absolutely continuous
distributions with respect to $\Pbase{}$. The set $\Gamma^{\ll}(b)$ is convex
by the same argument used for $\Gamma(b)$, as mixtures of distributions
that are absolutely continuous with respect to $\Pbase{}$ are absolutely
continuous themselves.

We now define two pairs of objects. On the primal side, for a divergence
budget $\delta\in\R_+$, we let
\begin{align}\label{eq: primal values ac}
    V_{\mathsf{P}}(\delta) &:= \sup_{P\in\calP_\phi(\rho,\delta)}\E_P[g(Z)],
    &
    V^{\ll}_{\mathsf{P}}(\delta) &:=
    \sup_{P\in\calP^{\ll}_\phi(\rho,\delta)}\E_P[g(Z)]
\end{align}
denote the value of the inner problem over the full robustness set and over
its absolutely continuous restriction, respectively. On the dual side, we let
\begin{equation}\label{eq: dual objective ac}
    f(\zeta,\eta,\lambda;\delta)
    :=\lambda_{12}^\top\rho+\zeta+\eta\delta+
    \E_{\Pbase{}}\big[(\phi^*)^\pi\big(g(Z)-\lambda^\top m(Z)-\zeta,\,
    \eta\big)\big]
\end{equation}
be the dual objective of Theorem~\ref{theorem: weak duality} and we let
\begin{equation}\label{eq: dual feasible set ac}
    \calF := \Big\{(\zeta,\eta,\lambda)\in\R\times\R_+\times\Rlam:\;
    \zeta+\eta\,\phi^\infty(1) \geq
    \sup_{z\in\calZ}\{g(z)-\lambda^\top m(z)\}\Big\}
\end{equation}
collect the dual variables that satisfy \eqref{eq: pointwise dual constraint}.
The two dual values of interest in this section are then
\begin{align}\label{eq: dual values ac}
    V_{\mathsf{D}}(\delta) &:= 
    \inf_{(\zeta,\eta,\lambda)\in\calF}f(\zeta,\eta,\lambda;\delta),
    &
    V^{\circ}_{\mathsf{D}}(\delta) &:= 
    \inf_{\substack{(\zeta,\eta,\lambda)\in\\
    \R\times\R_+\times\Rlam}}
    f(\zeta,\eta,\lambda;\delta),
\end{align}
that is, the value of the dual of Theorem~\ref{theorem: weak duality} and the
value of the same program with \eqref{eq: pointwise dual constraint} deleted. In
Section~\ref{sec:feasibility-inner} we reinstate the dependence on the
parameter and write $V_{\mathsf{P}}(\theta;\delta)$ and
$V_{\mathsf{D}}(\theta;\delta)$ for these same two values.

\begin{thm}[Restriction to absolutely continuous distributions]
\label{thm: ignoring dual constraint}
Let $\phi$ be an entropy function in the sense of
Definition~\ref{defi: entropy function} and let $\delta\in\R_+$.
Suppose that at least one of the following two conditions holds:
\begin{enumerate}[label=(\alph*),leftmargin=2em,topsep=2pt,itemsep=2pt]
    \item $\phi^\infty(1)=+\infty$;
    \item $\delta>0$, $\phi$ is continuous at $1$,
    $\int_{\calZ}|g(z)|\,\de\nu(z)<\infty$ for every
    $\nu\in\calM_+(\calZ)$ with
    $D_\phi(\nu\,\|\,\Pbase{})<\infty$,
    \[
    \calM(\rho)\cap\rint(\Gamma^{\ll}(\delta))
    \neq\emptyset,
    \]
    and, for every $\lambda\in\Rlam$,
    \[
    \sup_{z\in\calZ}\{g(z)-\lambda^\top m(z)\}
    =
    \Pbase{}\text{-}\esssup_{z\in\calZ}
    \{g(z)-\lambda^\top m(z)\}.
    \]
\end{enumerate}
Then
\[
V_{\mathsf P}(\delta)=V_{\mathsf P}^{\ll}(\delta).
\]
\end{thm}
\begin{flushright}
[\hyperref[proof: ignoring dual constraint]{Proof}]
\end{flushright}

Theorem~\ref{thm: ignoring dual constraint} provides two sets of sufficient conditions 
that allow us to discard from the robustness set all the distributions that are not
absolutely
continuous with respect to the baseline while leaving the value of the
inner primal
problem unchanged.
This result alone is not constructive from a practical standpoint and does not answer the 
central question of this subsection: when can we ignore
\eqref{eq: pointwise dual constraint}?
The next proposition addresses this question and shows that the
restricted primal
problem admits a strong dual in which the constraint
\eqref{eq: pointwise dual constraint} is absent, so that the restricted
bound can be computed by an unconstrained convex program.

\begin{prop}
\label{cor: ac unconstrained dual}
Let $\phi$ be an entropy function in the sense of
Definition~\ref{defi: entropy function}, let
$\int_{\calZ}|g(z)|\,\de\nu(z)<\infty$ for every $\nu\in\calM_+(\calZ)$ satisfying
$D_\phi(\nu\,\|\,\Pbase{})<\infty$, and suppose that $\delta>0$, $\phi$ is
continuous at $1$, and
\[
\calM(\rho)\cap\rint(\Gamma^{\ll}(\delta))
\neq\emptyset.
\]
Then
\[
V_{\mathsf P}^{\ll}(\delta)=V_{\mathsf D}^{\circ}(\delta).
\]
\end{prop}
\begin{flushright}
[\hyperref[proof: ac unconstrained dual]{Proof}]
\end{flushright}

Two comments are in order. First, the comparison between the
constrained and unconstrained
dual objectives is pointwise in $\lambda$. Therefore, the two dual programs
agree on the multiplier set $\Rlam$ generated by the maintained
moment equalities and inequalities. Second, we stress that full support
and lower semicontinuity together
suffice to verify the essential supremum identity of condition (b). Indeed, whenever
$\supp(\Pbase{})=\calZ$ and the map
$z\mapsto g(z)-\lambda^\top m(z)$ is lower semicontinuous for every
$\lambda\in\Rlam$, as happens when $g$ is lower semicontinuous and
$m$ is continuous, the displayed essential-supremum identity of
condition (b) is automatically satisfied. We illustrate a scenario in
which condition~(b) holds in Section~\ref{subsubsec: gumbel model verification}.

\subsection{Total Variation and Contamination Neighborhoods}
\label{subsec: contamination tv}
We close this section by specializing the entropy function to the total
variation one, that is $\phi=\phi_{\TV}$, and by introducing the robustness sets on
which the remainder of this supplement relies, namely the total variation
neighborhood and the two neighborhoods generated by the mixture restrictions of
Section~\ref{subsec:mixture-neighborhoods}. We then show that each of the latter
amounts to intersecting the total variation neighborhood with a one-sided
contamination neighborhood.

Fix a sensitivity parameter $\delta\in[0,1]$ and a mixing weight
$\kappa\in[0,1]$. We denote the
set of distributions satisfying the moment restrictions by
\[\calE(\rho)
:=\bigl\{P\in{\calP(\calZ)}:\ \E_P[m(Z)]\in\calM(\rho)\bigr\},\]
and we define the following three sets:
\begin{align*}
\calT_\delta(\Pbase{}) &:= \bigl\{P\in\Delta(\calZ):\ \TV(P\,\|\,\Pbase{})\leq
\delta\bigr\},
\\
\calC_\kappa(\Pbase{}) &:= \bigl\{P\in\Delta(\calZ):\ P = \kappa\Pbase{}+(1
-\kappa)R,\ R\in\Delta(\calZ)\bigr\},
\\
\calC^{\ll}_\kappa(\Pbase{}) &:= \bigl\{P\in\Delta(\calZ):\ \Pbase{} = \kappa P+(1
-\kappa)R,\ R\in\Delta(\calZ)\bigr\}.
\end{align*}
The first one is a total variation neighborhood. The second one is a
contamination neighborhood in the spirit of
\cite{huber1964RobustEstimationLocation} in which the baseline distribution
is a mixing component of the candidate distribution with weight $\kappa$,
so that $1-\kappa$ is the contamination share. 
In the third one, the roles are inverted and the candidate
distribution $P$ is a mixing component of the baseline with the same weight
$\kappa$. We refer to it as the decomposition neighborhood, since each of
its elements is obtained by decomposing the baseline population into a
retained subpopulation, distributed according to $P$, and a discarded one,
distributed according to $R$. The two-parameter families
\begin{equation}\label{eq: two parameter contamination}
\calC_{\kappa,\delta}(\Pbase{})
:=\calT_{\delta}(\Pbase{})\cap\calC_{\kappa}(\Pbase{})
\qquad\text{and}\qquad
\calC^{\ll}_{\kappa,\delta}(\Pbase{})
:=\calT_{\delta}(\Pbase{})\cap\calC^{\ll}_{\kappa}(\Pbase{})
\end{equation}
are of particular interest to us. These families nest the three sets above: at $\kappa=0$
both mixture constraints are vacuous, so that
$\calC_{0,\delta}(\Pbase{})=\calC^{\ll}_{0,\delta}(\Pbase{})=\calT_{\delta}(\Pbase{})$,
whereas at $\delta=1$ the total variation constraint is vacuous, so that
$\calC_{\kappa,1}(\Pbase{})=\calC_{\kappa}(\Pbase{})$ and
$\calC^{\ll}_{\kappa,1}(\Pbase{})=\calC^{\ll}_{\kappa}(\Pbase{})$. They also
coincide, for $\phi=\phi_{\TV}$, with the robustness sets of
Section~\ref{subsec:mixture-neighborhoods}. Indeed,
${\calP}_{\mathsf{mix}}(\rho,\delta)$ of
\eqref{eq: P mix baseline component} and
$\calP^{\ll}_{\mathsf{mix}}(\rho,\delta)$ of
\eqref{eq: P mix baseline mixture} can be rewritten as
\begin{equation}\label{eq: mixture sets as contamination}
{\calP}_{\mathsf{mix}}(\rho,\delta)
\;=\;
\calC_{\kappa,\delta}(\Pbase{})\cap\calE(\rho)
\qquad\text{and}\qquad
\calP^{\ll}_{\mathsf{mix}}(\rho,\delta)
\;=\;
\calC^{\ll}_{\kappa,\delta}(\Pbase{})\cap\calE(\rho),
\end{equation}
with $\delta$ free and $\kappa$ fixed. 

The next lemma leverages the equivalence between mixture constraints and dominance
constraints established in
Lemma~\ref{lemma: mixture density bounds} to provide a characterization of the sets in
\eqref{eq: two parameter contamination}.
\begin{lemma}[Contamination as constrained TV]
\label{lemma: contamination as constrained tv}
Let $\kappa\in[0,1]$ and $\delta\in[0,1]$. If $\kappa<1$, then for every
$P\in\calC_\kappa(\Pbase{})$ the mixing distribution is uniquely determined
as $R=(P-\kappa\Pbase{})/(1-\kappa)$, for every
$P\in\calC^{\ll}_\kappa(\Pbase{})$ as $R=(\Pbase{}-\kappa P)/(1-\kappa)$,
and the following characterizations hold:
\begin{enumerate}[label={\textbf{(\roman*)}},ref=\thelemma(\roman*),leftmargin=*,itemsep=10pt,topsep=6pt]
\item\label{lemma: contamination as constrained tv-i} \uline{\textbf{Baseline
as the mixture.}}
\begin{equation}\label{eq:reverse-contam-as-tv}
\begin{aligned}
\calC^{\ll}_{\kappa,\delta}(\Pbase{})
&\;=\;\Bigl\{P\in\Delta(\calZ):\ \Pbase{}=\kappa P+(1-\kappa)R,\
R\in\Delta(\calZ),\ \TV(P\,\|\,R)\leq\frac{\delta}{1-\kappa}\Bigr\}\\
&\;=\;\calT_{\delta}(\Pbase{})\cap\bigl\{P:\ \kappa P\leq \Pbase{}\bigr\},
\end{aligned}
\end{equation}
and the mixing distribution of every $P\in\calC^{\ll}_\kappa(\Pbase{})$
satisfies $\TV(P\,\|\,\Pbase{})=(1-\kappa)\TV(P\,\|\,R)$ and
$\TV(R\,\|\,\Pbase{})=\kappa\,\TV(P\,\|\,R)$;
\item\label{lemma: contamination as constrained tv-ii} \uline{\textbf{Baseline
as a mixing component.}}
\begin{equation}\label{eq:contam-as-tv}
\begin{aligned}
\calC_{\kappa,\delta}(\Pbase{})
&\;=\;\Bigl\{\kappa\Pbase{}+(1-\kappa)R:\ R\in\Delta(\calZ),\
\TV(R\,\|\,\Pbase{})\leq\frac{\delta}{1-\kappa}\Bigr\}\\
&\;=\;\calT_{\delta}(\Pbase{})\cap\bigl\{P:\ P\geq\kappa\Pbase{}\bigr\},
\end{aligned}
\end{equation}
and the mixing distribution of every $P\in\calC_\kappa(\Pbase{})$ satisfies
$\TV(P\,\|\,\Pbase{})=(1-\kappa)\TV(R\,\|\,\Pbase{})$.
\end{enumerate}
Both dominance constraints are interpreted as measure inequalities, that is,
$\kappa P\leq\Pbase{}$ means $\kappa P(A)\leq\Pbase{}(A)$ for every
$A\in\mathscr{F}$, and analogously for $P\geq\kappa\Pbase{}$. Moreover,
the rescaling identities imply
$\calC_{\kappa,\delta}(\Pbase{})\cup\calC^{\ll}_{\kappa,\delta}(\Pbase{})
\subseteq\calT_{\min\{\delta,1-\kappa\}}(\Pbase{})$, and whenever
$\delta\geq1-\kappa$ the total variation constraint is redundant, in the
sense that $\calC_{\kappa,\delta}(\Pbase{})=\calC_{\kappa}(\Pbase{})$ and
$\calC^{\ll}_{\kappa,\delta}(\Pbase{})=\calC^{\ll}_{\kappa}(\Pbase{})$.
Finally, if $\kappa=1$, then
$\calC_{1,\delta}(\Pbase{})=\calC^{\ll}_{1,\delta}(\Pbase{})=\{\Pbase{}\}$
for every $\delta\in[0,1]$.
\end{lemma}
\begin{flushright}
[\hyperref[proof: contamination as constrained tv]{Proof}]
\end{flushright}

Lemma~\ref{lemma: contamination as constrained tv} shows that intersecting a
total variation neighborhood with a contamination neighborhood produces
again a contamination class in which the mixing distribution is itself
disciplined in total variation. The two parameters thus play distinct
roles: $1-\kappa$ caps the share of the population that may deviate from
the baseline, while $\delta$ caps the total probability mass that may be
misplaced in doing so. Since, by
Lemma~\ref{lemma: contamination as constrained tv}, the total variation constraint is
redundant whenever $\delta\geq1-\kappa$, \eqref{eq: mixture sets as contamination} shows
that, for $\phi=\phi_{\TV}$ and $\delta=1-\kappa$, the robustness sets of
Section~\ref{subsec:mixture-neighborhoods} are pure contamination neighborhoods, that is,
\begin{equation}\label{eq: pure contamination case}
{\calP}_{\mathsf{mix}}(\rho,1-\kappa)
\;=\;
\calC_\kappa(\Pbase{})\cap\calE(\rho)
\qquad\text{and}\qquad
\calP^{\ll}_{\mathsf{mix}}(\rho,1-\kappa)
\;=\;
\calC^{\ll}_\kappa(\Pbase{})\cap\calE(\rho).
\end{equation}

We now map the sets defined above into the two robustness sets of the
main text. There, all moment restrictions are equalities, that is,
$\mathsf d_1=\mathsf d_3=0$ and $\rho_0$ reduces to its equality block, so that
$\calM(\rho_0)=\{\rhopad_0\}$ and $\calE(\rho_0)$ collects the distributions
satisfying $\E_P[m(Z;\theta)]=\rhopad_0$. The total variation
robustness set in equation~\textup{(3.4)} of the main text
requires the
baseline to be a mixture,
$\Pbase{}=\kappa P+(1-\kappa)R$ with $\kappa\in(0,1)$, and the two mixing components to
satisfy $\TV(P\,\|\,R)\leq\delta$. Therefore,
\begin{equation}\label{eq: paper tv set mapping}
\calP_{\TV}(\theta;\delta)
\;=\;
\calC^{\ll}_{\kappa,(1-\kappa)\delta}(\Pbase{})\cap\calE(\rho_0)
\;=\;
\calP^{\ll}_{\mathsf{mix}}\bigl(\rho_0,(1-\kappa)\delta\bigr),
\end{equation}
where the first equality follows from the first equality in
\eqref{eq:reverse-contam-as-tv} applied with budget $(1-\kappa)\delta$, and the second
one from \eqref{eq: mixture sets as contamination}. In words, a budget $\delta$ on
the distance between the two groups is a budget $(1-\kappa)\delta$ on the distance
between the candidate and the baseline distributions, in view of the rescaling
identity of Lemma~\ref{lemma: contamination as constrained tv-i}. The contamination
robustness set in equation~\textup{(3.5)} of the main text
instead ties the
mixing weight to the
sensitivity parameter, $\kappa=1-\delta$, and leaves the mixing distribution
unrestricted. Hence, for every $\delta\in(0,1)$,
\begin{equation}\label{eq: paper contamination set mapping}
\calP_{\mathsf{C}}(\theta;\delta)
\;=\;
\calC_{1-\delta}(\Pbase{})\cap\calE(\rho_0)
\;=\;
\calP_{\mathsf{mix}}(\rho_0,\delta)
\quad\text{with}\quad\kappa=1-\delta,
\end{equation}
where the first equality holds by definition of $\calC_{1-\delta}(\Pbase{})$, and the
second one follows from \eqref{eq: pure contamination case}. In both
sets, the paper
draws the mixing distribution $R$ from the integrability class $\calP(\calZ)$, which is
without loss here, because $(1-\kappa)R\leq\Pbase{}$ in the first and
$\delta R\leq P$ in the second make $R$ inherit the integrability of
$\Pbase{}\in\calP(\calZ)$ and of
$P\in\calE(\rho_0)\subseteq\calP(\calZ)$, respectively. Both robustness
sets of
the main text are thus members of the two-parameter families in
\eqref{eq: two parameter contamination}, intersected with $\calE(\rho_0)$.

\begin{rem}[Different interpretations]\label{rem: different interpretations tv}
    One key implication of the difference between the total variation
    neighborhood $\calT_\delta(\Pbase{})$ and the two mixture-based sets
    lies in their interpretation. If $P\in\calC_{1-\delta}(\Pbase{})$, not
    only can we say that $P$ and $\Pbase{}$ share at least a fraction
    $1-\delta$ of their mass, but also, in view of the mixture structure,
    that a $(1-\delta)$-fraction of the population behaves exactly as
    $\Pbase{}$. Similarly, if $P\in\calC^{\ll}_{1-\delta}(\Pbase{})$, then
    $P$ and $\Pbase{}$ share at least a fraction $1-\delta$ of their mass,
    and a $(1-\delta)$-fraction of the population behaves exactly as $P$.
    If instead all we know is that $\TV(P\,\|\,\Pbase{})=\delta$, with
    $\delta\in(0,1)$, then both distributions can be written as mixtures
    over a common component,
    \[
    P=(1-\delta)\,Q+\delta\,R_P
    \qquad\text{and}\qquad
    \Pbase{}=(1-\delta)\,Q+\delta\,R_{*},
    \]
    where $Q\in\Delta(\calZ)$ has density
    $\min\{\de P/\de\varrho,\de\Pbase{}/\de\varrho\}/(1-\delta)$ with
    respect to the dominating measure $\varrho:=(P+\Pbase{})/2$, and
    $R_P,R_{*}\in\Delta(\calZ)$ carry the mass on which the two
    distributions disagree. Hence $P$ and $\Pbase{}$ share exactly a
    fraction $1-\delta$ of their mass, but the shared component $Q$
    need not coincide with either $P$ or $\Pbase{}$. In general, neither
    distribution need be a mixture component of the other. Indeed, not every
    distribution within total variation distance $\delta$ of the baseline
    admits such a mixture representation. For example, for
    $\calZ=\{0,1\}$, $\Pbase{}=\mathsf{Be}(1/2)$, and $P(\{1\})=0.7$, so that
    $\TV(P\,\|\,\Pbase{})=0.2$, no $R\in\Delta(\calZ)$ satisfies $P=0.8\,\Pbase{}+0.2\,R$
    or $\Pbase{}=0.8\,P+0.2\,R$, as either would give $R$ a negative mass.
    Consequently, $P$ belongs to $\calT_{0.2}(\Pbase{})$ but to neither
    $\calC_{0.8}(\Pbase{})$ nor $\calC^{\ll}_{0.8}(\Pbase{})$.
\end{rem}

\clearpage
\section{Existence of Solutions to the Inner and Outer Problems}
\label{sec:feasibility-inner}
This section provides sufficient conditions for the dual inner problem and the
outer problem of \eqref{eq: general optimization problem} to admit solutions.
Accordingly, we make dependence of quantities on $\theta$ explicit again.
Section~\ref{subsec:existence-dual} treats the dual inner problem holding the
parameter $\theta$ fixed.
Section~\ref{subsec:existence-outer} then turns to the outer problem, which
optimizes the inner value over $\theta$. 
In this section, we keep suppressing the dependence
of the robustness set on the baseline distribution $\Pbase{}$
for notational convenience.

\subsection{Existence of Dual Solutions}\label{subsec:existence-dual}
We first establish sufficient conditions for the inner problem of
\eqref{eq: general optimization problem} to have solutions in its dual form,
holding $\theta\in\Theta$ fixed. We write
$V_{\mathsf P}(\theta;\delta)$ for the value
$\sup_{P\in\calP_\phi(\theta;\rho,\Pbase{},\delta)}\E_P[g(Z;\theta)]$ of the inner
problem and $V_\mathsf{D}(\theta;\delta)$ for
the value of the associated dual inner problem, displayed in
\eqref{eq: dual with moments} below, where we recall the convention of
Theorem~\ref{theorem: weak duality} that the product between $\eta$ and
$\phi^\infty(1)$ evaluates to $+\infty$ if $\eta=0$ and $\phi^\infty(1)=+\infty$.
Throughout this section, we write
$\calN(\theta):=\{m(z;\theta):z\in\calZ\}\subseteq\R^{\dimm}$ for the image of the moment
function at $\theta$ and, whenever $\dom(\phi)$ is bounded above,
$\bar{x}:=\sup\dom(\phi)$, and we recall the padded target $\rhopad$ of
Section~\ref{subsec: setup}.

We start by introducing the conditions on the primitives of the problem that
are maintained throughout this subsection. We rely on two
sets of conditions, which handle different scenarios that might occur in practice
and differ according to the properties of the recession of the
entropy function. Formally, Assumption~\ref{assump: dual primitives} covers entropies
with finite recession ($\phi^\infty(1)<\infty$), for which the pointwise
constraint in
\eqref{eq: dual with moments} is not vacuous, whereas
Assumption~\ref{assump: dual primitives superlinear} covers superlinear
entropies (with $\phi^\infty(1)=+\infty$), for which the constraint is vacuous. 
\begin{assump}[Primitives I: finite recession entropies]\label{assump: dual primitives}
    Let $\theta\in\Theta$. Suppose:
    \begin{enumerate}[label=(\roman*),ref=\theassump(\roman*),leftmargin=2em,topsep=2pt]
        \item\label{assump: dp-reg} (Boundedness from above.)
        $\bar{g}:=\sup_{z\in\calZ}g(z;\theta)<\infty$.
        \item\label{assump: dp-entropy} (Entropy.) $\phi$ is an entropy
        function in the sense of Definition~\ref{defi: entropy function} with
        $\phi^\infty(1)\in(0,+\infty)$ and $\phi(0)<\infty$.
        \item\label{assump: dp-feas} (Feasibility.) There
        exists a distribution $P^\circ\in\calP(\calZ)$ with
        $\E_{P^\circ}[m(Z;\theta)]\in\calM(\rho)$ and
        $D_\phi(P^\circ\,\|\,\Pbase{})<\delta$.
        \item\label{assump: dual slater} (Slater condition.)
        If $\dimm\geq1$, then
        \[0\in\interior\bigl(\mathrm{co}(\calN(\theta))-\rhopad\bigr).\]
        
    \end{enumerate}
\end{assump}

\begin{assump}[Primitives II: superlinear
entropies]\label{assump: dual primitives superlinear}
    Let $\theta\in\Theta$. Suppose:
    \begin{enumerate}[label=(\roman*),ref=\theassump(\roman*),leftmargin=2em,topsep=2pt]
        \item\label{assump: dps-entropy} (Entropy.) $\phi$ is an entropy
        function in the sense of Definition~\ref{defi: entropy function} with
        $\phi^\infty(1)=+\infty$ and $0\in\dom(\phi)$, and either
        \textup{(a)} $[0,\infty)\subseteq\dom(\phi)$, or
        \textup{(b)} $\bar{x}\in(1,\infty)$ with
        $\phi(\bar{x})<\infty$.
        \item\label{assump: dps-anchor} (Dual anchor.) There exists
        $\xi^{*}=(\zeta^{*},\eta^{*},\lambda^{*})\in\R\times\R_{+}\times\Rlam$
        with
        \[\E_{\Pbase{}}\big[(\phi^*)^\pi\big(g(Z;\theta)-\lambda^{*\top}
        m(Z;\theta)-\zeta^{*},\,\eta^{*}\big)\big]<\infty.\]
        \item\label{assump: dps-feas} (Feasibility.) There
        exists $P^\circ\in\calP(\calZ)$ with
        $\E_{P^\circ}[m(Z;\theta)]\in\calM(\rho)$ and
        $D_\phi(P^\circ\,\|\,\Pbase{})<\delta$.
        \item\label{assump: dps-int} (Integrability.)
        $\int_{\calZ}|g(z;\theta)|\,\de\nu(z)<\infty$ for every $\nu\in\calM_{+}(\calZ)$
        with $D_\phi(\nu\,\|\,\Pbase{})<\infty$.
        \item\label{assump: dps-support} (Support and continuity.) In case
        \textup{(a)} of Assumption~\ref{assump: dps-entropy}, $g(\cdot;\theta)$ and
        $m(\cdot;\theta)$ are continuous on $\calZ$ and
        $\supp(\Pbase{})=\calZ$.
        \item\label{assump: dps-ceiling} (Slater condition.)
        If $\dimm\geq1$, the following holds:
        In case \textup{(a)} of Assumption~\ref{assump: dps-entropy},
        $0\in\interior\bigl(\mathrm{co}(\calN(\theta))-\rhopad\bigr)$.
        In case \textup{(b)} of Assumption~\ref{assump: dps-entropy}, letting
        \[\calN_{\bar{x}}(\theta):=\Big\{\E_{Q}[m(Z;\theta)]:
        Q\in\Delta(\calZ),\ Q\ll\Pbase{},\
        \tfrac{\de Q}{\de\Pbase{}}\leq\bar{x}\ \ \Pbase{}\text{-a.s.}\Big\}
        \subseteq\R^{\dimm},\]
        \[0\in\interior\big(\calN_{\bar{x}}(\theta)-\rhopad\big).\]
    \end{enumerate}
\end{assump}
Cases (a) and (b) in Assumption~\ref{assump: dps-entropy} cover the two ways a
superlinear 
entropy can handle large density ratios. Case (a) allows arbitrarily large likelihood
ratios at finite cost, as with the Kullback--Leibler or $\chi^2$ divergences.
Case (b) accommodates entropies whose domain bounds the likelihood ratio at some 
$\bar{x}\in(1,\infty)$, as with the modified entropy $\overline{\phi}$ induced by 
the baseline-as-mixture restriction of Section~\ref{subsec:mixture-neighborhoods}, 
whose cap is $\bar{x}=1/\kappa$.

As already highlighted, the two sets of assumptions differ mainly in the
recession of the entropy function: under Assumption~\ref{assump: dual
primitives} the pointwise constraint in \eqref{eq: dual with moments} is
not vacuous, whereas under
Assumption~\ref{assump: dual primitives superlinear} it is vacuous. They also differ in
how they control the objective. Assumption~\ref{assump: dual primitives} requires
$g(\cdot;\theta)$ to be bounded above, while
Assumption~\ref{assump: dual primitives superlinear} removes every boundedness requirement
at the price of the stronger integrability condition
of Assumption~\ref{assump: dps-int} and of the support and continuity
condition of Assumption~\ref{assump: dps-support}. The feasibility
margins of Assumptions~\ref{assump: dp-feas}
and~\ref{assump: dps-feas} coincide as they both ask for a
distribution that satisfies the moment restrictions strictly inside the divergence budget,
which holds at $P^\circ=\Pbase{}$ whenever $\delta>0$ and
$\E_{\Pbase{}}[m(Z;\theta)]\in\calM(\rho)$, as
$D_\phi(\Pbase{}\,\|\,\Pbase{})=\phi(1)=0$, but also accommodates a
misspecified baseline. Finally, both sets close with a
Slater condition, which requires room to perturb the moments in every
direction simultaneously and is generically satisfied when the moment
functions are sufficiently rich on $\calZ$.

We now turn to the main results of this section. Define the dual
support function
\[\sigma(\lambda,\theta):=\sup_{z\in\calZ}\{g(z;\theta)-\lambda^\top
m(z;\theta)\},\qquad \lambda\in\Rlam,\]
and, within the ambient cone $\Lbb:=\R\times\R_+\times\Rlam$ of dual
multipliers, the dual feasibility set
\[\calF(\theta):=\Big\{(\zeta,\eta,\lambda)\in\Lbb
\,:\, \zeta+\eta\phi^\infty(1)\geq\sigma(\lambda,\theta)\Big\},\]
and, gathering the multipliers into $\xi:=(\zeta,\eta,\lambda)$, also define the dual
objective as
\[f(\xi,\theta):=\lambda_{12}^\top\rho+\zeta+\eta\delta
+\E_{\Pbase{}}\big[(\phi^*)^\pi\big(g(Z;\theta)-\lambda^\top m(Z;\theta)
-\zeta,\,\eta\big)\big].\]
As we show in the proof of Lemma~\ref{lemma: slater dual form} below,
the function $f(\cdot,\theta)$ is proper, convex, and lower semicontinuous, so that
its recession function $f^\infty(\cdot,\theta)$ is defined as in
Definition~\ref{defi: recession function}, and the set $\calF(\theta)$ is closed and
convex. The recession cone of a nonempty closed convex set $C\subseteq\R^k$ is
$\mathrm{rec}(C):=\{d\in\R^k:C+\tau d\subseteq C\text{ for all }\tau\geq0\}$, and we
call its elements the recession directions of $C$.

The next lemma presents some auxiliary results
that will be useful in the arguments that follow. In particular, it establishes (i)
the existence of a pair of distributions $(Q_-,Q_+)$ at which
the moment inequalities are strictly slack and the moment equalities are perturbed
to either side of their targets, (ii) the existence of a strictly
feasible point of $\calF(\theta)$, and (\textit{iii}) the strict
positivity of the recession function of the dual objective along every nonzero
recession direction of $\calF(\theta)$.

\begin{lemma}[Slater in dual form]\label{lemma: slater dual form}
    Suppose either Assumption~\ref{assump: dual primitives} or
    Assumption~\ref{assump: dual primitives superlinear} holds at
    $\theta\in\Theta$. Then:
    \begin{enumerate}[label=(\roman*),ref=\thelemma(\roman*)]
        \item\label{lemma: slater dual form-i} (Perturbed-moment
        distributions.) When $\dimm\geq1$, there
        exist $Q_-,Q_+\in\Delta(\calZ)$ and $\varepsilon>0$ such that the
        inequality blocks are strictly slack at both distributions, that is
        \[
        \E_{Q_\pm}[m_1(Z;\theta)]<\rho_1, \quad \E_{Q_\pm}[m_3(Z;\theta)]<0,
        \]
        while the
        equality blocks are perturbed to either side of their targets, that is
        \[\E_{Q_\pm}[m_2(Z;\theta)]=\rho_2\pm 1_{\mathsf{d}_2}\varepsilon\quad
        \text{and}\quad
        \E_{Q_\pm}[m_4(Z;\theta)]=\pm 1_{\mathsf{d}_4}\varepsilon.\]
        \item\label{lemma: slater dual form-ii} (Strictly feasible point.)
        Under Assumption~\ref{assump: dual primitives},
        $\calF(\theta)$ admits a strictly feasible point, that is, there
        exists $\xi^\circ=(\zeta^\circ,\eta^\circ,\lambda^\circ)$ in
        $\R\times\R_{++}\times\Rlam$ such that
        \[\zeta^\circ+\eta^\circ\phi^\infty(1)>
        \sigma(\lambda^\circ,\theta).\]
        Under Assumption~\ref{assump: dual primitives
        superlinear}, the pointwise constraint is vacuous and
        $\calF(\theta)=\R\times\R_+\times\Rlam$.
        \item\label{lemma: slater dual form-iii} (Positive recession function.) The
        recession function $f^\infty(\cdot,\theta)$ is
        strictly positive on every nonzero recession direction of $\calF(\theta)$.
    \end{enumerate}
\end{lemma}
\begin{flushright}
[\hyperref[proof: slater dual form]{Proof}]
\end{flushright}

The three parts of Lemma~\ref{lemma: slater dual form} play distinct
roles in what follows. Part (i) is the device behind part (iii): the two
distributions perturb the moment equalities to either side of their targets,
which is what makes the recession function of the dual objective strictly
positive along every nonzero recession direction of the feasible set. Part
(ii) provides a strictly feasible dual point, which
Lemma~\ref{lemma: outer value regularity} uses to show that the dual value is
finite and upper semicontinuous in $\theta$. Part (iii) is what
Theorem~\ref{thm: existence of dual solutions} invokes: by
Proposition~3.2.2 in
\citet{bertsekas2009ConvexOptimizationTheory}, it is equivalent to the set of
minimizers being nonempty and compact.

\begin{thm}[Existence of dual solutions]\label{thm: existence of dual solutions}
    Suppose either Assumption~\ref{assump: dual primitives} or
    Assumption~\ref{assump: dual primitives superlinear} holds at
    $\theta\in\Theta$. Then the set of minimizers of
    \begin{align}\label{eq: dual with moments}
    \begin{split}
    &\inf_{\substack{(\zeta,\eta,\lambda)\in\\\R\times\R_+\times\Rlam}}\,\lambda_{12}^\top\rho
    + \zeta + \eta\delta +\E_{\Pbase{}}\big[(\phi^*)^\pi\big(
    g(Z;\theta)-\lambda^\top
    m(Z;\theta)-\zeta,\,\eta\big)\big] \\
    &\quad\text{s.t.}\quad \zeta + \eta\phi^\infty(1) \geq
    \sup_{z\in\calZ}\{g(z;\theta)-\lambda^\top m(z;\theta)\},
    \end{split}
    \end{align}
    is nonempty, compact, and convex.
\end{thm}
\begin{flushright}
[\hyperref[proof:thm: existence of dual solutions]{Proof}]
\end{flushright}

We conclude by showing that the feasibility and Slater conditions of
Assumptions~\ref{assump: dual
primitives} and~\ref{assump: dual primitives superlinear} can be packaged,
when all moment restrictions are equalities, into a single interior
condition on the target, which is the form the main text adopts. The next
lemma records the implication, together with a primitive sufficient
condition.

\begin{lemma}[Constraint qualification with moment equalities]
\label{lemma: constraint qualification}
Let $\theta\in\Theta$ and $\delta>0$, let $\phi$ be an entropy function in
the sense of Definition~\ref{defi: entropy function}, and suppose that all moment
restrictions are equalities, so that $\calM(\rho)=\{\rhopad\}$ and
$\Rlam=\R^{\dimm}$. Let
$\Gamma(\delta)$ be the attainable moment set of \eqref{eq: attainable moment set} at
$\theta$, and, whenever $\bar{x}<\infty$, let $\calN_{\bar{x}}(\theta)$ be the set of
Assumption~\ref{assump: dps-ceiling}.
\begin{enumerate}[label=(\roman*),ref=\thelemma(\roman*),leftmargin=2em,topsep=2pt]
    \item\label{lemma: constraint qualification-i} (Consequences.) If
    $\rhopad\in\interior(\Gamma(\delta))$, then \textup{(a)} there exists
    $P^{\circ}\in\calP(\calZ)$ with $\E_{P^{\circ}}[m(Z;\theta)]=\rhopad$ and
    $D_\phi(P^{\circ}\,\|\,\Pbase{})<\delta$; \textup{(b)}
    $0\in\interior(\mathrm{co}(\calN(\theta))-\rhopad)$; and
    \textup{(c)} if $\bar{x}<\infty$, then
    $0\in\interior(\calN_{\bar{x}}(\theta)-\rhopad)$.
    \item\label{lemma: constraint qualification-ii} (A sufficient
    condition.) Suppose that $\phi(0)<\infty$ and $\phi^\infty(1)<\infty$. If
    there exists $P^{\circ}\in\calP(\calZ)$ with
    $\E_{P^{\circ}}[\|m(Z;\theta)\|^2]<\infty$, $\E_{P^{\circ}}[m(Z;\theta)]=\rhopad$,
    $D_\phi(P^{\circ}\,\|\,\Pbase{})<\delta$, and a nonsingular covariance matrix
    $\V_{P^{\circ}}[m(Z;\theta)]$, then
    $\rhopad\in\interior(\Gamma(\delta))$. If, moreover, $P^{\circ}\ll\Pbase{}$,
    the same conclusion holds for the set obtained from $\Gamma(\delta)$ by
    restricting to distributions dominated by $\Pbase{}$.
\end{enumerate}
\end{lemma}
\begin{flushright}
[\hyperref[proof: constraint qualification]{Proof}]
\end{flushright}

\subsection{Existence of Solutions to the Outer Problem}\label{subsec:existence-outer}
The previous subsection established existence of dual solutions to the inner
problem of \eqref{eq: general optimization problem} for each fixed value of the
parameter $\theta\in\Theta$. We now turn to the outer problem.
We first introduce the additional regularity conditions used for upper
semicontinuity.

\begin{assump}[Outer regularity]\label{assump: outer regularity}
Suppose that either
Assumption~\ref{assump: dual primitives} holds at every $\theta\in\Theta$ or
Assumption~\ref{assump: dual primitives superlinear} holds at every
$\theta\in\Theta$, and that:
\begin{enumerate}[label=(\roman*),ref=\theassump(\roman*),leftmargin=2em,topsep=2pt]
    \item\label{assump: or-cont} (Continuity in the parameter.) For each
    $\theta\in\Theta$, the map
    $\theta'\mapsto(g(z;\theta'),m(z;\theta'))$ is continuous at $\theta$
    for $\Pbase{}$-almost every $z\in\calZ$.
    \item\label{assump: or-usc sigma} (Upper semicontinuity of the dual
    support function.) If $\phi^\infty(1)<\infty$, then, for every
    $\lambda\in\Rlam$, the map $\theta\mapsto\sigma(\lambda,\theta)$ is
    upper semicontinuous at every $\theta\in\Theta$ at which it is finite.
    \item\label{assump: or-dom} (Local domination.) If
    $\phi^\infty(1)=+\infty$, then, in case (a) of
    Assumption~\ref{assump: dps-entropy}, for every
    $(\zeta,\eta,\lambda)\in\R\times\R_{++}\times\Rlam$ and every
    $\theta\in\Theta$ there exist a radius $r>0$ and a
    function $G\in L_1(\Pbase{})$ such that
    \[\eta\,\phi^*\Big(\frac{g(z;\theta')-\lambda^\top
    m(z;\theta')-\zeta}{\eta}\Big)\leq G(z)\]
    for all $\theta'\in\calB_r(\theta)\cap\Theta$ and
    $\Pbase{}$-almost every $z\in\calZ$.
    In case (b) of Assumption~\ref{assump: dps-entropy}, for each
    $\theta\in\Theta$ there exist a radius $r>0$
    and functions $G_g,G_m\in L_1(\Pbase{})$ with
    \[
    \sup_{\theta'\in\calB_r(\theta)\cap\Theta}|g(z;\theta')|\leq G_g(z),
    \qquad
    \sup_{\theta'\in\calB_r(\theta)\cap\Theta}\|m(z;\theta')\|
    \leq G_m(z),
    \]
    for $\Pbase{}$-almost every $z\in\calZ$.
\end{enumerate}
\end{assump}

The three conditions play distinct roles, and we comment on each in turn.
Assumption~\ref{assump: or-cont} asks for continuity in $\theta$
only at $\Pbase{}$-almost every point of the support, which accommodates,
for example, indicator-type moment functions whose discontinuity locus is
$\Pbase{}$-null, as in the Gumbel marital-premium example of
Section~\ref{subsec: structural models}. Assumption~\ref{assump: or-usc sigma}
is stated on the derived support function $\sigma$ rather than on the
primitives $(g,m)$. A sufficient condition on the primitives is
that $\calZ$ be compact and $(z,\theta)\mapsto(g(z;\theta),m(z;\theta))$ be jointly
continuous, in which case $\theta\mapsto\sigma(\lambda,\theta)$ is
continuous by Berge's maximum theorem. Finally, Assumption~\ref{assump: or-dom} is the
price of superlinearity: the constraint is then vacuous and the dual integrand must be
controlled by a local domination condition. It has no counterpart under finite recession,
where the dual constraint itself supplies the dominating bound.

\begin{lemma}[Regularity of the dual value function]
\label{lemma: outer value regularity}
\begin{enumerate}[label=(\roman*),ref=\thelemma(\roman*),leftmargin=2em,topsep=2pt]
    \item\label{lemma: outer value regularity-i} (Finiteness.) If either
    Assumption~\ref{assump: dual primitives} or
    Assumption~\ref{assump: dual primitives superlinear} holds at
    $\theta\in\Theta$, then
    $V_{\mathsf{P}}(\theta;\delta)\in\R$ and
    $V_{\mathsf{D}}(\theta;\delta)\in\R$;
    \item\label{lemma: outer value regularity-ii} (Upper semicontinuity.)
    Under Assumption~\ref{assump: outer regularity}, the dual value function
    $V_{\mathsf{D}}(\cdot\,;\delta)$ defined in
    \eqref{eq: dual with moments} is upper
    semicontinuous on $\Theta$.
\end{enumerate}
\end{lemma}
\begin{flushright}
[\hyperref[proof:lemma: outer value regularity]{Proof}]
\end{flushright}

Combining the preceding auxiliary results with
Theorem~\ref{thm: existence of dual solutions} gives the following theorem.

\begin{thm}[Existence of solutions and strong duality]\label{thm: existence full problem}
    Let $\Theta\subseteq\R^{\dimt}$ be nonempty and compact and let
    Assumption~\ref{assump: outer regularity} hold. Then the following hold:
    \begin{enumerate}[label=(\roman*),ref=\thethm(\roman*),leftmargin=2em,topsep=2pt]
        \item\label{thm: existence full problem-i} (Existence of solutions.) The set of
        maximizers 
        \[\argsup_{\theta\in\Theta}V_{\mathsf{D}}(\theta;\delta)\]
        is nonempty and compact and, at every maximizer,
        the set of minimizers of the dual inner problem
        \eqref{eq: dual with moments} is nonempty, compact, and convex.
        \item\label{thm: existence full problem-ii} (Strong duality.) If, in
        addition, $\int_{\calZ}|g(z;\theta)|\,\de\nu(z)<\infty$ for every
        $\theta\in\Theta$ and every $\nu\in\calM_+(\calZ)$ satisfying
        $D_\phi(\nu\,\|\,\Pbase{})<\infty$, then
        $V_{\mathsf{D}}(\theta;\delta)=V_{\mathsf{P}}(\theta;\delta)$ for
        every $\theta\in\Theta$ and, consequently,
        \[\argsup_{\theta\in\Theta}V_{\mathsf{D}}(\theta;\delta)
        =\argsup_{\theta\in\Theta}V_{\mathsf{P}}(\theta;\delta).\]
    \end{enumerate}
\end{thm}
\begin{flushright}
[\hyperref[proof:thm: existence full problem]{Proof}]
\end{flushright}

\begin{rem}[Dual feasibility under mixture constraints]
\label{remark: feasibility mixture}
The results for the dual problem of this section cover the
baseline-as-mixture sets whenever $0,1/\kappa\in\dom(\phi)$. Indeed,
$\overline{\phi}$ has infinite recession,
$\bar{x}=\sup\dom(\overline{\phi})=1/\kappa\in(1,\infty)$, and both
$\overline{\phi}(0)=\phi(0)$ and
$\overline{\phi}(1/\kappa)=\phi(1/\kappa)$ are finite, so it falls under case
(b) of Assumption~\ref{assump: dps-entropy}.
Lemma~\ref{lemma: outer value regularity} and
Theorem~\ref{thm: existence full problem} then apply whenever the remaining conditions of
Assumption~\ref{assump: dual primitives superlinear} and of
Assumption~\ref{assump: outer regularity} hold, with the local domination condition of
Assumption~\ref{assump: or-dom} in its case-(b) form. The additional
integrability condition in Theorem~\ref{thm: existence full problem-ii} is
automatic in this case as any $\nu\in\calM_{+}(\calZ)$ with
$D_{\overline{\phi}}(\nu\,\|\,\Pbase{})<\infty$ admits a density with respect to
$\Pbase{}$ bounded by $1/\kappa$, so the condition follows from
$\E_{\Pbase{}}[|g(Z;\theta)|]<\infty$, which holds as
$\Pbase{}\in\calP(\calZ)$. The contamination neighborhoods are 
outside the scope of these statements because the induced entropy $\underline{\phi}$
has $\underline{\phi}(0)=+\infty$. Since
$\underline{\phi}^\infty(1)=\phi^\infty(1)$ by
Proposition~\ref{prop: induced entropy-ii}, this violates
Assumption~\ref{assump: dp-entropy} when $\phi^\infty(1)<\infty$, and
Assumption~\ref{assump: dps-entropy}, which requires $0$ to lie in the domain, when
$\phi^\infty(1)=+\infty$.
\end{rem}

\clearpage
\section{Large Sample Properties}
\label{sec:estimation-inference}
\label{subsec:caseB-EB}
In this section, we develop the asymptotic theory for the plug-in estimators of
the sensitivity bounds. In particular, we (i) establish consistency,
(ii) characterize the asymptotic distribution, and (iii)
propose a bootstrap procedure to construct valid confidence intervals. 
We work with a family of robustness sets which contains
the two robustness sets \textup{(3.4)} and
\textup{(3.5)} of the main text and is slightly less general than those
defined via $\phi$-divergences in the previous sections. Throughout, the
baseline distribution $\Pbase{}$ is an identified feature of the law $P_0$ of the observed
data introduced in Remark~\ref{rem: baseline pseudo true}, and it is estimated
nonparametrically by the empirical measure
$\widehat P_{n}=n^{-1}\sum_{i=1}^{n}\delta^{\mathtt{Dirac}}_{Z_{i}}$ of an i.i.d.\ sample
$Z_{1},\ldots,Z_{n}$ from $\Pbase{}$. The case in
which $\Pbase{}$ is known, as in the structural model of
Section~\ref{subsec: structural models}, is treated as a special case (see
Remark~\ref{rem:known-baseline}). Sampling uncertainty comes from the estimator of
the moment targets $\widehat\rho_{n}$ and the empirical baseline $\widehat P_{n}$.

\textbf{Roadmap.} Section~\ref{subsec:ls-setup} introduces the family of
robustness sets, the common form of their duals, and the two auxiliary
objects on which the theory rests: a constraint qualification, which packages
the Slater and feasibility-margin conditions of
Section~\ref{subsec:existence-dual}, and a minimum-divergence function, which
governs which parameter values are feasible within the budget.
Section~\ref{subsec:ls-assumptions} records the regularity conditions.
Section~\ref{subsec:ls-auxiliary} proves that the integrands through which
the baseline enters the dual form a Donsker class, and that the bound is
continuous in $(\rho,P)$ at the population value. From these we
deduce, in turn, consistency, Hadamard
directional differentiability, and the asymptotic distribution, the latter
obtained from the joint empirical process limit through an appropriate directional delta
method (Section~\ref{subsec:ls-hd}), and the validity of our
bootstrap procedure (Section~\ref{subsec:ls-bootstrap}). 
Throughout, the lower bound is obtained from the upper bound by substituting $g$ with
$-g$.

\subsection{A Family of Robustness Sets}\label{subsec:ls-setup}

Fix $\underline{u}$ and $\overline{u}$ with $\underline{u}\in[0,1)$ and
$\overline{u}\in(1,+\infty]$
and write $\mathbb{U}:=[\underline{u},\overline{u}]$, understood
as $[\underline{u},+\infty)$ when $\overline{u}=+\infty$. The
endpoint $\overline{u}=1$ is excluded, as it would force $Q=P$ in the set
defined next. Hence $1$ is an interior point of $\mathbb{U}$, which makes the
entropy function of Lemma~\ref{lemma: family entropy} continuous at $1$ and
the set of dual minimizers in Lemma~\ref{lemma: family duality-iii}
compact. For a distribution
$P\in\Delta(\calZ)$, let
\begin{equation*}
\calP_{\mathbb U}(P):=\bigl\{Q\in\Delta(\calZ):\
\underline{u}\,P\leq Q\leq\overline{u}\,P\bigr\},
\end{equation*}
where the upper inequality is omitted when $\overline{u}=+\infty$.
By the equivalence of~(2) and~(3) in
Lemma~\ref{lemma: mixture density bounds}, which holds for every $\kappa>0$, applied with
$(P,Q)$ in place of $(\Pbase{},P)$, with $\kappa=\underline{u}$ in part~(ii) when
$\underline{u}>0$, and with $\kappa=1/\overline{u}$ in part~(i) when
$\overline{u}<\infty$, a
distribution $Q$ belongs to $\calP_{\mathbb U}(P)$ if and only if the density $w$
of the absolutely
continuous part of $Q$ with respect to $P$ satisfies $w\geq\underline{u}$
$P$-almost surely and, when $\overline{u}<\infty$, $Q\ll P$ with
$w\leq\overline{u}$ $P$-almost surely. When $\overline{u}=+\infty$ the
singular part of $Q$ is unrestricted. For $\theta\in\Theta$,
$\rho\in\R^{\dime}$, and a budget $\delta\in(0,1]$ (larger budgets add
nothing, as $\TV\leq1$), the robustness set is
\begin{equation}\label{eq:robustness family}
\calP_{\mathbb U}(\theta;\rho,P,\delta):=\bigl\{Q\in\calP_{\mathbb U}(P):\
\E_{Q}[m(Z;\theta)]\in\calM(\rho),\ \TV(Q\,\|\,P)\leq\delta\bigr\}.
\end{equation}
Three members of the family are of interest:
\begin{enumerate}
    \item when $\mathbb{U}=[0,+\infty)$ no weight is restricted and 
    \eqref{eq:robustness family} is the plain total variation robustness set
    \eqref{eq: robustness set definition} with $\phi=\phi_{\TV}$;
    \item when $\mathbb{U}=[0,1/\kappa]$ with $\kappa\in(0,1)$, it is the set
    $\calP^{\ll}_{\mathsf{mix}}(\rho,\delta)$ of \eqref{eq: P mix baseline mixture} with
    $\phi=\phi_{\TV}$,
    which coincides with the decomposition neighborhood of
    Section~\ref{subsec: contamination tv}. At budget $(1-\kappa)\delta$,
    it is the total variation robustness set \textup{(3.4)} of the main text
    at budget $\delta$, by \eqref{eq: paper tv set mapping};
    \item when $\mathbb{U}=[1-\delta,+\infty)$, it is the contamination set
    $\calC_{1-\delta}(P)\cap\calE(\rho)$ of Section~\ref{subsec: contamination tv} and of
    the main text.
\end{enumerate}
The next lemma identifies the family with the $\phi$-divergence
neighborhoods of Section~\ref{sec:convex-duality}, which is what lets us
import the duality theory developed there. Its proof applies
Proposition~\ref{prop: induced entropy} once for each endpoint of $\mathbb{U}$ and
computes the conjugate directly.

\begin{lemma}\label{lemma: family entropy}
Let $\phi_{\mathbb U}:=\phi_{\TV}+\iota_{\mathbb U}$. Then $\phi_{\mathbb U}$
is an entropy function in the sense of Definition~\ref{defi: entropy
function}, continuous at $1$, with recession
$\phi_{\mathbb U}^{\infty}(1)=1/2$ if $\overline{u}=+\infty$ and
$\phi_{\mathbb U}^{\infty}(1)=+\infty$ if $\overline{u}<\infty$. For every
$P\in\Delta(\calZ)$ and $Q\in\Delta(\calZ)$,
\[
D_{\phi_{\mathbb U}}(Q\,\|\,P)=
\begin{cases}
\TV(Q\,\|\,P), & Q\in\calP_{\mathbb U}(P),\\
+\infty, & \text{otherwise},
\end{cases}
\]
so that, whenever either $g(\cdot;\theta)$ and $m(\cdot;\theta)$ are
bounded or $\overline{u}<\infty$ and $P\in\calP(\calZ)$,
$\calP_{\mathbb U}(\theta;\rho,P,\delta)=\{Q\in\calP(\calZ):
\E_{Q}[m(Z;\theta)]\in\calM(\rho),\ D_{\phi_{\mathbb U}}(Q\,\|\,P)\leq\delta\}$.
Finally, the perspective of the conjugate of $\phi_{\mathbb U}$ is the
function $\mathsf{r}:\R\times\R_{+}\to\overline{\R}$ given by
\begin{equation}\label{eq:integrand family}
\mathsf{r}(x,\eta):=
\begin{cases}
\max\bigl\{\underline{u}\,x-\tfrac{1-\underline{u}}{2}\,\eta,\ x,\
\overline{u}\,x-\tfrac{\overline{u}-1}{2}\,\eta\bigr\}, &
\overline{u}<\infty,\\[4pt]
\max\bigl\{\underline{u}\,x-\tfrac{1-\underline{u}}{2}\,\eta,\ x\bigr\}, &
\overline{u}=+\infty\ \text{and}\ x\leq\eta/2,\\[4pt]
+\infty, & \overline{u}=+\infty\ \text{and}\ x>\eta/2,
\end{cases}
\end{equation}
that is, $(\phi_{\mathbb U}^{*})^{\pi}(x,\eta)=\mathsf{r}(x,\eta)$ for every
$x\in\R$ and $\eta\geq0$.
\end{lemma}
\begin{flushright}
[\hyperref[proof:lemma: family entropy]{Proof}]
\end{flushright}

For the three cases above, the integrand $\mathsf{r}$ 
reads $\max\{x,-\eta/2\}$ for the plain total variation robustness set, 
$\max\{-\eta/2,\ x,\ x/\kappa-(1-\kappa)\eta/(2\kappa)\}$
for the decomposition neighborhood, and
$\max\{(1-\delta)x-\delta\eta/2,\ x\}$ for the contamination robustness
set. In the first and third cases $\overline{u}=+\infty$, so that these
expressions apply for $x\leq\eta/2$, whereas $\mathsf{r}(x,\eta)=+\infty$ for
$x>\eta/2$. The last two are the integrands $\mathsf{r}_{\TV}$ and
$\mathsf{r}_{\mathsf{C}}$ in \textup{(4.4)} of the main text.  

Write $u^{\dagger}:=\overline{u}$ if $\overline{u}<\infty$ and $u^{\dagger}:=1$
if $\overline{u}=+\infty$ for the largest finite slope. Four properties of
$\mathsf{r}$ are used repeatedly and follow directly
from~\eqref{eq:integrand family}:
\begin{enumerate}[label=(r\arabic*),ref=(r\arabic*),leftmargin=2.6em,
  topsep=2pt,itemsep=2pt]
\item\label{prop:r-sup} $\mathsf{r}(x,\eta)=\sup_{u\in\mathbb U}
\{ux-\eta|u-1|/2\}$ for every $x\in\R$ and $\eta\geq0$;
\item\label{prop:r-hom} $\mathsf{r}$ is jointly convex and positively
homogeneous of degree one in $(x,\eta)$, and $\mathsf{r}(x,\cdot)$ is
nonincreasing;
\item\label{prop:r-lip} on its effective domain, $\mathsf{r}$ is Lipschitz:
$|\mathsf{r}(x,\eta)-\mathsf{r}(x',\eta')|\leq u^{\dagger}|x-x'|
+\tfrac{u^{\dagger}}{2}|\eta-\eta'|$, and
$|\mathsf{r}(x,\eta)|\leq u^{\dagger}|x|+\tfrac{u^{\dagger}}{2}\eta$;
\item\label{prop:r-lb} $\mathsf{r}(x,\eta)\geq x$ and
$\mathsf{r}(x,\eta)\geq\underline{u}\,x-\tfrac{1-\underline{u}}{2}\,\eta$. If
$\overline{u}<\infty$, also
$\mathsf{r}(x,\eta)\geq\overline{u}\,x-\tfrac{\overline{u}-1}{2}\,\eta$.
\end{enumerate}

\textbf{The dual.} For $\xi=(\zeta,\eta,\lambda)\in\R\times\R_{+}\times
\R^{\dimm}$ and $\theta\in\Theta$, define the dual integrand
\begin{equation}\label{eq:ell-EB}
\ell_{\xi,\theta}(z):=\mathsf{r}\bigl(g(z;\theta)-\lambda^{\top}m(z;\theta)
-\zeta,\ \eta\bigr),\qquad z\in\calZ,
\end{equation}
the dual objective
\begin{equation}\label{eq:Phi-def}
\Jdual(\xi,\theta;\rho,P):=\zeta+\eta\delta+\lambda_{12}^{\top}\rho
+P\ell_{\xi,\theta},
\end{equation}
and, with $\sigma(\lambda,\theta)$ the dual support function of
Section~\ref{subsec:existence-dual}, the dual feasible set
\begin{equation}\label{eq:F-def}
\calF(\theta):=
\begin{cases}
\bigl\{\xi\in\R\times\R_{+}\times\Rlam:\
\zeta+\tfrac{\eta}{2}\geq\sigma(\lambda,\theta)\bigr\}, &
\overline{u}=+\infty,\\[3pt]
\R\times\R_{+}\times\Rlam, & \overline{u}<\infty.
\end{cases}
\end{equation}
In view of Lemma~\ref{lemma: family
entropy}, the program
$\inf_{\xi\in\calF(\theta)}\Jdual(\xi,\theta;\rho,P)$
is the dual of Theorem~\ref{thm: strong duality} for the entropy
$\phi_{\mathbb U}$ and the baseline $P$: the pointwise constraint
$\zeta+\eta\phi_{\mathbb U}^{\infty}(1)\geq\sigma(\lambda,\theta)$ is
present when $\overline{u}=+\infty$, because singular mass is then admissible
and priced at the finite recession $1/2$, and vacuous when
$\overline{u}<\infty$, because its left-hand side is then $+\infty$ for every
$\eta\geq0$ under the convention of that theorem. On $\calF(\theta)$ the
integrand \eqref{eq:ell-EB} is finite, since the constraint forces
$g-\lambda^{\top}m-\zeta\leq\eta/2$ on $\calZ$ when $\overline{u}=+\infty$.
For the total variation robustness set of the main text, the budget in
\eqref{eq:Phi-def} is $(1-\kappa)\delta$ by the second item of the list above, so that
the term $(1-\kappa)\eta\delta$ is the term $c_{\TV}\,\eta\delta$ of
\textup{(4.2)} of the main text, with $c_{\TV}=1-\kappa$.
We call
\begin{equation}\label{eq:dual}
\Kmax(\theta;\rho,P):=\inf_{\xi\in\calF(\theta)}\Jdual(\xi,\theta;\rho,P)
\;\in\;[-\infty,+\infty)
\end{equation}
the criterion function. It is well defined at every $(\theta,\rho,P)$ at
which $g(\cdot;\theta)$ is bounded above, because $\calF(\theta)$ then
contains the point $(\sup_{z\in\calZ}g(z;\theta),0,0)$, at which
$\Jdual$ is finite by \ref{prop:r-lip}. 
The representation \eqref{eq:dual} has two structural properties on which
every large sample argument below relies:
\begin{description}[leftmargin=2em,topsep=2pt,itemsep=2pt]
  \item[(L)]\phantomsection\label{prop:linearity-EB}
        \uline{\textbf{Joint linearity.}} For each $\xi$ and $\theta$,
        $\Jdual(\xi,\theta;\rho,P)$ is affine in $\rho$ and linear in
        $P$. Consequently, by \hyperref[prop:independence-EB]{(I)},
        $(\rho,P)\mapsto\Kmax(\theta;\rho,P)$ is concave, as an
        infimum over the fixed set $\calF(\theta)$ of jointly affine maps, with
        supergradient
        $(\lambda_{12},\ell_{\xi,\theta})$ at any minimizer $\xi$.
  \item[(I)]\phantomsection\label{prop:independence-EB}
        \uline{\textbf{Independence.}} The feasible set $\calF(\theta)$
        depends only on $g,m,\theta,\calZ$, and neither on $\rho$ nor on $P$.
\end{description}

\textbf{Attainable moments and the constraint qualification.} Let
\begin{equation}\label{eq:attainable family}
\Gamma_{\mathbb U}(\theta;P,\delta):=\bigl\{\E_{Q}[m(Z;\theta)]:\
Q\in\calP_{\mathbb U}(P),\ \TV(Q\,\|\,P)\leq\delta\bigr\}\subseteq\R^{\dimm}
\end{equation}
be the set of moment vectors attainable within the budget, a convex set
because $\calP_{\mathbb U}(P)$ is convex and $\TV(\cdot\,\|\,P)$ is convex.
By Lemma~\ref{lemma: family entropy}, it is the counterpart of the
attainable moment set \eqref{eq: attainable moment set} for the entropy $\phi_{\mathbb U}$
and the baseline $P$. We say that the constraint qualification holds at $(\theta,\rho,P)$
with radius $r>0$ if there is $w_{0}\in\calM(\rho)$ with
\begin{equation}\label{eq:cq}
\calB_{r}(w_{0})\subseteq\Gamma_{\mathbb U}(\theta;P,\delta),
\end{equation}
and that it holds at $(\theta,\rho,P)$ if it does so for some
radius. Equivalently,
$\calM(\rho)\cap\interior(\Gamma_{\mathbb U} (\theta;P,\delta))\neq\emptyset$. When no
moment restrictions are present, the condition is vacuous, as we then set
$\Gamma_{\mathbb U}(\theta;P,\delta):=\{0\}$ in line with the conventions of
Section~\ref{subsec: setup}. In words, the moment restrictions must be
attainable within the budget with room to perturb the attained moment
vector in every direction.
Condition \eqref{eq:cq} and the feasibility and Slater conditions of
Section~\ref{subsec:existence-dual} differ in one respect. With inequality restrictions,
the Slater conditions require the specific point $\rhopad$ to be interior to the
budget-free attainable set, whereas \eqref{eq:cq} only requires some point of
$\calM(\rho)$ to be interior to the attainable set within the budget, which is weaker.
With equality restrictions only, $\calM(\rho)=\{\rhopad\}$ and the two coincide.

\textbf{Feasible parameters and the objects of interest.} Define the
minimum-divergence function, that is, the smallest distance
from the baseline $P$ to a candidate distribution whose moments satisfy the restrictions
at $\theta$,
\begin{equation}\label{eq:mindiv-primal}
\mathsf{D}(\theta;\rho,P):=\inf\bigl\{\TV(Q\,\|\,P):\ Q\in\calP_{\mathbb U}(P),\
\E_{Q}[m(Z;\theta)]\in\calM(\rho)\bigr\},
\end{equation}
with $\mathsf{D}(\theta;\rho,P)=+\infty$ when no such $Q$ exists. Its dual
representation, given in Lemma~\ref{lemma: mindiv dual} below, carries a
pointwise restriction on its test functions when the weights are unbounded
above, which we record through the set
\begin{equation*}
\calF_{\mathsf D}(\theta):=
\begin{cases}
\bigl\{(\zeta,\lambda)\in\R\times\Rlam:\
\zeta+\tfrac12\geq\sigma^{\infty}(\lambda,\theta)\bigr\}, &
\overline{u}=+\infty,\\[3pt]
\R\times\Rlam, & \overline{u}<\infty,
\end{cases}
\qquad
\sigma^{\infty}(\lambda,\theta):=\sup_{z\in\calZ}\{-\lambda^{\top}
m(z;\theta)\},
\end{equation*}
the analog for the projection problem of the constraint defining
$\calF(\theta)$, obtained by setting $g\equiv0$ and $\eta=1$ there. We then
define the strictly feasible parameter set, the bound, the outer argmax, the
dual optimizer set, and the set of optimal estimated-block multipliers as
\begin{align*}
  \Theta_{\delta}(\rho,P)&:=\{\theta\in\Theta:\mathsf
  D(\theta;\rho,P)<\delta\},
  & \bmax(\delta,\rho,P)&:=\sup_{\theta\in\Theta_{\delta}(\rho,P)}
  \Kmax(\theta;\rho,P),\\
  \Tmax(\rho,P)&:=\argsup_{\theta\in\Theta}\Kmax(\theta;\rho,P),
  & \Xmax(\theta;\rho,P)&:=\arginf_{\xi\in\calF(\theta)}
  \Jdual(\xi,\theta;\rho,P),
\end{align*}
and $\Lmax(\theta;\rho,P):=\Pi_{12}(\Xmax(\theta;\rho,P))$, where
$\Pi_{12}(\zeta,\eta,\lambda)=\lambda_{12}$ is the projection onto the
estimated-block multipliers, that is, the multipliers $\lambda_{12}$ of
Section~\ref{subsec: setup}, which are attached to the moments whose targets
$\rho=(\rho_1^\top,\rho_2^\top)^\top$ are estimated. The
lower bound objects $\Kmin,\bmin,\Tmin,\Xmin,\Lmin$ are obtained by using
$-g$ in place of $g$ and, writing
$\widetilde{K},\widetilde{\mathsf{b}},\widetilde{\Theta}{}^{*},
\widetilde{\Xi}{}^{*},\widetilde{\Lambda}{}^{*}$ for
$\Kmax,\bmax,\Tmax,\Xmax,\Lmax$, setting
$\Kmin:=-\widetilde{K}$ and $\bmin:=-\widetilde{\mathsf{b}}$, so that
$\bmin(\delta,\rho,P)=\inf_{\theta\in\Theta_{\delta}(\rho,P)}\Kmin(\theta;\rho,P)$
is the lower sensitivity bound. Finally, the plug-in estimators of the bounds are
\[
  \bmaxhat(\delta):=\bmax(\delta,\widehat\rho_{n},\widehat P_{n}),\qquad
  \bminhat(\delta):=\bmin(\delta,\widehat\rho_{n},\widehat P_{n}).
\]

\textbf{Two auxiliary results.} The first lemma specializes the duality
theory of Section~\ref{sec:convex-duality} to the family of robustness sets
considered in this section.

\begin{lemma}[Duality for the family]\label{lemma: family duality}
Fix $\theta\in\Theta$, let $\calZ$ be compact, let $g(\cdot;\theta)$ and
$m(\cdot;\theta)$ be
continuous on $\calZ$, and let $\delta\in(0,1]$, $\rho\in\R^{\dime}$, and
$P\in\Delta(\calZ)$.
\begin{enumerate}[label=(\roman*),ref=\thelemma(\roman*),leftmargin=2em,
  topsep=2pt,itemsep=3pt]
\item\label{lemma: family duality-i} (Weak duality with slack.) For every
$Q\in\calP_{\mathbb U}(P)$ and every $\xi=(\zeta,\eta,\lambda)\in\calF(\theta)$,
\[
\Jdual(\xi,\theta;\rho,P)-\E_{Q}[g(Z;\theta)]
\;\geq\;\eta\bigl(\delta-\TV(Q\,\|\,P)\bigr)
+\bigl(\lambda_{12}^{\top}\rho-\lambda^{\top}\E_{Q}[m(Z;\theta)]\bigr),
\]
and both terms on the right are nonnegative when
$Q\in\calP_{\mathbb U}(\theta;\rho,P,\delta)$. In particular,
$\Kmax(\theta;\rho,P)$ is at least the upper bound over
\eqref{eq:robustness family}, and it is finite whenever
\eqref{eq:robustness family} is nonempty.
\item\label{lemma: family duality-ii} (Strong duality.) If the constraint
qualification holds at $(\theta,\rho,P)$, then
\[
\Kmax(\theta;\rho,P)=\sup_{Q\in\calP_{\mathbb U}(\theta;\rho,P,\delta)}
\E_{Q}[g(Z;\theta)].
\]
\item\label{lemma: family duality-iii} (Existence of dual solutions.) The
recession cone of $\calF(\theta)$ is
$0^{+}\calF(\theta)=\{d=(\zeta_{d},\eta_{d},\lambda_{d}):\eta_{d}\geq0,\
\lambda_{d}\in\Rlam,\ \zeta_{d}+\eta_{d}/2\geq\sigma^{\infty}(\lambda_{d},\theta)\}$
if $\overline{u}=+\infty$ and $0^{+}\calF(\theta)=\R\times\R_{+}\times\Rlam$ if
$\overline{u}<\infty$, and the recession function of
$\Jdual(\cdot,\theta;\rho,P)$ on $\calF(\theta)$ is
\[
\Jdual^{\infty}(d,\theta;\rho,P)=\zeta_{d}+\eta_{d}\delta
+\lambda_{d,12}^{\top}\rho
+P\,\mathsf{r}\bigl(-\lambda_{d}^{\top}m(\cdot;\theta)-\zeta_{d},\eta_{d}\bigr),
\qquad d\in0^{+}\calF(\theta).
\]
If the constraint qualification holds at $(\theta,\rho,P)$, then
$\Jdual^{\infty}(d,\theta;\rho,P)>0$ for every nonzero
$d\in0^{+}\calF(\theta)$, and the set $\Xmax(\theta;\rho,P)$ of minimizers
of \eqref{eq:dual} is nonempty, compact, and convex.
\item\label{lemma: family duality-iv} (Equivalent form.) The constraint
qualification holds at $(\theta,\rho,P)$ if and only if there exists
$Q^{\circ}\in\calP_{\mathbb U}(P)$ with $\E_{Q^{\circ}}[m(Z;\theta)]\in
\calM(\rho)$ and $\TV(Q^{\circ}\,\|\,P)<\delta$, and
$\calM(\rho)\cap\interior(\Gamma_{\mathbb U}(\theta;P,1))\neq\emptyset$.
In particular, it implies $\mathsf{D}(\theta;\rho,P)<\delta$, that is,
$\theta\in\Theta_{\delta}(\rho,P)$.
\end{enumerate}
\end{lemma}
\begin{flushright}
[\hyperref[proof:lemma: family duality]{Proof}]
\end{flushright}

Two comments are in order. First, part~\ref{lemma: family duality-iii} is
the counterpart of Theorem~\ref{thm: existence of dual
solutions}. The latter cannot be invoked directly for every member because
the entropy $\phi_{\mathbb U}$ is infinite at zero when $\underline{u}>0$,
which violates Assumption~\ref{assump: dp-entropy}, whereas the direct
argument given in the proof only uses the slack inequality of
part~\ref{lemma: family duality-i} together with the room provided by the
constraint qualification. Second, part~\ref{lemma: family duality-iv}
connects the constraint qualification to the primitives.

The second lemma collects the properties of the minimum-divergence function
on which the theory relies: its dual representation, how it relates to
the feasibility of the criterion function, and its stability
under perturbations of $(\theta,\rho,P)$. The last property is what keeps
the population optimizers feasible when the inputs are replaced by their
sample analogs. To measure the baseline perturbation we use the class of recession
integrands
\begin{equation}\label{eq:recession class}
\calH^{\infty}_{\calA}:=\bigl\{z\mapsto\mathsf{r}\bigl(-\lambda^{\top}
m(z;\theta)-\zeta,\eta\bigr):\ \theta\in\Theta,\
\xi=(\zeta,\eta,\lambda)\in \calA,\
\text{and }\zeta+\tfrac{\eta}{2}\geq\sigma^{\infty}(\lambda,\theta)
\text{ when }\overline{u}=+\infty\bigr\},
\end{equation}
indexed by a compact set $\calA\subseteq\R\times\R_{+}\times\R^{\dimm}$,
in which
the multiplier $\lambda$ is not restricted to $\Rlam$. When $\overline{u}=+\infty$, the
restriction on the index places the argument of $\mathsf{r}$ in its domain at
every $z$. When $\overline{u}<\infty$, the function $\mathsf{r}$ is finite
everywhere by \eqref{eq:integrand family}, and no restriction is needed.
Lemma~\ref{lem:donsker-EB} below
shows that the class is $\Pbase{}$-Donsker.

\begin{lemma}[The minimum-divergence function]\label{lemma: mindiv dual}
Fix $\theta\in\Theta$ and $\delta\in(0,1]$, let $\calZ$ be compact and
$m(\cdot;\theta)\in C(\calZ;\R^{\dimm})$, and
let $\rho\in\R^{\dime}$ and $P\in\Delta(\calZ)$. Whenever the
moment-constrained set in~\eqref{eq:mindiv-primal} is nonempty, the infimum
defining $\mathsf D(\theta;\rho,P)$ is attained.
\begin{enumerate}[label=(\roman*),ref=\thelemma(\roman*),leftmargin=2em,
  topsep=2pt,itemsep=3pt]
  \item\label{lemma: mindiv dual-i} (Dual representation.) If there is
        $w_{0}\in\calM(\rho)$ with $\calB_{r}(w_{0})\subseteq
        \Gamma_{\mathbb U}(\theta;P,1)$ for some $r>0$, in particular if
        the constraint qualification holds at $(\theta,\rho,P)$ with radius
        $r$, then
        \begin{align}\label{eq:mindiv}
          \mathsf{D}(\theta;\rho,P)
            =\max_{(\zeta,\lambda)\in\calF_{\mathsf D}(\theta)}
            v_{P}(\zeta,\lambda),\qquad
          v_{P}(\zeta,\lambda):=-\E_{P}\bigl[\mathsf{r}\bigl(-\zeta
                -\lambda^{\top}m(Z;\theta),1\bigr)\bigr]-\zeta
                -\lambda_{12}^{\top}\rho.
        \end{align}
        Moreover, every maximizer
        $(\zeta^{*},\lambda^{*})$ satisfies $\|\lambda^{*}\|\leq1/r$ and
        $|\zeta^{*}|\leq\tfrac12+\sup_{z\in\calZ}\|m(z;\theta)\|/r$.
  \item\label{lemma: mindiv dual-ii} (Criterion dichotomy.) Let
        $g(\cdot;\theta)\in C(\calZ;\R)$. If $\mathsf{D}(\theta;\rho,P)\leq
        \delta$, then $\Kmax(\theta;\rho,P)\in\R$. If the hypothesis
        of~(i) holds and $\mathsf{D}(\theta;\rho,P)>\delta$, then
        $\Kmax(\theta;\rho,P)=-\infty$.
  \item\label{lemma: mindiv dual-iii} (Persistence and local Lipschitz
        continuity.) Suppose $m$ is jointly continuous on $\calZ\times\Theta$
        with $\Theta$ compact, and that the constraint qualification holds
        at $(\theta,\rho,P)$ with radius $r$. Then there exist
        $\varepsilon>0$, a constant $C<\infty$, and a compact set
        $\calA_{\mathsf D}\subseteq\R\times\R_{+}\times\R^{\dimm}$ such that the
        constraint qualification holds, with radius $r/3$, at every
        $(\theta',\rho',P')\in\Theta\times\R^{\dime}\times\Delta(\calZ)$
        with
        \[
        \|\theta'-\theta\|+\|\rho'-\rho\|
        +\|P'-P\|_{\calH^{\infty}_{\calA_{\mathsf D}}}<\varepsilon,
        \]
        and, for all such triples $(\theta_{1},\rho_{1},P_{1})$ and
        $(\theta_{2},\rho_{2},P_{2})$,
        \[
          \bigl|\mathsf D(\theta_{1};\rho_{1},P_{1})-\mathsf
          D(\theta_{2};\rho_{2},P_{2})\bigr|
          \;\leq\;
          C\Bigl(\|\rho_{1}-\rho_{2}\|
          +\sup_{z\in\calZ}\|m(z;\theta_{1})-m(z;\theta_{2})\|
          +\|P_{1}-P_{2}\|_{\calH^{\infty}_{\calA_{\mathsf D}}}\Bigr).
        \]
        The constant $C$ and the set $\calA_{\mathsf D}$ depend only on
        $r$, $\delta$, $\mathbb{U}$, and
        $\sup_{(z,\theta')\in\calZ\times\Theta}\|m(z;\theta')\|$, while
        $\varepsilon$ depends in addition on the modulus of continuity of
        $m$ on $\calZ\times\Theta$.
\end{enumerate}
\end{lemma}
\begin{flushright}
[\hyperref[proof:lemma: mindiv dual]{Proof}]
\end{flushright}

\begin{rem}[A specification statistic in probability units]
\label{rem: mindiv specification}
The minimum-divergence function of
\eqref{eq:mindiv-primal} might be a quantity of interest in itself. Indeed, it measures
the smallest probability mass that must be reallocated from the baseline,
within the family of distributions selected by $\mathbb{U}$, for the moment
restrictions to hold at $\theta$. Its infimum over $\theta$ is, under total variation,
the smallest divergence needed to reconcile the baseline with the model
(Section~\ref{subsec: setup}). This is the
total variation analogue of the optimally-transported generalized method of
moments criterion of
\citet{schennach2026OptimallytransportedGeneralizedMethod}, which reconciles
model and data through the least mean squared transport cost. 
\end{rem}

\subsection{Regularity Conditions}\label{subsec:ls-assumptions}

The following assumption collects the regularity conditions of the
large sample theory. It is the analog of Assumption~M in
\citet{christensen2023CounterfactualSensitivityRobustness}
(\textsf{CC23}, henceforth), modified to accommodate the non-smoothness of the
conjugate, the estimation of the baseline, and the family of robustness sets.

\begin{assump}[Estimation regularity]\label{assump: est-EB}
The following conditions hold.
\begin{enumerate}[label=(\roman*),ref=\theassump(\roman*),leftmargin=2.6em]
  \item\label{assump: reg-EB} {(Primitive regularity.)} $\calZ$ is
        compact; $\Theta$ is a compact subset of $\R^{\dimt}$; the maps
        $(z,\theta)\mapsto g(z;\theta)$ and $(z,\theta)\mapsto m(z;\theta)$
        are jointly continuous on $\calZ\times\Theta$ and Lipschitz in
        $\theta$ uniformly in $z$: there is $L<\infty$ with
        $\sup_{z\in\calZ}|g(z;\theta)-g(z;\theta')|\leq L\|\theta-\theta'\|$
        and $\sup_{z\in\calZ}\|m(z;\theta)-m(z;\theta')\|\leq
        L\|\theta-\theta'\|$ for all $\theta,\theta'\in\Theta$.
  \item\label{assump: slater-EB} {(Constraint qualification.)} The
        constraint qualification \eqref{eq:cq} holds at
        $(\theta,\rho_{0},\Pbase{})$ for each
        $\theta\in\Tmin(\rho_{0},\Pbase{})\cup\Tmax(\rho_{0},\Pbase{})$.
  \item\label{assump: feas-EB} {(Feasibility and strictly feasible
        optimizers.)}
        $\Tmin(\rho_{0},\Pbase{})\cup\Tmax(\rho_{0},\Pbase{})\subseteq
        \Theta_{\delta}(\rho_{0},\Pbase{})\ne\emptyset$.
  \item\label{assump: clt-EB} {(Asymptotic linearity.)}
        The estimator $\widehat\rho_{n}$ of $\rho_{0}$ is
        asymptotically linear,
        $\widehat\rho_{n}=\rho_{0}+\widehat P_{n}\psi+\op(n^{-1/2})$, with
        influence function $\psi:\calZ\to\R^{\dime}$ measurable and
        satisfying $\Pbase{}\psi=0$ and $\Pbase{}\|\psi\|^{2}<\infty$. The
        known block carries no sampling uncertainty.
  \item\label{assump: mult-EB} {(Dual optimizer lower hemicontinuity.)}
        The maps $\theta\mapsto\Xmin(\theta;\rho_{0},\Pbase{})$ and
        $\theta\mapsto\Xmax(\theta;\rho_{0},\Pbase{})$ are lower
        hemicontinuous at each $\theta\in\Tmin(\rho_{0},\Pbase{})$ and
        $\theta\in\Tmax(\rho_{0},\Pbase{})$, respectively. 
\end{enumerate}
\end{assump}

We now comment on each piece in turn.

\uline{\textbf{On~\ref{assump: reg-EB}.}} Compactness of $\calZ$ is the
substantive requirement, with no counterpart in \textsf{CC23}, who impose compactness
only on $\Theta$ (their Assumption~M(v)). It lets us control the support supremum
$\sigma(\lambda,\theta)$ in the pointwise constraint
through Berge's maximum theorem, and it bounds the dual integrand \eqref{eq:ell-EB}
uniformly over compact sets of multipliers. Proposition~\ref{prop: attrition inference}
relaxes it in the RCT with attrition application. Continuity of $g$ and $m$ is the analog
of \textsf{CC23}'s Assumption~M(i)--(ii), and the Lipschitz requirement makes the
integrand class Donsker (Lemma~\ref{lem:donsker-EB}).

\uline{\textbf{On~\ref{assump: slater-EB}.}} This condition is the analog of
\textsf{CC23}'s Condition~S$'$ (their Assumption~M(iii)), except that we impose it only
at the population optimizers rather than on all of $\Theta_{\delta}$. Its
validity at nearby triples $(\theta,\rho,P)$ is supplied by
Lemma~\ref{lemma: mindiv dual-iii}. By Lemma~\ref{lemma: family duality-iv} it
is a classical strict-feasibility condition, and by
Lemma~\ref{lemma: family duality-iii} it makes the dual objective coercive, so
that $\Xmin(\theta;\rho_{0},\Pbase{})$ and $\Xmax(\theta;\rho_{0},\Pbase{})$
are nonempty, compact, and convex. 

\uline{\textbf{On~\ref{assump: feas-EB}.}} The inclusion is implied by
\ref{assump: slater-EB} through Lemma~\ref{lemma: family duality-iv} and is
recorded for ease of reference. Nonemptiness of
$\Theta_{\delta}(\rho_{0},\Pbase{})$ ensures that the criterion is not
identically $-\infty$. Strictness of the budget inequality at the optimizers
keeps them feasible under small perturbations of $(\rho,P)$
(Lemma~\ref{lem:aux-EB-i}), on which the consistency and directional-differentiability
arguments rely. It is the analog of \textsf{CC23}'s
Assumption~M(vi).

\uline{\textbf{On~\ref{assump: clt-EB}.}} Under this condition,
$\sqrt n(\widehat\rho_{n}-\rho_{0})=\mathbb G_{n}\psi+\op(1)$, so that
$\mathbb Z:=\mathbb G\psi\sim\normal(0,\Pbase{}\psi\psi^{\top})$ and the process limit
$\mathbb G$ are jointly Gaussian with cross-covariance
$\E[\mathbb G f\,\mathbb Z]=\Pbase{}(f\psi)$. We impose the linear expansion rather
than marginal asymptotic normality because the delta method of
Theorem~\ref{thm:HD-EB} uses the joint law of
$\bigl(\sqrt n(\widehat\rho_{n}-\rho_{0}),\sqrt n(\widehat P_{n}
-\Pbase{})\bigr)$, which determines the asymptotic variance of the bounds
(Remark~\ref{rem:singleton-EB}) and the consistency of the bootstrap
(Theorem~\ref{thm:boot-EB}). It is the counterpart of the $\sqrt n$-Gaussian
hypothesis that \textsf{CC23}'s Theorem~6.2 places on the estimated vector
of targeted moments, the analog of $\widehat\rho_{n}$.

\uline{\textbf{On~\ref{assump: mult-EB}.}} This condition is the counterpart
of \textsf{CC23}'s Assumption~M(vii), and its strengthening for the bootstrap
(Assumption~\ref{assump: M-joint-EB}) is the counterpart of their
Assumption~M(vii$'$). It is imposed on $\Xmax$ rather than on $\Lmax$ because
the derivative depends on the entire dual optimizer $\xi$ through the
integrand $\ell_{\xi,\theta}$. We use it only to establish the directional
differentiability of the value function. It holds automatically when the dual
optimizer is unique on $\Tmin\cup\Tmax$, since the optimizer sets are
nonempty and locally bounded (Lemma~\ref{lem:aux-EB-iii}) and upper
hemicontinuous in $\theta$ (Lemma~\ref{lem:aux-EB-iv}).

\subsection{Auxiliary Results}\label{subsec:ls-auxiliary}

We first record the Donsker property of the integrand classes through which
the baseline enters the program. The objective $\Jdual$, the
minimum-divergence function $\mathsf D$ in its dual form \eqref{eq:mindiv},
and the recession function of Lemma~\ref{lemma: family duality-iii} each
depend on $P$ only through an expectation, respectively of the dual
integrands $\ell_{\xi,\theta}$ of \eqref{eq:ell-EB}, of the divergence
integrands $\mathsf{r}(-\zeta-\lambda^{\top}m(z;\theta),1)$, and of
the recession integrands \eqref{eq:recession class}. All are of the form
$z\mapsto\mathsf{r}(a(z;\theta,\zeta,\lambda),\eta)$, where
\[
a(z;\theta,\zeta,\lambda):=g(z;\theta)-\lambda^{\top}m(z;\theta)-\zeta,
\]
and $g\equiv0$ for the last two classes, so a single lemma covers all of
them.

\begin{lemma}[Donsker property]\label{lem:donsker-EB}
Let Assumption~\ref{assump: reg-EB} hold and let
$\calA\subseteq\R\times\R_{+}\times\R^{\dimm}$ be compact. Then each of the
classes
\begin{align*}
  \calH^{\Jdual}_{\calA}&:=\bigl\{\ell_{\xi,\theta}:\theta\in\Theta,\
    \xi\in \calA\cap\calF(\theta)\bigr\},\\
  \calH^{\mathsf D}_{\calA}&:=\bigl\{z\mapsto\mathsf{r}\bigl(-\zeta
  -\lambda^{\top}m(z;\theta),1\bigr):
    \theta\in\Theta,\ (\zeta,\lambda)\in\Pi_{\zeta\lambda}(\calA)\cap
    \calF_{\mathsf D}(\theta)\bigr\},
\end{align*}
with $\Pi_{\zeta\lambda}(\calA)$ the projection of $\calA$ onto its
$(\zeta,\lambda)$ coordinates, as well as the class of recession integrands
$\calH^{\infty}_{\calA}$ of \eqref{eq:recession class}, is $\Pbase{}$-Donsker
with bounded envelope, hence $\Pbase{}$-Glivenko--Cantelli.
\end{lemma}
\begin{flushright}
[\hyperref[proof:lem:donsker-EB]{Proof}]
\end{flushright}

Lemma~\ref{lemma: family duality-iii} gives, at each population optimizer
and at the population value $(\rho_{0},\Pbase{})$, a nonempty compact set of dual
minimizers, and Lemma~\ref{lemma: mindiv dual-iii} shows how the
constraint qualification and of the strict feasibility of the optimizers are maintained
locally. With an
estimated baseline, these conclusions must hold uniformly over joint perturbations of
$(\rho,P)$, which index the Donsker class used below. Properties
\hyperref[prop:linearity-EB]{(L)} and \hyperref[prop:independence-EB]{(I)} reduce this
requirement to measuring changes in $P$ in a seminorm under which $P\mapsto P\ell$ is
continuous on the relevant integrands $\ell$.

The estimated baseline enters the large sample analysis through the
empirical process $\mathbb G_{n}=\sqrt n(\widehat P_{n}-\Pbase{})$, which we
view as a random element of $\ell^{\infty}(\calH)$ for a class $\calH$ of
functions. We take
\[
\calH:=\calH^{\Jdual}_{\calA}\cup\calH^{\Jdual,-}_{\calA}
\cup\calH^{\mathsf D}_{\calA}\cup\calH^{\infty}_{\calA},
\]
where $\calA\subseteq\R\times\R_{+}\times\R^{\dimm}$ is a compact index set
chosen below, $\calH^{\Jdual}_{\calA}$, $\calH^{\mathsf D}_{\calA}$, and
$\calH^{\infty}_{\calA}$ are the classes of Lemma~\ref{lem:donsker-EB}, and
$\calH^{\Jdual,-}_{\calA}$ is the class $\calH^{\Jdual}_{\calA}$ formed with
$-g$ in place of $g$, the substitution that defines the lower bound objects.
Whatever the compact set $\calA$, the class $\calH$ is a finite union of the
$\Pbase{}$-Donsker classes of Lemma~\ref{lem:donsker-EB} and so itself
$\Pbase{}$-Donsker. For any class of functions $\calG$, with 
$\|P-P'\|_{\calG} = \sup_{f\in\calG}|Pf-P'f|$, define
\[
\vertiii{(\rho,P)-(\rho',P')}_{\calG}:=\|\rho-\rho'\|+\|P-P'\|_{\calG}
\]
on the pair $(\rho,P)$. We equip the pair $(\rho,P)$ with the seminorm
$\vertiii{\cdot}_{\calH}$, and throughout the rest of the section the
convergence $(\rho,P)\to(\rho_{0},\Pbase{})$ is understood as
$\vertiii{(\rho,P)-(\rho_{0},\Pbase{})}_{\calH}\to0$.

It remains to choose the index set $\calA$, which has to serve two purposes.
First, the integrand $\ell_{\xi,\theta}$ belongs to $\calH$ only if
$\xi\in\calA$, so $\calA$ must contain the dual optimizers of the upper- and
lower bound programs realized near the population value. Second, the
baseline enters the minimum-divergence function only through expectations of
recession integrands, and Lemma~\ref{lemma: mindiv dual-iii} states the
persistence of the constraint qualification and of the strict feasibility of
the optimizers in a norm over compact families of such integrands, so
$\calA$ must contain the indices of these families. Accordingly, we fix any compact set
\[
\calA\subseteq\R\times\R_{+}\times\R^{\dimm}
\qquad\text{with}\qquad
\calA\supseteq\Xi\cup\Xi^{-}\cup \calA_{0},
\]
where $\Xi$ and $\Xi^{-}$ are bounded sets collecting the dual optimizers
of the two programs and $\calA_{0}$ is a compact set of indices of recession
integrands. We define these sets in turn.

\uline{\textbf{The sets $\Xi$ and $\Xi^{-}$.}} Let
$\calH_{0}:=\calH^{\infty}_{\calA_{0}}$, with $\calA_{0}$ as defined below. For
each $\theta\in\Tmax(\rho_{0},\Pbase{})$ there is a radius
$\varepsilon_{\theta}>0$ such that, over all triples $(\theta',\rho',P')$ with
$\|\theta'-\theta\|+\vertiii{(\rho',P')-(\rho_{0},\Pbase{})}_{\calH_{0}}
<\varepsilon_{\theta}$, the sets $\Xmax(\theta';\rho',P')$ are nonempty and
compact and their union is bounded, as we verify in
Section~\ref{subsec:ls-construction}. The balls
$\calB_{\varepsilon_{\theta}}(\theta)$ cover the compact set
$\Tmax(\rho_{0},\Pbase{})$, so we can extract a finite subcover, with centers
$\theta_{k}$ and radii $\varepsilon_{k}:=\varepsilon_{\theta_{k}}$, collect
the associated sets of triples into the set
\[
\calN_{0}:=\bigcup_{k}\bigl\{(\theta',\rho',P')
\in\Theta\times\R^{\dime}\times\Delta(\calZ):
\ \|\theta'-\theta_{k}\|+\vertiii{(\rho',P')-(\rho_{0},\Pbase{})}_{\calH_{0}}
<\varepsilon_{k}\bigr\},
\]
and let
\[
  \Xi \;:=\; \bigcup_{(\theta',\rho',P')\in\calN_{0}}\Xmax(\theta';\rho',P')
  \ \subseteq\ \R\times\R_{+}\times\R^{\dimm},
\]
which is bounded as a finite union of bounded sets. Repeating the
construction with $-g$ in place of $g$ yields a bounded set $\Xi^{-}$
containing the optimizers of the lower  program over an analogous
cover of $\Tmin(\rho_{0},\Pbase{})$.

\uline{\textbf{The set $\calA_{0}$.}} This set fixes the reference points at
which the minimum-divergence function is controlled. By
Lemma~\ref{lemma: mindiv dual-iii}, each
$\theta\in\Tmax(\rho_{0},\Pbase{})$ admits a radius
$\varepsilon^{\mathsf D}_{\theta}>0$, a constant $C_{\theta}<\infty$, and a
compact index set $\calA_{\mathsf D,\theta}$ such that the constraint
qualification, with radius $r_{\theta}/3$, and the Lipschitz estimate of
that lemma are available at every $(\theta',\rho',P')$ with
$\|\theta'-\theta\|+\|\rho'-\rho_{0}\|
+\|P'-\Pbase{}\|_{\calH^{\infty}_{\calA_{\mathsf D,\theta}}}
<\varepsilon^{\mathsf D}_{\theta}$, where $r_{\theta}>0$ is a radius
with which the constraint qualification of Assumption~\ref{assump: slater-EB} holds at
$(\theta,\rho_{0},\Pbase{})$, fixed once and for all. Since
$\Tmax(\rho_{0},\Pbase{})$ is compact, finitely many of the balls
$\calB_{\varepsilon^{\mathsf D}_{\theta}/2}(\theta)$, with centers
$\theta_{1},\dots,\theta_{J}$, cover it, and we fix such a subcover once and
for all. Let
\[
\calA_{0}:=\bigcup_{j=1}^{J}\calA_{\mathsf D,\theta_{j}},
\]
which is compact as a finite union of compact subsets of
$\R\times\R_{+}\times\R^{\dimm}$. Since $\calH^{\infty}_{\calA}$ grows with
its index set $\calA$, the single class $\calH_{0}$ contains
$\calH^{\infty}_{\calA_{\mathsf D,\theta_{j}}}$ for every $j\in[J]$, and thus
controls the perturbation of the baseline at every center of the subcover.
For the same reason $\calA_{0}\subseteq\calA$ gives $\calH_{0}\subseteq\calH$,
so that $\vertiii{\cdot}_{\calH_{0}}\leq\vertiii{\cdot}_{\calH}$ and every
property established on a $\vertiii{\cdot}_{\calH_{0}}$-neighborhood of
$(\rho_{0},\Pbase{})$ holds on the $\vertiii{\cdot}_{\calH}$-neighborhood of
the same radius.

The next lemma collects the regularity properties of the dual program and
of the bound on which the consistency and directional-differentiability
arguments rest. The large sample analysis perturbs the pair $(\rho,P)$
around its population value and propagates the perturbation through three
nested objects: the feasible set $\Theta_{\delta}(\rho,P)$ of the outer
program, the dual optimizers $\Xmax(\theta;\rho,P)$, and the value $\bmax(\delta,\rho,P)$
of the bound. For the directional functional delta method we need each of these objects
to behave continuously in $(\rho,P)$ near $(\rho_{0},\Pbase{})$. Part~(i)
guarantees that no optimizer is lost from the feasible set under small
perturbations. Parts~(ii)--(iv) show that the
optimizer set is nonempty, compact, and locally uniformly bounded, varies
upper hemicontinuously in $(\rho,P)$, and delivers an inner value
continuous in $\theta$. Part~(v) proves continuity of the bound itself,
which is what consistency requires. Part~(vi) shows that the
near-optimizers of every perturbed program must belong 
to any prescribed neighborhood of the population argmax, which is 
what lets the directional-differentiability and bootstrap arguments 
focus on compact neighborhoods.

\begin{lemma}[Regularity of the empirical-baseline program]\label{lem:aux-EB}
Let Assumptions~\ref{assump: reg-EB}--\ref{assump: feas-EB} hold. Then:
\begin{enumerate}[label=(\roman*),ref=\thelemma(\roman*),leftmargin=2em,topsep=2pt,%
itemsep=3pt]
  \item\label{lem:aux-EB-i} (Feasibility persistence at the optimizers.)
        There is $r>0$ such that
        $\Tmax(\rho_{0},\Pbase{})\subseteq\Theta_{\delta}(\rho,P)$
        for every $(\rho,P)\in\calB_{r}(\rho_{0},\Pbase{})$, where the
        ball is taken with respect to $\vertiii{\cdot}_{\calH}$. In particular, for every
        sequence $(\rho_{n},P_{n})\to(\rho_{0},\Pbase{})$,
        $\Tmax(\rho_{0},\Pbase{})\subseteq\Theta_{\delta}(\rho_{n},P_{n})$ for all
        $n$ large.
  \item\label{lem:aux-EB-iii} (Nonemptiness, compactness, and local
        boundedness.) For any $\theta\in\Tmax(\rho_{0},\Pbase{})$ there is
        $\varepsilon>0$ such that $\Xmax(\theta';\rho',P')$ is nonempty and
        compact, and their union is bounded, over all $(\theta',\rho',P')$
        in the ball $\calB_{\varepsilon}(\theta,\rho_{0},\Pbase{})$
        of points with $\|\theta'-\theta\|+\vertiii{(\rho',P')
        -(\rho_{0},\Pbase{})}_{\calH}<\varepsilon$.
  \item\label{lem:aux-EB-iv} (Joint upper hemicontinuity.) For any
        $\theta\in\Tmax(\rho_{0},\Pbase{})$, the correspondence
        $(\theta',\rho,P)\mapsto\Xmax(\theta';\rho,P)$, and hence its
        $\lambda_{12}$-projection $\Lmax(\theta';\rho,P)$, is upper
        hemicontinuous at $(\theta,\rho_{0},\Pbase{})$.
  \item\label{lem:aux-EB-v} (Continuity of the inner value.) If, in
        addition, Assumption~\ref{assump: mult-EB} holds, then for each
        $h\in\R^{\dime}$ and each $\gamma\in UC(\calH,e_{\Pbase{}})$ the map
        $\theta\mapsto v(\theta;h,\gamma):=
        \min_{\xi\in\Xmax(\theta;\rho_{0},\Pbase{})}
        \{\lambda_{12}^{\top}h+\gamma\ell_{\xi,\theta}\}$, well defined
        for $\theta\in\calN\cap\Theta$ with $\calN$ as in part~\textup{(v)}, where
        $\Xmax(\theta;\rho_{0},\Pbase{})$ is nonempty, compact, and contained in
        $\Xi\subseteq\calA$, is continuous
        at every $\theta\in\Tmax(\rho_{0},\Pbase{})$.
  \item\label{lem:aux-EB-vi} (Continuity of the bound.) The map
        $(\rho,P)\mapsto\bmax(\delta,\rho,P)$ is continuous at $(\rho_{0},\Pbase{})$
        along every sequence $(\rho_{n},P_{n})\to(\rho_{0},\Pbase{})$ such that
        $\|P_{n}-\Pbase{}\|_{\calH^{\Jdual}_{K}}\to0$ for every compact
        $K\subseteq\R\times\R_{+}\times\R^{\dimm}$. Moreover, there
        are a compact neighborhood $\calN$ of $\Tmax(\rho_{0},\Pbase{})$,
        $r>0$, and $M<\infty$ such that, for all
        $\theta\in\calN\cap\Theta$ and all
        $(\rho,P)$ and $(\rho',P')$ in the $\vertiii{\cdot}_{\calH}$-ball
        $\calB_{r}(\rho_{0},\Pbase{})$, the set $\Xmax(\theta;\rho,P)$ is
        nonempty, compact, and contained in $\Xi\subseteq\calA$, every
        $\lambda_{12}\in\Lmax(\theta;\rho,P)$ satisfies $\|\lambda_{12}\|\leq M$,
        and
        \[|\Kmax(\theta;\rho,P)-\Kmax(\theta;\rho',P')|\leq
        M\,\vertiii{(\rho,P)-(\rho',P')}_{\calH}.\]
  \item\label{lem:aux-EB-vii} (Localization of near-optimizers.)
        Let $(\rho_{n},P_{n})\to(\rho_{0},\Pbase{})$ be such that
        $\|P_{n}-\Pbase{}\|_{\calH^{\Jdual}_{K}}\to0$ for every compact
        $K\subseteq\R\times\R_{+}\times\R^{\dimm}$, and let $\nu_{n}\geq0$ with
        $\nu_{n}\to0$. Then, for every open set
        $\calU\supseteq\Tmax(\rho_{0},\Pbase{})$, there is $n_{0}$ such that,
        for all $n\geq n_{0}$, every $\theta\in\Theta_{\delta}(\rho_{n},P_{n})$
        with $\Kmax(\theta;\rho_{n},P_{n})\geq\bmax(\delta,\rho_{n},P_{n})-\nu_{n}$
        belongs to $\calU$.
\end{enumerate}
The same conclusions hold for the lower bound, with
$\Kmin,\bmin,\Tmin,\Xmin,\Lmin$ in place of
$\Kmax,\bmax,\Tmax,\Xmax,\Lmax$.
\end{lemma}
\begin{flushright}
[\hyperref[proof:lem:aux-EB]{Proof}]
\end{flushright}

\subsection{Consistency, Directional Differentiability, and Distributional Approximation}
\label{subsec:ls-hd}

We now establish the consistency of $\bminhat(\delta)$ and $\bmaxhat(\delta)$ via 
the joint continuity of $(\rho,P)\mapsto\bmax(\delta,\rho,P)$ at $(\rho_{0},\Pbase{})$
(Lemma~\ref{lem:aux-EB-vi}), together with the convergence of the
empirical baseline uniformly over the integrand classes that the bound
sees, which is what the Glivenko--Cantelli half of
Lemma~\ref{lem:donsker-EB} provides.

\begin{theorem}[Consistency]\label{thm:cons-EB}
Under Assumptions~\ref{assump: reg-EB}--\ref{assump: clt-EB},
\[
  \bminhat(\delta) = \bmin(\delta,\widehat\rho_{n},\widehat
  P_{n})\cprob\bmin(\delta,\rho_{0},\Pbase{})\qquad\text{and}\qquad
  \bmaxhat(\delta) = \bmax(\delta,\widehat\rho_{n},\widehat P_{n})
  \cprob\bmax(\delta,\rho_{0},\Pbase{}).
\]
\end{theorem}
\begin{flushright}
[\hyperref[proof:thm:cons-EB]{Proof}]
\end{flushright}

We next provide an asymptotically valid distributional approximation
for the plug-in estimators, which
rests on the Hadamard directional differentiability of the bounds at the
population value. Recall that a map $f:\mathbb D\to\R$ on a normed space
$\mathbb D$ is Hadamard directionally differentiable at $x_{0}$
tangentially to a set $\mathbb D_{0}\subseteq\mathbb D$ if there is a 
continuous map
$df_{x_{0}}:\mathbb D_{0}\to\R$ with
$\{f(x_{0}+t_{n}h_{n})-f(x_{0})\}/t_{n}\to df_{x_{0}}[h]$ for every
$h\in\mathbb D_{0}$ and all sequences $t_{n}\downarrow0$ and $h_{n}\to h$
in $\mathbb D$. For $f$ to be Hadamard differentiable, $df_{x_{0}}$ would
in addition have to be linear. See also
\cite{shapiro1990ConceptsDirectionalDifferentiability}.
The reason why the bounds are in general directionally, rather than fully,
differentiable lies in their nested variational form. For fixed $\theta$ and
$\xi$, the dual objective
$\zeta+\eta\delta+\lambda_{12}^{\top}\rho+P\ell_{\xi,\theta}$ is affine in
$(\rho,P)$, so that $\Kmax(\theta;\rho,P)$ is a pointwise infimum of affine
maps of $(\rho,P)$, and $\bmax(\delta,\rho,P)$ is a supremum over $\theta$ of
such infima. Under the regularity of Lemma~\ref{lem:aux-EB}, value functions
of this kind are directionally differentiable, with derivative obtained by
differentiating the criterion at the optimizers only, in the spirit of 
Danskin's theorem. At a direction
$(h,\gamma)$, the derivative of the inner infimum is the minimum over
$\xi^{*}\in\Xmax(\theta;\rho_{0},\Pbase{})$ of the linear map
$(h,\gamma)\mapsto(\lambda_{12}^{*})^{\top}h+\gamma\ell_{\xi^{*},\theta}$, and
the derivative of the outer supremum is the maximum of these inner
derivatives over $\theta^{*}\in\Tmax(\rho_{0},\Pbase{})$. A minimum (maximum) 
of linear maps over a set is positively homogeneous and continuous,
and it is linear whenever the set is a singleton. It fails in general to be
additive otherwise, because the optimizer selected by the extremum may vary
with the direction. Hence, when the outer or the dual optimizer set at
the population value contains more than one element, the derivative is in
general nonlinear. Neither
form of nonuniqueness is exceptional in our setting. The outer program is a
nonconvex program in $\theta$ whose supremum may be attained at several
points, and the dual program, although convex in $\xi$, need not be strictly
convex, so that its minimizers may form a nontrivial compact set
(Lemma~\ref{lem:aux-EB-iii}).

To derive the asymptotic distribution of $\bmaxhat(\delta)$, it is useful to
consider the bound as a functional on the normed space
$\mathbb D:=\R^{\dime}\times\ell^{\infty}(\calH)$. 
Two localizations make this extension
well behaved. First, because an element $\mathsf b$ evaluates only
integrands with index in $\calA$, the extended criterion restricts the dual
program to $\calF(\theta)\cap \calA$, and the extended minimum-divergence
function is defined through its dual form \eqref{eq:mindiv} restricted to
the indices of $\calH^{\mathsf D}_{\calA}$. Second, the outer supremum is
restricted to a fixed compact neighborhood of the population argmax
within $\Theta$. The union of the finitely many $\theta$-balls extracted in
the construction of $\Xi$ in Section~\ref{subsec:ls-auxiliary} is an open set that
contains the compact set $\Tmax(\rho_{0},\Pbase{})$, so we can fix
$\varepsilon_{\Theta}>0$ such that the compact set
\[
\calN_{\Theta}:=\bigl\{\theta\in\Theta:\ d\bigl(\theta,\Tmax(\rho_{0},\Pbase{})\bigr)
\leq\varepsilon_{\Theta}\bigr\}
\]
is contained in that union. We take the neighborhood relative to $\Theta$, rather
than requiring $\Tmax(\rho_{0},\Pbase{})$ to lie in the interior of
$\calN_{\Theta}$, so as to allow population optimizers on the boundary of
$\Theta$. By compactness of
$\calN_{\Theta}$, the continuous map
$\theta\mapsto\max_{k}\{\varepsilon_{k}-\|\theta-\theta_{k}\|\}$, with
$\theta_{k}$ and $\varepsilon_{k}$ the centers and radii of that finite
subcover, is bounded away from zero on $\calN_{\Theta}$, so there is
$\varepsilon_{*}>0$, which we shrink further so that the persistence
neighborhoods of Lemma~\ref{lemma: mindiv dual-iii} at the centers
$\theta_{1},\dots,\theta_{J}$ also apply. Therefore, whenever
$\theta\in\calN_{\Theta}$ and
$\vertiii{(\rho,P)-(\rho_{0},\Pbase{})}_{\calH_{0}}<\varepsilon_{*}$, the
optimizer set $\Xmax(\theta;\rho,P)$ is nonempty and contained in
$\Xi\subseteq \calA$, and every maximizer of the dual \eqref{eq:mindiv} at
$(\theta,\rho,P)$ lies in $\Pi_{\zeta\lambda}(\calA_{0})$, by the bounds of
Lemma~\ref{lemma: mindiv dual-i} at the persisting radius $r_{\theta_{j}}/3$ and
the construction of $\calA_{0}$. Concretely, we set
\begin{align*}
  \Kmax(\theta;\rho,\mathsf b)&:=\inf_{\xi\in\calF(\theta)\cap \calA}
   \{\zeta+\eta\delta+\lambda_{12}^{\top}\rho+\mathsf b(\ell_{\xi,\theta})\},\\
  \mathsf D(\theta;\rho,\mathsf b)&:=\sup\bigl\{-\mathsf
  b\bigl(\mathsf{r}(-\zeta-\lambda^{\top}m(\cdot;\theta),1)\bigr)
  -\zeta-\lambda_{12}^{\top}\rho:
  (\zeta,\lambda)\in\Pi_{\zeta\lambda}(\calA)\cap\calF_{\mathsf
  D}(\theta)\bigr\},\\
  \bmax(\delta,\rho,\mathsf b)&:=\sup\,\bigl\{\Kmax(\theta;\rho,\mathsf
  b):\theta\in\calN_{\Theta},\ \mathsf D(\theta;\rho,\mathsf b)<\delta\bigr\},
\end{align*}
with the analogous localization around $\Tmin(\rho_{0},\Pbase{})$ for the
lower bound. Note that for $\theta\in\calN_{\Theta}$ and
$(\rho,P)$ as above, restricting the infimum to $\calF(\theta)\cap \calA$ does
not change its value and, similarly, restricting the
dual of $\mathsf D$ to $\Pi_{\zeta\lambda}(\calA)$ does not change its value
either, so, at $\mathsf b=(P\ell)_{\ell\in\calH}$,
$\Kmax(\theta;\rho,\mathsf b)=\Kmax(\theta;\rho,P)$ and
$\mathsf D(\theta;\rho,\mathsf b)=\mathsf D(\theta;\rho,P)$ for every
$\theta\in\calN_{\Theta}$. In particular, writing
$\mathsf b_{0}:=(\Pbase{}\ell)_{\ell\in\calH}$ for the element of
$\ell^{\infty}(\calH)$ induced by the population baseline $\Pbase{}$, we
obtain these two identities at $(\rho_{0},\mathsf b_{0})$ for every
$\theta\in\calN_{\Theta}$, and hence, since
$\Tmax(\rho_{0},\Pbase{})\subseteq\calN_{\Theta}$,
$\bmax(\delta,\rho_{0},\mathsf b_{0})=\bmax(\delta,\rho_{0},\Pbase{})$ and
$\Xmax(\theta;\rho_{0},\mathsf b_{0})=\Xmax(\theta;\rho_{0},\Pbase{})$ for
every $\theta\in\calN_{\Theta}$. The main result of this subsection is the
following.

\begin{theorem}[Distributional approximation]
\label{thm:HD-EB}
Under Assumption~\ref{assump: est-EB}, the map $(\rho,P)\mapsto
(\bmin(\delta,\rho,P),\bmax(\delta,\rho,P))$ is Hadamard
directionally differentiable at $(\rho_{0},\Pbase{})$ tangentially to
$\R^{\dime}\times\allowbreak UC(\calH,e_{\Pbase{}})$. At a direction
$(h,\gamma)\in\R^{\dime}\times UC(\calH,e_{\Pbase{}})$, the upper and lower bound
derivatives are
\begin{equation}\label{eq:HD-EB-formula}
\begin{aligned}
  d\bmaxAt{(\rho_{0},\Pbase{})}[h,\gamma]
   &=\max_{\theta^{*}\in\Tmax(\rho_{0},\Pbase{})}\,
    \min_{\xi^{*}\in\Xmax(\theta^{*};\rho_{0},\Pbase{})}
    \bigl\{(\lambda_{12}^{*})^{\top}h+\gamma\,\ell_{\xi^{*},\theta^{*}}\bigr\},\\
  d\bminAt{(\rho_{0},\Pbase{})}[h,\gamma]
   &=\min_{\theta^{*}\in\Tmin(\rho_{0},\Pbase{})}\,
    \max_{\xi^{*}\in\Xmin(\theta^{*};\rho_{0},\Pbase{})}
    \bigl\{-(\lambda_{12}^{*})^{\top}h-\gamma\,\ell^{-}_{\xi^{*},\theta^{*}}\bigr\},
\end{aligned}
\end{equation}
where $\ell^{-}_{\xi,\theta}(z):=\mathsf{r}(-g(z;\theta)-\lambda^{\top}
m(z;\theta)-\zeta,\eta)$ is the dual integrand \eqref{eq:ell-EB} of the
program obtained by the substitution $g\mapsto-g$. With
$(\mathbb Z,\mathbb G)$ the joint Gaussian limit of $\sqrt
n\bigl((\widehat\rho_{n}-\rho_{0}),\,(\widehat P_{n}-\Pbase{})\bigr)$ in
$\R^{\dime}\times\ell^{\infty}(\calH)$, so that
$\mathbb G$ is the $\Pbase{}$-Brownian bridge indexed by $\calH$ and
$\mathbb Z=\mathbb G\psi$, we have
\[
  \sqrt n\!\begin{pmatrix}\bminhat(\delta)-\bmin(\delta,\rho_{0},\Pbase{})\\
   \bmaxhat(\delta)-\bmax(\delta,\rho_{0},\Pbase{})\end{pmatrix}
   \;\rightsquigarrow\;
   \begin{pmatrix}d\bminAt{(\rho_{0},\Pbase{})}[\mathbb Z,\mathbb G]\\
   d\bmaxAt{(\rho_{0},\Pbase{})}[\mathbb Z,\mathbb G]\end{pmatrix}.
\]
\end{theorem}
\begin{flushright}
[\hyperref[proof:thm:HD-EB]{Proof}]
\end{flushright}

The limiting law carries the bridge term $\mathbb G\ell_{\xi^{*},
\theta^{*}}$ alongside the finite-dimensional term
$(\lambda_{12}^{*})^{\top}\mathbb Z$, and the two are correlated, as both
are functionals of the same sample. When the
sets of outer maximizers and dual minimizers are singletons the derivative
is linear and the limit is Gaussian. We return to this case in
Remark~\ref{rem:singleton-EB}.

\subsection{Bootstrap}\label{subsec:ls-bootstrap}

Because the derivative \eqref{eq:HD-EB-formula} is generally nonlinear in
$(h,\gamma)$, the naive bootstrap is not consistent 
\citep[see Theorem~3.1
in][]{fang2019InferenceDirectionallyDifferentiable}, 
and we thus specialize the general resampling procedure 
of \citet{fang2019InferenceDirectionallyDifferentiable} to our setting. 
In particular, we draw $Z_{1}^{*},\ldots,Z_{n}^{*}$ i.i.d.\ from $\widehat P_{n}$,
yielding $\widehat P_{n}^{*}=n^{-1}\sum_{i}\delta^{\mathtt{Dirac}}_{Z_{i}^{*}}$, and
recompute the estimator on the resample, which gives
$\widehat\rho_{n}^{*}$. In the leading case in which $\widehat\rho_{n}$ is the sample
mean of a function of the data, the influence function $\psi$ of
Assumption~\ref{assump: clt-EB} is that function minus its mean $\rho_{0}$ under
$\Pbase{}$, so that $\widehat\rho_{n}=\widehat P_{n}(\rho_{0}+\psi)$ and
$\widehat\rho_{n}^{*}=\widehat P_{n}^{*}(\rho_{0}+\psi)$. Hence, 
$\sqrt n(\widehat\rho_{n}^{*}-\widehat\rho_{n})=\mathbb G_{n}^{*}\psi$, where
$\mathbb G_{n}^{*}:=\sqrt n(\widehat P_{n}^{*}-\widehat P_{n})$ is the bootstrapped
empirical process. For a general asymptotically linear $\widehat\rho_{n}$, we instead
require the bootstrap linearization
$\sqrt n(\widehat\rho_{n}^{*}-\widehat\rho_{n})=\mathbb G_{n}^{*}\psi+\op(1)$, which is
the hypothesis imposed by Theorem~\ref{thm:boot-EB} below.
The estimated derivative replaces
$\Tmax(\rho_{0},\Pbase{})$ and $\Xmax(\theta;\rho_{0},\Pbase{})$ by their
sample analogs
\[
  \widehat{d\bmax}_{n}[h,\gamma]
   :=\sup_{\theta\in\Tmaxhat}\,\inf_{\xi\in\Xmax(\theta;\widehat\rho_{n},\widehat
   P_{n})}
    \bigl\{\lambda_{12}^{\top}h+\gamma\,\ell_{\xi,\theta}\bigr\},
\]
with the estimated argmax set
\[
  \Tmaxhat:=\bigl\{\theta\in\Theta_{\delta}(\widehat\rho_{n},\widehat P_{n}):
   \Kmax(\theta;\widehat\rho_{n},\widehat P_{n})\geq\bmaxhat(\delta)-\widehat\nu_{n}
   \sqrt{\tfrac{\log n}{n}}\bigr\},\qquad\widehat\nu_{n}\geq0,\ \
   \widehat\nu_{n}=\Op(1),\ \ \widehat\nu_{n}^{-1}=\Op(1),
\]
where $\widehat\nu_{n}$ is a tuning sequence whose choice is discussed in
Remark~\ref{rem:nu-choice} below, and
$\Lmaxhat(\theta):=\Pi_{12}(\Xmax(\theta;\widehat\rho_{n},\widehat P_{n}))$.
The resulting resampling procedure and the confidence sets it
delivers are summarized in Algorithm~\ref{alg:FS-EB}, which reproduces
Algorithm~1 of the main text in the notation of this
section,
and whose
validity we establish in Theorem~\ref{thm:boot-EB} below.
\begin{algorithm}[t]
\caption{Confidence sets for the sensitivity bounds}
\label{alg:FS-EB}
\begin{algorithmic}[1]
\State Compute $\widehat\rho_{n}$, $\widehat P_{n}$, the bounds
$\bminhat(\delta),\bmaxhat(\delta)$, the near-argmax sets
$\Tminhat,\Tmaxhat$, and the dual minimizer sets
$\Xmax(\theta;\widehat\rho_{n},\widehat P_{n})$ and
$\Xmin(\theta;\widehat\rho_{n},\widehat P_{n})$.
\For{$b=1,\ldots,B$}
\State Draw an i.i.d.\ sample $\{Z_{i}^{*,b}\}_{i=1}^{n}$ from
$\widehat P_{n}$ and form $\widehat P_{n}^{*,b}$ and
$\widehat\rho_{n}^{*,b}$.
\State Set
$\widehat T_{n}^{*,b,\mathrm{up}}
:=\widehat{d\bmax}_{n}\big[\sqrt{n}(\widehat\rho_{n}^{*,b}
-\widehat\rho_{n}),\,
\sqrt{n}(\widehat P_{n}^{*,b}-\widehat P_{n})\big]$ and the analog
$\widehat T_{n}^{*,b,\mathrm{low}}$.
\EndFor
\State For $u\in(0,1)$, let
$\widehat{\mathfrak{c}}_{u}^{\mathrm{low}}$ and
$\widehat{\mathfrak{c}}_{u}^{\mathrm{up}}$ be the empirical $u$-quantiles
of $\{\widehat T_{n}^{*,b,\mathrm{low}}\}_{b=1}^{B}$ and
$\{\widehat T_{n}^{*,b,\mathrm{up}}\}_{b=1}^{B}$, respectively. Valid
lower and upper $100(1-\alpha)\%$ confidence sets are
\[
[\bminhat(\delta)-\widehat{\mathfrak{c}}_{1-\alpha}^{\mathrm{low}}/\sqrt{n},
+\infty)
\quad\text{and}\quad
(-\infty,\bmaxhat(\delta)-\widehat{\mathfrak{c}}_{\alpha}^{\mathrm{up}}/\sqrt{n}].
\]
\end{algorithmic}
\end{algorithm}
\begin{rem}[Choice of the tuning sequence $\widehat\nu_{n}$]\label{rem:nu-choice}
The sequence $\widehat\nu_{n}$ governs the slack used to form the
estimated argmax set $\Tmaxhat$, and such slack, namely $\widehat\nu_{n}\sqrt{\log n/n}$,
must vanish slowly enough that $\Tmaxhat$ contains the population set
$\Tmax(\rho_{0},\Pbase{})$ with probability approaching one, yet fast
enough that $\Tmaxhat$ excludes strictly suboptimal $\theta$ in the limit.
Both requirements are made precise in the proof of
Theorem~\ref{thm:boot-EB}. For concrete guidance on this choice we refer
the reader to \citet[Section~6.2]{christensen2023CounterfactualSensitivityRobustness},
who discuss the selection of the analogous tuning sequence in their
setting. Following their implementation, we set
$\widehat\nu_{n}=\min_{1\leq j\leq\dimm}(\widehat\Sigma_{m})_{jj}$, the
smallest diagonal entry of the covariance matrix of the moment function under the
empirical baseline
\[
\widehat\Sigma_{m}
  := \widehat P_{n}\bigl[\bigl(m(\cdot;\widehat\theta_{n})-\widehat P_{n}
  m(\cdot;\widehat\theta_{n})\bigr)\bigl(m(\cdot;\widehat\theta_{n})
  -\widehat P_{n}m(\cdot;\widehat\theta_{n})\bigr)^{\top}\bigr]
\]
evaluated at an approximate maximizer
$\widehat\theta_{n}\in\Theta_{\delta}(\widehat\rho_{n},\widehat P_{n})$ of the estimated
outer problem, that is, one with
$\Kmax(\widehat\theta_{n};\widehat\rho_{n},\widehat P_{n})\geq\bmaxhat(\delta)-1/n$.
This choice meets the requirement stated with $\Tmaxhat$. It is bounded, as $m$ is
bounded on the compact set $\calZ\times\Theta$ under Assumption~\ref{assump: reg-EB}.
Moreover, Lemma~\ref{lem:aux-EB-vii}, applied along almost surely convergent
subsequences as in the proof of Theorem~\ref{thm:boot-EB}, gives
$d(\widehat\theta_{n},\Tmax(\rho_{0},\Pbase{}))\cprob0$. By a uniform law of large
numbers over the Lipschitz class $\{m(\cdot;\theta):\theta\in\Theta\}$ and the
compactness of $\Tmax(\rho_{0},\Pbase{})$, $\widehat\nu_{n}$ is then bounded away from
zero in probability, provided that no component of $m(Z;\theta)$ is
$\Pbase{}$-almost surely constant at any $\theta\in\Tmax(\rho_{0},\Pbase{})$. A
constant component, such as a normalization, must therefore be excluded from the
minimum.
\end{rem}

Bootstrap consistency requires a joint strengthening of the pointwise
lower hemicontinuity of Assumption~\ref{assump: mult-EB}, needed for the
convergence of the estimated derivative at every direction of the tangent set.
Like that assumption, it is imposed on $\Xmax$ rather than on $\Lmax$, for
the reason given in Section~\ref{subsec:ls-assumptions}.

\begin{assump}[Joint lower hemicontinuity]\label{assump: M-joint-EB}
The correspondences
$(\theta,\rho,P)\mapsto\Xmin(\theta;\rho,P)$ and
$(\theta,\rho,P)\mapsto\Xmax(\theta;\rho,P)$ are lower hemicontinuous
jointly at every $(\theta,\rho_{0},\Pbase{})$ with
$\theta\in\Tmin(\rho_{0},\Pbase{})$ (resp.\
$\theta\in\Tmax(\rho_{0},\Pbase{})$). 
\end{assump}

Theorem~\ref{thm:boot-EB} establishes the validity of
Algorithm~\ref{alg:FS-EB}.
\begin{theorem}[Bootstrap consistency]\label{thm:boot-EB}
Suppose Assumptions~\ref{assump: est-EB} and~\ref{assump: M-joint-EB} hold,
and that the bootstrap estimator satisfies the linearization
$\sqrt n(\widehat\rho_{n}^{*}-\widehat\rho_{n})
=\mathbb G_{n}^{*}\psi+\op(1)$, with $\mathbb G_{n}^{*}:=\sqrt n(\widehat
P_{n}^{*}-\widehat P_{n})$. Then, conditionally on the data,
\[
  \widehat{d\bmax}_{n}\!\bigl[\sqrt n(\widehat\rho_{n}^{*}-\widehat\rho_{n}),\,
   \sqrt n(\widehat P_{n}^{*}-\widehat P_{n})\bigr]
   \;\rightsquigarrow\; d\bmaxAt{(\rho_{0},\Pbase{})}[\mathbb Z,\mathbb G].
\]
If, in addition, the distribution function of
$d\bmaxAt{(\rho_{0},\Pbase{})}[\mathbb Z,\mathbb G]$ is continuous and strictly
increasing at its $\alpha$- and $(\alpha/2)$-quantiles and the
distribution function of
$d\bminAt{(\rho_{0},\Pbase{})}[\mathbb Z,\mathbb G]$ is continuous and strictly increasing
at its $(1-\alpha)$- and $(1-\alpha/2)$-quantiles, then the one-sided confidence sets of
Algorithm~\ref{alg:FS-EB}, as well as the two-sided set
$[\bminhat(\delta)-\widehat{\mathfrak{c}}_{1-\alpha/2}^{\mathrm{low}}/\sqrt n,
\bmaxhat(\delta)-\widehat{\mathfrak{c}}_{\alpha/2}^{\mathrm{up}}/\sqrt n]$
formed from its critical values, have asymptotic coverage at least
$1-\alpha$. In particular
\[\liminf_{n\to\infty}\PP\bigl(\bmax(\delta,\rho_{0},\Pbase{})\in(-\infty,\bmaxhat(\delta)
-\widehat{\mathfrak{c}}_{\alpha}^{\mathrm{up}}/\sqrt n]\bigr)\geq1-\alpha,\]
with analogous statements for the lower and two-sided sets.
\end{theorem}
\begin{flushright}
[\hyperref[proof:thm:boot-EB]{Proof}]
\end{flushright}

\begin{rem}[Singleton dual]\label{rem:singleton-EB}
Suppose that
\[
\Tmax(\rho_{0},\Pbase{})=\{\theta^{*}\},\qquad
\Xmax(\theta^{*};\rho_{0},\Pbase{})=\{\xi^{*}\}.
\]
Then the derivative
$d\bmaxAt{(\rho_{0},\Pbase{})}[h,\gamma]=(\lambda_{12}^{*})^{\top}h
+\gamma\ell_{\xi^{*},\theta^{*}}$ is linear, so $\bmaxhat(\delta)$ is
asymptotically Gaussian,
\[
  \sqrt n(\bmaxhat(\delta)-\bmax(\delta,\rho_{0},\Pbase{}))
  \;\rightsquigarrow\;\normal\bigl(0,\;\sigma^{2}_{*}\bigr),\qquad
  \sigma^{2}_{*}=\V\bigl((\lambda_{12}^{*})^{\top}\mathbb Z+\mathbb
  G\ell_{\xi^{*},\theta^{*}}\bigr),
\]
where, by Assumption~\ref{assump: clt-EB}, $\sigma^{2}_{*}=
\V_{\Pbase{}}\bigl((\lambda_{12}^{*})^{\top}\psi+\ell_{\xi^{*},\theta^{*}}\bigr)$ combines
the variance of the
estimated moments, the variance of the empirical baseline, and their
covariance. The bootstrap of Algorithm~\ref{alg:FS-EB}
reduces to the standard nonparametric bootstrap in this case. 
\end{rem}

\begin{rem}[Uncertainty in external nuisance
parameters]\label{rem:nuisance-EB}
Throughout this section the only objects carrying sampling uncertainty are
the estimated moment block $\widehat\rho_{n}$ and the empirical baseline
$\widehat P_{n}$, and the large sample theory of
Theorems~\ref{thm:cons-EB}--\ref{thm:boot-EB} is stated for the bound as a
functional of the pair $(\rho,P)$ alone. Some applications, however, also
feature an auxiliary parameter $\varsigma$
that must itself be estimated and that
enters the program through $g$, $m$, or the geometry of the neighborhood,
generally in a nonlinear fashion. We
derive the asymptotic distribution assuming $\varsigma$ is known and
suppressing the dependence of all quantities on it, so that the limit law
and the bootstrap of Algorithm~\ref{alg:FS-EB} reflect only the sampling
uncertainty of $(\widehat\rho_{n},\widehat P_{n})$. 
We leave to future work accommodating sampling
variation in $\widehat\varsigma$ in a tractable and general manner.

In specific applications, however, it is possible to leverage additional
structure to quantify the uncertainty from estimating these parameters.
The RCT-with-attrition application of Section~\ref{subsec: rct attrition}
is a leading instance. There the complier share $\pi$, the share of
compliers among the observed treated, is exactly such an auxiliary
parameter: it enters the bound nonlinearly, through the cap $1/\pi$ and
the effective budget $(1-\pi)\delta$, as well as through the affine map
relating the complier mean to the always observed mean, and it is
estimated from the same sample as the baseline distribution.
Section~\ref{subsubsec: rct attrition inference} takes up exactly this
point, exploiting the explicit structure of the bound to account for the
sampling uncertainty in $\widehat{\pi}$ that the general theory holds
fixed. Proposition~\ref{prop: attrition inference} there shows that
estimating $\pi$ adds the term $\varkappa(\delta)\psi_\pi$ to the influence
function of the plug-in bound, and that the nonparametric bootstrap accounts
for it once $\widehat{\pi}$ is recomputed in each resample.
\end{rem}

\begin{rem}[Known baseline distribution]\label{rem:known-baseline}
When $\Pbase{}$ is known, as with the Gumbel shocks of
Section~\ref{subsec: structural models}, expectations against it carry no sampling
uncertainty, as they can be computed exactly or by Monte Carlo integration with an
arbitrarily large number of draws, and the data enter only through $\widehat\rho_{n}$.
Writing $\Kmax(\theta;\rho):=\Kmax(\theta;\rho,\Pbase{})$, and likewise for
$\mathsf D$, $\Theta_{\delta}$, $\bmax$, $\Tmax$, $\Xmax$, and $\Lmax$, the
plug-in estimator is $\bmaxhat(\delta)=\bmax(\delta,\widehat\rho_{n},\Pbase{})$,
and Theorems~\ref{thm:cons-EB}--\ref{thm:boot-EB} hold with the baseline
slot frozen under a weakened version of Assumption~\ref{assump: est-EB}: the
Lipschitz requirement in $\theta$ of Assumption~\ref{assump: reg-EB} is
dropped; Assumption~\ref{assump: clt-EB}
is replaced by $\sqrt n(\widehat\rho_{n}-\rho_{0})\rightsquigarrow
\normal(0,\Sigma)$ for some positive semidefinite
$\Sigma\in\R^{\dime\times\dime}$, so that $\widehat\rho_{n}$ may be computed from data on
a random vector other than $Z$; and Assumptions~\ref{assump: mult-EB}
and~\ref{assump: M-joint-EB} may be imposed on the multiplier projections
$\Lmax$ rather than on $\Xmax$. The derivative becomes
\[
  d\bmaxAt{\rho_{0}}[h]=\max_{\theta^{*}\in\Tmax(\rho_{0})}\,
   \min_{\lambda^{*}\in\Lmax(\theta^{*};\rho_{0})}(\lambda^{*})^{\top}h,
\]
which coincides with that of
\citet[Theorem~6.2]{christensen2023CounterfactualSensitivityRobustness},
so that $\sqrt n(\bmaxhat(\delta)-\bmax(\delta,\rho_{0}))\rightsquigarrow
d\bmaxAt{\rho_{0}}[\mathbb Z]$ with $\mathbb Z\sim\normal(0,\Sigma)$, and
Algorithm~\ref{alg:FS-EB} reduces to resampling $\widehat\rho_{n}$ alone,
by any bootstrap satisfying Assumption~3 of
\citet{fang2019InferenceDirectionallyDifferentiable}, such as the Gaussian
draw $\widehat\rho_{n}^{*}\sim\normal(\widehat\rho_{n},\widehat\Sigma/n)$
with $\widehat\Sigma\cprob\Sigma$, and evaluating
$\sup_{\theta\in\Tmaxhat}\inf_{\lambda\in\Lmax(\theta;\widehat\rho_{n})}
\lambda^{\top}h$ at $h=\sqrt n(\widehat\rho_{n}^{*}-\widehat\rho_{n})$, which
is the procedure of
\citet[Section~6.2]{christensen2023CounterfactualSensitivityRobustness}. The
tuning sequence of Remark~\ref{rem:nu-choice} is computed from the Monte Carlo draws
used to evaluate $\E_{\Pbase{}}[\cdot]$.
\end{rem}

\clearpage
\section{Summary Measures for Sensitivity Analysis}
\label{sec: summary measures}

Our framework outputs a map
$\delta\mapsto[\underline{\tau}(\delta),\overline{\tau}(\delta)]$ that traces out the
range of the estimand as the sensitivity parameter varies over $[0,1]$. Reporting the
bounds alone is not always the most effective way of communicating the results, and it
may hide information that is relevant for decision making. This section develops
the four summary measures of Section~5.2 of the main
text, which are computed directly from the bounds. The first is the \textit{breakdown
budget}, the largest $\delta$ at which a conclusion still holds.
The second is the \textit{certified breakdown budget}, the largest
$\delta$ at which it also holds once we account for sampling uncertainty at level
$1-\alpha$. The third is the \textit{shadow price of robustness}, the
marginal widening of either bound with respect to $\delta$, which
Proposition~\ref{thm: shadow price budget} identifies up to a scalar with the optimal
multiplier on the divergence constraint. The fourth is the
\textit{certification frontier}, the sample size a study would need to certify robustness
at a given budget. Proposition~\ref{prop: exchange rate} computes its semi-elasticity with
respect to $\delta$, that is, the proportional increase in sample size needed to certify
one additional percentage point of misspecified probability mass. We study
the shadow price first, in Section~\ref{subsec: shadow price budget}, because the
asymptotic analysis of the certified breakdown budget in
Section~\ref{subsec: breakdown budget} rests on it.

Throughout, we work in a simplified version of the framework of
Section~\ref{subsec: setup} in which the integrand does not depend on a structural
parameter, no moment restriction is imposed, and the robustness set is the total
variation robustness set of the main text. Formally, we let
$\calZ\subseteq\R^{\dimz}$ be
closed, let $\Pbase{}\in\Delta(\calZ)$ be the baseline
distribution, let $g:\calZ\to\R$ be such that $\E_{\Pbase{}}[|g(Z)|]<\infty$, and
we let $\dimm=0$ for simplicity. For
a known mixing weight $\kappa\in(0,1)$ and a budget $\delta\geq0$, we
consider the
robustness set
\begin{align}
    \calP_{\TV}(\delta)
    :=&\bigl\{P\in\calP(\calZ):\ \TV(P\,\|\,R)\leq\delta,\
    \Pbase{}=\kappa P+(1-\kappa)R\ \text{for some}\ R\in\calP(\calZ)\bigr\}\nonumber\\
    =&\bigl\{P\in\calP(\calZ):\ \kappa P\leq\Pbase{},\
    \TV(P\,\|\,\Pbase{})\leq(1-\kappa)\delta\bigr\},
    \label{eq: tv set summary}
\end{align}
where the second equality follows by Lemma~\ref{lemma: contamination as constrained tv-i}
for $\delta\leq1$. For $\delta>1$, both sets equal
$\{P\in\calP(\calZ):\kappa P\leq\Pbase{}\}$, as $\TV(P\,\|\,R)\leq1$ for all $P$ and $R$,
and as $\kappa P\leq\Pbase{}$ implies $\TV(P\,\|\,\Pbase{})\leq1-\kappa$ by the same
lemma. The set thus does not change for $\delta\geq1$, and we allow $\delta>1$ only to
let the one-sided derivatives at $\delta=1$ be well defined.
The above set coincides with the robustness set 
$\calP^{\ll}_{\mathsf{mix}}((1-\kappa)\delta)$
of \eqref{eq: P mix baseline mixture} with entropy $\phi_{\TV}$, which
Proposition~\ref{prop: induced entropy-i} identifies with the neighborhood generated by
the upper-truncated entropy $\overline{\phi}_{\TV}$ of
Section~\ref{subsec: correspondence with the paper} at
budget $(1-\kappa)\delta$. The sensitivity bounds are thus
\begin{align}
    \overline{\tau}(\delta):=\sup_{P\in\calP_{\TV}(\delta)}\E_P[g(Z)]
    \qquad\text{and}\qquad
    \underline{\tau}(\delta):=\inf_{P\in\calP_{\TV}(\delta)}\E_P[g(Z)].
    \label{eq: bounds summary}
\end{align}
This is the scenario of the RCT with attrition
application of Section~\ref{subsec: rct attrition}.
Absent a structural parameter, the dual problem is only
two-dimensional and uniqueness of the optimal budget multiplier can be
verified directly. A structural parameter
raises difficulties of a different nature, which we leave to future work. For example, 
the bounds
become upper envelopes over the parameter space of the values of the inner problems,
and an envelope of concave functions of the budget need not be concave, so that the
shadow price is defined only at each outer maximizer and may jump when the maximizer
changes with the budget. 

\subsection{The Shadow Price of Robustness}
\label{subsec: shadow price budget}
In this simplified scenario, the dual of the inner problem at budget
$\delta>0$ is, by Theorem~\ref{theorem: strong duality with mixture constraint-i} and
Lemma~\ref{lemma: family entropy},
\begin{align}
    \overline{\tau}(\delta)
    =\inf_{(\zeta,\eta)\in\R\times\R_+}\Bigl\{\zeta+\eta(1-\kappa)\delta
    +\E_{\Pbase{}}\bigl[\mathsf{r}_{\TV}\bigl(g(Z)-\zeta,\eta\bigr)\bigr]\Bigr\},
    \qquad
    \mathsf{r}_{\TV}(x,\eta):=\max\Bigl\{-\frac{\eta}{2},\ x,\
    \frac{x}{\kappa}-\frac{(1-\kappa)\eta}{2\kappa}\Bigr\}.
    \label{eq: dual summary tv}
\end{align}
The budget enters only through
the term $\eta(1-\kappa)\delta$, so that $\eta$ prices the total variation constraint as
a Lagrange multiplier prices a resource constraint. The next result shows that
$\overline{\tau}$ is concave and
$\underline{\tau}$ convex in $\delta$, so that their one-sided derivatives exist, and
that these derivatives equal, up to the factor $1-\kappa$, the optimal multipliers,
whence the name shadow price.

\begin{prop}[The shadow price of the robustness budget]
\label{thm: shadow price budget}
For $\delta>0$, let $\Xmax$ and
$\Xmin$ denote the solution sets of the dual problem \eqref{eq: dual summary tv} with
budget $\delta$, computed at $g$ and at $-g$, respectively. Let
\begin{alignat*}{3}
\overline{\eta}^{\min}_{\delta}&:=\min\{\eta:(\zeta,\eta)\in\Xmax\},
\qquad
&\overline{\eta}^{\max}_{\delta}&:=\max\{\eta:(\zeta,\eta)\in\Xmax\},\\
\underline{\eta}^{\min}_{\delta}&:=\min\{\eta:(\zeta,\eta)\in\Xmin\},
\qquad
&\underline{\eta}^{\max}_{\delta}&:=\max\{\eta:(\zeta,\eta)\in\Xmin\}
\end{alignat*}
denote the smallest and the largest optimal budget multiplier of each program, which
are well defined because each solution set is nonempty and compact. Then the following
hold:
\begin{enumerate}[label=(\roman*),ref=\theprop(\roman*),leftmargin=2em,topsep=2pt]
    \item\label{thm: shadow price budget-i} (Regularity.) On $[0,\infty)$ the map
    $\overline{\tau}$ is finite, nondecreasing, concave, and continuous, and the map
    $\underline{\tau}$ is finite, nonincreasing, convex, and continuous. Both admit
    finite one-sided derivatives at every $\delta\in(0,\infty)$, both are
    differentiable at all but at most countably many budgets, and both are constant on
    $[1,\infty)$.
    \item\label{thm: shadow price budget-ii} (Shadow prices.) For every
    $\delta\in(0,\infty)$,
    \[
    \partial_{+}\overline{\tau}(\delta)=(1-\kappa)\,\overline{\eta}^{\min}_{\delta},
    \qquad
    \partial_{-}\overline{\tau}(\delta)=(1-\kappa)\,\overline{\eta}^{\max}_{\delta},
    \qquad
    -\partial_{+}\underline{\tau}(\delta)=(1-\kappa)\,\underline{\eta}^{\min}_{\delta},
    \qquad
    -\partial_{-}\underline{\tau}(\delta)=(1-\kappa)\,\underline{\eta}^{\max}_{\delta}.
    \]
    \item\label{thm: shadow price budget-iii} (Declining prices.) For every
    $0<\delta_1<\delta_2<\infty$, both
    $\overline{\eta}^{\min}_{\delta_1}\geq\overline{\eta}^{\max}_{\delta_2}$ and
    $\underline{\eta}^{\min}_{\delta_1}\geq\underline{\eta}^{\max}_{\delta_2}$.
    \item\label{thm: shadow price budget-iv} (A single shadow price.) The bound
    $\overline{\tau}$ is differentiable at $\delta\in(0,\infty)$ if and only if
    $\overline{\eta}^{\min}_{\delta}=\overline{\eta}^{\max}_{\delta}$, and
    $\underline{\tau}$ is differentiable at $\delta$ if and only if
    $\underline{\eta}^{\min}_{\delta}=\underline{\eta}^{\max}_{\delta}$, that is,
    exactly when the optimal budget multiplier of the corresponding program is unique.
    At every such budget we write
    \begin{align*}
        \overline{\eta}(\delta):=\overline{\tau}'(\delta)
        =(1-\kappa)\,\overline{\eta}^{\min}_{\delta},
        \qquad
        \underline{\eta}(\delta):=-\underline{\tau}'(\delta)
        =(1-\kappa)\,\underline{\eta}^{\min}_{\delta}.
    \end{align*}
    Both $\overline{\eta}$ and $\underline{\eta}$ are nonnegative, defined at all but at
    most countably many $\delta\in(0,\infty)$, and nonincreasing on the set of budgets
    at which they are defined.
\end{enumerate}
\end{prop}
\begin{flushright}
[\hyperref[proof: shadow price budget]{Proof}]
\end{flushright}

Beyond $\delta=1$ the bounds are constant, so that $\partial_{+}\overline{\tau}(1)=0$ and
the smallest optimal multiplier at $\delta=1$ vanishes, whereas the largest need not.

\subsection{Breakdown and Certified Breakdown}
\label{subsec: breakdown budget}

Let $\tau_\star\in\R$ be a reference value for the estimand, typically $\tau_\star=0$
when the sign of a treatment effect is of interest. The bounds in
\eqref{eq: bounds summary} are
finite and continuous on $[0,1]$ by Proposition~\ref{thm: shadow price budget-i}, and
the breakdown budget for $\tau_\star$ is
\begin{align}
    \delta_{\mathsf{b}}\equiv\delta_{\mathsf{b}}(\tau_\star)
    :=
    \inf\bigl\{\delta\in[0,1]:\;\underline{\tau}(\delta)\leq\tau_\star
    \leq\overline{\tau}(\delta)\bigr\},
    \label{eq: breakdown point}
\end{align}
with the convention that the infimum over an empty set equals $1$. Since the bounds are
monotone in $\delta$, at most one of the two inequalities fails at
$\delta=0$, and the same one fails for every budget below the breakdown. If
$\underline{\tau}(0)>\tau_\star$, the case on which we focus, \eqref{eq: breakdown point}
reduces to $\inf\{\delta\in[0,1]:\underline{\tau}(\delta)\leq\tau_\star\}$. The case
$\overline{\tau}(0)<\tau_\star$ is symmetric, and if
$\tau_\star\in[\underline{\tau}(0),\overline{\tau}(0)]$, then $\delta_{\mathsf{b}}=0$.

The breakdown budget ignores sampling uncertainty. For $\alpha\in(0,1)$ and a sample
size $n$ held fixed, the population certified breakdown budget is
\begin{align}
    \delta_{\mathsf{cb}}(\alpha;n)
    :=
    \inf\Bigl\{\delta\in[0,1]:\;
    \underline{\tau}(\delta)
    -\frac{\mathfrak{c}_{1-\alpha/2}\,\sigma(\delta)}{\sqrt{n}}
    \leq\tau_\star\Bigr\},
    \label{eq: population certified breakdown}
\end{align}
where the infimum over an empty set equals $1$ as in
\eqref{eq: breakdown point}, $\sigma(\delta)$ is the standard deviation at $\delta$ of the
Gaussian limit of $\sqrt n(\widehat{\underline{\tau}}_n-\underline{\tau})$, with
$\widehat{\underline{\tau}}_n$ the estimated lower bound of
\eqref{eq: certified breakdown}, and
$\mathfrak{c}_{1-\alpha/2}$ is the $(1-\alpha/2)$-th
quantile of the standard normal distribution. It is the smallest budget at which a study
of size $n$ would fail to certify the conclusion, were the bounds and their standard
deviations known. When $\underline{\tau}(0)>\tau_\star$,
$\underline{\tau}$ is nonincreasing and continuous with
$\delta_{\mathsf{b}}\in(0,1)$,
and $\sigma$ is continuous and strictly positive on $[0,1]$, as
Assumption~\ref{assump: breakdown regularity} below imposes,
$\delta_{\mathsf{cb}}(\alpha;n)\leq\delta_{\mathsf{b}}$ for every $n$, because
$\underline{\tau}(\delta_{\mathsf{b}})=\tau_\star$ places $\delta_{\mathsf{b}}$ in the set
in \eqref{eq: population certified breakdown}, the map
$n\mapsto\delta_{\mathsf{cb}}(\alpha;n)$ is nondecreasing, and
$\delta_{\mathsf{cb}}(\alpha;n)\to\delta_{\mathsf{b}}$ as $n\to\infty$. The sample
analogue replaces the population lower bound and its standard deviation by estimators,
\begin{align}
    \widehat{\delta}_{\mathsf{cb}}(\alpha)
    :=
    \inf\Bigl\{\delta\in[0,1]:\;
    \widehat{\underline{\tau}}_n(\delta)
    -\frac{\mathfrak{c}_{1-\alpha/2}\,\widehat{\sigma}_n(\delta)}{\sqrt{n}}
    \leq\tau_\star\Bigr\},
    \label{eq: certified breakdown}
\end{align}
where $\widehat{\underline{\tau}}_n(\delta)$ and $\widehat{\sigma}_n(\delta)$ estimate
$\underline{\tau}(\delta)$ and $\sigma(\delta)$, for instance through the bootstrap of
Section~\ref{sec:estimation-inference} or the influence-function estimator of
Proposition~\ref{prop: attrition inference}. It is the infimum of the budgets at
which $\tau_\star$ is not rejected at level $\alpha$. In the case on which we
focus, the plug-in estimator of the breakdown budget is
\[
\widehat{\delta}_{\mathsf{b}}
:=\inf\bigl\{\delta\in[0,1]:\widehat{\underline{\tau}}_n(\delta)\leq\tau_\star\bigr\}.
\]

\begin{rem}[Coverage of the identified set, not of a point effect]
\label{rem: attrition IM interval}
In the literature, there are at least two approaches
to constructing confidence intervals for a partially identified
parameter. The first, and the one used throughout this work, covers the entire
identified interval by placing a two-sided critical value at each estimated
endpoint. The second, following \citet{imbens2004ConfidenceIntervalsPartially},
covers a single point of that interval, such as the ATE $\tau_0$ in the
attrition example of Section~\ref{subsec: rct attrition},
through a critical value that lies between the one- and two-sided normal quantiles and
is therefore narrower at the same nominal level. \citet{lee2009TrainingWagesSample} favors
the latter because his object of interest is the treatment effect, a point inside
the identified set, and not the region of all rationalizable effects. Here the
object of interest is precisely that region: the sensitivity analysis reports
the identified set $[\underline{\tau}(\delta),\overline{\tau}(\delta)]$
as a function of the budget $\delta$, and it is the
bounds delimiting this set whose sampling uncertainty we quantify.
\end{rem}

\begin{assump}[Regularity at the breakdown]\label{assump: breakdown regularity}
For the reference value $\tau_\star\in\R$ of \eqref{eq: breakdown point}
the following hold:
\begin{enumerate}[label=(\roman*),ref=\theassump(\roman*),leftmargin=2em,topsep=2pt]
    \item\label{assump: breakdown regularity-i} (Population bound.)
    $\underline{\tau}:[0,1]\to\R$ is nonincreasing and continuous,
    $\underline{\tau}(0)>\tau_\star$,
    $\delta_{\mathsf{b}}\in(0,1)$, and $\underline{\tau}$ is
    differentiable at $\delta_{\mathsf{b}}$ with
    $\underline{\eta}(\delta_{\mathsf{b}})=-\underline{\tau}'(\delta_{\mathsf{b}})\in(0,\infty)$.
    \item\label{assump: breakdown regularity-ii} (Estimator.)
    $\sqrt{n}\,(\widehat{\underline{\tau}}_n-\underline{\tau})
    \rightsquigarrow\mathbb{G}$ in $\ell^{\infty}([0,1])$, where
    $\mathbb{G}$ is a tight zero-mean Gaussian process with continuous
    sample paths on $[0,1]$ whose standard deviation function
    $\sigma(\delta):=\{\E[\mathbb{G}(\delta)^2]\}^{1/2}$ is continuous and
    strictly positive. Moreover,
    $\delta\mapsto\widehat{\underline{\tau}}_n(\delta)$ has continuous
    sample paths.
    \item\label{assump: breakdown regularity-iii} (Standard errors.)
    $\sup_{\delta\in[0,1]}
    |\widehat{\sigma}_n(\delta)-\sigma(\delta)|=\op(1)$, and
    $\delta\mapsto\widehat{\sigma}_n(\delta)$ has continuous sample paths.
\end{enumerate}
\end{assump}

In this scenario, the monotonicity and continuity of
$\underline{\tau}$ and its differentiability off a countable set are automatic by
Proposition~\ref{thm: shadow price budget-i}, so that
Assumption~\ref{assump: breakdown regularity-i} only requires an interior breakdown
budget at which $\underline{\tau}$ is differentiable, that is, by
Proposition~\ref{thm: shadow price budget-iv}, at which the optimal multiplier is
unique, and a strictly positive shadow price there, which rules out a bound path that
is flat at the crossing. Assumption~\ref{assump: breakdown regularity-ii} imposes a
Gaussian limit uniformly over $[0,1]$, which we discuss in the next remark.

\begin{rem}[Uniqueness and the certified breakdown budget]
\label{rem: certified breakdown uniqueness}
The certified breakdown budgets \eqref{eq: population certified breakdown} and
\eqref{eq: certified breakdown} build the lower confidence bound on
$\underline{\tau}(\delta)$ from a normal critical value, which presumes that
$\sqrt{n}(\widehat{\underline{\tau}}_n(\delta)-\underline{\tau}(\delta))$ has a Gaussian
limit. As there is no outer maximization in this scenario,
Remark~\ref{rem:singleton-EB} delivers such a limit, with $(1-\alpha/2)$-quantile
$\mathfrak{c}_{1-\alpha/2}\,\sigma(\delta)$, whenever the dual solution set $\Xmin$ is a
singleton. Non-uniqueness of the budget multiplier is one way in which this property
can fail, and by Proposition~\ref{thm: shadow price budget-iv} it occurs exactly at the
budgets at which $\underline{\tau}$ has a kink. The RCT application of
Section~\ref{subsec: rct attrition} is not affected. There, the baseline has a density
that is positive at the trimming quantiles of the lower bound, under the
version of Assumption~\ref{assump: attrition inference} for the lower
bound given in Section~\ref{subsubsec: rct attrition inference}, so that the optimizer
set is a singleton
and the plug-in bound is asymptotically normal
(Proposition~\ref{prop: attrition inference}). When the dual solution is not unique,
the limit is in general not Gaussian, and $\mathfrak{c}_{1-\alpha/2}\,\sigma(\delta)$
should be replaced by the $(1-\alpha/2)$-quantile of the limit, estimated by the
Fang--Santos bootstrap as in
Remark~\ref{rem: certified breakdown bootstrap}.
\end{rem}

\begin{rem}[Bootstrap critical values]\label{rem: certified breakdown bootstrap}
The certified breakdown budget can also be built from the quantiles delivered by
Algorithm~\ref{alg:FS-EB}, without estimating $\sigma(\delta)$. Letting
$\widehat{\mathfrak{c}}^{\mathrm{low}}_{1-\alpha/2}(\delta)$ denote the critical value of
Algorithm~\ref{alg:FS-EB} for the lower bound at budget $\delta$, one replaces
$\mathfrak{c}_{1-\alpha/2}\,\widehat\sigma_n(\delta)$ in
\eqref{eq: certified breakdown} by
$\widehat{\mathfrak{c}}^{\mathrm{low}}_{1-\alpha/2}(\delta)$, that is,
\[
\widehat{\delta}_{\mathsf{cb}}(\alpha)
:=\inf\Bigl\{\delta\in[0,1]:\;
\widehat{\underline{\tau}}_n(\delta)
-\frac{\widehat{\mathfrak{c}}^{\mathrm{low}}_{1-\alpha/2}(\delta)}{\sqrt{n}}
\leq\tau_\star\Bigr\}.
\]
At each fixed budget this critical value is consistent even when the dual solution is not
unique (Theorem~\ref{thm:boot-EB}). When the dual solution is unique, it estimates the
same quantity as $\mathfrak{c}_{1-\alpha/2}\,\widehat\sigma_n(\delta)$, and so does the
distance between $\widehat{\underline{\tau}}_n(\delta)$ and the lower endpoint of a
percentile bootstrap band, the $\alpha/2$-quantile of the bootstrap distribution of
$\widehat{\underline{\tau}}_n(\delta)$. The certified breakdown budgets reported in
Section~\ref{sec: empirical applications details} are computed from such a percentile
band.
\end{rem}

We define the robustness standard error as
\begin{align}
    \varsigma_{\mathsf{b}} :=
    \frac{\sigma(\delta_{\mathsf{b}})}{\underline{\eta}(\delta_{\mathsf{b}})},
    \label{eq: robustness se}
\end{align}
which converts the standard error of the lower bound at the breakdown from estimand
units into budget units through the shadow price. The next result shows that the
certified breakdown budget is consistent for $\delta_{\mathsf{b}}$, that it has a
Gaussian limit whose location and scale are governed by $\varsigma_{\mathsf{b}}$, and that
$[\widehat{\delta}_{\mathsf{cb}}(\alpha),1]$ is an asymptotically valid lower confidence
interval for $\delta_{\mathsf{b}}$ with coverage $1-\alpha/2$.

\begin{prop}[The certified breakdown budget]\label{prop: certified breakdown}
Let Assumption~\ref{assump: breakdown regularity} hold and
$\alpha\in(0,1)$. Then, as $n\to\infty$:
\begin{enumerate}[label=(\roman*),ref=\theprop(\roman*),leftmargin=2em,topsep=2pt]
    \item\label{prop: certified breakdown-i}
    $\widehat{\delta}_{\mathsf{cb}}(\alpha)
    \cprob\delta_{\mathsf{b}}$;
    \item\label{prop: certified breakdown-ii}
    $\sqrt{n}\,\bigl(\widehat{\delta}_{\mathsf{cb}}(\alpha)
    -\delta_{\mathsf{b}}\bigr)\rightsquigarrow
    \normal\bigl(-\mathfrak{c}_{1-\alpha/2}\,\varsigma_{\mathsf{b}},
    \ \varsigma_{\mathsf{b}}^{2}\bigr)$;
    \item\label{prop: certified breakdown-iii}
    $\lim_{n\to\infty}
    \PP\bigl(\widehat{\delta}_{\mathsf{cb}}(\alpha)
    \leq\delta_{\mathsf{b}}\bigr)=1-\alpha/2$.
\end{enumerate}
\end{prop}
\begin{flushright}
[\hyperref[proof: certified breakdown]{Proof}]
\end{flushright}

By Proposition~\ref{prop: certified breakdown-ii},
$\delta_{\mathsf{b}}-\widehat{\delta}_{\mathsf{cb}}(\alpha)
=\varsigma_{\mathsf{b}}\,(\mathfrak{c}_{1-\alpha/2}+\mathbb Z_n)/\sqrt{n}+\op(n^{-1/2})$
with
$\mathbb Z_n\rightsquigarrow\normal(0,1)$, so that the first-order certification gap
$\mathfrak{c}_{1-\alpha/2}\varsigma_{\mathsf{b}}/\sqrt{n}$ increases with the confidence
level and with the robustness standard error and decreases with the sample size.
Proposition~\ref{prop: certified breakdown-iii} says that the certified budget is a
one-sided $1-\alpha/2$ confidence bound on the population breakdown.

\subsection[The Certification Frontier]{The Certification Frontier}
\label{subsec: exchange rate}

For $\delta\in[0,\delta_{\mathsf{b}})$, the certification frontier
\begin{align}
    n^{\star}(\delta;\alpha)
    := \frac{\mathfrak{c}_{1-\alpha/2}^{2}\,\sigma^{2}(\delta)}
    {\bigl(\underline{\tau}(\delta)-\tau_\star\bigr)^{2}}
    \label{eq: certification frontier}
\end{align}
is the sample size at which the lower confidence bound in
\eqref{eq: population certified breakdown} equals the reference value at budget
$\delta$. We treat it as a real valued
quantity. The smallest certifying integer sample size is
$\lfloor n^{\star}(\delta;\alpha)\rfloor+1$. The next result computes how the frontier
grows with the robustness target, that is, the marginal rate of substitution between
observations and misspecified probability mass along the frontier.

\begin{prop}[Semi-elasticity of the certification frontier]
\label{prop: exchange rate}
Let Assumption~\ref{assump: breakdown regularity} hold, and suppose in
addition that $\underline{\tau}$ is convex on $[0,1]$ and that $\sigma$ is
continuously differentiable on $[0,1]$. Fix $\alpha\in(0,1)$.
Then, at
every $\delta\in(0,\delta_{\mathsf{b}})$ at which $\underline{\tau}$ is
differentiable, $n^{\star}(\cdot\,;\alpha)$ is differentiable with
\begin{align}
    \frac{\de\log n^{\star}(\delta;\alpha)}{\de\delta}
    = 2\left[
    \frac{\underline{\eta}(\delta)}
    {\underline{\tau}(\delta)-\tau_\star}
    +\frac{\sigma'(\delta)}{\sigma(\delta)}
    \right],
    \label{eq: exchange rate formula}
\end{align}
and as $\delta\uparrow\delta_{\mathsf{b}}$
\[\frac{\de\log n^{\star}(\delta;\alpha)}{\de\delta}
=\frac{2}{\delta_{\mathsf{b}}-\delta}\{1+o(1)\}.\]
\end{prop}
\begin{flushright}
[\hyperref[proof: exchange rate]{Proof}]
\end{flushright}

The convexity of $\underline{\tau}$ is automatic
in this scenario by
Proposition~\ref{thm: shadow price budget-i}. Formula
\eqref{eq: exchange rate formula} decomposes the rate into the ratio of the shadow
price to the slack $\underline{\tau}(\delta)-\tau_\star$, which dominates near the
breakdown, and the precision correction $\sigma'/\sigma$, which stays bounded while the
slack vanishes. Near the breakdown, one additional percentage point of certified
robustness thus costs approximately $2/(\delta_{\mathsf{b}}-\delta)$ percent additional
sample, so that closing the certification gap from four to three percentage points
requires multiplying the sample size by approximately $1.78$. 

\clearpage
\section{Applications of the Framework}
\label{sec: applications of tv sensitivity}
In this section we show how the framework covers several settings that are
common in applied economics.
In Section~6 of the main text we develop three
applications in detail: randomized
experiments with attrition (Section~\ref{subsec: rct attrition}),
counterfactuals in structural models (Section~\ref{subsec: structural
models}), and formula and shift-share instruments
(Section~\ref{subsec: formula instruments}). Here we provide further details
on these applications together with formal results.
Section~\ref{subsec: other applications} then collects a series of further 
applications in a deliberately succinct format. 
A closing subsubsection shows that the sensitivity bounds at the endpoint $\delta=1$
coincide with the nonparametric sharp bounds.

\subsection{RCTs with Attrition}\label{subsec: rct attrition}

This section maintains the setup of Example~2.1 of the
main text, which
we briefly
recall, and expands on it in three directions. First, we show
that the sensitivity bounds have a closed form and coincide with the
trimming bounds of \citet{lee2009TrainingWagesSample} at
$\delta=1$. Section~\ref{subsec: role of covariates RCT} shows how to add pre-treatment
covariates. Section~\ref{subsubsec: rct attrition inference} develops
inference on the sensitivity bounds that accounts for the estimation of the
nuisance parameter $\pi$.

\textbf{Setup.} Let $D\in\{0,1\}$ be a binary treatment,
$Y(d)\in\calY\subseteq \R,d\in\{0,1\}$ be potential outcomes, and
$S(d)\in\{0,1\},d\in\{0,1\}$ be potential response indicators, with
\[Y=DY(1)+(1-D)Y(0),\qquad S=DS(1)+(1-D)S(0).\]
The outcome is recorded only when $S=1$, so the observed data are $(YS,S,D)$.
The distribution $P_0$ of the population vector $(Y(0),Y(1),S(0),S(1),D)$ satisfies random
treatment assignment, that is, $(S(0),S(1),Y(0),Y(1))\independent D$. The potential
response indicators partition the population into four groups, which we index by the
latent type
\begin{equation*}
G:=\begin{cases}
\mathsf{AO} & \text{if } (S(0),S(1))=(1,1),\\
\mathsf{C} & \text{if } (S(0),S(1))=(0,1),\\
\mathsf{NO} & \text{if } (S(0),S(1))=(0,0),\\
\mathsf{DF} & \text{if } (S(0),S(1))=(1,0),
\end{cases}
\end{equation*}
namely the always observed, the compliers, the never observed, and the defiers.
The causal estimand of interest is the average treatment effect (ATE) on the
always observed units
\citep{lee2009TrainingWagesSample,semenova2025GeneralizedLeeBounds},\footnote{
A different estimand, the average treatment effect for the whole population
$\E_{P_0}[Y(1)-Y(0)]$, can be
accommodated as well. Because in that case the baseline is a mixing component of the
candidate distribution rather than the mixture itself, we defer its analysis to the
missing data case analyzed in Section~\ref{subsec: missing data}.}
\[
\tau_0 := \E_{P_0}[Y(1)-Y(0)\mid G=\mathsf{AO}].
\]
Two assumptions are routinely imposed in this setting. The first, which dates
back to \cite{rubin1976InferenceMissingData}, is that observations are missing
completely at random (MCAR, henceforth), that is
\begin{equation*}\tag{\textsf{MCAR}}\label{assump: MCAR}
(S(0),S(1))\independent (Y(0),Y(1)),
\end{equation*}
which point identifies $\tau_0$. The second, proposed in
\cite{lee2009TrainingWagesSample}, is monotonicity of selection,
\begin{equation*}\tag{\textsf{Mono}}\label{assump: Mono}
S(1)\geq S(0)\quad\text{a.s.},
\end{equation*}
which rules out defiers, $P_0(G=\mathsf{DF})=0$, and delivers sharp
nonparametric bounds for $\tau_0$. We maintain \ref{assump: Mono} throughout
and relax \ref{assump: MCAR}.

Under \ref{assump: Mono} a control unit is observed only if it is
always observed, so that
\begin{equation}\label{eq: mu0 AO identified}
\mu^{\mathsf{AO}}(0):=\E_{P_0}[Y(0)\mid G=\mathsf{AO}] = \E_{P_0}[Y\mid D=0, S=1]
\end{equation}
is identified, whereas the treated respondents comprise both compliers and
always observed, so that $\mu^{\mathsf{AO}}(1):=\E_{P_0}[Y(1)\mid G=\mathsf{AO}]$
is not. Let $\PT$ denote the conditional distribution of $Y$ given $D=1$ and
$S=1$ under $P_0$, with mean $\mu_{\mathsf{T}}:=\E_{\PT}[Y]$, let
$P_{\mathsf{C}}$ and $P_{\mathsf{AO}}$ denote the conditional distributions of
$Y(1)$ given $G=\mathsf{C}$ and given $G=\mathsf{AO}$ under $P_0$,
respectively, and let
\[\pi := P_0(G=\mathsf{C}\mid D=1,S=1)=
\frac{P_0(S=1\mid D=1)-P_0(S=1\mid D=0)}{P_0(S=1\mid D=1)}\]
be the complier share among treated respondents, where the second equality
follows from random assignment and \ref{assump: Mono}. Since $S(1)=1$ if and
only if $G\in\{\mathsf{AO},\mathsf{C}\}$, random assignment yields the mixture
decomposition
\begin{align}\label{eq: mixture AT and C}
    \PT = \pi P_{\mathsf{C}} +
    (1-\pi)P_{\mathsf{AO}},
\end{align}
in which only $\PT$ is identified. In the treated arm, \ref{assump: MCAR}
implies, as $G$ is a function of $(S(0),S(1))$, that
\begin{equation*}
P_{\mathsf{C}}=P_{\mathsf{AO}}=\PT,
\end{equation*}
that is, $\TV(P_{\mathsf{C}}\,\|\,P_{\mathsf{AO}}) = 0$, whereas
\ref{assump: Mono} leaves the distance between the two distributions
unrestricted. We interpolate between the two by imposing
$\TV(P_{\mathsf{C}}\,\|\,P_{\mathsf{AO}})\leq \delta \in[0,1]$. In view of
\eqref{eq: mixture AT and C}, we get
\[\TV(P_{\mathsf{C}}\,\|\,\PT) =
(1-\pi)\TV(P_{\mathsf{C}}\,\|\,P_{\mathsf{AO}}),\]
so that the restriction can be centered at the identified law $\PT$ and only
one distribution is left to vary.\footnote{Wasserstein-1 would also have
this property, whereas for the Kullback--Leibler divergence convexity
yields only an inequality. In general, any metric that is
norm-induced would preserve the equality, whereas $\phi$-divergences will not in general.
Total variation enjoys this property
because it is not only a $\phi$-divergence but also an optimal transport metric
\citep[see, e.g., Proposition~2.24 in][]{kuhn2025DistributionallyRobustOptimization}.}

\textbf{Sensitivity bounds.} For any candidate complier distribution
$Q\in\Delta(\calY)$, the mixture identity \eqref{eq: mixture AT and C} forces
the always observed distribution to be $(\PT-\pi Q)/(1-\pi)$. Subtracting the
identified control mean \eqref{eq: mu0 AO identified}, we obtain the map
\begin{equation}\label{eq: tau map attrition}
\tau(Q):=\frac{\mu_{\mathsf{T}}-\pi\,\E_Q[Y]}{1-\pi}-\mu^{\mathsf{AO}}(0),\qquad
Q\in\Delta(\calY)\ \text{with}\ \E_Q[|Y|]<\infty,
\end{equation}
in which $\E_Q[Y]$ is the only quantity not identified from the data, with
$\tau_0=\tau(P_{\mathsf{C}})$ and, under \ref{assump: MCAR},
$\tau_0=\tau(\PT)=\mu_{\mathsf{T}}-\mu^{\mathsf{AO}}(0)$.
For $\delta\in[0,1]$, the set of candidate complier distributions is
\begin{align*}
\calP_{\mathsf{RCT}}(\delta)
:&=\{Q\in\Delta(\calY):\TV(Q\,\|\,\PT)\leq (1-\pi)\delta,\, \exists\,R\in\Delta(\calY) :
\PT = \pi Q + (1-\pi)R\} \\
&=\{Q\in\Delta(\calY):\TV(Q\,\|\,\PT)\leq (1-\pi)\delta,\, \pi
Q\leq \PT\},
\end{align*}
where the second equality follows from Lemma \ref{lemma: mixture density bounds-i}.
When $\E_{\PT}[|Y|]<\infty$, every $Q\in\calP_{\mathsf{RCT}}(\delta)$ has
$\E_Q[|Y|]\leq\pi^{-1}\E_{\PT}[|Y|]<\infty$ because $\pi Q\leq\PT$, so that $\tau$ is
well defined on $\calP_{\mathsf{RCT}}(\delta)$, and $\calP_{\mathsf{RCT}}(\delta)$ is the
set $\calP_{\TV}(\delta)$ of \eqref{eq: tv set summary} with
$(\kappa,\Pbase{})=(\pi,\PT)$.
The sensitivity bounds on the complier mean are
\[\underline{\mu}^{\mathsf{C}}(1;\delta)
:=\inf_{Q\in\calP_{\mathsf{RCT}}(\delta)}\E_Q[Y],\qquad
\overline{\mu}^{\mathsf{C}}(1;\delta)
:=\sup_{Q\in\calP_{\mathsf{RCT}}(\delta)}\E_Q[Y],\]
and, through the mixture identity, they translate into bounds on the
always observed mean $\mu^{\mathsf{AO}}(1)$,
\begin{equation}\label{eq: at bounds def}
\underline{\mu}^{\mathsf{AO}}(1;\delta):= \frac{\mu_{\mathsf{T}} -
\pi\overline{\mu}^{\mathsf{C}}(1;\delta)}{1-\pi},\qquad
\overline{\mu}^{\mathsf{AO}}(1;\delta):= \frac{\mu_{\mathsf{T}} -
\pi\underline{\mu}^{\mathsf{C}}(1;\delta)}{1-\pi},
\end{equation}
where the first argument records the treatment arm and the second the budget.
At $\delta=0$ we have $\calP_{\mathsf{RCT}}(0)=\{\PT\}$ and the two bounds
coincide. In
this application $g(y)=y$ and there is no moment function, so that $\dimm=0$
and the moment restriction is vacuous. Since $\tau(Q)$ is a decreasing affine
function of $\E_Q[Y]$, the sensitivity bounds on the estimand of interest are
\begin{equation}\label{eq: tau bounds def}
\underline{\tau}(\delta):=\inf_{Q\in\calP_{\mathsf{RCT}}(\delta)}\tau(Q)=
\underline{\mu}^{\mathsf{AO}}(1;\delta)-\mu^{\mathsf{AO}}(0),
\qquad
\overline{\tau}(\delta):=\sup_{Q\in\calP_{\mathsf{RCT}}(\delta)}\tau(Q)=
\overline{\mu}^{\mathsf{AO}}(1;\delta)-\mu^{\mathsf{AO}}(0).
\end{equation}
The strong duality result of
Theorem~\ref{theorem: strong duality with mixture constraint-i} applies, and the
next theorem shows that in this particular case the dual problem admits a
closed form solution.

\begin{thm}[$\TV$ sensitivity bounds for the average treatment effect on the
always observed]\label{theorem: tv dual recovers lee bound}
Assume \ref{assump: Mono} holds and $\pi\in(0,1)$. Fix
$\delta\in(0,1]$, and suppose
$\E_{\PT}[|Y|]<\infty$ and that $\mu^{\mathsf{AO}}(0)$ is finite. Define
the trimming quantiles
\[
t_\mathsf{L}(\delta):=F_{\PT}^{-1}\big((1-\pi)\delta\big),
\quad
t_\mathsf{U}(\delta):=F_{\PT}^{-1}\big(1-\pi\delta\big),
\quad
s_\mathsf{L}(\delta):=F_{\PT}^{-1}\big(\pi\delta\big),
\quad
s_\mathsf{U}(\delta):=F_{\PT}^{-1}\big(1-(1-\pi)\delta\big),
\]
and assume
$\PT$ has no atoms at any of them. Then, the sensitivity bounds on
$\tau_0$ defined in \eqref{eq: tau bounds def} are
\[
\underline{\tau}(\delta)
=\E_{\PT}\big[Y\indic\{t_\mathsf{L}(\delta)<Y<t_\mathsf{U}(\delta)\}\big]
+\frac{1}{1-\pi}\E_{\PT}\big[Y\indic\{Y\leq t_\mathsf{L}(\delta)\}\big]
-\mu^{\mathsf{AO}}(0),
\]
\[
\overline{\tau}(\delta)
=\E_{\PT}\big[Y\indic\{s_\mathsf{L}(\delta)<Y<s_\mathsf{U}(\delta)\}\big]
+\frac{1}{1-\pi}\E_{\PT}\big[Y\indic\{Y\geq s_\mathsf{U}(\delta)\}\big]
-\mu^{\mathsf{AO}}(0).
\]
Moreover, both bounds are attained, and the set of values of $\tau_0$ consistent with
the observed distribution $\PT$, the complier share $\pi$, the
identified control mean $\mu^{\mathsf{AO}}(0)$, and the sensitivity restriction
$\TV(P_{\mathsf{C}}\,\|\,P_{\mathsf{AO}})\leq\delta$ is exactly the closed interval
$\big[\underline{\tau}(\delta),\overline{\tau}(\delta)\big]$.
\end{thm}
\begin{flushright}
[\hyperref[proof:theorem: tv dual recovers lee bound]{Proof}]
\end{flushright}

When $\delta = 1$, the trimming quantiles coincide,
$t_\mathsf{L}(1)=t_\mathsf{U}(1)=: y_{1-\pi}$,
the upper bound of the mean for compliers becomes $\overline{\mu}^{\mathsf{C}}(1;1) =
\frac{1}{\pi}\E_{\PT}[Y\indic\{Y\geq
y_{1-\pi}\}]$, and so the lower bound for always observed is
\[\underline{\mu}^{\mathsf{AO}}(1;1) = \frac{\mu_{\mathsf{T}} -
\pi\overline{\mu}^{\mathsf{C}}(1;1)}{1-\pi} = \E_{\PT}[Y\mid Y\leq
y_{1-\pi}].\]
Symmetrically, at $\delta=1$ we have
$s_\mathsf{L}(1)=s_\mathsf{U}(1)=: y_{\pi}$ and the upper bound reads
$\overline{\mu}^{\mathsf{AO}}(1;1) = \E_{\PT}[Y\mid Y\geq y_{\pi}]$. After
subtracting the identified control mean as in \eqref{eq: tau bounds def}, the
sensitivity bounds on $\tau_0$ at $\delta=1$ read
\begin{align*}
    \underline{\tau}(1) &= \E_{\PT}\big[Y\mid Y\leq y_{1-\pi}\big] -
    \E_{P_0}[Y\mid D=0,S=1]=: \underline{\tau}_{\mathsf{Lee}}, \\
    \overline{\tau}(1) &= \E_{\PT}\big[Y\mid Y\geq y_{\pi}\big] -
    \E_{P_0}[Y\mid D=0,S=1] =: \overline{\tau}_{\mathsf{Lee}},
\end{align*}
which are precisely the sharp bounds on the average treatment effect for the
always observed
of Proposition 1a in \cite{lee2009TrainingWagesSample} under \ref{assump: Mono}.
We note that the requirement that
$\PT$ place no atom at the trimming quantiles is weaker than the continuity of the
potential outcomes maintained there.

Total variation thus interpolates between the point identified estimand
under \ref{assump: MCAR} at $\delta=0$ and the Lee bounds at $\delta=1$. For an
intermediate budget, say $\delta=0.3$, the bounds allow at most $30\%$ of the treated
respondents to depart from \ref{assump: MCAR}. This population reading of $\delta$ rests
on the mixture structure \eqref{eq: mixture AT and C}, since on its own a total
variation neighborhood of radius $\delta$ bounds the probability mass that is moved,
not a share of a population (see Remark~\ref{rem: different interpretations tv}).
Finally, the fact that
$[\underline{\tau}(1),\overline{\tau}(1)]$ coincides exactly with the sharp
nonparametric bounds, and does not merely approximate them, is an instance
of the general result of Section~\ref{subsubsec: nonparametric sharp bounds}.

\begin{rem}[Relaxing monotonicity]\label{rem: relaxing monotonicity attrition}
We have maintained \ref{assump: Mono} throughout, but the approach does not hinge
on it. Once defiers are allowed, the control respondents are a mixture of
always observed and defiers, and the control arm can be handled exactly as the treated
arm, with a second sensitivity parameter bounding the total variation distance between
the outcome distributions of defiers and always observed. The reduction to a single
distribution and the closed form trimmed means of
Theorem~\ref{theorem: tv dual recovers lee bound} then carry over to each arm, and
bounds on $\tau_0$ follow by combining the bounds on $\mu^{\mathsf{AO}}(1)$ and
$\mu^{\mathsf{AO}}(0)$, with one budget per arm. The only new ingredient is that the
response rates of the two arms no longer pin down the shares of the latent types. The
share of defiers can be anything between zero and the minimum between the control response
rate and the treated nonresponse rate. For
bounds on $\tau_0$ that do not require monotonicity, see also
\cite{zhang2003EstimationCausalEffects} and \cite{imai2008SharpBoundsCausal}.
\end{rem}

\subsubsection{The role of covariates}
\label{subsec: role of covariates RCT}
Suppose we also observe a vector $X$ of pre-treatment covariates with finite
support $\calX$.
With a slight abuse of notation we now use $P_0$ as the distribution of the
vector $(Y(0),Y(1),S(0),S(1),D,X)$, and we maintain that random assignment
holds jointly with the covariates, $D\independent(Y(0),Y(1),S(0),S(1),X)$.

Covariates play two distinct roles. They change the point identifying
assumption imposed at $\delta=0$, and they may tighten the sensitivity bounds for
$\delta\in(0,1]$, hence deliver different nonparametric sharp bounds at $\delta=1$.
We now discuss each of the two phenomena in detail.

\textbf{Change of identifying assumption.} Covariates let us change the
point identifying assumption that pins down the causal estimand at $\delta=0$.
In place of \ref{assump: MCAR}, we may ask that potential outcomes and
attrition be independent only once we condition on $X$, that is
\begin{equation*}\tag{\textsf{MAR}}\label{assump: MAR}
(S(0),S(1))\independent (Y(0),Y(1))\mid X,
\end{equation*}
the so called ``missing at random'' (MAR, henceforth) assumption.
MAR identifies $\tau_0$ through a covariate-weighted difference in means,
\begin{equation*}
\tau_{\mathsf{MAR}}(P_0) := \E_{P_0}[\E_{P_0}[Y\mid X,D=1,S=1]
-\E_{P_0}[Y\mid X,D=0,S=1]\;|\;G=\mathsf{AO}].
\end{equation*}

Under \ref{assump: MAR}, the inner difference equals
$\E_{P_0}[Y(1)-Y(0)\mid X,G=\mathsf{AO}]$, so that
$\tau_{\mathsf{MAR}}(P_0)=\tau_0$ by the law of iterated
expectations. The difference-in-means estimand
\[\tau_{\mathsf{MCAR}}(P_0):=\tau(\PT)=\E_{P_0}[Y\mid D=1,S=1]-\E_{P_0}[Y\mid D=0,S=1],\]
with $\tau(\cdot)$ the map \eqref{eq: tau map attrition},
identifies the same target $\tau_0$,
but under \ref{assump: MCAR} in place of \ref{assump: MAR}.
Let $P_{X\mid\mathsf{C}}$, $P_{X\mid\mathsf{AO}}$, and $P_{X\mid D=1,S=1}$
denote the conditional distributions of $X$ given $G=\mathsf{C}$, given $G=\mathsf{AO}$,
and given $D=1,S=1$ under $P_0$. Any
discrepancy between the two estimands comes from the treated arm, as they differ only
through the covariate distribution against which the treated stratum means
$\E_{P_0}[Y\mid X=x,D=1,S=1]$ are averaged, namely
$P_{X\mid D=1,S=1}=\pi P_{X\mid\mathsf{C}}+(1-\pi)P_{X\mid\mathsf{AO}}$ for
$\tau_{\mathsf{MCAR}}(P_0)$ and $P_{X\mid\mathsf{AO}}$ for $\tau_{\mathsf{MAR}}(P_0)$.
They therefore coincide when compliers and always observed share the same covariate
distribution, as is the case when the share of compliers does not depend on $x$, or
when the treated stratum means do not depend on $x$. Under \ref{assump: MAR} their
difference equals
$\pi\big(\E_{P_0}[Y(1)\mid G=\mathsf{C}]-\E_{P_0}[Y(1)\mid G=\mathsf{AO}]\big)$, which
is nonzero only when $Y(1)$ has a different mean for compliers and always observed,
itself a violation of \ref{assump: MCAR}, and in this case only
$\tau_{\mathsf{MAR}}(P_0)$ recovers $\tau_0$.

\textbf{Bounds for $\delta>0$.} In the literature on RCTs with attrition it is
known that covariates can be used to tighten bounds at $\delta=1$ (see
Proposition~1b in \citet{lee2009TrainingWagesSample} and Lemma~3 in
\citet{semenova2025GeneralizedLeeBounds}). Whether covariates also tighten the
bounds at interior budgets $\delta\in(0,1)$ is a more delicate matter and it is
not true in general, because the addition of covariates changes the
point identifying assumption maintained at $\delta=0$, thus precluding a direct
comparison of the two sets of bounds. We also note that the distribution of $X$
among the always observed is identified from the control group by that of
$X\mid S=1,D=0$, so that only the distribution of the potential outcomes given
$X$ is left unrestricted by the data.

There are at least two ways to introduce covariates.
To state them, fix $x\in\calX$, let $P_{\mathsf{C}}(x)$ and
$P_{\mathsf{AO}}(x)$ denote the conditional distributions of $Y(1)$ given $X=x$ and
$G=\mathsf{C}$ and given $X=x$ and $G=\mathsf{AO}$ under $P_0$, and let $\PTc{x}$ denote
the conditional distribution of $Y$ given $D=1$, $S=1$, and $X=x$ under $P_0$, that is,
the outcome distribution of the observed treated units within the stratum. With a slight
abuse of notation, in this section $\PT$ denotes the joint distribution of $(Y,X)$ given
$D=1$ and $S=1$ under $P_0$, of which $\PTc{x}$ is the conditional given $X=x$.
We maintain that $\PTc{x}$ has a continuous distribution function
for every stratum under consideration.
Repeating the argument leading to \eqref{eq: mixture AT and C} conditionally
on $X=x$ gives
\begin{equation}\label{eq: mixture AT and C within stratum}
\begin{split}
\PTc{x} &= \pi(x)P_{\mathsf{C}}(x) + (1-\pi(x))P_{\mathsf{AO}}(x), \\[0.3em]
\pi(x) &= \frac{P_0(S=1\mid D=1,X=x)-P_0(S=1\mid D=0,X=x)}{P_0(S=1\mid D=1,X=x)}.
\end{split}
\end{equation}
The within-stratum complier share $\pi(x)$ is identified from the data
exactly as $\pi$ is. We maintain that $\pi(x)\in(0,1)$ for every
$x\in\calX$, so that every stratum contains both compliers and always observed units,
and that $\E_{\PT}[|Y|]$ and $\E_{P_0}[|Y|\mid D=0,S=1]$ are finite.
On the one hand, we can replicate the exercise above
and impose the sensitivity restriction within each stratum $x\in\calX$ as
\begin{equation}\label{eq: within stratum tv restriction}
\TV(P_{\mathsf{C}}(x)\,\|\,P_{\mathsf{AO}}(x))\leq \delta.
\end{equation}
This set of restrictions induces the collection of
within-stratum robustness sets on the conditional distributions of the outcome given the
covariates
\begin{align*}
\calP^{\mathsf{pw}}_{\mathsf{RCT}}(\delta;x)
:=\{Q\in\Delta(\calY):\TV(Q\,\|\,\PTc{x})\leq \delta(1-\pi(x)),\, \pi(x)
Q\leq \PTc{x}\}.
\end{align*}
On the other hand, we can impose
\begin{equation}\label{eq: average stratum tv restriction}
\int_{\calX} \TV\big(P_{\mathsf{C}}(x)\,\|\,P_{\mathsf{AO}}(x)\big)\,\de
P_{X\mid D=0,S=1}(x) =
\int_{\calX} \TV\big(P_{\mathsf{C}}(x)\,\|\,P_{\mathsf{AO}}(x)\big)\,\de
P_{X\mid \mathsf{AO}}(x)\leq \delta,
\end{equation}
where $P_{X\mid D=0,S=1}$ denotes the conditional distribution of $X$ given $D=0$ and
$S=1$ under $P_0$,
and where the first equality comes from the fact that observed control units are
always observed units under \ref{assump: Mono}. For
$Q\in\Delta(\calY\times\calX)$, we write $Q_X$ for its marginal on $\calX$ and
$Q(\cdot\mid x)$ for the conditional distribution of the outcome given $X=x$, and
likewise for $R$. The restriction
\eqref{eq: average stratum tv restriction} induces a single pooled
robustness set on the joint distributions of outcome and covariates,
\begin{align}\label{eq: rct covariate robustness set}
\calP_{\mathsf{RCT},X}(\delta)
:&=\Big\{Q\in\Delta(\calY\times\calX):\,
\exists\,R\in\Delta(\calY\times\calX)\ \text{s.t.}\
\PT = \pi Q + (1-\pi)R,\nonumber\\
&\quad R_X = P_{X\mid D=0,S=1},\
\int_{\calX} \TV\big(Q(\cdot\mid x)\,\|\,R(\cdot\mid x)\big)\,\de
P_{X\mid\mathsf{AO}}(x)\leq\delta\Big\}\nonumber\\
&=\Big\{Q\in\Delta(\calY\times\calX):\, \pi Q\leq\PT,\
\E_Q[\indic\{X = x\}]=\tfrac{\pi(x)}{\pi}\,\E_{\PT}[\indic\{X=x\}],\
x\in\calX,\nonumber\\
&\quad\int_{\calX}\TV\big(Q(\cdot\mid x)\,\|\,\PTc{x}\big)\,\de
P_{X\mid D=1,S=1}(x)\leq(1-\pi)\delta\Big\},
\end{align}
where the second equality uses the fact that $X$ has discrete support and
Lemma~\ref{lemma: mixture density bounds-i}.
The mixture constraint and the $|\calX|$ constraints on the marginal in $X$
encode what the data identify, whereas the budget constraint involves the only
component the data leave unrestricted, the conditional distribution of $Y(1)$
given $X$. Unlike \eqref{eq: within stratum tv restriction}, the restriction in
\eqref{eq: average stratum tv restriction} allows deviations from the
conditional treated outcome equality implied by \ref{assump: MAR} to be
heterogeneous across strata and involves one robustness set rather than
$|\calX|$. It is therefore weaker, hence the bounds it induces are weakly wider.
Under \eqref{eq: within stratum tv restriction} the population reading
of $\delta$ holds stratum by stratum, whereas under
\eqref{eq: average stratum tv restriction} it applies to the always observed as a
whole. For instance, $\delta=0.3$ then reads as ``at most $30\%$ of the
always observed depart from \ref{assump: MAR}'', a share that can be allocated freely
across strata, so that the share within a stratum may exceed $\delta$ as long as it
falls below $\delta$ in other strata.

\textbf{Sensitivity bounds.}
Even without \ref{assump: MAR}, $\mu^{\mathsf{AO}}(0)$ remains identified
as in \eqref{eq: mu0 AO identified}.
For the other mean potential outcome, we apply the results above stratum by
stratum, for a fixed $x\in\calX$.
Since \eqref{eq: mixture AT and C within stratum} and
\eqref{eq: within stratum tv restriction}
reproduce \eqref{eq: mixture AT and C} and the restriction underlying
$\calP_{\mathsf{RCT}}(\delta)$ upon replacing $(\PT,\pi)$ with
$(\PTc{x},\pi(x))$, the within-stratum robustness set is
$\calP^{\mathsf{pw}}_{\mathsf{RCT}}(\delta;x)$, and Theorem~\ref{theorem: tv dual
recovers lee bound} applies within each stratum and yields the bounds
$\underline{\mu}^{\mathsf{AO},\mathsf{pw}}(1;\delta,x)$ and
$\overline{\mu}^{\mathsf{AO},\mathsf{pw}}(1;\delta,x)$, defined as in
\eqref{eq: at bounds def} upon the same replacement, on
$\E_{P_0}[Y(1)\mid X=x,G=\mathsf{AO}]$.

We then aggregate the above bounds using the distribution of $X$ given $G=\mathsf{AO}$,
which, as noted above, is identified from the observed control group, and so
\begin{equation}\label{eq: tau bounds def covariates}
\begin{aligned}
\underline{\tau}_{X}^{\mathsf{pw}}(\delta) &:=
\E_{P_0}\big[\underline{\mu}^{\mathsf{AO},\mathsf{pw}}(1;\delta,X)\mid G=\mathsf{AO}\big]
-
\mu^{\mathsf{AO}}(0), \\
\overline{\tau}_{X}^{\mathsf{pw}}(\delta) &:=
\E_{P_0}\big[\overline{\mu}^{\mathsf{AO},\mathsf{pw}}(1;\delta,X)\mid G=\mathsf{AO}\big] -
\mu^{\mathsf{AO}}(0).
\end{aligned}
\end{equation}
At $\delta=0$ the budget forces $P_{\mathsf{C}}(x)=P_{\mathsf{AO}}(x)=\PTc{x}$ in
every stratum, so the two bounds collapse to $\tau_{\mathsf{MAR}}(P_0)$.
At $\delta=1$ the sensitivity bounds with covariates coincide with the Lee bounds
\citep[][Proposition~1b]{lee2009TrainingWagesSample}, which are
no wider than the bounds without covariates of
Theorem~\ref{theorem: tv dual recovers lee bound} and are given by
\begin{equation*}
\begin{aligned}
\underline{\tau}^{\mathsf{pw}}_{X}(1)&=
\E_{P_0}\big[\E_{P_0}\big[Y\mid X,D=1,S=1,Y\leq y_{1-\pi(X)}(X)\big]\,\big|\,D=0,S=1\big]
-\mu^{\mathsf{AO}}(0)=:\underline{\tau}_{\mathsf{Lee},X}, \\
\overline{\tau}_{X}^{\mathsf{pw}}(1)&=
\E_{P_0}\big[\E_{P_0}\big[Y\mid X,D=1,S=1,Y\geq y_{\pi(X)}(X)\big]\,\big|\,D=0,S=1\big]
-\mu^{\mathsf{AO}}(0)=:\overline{\tau}_{\mathsf{Lee},X},
\end{aligned}
\end{equation*}
where, for $u\in(0,1)$ and $x\in\calX$, $y_u(x):=F_{\PTc{x}}^{-1}(u)$ denotes
the $u$-quantile of $\PTc{x}$.

If we instead use the joint robustness set of \eqref{eq: rct covariate robustness set} 
the sensitivity bounds for the complier mean are defined as
\begin{equation}\label{eq: complier bounds joint}
\underline{\mu}_{X}^{\mathsf{C}}(1;\delta)
:=\inf_{Q\in\calP_{\mathsf{RCT},X}(\delta)}\E_Q[Y],
\qquad
\overline{\mu}_{X}^{\mathsf{C}}(1;\delta)
:=\sup_{Q\in\calP_{\mathsf{RCT},X}(\delta)}\E_Q[Y].
\end{equation}
Sensitivity bounds on $\tau_0$ can then be formed as before through
\eqref{eq: at bounds def} and \eqref{eq: tau bounds def}, with
$\underline{\mu}_{X}^{\mathsf{C}}(1;\delta)$ and
$\overline{\mu}_{X}^{\mathsf{C}}(1;\delta)$ in place of
$\underline{\mu}^{\mathsf{C}}(1;\delta)$ and
$\overline{\mu}^{\mathsf{C}}(1;\delta)$. We denote such bounds
with $\underline{\tau}_{X}(\delta)$ and
$\overline{\tau}_{X}(\delta)$. We stress that these
bounds share the endpoints of the pointwise ones defined in
\eqref{eq: tau bounds def covariates}. At $\delta=0$ they also collapse on
$\tau_{\mathsf{MAR}}(P_0)$, whereas at $\delta=1$ they coincide with the
covariate based Lee bounds
$\underline{\tau}_{\mathsf{Lee},X}$ and $\overline{\tau}_{\mathsf{Lee},X}$ 
because the total variation
constraint is implied by the other constraints. For $\delta\in(0,1)$,
the joint bounds are weakly wider than the pointwise one in
\eqref{eq: tau bounds def covariates}
because \eqref{eq: within stratum tv restriction} implies
\eqref{eq: average stratum tv restriction} but not vice versa.

Obtaining the dual of the optimization problems in \eqref{eq: complier bounds joint}
is not immediate, but the results in Section~\ref{sec:convex-duality} still 
help us reduce them to a single
finite-dimensional convex program, as we now show. In what follows, we maintain 
the assumptions that $\calX$ is
finite and that every stratum satisfies
$\E_{\PT}[\indic\{X=x\}]>0$.

Because of the constraints on the marginal of $X$, a
candidate $Q\in\calP_{\mathsf{RCT},X}(\delta)$ is
determined entirely by its conditional distribution. Furthermore,
we can rewrite the estimand as
\[
\E_Q[Y]=\E_{\PT}\Big[\tfrac{\pi(X)}{\pi}\,\E_{Q(\cdot\mid X)}[Y]\Big].
\]
These two facts suggest that we can focus on each stratum first.
In this spirit, define the value function
\[
V(t;x):=\sup\big\{\E_Q[Y]\,:\,Q\in\Delta(\calY),\ \pi(x)Q\leq\PTc{x},\
\TV(Q\,\|\,\PTc{x})\leq t\big\}.
\]
By Lemma~\ref{lemma: mixture density bounds-i}, the feasible set of $V(t;x)$ is
the mixture-restricted neighborhood $\calP^{\ll}_{\mathsf{mix}}(\rho,t)$ of
\eqref{eq: P mix baseline mixture}, evaluated at baseline $\PTc{x}$ and mixing
weight $\kappa=\pi(x)$, with vacuous moment restriction. Moreover, the 
hypotheses of Theorem~\ref{theorem: strong duality with mixture
constraint-i} hold here, since $\E_{\PTc{x}}[|Y|]<\infty$ follows from
$\E_{\PT}[|Y|]<\infty$ and because each stratum has positive mass, while with
$\dimm=0$, both $\calM(\rho)=\{0\}$ and
$\Gamma_{\overline{\phi}_{\TV,x}}(t)=\{0\}$ per our convention, the latter nonempty
because the baseline itself is feasible. Therefore, for every $t>0$,
\begin{equation}\label{eq: stratum dual rct covariates}
V(t;x)=\inf_{\eta_x\geq0}\big\{\eta_x t+\Upsilon(\eta_x;x)\big\},
\qquad
\Upsilon(\eta_x;x):=\inf_{\zeta\in\R}\Big\{\zeta
+\E_{\PTc{x}}\big[(\overline{\phi}_{\TV,x}^*)^\pi(Y-\zeta,\eta_x)\big]\Big\},
\end{equation}
where $\overline{\phi}_{\TV,x}:=\phi_\TV+\iota_{[0,1/\pi(x)]}$ is the induced
entropy of Proposition~\ref{prop: induced entropy-i} with $\kappa=\pi(x)$.
Specializing Example~\ref{ex: tv truncated duals} to $\kappa=\pi(x)$,
the perspective appearing in \eqref{eq: stratum dual rct covariates} is
\begin{equation*}
(\overline{\phi}_{\TV,x}^*)^\pi(z,\eta)=\max\Big\{-\frac{\eta}{2},\ z,\
\frac{z}{\pi(x)}-\eta\,\frac{1-\pi(x)}{2\pi(x)}\Big\},
\qquad z\in\R,\ \eta\geq0,
\end{equation*}
which at $\eta=0$ reduces to the recession value $\max\{z,0\}/\pi(x)$.

With these ingredients, the upper bound in \eqref{eq: complier bounds joint}
becomes an exact budget-allocation program across strata,
\begin{equation}\label{eq: rct covariate budget allocation}
\overline{\mu}_{X}^{\mathsf{C}}(1;\delta)
=\sup\Big\{\E_{\PT}\Big[\tfrac{\pi(X)}{\pi}\,V(t(X);X)\Big]\,:\,
t:\calX\to\R_+,\ \E_{\PT}[t(X)]\leq(1-\pi)\delta\Big\}.
\end{equation}
The program \eqref{eq: rct covariate budget allocation}
is finite-dimensional and concave: each $V(\cdot;x)$
is nondecreasing and concave, because if $Q_0$ and $Q_1$ are feasible at
budgets $t_0$ and $t_1$ then any mixture of the two is feasible at the mixed
budget, by convexity of $\TV(\cdot\,\|\,\PTc{x})$, while the objective is linear.
Moreover, each $V(\cdot;x)$ is constant on $[1-\pi(x),\infty)$, because the ceiling
alone forces
$\TV(Q\,\|\,\PTc{x})\leq1-\pi(x)$, in view of the rescaling identity of
Lemma~\ref{lemma: contamination as constrained tv-i}.

Since $\delta>0$, the allocation $t\equiv(1-\pi)\delta/2$ is strictly
feasible, so Slater's condition holds and we can introduce a multiplier
for the budget constraint to get
\[
\overline{\mu}_{X}^{\mathsf{C}}(1;\delta)
=\inf_{\eta\geq0}\Big\{\eta(1-\pi)\delta
+\E_{\PT}\Big[\tfrac{\pi(X)}{\pi}\,\sup_{t\geq0}\Big\{V(t;X)
-\tfrac{\pi\,\eta}{\pi(X)}\,t\Big\}\Big]\Big\}.
\]
For each $x\in\calX$ and $c\geq0$, the inner supremum equals $\Upsilon(c;x)$.
Indeed, $\Upsilon(\cdot;x)$ is convex on $\R_+$, as the partial minimum over $\zeta$ of
a jointly convex function, nonincreasing, because the perspective displayed above is
nonincreasing in $\eta$, and finite, as $\E_{\PTc{x}}[|Y|]<\infty$. It is also lower
semicontinuous at $0$, because the perspective exceeds its value at $\eta=0$ minus
$\eta/(2\pi(x))$, so that $\Upsilon(\eta;x)\geq\Upsilon(0;x)-\eta/(2\pi(x))$. The
extension of $\Upsilon(\cdot;x)$ by $+\infty$ to $(-\infty,0)$ is therefore closed,
proper, and convex. Writing $\Upsilon^*(\cdot;x)$ for its convex conjugate,
Theorem~12.2 in \citet{rockafellar1970ConvexAnalysis} gives
$\Upsilon(c;x)=\sup_{s\in\R}\{cs-\Upsilon^*(s;x)\}$. For $t>0$,
$-\Upsilon^*(-t;x)=\inf_{\eta\geq0}\{\eta t+\Upsilon(\eta;x)\}=V(t;x)$ by
\eqref{eq: stratum dual rct covariates}. For $t<0$, the same infimum is $-\infty$, as
$\Upsilon(\cdot;x)\leq\Upsilon(0;x)$. The term at $s=0$ equals
$\inf_{\eta\geq0}\Upsilon(\eta;x)$, which is at most $V(t;x)$ for every $t>0$ and
hence, letting $t\downarrow0$, at most $\sup_{t>0}\{V(t;x)-ct\}$. Since $V(\cdot;x)$ is
nondecreasing, the value at $t=0$ does not affect the supremum either, so that
$\sup_{t\geq0}\{V(t;x)-ct\}=\Upsilon(c;x)$. Substituting $c=\pi\eta/\pi(x)$ in each
stratum and minimizing jointly over the intercepts of the strata, we obtain
\[
\overline{\mu}_{X}^{\mathsf{C}}(1;\delta)
=\inf_{\substack{\eta\geq0,\\(\zeta_x)_{x\in\calX}\in\R^{|\calX|}}}
\Big\{\eta(1-\pi)\delta+\sum_{x\in\calX}\E_{\PT}[\indic\{X=x\}]\,
\tfrac{\pi(x)}{\pi}\Big(\zeta_x
+\E_{\PTc{x}}\Big[(\overline{\phi}_{\TV,x}^*)^\pi\Big(Y-\zeta_x,
\tfrac{\pi\,\eta}{\pi(x)}\Big)\Big]\Big)\Big\},
\]
which is a single finite-dimensional convex program in the $|\calX|+1$ variables
$(\eta,(\zeta_x)_{x\in\calX})$, as the perspective is jointly convex by
Lemma~\ref{lemma: entropy function proper cvx lsc} and its arguments are affine in these
variables.

\subsubsection{Inference on the sensitivity bounds}
\label{subsubsec: rct attrition inference}

We now quantify the sampling uncertainty in the estimated bounds. The
large sample theory of Section~\ref{sec:estimation-inference} holds nuisance
parameters such as the complier share $\pi$ fixed (Remark~\ref{rem:nuisance-EB}).
Here no structural parameter enters, the bound is an explicit functional of
$(\PT,\pi,\mu^{\mathsf{AO}}(0))$, and the dual optimizer is unique whenever the density of
$\PT$ is continuous and strictly positive at the trimming quantiles. We therefore work
directly with the trimmed-mean representation of the bound: conditional on
$\pi$, standard quantile and truncated-moment expansions deliver its influence
function, to which we add the first-order contribution of estimating $\pi$ from
the same sample. This route does not impose compactness of the outcome
support. By Remark~\ref{rem:singleton-EB}, the bootstrap of
Algorithm~\ref{alg:FS-EB} reduces here to the ordinary nonparametric
bootstrap, in which $\pi$, the control-observed mean, and the trimming quantiles
are recomputed in each resample.

Throughout, the data $\{(Y_iS_i,S_i,D_i)\}_{i=1}^{n}$ are an i.i.d.\ sample generated
under
$P_0$ and we only discuss the upper bound. Write
$p:=P_0(D=1)$, $r_1:=P_0(S=1\mid D=1)$, and $r_0:=P_0(S=1\mid D=0)$ for the
assignment and arm-specific response probabilities, so that the complier share
of \eqref{eq: mixture AT and C} is
\[
  \pi \;=\; 1-\frac{r_0}{r_1}.
\]
The bound on $\tau_0$ is an explicit functional
of $(\PT,\pi,\mu^{\mathsf{AO}}(0))$ as shown above, with upper-tail trimming
points $s_\mathsf{L}(\delta)=F_{\PT}^{-1}(\pi\delta)$ and
$s_\mathsf{U}(\delta)=F_{\PT}^{-1}(1-(1-\pi)\delta)$. Its
plug-in estimator replaces $r_1,r_0,p$ by the corresponding sample frequencies, $\PT$
by the empirical distribution $\widehat P_{\mathsf T,n}$ of the treated observed outcomes
$\{Y_i:D_i=1,S_i=1\}$, the trimming points by the empirical quantiles
$\widehat s_\mathsf{L}(\delta),\widehat s_\mathsf{U}(\delta)$ of
$\widehat P_{\mathsf T,n}$, and $\mu^{\mathsf{AO}}(0)$ by the control observed sample
mean.

Write $\widehat{\overline\tau}(\delta)$ for this plug-in estimator of the bound
$\overline\tau(\delta)$, and let
$\{\widehat{\overline\tau}{}^{(b)}(\delta)\}_{b=1}^{B}$ be its values on $B=1000$
nonparametric bootstrap samples, with mean
\[{\widehat{\overline\tau}}_{\mathsf{boot}}(\delta)
=\frac{1}{B}\sum_{b=1}^B\widehat{\overline\tau}{}^{(b)}%
(\delta).\]
The limiting law of $\sqrt
n\bigl(\widehat{\overline\tau}(\delta)-\overline\tau(\delta)\bigr)$ is
governed by three first-order influence functions. Writing
$W=(YS,S,D)$, $\overline\mu(\delta):=\overline\mu^{\mathsf{AO}}(1;\delta)$, and
$\vartheta_+(y;\delta):=y\,\indic\{s_\mathsf{L}(\delta)<y<s_\mathsf{U}(\delta)\}
+\tfrac1{1-\pi}\,y\,\indic\{y\geq s_\mathsf{U}(\delta)\}$ so
that $\E_{\PT}[\vartheta_+(Y;\delta)]=\overline\mu(\delta)$, these influence functions are
\begin{align*}
  \psi_{\mathsf{T}}(W;\delta) &=
  \frac{DS}{p\,r_1}\Bigl\{\vartheta_+(Y;\delta)-\overline\mu(\delta)
     + s_\mathsf{L}(\delta)\bigl(\indic\{Y\leq s_\mathsf{L}(\delta)\}-\pi\delta\bigr)
     + \tfrac{\pi}{1-\pi}\,s_\mathsf{U}(\delta)\bigl(\indic\{Y\leq
     s_\mathsf{U}(\delta)\}-1+(1-\pi)\delta\bigr)\Bigr\},\\
  \psi_{\pi}(W) &= \frac{r_0}{r_1^{2}}\,\frac{D}{p}\,(S-r_1)
     -\frac{1}{r_1}\,\frac{1-D}{1-p}\,(S-r_0),\qquad
  \psi_0(W) = \frac{(1-D)\,S}{(1-p)\,r_0}\bigl(Y-\mu^{\mathsf{AO}}(0)\bigr).
\end{align*}
Only $\psi_{\mathsf{T}}(\cdot\,;\delta)$ varies with the sensitivity
level. The other two, the influence functions of
the estimated complier share $\widehat{\pi}=1-\widehat r_0/\widehat r_1$
and of the control mean $\mu^{\mathsf{AO}}(0)$, are free of $\delta$.
Denote the total derivative of the upper bound with respect to the complier share as
\[
  \varkappa(\delta)\;:=\;-\delta\Bigl(s_\mathsf{L}(\delta)
  +\tfrac{\pi}{1-\pi}\,s_\mathsf{U}(\delta)\Bigr)
  +\tfrac{1}{(1-\pi)^2}\,\E_{\PT}\bigl[Y\,\indic\{Y\geq s_\mathsf{U}(\delta)\}\bigr]
\]
and define the second-step influence function
\[\psi_2(W;\delta):=\psi_{\mathsf{T}}(W;\delta)-\psi_0(W),\] which holds $\pi$ fixed,
and the full influence function
\[\psi_{\mathsf{full}}(W;\delta):=\psi_{\mathsf{T}}(W;\delta)
+\varkappa(\delta)\,\psi_{\pi}(W)-\psi_0(W),\]
which also takes $\widehat{\pi}$ into account. Because $\psi_{\mathsf{full}}$ is the
influence function of the upper bound
functional, the moment function $\overline\tau(\delta)+\psi_{\mathsf{full}}(W;\delta)$
has zero pathwise derivative with respect to $(\PT,\pi,\mu^{\mathsf{AO}}(0))$ at their
population values, that is, it is Neyman orthogonal.
We form four estimators of the sampling variability of
$\widehat{\overline\tau}(\delta)$
\[
\widehat\sigma_{\mathsf{naive}}^{2}(\delta)=\widehat V_2(\delta),\quad
\widehat\sigma_{\mathsf{NO}}^{2}(\delta)=\widehat V_{\mathsf{full}}(\delta),\quad
\widehat\sigma_{\mathsf{boot}}^{2}(\delta)=\frac{n}{B-1}\sum_{b=1}^{B}\bigl(\widehat{%
\overline\tau}{}^{(b)}(\delta)
-{\widehat{\overline\tau}}_{\mathsf{boot}}(\delta)\bigr)^{2},\quad
\widehat\sigma_{\mathsf{Lee}}^{2}=\widehat V_{\mathsf{Lee},\mathsf{U}}+\widehat
V_{\mathsf{Lee},\mathsf{C}},
\]
where $\widehat V_2(\delta):=n^{-1}\sum_{i}\widehat\psi_2(W_i;\delta)^2$ and
$\widehat V_{\mathsf{full}}(\delta):=n^{-1}\sum_{i}\widehat\psi_{\mathsf{full}}(W_i;\delta)^2$
are the
influence-function variances. The remaining two are the ingredients of
the trimming bound variance of \citet[Proposition~3]{lee2009TrainingWagesSample}:
$V_{\mathsf{Lee},\mathsf{U}}$ is the asymptotic variance of the estimated treated trimmed
mean $\E_{\PT}[Y\mid Y\geq F_{\PT}^{-1}(\pi)]$ and $V_{\mathsf{Lee},\mathsf{C}}$ that of
the estimated control mean $\mu^{\mathsf{AO}}(0)$, with sample analogs $\widehat
V_{\mathsf{Lee},\mathsf{U}}$ and $\widehat V_{\mathsf{Lee},\mathsf{C}}$.
The naive estimator $\widehat\sigma_{\mathsf{naive}}^{2}(\delta)$ treats the estimated
complier share $\widehat{\pi}$ as known,
whereas the variance estimator $\widehat\sigma^2_{\mathsf{NO}}(\delta)$ is based on the
Neyman-orthogonal score and therefore uses the full influence function, including
the channel through which $\widehat{\pi}$ contributes to sampling uncertainty.

We now state some standard regularity conditions before presenting the main
inferential result of this section.
\begin{assump}[Inference regularity]\label{assump: attrition inference}
The sample $\{(Y_iS_i,S_i,D_i)\}_{i=1}^{n}$ is i.i.d.\ under $P_0$, $p\in(0,1)$,
$r_0,r_1\in(0,1]$ with $\pi=1-r_0/r_1\in(0,1)$,
$\delta\in(0,1]$, $\E_{P_0}[Y^2\mid D=d,S=1]<\infty,d\in\{0,1\}$,
and the baseline $\PT$ admits a density $f_{\PT}$ that is continuous and
strictly positive at the trimming quantiles $s_\mathsf{L}(\delta)$ and
$s_\mathsf{U}(\delta)$.
\end{assump}

\begin{prop}[Nuisance parameter uncertainty
quantification]\label{prop: attrition inference}
Let Assumption~\ref{assump: attrition inference} hold. Then:
\begin{enumerate}[label=(\roman*),ref=\theprop(\roman*),leftmargin=2.4em,
  itemsep=5pt,topsep=4pt]
\item\label{prop: attrition inference-i}
The plug-in is asymptotically normal,
\[
  \sqrt
  n\bigl(\widehat{\overline\tau}(\delta)-\overline\tau(\delta)\bigr)
  \;\rightsquigarrow\;\normal\bigl(0,\sigma_{\mathsf{full}}^2(\delta)\bigr),\qquad
  \sigma_{\mathsf{full}}^2(\delta)=\E\bigl[\psi_{\mathsf{full}}(W;\delta)^2\bigr]
  .
\]

\item\label{prop: attrition inference-ii}
Writing $\sigma_2^2(\delta):=\E[\psi_2(W;\delta)^2]$ for the asymptotic variance
when $\pi$ is known and $V_{\pi}:=\E[\psi_{\pi}^2]$, as $n\to\infty$
\[
\widehat\sigma_{\mathsf{NO}}^2(\delta)\;\cprob\; \sigma_{\mathsf{full}}^2(\delta),\qquad 
\widehat\sigma_{\mathsf{boot}}^2(\delta)\;\cprob\; \sigma_{\mathsf{full}}^2(\delta),
\qquad
  \widehat\sigma_{\mathsf{naive}}^2(\delta)
     \;\cprob\; \sigma_2^2(\delta),
\]
where $\sigma_2^2(\delta)=\sigma_{\mathsf{full}}^2(\delta)-\varkappa(\delta)^2\,V_{\pi}
\;\leq\;\sigma_{\mathsf{full}}^2(\delta)$.
If, in addition, $r_1<1$ and $f_{\PT}$ is continuous and strictly positive
at $F_{\PT}^{-1}(\pi)$, then
$\widehat\sigma_{\mathsf{Lee}}^2\cprob\sigma_{\mathsf{full}}^2(1)$.

\end{enumerate}
\end{prop}

\begin{flushright}
[\hyperref[proof:prop: attrition inference]{Proof}]
\end{flushright}
The term $\varkappa\,\psi_{\pi}$ is exactly the contribution of estimating
$\pi$. Discarding it, that is, treating $\pi$ as known as the general theory
does, understates the variance. The understatement vanishes at
$\delta=0$, where $\varkappa\to0$ and the bound is the difference in means, which does
not require estimation of $\pi$. Because the plug-in is asymptotically
normal (part~(i)) with a variance that $\widehat\sigma_{\mathsf{NO}}(\delta)$ and
$\widehat\sigma_{\mathsf{boot}}(\delta)$ estimate consistently (part~(ii)), the Wald
intervals based on these two standard errors attain asymptotic coverage,
whereas the naive interval under-covers whenever $\varkappa(\delta)\neq0$.
At the endpoint $\delta=1$, the bound reduces to Lee's trimmed mean and
$\sigma_{\mathsf{full}}^2(1)=V_{\mathsf{Lee},\mathsf{U}}+V_{\mathsf{Lee},\mathsf{C}}$ is
the trimming bound variance of \citet[Proposition~3]{lee2009TrainingWagesSample}.

\subsection{Counterfactuals in Structural Models}
\label{subsec: structural models}
Counterfactual predictions from structural models depend on the distribution
specified for latent variables such as taste shocks, unobserved heterogeneity,
or productivity shocks, which is typically not nonparametrically identifiable
and is therefore chosen for tractability.
\citet{christensen2023CounterfactualSensitivityRobustness} (\textsf{CC23})
propose to assess how sensitive counterfactuals are to this choice
by letting the latent distribution range over a $\phi$-divergence neighborhood
of the parametric specification. We conduct the same exercise with a
sensitivity parameter that admits a population share interpretation, a case
that is not covered by the theory of \textsf{CC23}, whose entropy functions
are superlinear (see the discussion of the pointwise dual constraint
at the beginning of Section~\ref{subsec: ignoring dual constraint}).

\textbf{A generic structural model.} Let $Z\in\calZ$ collect the latent variables, let
$\Pbase{}\in\calP(\calZ)$ be the distribution postulated by the econometrician,
which is known, and let $\theta\in\Theta$ collect the structural parameters.
The model rationalizes the observed data through the moment conditions
$\E_P[m(Z;\theta)]\in\calM(\rho_0)$, where $\rho_0$ collects the moments that
are estimated from the data, and the counterfactual of interest is the
expectation of a known function, $\E_P[g(Z;\theta)]$. The relevant robustness set is
the contamination robustness set of the main text, recalled in
\eqref{eq: paper contamination set mapping},
\[
\calP_{\mathsf{C}}(\theta;\delta)=\big\{P\in\calP(\calZ):\
P=(1-\delta)\Pbase{}+\delta R\ \text{for some}\ R\in\calP(\calZ),\
\E_P[m(Z;\theta)]\in\calM(\rho_0)\big\},
\]
so that $\delta$ is the share of the population whose latent variables are not
drawn from $\Pbase{}$. This set is the member $\mathbb{U}=[1-\delta,+\infty)$
of the family of Section~\ref{subsec:ls-setup}, that is, the contamination
neighborhood $\calC_{1-\delta}(\Pbase{})\cap\calE(\rho_0)$ of
Section~\ref{subsec: contamination tv}. The sensitivity bounds on the
counterfactual are
\[
\underline{\mathsf{b}}_{\mathsf{C}}(\delta)
:=\inf_{\theta\in\Theta}\ \inf_{P\in\calP_{\mathsf{C}}(\theta;\delta)}\E_P[g(Z;\theta)],
\qquad
\overline{\mathsf{b}}_{\mathsf{C}}(\delta)
:=\sup_{\theta\in\Theta}\ \sup_{P\in\calP_{\mathsf{C}}(\theta;\delta)}\E_P[g(Z;\theta)],
\]
with the understanding that the inner infimum and supremum are $+\infty$ and
$-\infty$, respectively, whenever $\calP_{\mathsf{C}}(\theta;\delta)$ is empty.
Their plug-in estimators replace $\rho_0$ by an estimator $\widehat\rho_n$, and,
as the baseline is known, their large sample properties are those of
Remark~\ref{rem:known-baseline}.

\subsubsection{The marital college premium}
\label{subsubsec: gumbel model verification}
We borrow the setting from Example~2.2 of the
main text,
which
revisits one of the
leading empirical applications of \textsf{CC23}: the marital college premium in
the marriage-matching model that
\citet*{chiappori2017PartnerChoiceInvestment} estimate on American Community
Survey data. Women belong to one of $J\in\N$ education groups, which we term
``types,'' $j=0$ denotes the outside option of remaining unmatched, and a
type-$b\in[J]$ woman receives utility $\nu_{0b}$ if unmatched and
$t_{jb}+\nu_{jb}$ if matched with a type-$j\in[J]$ man, where we set
$t_{0b}:=0$ for the outside option. As in the main text,
the distribution of the preference shocks does not depend on the type of the
woman (see also Remark~\ref{rem: non vacuous bounds in Gumbel}), so that we
consider a single random vector $U=(\nu_0,\ldots,\nu_J)^\top$, whose postulated
distribution $\widetilde{P}$ makes $\nu_0,\ldots,\nu_J$ independent Gumbel
random variables. We fix two female types $b,b'\in[J]$,
let $\varrho\in\R^{2J}$ collect their observed marriage probabilities with each
male type, and let $\vartheta:=(t_{jb},t_{jb'},j\in[J])$ range over a parameter
space $\Theta\subseteq\R^{2J}$. The moment function is
$s(U;\vartheta)=(s_1(U;\vartheta)^\top,s_2(U;\vartheta)^\top)^\top$, where
\[s_1(U; \vartheta)=
\begin{bmatrix}
\Big(\indic\Big\{j=\argmax\limits_{j'\in\{0, \ldots, J\}}\Big(t_{j'
b}+\nu_{j'}\Big)\Big\}\Big)_{j=1}^J \\
\Big(\indic\Big\{j=\argmax\limits_{j'\in\{0, \ldots, J\}}\Big(t_{j'
b'}+\nu_{j'}\Big)\Big\}\Big)_{j=1}^J
\end{bmatrix},
\qquad
s_2(U;\vartheta) = (\nu_j, \nu_j^2-\pi^2/6)_{j=0}^J,\]
with ties broken toward the lowest indexed alternative. The moment conditions
$\E_{P}[s_1(U;\vartheta)]=\varrho$ equate the observed marriage probabilities to
the model implied choice probabilities, and
$\E_{P}[s_2(U;\vartheta)]=0_{2(J+1)}$ imposes that each shock has mean zero and
variance $\pi^2/6$, the variance of a standard Gumbel distribution. The
counterfactual is the marital college premium for female types $b$ and $b'$,
$\E_P[f(U;\vartheta)]$, where
\[
f(U;\vartheta)
=
\max_{j\in\{0,\dots,J\}}\{t_{jb}+\nu_{j}\} -
\max_{j\in\{0,\dots,J\}}\{t_{jb'}+\nu_{j}\}.
\]
The robustness set is $\calP_{\mathsf{STR}}(\vartheta;\delta)$ of
\textup{(2.7)} in Example~2.2 of the
main text, so that $\delta$ is the share of women whose shocks are not drawn from the
Gumbel distribution, and the sensitivity bounds $\underline{\mathsf{k}}(\delta)$ and
$\overline{\mathsf{k}}(\delta)$ on the premium are the infimum and the supremum of
$\E_{P}[f(U;\vartheta)]$ over $\vartheta\in\Theta$ and
$P\in\calP_{\mathsf{STR}}(\vartheta;\delta)$.
This is the generic exercise described above with $Z=U$, $\calZ=\R^{J+1}$,
$\theta=\vartheta$, $g=f$, $m=s$, $\rho_0=\varrho$, the block $s_2$ having a
known target, and $\Pbase{}=\widetilde{P}$, so that
$\calP_{\mathsf{STR}}(\vartheta;\delta)=\calP_{\mathsf{C}}(\vartheta;\delta)$,
$\underline{\mathsf{k}}(\delta)=\underline{\mathsf{b}}_{\mathsf{C}}(\delta)$, and
$\overline{\mathsf{k}}(\delta)=\overline{\mathsf{b}}_{\mathsf{C}}(\delta)$.
The first identity uses that every
$P\in\calP_{\mathsf{STR}}(\vartheta;\delta)$ lies in the admissible class
$\calP(\R^{J+1})$ of \eqref{eq: admissible P}. Indeed,
$|f(u;\vartheta)|\leq\max_{j\in\{0,\ldots,J\}}|t_{jb}-t_{jb'}|$ and $s_1$ takes
values in $\{0,1\}^{2J}$, while the normalization $\E_P[s_2(U;\vartheta)]=0_{2(J+1)}$
makes $s_2(U;\vartheta)$ integrable under $P$. The
observed marriage probabilities are estimated by the corresponding empirical
frequencies $\widehat{\varrho}_n$. In the remainder of this section we state
the results in the notation of the framework.

\begin{rem}[Vacuous bounds]\label{rem: non vacuous bounds in Gumbel}
If $g$ is unbounded above or below, then the robustness set alone is too
large, as the contaminating distribution $R$ can place its mass on regions where
$g$ is arbitrarily large or arbitrarily negative, implying that
\[
\overline{\mathsf{k}}(\delta)=+\infty
\qquad\text{and}\qquad
\underline{\mathsf{k}}(\delta)=-\infty
\]
for every $\delta>0$. This issue arises, for example, if one writes the premium
using two different collections of shocks for types $b$ and $b'$, and the shocks
$\nu_{jb}$ and
$\nu_{jb'}$ have unbounded support, as under Gaussian or Gumbel specifications.
There are at least two ways to make such bounds non-trivial:
\begin{enumerate}
    \item impose a common-shock representation, so that the premium becomes bounded;
    \item intersect the robustness set with moment normalizations which impose structure
    on the tails of the distribution.
\end{enumerate}
In this section we adopt both for simplicity, but the moment
normalizations contained in $s_2$ would have sufficed to bound the premium uniformly over
the robustness set.
\end{rem}

Throughout the remainder of this application, we consider only
distributions that are absolutely continuous with respect to the baseline. This is
without loss of generality for the sensitivity bounds, because condition~(b) of
Theorem~\ref{thm: ignoring dual constraint} holds for the entropy
$\underline{\phi}_{\TV}=\phi_\TV+\iota_{[1-\delta,\infty)}$ of
Proposition~\ref{prop: induced entropy-ii}, whose neighborhood is the
contamination set, as we show in the next lemma. Moreover,
Proposition~\ref{cor: ac unconstrained dual} allows us to remove the
pointwise
constraint from the dual.

For the next result, define
$\rhopad_0:=(\rho_{0}^{\top},0_{2(J+1)}^{\top})^{\top}$ and let
$\Gamma^{\ll}_{\underline{\phi}_{\TV}}(\theta;\delta)$ be the attainable moment set of
\eqref{eq: attainable moment set} for this entropy, restricted to distributions dominated
by $\Pbase{}$, that is, by Proposition~\ref{prop: induced entropy-ii}, the set of the
$\E_{P}[m(U;\theta)]$ with $P\in\calP(\R^{J+1})$, $P\ll\Pbase{}$, and
$P=(1-\delta)\Pbase{}+\delta R$ for some $R\in\calP(\R^{J+1})$.
At every $\theta$ in a neighborhood of the population optimizers, we maintain
the interior condition
$\rhopad_0\in\rint(\Gamma^{\ll}_{\underline{\phi}_{\TV}}(\theta;\delta))$.
A primitive sufficient condition is obtained as in
Lemma~\ref{lemma: constraint qualification-ii}: it suffices that the moment
restrictions be satisfied by some $\bar P\ll\Pbase{}$ whose density with respect
to $\Pbase{}$ is bounded below by a constant strictly larger than $1-\delta$
and under which $m(U;\theta)$ has a finite, full rank covariance matrix (see
the discussion following Lemma~\ref{lemma: family duality}).

\begin{lemma}[Absolutely continuous Gumbel formulation]\label{lem:gumbel-coercive}
For every $\theta\in\Theta$ and every $\lambda\in\Rlam$,
\[
\sup_{u\in\R^{J+1}}\{g(u;\theta)-\lambda^{\top}m(u;\theta)\}
=\Pbase{}\text{-}\esssup_{u\in\R^{J+1}}\{g(u;\theta)-\lambda^{\top}m(u;\theta)\},
\]
and the same identity holds with $-g$ in place of $g$. Consequently, for every
$(\theta,\delta)$ such that $\delta>0$ and
$\rhopad_0\in\rint(\Gamma^{\ll}_{\underline{\phi}_{\TV}}(\theta;\delta))$,
condition~(b) of Theorem~\ref{thm: ignoring dual constraint} holds for the entropy
$\underline{\phi}_{\TV}=\phi_\TV+\iota_{[1-\delta,\infty)}$, for $g(\cdot;\theta)$
and for $-g(\cdot;\theta)$.
\end{lemma}
\begin{flushright}
[\hyperref[proof:lem:gumbel-coercive]{Proof}]
\end{flushright}

\subsection{Formula and Shift-Share Instruments}
\label{subsec: formula instruments}
A growing body of empirical work identifies causal effects by means of formula
instruments, that is, instruments obtained by applying a known formula to a
vector of exogenous shocks and to a vector of predetermined covariates.
We refer the
reader to \citet*{borusyak2023DesignBasedIdentificationFormula} for a review.
\citet{borusyak2023NonrandomExposureExogenous} (\textsf{BH23}, henceforth) point out that
exogeneity of the shocks alone does not deliver a consistent estimator.
In the main text we show that this scenario fits
into the framework of Section~\ref{subsec: setup}.
In this subsection we redefine all the relevant quantities and explain in detail how
the sensitivity bounds can be computed.

We follow the notation of \textsf{BH23} throughout. 
Let $i\in[n], n\in\N$ index the units of the sample and consider the causal or structural
model
\[
y_i=\beta x_i+\varepsilon_i,
\]
where $y_i\in\R$ is a demeaned outcome, $x_i\in\R$ is a demeaned endogenous regressor,
$\varepsilon_i\in\R$ is
an unobserved residual, and $\beta\in\R$ is the parameter of interest.
Let $v=(v_1,\ldots,v_K)\in\calV$, $K\in\N$, denote a vector of
exogenous shocks, where $\calV$ is the set of shock configurations the design
allows for, a subset of $\R^K$ when the shocks are real valued, and let $w$ collect
predetermined covariates of unrestricted
dimension. The candidate instrument of unit $i\in[n]$ is
\[
z_i:=f_i(v;w),
\]
where $f_i(\cdot\,;w):\calV\to\R$ are known nonstochastic functions on $\calV$.
\textsf{BH23} assume that the shocks $v$ are exogenous, that is
\[v\independent(\varepsilon_1,\ldots,\varepsilon_n)\mid w,\]
and show that this alone does not make $z_i$ orthogonal to $\varepsilon_i$.
Their proposed solution is to recenter $z_i$ at its expected value under the assignment
process of the shocks. Computing that expected value requires the
researcher to specify a distribution for the shocks, an assumption that economic theory
rarely pins down. Exogeneity is maintained throughout the sensitivity
exercise. What the sensitivity exercise relaxes is the assumption that the
law of $v$ given $w$ is known. We let $\widetilde{P}\in\Delta(\calV)$
denote the assignment process postulated by the econometrician, and we assume
for simplicity that the distribution of $v$ given $w$ does not depend on $w$,
as is the case in the application we consider.

The postulated assignment process defines the expected instruments
$\mu_i\equiv \mu_i(\widetilde{P}),$ where for $i\in[n]$
\[
\mu_i(P):=\E_{P}\bigl[f_i(v;w)\bigm| w\bigr]
=\int_{\calV}f_i\bigl(v';w\bigr)\,\de P(v'),
\]
and the recentered instrument of unit $i$ is $\widetilde{z}_i(P):=z_i-\mu_i(P)$, 
with $\widetilde{z}_i\equiv \widetilde{z}_i(\widetilde{P})$.
The recentered instrumental variable estimator of
\textsf{BH23} is
\begin{align}
\widehat{\beta}_*
:=\frac{\sum_{i=1}^{n}\widetilde{z}_i\,y_i}
       {\sum_{i=1}^{n}\widetilde{z}_i\,x_i}.
\label{eq: riv estimator}
\end{align}

We use a ``design based'' setup in which we hold the realized sample fixed (outcomes,
treatments, covariates, and the realized shocks $\check{v}$, at which
$z_i=f_i(\check{v};w)$ for every $i\in[n]$), so
that the recentered estimate is a function of the assignment process alone,
and we ask how much it moves as that process ranges over a neighborhood of
$\widetilde{P}$. Because \eqref{eq: riv estimator} depends on the postulated assignment
process $\widetilde{P}$ only through the expected instruments, the sensitivity exercise
amounts to replacing $\widetilde{P}$ by candidate assignment processes in some robustness
set. We accordingly define
\begin{align}
\widehat\beta(P)
:=\frac{\sum_{i=1}^{n}\widetilde{z}_i(P)\,y_i}
       {\sum_{i=1}^{n}\widetilde{z}_i(P)\,x_i}
=\frac{G_y(P)}{G_x(P)},
\qquad
\begin{aligned}
G_y(P)&:=g_y(\check{v})-\E_P[g_y],\\
G_x(P)&:=g_x(\check{v})-\E_P[g_x],
\end{aligned}
\label{eq: riv estimator candidate}
\end{align}
where 
\[
g_y(\cdot):=\sum_{i=1}^{n}y_i f_i(\cdot\,;w),
\qquad
g_x(\cdot):=\sum_{i=1}^{n}x_i f_i(\cdot\,;w).
\]
By construction $\widehat\beta_*=\widehat\beta(\widetilde{P})$, so the exercise is
centered around the reported estimate.

We can conduct a sensitivity exercise on $\widehat{\beta}_*$ by letting
the assignment process of $v$ vary over the robustness set
\[
\calP_{\mathsf{FI}}(\delta):=\{P\in\Delta(\calV):\ P=(1-\delta)\widetilde{P}+\delta
R\ \text{for some}\ R\in\Delta(\calV)\}.
\]
Also, define the set
\[
\calP_{\neq0}
:=\bigl\{P\in\Delta(\calV):\ G_x(P)\neq0\bigr\},
\]
which excludes the distributions in the robustness set at which
the recentered first stage equals zero and the estimator is therefore
undefined.

We assume throughout that the recentered first stage is positive at the
postulated law, $G_x(\widetilde{P})>0$, so that the reported estimate in
\eqref{eq: riv estimator} is well defined. This is without loss of generality,
because replacing $(g_y,g_x)$ by $(-g_y,-g_x)$ leaves $\widehat\beta(\cdot)$
unchanged and reverses the sign of $G_x(\widetilde{P})$. The space $\calV$ of
shock realizations is otherwise arbitrary, and we maintain instead a condition
on the formula itself, namely that it is bounded uniformly over the sample,
\begin{align}
\|f\|_{\infty}:=\max_{i\in[n]}\ \sup_{v'\in\calV}\bigl|f_i(v';w)\bigr|<\infty.
\label{eq: fi bounded criteria}
\end{align}
Since the exercise is conducted conditionally on the realized sample,
\eqref{eq: fi bounded criteria} implies
\[
\sup_{v'\in\calV}\bigl|g_y(v')\bigr|\leq\|f\|_{\infty}\sum_{i=1}^{n}|y_i|,
\qquad
\sup_{v'\in\calV}\bigl|g_x(v')\bigr|\leq\|f\|_{\infty}\sum_{i=1}^{n}|x_i|.
\]
This condition keeps the sensitivity bounds informative. Without it, the
robustness set can contain assignment processes that place mass where $g_x$ or $g_y$
is arbitrarily large, and the bounds can be infinite
(Remark~\ref{rem: formula instruments unbounded}). Condition
\eqref{eq: fi bounded criteria} does not restrict $\calV$, and it holds in each of the
three examples that motivate \textsf{BH23}, which we discuss next.

\begin{ex}[\textsf{BH23} motivating examples]
In their network spillover example, a stylized version of the setting of
\citet*{miguel2004WormsIdentifyingImpacts}, the treatment of student $i$ counts
the number of her neighbors who have been dewormed in an intervention, so that
\[
f_i(v';w)=\sum_{k=1}^{K}\mathsf{Neighbor}_{ik}\,v'_k,
\]
where $v'_k\in\{0,1\}$ indicates that student $k$ is selected for the deworming
intervention, $\mathsf{Neighbor}_{ik}\in\{0,1\}$ indicates that $i$ and $k$ are
connected by an observed network link, and the shocks are assigned at the level
of the $K=n$ students. Then, the formula is bounded by the largest number of neighbors
that a student has. In their market access example, which follows
\citet*{donaldson2016RailroadsAmericanEconomic}, the treatment of region $i$ is
the growth of its market access between two dates due to improvements to the
interregional railroad network, that is
\[
f_i(v';w)
=\log\mathsf{MA}_{i}\bigl(\mathsf{Network}_1(v')\bigr)
-\log\mathsf{MA}_{i}\bigl(\mathsf{Network}_0\bigr),
\qquad
\mathsf{MA}_{i}(\mathsf{Network})
:
=\sum_{j=1}^{n}\frac{\mathsf{Pop}_j}{\mathsf{Time}\bigl(\mathsf{Network},\mathsf{Loc}_i,\mathsf{Loc}_j\bigr)},
\]
where $\mathsf{Pop}_j$ is the time invariant population of region $j$,
$\mathsf{Loc}_j$ is its location, $\mathsf{Time}(\cdot)$ gives the travel time between two
regions, and $\mathsf{Network}$ is the set of railway lines and other types of
transit in operation. The shocks enter through the network in operation at the
later date: $v'_k\in\{0,1\}$ indicates that planned line $k$ faces no delays, so
that
\[
\mathsf{Network}_1(v')
=\mathsf{Network}_0\cup\bigl\{k\in[K]:v'_k=1\bigr\},
\]
where $\mathsf{Network}_0$ denotes the initial network. 
Since the shocks are indicators, the
formula takes finitely many values and is therefore bounded. The
application in Section~6.3 of the main text takes up this design,
in the version that \textsf{BH23} apply to the construction of the Chinese high-speed
rail network. In their Medicaid eligibility
example, which comes from \citet*{currie1996HealthInsuranceEligibility}, the
treatment of individual $i$ indicates her eligibility for Medicaid, so that
\[
f_i(v';w)=\indic\bigl\{\mathsf{IncDem}_i\in v'_{\mathsf{State}_i}\bigr\},
\]
where $\mathsf{IncDem}_i$ is a vector of individual income and demographics,
$\mathsf{State}_i\in[K]$ indexes $i$'s state of residence, and $v'_k$ is state
$k$'s eligibility policy, that is, the set of income and demographic groups
eligible for Medicaid in that state. Here $\|f\|_{\infty}=1$ whatever the space
of policy profiles.

Condition \eqref{eq: fi bounded criteria} fails, on the other hand, for
a shift-share formula 
\[f_i(v';w)=\sum_{k\in[K]}w_{ik}v'_k\] 
with shocks ranging over an unbounded set, such as $\calV=\R^K$, and
Remark~\ref{rem: formula instruments unbounded} describes the two
ways of rescuing the exercise in that case.
\end{ex}

The objects of interest are the largest and the smallest value that the
recentered estimator takes as the law of $v$ ranges over the
robustness set, that is
\begin{align}
\underline{\beta}(\delta)
:=\inf_{P\in\calP_{\mathsf{FI}}(\delta)\cap\calP_{\neq0}}\widehat\beta(P),
\qquad
\overline{\beta}(\delta)
:=\sup_{P\in\calP_{\mathsf{FI}}(\delta)\cap\calP_{\neq0}}\widehat\beta(P).
\label{eq: fi cont bounds}
\end{align}

The estimator in \eqref{eq: riv estimator candidate} is not linear in the
candidate distribution $P$, so we cannot apply our framework directly to
the two optimization programs above. Its numerator and denominator are,
however, both affine in $P$, so that the estimator depends on $P$ only through the two
scalars $\E_P[g_y]$ and $\E_P[g_x]$, so that the exercise takes place in a
two-dimensional set no matter the dimensionality of $\calV$. Moreover, 
given $b\in\R$, the requirement that the candidate
estimate equals $b$ reduces to a single linear restriction on the candidate
$P\in\calP_{\neq0}$. Indeed, $\widehat\beta(P)=b$ if and only if
$G_y(P)-b\,G_x(P)=0$ and
\[
G_y(P)-b\,G_x(P)
=\bigl\{g_y(\check{v})-b\,g_x(\check{v})\bigr\}-\E_P\bigl[g_y-b\,g_x\bigr]
=g_b(\check{v})-\E_P[g_b],
\]
where the first equality follows from the definitions of $G_y$ and $G_x$ in
\eqref{eq: riv estimator candidate} and
the second from setting $g_b:=g_y-b\,g_x$. That is, the estimator attains
the value $b$ under $P$ exactly
when the sample orthogonality condition
\[\sum_{i=1}^{n}\widetilde{z}_i(P)\,(y_i-b\,x_i)=0\] 
of the recentered instrument holds at $b$ under $P$.
Once $b$ is fixed, $g_b$ is a known function on $\calV$ that does not vary with
$P$, so that the exercise fits the framework of Section~\ref{subsec: setup} with
$g=g_b$ and no moment restriction, as explained in the main text. 
The next proposition collects the
properties of the sensitivity exercise.

\begin{prop}[Sensitivity bounds under a contamination neighborhood]
\label{prop: formula instruments contamination}
Let $\delta\in[0,1]$.
Write $\overline{g}_y^*:=\E_{\widetilde{P}}[g_y]$ and
$\overline{g}_x^*:=\E_{\widetilde{P}}[g_x]$, so that
$G_y(\widetilde{P})=g_y(\check{v})-\overline{g}_y^*$ and
$G_x(\widetilde{P})=g_x(\check{v})-\overline{g}_x^*$. Also, define
\[
\begin{aligned}
G_y(v';\delta)&:=G_y\bigl((1-\delta)\widetilde{P}
+\delta\,\delta^{\mathtt{Dirac}}_{v'}\bigr)
=\,G_y(\widetilde{P})-\delta\bigl\{g_y(v')-\overline{g}_y^*\bigr\},\\
G_x(v';\delta)&:=G_x\bigl((1-\delta)\widetilde{P}
+\delta\,\delta^{\mathtt{Dirac}}_{v'}\bigr)
=\,G_x(\widetilde{P})-\delta\bigl\{g_x(v')-\overline{g}_x^*\bigr\},
\end{aligned}
\]
for $v'\in\calV$, the values of $G_y$ and $G_x$ when the postulated law is
contaminated by a point mass at $v'$. Then the following statements hold.
\begin{enumerate}[label=(\roman*),ref=\theprop(\roman*)]
    \item\label{prop: formula instruments contamination-i} If
    $\inf_{v'\in\calV}G_x(v';\delta)>0$, then the estimator
    \eqref{eq: riv estimator candidate} is well defined at every
    $P\in\calP_{\mathsf{FI}}(\delta)$ and both bounds in
    \eqref{eq: fi cont bounds} are finite. If instead
    $\inf_{v'\in\calV}G_x(v';\delta)<0$, the neighborhood contains assignment
    laws at which the recentered first stage vanishes.  The former holds exactly
    on
    \[
    \calI_{\mathsf{FS}}
    :=\Bigl\{\delta\in[0,1]:\ \delta\,\sup_{v'\in\calV}\{g_x(v')-\overline{g}_x^*\}
    <G_x(\widetilde{P})\Bigr\},
    \]
    an interval containing the origin whose right endpoint is the first-stage
    breakdown budget
    \[
    \delta_{\mathsf{FS}}
    :=\min\Bigl\{1,\
    \frac{G_x(\widetilde{P})}{\sup_{v'\in\calV}\{g_x(v')-\overline{g}_x^*\}}\Bigr\},
    \]
    with the convention that the ratio equals $+\infty$ when the supremum in its
    denominator vanishes.
    \item\label{prop: formula instruments contamination-ii} For every
    $\delta\in\calI_{\mathsf{FS}}$ the bounds in
    \eqref{eq: fi cont bounds} are given in closed form by
    \[
    \underline{\beta}(\delta)
    =\inf_{v'\in\calV}\frac{G_y(v';\delta)}{G_x(v';\delta)},
    \qquad
    \overline{\beta}(\delta)
    =\sup_{v'\in\calV}\frac{G_y(v';\delta)}{G_x(v';\delta)}.
    \]
    Every value strictly between the two bounds is
    taken by the estimator at some
    $P\in\calP_{\mathsf{FI}}(\delta)$.
\end{enumerate}
\end{prop}
\begin{flushright}
[\hyperref[proof: formula instruments contamination]{Proof}]
\end{flushright}

\begin{rem}[Unbounded formulas]\label{rem: formula instruments unbounded}
When $g_x$ is unbounded above on $\calV$, as it is for a shift-share formula
with shocks ranging over $\calV=\R^K$ whenever $\sum_{i=1}^{n}x_iw_{ik}\neq0$ for some
$k\in[K]$, the exercise as stated is vacuous. Indeed,
$\sup_{v'\in\calV}\{g_x(v')-\overline{g}_x^*\}$ is then $+\infty$, so that at every
$\delta>0$ the robustness set contains assignment processes that place a share
$\delta$ of mass where $g_x$ is arbitrarily large, the first-stage breakdown
budget of part~\ref{prop: formula instruments contamination-i} equals zero, and
both bounds on the estimate are infinite. Remark~\ref{rem: non vacuous bounds in Gumbel}
records the same phenomenon in the marital college premium of
Section~\ref{subsubsec: gumbel model verification}. Here we propose two alternatives
to make the sensitivity exercise informative again.

\uline{\textbf{Normalizing the moments of the shocks.}} A first solution is to leave
$\calV$ unrestricted and impose that the candidate distributions must match the
second moments under $\widetilde{P}$, a moment restriction of the kind the
framework of Section~\ref{subsec: setup} accommodates. The closed forms of
Proposition~\ref{prop: formula instruments contamination} do not survive the restriction,
and the bounds are then computed from the dual representations of
Section~\ref{sec:convex-duality}.

\uline{\textbf{Restricting the configuration space.}} The second route keeps the
formula as it is and takes $\calV$ to be a finite set of possible shock 
configurations that the design allows for. For example, for the
shift-share formula of the \textsf{BH23} examples above one may take $\calV$ to be the
finite set of permutations of the realized shocks $\check{v}$, on which the formula takes
finitely many values and is therefore bounded. In the application of
Section~6.3 of the main text, the permutations of line completion
status of \textsf{BH23} likewise define a finite $\calV$, although there the formula is
bounded on the whole of $\{0,1\}^K$.
\end{rem}

\begin{rem}[Inference]\label{rem: formula instruments inference}
The bounds of this section carry no sampling uncertainty. As in the
design-based setup above, the sample is held fixed and only the postulated assignment
process varies. Unlike in Sections~\ref{subsec: rct attrition}
and~\ref{subsec: structural models}, where the baseline or the moment targets are
estimated, the bounds of this section are therefore bounds for the estimate
$\widehat{\beta}_*$ rather than for an estimand.
\end{rem}

\subsection{Other Applications}\label{subsec: other applications}
Our framework accommodates a considerably wider range of exercises than the
RCTs with attrition of Section~\ref{subsec: rct attrition}, the counterfactuals in
structural models of Section~\ref{subsec: structural models}, and the formula
instruments of Section~\ref{subsec: formula instruments}. This section collects several
other examples that are relevant for economic applications.
We then close by discussing
how the framework recovers exact nonparametric sharp bounds in all these
applications, beyond sensitivity analysis. For expositional
purposes, we avoid including moment conditions in the majority of these applications,
but they can be incorporated in either of the two ways described in
Section~\ref{subsec: role of covariates RCT}.

Before turning to the individual applications, we present a lemma
that we invoke repeatedly in what follows. 
\begin{lemma}
\label{lemma: expectation decomposition in a TV neighborhood}
Let $(\calY,\mathscr{F})$ be a measurable space, and let $Q$ be a
probability measure on $(\calY,\mathscr{F})$. Then:
\begin{enumerate}[label=(\roman*),ref=\thelemma(\roman*)]
    \item\label{lemma: expectation decomposition in a TV neighborhood-i}
    if $P$ is a probability measure on $(\calY,\mathscr{F})$ such that
    $\TV(P\,\|\,Q)\leq \delta\in[0,1]$, there exist two
    mutually singular nonnegative measures $\nu^+$ and
    $\nu^-$ with $P-Q=\nu^+-\nu^-$, $\nu^+\leq P$, and $\nu^-\leq Q$, together
    with a set $\calY^+\in\mathscr{F}$ carrying $\nu^+$ in the sense
    that $\nu^+(\calY\setminus\calY^+)=\nu^-(\calY^+)=0$, such that
    \[\nu^+(\calY)=\nu^-(\calY)=\TV(P\,\|\,Q)\leq \delta.\]
    Moreover, the supremum defining the total variation is attained at $\calY^+$,
    that is,
    \[\TV(P\,\|\,Q)=\sup_{A\in\mathscr{F}}|P(A)-Q(A)|
    =|P(\calY^+)-Q(\calY^+)|;\]
    \item\label{lemma: expectation decomposition in a TV neighborhood-ii} if in addition
    $\calY=[\underline{Y},\overline{Y}]$, then $\E_P[Y] \in
    [\underline{\mu}_{Q}(\delta),\overline{\mu}_{Q}(\delta)]$, where $Y$ denotes the
    identity map on $\calY$, $q:=F_Q^{-1}$ denotes the quantile function
    of $Q$, and
    \[
    \begin{aligned}
    \overline{\mu}_{Q}(\delta) &=
    \E_{Q}[Y]+\delta\,\overline{Y}-\int_0^\delta q(u)\,\de u =
    \delta\,\overline{Y}+\int_\delta^1 q(u)\,\de u, \\
    \underline{\mu}_{Q}(\delta) &=
    \E_{Q}[Y]+\delta\,\underline{Y}-\int_{1-\delta}^1 q(u)\,\de
    u = \delta\,\underline{Y}+\int_0^{1-\delta} q(u)\,\de u.
    \end{aligned}
    \]
    Moreover, both endpoints are attained in the sense that there
    exist probability measures $P_+$ and $P_-$ on $(\calY,\mathscr{F})$
    with $\TV(P_\pm\,\|\,Q)\leq\delta$,
    $\E_{P_+}[Y]=\overline{\mu}_{Q}(\delta)$, and
    $\E_{P_-}[Y]=\underline{\mu}_{Q}(\delta)$;
    \item\label{lemma: expectation decomposition in a TV neighborhood-iii} if instead
    $\calY\subseteq\R$ and $\E_Q[|Y|]<\infty$, then, for any
    $\kappa\in(0,1]$, the worst case means over the decomposition
    neighborhood $\calC^{\ll}_{\kappa}(Q)$ are 
    \[
    \sup_{\tilde P\in\calC^{\ll}_{\kappa}(Q)}\E_{\tilde P}[Y]
    =\frac{1}{\kappa}\int_{1-\kappa}^{1} q(u)\,\de u,
    \qquad
    \inf_{\tilde P\in\calC^{\ll}_{\kappa}(Q)}\E_{\tilde P}[Y]
    =\frac{1}{\kappa}\int_{0}^{\kappa} q(u)\,\de u,
    \]
    with $Y$ and $q$ as in part (ii). Moreover, both the supremum and the
    infimum are attained by distributions in $\calC^{\ll}_{\kappa}(Q)$.
\end{enumerate}
\end{lemma}
\begin{flushright}
[\hyperref[proof:lemma: expectation decomposition in a TV neighborhood]{Proof}]
\end{flushright}

Each part of
Lemma~\ref{lemma: expectation decomposition in a TV neighborhood} is put to
work at a specific point of this section. Part (i) underlies the equivalence
between corrupted sampling and the total variation neighborhood established
in Section~\ref{subsec: contaminated data}. Part (ii) delivers the
closed form sensitivity bounds of the external validity exercise of
Section~\ref{subsec: rct external validity}, the missing-data bounds of
Section~\ref{subsec: missing data}, and the closed
forms of the discrete choice exercise of
Section~\ref{subsec: discrete choice}. Part (iii) delivers the
trimmed-mean bounds of the departures-from-randomization exercise of
Section~\ref{subsec: randomization}.

\subsubsection{RCT external validity}\label{subsec: rct external validity}
Suppose an RCT is conducted in one location, indexed by $S=1$, and the object
of interest is the average treatment effect in a different location, indexed
by $S=0$. This is a common situation in which an experiment run in one site, say
a microcredit program evaluated in one country or a cash transfer piloted in
selected municipalities, is used to inform policy elsewhere. Let
$Y(0),Y(1)\in[\underline Y,\overline Y]=:\calY$ denote potential outcomes, let
$D\in\{0,1\}$ denote treatment assignment, let $Y=DY(1)+(1-D)Y(0)$ be the
observed outcome, and let $P_0$ denote the population distribution of
$(Y(0),Y(1),D,S)$. Treatment is randomly assigned in the
experimental location, that is  $D\independent(Y(0),Y(1))$ given $S=1$, with
$0<P_0(D=1\mid S=1)<1$. Therefore, the source-site effect
\[
\tau_{1}(P_0):=\E_{P_0}[Y(1)-Y(0)\mid S=1]
=\E_{P_0}[Y\mid D=1,S=1]-\E_{P_0}[Y\mid D=0,S=1]
\]
is identified, whereas the target parameter is
$\tau_0(P_0)=\E_{P_0}[Y(1)-Y(0)\mid S=0]$, which need not equal $\tau_{1}(P_0)$
if the distribution of potential outcomes differs across locations. The
standard assumption equating the two causal estimands is
transportability, that is
\[(Y(0),Y(1))\mid S=0\overset{d}{=}(Y(0),Y(1))\mid S=1.\] 
This restriction is strong and untestable, so we treat it as a reference
point and ask how departures from this assumption affect the target
estimand.

Because only the marginals of $(Y(0),Y(1))\mid S=1$ are identified under
random assignment and the average treatment effect depends only on those
marginals, we conduct a separate sensitivity exercise for each arm. For
$d\in\{0,1\}$, let $\Pbasesub{d}(\cdot):=P_0(Y(d)\in\cdot\mid S=1)$ denote the
baseline marginal distribution implied by transportability, and define
\[
\calP_{\mathsf{EV}}^d(\delta,\Pbasesub{d}):
=\big\{P\in\Delta(\calY):\TV(P\,\|\,\Pbasesub{d})\leq\delta\big\},
\]
and
\[\underline{\mu}_d(\delta):
=\inf_{P\in\calP_{\mathsf{EV}}^d(\delta,\Pbasesub{d})}\E_P[Y]
=\delta\,\underline{Y}+\int_0^{1-\delta} q_d(u)\,\de u,
\qquad
\overline{\mu}_d(\delta):=\sup_{P\in\calP_{\mathsf{EV}}^d(\delta,\Pbasesub{d})}\E_P[Y]
=\delta\,\overline{Y}+\int_\delta^1 q_d(u)\,\de u,\]
where the second equalities follow from 
Lemma~\ref{lemma: expectation decomposition in a TV neighborhood-ii} and
$q_d$ denotes the quantile function of $\Pbasesub{d}$.
The sensitivity bounds for $\tau_0(P_0)$ are then
\[
\underline{\tau}_{\TV}(\delta)
=\underline{\mu}_1(\delta)-\overline{\mu}_0(\delta),
\qquad
\overline{\tau}_{\TV}(\delta)
=\overline{\mu}_1(\delta)-\underline{\mu}_0(\delta).
\]
At $\delta=0$ both bounds equal $\tau_{1}(P_0)$, whereas 
at $\delta=1$ the closed forms return
$\underline{\tau}_{\TV}(1)=\underline{Y}-\overline{Y}$ and
$\overline{\tau}_{\TV}(1)=\overline{Y}-\underline{Y}$, the sharp nonparametric
bounds implied by the bounded support alone
\citep{manski2003PartialIdentificationProbability}. The exercise thus interpolates
between exact transportability and the fully agnostic bounds, with $\delta$ 
measuring how much probability mass of the target site distribution 
differs from that of the experimental site. In Section
\ref{subsubsec: nonparametric sharp bounds}
we show that the recovery of exact nonparametric sharp bounds is not specific to this
application but a feature of our framework.

\subsubsection{Contaminated and corrupted data}\label{subsec: contaminated data}
Suppose we are interested in a linear functional of the distribution of a
random variable $Y_1\in\calY$,
but we are concerned that the population we sample from might contain data errors, that
is, we observe
\[
Y=(1-E)Y_0+EY_1,
\]
where $E\in\{0,1\}$ is unobserved and indicates whether an observation is error free
($E=1$) or not $(E=0)$. Realizations of $Y$ corresponding to $Y_0$ are termed
``contaminated'' or ``corrupted''. A leading occurrence is survey data in which a share of
the responses is misreported or miscoded, as with income measures in household surveys.
Let $P$ denote the law of the observed outcome $Y$, let $P_1$ denote the
marginal law of $Y_1$, our object of interest, let $\widetilde{P}$ denote the
conditional law of $Y_0$ given $E=0$, and write $\delta:=\PP(E=0)\in(0,1)$ for the error
probability. \cite*{horowitz1995IdentificationRobustnessContaminated} study this
model under two sampling schemes. First, under ``contaminated sampling'' the
occurrence of errors is independent of the error-free outcome,
$E\independent Y_1$. Second, under ``corrupted sampling'' the dependence between $E$ and
$Y_1$ is left unrestricted. By setting the baseline distribution as
$\Pbase{}=P$, our framework covers both settings described above. Indeed,
contaminated sampling is based on the
decomposition neighborhood of Section~\ref{subsec: contamination tv}, whereas
corrupted sampling is based on the pure total variation
neighborhood. 

\uline{\textbf{Contaminated sampling.}} Under independence the distribution of
$Y_1$ among error-free observations coincides with $P_1$, so the sampling
distribution is the mixture
\[
P=\delta \widetilde{P}+(1-\delta)P_1.
\]
Therefore, the candidate distributions are the distributions that can arise 
as the error-free component of the observed mixture, namely those
$Q\in\Delta(\calY)$ for which $P=\delta R+(1-\delta)Q$ for some
$R\in\Delta(\calY)$ or, equivalently by
Lemma~\ref{lemma: mixture density bounds-i} with $\kappa=1-\delta$, those
satisfying the domination constraint $(1-\delta)Q\leq P$. 
The robustness set is thus
\[
\calP_{\mathsf{CON}}(\delta,P) :=
\big\{Q\in\Delta(\calY):\ (1-\delta)Q\leq P\big\}.
\]
Sensitivity bounds can then be
obtained from the dual of
Theorem~\ref{theorem: strong duality with mixture constraint-i}.

\uline{\textbf{Corrupted sampling.}} Suppose now that the independence
assumption is dropped, so that errors may occur selectively. Writing
$P_{1|e}$ for the conditional distribution of $Y_1$ given $E=e$, the observed distribution
and the distribution of interest decompose as
\[
P=\delta \widetilde{P}+(1-\delta)P_{1|1},
\qquad
P_1=\delta P_{1|0}+(1-\delta)P_{1|1},
\]
where $P_{1|1}$ need no longer coincide with $P_1$ and $P_{1|0}$ is
entirely unrestricted. In this case the robustness set coincides with
the total variation neighborhood of
Section~\ref{subsec: contamination tv} centered at the baseline
distribution $P$, namely
\[
\calP_{\mathsf{COR}}(\delta,P) :=
\big\{Q\in\Delta(\calY):\ \TV(Q\,\|\,P)\leq\delta\big\}.
\]
To see that $P_1$ must satisfy
$\TV(P_1\,\|\,P)\leq\delta$, fix a measurable event $A$ and decompose both $P_1(A)$ and
$P(A)$
over $\{E=1\}$ and $\{E=0\}$. The contributions on $\{E=1\}$ cancel,
because conditionally on this event we get $Y=Y_1$, and so
\[
P_1(A)-P(A)=\PP(Y_1\in A,E=0)-\PP(Y_0\in A,E=0),
\]
which is a difference of two numbers lying in $[0,\delta]$, so that
$|P_1(A)-P(A)|\leq\delta$. Taking the supremum over $A$ yields
$\TV(P_1\,\|\,P)\leq\delta$.

Conversely, every distribution in the total variation neighborhood arises as
the distribution of $Y_1$ under some corrupted-sampling process with error
probability $\delta$, so the inclusion just derived above is in fact an equality.
To see it, fix
$Q\in\calP_{\mathsf{COR}}(\delta,P)$ and set $t:=\TV(Q\,\|\,P)\leq\delta$.
In view of
Lemma~\ref{lemma: expectation decomposition in a TV neighborhood-i} and writing
$Q-P=\nu^+-\nu^-$ with $\nu^+\leq Q$, $\nu^-\leq P$, and
$\nu^+(\calY)=\nu^-(\calY)=t$, the two distributions share the common
component $C:=P-\nu^-=Q-\nu^+$, which is a nonnegative measure of mass
$1-t\geq1-\delta$. Let a unit be error free ($E=1$) with probability
$1-\delta$, in which case its true and recorded outcomes coincide and are
drawn from the normalized common component $C/(1-t)$. Instead, with
probability $\delta$ the unit is erroneous ($E=0$), its recorded outcome
$Y_0$ is drawn from
$\bigl(P-\tfrac{1-\delta}{1-t}\,C\bigr)/\delta$ and its true outcome $Y_1$
from $\bigl(Q-\tfrac{1-\delta}{1-t}\,C\bigr)/\delta$. 
Both numerators are nonnegative measures of total mass $\delta$, because
$\tfrac{1-\delta}{1-t}\leq1$ and $C$ is dominated by both $P$ and $Q$.
Hence, after division by $\delta$, the displayed measures are probability laws.
By construction $\PP(E=0)=\delta$, the recorded outcome
has distribution $P$, and the true outcome has distribution $Q$, as desired.
When $t<\delta$, part of the erroneous units simply carry correct values,
reflecting that the corruption budget need not be exhausted.
Sensitivity bounds can thus be obtained via
Theorem~\ref{thm: strong duality}. 

\subsubsection{Missing data}\label{subsec: missing data}
Consider an outcome $Y$ with support $\calY=[\underline{Y},\overline{Y}]$
that is observed only when an indicator $S\in\{0,1\}$ equals one -- as with item
nonresponse in a survey or dropout in a panel. Write $p:=\PP(S=1)$, with
$p\in(0,1)$, and let
$\Pbase{}(\cdot):=\PP(Y\in\cdot\mid S=1)$ and
$R(\cdot):=\PP(Y\in\cdot\mid S=0)$ denote the outcome distributions of the
observed and of the missing units, respectively. The object of interest is
the linear functional of the whole population
\[\mu_{\mathsf{MD}}(P) := \E_P[g(Y)],\qquad
P=p\,\Pbase{}+(1-p)\,R.
\]
While $p$ and $\Pbase{}$ are identified from the data, $R$, the law of
$Y$ among the missing, is not. The main difference from the sensitivity exercise
we conducted in Section~\ref{subsec: rct attrition} is that the baseline distribution
$\Pbase{}$ has to be a mixing component of any candidate distribution $P$, 
rather than being the mixture itself.

We can point identify $\mu_{\mathsf{MD}}(P)$ under an
MCAR-type assumption, which would imply that $R=\Pbase{}$. The sensitivity exercise
would then relax this benchmark by imposing that $\TV(R\,\|\,\Pbase{})\leq\delta$.
The mixture displays $P$ as a contamination of $\Pbase{}$ with
mixing weight $p$ and mixing distribution $R$, in the sense of
Section~\ref{subsec: contamination tv}, so, in view of the rescaling
identity $\TV(P\,\|\,\Pbase{})=(1-p)\TV(R\,\|\,\Pbase{})$ of
Lemma~\ref{lemma: contamination as constrained tv-ii}, the requirement is
equivalent to the pair of constraints $\TV(P\,\|\,\Pbase{})\leq(1-p)\delta$ and
$p\,\Pbase{}\leq P$. Therefore,  the robustness set is
\[
\calP_{\mathsf{MD}}(\delta,\Pbase{}):=\big\{P\in\Delta(\calY):\TV(P\,\|\,\Pbase{})\leq(1
-p)\delta,\ p\,\Pbase{}\leq P\big\},
\]
with $\delta$ bounding the fraction of the missing subpopulation that does 
not satisfy the missing completely at random assumption.
The sensitivity bounds for $\delta\in[0,1]$ are then defined as
\[\underline{\mu}_{\mathsf{MD}}(\delta):
=\inf_{P\in\calP_{\mathsf{MD}}(\delta,\Pbase{})}\E_P[g(Y)],\qquad
\overline{\mu}_{\mathsf{MD}}(\delta):
=\sup_{P\in\calP_{\mathsf{MD}}(\delta,\Pbase{})}\E_P[g(Y)],\]
and their duals are readily obtained from 
Theorem~\ref{theorem: strong duality with mixture constraint-ii}. 
Were the support unbounded, the moment
restrictions $\E_P[m(Y)]\in\calM(\rho)$ would again be what keeps the bounds finite
(cf.~Section~\ref{subsec: structural models} and
Remark~\ref{rem: non vacuous bounds in Gumbel}). The
closest exercise in the literature is that of
\cite*{ober-reynolds2026RobustnessMissingData}, who studies GMM breakdown
points in the presence of missing data under Hellinger neighborhoods. 
The neighborhoods there are defined through densities with respect to the baseline, 
which restricts the candidate distributions to be absolutely continuous with 
respect to it, and the results require
strict convexity of the entropy function, so they do not cover $\TV$. We stress
that our framework covers the Hellinger divergence.

The same setting described above can also be used to perform sensitivity analysis
around a causal estimand that the randomized
experiment with attrition of Section~\ref{subsec: rct attrition} set aside, namely
the ATE for the whole population,
\[\tau_{\mathsf{MB}}(P_0):=\E_{P_0}[Y(1)-Y(0)],\]
where, as in Section~\ref{subsec: rct attrition}, $P_0$ is the distribution of the
population vector $(Y(0),Y(1),S(0),S(1),D)$. 
Formally, we set $g(y)=y$ in the above setting and
focus on the treated arm. The law of $Y(1)$ is the mixture
\[P_{Y(1)} = \pi P_{Y(1)|\mathsf{NA}} + (1-\pi)P_{Y(1)|\mathsf{A}},\]
where $P_{Y(1)|\mathsf{NA}}$ and $P_{Y(1)|\mathsf{A}}$ denote the distribution of
$Y(1)$ for the non-attritors $(S(1)=1)$ and the attritors $(S(1)=0)$,
respectively, and $\pi:=P_0(S(1)=1)\in(0,1)$ is the non-attritor share, identified
as $P_0(S=1\mid D=1)$ by the random assignment of $D$. Under \ref{assump: MCAR},
$P_{Y(1)|\mathsf{NA}}=P_{Y(1)|\mathsf{A}}$. Absent \ref{assump: MCAR},
the two laws need not coincide, and $\delta$ controls their distance.
The baseline of this exercise is the identified outcome
distribution of the non-attritors in the treated arm,
$\PT=P_0(Y\in\cdot\mid D=1,S=1)$, which equals $P_{Y(1)\mid\mathsf{NA}}$ by the
random assignment of $D$.
The robustness set is thus
\[\calP_{\mathsf{MB}}(\delta,\PT):=
\big\{P\in\Delta(\calY):\TV(P\,\|\,\PT)\leq(1-\pi)\delta,\ \pi\PT\leq P\big\}.\]
The upper sensitivity
bound $\overline{\mu}_1^{\mathsf{MB}}(\delta)$ is then the supremum of
$\E_P[Y(1)]$ over $P\in\calP_{\mathsf{MB}}(\delta,\PT)$.

Applying
Lemma~\ref{lemma: expectation decomposition in a TV neighborhood-ii} to the attritors'
distribution, with baseline $\PT$ and budget $\delta\in[0,1]$, we get
\[\overline{\mu}_1^{\mathsf{MB}}(\delta) = \pi\E_{\PT}[Y] +
(1-\pi)\left(\E_{\PT}[Y]+\delta\overline{Y}-\int_0^\delta q(u)\,\de u\right),\]
where $q$ denotes the quantile function of $\PT$. In words, the upper bound is obtained by
trimming from the identified average the low
$\delta$-mass tail of $\PT$ and relocating it to $\overline{Y}$. At
$\delta=0$ the two subpopulation distribution are forced to coincide, as they would under
\ref{assump: MCAR}, and the display reduces to $\E_{\PT}[Y]$, so that $\E_{P_0}[Y(1)]$ is
point identified. At $\delta=1$ the attritors' distribution is unrestricted, and the
display
allocates their entire mass $1-\pi$ to $\overline{Y}$, which delivers the worst case
bounds of \citet{manski2003PartialIdentificationProbability}. In contrast
with the Lee bounds case, the mixture constraint is reversed, so that
$\PT\ll P$ but not conversely, and the elements of
$\calP_{\mathsf{MB}}(\delta,\PT)$ need not be dominated by $\PT$, so that
the relocated mass may be singular with respect to $\PT$.

The same construction applies to $Y(0)$ upon replacing $\pi$ with $P_0(S(0)=1)$
and $\PT$ with $P_0(Y\in\cdot\mid D=0,S=1)$, and the lower bounds
$\underline{\mu}_d^{\mathsf{MB}}(\delta)$ follow from the symmetric closed form
$\underline{\mu}_{Q}(\delta)$ of
Lemma~\ref{lemma: expectation decomposition in a TV neighborhood-ii}, so that
\[\tau_{\mathsf{MB}}(P_0)\in\left[\underline{\mu}_1^{\mathsf{MB}}(\delta)
-\overline{\mu}_0^{\mathsf{MB}}(\delta),\:
\overline{\mu}_1^{\mathsf{MB}}(\delta)-\underline{\mu}_0^{\mathsf{MB}}(\delta)\right].\]
The interval is sharp because no restriction links the law of $Y(1)$ on
$\{S(1)=0\}$ to that of $Y(0)$ on $\{S(0)=0\}$.

\subsubsection{Departures from randomization}\label{subsec: randomization}
Consider an observational study in which treatment effects are identified by
assuming that assignment is as good as random, that is $D\independent (Y(0),Y(1))$,
where $Y(d)\in [\underline{Y},\overline{Y}]=:\calY$ almost surely.
In some situations the assumption may fail for part of the population.
For example, in a job-training evaluation, some individuals may select into the program on
unobserved motivation, while for the remainder enrollment is as good as random. Suppose
that the distribution $P_0$ from which the data are sampled is the mixture
\[
P_0=\delta P_{\mathsf{C}}+(1-\delta)P_{\mathsf{U}},
\]
where $P_{\mathsf{U}}$ is a distribution in which assignment is random, 
$P_{\mathsf{C}}$ describes the ``confounded'' population in which
$D\not\independent Y(d)$, and $\delta$ is the proportion of confounded units. 
The causal estimand of interest is
\[
\tau_{\mathsf{U}}(P_{\mathsf{U}}):=\E_{P_{\mathsf{U}}}[Y(1)-Y(0)],
\]
the average treatment effect on the unconfounded subpopulation.

Let $\varpi_d:=P_0(D=d)$, $d\in\{0,1\}$, and $\pi:=P_{\mathsf{U}}(D=1)$. Among treated
units, a
fraction $\alpha_1:=(1-\delta)\pi/\varpi_1$ comes from the unconfounded component,
and similarly a fraction $\alpha_0:=(1-\delta)(1-\pi)/\varpi_0$ among untreated
units, so the observed outcome distributions
$\Pbasesub{d}(\cdot):=P_0(Y\in\cdot\mid D=d)$ are the mixtures
\[
\Pbasesub{d}=\alpha_d P_{\mathsf{U},d}+(1-\alpha_d)P_{\mathsf{C},d},
\qquad d\in\{0,1\},
\]
with $P_{\mathsf{U},d}$ and $P_{\mathsf{C},d}$ the conditional laws of $Y$
given $D=d$ under $P_{\mathsf{U}}$ and $P_{\mathsf{C}}$, respectively. The
baseline distribution of the exercise is thus $\Pbasesub{d}$,
which is identified from the data and plays the role of the mixture as in
the RCT with attrition example of Section \ref{subsec: rct attrition}.

The candidate distributions are the
possible unconfounded components $P_{\mathsf{U},d}$, which satisfy the
domination constraint $\alpha_d P_{\mathsf{U},d}\leq \Pbasesub{d}$ by
Lemma~\ref{lemma: mixture density bounds-i}. The domination constraint
already implies $\TV(P_{\mathsf{U},d}\,\|\,\Pbasesub{d})\leq 1-\alpha_d$, in view
of the rescaling identity of
Lemma~\ref{lemma: contamination as constrained tv-i}, so the robustness set
of this exercise is, for $d\in\{0,1\}$, the set
$\calP_{\mathsf{DR}}^{d}(1-\alpha_d,\Pbasesub{d})$, where, for
$\delta\in[0,1)$,
\[
\calP_{\mathsf{DR}}^{d}(\delta,\Pbasesub{d}):=\big\{P\in\Delta(\calY):\
\Pbasesub{d}=(1-\delta)\,P+\delta\,R\ \text{for some }R\in\Delta(\calY)\big\}.
\]
We can obtain sensitivity bounds by means of the dual in
Theorem~\ref{theorem: strong duality with mixture constraint-i}, applied
arm by arm. One caveat is in order: the data identify
$\varpi_d$ but not $\pi$, so the weights $\alpha_d$ are not identified. We
sidestep this issue by imposing equal treatment rates across the two
components, $P_{\mathsf{C}}(D=1)=\pi$, which yields $\pi=\varpi_1$ and
$\alpha_d=1-\delta$ in both arms, so that the robustness set of
each arm becomes $\calP_{\mathsf{DR}}^{d}(\delta,\Pbasesub{d})$. Alternatively, one could
profile the bounds over the values of $\pi$ for which $\alpha_0,\alpha_1\in[0,1]$.

Since assignment is random under $P_{\mathsf{U}}$, we have
$\E_{P_{\mathsf{U}}}[Y(d)]=\E_{P_{\mathsf{U},d}}[Y]$ for $d\in\{0,1\}$, so
it suffices to bound each arm mean as $P_{\mathsf{U},d}$ ranges over the
robustness set $\calP_{\mathsf{DR}}^{d}(\delta,\Pbasesub{d})$ defined
above. Fix $\delta\in[0,1)$ and let $q_d$
denote the quantile function of $\Pbasesub{d}$. In view of
Lemma~\ref{lemma: expectation decomposition in a TV neighborhood-iii},
applied with $\kappa=1-\delta$ and baseline $\Pbasesub{d}$,
the sensitivity bounds for the mean of each arm $d\in\{0,1\}$ are
\[
\underline{\mu}^{\mathsf{U}}_d(\delta)
:=\inf_{P\in{\calP_{\mathsf{DR}}^{d}(\delta,\Pbasesub{d})}}\E_P[Y]
=\frac{1}{1-\delta}\int_0^{1-\delta}q_d(u)\,\de u,
\qquad
\overline{\mu}^{\mathsf{U}}_d(\delta)
:=\sup_{P\in{\calP_{\mathsf{DR}}^{d}(\delta,\Pbasesub{d})}}\E_P[Y]
=\frac{1}{1-\delta}\int_\delta^1 q_d(u)\,\de u,
\]
the averages of $Y$ over the lower and upper $(1-\delta)$-fraction of each
arm. Combining the two arms then yields
\[
\underline{\mu}^{\mathsf{U}}_1(\delta)-\overline{\mu}^{\mathsf{U}}_0(\delta)
\;\leq\;\tau_{\mathsf{U}}(P_\mathsf{U})\;\leq\;
\overline{\mu}^{\mathsf{U}}_1(\delta)-\underline{\mu}^{\mathsf{U}}_0(\delta).
\]
In this application, $\delta$ thus traces out the interpolation between
point identification and worst case selection and bounds
the fraction of the population that does not satisfy unconfoundedness. 
At $\delta=0$ the trimmed means reduce to the arm means and both bounds collapse to the
difference in means of the two arms, whereas as $\delta\to1$ each arm is trimmed down
to its most extreme $(1-\delta)$-fraction, so that
$\overline{\mu}^{\mathsf{U}}_d(\delta)$ and
$\underline{\mu}^{\mathsf{U}}_d(\delta)$ converge to the essential supremum
and the essential infimum of the observed distribution $\Pbasesub{d}$.

The exercise accommodates the more general case of 
unconfoundedness or selection on observables, which relaxes randomization as
\[D\independent (Y(0),Y(1))\mid X.\]
It suffices to posit the mixture decomposition, with the same
confounded share $\delta$, within each covariate cell and
repeat the argument conditionally on $X=x$, which bounds
$\E_{P_{\mathsf{U}}}[Y(d)\mid X=x]$. The robustness set of this
extension is, within the covariate cell $X=x$, the set
$\calP_{\mathsf{DR}}^{d}(\delta,\Pbasesub{d}(x))$, where
$\Pbasesub{d}(x):=P_0(Y\in\cdot\mid D=d,X=x)$ denotes the observed
outcome distribution of arm $d$ within the cell. Since a constant share
implies that $X$ has the same distribution under $P_{\mathsf{U}}$ and
$P_0$, averaging these bounds over the observed distribution
of $X$ then bounds the average treatment effect for the unconfounded
subpopulation, in the spirit of
\cite*{bonvini2022SensitivityAnalysisProportion}. We stress that our target
parameter differs from theirs, as they bound the average treatment effect for the
entire population, which corresponds to $P_0$ in our notation, whereas we
target the unconfounded subpopulation $P_{\mathsf{U}}$. 

\subsubsection{Discrete choice models}\label{subsec: discrete choice}
The linearity of the estimand $\E_P[g(Z;\theta)]$ in $P$ is what makes the
dualization behind Theorem~\ref{theorem: weak duality} possible. The
estimands of interest in discrete choice models do not possess this structure. In this
section, we explain why discrete choice models cannot, in general, be
accommodated within our framework and propose three alternatives. For simplicity, we
illustrate the argument on the canonical binary logit, but the same reasoning extends to
multinomial logit, probit, and ordered choice.

Let $X\in\R^k$ be a vector of regressors and
$Y\in\{0,1\}$ a binary outcome, whose conditional choice probability we assume to be
$\PP(Y=1\mid X=x)=\Lambda(x^\top\beta)$, where 
$\Lambda(t):=(1+e^{-t})^{-1}$ is the logistic link function. 
The population analog of the MLE is the solution $\beta(P)$ of the nonlinear system 
of moment equations
\begin{align}
    \E_P\![(Y-\Lambda(X^\top\beta(P)))X] \;=\; 0.
    \label{eq: logit pop FOC}
\end{align}
The functional $\beta(P)$ is not of the form $\E_P[f(Z)]$ for any known $f$. In this
particular case the link is nonlinear in $\beta$, so \eqref{eq: logit pop FOC} 
cannot be solved in closed form. The coefficient $\beta(P)$ is, however,
seldom the target of applied work. Researchers are typically interested, for example, in
the average marginal effect
\[\mathsf{AME}_j(P):=\E_{P_{X}}[\lambda(X^\top\beta(P))\beta_j(P)],\qquad \beta_j(P)
=e_j^\top\beta(P),\]
where $\lambda(t):=\Lambda'(t)=\Lambda(t)(1-\Lambda(t))$. Unlike in a linear model,
where fixing the regressor marginal renders the estimand an expectation under
the conditional law of the error, conditioning on $X$ does not restore
linearity here. Although the framework does not apply to $\beta(P)$ directly, 
we propose three ways in which some version of the sensitivity
analysis exercise can still be performed in a discrete choice model.

\uline{\textbf{Transportability of $P_{X}$.}} Fix $\beta$ at the
population value $\beta(P_0)$ solving \eqref{eq: logit pop FOC} and the link to be the
logit one, and let the covariate distribution vary. The AME reduces to the linear
functional
\[\mathsf{AME}_j(P_{X})=\beta_j(P_0)\,\E_{P_{X}}\big[\lambda(X^\top\beta(P_0))\big],\]
which is bounded because $\lambda\in[0,1/4]$. In this case,
sensitivity to contamination or corruption in $P_{X}$ can be readily
assessed following Section~\ref{subsec: contaminated data}.
Under corrupted sampling the robustness set is
$\calP_{\mathsf{COR}}(\delta,P_{X})$, whereas under contaminated sampling it
is $\calP_{\mathsf{CON}}(\delta,P_{X})$, the two sets of
Section~\ref{subsec: contaminated data} evaluated at baseline $P_{X}$.

\uline{\textbf{Relaxation of the logit link.}} This route
holds the covariate marginal $P_{X}$ fixed and perturbs the conditional
distribution of the outcome, retaining the logit specification
as the center of the robustness set. Because $Y$ is binary, for any measurable
conditional choice probability $P:\R^k\to[0,1]$ let
$P_X\otimes\Be(P)$ denote the joint law under which $X\sim P_X$ and
$Y$ given $X=x$ is distributed as the Bernoulli law $\Be(P(x))$.
Writing $\Pbase{}(x):=\Lambda(x^\top\beta_0)$ for the logit
specification, with $\beta_0:=\beta(P_0)$ solving \eqref{eq: logit pop FOC}
at the law $P_0$ from which the data are sampled, so that
$\Be(\Pbase{}(x))$ is the implied baseline
conditional, the robustness set for each covariate cell $x$ reads
\begin{align*}
    \calP_{\mathsf{DC}}(\delta,\Be(\Pbase{}(x)))
    :=\big\{Q\in\Delta(\{0,1\})\,:\,\TV(Q\,\|\,\Be(\Pbase{}(x)))\leq\delta\big\}
    \qquad\text{for $P_{X}$-almost every }x,
\end{align*}
which contains all conditional distributions that are at most $\delta$ away from 
the logit specification in total variation.

The within-cell sets do not restrict the derivatives of $P$, so
derivative-based functionals such as the $\mathsf{AME}_j$ admit no
informative bounds and the natural targets are averaged discrete
contrasts. For a binary regressor of interest $X_j\in\{0,1\}$, write
$X=(X_j,X_{-j})$ and $x=(x_j,x_{-j})$, where $X_{-j}$
collects the remaining covariates, with marginal $P_{X_{-j}}$
fixed by $P_{X}$, and, with a slight abuse of notation, let
$P(t,x_{-j})$ denote the value of $P$ when the regressor of
interest equals $t\in\{0,1\}$. The target estimand is the average
effect of switching $X_j$ from $0$ to $1$,
\begin{align*}
    \tau_{\mathsf{DC}}(P):=\E_{P_{X_{-j}}}\big[P(1,X_{-j})-P(0,X_{-j})\big],
\end{align*}
which is linear in $P$. Sensitivity bounds can be obtained cell by
cell. Since $\TV(Q\,\|\,\Be(p))=|Q(\{1\})-p|$ for every
$Q\in\Delta(\{0,1\})$ and $p\in[0,1]$, the choice probabilities $Q(\{1\})$ of the
distributions $Q\in\calP_{\mathsf{DC}}(\delta,\Be(\Pbase{}(x)))$ form exactly the
interval
$[\underline{P}(x),\overline{P}(x)]$, where
$\underline{P}(x):=\max\{\Pbase{}(x)-\delta,0\}$ and
$\overline{P}(x):=\min\{\Pbase{}(x)+\delta,1\}$. 
Aggregating over the fixed marginal $P_{X_{-j}}$ yields
\begin{align*}
    \underline{\tau}_{\mathsf{DC}}(\delta)
    &:=\E_{P_{X_{-j}}}\big[\underline{P}(1,X_{-j})\big]
    -\E_{P_{X_{-j}}}\big[\overline{P}(0,X_{-j})\big], \\
    \overline{\tau}_{\mathsf{DC}}(\delta)
    &:=\E_{P_{X_{-j}}}\big[\overline{P}(1,X_{-j})\big]
    -\E_{P_{X_{-j}}}\big[\underline{P}(0,X_{-j})\big].
\end{align*}
At $\delta=0$ the interval collapses to
the logit implied contrast $\tau_{\mathsf{DC}}(\Pbase{})$, whereas at $\delta=1$
it widens to the trivial bounds $[-1,1]$.

An alternative robustness set is worth recording. It replaces the
within-cell restriction with its integrated version,
$\E_{P_{X}}[\,|P(X)-\Pbase{}(X)|\,]\leq\delta$. Since two joint distributions sharing a
marginal have total variation equal to the average of their conditional total variations
\citep[see,
e.g.,][Proposition~7.2]{polyanskiy2025InformationTheoryCoding}, this
restriction is equivalent to
\[
\calP_{\mathsf{DC}}^{\mathsf{int}}(\delta,\Pbase{}):
=\big\{P_X\otimes\Be(P')\in\Delta(\R^k\times\{0,1\}):\
P':\R^k\to[0,1]\ \text{measurable},\
\TV(P_X\otimes\Be(P')\,\|\,P_X\otimes\Be(\Pbase{}))\leq\delta\big\},
\]
so that the budget $\delta$ is allocated across covariate cells
rather than imposed cell by cell. Provided the propensity score
$\pi(x_{-j}):=\PP(X_j=1\mid X_{-j}=x_{-j})$ lies in $(0,1)$ for
$P_{X_{-j}}$-almost every $x_{-j}$, the law of iterated
expectations gives
$\tau_{\mathsf{DC}}(P)=\E_{P_X\otimes\Be(P)}[\omega(X)\,Y]$ for the
inverse-probability weight
$\omega(x):=x_j/\pi(x_{-j})-(1-x_j)/(1-\pi(x_{-j}))$, so that,
when $X$ takes finitely many values, bounds are delivered by
the dual of Theorem~\ref{thm: strong duality} applied with baseline
$P_X\otimes\Be(\Pbase{})$ and with the fixed covariate marginal
imposed through one moment equality per covariate cell. 

\uline{\textbf{Unobserved heterogeneity.}} Keep the logit specification at
the individual level and augment the latent utility index with a unit-level
intercept 
\[Y=\indic\{X^\top\beta+A+\varepsilon>0\}\] with
$\varepsilon\sim\mathsf{Logistic}(0,1)$ independent of $(X,A)$ and $A\sim Q\in\Delta(\R)$
independent of $X$. The average marginal effect is thus
\[\mathsf{AME}_j(Q)=\beta_j\int\psi(a)\,\de Q(a),\qquad \psi(a):
=\E_{P_{X}}[\lambda(X^\top\beta+a)],\]
which is bounded since $\lambda\leq1/4$ and linear in $Q$.
We center the robustness set at a Gaussian benchmark,
$Q_{\star}:=\normal(\mu_0,\sigma_0^2)$ for given $\mu_0\in\R$ and
$\sigma_0>0$, a classic random effects logit specification, and take
as robustness set the total variation ball
\begin{align*}
    \calP_{\mathsf{DC}}^{\mathsf{het}}(\delta,Q_{\star}):
    =\big\{Q\in\Delta(\R):\TV(Q\,\|\,Q_{\star})\leq\delta\big\},
\end{align*}
the set of heterogeneity distributions deviating from normality by
at most $\delta$ in total variation.
Let $q_{\psi}$ denote the quantile function of $\psi(A)$ when $A\sim Q_{\star}$, and set
\[
\underline\psi:=\inf_{a\in\R}\psi(a),
\qquad
\overline\psi:=\sup_{a\in\R}\psi(a).
\]
The lower and upper values of the average derivative are
\begin{align*}
\underline\mu_{\psi}(\delta)
&:=\inf_{Q\in\calP_{\mathsf{DC}}^{\mathsf{het}}(\delta,Q_{\star})}
\int\psi(a)\,\de Q(a)
=\delta\,\underline\psi+\int_{0}^{1-\delta}q_{\psi}(u)\,\de u,\\
\overline\mu_{\psi}(\delta)
&:=\sup_{Q\in\calP_{\mathsf{DC}}^{\mathsf{het}}(\delta,Q_{\star})}
\int\psi(a)\,\de Q(a)
=\delta\,\overline\psi+\int_{\delta}^{1}q_{\psi}(u)\,\de u.
\end{align*}
These closed forms follow from
Lemma~\ref{lemma: expectation decomposition in a TV neighborhood-ii} applied
to the law of $\psi(A)$ under $Q_{\star}$. Therefore, the sensitivity bounds
on the average marginal effect are
\begin{align*}
\underline{\mathsf{AME}}_j(\delta)
&:=\inf_{Q\in\calP_{\mathsf{DC}}^{\mathsf{het}}(\delta,Q_{\star})}
\mathsf{AME}_j(Q)
=\min\big\{\beta_j\underline\mu_{\psi}(\delta),
\beta_j\overline\mu_{\psi}(\delta)\big\},\\
\overline{\mathsf{AME}}_j(\delta)
&:=\sup_{Q\in\calP_{\mathsf{DC}}^{\mathsf{het}}(\delta,Q_{\star})}
\mathsf{AME}_j(Q)
=\max\big\{\beta_j\underline\mu_{\psi}(\delta),
\beta_j\overline\mu_{\psi}(\delta)\big\}.
\end{align*}
The upper value $\overline\mu_{\psi}(\delta)$ is attained whenever there
exists $a_{+}\in\R$ with $\psi(a_{+})=\overline\psi$, and the lower value
$\underline\mu_{\psi}(\delta)$ is attained whenever there exists
$a_{-}\in\R$ with $\psi(a_{-})=\underline\psi$.
By appropriately changing the baseline distribution, we can also study
sensitivity to departures from a no heterogeneity benchmark. In particular,
we can choose the baseline distribution to be $\delta^{\mathtt{Dirac}}_{a_0}$,
the Dirac measure at some $a_0\in\R$, so that the robustness set is
\[\calP_{\mathsf{DC}}^{\mathsf{het}}(\delta,\delta^{\mathtt{Dirac}}_{a_0})
=\big\{Q\in\Delta(\R):\TV(Q\,\|\,\delta^{\mathtt{Dirac}}_{a_0})\leq\delta\big\}.\]
Finally, we can also allow $A$ to depend on $X$ by constructing robustness
sets that vary the conditional distribution of $A$ given $X$ while
keeping the identified marginal of $X$ fixed.

\subsubsection{Just-identified method of moments}
\label{subsec: just identified mom}

Let $Z\sim P_0\in\Delta(\calZ)$ denote
the law from which the data are drawn, let $\Theta\subseteq\R^{\dimt}$, and let 
$m:\calZ\times\Theta\to\R^{\dimm}$ be a moment
function with as many elements as parameters $(\dimm=\dimt)$ 
that defines the moment condition
\begin{align}
  \E_{P_0}\bigl[m(Z;\theta)\bigr]=0.
  \label{eq: mm moment}
\end{align}
We assume that the moment condition in
\eqref{eq: mm moment} has a unique solution $\theta\in\Theta$. This assumption is
standard in much of the literature
on generalized method of moments estimators
\citep[see, e.g.,][Section~2.2.3]{newey1994Chapter36Large}.
The formulation covers just-identified moment estimators and the
first-order
conditions of $M$-estimators. 
We consider the case in which $\dimm>\dimt$ at the end of the section.

Suppose that an econometrician has as target estimand $\theta_j=e_j^\top\theta$
and is concerned that the sampling distribution $P_0$ differs from the
population of interest $P_1$ because a share $\delta$ of observations is
erroneous. Under contaminated sampling,
\[P_0=\delta\widetilde P+(1-\delta)P_1,\]
where $\widetilde P$ is the distribution of the erroneous observations,
whereas under corrupted sampling $P_1$ need not coincide with the error-free
mixture component. As established in Section~\ref{subsec: contaminated data},
the two sampling schemes instead imply, respectively, the domination
restriction $(1-\delta)P_1\leq P_0$ and the total variation restriction
$\TV(P_1\,\|\,P_0)\leq\delta$. In both settings, the baseline distribution is
$\Pbase{}=P_0$, whereas the distribution of interest is $P_1$. Accordingly,
the robustness set can be defined as either
\[
\calP_{\mathsf{MM}}^{\mathsf{con}}(\theta;\delta,\Pbase{}):=\big\{P\in\Delta(\calZ):\
(1-\delta)P\leq \Pbase{},\ \E_P[m(Z;\theta)]=0\big\}
\]
in the case of contamination, or, in the case of corruption, as
\[
\calP_{\mathsf{MM}}^{\mathsf{cor}}(\theta;\delta,\Pbase{}):=\big\{P\in\Delta(\calZ):\
\TV(P\,\|\,\Pbase{})\leq\delta,\ \E_P[m(Z;\theta)]=0\big\}.
\]
Sensitivity bounds can then be obtained by optimizing over the
parameter values that are consistent with at least one distribution in the
robustness set, that is,
\begin{align*}
  \overline{\theta_j}(\delta)\;=\;\sup_{\theta\in\Theta_{\mathsf{MM}}(\delta)}
  e_j^\top\theta,\qquad
  \underline{\theta_j}(\delta)\;=\;\inf_{\theta\in\Theta_{\mathsf{MM}}(\delta)}
  e_j^\top\theta,\qquad
  \Theta_{\mathsf{MM}}(\delta):=\big\{\theta\in\Theta:\
  \calP_{\mathsf{MM}}^{\bullet}(\theta;\delta,\Pbase{})\neq\emptyset\big\},
\end{align*}
where $\bullet\in\{\mathsf{con},\mathsf{cor}\}$.

If instead the model is over-identified, i.e., $\dimm>\dimt$, there might not
exist a $\theta\in\Theta$ for which the sample analog
$\E_{\widehat P_n}[m(Z;\theta)]=0$ of \eqref{eq: mm moment} holds, even
when \eqref{eq: mm moment} itself has a solution. In this case, we refer the
reader to the two solutions proposed
to circumvent this issue in Section~3.3 of
\cite{christensen2023CounterfactualSensitivityRobustness}.

\subsubsection{Nonparametric sharp bounds}
\label{subsubsec: nonparametric sharp bounds}
A recurring feature of the applications of this section is the behavior of
the bounds at the endpoint $\delta=1$. For this value of the sensitivity parameter,
the total variation constraint
becomes vacuous, and the sensitivity bounds coincide with the sharp
nonparametric bounds implied by whatever support and moment
restrictions are maintained. So far this coincidence was verified case by
case. The next result, together with the discussion that follows it, shows
that this is a general feature of our framework. In particular, we show that it 
holds for every entropy function that is finite at the endpoints
of its effective domain, including the induced entropies of
Proposition~\ref{prop: induced entropy}, for which the sharp bounds are those
implied by the maintained mixture restriction.
Throughout, $\calP_{\phi}(\theta;\rho,\Pbase{},\delta)$ denotes the
robustness set of \eqref{eq: robustness set definition} with entropy $\phi$, and
$\calE(\theta;\rho):=\{P\in\calP(\calZ):\E_P[m(Z;\theta)]\in\calM(\rho)\}$ denotes the
collection of distributions compatible with the moment constraints at $\theta$, as in
Section~\ref{subsec: contamination tv}.

\begin{thm}[Exact nonparametric sharp bounds]
\label{thm: endpoint sharp bounds}
Let $\Pbase{}\in\Delta(\calZ)$, and let $\phi$ be an entropy
function in the sense of Definition~\ref{defi: entropy function} whose
effective domain has endpoints
$\underline{x}:=\inf\dom(\phi)$ and $\bar{x}:=\sup\dom(\phi)$, with
$\underline{x}<\bar{x}$. Assume that $\phi(\underline{x})<\infty$, that
$\phi(\bar{x})<\infty$ if $\bar{x}<\infty$, and that $\phi^\infty(1)<\infty$
if $\bar{x}=\infty$. Define the domination set and the threshold
\[
\calD_\phi(\Pbase{}):=\big\{P\in\Delta(\calZ):\
\underline{x}\,\Pbase{}\leq P\leq\bar{x}\,\Pbase{}\big\},
\qquad
\overline{\delta}_\phi:=
\frac{(\bar{x}-1)\,\phi(\underline{x})+(1-\underline{x})\,\phi(\bar{x})}{\bar{x}-\underline{x}},
\]
if $\bar{x}<\infty$, and
\[
\calD_\phi(\Pbase{}):=\big\{P\in\Delta(\calZ):\
\underline{x}\,\Pbase{}\leq P\big\},
\qquad
\overline{\delta}_\phi:=\phi(\underline{x})+(1-\underline{x})\,\phi^\infty(1),
\]
if $\bar{x}=\infty$. Then, for every $\delta\geq\overline{\delta}_\phi$
and every $\theta\in\Theta$,
\[\calP_{\phi}(\theta;\rho,\Pbase{},\delta)
=\calD_\phi(\Pbase{})\cap\calE(\theta;\rho),\]
and consequently the sensitivity bounds of \eqref{eq: general optimization problem}
satisfy
\begin{align*}
\inf_{\theta\in\Theta}\
\inf_{P\in\calP_{\phi}(\theta;\rho,\Pbase{},\delta)}\E_P[g(Z;\theta)]
&=\inf_{\theta\in\Theta}\
\inf_{P\in\calD_\phi(\Pbase{})\cap\calE(\theta;\rho)}\E_P[g(Z;\theta)],\\
\sup_{\theta\in\Theta}\
\sup_{P\in\calP_{\phi}(\theta;\rho,\Pbase{},\delta)}\E_P[g(Z;\theta)]
&=\sup_{\theta\in\Theta}\
\sup_{P\in\calD_\phi(\Pbase{})\cap\calE(\theta;\rho)}\E_P[g(Z;\theta)].
\end{align*}
\end{thm}
\begin{flushright}
[\hyperref[proof:thm: endpoint sharp bounds]{Proof}]
\end{flushright}

Three cases deserve to be singled out. First, when
$\dom(\phi)=[0,\infty)$ the domination set is the whole of $\Delta(\calZ)$,
the threshold is $\overline{\delta}_\phi=\phi(0)+\phi^\infty(1)$, and the
theorem delivers exactly the nonparametric sharp bounds implied by the moment
constraints. In terms of Table~\ref{tab: phi conjugate recession}, this
covers the total variation entropy, for which
$\phi_\TV(0)=\phi_\TV^\infty(1)=1/2$ and thus
$\overline{\delta}_{\phi_\TV}=1$, as well as the squared Hellinger entropy
($\overline{\delta}_\phi=2$) and the Jensen--Shannon entropy
($\overline{\delta}_\phi=2\log2$). It excludes the Kullback--Leibler and
Pearson $\chi^2$ entropies, for which $\phi^\infty(1)=+\infty$, and the
likelihood and Neyman $\chi^2$ entropies, for which $\phi(0)=+\infty$.
Second, for the upper-truncated entropy
$\overline{\phi}=\phi+\iota_{[0,1/\kappa]}$ of
Proposition~\ref{prop: induced entropy-i} we have $\underline{x}=0$ and
$\bar{x}=1/\kappa$, so that the domination set is the baseline-as-mixture
restriction $\{P\in\Delta(\calZ):\kappa P\leq\Pbase{}\}$ and
$\overline{\delta}_{\overline{\phi}}=(1-\kappa)\phi(0)+\kappa\phi(1/\kappa)$,
provided $\phi(0)$ and $\phi(1/\kappa)$ are finite. Third, for the
lower-truncated entropy $\underline{\phi}=\phi+\iota_{[\kappa,\infty)}$ of
Proposition~\ref{prop: induced entropy-ii} we have $\underline{x}=\kappa$ and
$\bar{x}=\infty$, so that the domination set is the baseline-as-mixing-component
restriction $\{P\in\Delta(\calZ):\kappa\Pbase{}\leq P\}$ and
$\overline{\delta}_{\underline{\phi}}=\phi(\kappa)+(1-\kappa)\phi^\infty(1)$,
provided $\phi^\infty(1)$ is finite. In the last two cases the theorem
delivers the sharp bounds implied by the mixture restriction together with
the moment constraints.

A similar result was derived in Theorem 2.1 in
\cite{christensen2023CounterfactualSensitivityRobustness},
which shows that divergence-based
sensitivity bounds \textit{approximate} the nonparametric sharp bounds as the
sensitivity parameter diverges. However, three differences are worth
highlighting. First, the assumptions of \textsf{CC23}'s Theorem 2.1 do
not allow for $\phi= \phi_{\TV}$. Second, the entropy functions \textsf{CC23} consider are
superlinear, that is, $\phi^\infty(1)=+\infty$, so the associated divergences have
unbounded range and no finite upper bound for the budget exists. Therefore, the sharp
bounds are recovered only in the limit as $\delta\to+\infty$. On the contrary, our result
shows that the framework presented here allows us to construct the \textit{exact}
nonparametric sharp bounds. Third, \textsf{CC23} recover the nonparametric sharp bounds
when only distributions that are absolutely continuous with respect to the baseline are
considered. This is worrisome, for example, when the latter has a positive Lebesgue
density, because distributions with mass points are automatically discarded. Our 
nonparametric sharp bounds are instead constructed by considering all these 
distributions as well. We recall that our Theorem~\ref{thm: ignoring dual constraint}
gives sufficient conditions for when ignoring such distributions is allowed.

Finally, in some applications the nonparametric sharp bounds have known
closed forms, such as the Lee bounds, to which the trimming bounds of
Theorem~\ref{theorem: tv dual recovers lee bound} reduce at $\delta=1$, and the Manski
bounds of Sections~\ref{subsec: rct external validity} and~\ref{subsec: missing data},
whereas in others, such as the marital college premium of
Section~6.2 of the main text, they do not.
Theorem~\ref{thm: endpoint sharp bounds} delivers them in either case.

\clearpage
\section{Additional Empirical Applications}
\label{sec: empirical applications details}

This section collects two empirical exercises that take the attrition framework
of Section~\ref{subsec: rct attrition} to the data and complement the one in
Section~6.1 of the main text, which revisits the
online
tutoring program of
\citet{carlana2025ApartConnectedOnline}, a randomized experiment in which
attrition is moderate in level but strongly differential across arms.
Differential attrition is ultimately what threatens a design that aims at identifying the
ATE for the always observed. Heuristically, 
when the two arms respond at nearly the same rate, the complier share
among treated respondents, $\pi$, of Section~\ref{subsec: rct attrition}, is close to
zero,
there are few compliers ``contaminating'' the treatment group, and their impact
on the causal estimand is limited. When treatment assignment itself induces
response, $\pi$ is large, and compliers can substantially move the estimand of
interest. Section~\ref{subsec: empirical cno} revisits the large experiment of
\citet{christensen2026DemandMobility}, which randomized price discounts
for transport services and sits at the opposite end of the spectrum relative to the
tutoring program: attrition is severe in level yet similar across arms,
so that the sensitivity paths widen slowly.
Section~\ref{subsec: empirical clinical} runs the same sensitivity exercise on
every eligible trial in a public collection of clinical randomized experiments
and asks how often attrition threatens the sign of the estimated effect.

\subsection{RCT with Attrition I:
\texorpdfstring{\citet*{christensen2026DemandMobility}}{Christensen, Nino, and Osman
(2026)}}
\label{subsec: empirical cno}

\textbf{Setup.} The experiment recruited $1{,}373$ existing Uber
riders in Cairo, Egypt, and randomized them into a $25\%$ price reduction 
($457$ participants), a $50\%$ price reduction ($461$), or standard prices 
($455$), applied directly in the app for twelve weeks. The outcome we consider is the
headline measure of total mobility of
\citet{christensen2026DemandMobility}: kilometers traveled per week, recorded by
the Google Maps Timeline application and collected in four follow up phone
surveys. For this particular outcome variable, the authors report
intention-to-treat effects on the inverse hyperbolic sine of weekly
kilometers of $0.10$ ($0.10$) for the $25\%$ reduction
and $0.40$ $(0.09)$ for the $50\%$
reduction, against a control mean of $205$ kilometers per week, in
column~(1) of
their Table~2. Attrition operates at the level of the individual survey
round, as we report in Table~\ref{tab: attrition rates cno}: response rates in
the control group fall from $82\%$ and $78\%$ in the
first two rounds to $40\%$ and $38\%$ in the last two. Treatment
assignment
plausibly induces differential attrition, as subsidized riders remain engaged with
the app, and the authors themselves report the bounds of
\citet{lee2009TrainingWagesSample} in their Table~B11.

\begin{table}[htbp]
\centering
\caption{Attrition by survey round in the Uber price-reduction experiment.}
\label{tab: attrition rates cno}
\begin{tabular}{lccc}
\toprule\toprule
 & Control & 25\% reduction & 50\% reduction \\
 & ($n=455$) & ($n=457$) & ($n=461$) \\
\midrule
Round 1 & 18.0\% & 18.8\% & 15.8\% \\
Round 2 & 22.2\% & 17.5\% & 14.1\% \\
Round 3 & 59.8\% & 56.0\% & 53.8\% \\
Round 4 & 62.2\% & 60.2\% & 54.0\% \\
\addlinespace
All rounds & 40.5\% & 38.1\% & 34.4\% \\
\bottomrule\bottomrule
\end{tabular}

\par
\begin{center}
\parbox[t]{\textwidth}{\footnotesize
\textit{Notes:} Computed from the replication data of
\citet{christensen2026DemandMobility}. Columns are the three randomized arms,
with $n$ the number of participants assigned to each. Cells are the share of
that arm's participants whose weekly-kilometers outcome is unobserved.}
\end{center}
\end{table}

We take the independent sampling unit to be the participant, of whom there are
$1{,}373$, and treat the four survey rounds as repeated measurements within
each participant. The sensitivity bounds pool the resulting participant--round
observations, while inference resamples participants and thereby preserves
dependence across rounds within each participant.
We let $Y$ be the inverse hyperbolic sine of weekly kilometers, $S$ the
indicator that the participant responded in the round in question,
and $D$ the treatment indicator of the contrast analyzed,
namely each price reduction ($25\%$ or $50\%$) against control (no reduction).
The baseline $\PT$ is estimated by the empirical distribution of the observed treated
units, while $\mu^{\mathsf{AO}}(0)$ is identified by
\eqref{eq: mu0 AO identified} and estimated by the mean among the observed
controls. The estimated complier
shares are $\widehat{\pi}=0.039$ for the $25\%$ arm and $0.093$ for the $50\%$
arm. Both estimates are an order of magnitude smaller than the complier shares
of the tutoring program, which are $0.47$ and $0.22$ in its two waves. 

\textbf{Sensitivity bounds.} Figure~\ref{fig: cno attrition empirical} traces
$(\widehat{\underline{\tau}}(\delta), \widehat{\overline{\tau}}(\delta))$ and 
$(\widehat{\underline{\tau}}_{X}(\delta), \widehat{\overline{\tau}}_{X}(\delta))$ over a
grid of $51$ values of $\delta$, separately for the two
comparisons. We stress two features of the analysis without covariates. 
First, the sensitivity
paths originate at the sample analog of $\tau_{\mathsf{MCAR}}(P_0)$, which is $0.103$
for the $25\%$ arm and $0.407$ for the $50\%$ arm, and end at
$\delta=1$ on the Lee bounds,
namely $[-0.018,\,0.307]$ and $[0.168,\,0.794]$,
respectively. Second, the small complier shares make the paths fan out
slowly, as the breakdown budgets reflect. Indeed,
$\widehat{\underline{\tau}}(\delta)$ stays above zero at every budget the
exercise considers in the $50\%$ arm, so that
$\widehat{\delta}_{\mathsf{b}}$ attains its largest possible value,
$\widehat{\delta}_{\mathsf{b}}=1$. 
In the $25\%$ arm the estimated
lower bound reaches zero at $\widehat{\delta}_{\mathsf{b}}=0.62$. When
sampling uncertainty is taken into account, the $95\%$ confidence
interval of $[\underline{\tau}(\delta),\overline{\tau}(\delta)]$
contains zero from $\delta=0.54$ onwards in the $50\%$ arm
and already at $\delta=0$ in the $25\%$ arm, in line with
the original point estimates in Table~2 of
\citet{christensen2026DemandMobility}.

We now comment on the sensitivity exercise which includes covariates. Formally,
we impose the aggregate sensitivity constraint across covariate strata as in
\eqref{eq: average stratum tv restriction} and compute the joint bounds through
the program in \eqref{eq: rct covariate budget allocation}. We take $X$ to be the
full interaction of an indicator for above median baseline travel distance 
and the two randomization stratifiers, gender and an
indicator for whether the participant was actively looking for a job. We retain a cell
only if it contains at least five treated and
five control respondents, and all eight cells satisfy this rule. The
control respondents keep the plug-in $\widehat{\pi}(x)$ strictly below one, as
Section~\ref{subsec: role of covariates RCT} maintains for $\pi(x)$.\footnote{When
$\widehat{\pi}(x)<0$, we set it to 0, since \ref{assump: Mono} implies that the
two arms' response rates are ordered within each cell and thus we attribute a
violation of that ordering in the sample to noise.} We recall that
the covariate-augmented paths originate at the sample analog of
$\tau_{\mathsf{MAR}}(P_0)$ 
rather than at the sample analog of $\tau_{\mathsf{MCAR}}(P_0)$. 
The introduction of covariates narrows the bounds at $\delta=1$
by $18.8\%$ in the $50\%$ arm, but widens them by $18.4\%$ in the
$25\%$ arm. Conditioning leaves the breakdown budget of the $50\%$
arm at $\widehat{\delta}_{\mathsf{b}}=1$, but cuts that of the $25\%$ arm
from $0.62$ to $0.10$. Covariates thus need not raise the breakdown budget, because
they also move the origin of the path. Here the covariate-augmented path of that
arm starts closer to zero, at $0.050$ against $0.103$, and its bounds are in fact wider
than the unconditional ones at every $\delta>0$, so that no tightening is available to
offset the shift in the origin. Finally, we stress that differences between
the estimated sensitivity bounds can also stem from differences in the
underlying samples, as the covariates are not observed for every unit for
which $Y$ and $D$ are observed.

\begin{figure}[htbp]
\caption{$\TV$ sensitivity bounds for the effect of Uber price reductions on
total mobility.}
\centering
\begin{minipage}[t]{0.495\textwidth}
\centering
\resizebox{\textwidth}{!}{%
  \adjustbox{trim=0pt 70pt 0pt 0pt,clip}{%
    \input{figures/TV_bounds_mobility_1.tex}}}
\end{minipage}\hfill
\begin{minipage}[t]{0.495\textwidth}
\centering
\resizebox{\textwidth}{!}{%
  \adjustbox{trim=0pt 70pt 0pt 0pt,clip}{%
    \input{figures/TV_bounds_mobility_2.tex}}}
\end{minipage}
\par\medskip
\resizebox{0.59\textwidth}{!}{% =====================================================================
% Shared legend for the two-panel RCT-with-attrition figure in main.tex.
% Extracted verbatim from the legend block of
% TV_bounds_performance_main.tex (tikzDevice output), so that the two
% panels of the merged figure can be shown without their own legends.
% Coordinates are the original picture coordinates; the bounding box is
% restricted to the legend block. Re-extracted 2026-08-22 after the legend
% moved from three rows of two to two rows of three (no covariates / with
% covariates); regenerate this file whenever the legend block changes.
% =====================================================================
\providecommand{\bgcolor}{white}% background inherited from the document
\begin{tikzpicture}[x=1pt,y=1pt]
\path[use as bounding box] (62.00, 8.50) rectangle (555.57, 62.91);
\begin{scope}
\path[clip] (  0.00,  0.00) rectangle (578.16,433.62);
\colorlet{fillColor}{\bgcolor}

\path[fill=fillColor] ( 62.00,  8.50) rectangle (555.57, 62.91);
\end{scope}
\begin{scope}
\path[clip] (  0.00,  0.00) rectangle (578.16,433.62);
\colorlet{fillColor}{\bgcolor}

\path[fill=fillColor] ( 70.50, 39.95) rectangle (113.18, 54.41);
\definecolor{drawColor}{RGB}{31,78,121}

\path[draw=drawColor,line width= 2.3pt,line join=round] ( 74.77, 47.18) -- (108.91, 47.18);
\end{scope}
\begin{scope}
\path[clip] (  0.00,  0.00) rectangle (578.16,433.62);
\colorlet{fillColor}{\bgcolor}

\path[fill=fillColor] (276.54, 39.95) rectangle (319.22, 54.41);
\definecolor{drawColor}{RGB}{31,78,121}

\path[draw=drawColor,line width= 2.3pt,dash pattern=on 4pt off 4pt ,line join=round] (280.80, 47.18) -- (314.95, 47.18);
\end{scope}
\begin{scope}
\path[clip] (  0.00,  0.00) rectangle (578.16,433.62);
\colorlet{fillColor}{\bgcolor}

\path[fill=fillColor] (402.47, 39.95) rectangle (445.14, 54.41);
\definecolor{drawColor}{RGB}{31,78,121}

\path[draw=drawColor,line width= 2.3pt,dash pattern=on 1pt off 3pt ,line join=round] (406.73, 47.18) -- (440.88, 47.18);
\end{scope}
\begin{scope}
\path[clip] (  0.00,  0.00) rectangle (578.16,433.62);
\colorlet{fillColor}{\bgcolor}

\path[fill=fillColor] ( 70.50, 17.00) rectangle (113.18, 31.45);
\definecolor{drawColor}{RGB}{213,94,0}

\path[draw=drawColor,line width= 2.3pt,line join=round] ( 74.77, 24.23) -- (108.91, 24.23);
\end{scope}
\begin{scope}
\path[clip] (  0.00,  0.00) rectangle (578.16,433.62);
\colorlet{fillColor}{\bgcolor}

\path[fill=fillColor] (276.54, 17.00) rectangle (319.22, 31.45);
\definecolor{drawColor}{RGB}{213,94,0}

\path[draw=drawColor,line width= 2.3pt,dash pattern=on 4pt off 4pt ,line join=round] (280.80, 24.23) -- (314.95, 24.23);
\end{scope}
\begin{scope}
\path[clip] (  0.00,  0.00) rectangle (578.16,433.62);
\colorlet{fillColor}{\bgcolor}

\path[fill=fillColor] (402.47, 17.00) rectangle (445.14, 31.45);
\definecolor{drawColor}{RGB}{213,94,0}

\path[draw=drawColor,line width= 2.3pt,dash pattern=on 1pt off 3pt ,line join=round] (406.73, 24.23) -- (440.88, 24.23);
\end{scope}
\begin{scope}
\path[clip] (  0.00,  0.00) rectangle (578.16,433.62);
\definecolor{drawColor}{RGB}{0,0,0}

\node[text=drawColor,anchor=base west,inner sep=0pt, outer sep=0pt, scale=  1.60] at (121.68, 41.67) {\legTVbounds};
\end{scope}
\begin{scope}
\path[clip] (  0.00,  0.00) rectangle (578.16,433.62);
\definecolor{drawColor}{RGB}{0,0,0}

\node[text=drawColor,anchor=base west,inner sep=0pt, outer sep=0pt, scale=  1.60] at (327.72, 41.67) {95\% CI};
\end{scope}
\begin{scope}
\path[clip] (  0.00,  0.00) rectangle (578.16,433.62);
\definecolor{drawColor}{RGB}{0,0,0}

\node[text=drawColor,anchor=base west,inner sep=0pt, outer sep=0pt, scale=  1.60] at (453.64, 41.67) {\legLeebound};
\end{scope}
\begin{scope}
\path[clip] (  0.00,  0.00) rectangle (578.16,433.62);
\definecolor{drawColor}{RGB}{0,0,0}

\node[text=drawColor,anchor=base west,inner sep=0pt, outer sep=0pt, scale=  1.60] at (121.68, 18.72) {\legTVboundsX};
\end{scope}
\begin{scope}
\path[clip] (  0.00,  0.00) rectangle (578.16,433.62);
\definecolor{drawColor}{RGB}{0,0,0}

\node[text=drawColor,anchor=base west,inner sep=0pt, outer sep=0pt, scale=  1.60] at (327.72, 18.72) {95\% CI};
\end{scope}
\begin{scope}
\path[clip] (  0.00,  0.00) rectangle (578.16,433.62);
\definecolor{drawColor}{RGB}{0,0,0}

\node[text=drawColor,anchor=base west,inner sep=0pt, outer sep=0pt, scale=  1.60] at (453.64, 18.72) {\legLeeboundX};
\end{scope}
\end{tikzpicture}}
\par
\begin{center}
\parbox[t]{\textwidth}{\footnotesize
\textit{Notes:}
Computed from the replication data of \citet{christensen2026DemandMobility}.
Each panel plots the estimated $\widehat{\underline{\tau}}(\delta)$ and
$\widehat{\overline{\tau}}(\delta)$ of \eqref{eq: tau bounds def} (solid blue lines)
over $\delta\in\{0,0.02,\ldots,1\}$, together with their joint covariate-augmented
counterparts $\widehat{\underline{\tau}}_{X}(\delta)$ and
$\widehat{\overline{\tau}}_{X}(\delta)$
obtained from \eqref{eq: complier bounds joint} (solid orange lines), and
the corresponding bounds of \citet{lee2009TrainingWagesSample} (dotted
horizontal lines). The dashed lines are the outer $95\%$ confidence
bands of the bounds, pointwise in $\delta$, formed by the $2.5\%$ quantile of the
bootstrapped lower bound and the $97.5\%$ quantile of the bootstrapped upper bound.
The colored vertical dotted lines mark the estimated breakdown
budgets $\widehat{\delta}_{\mathsf{b}}$ of Section~\ref{sec: summary measures}
and the annotations beside them report
their values. The remaining annotation reports the reduction in the width
of the sensitivity bounds at $\delta=1$ achieved by the use of additional
covariates, so that a negative value means that the covariates widen the
bounds.
The confidence intervals are computed using $1{,}000$ iterations of
a nonparametric percentile bootstrap clustered at the individual level. The corresponding
standard errors are reported in
Table~\ref{tab: cno attrition se}.}
\end{center}
\label{fig: cno attrition empirical}
\end{figure}

\begin{table}[t]
\centering
\caption{$\TV$ sensitivity bounds for the effect of Uber price reductions.}
\label{tab: cno attrition se}
{\small\setlength{\tabcolsep}{4pt}
\begin{tabular}{lcccccc}
\toprule\toprule
 & \multicolumn{3}{c}{\textit{25\% reduction}} & \multicolumn{3}{c}{\textit{50\% reduction}} \\
\cmidrule(lr){2-4}\cmidrule(lr){5-7}
$\delta$ & $\widehat{\underline{\tau}}(\delta)$ & $\widehat{\overline{\tau}}(\delta)$ & Width & $\widehat{\underline{\tau}}(\delta)$ & $\widehat{\overline{\tau}}(\delta)$ & Width \\
\midrule
\multicolumn{7}{l}{\uline{\textit{Panel A: Without covariates}}} \\[0.3em]
\multirow{2}{*}{\textbf{0.00}} & $0.103$ & $0.103$ & $0.000$ & $0.407$ & $0.407$ & $0.000$ \\
 & ($0.108$) & ($0.108$) & ($0.000$) & ($0.101$) & ($0.101$) & ($0.000$) \\[0.25em]
\multirow{2}{*}{\textbf{0.25}} & $0.041$ & $0.171$ & $0.130$ & $0.284$ & $0.571$ & $0.287$ \\
 & ($0.115$) & ($0.117$) & ($0.084$) & ($0.107$) & ($0.113$) & ($0.080$) \\[0.25em]
\multirow{2}{*}{\textbf{0.50}} & $0.011$ & $0.228$ & $0.217$ & $0.222$ & $0.704$ & $0.482$ \\
 & ($0.122$) & ($0.137$) & ($0.139$) & ($0.112$) & ($0.121$) & ($0.117$) \\[0.25em]
\multirow{2}{*}{\textbf{0.75}} & $-0.009$ & $0.278$ & $0.287$ & $0.183$ & $0.769$ & $0.585$ \\
 & ($0.128$) & ($0.159$) & ($0.183$) & ($0.116$) & ($0.120$) & ($0.128$) \\[0.25em]
\multirow{2}{*}{\textbf{1.00}} & $-0.018$ & $0.307$ & $0.325$ & $0.168$ & $0.794$ & $0.626$ \\
 & ($0.130$) & ($0.178$) & ($0.210$) & ($0.117$) & ($0.118$) & ($0.129$) \\
\addlinespace[0.9em]
\multicolumn{7}{l}{\uline{\textit{Panel B: With covariates}}} \\[0.3em]
\multirow{2}{*}{\textbf{0.00}} & $0.050$ & $0.050$ & $0.000$ & $0.388$ & $0.388$ & $0.000$ \\
 & ($0.109$) & ($0.109$) & ($0.000$) & ($0.097$) & ($0.097$) & ($0.000$) \\[0.25em]
\multirow{2}{*}{\textbf{0.25}} & $-0.040$ & $0.173$ & $0.213$ & $0.258$ & $0.574$ & $0.316$ \\
 & ($0.113$) & ($0.117$) & ($0.070$) & ($0.100$) & ($0.102$) & ($0.057$) \\[0.25em]
\multirow{2}{*}{\textbf{0.50}} & $-0.077$ & $0.244$ & $0.321$ & $0.211$ & $0.646$ & $0.435$ \\
 & ($0.115$) & ($0.128$) & ($0.106$) & ($0.103$) & ($0.109$) & ($0.086$) \\[0.25em]
\multirow{2}{*}{\textbf{0.75}} & $-0.093$ & $0.291$ & $0.384$ & $0.190$ & $0.693$ & $0.503$ \\
 & ($0.117$) & ($0.137$) & ($0.126$) & ($0.105$) & ($0.117$) & ($0.105$) \\[0.25em]
\multirow{2}{*}{\textbf{1.00}} & $-0.094$ & $0.291$ & $0.385$ & $0.189$ & $0.697$ & $0.509$ \\
 & ($0.117$) & ($0.138$) & ($0.128$) & ($0.105$) & ($0.119$) & ($0.110$) \\
\bottomrule\bottomrule
\end{tabular}
}
\par
\begin{center}
\parbox[t]{\textwidth}{\footnotesize
\textit{Notes:} Plug-in estimates of the bounds
$\widehat{\underline{\tau}}(\delta)$ and $\widehat{\overline{\tau}}(\delta)$ of
\eqref{eq: tau bounds def} (Panel A) and of their joint covariate-augmented counterparts
$\widehat{\underline{\tau}}_{X}(\delta)$ and $\widehat{\overline{\tau}}_{X}(\delta)$ of
\eqref{eq: complier bounds joint} (Panel B) at five budgets. Width is the difference
between the upper and the lower bound. Bootstrap standard errors are in parentheses, where
the standard error of the width is computed from the bootstrap distribution of the width
itself. All other technical details
can be found in the notes of Figure~\ref{fig: cno attrition empirical}.}
\end{center}
\end{table}

\textbf{Summary measures.} Table~\ref{tab: cno sa9} reports the
measures of Section~\ref{sec: summary measures} for the two comparisons.
The shadow prices are small,
$0.08$ at $\widehat{\delta}_{\mathsf{b}}=0.62$ in the $25\%$ arm and
$0.18$ at $\widehat{\delta}_{\mathsf{cb}}=0.54$ in the $50\%$ arm, which is what the
small complier shares lead one to expect in
view of the trimming quantiles of
Theorem~\ref{theorem: tv dual recovers lee bound}.

The flatness of the bounds, which is what makes the present application
robust, also makes certifying additional robustness cheap at the margin, because
by \eqref{eq: exchange rate formula} the marginal sample requirement is governed
mainly by the ratio of the shadow price to the slack
$\underline{\tau}(\delta)-\tau_\star$, with reference value
$\tau_\star=0$, and this slack is large in the $50\%$ arm. The precision correction
$\sigma'/\sigma$ in \eqref{eq: exchange rate formula} is comparatively small. Certifying
five additional percentage
points would require $167$ additional participants in the $50\%$
contrast. At fifty
dollars per participant, this requirement amounts to approximately \$1,670 per
additional percentage point of certified robustness, well below the \$12k to
\$80k per percentage point of the tutoring program reported in
Section~6.1 of the main text. On the other hand,
the $25\%$ contrast illustrates a case in
which the certified breakdown budget is $0$: signing the effect at
$\delta=0$ alone would require $2{,}941$ additional participants, and certifying a budget
of five percentage points $4{,}702$.

\begin{table}[t]
\centering
\caption{Summary measures for the effect of Uber price reductions on total
mobility.}
\label{tab: cno sa9}
{\small\setlength{\tabcolsep}{4pt}
\begin{tabular}{lccccccc}
\toprule\toprule
 & \multicolumn{2}{c}{\textit{Breakdown}} & \textit{Price of robustness} & \multicolumn{4}{c}{\textit{Certification frontier}} \\
\cmidrule(lr){2-3}\cmidrule(lr){4-4}\cmidrule(lr){5-8}
 & $\widehat{\delta}_{\mathsf{b}}$ & $\widehat{\delta}_{\mathsf{cb}}$ & $\widehat{\underline{\eta}}_{\mathsf{b}}$ & $n$ & $n^{\star}$ & $\Delta n$ & \$/pp \\
\midrule
$25\%$ reduction & $0.621$ & $0.000$ & $0.08$ & 912 & 5,614 & $+$4,702 & $47$k \\
$50\%$ reduction & $1$ & $0.543$ & $0.18$ & 916 & 1,083 & $+$167 & $2$k \\
\bottomrule\bottomrule
\end{tabular}
}
\par
\begin{center}
\parbox[t]{\textwidth}{\footnotesize
\textit{Notes:} All measures are computed without covariates and with
reference value $\tau_\star=0$. $\widehat{\delta}_{\mathsf{b}}$ is the plug-in breakdown
budget of \eqref{eq: breakdown point}, and $\widehat{\delta}_{\mathsf{cb}}$ is the
certified breakdown budget of \eqref{eq: certified breakdown} at $\alpha=0.05$, computed
from the percentile bootstrap band as in
Remark~\ref{rem: certified breakdown bootstrap}.
$\widehat{\underline{\eta}}_{\mathsf{b}}$ is the plug-in shadow
price, evaluated at $\widehat{\delta}_{\mathsf{b}}$, or at
$\widehat{\delta}_{\mathsf{cb}}$ when $\widehat{\delta}_{\mathsf{b}}=1$, $n^{\star}$ is
the certification
frontier \eqref{eq: certification frontier} at
$\widehat{\delta}_{\mathsf{cb}}+0.05$ and rounded up to the nearest
integer, and
$\Delta n=n^{\star}-n$ is the additional sample the study would
have to collect, measured against the realized sample size $n$ rather than
against the frontier at the certified budget. The
last column converts $\Delta n$ into dollars per additional percentage point
of certified budget at a marginal cost of fifty dollars per participant
enrolled and surveyed, the midpoint of the range implied by the survey costs
of the impact-evaluation portfolio of
\citet{puri2019ChallengesRealWorldImpact}.}
\end{center}
\end{table}

\textbf{Discussion.} The exercise is the mirror image of the tutoring
application of Section~6.1 of the main text. There, attrition is
moderate in level but strongly differential across arms, and a small share of treated
respondents departing from \ref{assump: MCAR} suffices to make the effect vanish. Here,
attrition is severe in level but similar across arms. In the $50\%$ arm no departure
from \ref{assump: MCAR} makes the effect vanish, and in the $25\%$ arm close to two
thirds of the treated respondents must depart from \ref{assump: MCAR} before the effect
loses its sign. The sensitivity exercise thus prices the differential in response
across arms, not its level, exactly as the trimming quantiles of
Theorem~\ref{theorem: tv dual recovers lee bound} predict.

\subsection{RCT with Attrition II:
\texorpdfstring{\citet*{shao2026HowShouldCovariates}}{Shao, Lyu, Yu, and Wang (2026)}}
\label{subsec: empirical clinical}

The case study of the previous subsection and the one in
Section~6.1 of the main text answer the question
``how fragile is this particular
conclusion'' for one experiment at a time. A different
question is how often attrition is a concern for robustness across a population of
published trials. To answer this question, we deploy our framework on every eligible trial
in RCT-Bench, a public collection of $125$ individually randomized clinical
trials with participant-level data -- $54{,}258$ participants in all,
published between 1991 and 2026 -- assembled by
\citet{shao2026HowShouldCovariates} to study covariate adjustment.\footnote{The
collection is available at
\url{https://bingkaiwang.github.io/RCT_Bench-main/website/}. For each trial we
take the first listed primary outcome (endpoint) as the outcome and treat a unit as
attrited when that outcome is missing in the participant-level file.}
Clinical trials are a natural population for the exercise as attrition is
endemic and pre-analysis plans typically commit researchers to use a specific outcome.

\textbf{Study selection.} The unit of analysis is a treatment-control
contrast rather than a trial: in trials with more than two arms, we compare each treated
arm with the common control group. 
The $125$ trials yield $160$ contrasts, $97$ from
two-arm trials and $63$ from the $28$ trials with three
or more arms.
We retain the contrasts with more than fifty units observed on the primary
outcome and with a strictly positive complier share
$\widehat{\pi}>0$, that is, with some differential attrition. The size screen leaves
$114$ contrasts, of which $68$
exhibit no differential attrition at all ($\widehat{\pi}=0$), 
leaving $46$ contrasts drawn from $37$
distinct trials. These
are the units displayed in Figure~\ref{fig: clinical forest}. 
They are small to medium size studies. Across the $46$ contrasts, the
median number of participants randomized to the two arms is $232$, and the median number
observed on the primary outcome is $195$. The largest contrast has $2{,}207$ randomized
participants and the smallest $56$. Figure~\ref{fig: clinical selection}
displays the histograms for the observed sample size and the estimated
complier share $\widehat{\pi}$ for the whole corpus of $160$ contrasts.

\begin{figure}[htbp]
\caption{Sample size and differential attrition across the clinical corpus.}
\label{fig: clinical selection}
\centering
\begin{subfigure}[t]{0.49\textwidth}
    \centering
    \caption{Observed sample size}
    \label{fig: clinical selection size}
    \resizebox{\textwidth}{!}{% Created by tikzDevice version 0.12.6 on 2026-09-29 13:37:44
% !TEX encoding = UTF-8 Unicode
\providecommand{\bgcolor}{white}% background inherited from the document
\begin{tikzpicture}[x=1pt,y=1pt]
\colorlet{fillColor}{\bgcolor}
\path[use as bounding box,fill=fillColor,fill opacity=0.00] (0,0) rectangle (433.62,310.76);
\begin{scope}
\path[clip] (  0.00,  0.00) rectangle (433.62,310.76);
\colorlet{drawColor}{\bgcolor}
\colorlet{fillColor}{\bgcolor}

\path[draw=drawColor,line width= 0.8pt,line join=round,line cap=round,fill=fillColor] (  0.00,  0.00) rectangle (433.62,310.76);
\end{scope}
\begin{scope}
\path[clip] ( 45.33, 68.15) rectangle (423.62,304.76);
\colorlet{fillColor}{\bgcolor}

\path[fill=fillColor] ( 45.33, 68.15) rectangle (423.62,304.76);
\colorlet{drawColor}{\bgcolor}
\definecolor{fillColor}{gray}{0.60}

\path[draw=drawColor,line width= 0.8pt,fill=fillColor] ( 54.48, 78.90) rectangle (109.39,264.34);

\path[draw=drawColor,line width= 0.8pt,fill=fillColor] (115.49,127.28) rectangle (170.41,219.99);

\path[draw=drawColor,line width= 0.8pt,fill=fillColor] (176.51,151.46) rectangle (231.42,240.15);

\path[draw=drawColor,line width= 0.8pt,fill=fillColor] (237.52,127.28) rectangle (292.44,163.56);

\path[draw=drawColor,line width= 0.8pt,fill=fillColor] (298.54, 91.00) rectangle (353.45,115.18);

\path[draw=drawColor,line width= 0.8pt,fill=fillColor] (359.55, 82.93) rectangle (414.47,115.18);
\definecolor{fillColor}{RGB}{31,78,121}

\path[draw=drawColor,line width= 0.8pt,fill=fillColor] ( 54.48, 78.90) rectangle (109.39, 78.90);

\path[draw=drawColor,line width= 0.8pt,fill=fillColor] (115.49, 78.90) rectangle (170.41,127.28);

\path[draw=drawColor,line width= 0.8pt,fill=fillColor] (176.51, 78.90) rectangle (231.42,151.46);

\path[draw=drawColor,line width= 0.8pt,fill=fillColor] (237.52, 78.90) rectangle (292.44,127.28);

\path[draw=drawColor,line width= 0.8pt,fill=fillColor] (298.54, 78.90) rectangle (353.45, 91.00);

\path[draw=drawColor,line width= 0.8pt,fill=fillColor] (359.55, 78.90) rectangle (414.47, 82.93);
\definecolor{drawColor}{RGB}{0,0,0}

\node[text=drawColor,anchor=base,inner sep=0pt, outer sep=0pt, scale=  1.48] at ( 81.94,268.92) {46};

\node[text=drawColor,anchor=base,inner sep=0pt, outer sep=0pt, scale=  1.48] at (142.95,224.58) {35};

\node[text=drawColor,anchor=base,inner sep=0pt, outer sep=0pt, scale=  1.48] at (203.97,244.74) {40};

\node[text=drawColor,anchor=base,inner sep=0pt, outer sep=0pt, scale=  1.48] at (264.98,168.14) {21};

\node[text=drawColor,anchor=base,inner sep=0pt, outer sep=0pt, scale=  1.48] at (326.00,119.77) {9};

\node[text=drawColor,anchor=base,inner sep=0pt, outer sep=0pt, scale=  1.48] at (387.01,119.77) {9};
\definecolor{drawColor}{RGB}{255,255,255}

\node[text=drawColor,anchor=base,inner sep=0pt, outer sep=0pt, scale=  1.42] at (142.95, 98.18) {\bfseries 12};

\node[text=drawColor,anchor=base,inner sep=0pt, outer sep=0pt, scale=  1.42] at (203.97,110.27) {\bfseries 18};

\node[text=drawColor,anchor=base,inner sep=0pt, outer sep=0pt, scale=  1.42] at (264.98, 98.18) {\bfseries 12};

\node[text=drawColor,anchor=base,inner sep=0pt, outer sep=0pt, scale=  1.42] at (326.00, 80.04) {\bfseries 3};
\definecolor{drawColor}{gray}{0.20}

\path[draw=drawColor,line width= 0.8pt,line join=round,line cap=round] ( 45.33, 68.15) rectangle (423.62,304.76);
\end{scope}
\begin{scope}
\path[clip] (  0.00,  0.00) rectangle (433.62,310.76);
\definecolor{drawColor}{gray}{0.30}

\node[text=drawColor,anchor=base east,inner sep=0pt, outer sep=0pt, scale=  1.40] at ( 38.13, 74.08) {0};

\node[text=drawColor,anchor=base east,inner sep=0pt, outer sep=0pt, scale=  1.40] at ( 38.13,114.39) {10};

\node[text=drawColor,anchor=base east,inner sep=0pt, outer sep=0pt, scale=  1.40] at ( 38.13,154.71) {20};

\node[text=drawColor,anchor=base east,inner sep=0pt, outer sep=0pt, scale=  1.40] at ( 38.13,195.02) {30};

\node[text=drawColor,anchor=base east,inner sep=0pt, outer sep=0pt, scale=  1.40] at ( 38.13,235.33) {40};

\node[text=drawColor,anchor=base east,inner sep=0pt, outer sep=0pt, scale=  1.40] at ( 38.13,275.64) {50};
\end{scope}
\begin{scope}
\path[clip] (  0.00,  0.00) rectangle (433.62,310.76);
\definecolor{drawColor}{gray}{0.20}

\path[draw=drawColor,line width= 0.8pt,line join=round] ( 41.33, 78.90) --
	( 45.33, 78.90);

\path[draw=drawColor,line width= 0.8pt,line join=round] ( 41.33,119.21) --
	( 45.33,119.21);

\path[draw=drawColor,line width= 0.8pt,line join=round] ( 41.33,159.53) --
	( 45.33,159.53);

\path[draw=drawColor,line width= 0.8pt,line join=round] ( 41.33,199.84) --
	( 45.33,199.84);

\path[draw=drawColor,line width= 0.8pt,line join=round] ( 41.33,240.15) --
	( 45.33,240.15);

\path[draw=drawColor,line width= 0.8pt,line join=round] ( 41.33,280.46) --
	( 45.33,280.46);
\end{scope}
\begin{scope}
\path[clip] (  0.00,  0.00) rectangle (433.62,310.76);
\definecolor{drawColor}{gray}{0.20}

\path[draw=drawColor,line width= 0.8pt,line join=round] ( 81.94, 64.15) --
	( 81.94, 68.15);

\path[draw=drawColor,line width= 0.8pt,line join=round] (142.95, 64.15) --
	(142.95, 68.15);

\path[draw=drawColor,line width= 0.8pt,line join=round] (203.97, 64.15) --
	(203.97, 68.15);

\path[draw=drawColor,line width= 0.8pt,line join=round] (264.98, 64.15) --
	(264.98, 68.15);

\path[draw=drawColor,line width= 0.8pt,line join=round] (326.00, 64.15) --
	(326.00, 68.15);

\path[draw=drawColor,line width= 0.8pt,line join=round] (387.01, 64.15) --
	(387.01, 68.15);
\end{scope}
\begin{scope}
\path[clip] (  0.00,  0.00) rectangle (433.62,310.76);
\definecolor{drawColor}{gray}{0.30}

\node[text=drawColor,anchor=base,inner sep=0pt, outer sep=0pt, scale=  1.40] at ( 81.94, 51.31) {$\leq 50$};

\node[text=drawColor,anchor=base,inner sep=0pt, outer sep=0pt, scale=  1.40] at (142.95, 51.31) {$(50,100]$};

\node[text=drawColor,anchor=base,inner sep=0pt, outer sep=0pt, scale=  1.40] at (203.97, 51.31) {$(100,250]$};

\node[text=drawColor,anchor=base,inner sep=0pt, outer sep=0pt, scale=  1.40] at (264.98, 51.31) {$(250,500]$};

\node[text=drawColor,anchor=base,inner sep=0pt, outer sep=0pt, scale=  1.40] at (326.00, 51.31) {$(500,10^3]$};

\node[text=drawColor,anchor=base,inner sep=0pt, outer sep=0pt, scale=  1.40] at (387.01, 51.31) {$>10^3$};
\end{scope}
\begin{scope}
\path[clip] (  0.00,  0.00) rectangle (433.62,310.76);
\definecolor{drawColor}{RGB}{0,0,0}

\node[text=drawColor,anchor=base,inner sep=0pt, outer sep=0pt, scale=  1.60] at (234.47, 33.56) {Units observed on the primary outcome $n_{\mathsf{obs}}$};
\end{scope}
\begin{scope}
\path[clip] (  0.00,  0.00) rectangle (433.62,310.76);
\definecolor{drawColor}{RGB}{0,0,0}

\node[text=drawColor,rotate= 90.00,anchor=base,inner sep=0pt, outer sep=0pt, scale=  1.60] at ( 17.02,186.45) {Number of contrasts};
\end{scope}
\begin{scope}
\path[clip] (  0.00,  0.00) rectangle (433.62,310.76);
\colorlet{fillColor}{\bgcolor}

\path[fill=fillColor] (136.66,  4.00) rectangle (332.28, 14.45);
\end{scope}
\begin{scope}
\path[clip] (  0.00,  0.00) rectangle (433.62,310.76);
\colorlet{fillColor}{\bgcolor}

\path[fill=fillColor] (136.66,  4.00) rectangle (151.12, 18.45);
\colorlet{drawColor}{\bgcolor}
\definecolor{fillColor}{RGB}{31,78,121}

\path[draw=drawColor,line width= 0.8pt,fill=fillColor] (137.70,  5.03) rectangle (150.08, 17.42);
\end{scope}
\begin{scope}
\path[clip] (  0.00,  0.00) rectangle (433.62,310.76);
\colorlet{fillColor}{\bgcolor}

\path[fill=fillColor] (225.64,  4.00) rectangle (240.10, 18.45);
\colorlet{drawColor}{\bgcolor}
\definecolor{fillColor}{gray}{0.60}

\path[draw=drawColor,line width= 0.8pt,fill=fillColor] (226.68,  5.03) rectangle (239.06, 17.42);
\end{scope}
\begin{scope}
\path[clip] (  0.00,  0.00) rectangle (433.62,310.76);
\definecolor{drawColor}{RGB}{0,0,0}

\node[text=drawColor,anchor=base west,inner sep=0pt, outer sep=0pt, scale=  1.50] at (159.12,  6.06) {Retained};
\end{scope}
\begin{scope}
\path[clip] (  0.00,  0.00) rectangle (433.62,310.76);
\definecolor{drawColor}{RGB}{0,0,0}

\node[text=drawColor,anchor=base west,inner sep=0pt, outer sep=0pt, scale=  1.50] at (248.10,  6.06) {Screened out};
\end{scope}
\end{tikzpicture}}
\end{subfigure}
\hfill
\begin{subfigure}[t]{0.49\textwidth}
    \centering
    \caption{Complier share $\widehat{\pi}$}
    \label{fig: clinical selection p0}
    \resizebox{\textwidth}{!}{% Created by tikzDevice version 0.12.6 on 2026-09-29 13:37:53
% !TEX encoding = UTF-8 Unicode
\providecommand{\bgcolor}{white}% background inherited from the document
\begin{tikzpicture}[x=1pt,y=1pt]
\colorlet{fillColor}{\bgcolor}
\path[use as bounding box,fill=fillColor,fill opacity=0.00] (0,0) rectangle (433.62,310.76);
\begin{scope}
\path[clip] (  0.00,  0.00) rectangle (433.62,310.76);
\colorlet{drawColor}{\bgcolor}
\colorlet{fillColor}{\bgcolor}

\path[draw=drawColor,line width= 0.8pt,line join=round,line cap=round,fill=fillColor] (  0.00,  0.00) rectangle (433.62,310.76);
\end{scope}
\begin{scope}
\path[clip] ( 52.32, 68.15) rectangle (423.62,304.76);
\colorlet{fillColor}{\bgcolor}

\path[fill=fillColor] ( 52.32, 68.15) rectangle (423.62,304.76);
\colorlet{drawColor}{\bgcolor}
\definecolor{fillColor}{gray}{0.60}

\path[draw=drawColor,line width= 0.8pt,fill=fillColor] ( 63.04, 78.90) rectangle (127.30,264.34);

\path[draw=drawColor,line width= 0.8pt,fill=fillColor] (134.44,137.56) rectangle (198.70,137.56);

\path[draw=drawColor,line width= 0.8pt,fill=fillColor] (205.84, 99.72) rectangle (270.10,118.64);

\path[draw=drawColor,line width= 0.8pt,fill=fillColor] (277.24, 84.58) rectangle (341.51, 94.04);

\path[draw=drawColor,line width= 0.8pt,fill=fillColor] (348.65, 80.79) rectangle (412.91, 82.69);
\definecolor{fillColor}{RGB}{31,78,121}

\path[draw=drawColor,line width= 0.8pt,fill=fillColor] ( 63.04, 78.90) rectangle (127.30, 78.90);

\path[draw=drawColor,line width= 0.8pt,fill=fillColor] (134.44, 78.90) rectangle (198.70,137.56);

\path[draw=drawColor,line width= 0.8pt,fill=fillColor] (205.84, 78.90) rectangle (270.10, 99.72);

\path[draw=drawColor,line width= 0.8pt,fill=fillColor] (277.24, 78.90) rectangle (341.51, 84.58);

\path[draw=drawColor,line width= 0.8pt,fill=fillColor] (348.65, 78.90) rectangle (412.91, 80.79);
\definecolor{drawColor}{RGB}{0,0,0}

\node[text=drawColor,anchor=base,inner sep=0pt, outer sep=0pt, scale=  1.48] at ( 95.17,268.92) {98};

\node[text=drawColor,anchor=base,inner sep=0pt, outer sep=0pt, scale=  1.48] at (166.57,142.15) {31};

\node[text=drawColor,anchor=base,inner sep=0pt, outer sep=0pt, scale=  1.48] at (237.97,123.23) {21};

\node[text=drawColor,anchor=base,inner sep=0pt, outer sep=0pt, scale=  1.48] at (309.38, 98.63) {8};

\node[text=drawColor,anchor=base,inner sep=0pt, outer sep=0pt, scale=  1.48] at (380.78, 87.27) {2};
\definecolor{drawColor}{RGB}{255,255,255}

\node[text=drawColor,anchor=base,inner sep=0pt, outer sep=0pt, scale=  1.42] at (166.57,103.32) {\bfseries 31};

\node[text=drawColor,anchor=base,inner sep=0pt, outer sep=0pt, scale=  1.42] at (237.97, 84.40) {\bfseries 11};
\definecolor{drawColor}{gray}{0.20}

\path[draw=drawColor,line width= 0.8pt,line join=round,line cap=round] ( 52.32, 68.15) rectangle (423.62,304.76);
\end{scope}
\begin{scope}
\path[clip] (  0.00,  0.00) rectangle (433.62,310.76);
\definecolor{drawColor}{gray}{0.30}

\node[text=drawColor,anchor=base east,inner sep=0pt, outer sep=0pt, scale=  1.40] at ( 45.12, 74.08) {0};

\node[text=drawColor,anchor=base east,inner sep=0pt, outer sep=0pt, scale=  1.40] at ( 45.12,111.93) {20};

\node[text=drawColor,anchor=base east,inner sep=0pt, outer sep=0pt, scale=  1.40] at ( 45.12,149.77) {40};

\node[text=drawColor,anchor=base east,inner sep=0pt, outer sep=0pt, scale=  1.40] at ( 45.12,187.61) {60};

\node[text=drawColor,anchor=base east,inner sep=0pt, outer sep=0pt, scale=  1.40] at ( 45.12,225.46) {80};

\node[text=drawColor,anchor=base east,inner sep=0pt, outer sep=0pt, scale=  1.40] at ( 45.12,263.30) {100};
\end{scope}
\begin{scope}
\path[clip] (  0.00,  0.00) rectangle (433.62,310.76);
\definecolor{drawColor}{gray}{0.20}

\path[draw=drawColor,line width= 0.8pt,line join=round] ( 48.32, 78.90) --
	( 52.32, 78.90);

\path[draw=drawColor,line width= 0.8pt,line join=round] ( 48.32,116.75) --
	( 52.32,116.75);

\path[draw=drawColor,line width= 0.8pt,line join=round] ( 48.32,154.59) --
	( 52.32,154.59);

\path[draw=drawColor,line width= 0.8pt,line join=round] ( 48.32,192.43) --
	( 52.32,192.43);

\path[draw=drawColor,line width= 0.8pt,line join=round] ( 48.32,230.28) --
	( 52.32,230.28);

\path[draw=drawColor,line width= 0.8pt,line join=round] ( 48.32,268.12) --
	( 52.32,268.12);
\end{scope}
\begin{scope}
\path[clip] (  0.00,  0.00) rectangle (433.62,310.76);
\definecolor{drawColor}{gray}{0.20}

\path[draw=drawColor,line width= 0.8pt,line join=round] ( 95.17, 64.15) --
	( 95.17, 68.15);

\path[draw=drawColor,line width= 0.8pt,line join=round] (166.57, 64.15) --
	(166.57, 68.15);

\path[draw=drawColor,line width= 0.8pt,line join=round] (237.97, 64.15) --
	(237.97, 68.15);

\path[draw=drawColor,line width= 0.8pt,line join=round] (309.38, 64.15) --
	(309.38, 68.15);

\path[draw=drawColor,line width= 0.8pt,line join=round] (380.78, 64.15) --
	(380.78, 68.15);
\end{scope}
\begin{scope}
\path[clip] (  0.00,  0.00) rectangle (433.62,310.76);
\definecolor{drawColor}{gray}{0.30}

\node[text=drawColor,anchor=base,inner sep=0pt, outer sep=0pt, scale=  1.40] at ( 95.17, 51.31) {$0$};

\node[text=drawColor,anchor=base,inner sep=0pt, outer sep=0pt, scale=  1.40] at (166.57, 51.31) {$(0,0.05]$};

\node[text=drawColor,anchor=base,inner sep=0pt, outer sep=0pt, scale=  1.40] at (237.97, 51.31) {$(0.05,0.10]$};

\node[text=drawColor,anchor=base,inner sep=0pt, outer sep=0pt, scale=  1.40] at (309.38, 51.31) {$(0.10,0.20]$};

\node[text=drawColor,anchor=base,inner sep=0pt, outer sep=0pt, scale=  1.40] at (380.78, 51.31) {$(0.20,0.50]$};
\end{scope}
\begin{scope}
\path[clip] (  0.00,  0.00) rectangle (433.62,310.76);
\definecolor{drawColor}{RGB}{0,0,0}

\node[text=drawColor,anchor=base,inner sep=0pt, outer sep=0pt, scale=  1.60] at (237.97, 33.56) {Complier share $\widehat{\pi}$};
\end{scope}
\begin{scope}
\path[clip] (  0.00,  0.00) rectangle (433.62,310.76);
\definecolor{drawColor}{RGB}{0,0,0}

\node[text=drawColor,rotate= 90.00,anchor=base,inner sep=0pt, outer sep=0pt, scale=  1.60] at ( 17.02,186.45) {Number of contrasts};
\end{scope}
\begin{scope}
\path[clip] (  0.00,  0.00) rectangle (433.62,310.76);
\colorlet{fillColor}{\bgcolor}

\path[fill=fillColor] (140.16,  4.00) rectangle (335.78, 14.45);
\end{scope}
\begin{scope}
\path[clip] (  0.00,  0.00) rectangle (433.62,310.76);
\colorlet{fillColor}{\bgcolor}

\path[fill=fillColor] (140.16,  4.00) rectangle (154.61, 18.45);
\colorlet{drawColor}{\bgcolor}
\definecolor{fillColor}{RGB}{31,78,121}

\path[draw=drawColor,line width= 0.8pt,fill=fillColor] (141.20,  5.03) rectangle (153.58, 17.42);
\end{scope}
\begin{scope}
\path[clip] (  0.00,  0.00) rectangle (433.62,310.76);
\colorlet{fillColor}{\bgcolor}

\path[fill=fillColor] (229.14,  4.00) rectangle (243.60, 18.45);
\colorlet{drawColor}{\bgcolor}
\definecolor{fillColor}{gray}{0.60}

\path[draw=drawColor,line width= 0.8pt,fill=fillColor] (230.18,  5.03) rectangle (242.56, 17.42);
\end{scope}
\begin{scope}
\path[clip] (  0.00,  0.00) rectangle (433.62,310.76);
\definecolor{drawColor}{RGB}{0,0,0}

\node[text=drawColor,anchor=base west,inner sep=0pt, outer sep=0pt, scale=  1.50] at (162.61,  6.06) {Retained};
\end{scope}
\begin{scope}
\path[clip] (  0.00,  0.00) rectangle (433.62,310.76);
\definecolor{drawColor}{RGB}{0,0,0}

\node[text=drawColor,anchor=base west,inner sep=0pt, outer sep=0pt, scale=  1.50] at (251.60,  6.06) {Screened out};
\end{scope}
\end{tikzpicture}}
\end{subfigure}
\par
\begin{center}
\parbox[t]{\textwidth}{\footnotesize
\textit{Notes:} Both panels use data from the $160$ treatment--control
contrasts of the corpus described in the text, shading in blue
the $46$ contrasts the exercise retains and in gray the remainder. Retention
requires (panel~(i)) more than fifty units observed on the primary
outcome and a strictly
positive estimated complier share $\widehat{\pi}$ (panel~(ii)).
The number atop each bar is the bin total. The number inside
the blue block is the count of retained contrasts.}
\end{center}
\end{figure}

\textbf{Results.} For each contrast we
trace $\widehat{\underline{\tau}}(\delta)$ and $\widehat{\overline{\tau}}(\delta)$
on a grid with step size $0.005$, with $B=2{,}000$ nonparametric bootstrap
iterations.
The estimated complier shares are small for
most contrasts: the median $\widehat{\pi}$ is $0.018$, the maximum is
$0.371$, and the smallest positive value is $7\times10^{-5}$.
We summarize each contrast by the certified breakdown budget
$\widehat{\delta}_{\mathsf{cb}}$ of \eqref{eq: certified breakdown},
computed from the percentile bootstrap band as in
Remark~\ref{rem: certified breakdown bootstrap}.
We set $\widehat{\delta}_{\mathsf{cb}}=1$ when zero is never reached on $[0,1]$.

For each contrast, Figure~\ref{fig: clinical forest} reports on a common standardized
scale the difference in means among respondents, the sample analog of
$\tau_{\mathsf{MCAR}}(P_0)$ (white dots), and its 95\% confidence intervals (black
lines),
together with $[\widehat{\underline{\tau}}(1),\widehat{\overline{\tau}}(1)]$
(non-transparent bands)
and their 95\% confidence intervals (shaded bands).
First, 25 of the 46 contrasts have $\widehat{\delta}_{\mathsf{cb}}=0$, meaning that the
95\% confidence interval
for $[\underline{\tau}(0),\overline{\tau}(0)]$ already contains 0. These are drawn in red 
in Figure~\ref{fig: clinical forest}. Put differently, the estimated treatment effect is
not signed even under \ref{assump: MCAR}, so that no positive budget is
needed to overturn the conclusion. Second, 13 contrasts,
drawn in green in Figure~\ref{fig: clinical forest},
are robust at every budget, meaning that the confidence intervals for
$[\underline{\tau}(1),\overline{\tau}(1)]$
do not contain 0. In other words, even if the whole population of treated respondents
departs from
\ref{assump: MCAR} and does so adversarially, the treatment effect would still be
statistically significant.
Third, the remaining 8 contrasts, depicted in amber in Figure~\ref{fig: clinical forest},
have a certified breakdown budget
in the interior of $[0,1]$, that is,
$\widehat{\delta}_{\mathsf{cb}}\in(0,1)$.

Before concluding, we stress that we did not use covariate-adjusted sensitivity bounds to
prioritize comparability across studies. Therefore, a
trial with $\widehat{\delta}_{\mathsf{cb}}=0$ may well report a significant
covariate-adjusted treatment effect in the primary analysis.

\begin{landscape}
\thispagestyle{empty}
\centering
\captionsetup{type=figure}
\caption{$\TV$ sensitivity bounds, contrast by contrast.}
\label{fig: clinical forest}
\resizebox{\linewidth}{!}{\input{figures/endpoints_forest.tex}}
\par
\begin{center}
\parbox[t]{\linewidth}{\footnotesize
\textit{Notes:} The $46$ treatment-control contrasts that the screens of this
subsection admit, drawn from $37$ distinct trials, one vertical bar each,
grouped by the value of the certified breakdown budget
$\widehat{\delta}_{\mathsf{cb}}$ of \eqref{eq: certified breakdown},
computed as in Remark~\ref{rem: certified breakdown bootstrap}, namely
$\widehat{\delta}_{\mathsf{cb}}=0$ (left), $\widehat{\delta}_{\mathsf{cb}}\in(0,1)$
(middle), and $\widehat{\delta}_{\mathsf{cb}}=1$ (right), the latter under the convention
that the infimum over an empty set equals one, drawn in red, amber, and
green, respectively,
and ordered within each group by increasing standardized effect.
A label ``\texttt{t$j$-$k$}'' denotes the $k$-th treated arm of trial $j$,
indexed in the alphabetical order of the arm names recorded in the metadata, whereas a
simple ``\texttt{t$j$}'' is a two-arm trial. 
Because the primary outcomes are on incomparable scales, every quantity is
divided by $\widehat\sigma$, the contrast's pooled-respondent outcome standard
deviation, so that the vertical axis is in standard-deviation units. Each bar nests four
quantities: the outer $95\%$
confidence band of the bounds at $\delta=1$ (shaded bands); the identified set at
$\delta=1$, whose endpoints are the Lee bounds (non-transparent bands);
the $95\%$ confidence interval of the difference in means among
respondents (black lines); and the difference in means
among respondents itself, the sample analog of $\tau_{\mathsf{MCAR}}(P_0)$ (white dots).
Confidence intervals use $B=2{,}000$ nonparametric bootstrap iterations
per contrast.}
\end{center}
\par
\vfill
\centerline{\thepage}
\end{landscape}

\clearpage
\section{Proofs}\label{sec: proofs}

\subsection{Proof of Lemma~\ref{lemma: divergence lebesgue decomposition}}
\begin{proof}\label{proof: divergence lebesgue decomposition}
    Choose $\varrho=\Pbase{}+\nu^\perp$ as the dominating measure and let
    $\calN\in\mathscr{F}$ be such that $\Pbase{}(\calN)=0$ and
    $\nu^\perp(\calZ\setminus \calN)=0$. Such $\calN$ exists because
    $\nu^\perp\perp \Pbase{}$. On $\calZ\setminus \calN$ we have
    $\de \Pbase{}/\de\varrho=1$
    and $\de\nu/\de\varrho=w$ $\varrho$-almost everywhere, so the integrand in
    Definition~\ref{defi: generalized phi divergence} equals
    $\phi^\pi(w,1)=\phi(w)$, whose integral may be taken over all of $\calZ$
    because $\Pbase{}(\calN)=0$. On $\calN$ we have $\de \Pbase{}/\de\varrho=0$ and
    $\de\nu/\de\varrho=1$ $\varrho$-almost everywhere, since $\varrho$ and $\nu$ both
    restrict to $\nu^\perp$ on $\calN$, so the integrand equals
    $\phi^\pi(1,0)=\phi^\infty(1)$ by definition of the perspective function,
    and it integrates to
    $\phi^\infty(1)\,\varrho(\calN)=\phi^\infty(1)\,\nu^\perp(\calZ)$. Summing the
    two contributions yields the claim. Finally, the argument does not depend on
    the choice of the dominating measure, in view of the positive homogeneity
    of the perspective function $\phi^\pi$.
\end{proof}

\subsection{Proof of Lemma~\ref{lem:tv_posneg}}
\begin{proof}\label{proof:lem:tv_posneg}
Since $\phi_{\TV}(s)=\frac{1}{2}|s-1|$ for
$s\geq0$ and $\phi_{\TV}^{\infty}(s)=s/2$ for $s\geq0$, the perspective
function satisfies $\phi^\pi_{\TV}(s,t)=\frac{1}{2}|s-t|$ for all
$s,t\geq0$: for $t>0$ this follows from
$t\phi_{\TV}(s/t)=\frac{t}{2}|s/t-1|=\frac{1}{2}|s-t|$, and for $t=0$ from
$\phi^\pi_{\TV}(s,0)=\phi^{\infty}_{\TV}(s)=s/2$. Hence, by
Definition~\ref{defi: generalized phi divergence},
\[
D_{\phi_{\TV}}(P\,\|\,Q)=\int\phi^\pi_{\TV}(p,q)\,\de\varrho
=\frac{1}{2}\int|p-q|\,\de\varrho.
\]
Next, for any $\calA\in\mathscr{F}$, since
$P(\calA)-Q(\calA)=\int_\calA(p-q)\,\de\varrho$, dropping the negative part
of the integrand and enlarging the domain of integration yields
$P(\calA)-Q(\calA)\leq\int[p-q]_+\,\de\varrho$, with equality at
$\calA=\{p\geq q\}$. Symmetrically,
$Q(\calA)-P(\calA)\leq\int[q-p]_+\,\de\varrho$. Since
$\int(p-q)\,\de\varrho=0$ in view of $P(\calZ)=Q(\calZ)=1$, the two bounds
coincide and equal $\frac{1}{2}\int|p-q|\,\de\varrho$, whence
\[
\TV(P\,\|\,Q)=\sup_{\calA\in\mathscr{F}}|P(\calA)-Q(\calA)|
=\frac{1}{2}\int|p-q|\,\de\varrho,
\]
attained at $\calA=\{p\geq q\}$.
\end{proof}

\subsection{Proof of Lemma~\ref{lemma: entropy function proper cvx lsc}}
\begin{proof}\label{proof: entropy function proper cvx lsc}
    By Definition \ref{defi: entropy function}, $\phi$ is a convex function.
    Furthermore, $\phi$ is proper: it is not identically $+\infty$ because
    $\phi(1)=0$, and it never takes the value $-\infty$ because $\phi\geq0$. Thus, by
    Theorem~12.2 in \citet{rockafellar1970ConvexAnalysis}, $\phi^*$ %
    is
    proper, convex, and closed. Closedness and lower semicontinuity are
    equivalent for any
    proper convex function \citep[p.~52]{rockafellar1970ConvexAnalysis}.

    Regarding the second part, take any proper closed convex function
    $f:\R^k\to\overline{\R}$ and define the function
    $\tilde f:\R^{k}\times\R\to\overline{\R}$
    \[\tilde f(z,t) =
    \begin{cases}
    tf(z/t),\quad &t>0,\\
    \iota_{\{0\}}(z), \quad &t=0,\\
    +\infty, \quad &t<0.
    \end{cases}\]     
    Then, $\tilde f$ is also proper and convex
    \citep[p.~35]{rockafellar1970ConvexAnalysis}.
    The closure of $\tilde f$ is equivalent to
    \[(\mathrm{cl}\, \tilde f)(z,t) =
    \begin{cases}
    tf(z/t),\quad &t>0,\\
    f^\infty(z), \quad &t=0,\\
    +\infty, \quad &t<0,
    \end{cases}\]      
    as highlighted on p.~67 in \citet{rockafellar1970ConvexAnalysis}.
    Note
    that
    $\mathrm{cl}\, \tilde f$ is closed and convex by construction and it is proper
    by Theorem~7.4 in \citet{rockafellar1970ConvexAnalysis}, because $\tilde f$
    is
    proper.
    Hence $\mathrm{cl}\, \tilde f$ is also lower semicontinuous. Finally, by definition
    of
    perspective function (Definition \ref{defi: perspective function}), we have
    $f^\pi(z,t)=(\mathrm{cl}\, \tilde f)(z,t)$. Taking $f=\phi$, which is proper and
    convex as argued above and lower semicontinuous by
    Definition~\ref{defi: entropy function}, and $f=\phi^*$ concludes.
\end{proof}

\subsection{Proof of Lemma~\ref{lemma: perspective larger than argument}}
\begin{proof}\label{proof: perspective larger than argument}
    Fix $z\in\R$. If $t<0$, the statement follows from the definition of
    perspective
    function as $(\phi^*)^\pi(z,t)=+\infty$. If $t>0$, then
    \[(\phi^*)^\pi(z,t) = t \phi^*(z/t) \geq \check{x}z-t\phi(\check{x}),\]
    for all $\check{x}\in\dom(\phi)$, where the inequality follows by
    multiplying the
    Young--Fenchel inequality
    \citep[p.~105]{rockafellar1970ConvexAnalysis},
    \[\phi^*(y) + \phi(x) \geq xy\:\:
    \forall\,x\in\dom(\phi),\,y\in\dom(\phi^*),\]
    taken at $y=z/t$ and $x=\check{x}$, by $t>0$. If
    $z/t\notin\dom(\phi^*)$, then $(\phi^*)^\pi(z,t)=t\phi^*(z/t)=+\infty$ and
    the claimed inequality holds trivially.
    If $t=0$, then, since $\phi$ is closed, proper, and convex, it coincides
    with its biconjugate, $\phi=\phi^{**}$, so the second assertion of
    Theorem~13.3 in \citet{rockafellar1970ConvexAnalysis}, applied with
    $f=\phi^*$, gives
    that
    \[(\phi^*)^\infty(z) = \iota_{\dom(\phi^{**})}^*(z) =
    \iota_{\dom(\phi)}^*(z) = \sup_{x\in\dom(\phi)}
    xz\geq
    \check{x}z = \check{x}z - t\phi(\check{x}).\]
\end{proof}

\subsection{Proof of Corollary~\ref{lemma: implicit regularization of entropy}}
\begin{proof}\label{proof: implicit regularization of entropy}
Lemma~\ref{lemma: lemma 4.12 kuhn} gives
$\sup\dom(\phi^*)=\phi^\infty(1)$. Hence (1) and
(2) are equivalent. Condition~(2) is equivalent to the existence of an
$M\in\R$ such that $\dom(\phi^*)\subseteq(-\infty,M]$, which is precisely
condition~(3). When these conditions hold, the same identity permits the
choice $M=\phi^\infty(1)$.
\end{proof}

\subsection{Proof of Theorem~\ref{theorem: weak duality}}
\begin{proof}\label{proof: weak duality}
    The proof follows the logic of
    \cite*{kuhn2025DistributionallyRobustOptimization} and proceeds in four
    steps:
    \begin{enumerate}
        \item Split $\calP_\phi({\rho},\delta)$ along its fibers over
        $\calM(\rho)$.
        \item Dualize the inner maximization problem. This step transforms the
        infinite-dimensional problem over $\calP_\phi({\rho},\delta)$ into a
        finite-dimensional problem in a Euclidean space.
        \item Using classical results from convex analysis, the optimal value
        function $h$
        of the primal of the inner maximization problem turns out to be convex.
        Its biconjugate then satisfies the inequality
        $h^{**}\leq h$.
        \item By the last step, it is possible to upper bound the value of the
        inner maximization
        problem.
    \end{enumerate}

    \uline{\textbf{Fiber decomposition and value function of the primal.}} Fix
    $\delta\in\R_+$ and let $\phi$ be an entropy function. Then,
    note that we
    can separate the set of probability distributions
    $\calP_\phi({\rho},\delta)$ along its fibers as follows
    \begin{align*}
        \calP_\phi({\rho},\delta) &= \left\{P\in{\calP(\calZ)}:
        \E_P[m(Z)] \in\calM(\rho), \:\:D_\phi(P\,\|\,\Pbase{})\leq \delta\right\} \\
        &=\bigcup_{\om\in\calM(\rho)}\left\{P\in{\calP(\calZ)}: \E_P[m(Z)]=\om,
        \:\:D_\phi(P\,\|\,\Pbase{})\leq \delta\right\}.
    \end{align*}
    The inner problem in \eqref{eq: general optimization problem} can then be
    rewritten as
    \begin{align}\label{aux eq: rewrite primal}
        \sup_{P\in\calP_\phi({\rho},\delta)}
        \E_P[g(Z)]=\sup_{\om\in\calM(\rho)}
        \sup_{\substack{P\in\calP(\calZ):\ \E_P[m(Z)]=\om,\\
        D_\phi(P\,\|\,\Pbase{})\leq\delta}}
        \E_P[g(Z)] = \sup_{\om\in\calM(\rho)}-h(1,\delta,\om),
    \end{align}
    where 
    $h$ is the value function \eqref{aux eq: function domain}.

    \uline{\textbf{Convexity of the value function.}} Consider the function
    on $\R^{\dimm+2}\times\calS$, where $\calS$ denotes the vector space of
    signed measures on $\calZ$ with finite $m$-moment, obtained from
    \eqref{aux eq: function domain} by adding to the objective
    $-\int g\,\de\nu$ the
    convex indicators of the constraints, namely those of the sets
    $\{(\nu,x):\int m\,\de\nu=x\}$, $\{(\nu,a):\int\de\nu=a\}$,
    $\{(\nu,b):D_\phi(\nu\,\|\,\Pbase{})\leq b\}$, and $\calM_{+,m}(\calZ)$. This
    function is jointly convex in $(a,b,x,\nu)$: the objective is linear in
    $\nu$; the first two indicator sets are linear subspaces; the third is
    convex because $\nu\mapsto D_\phi(\nu\,\|\,\Pbase{})$ is convex on
    $\calM_+(\calZ)$ -- for any two measures, fix a common dominating measure
    and use the joint convexity of the perspective $\phi^\pi$ (see
    Theorem~7.5(b) in \citet{polyanskiy2025InformationTheoryCoding} for
    the
    case of probability measures) -- and $\calM_{+,m}(\calZ)$ is a convex
    cone. As $h$ is the infimum of this function over $\nu\in\calS$,
    Theorem~1 in \citet{rockafellar1974ConjugateDualityOptimization} shows that $h$ is
    jointly convex in its three arguments. Its biconjugate $h^{**}$ is
    therefore convex
    and closed and satisfies $h\geq
    h^{**}$. We first work
    out the convex conjugate of $h$, $h^*$, and then the biconjugate $h^{**}$.

    \uline{\textbf{The convex conjugate.}} The convex conjugate of $h$ can be
    obtained as
    \begin{align*}
        h^*(-\zeta,-\eta,-\lambda)
        &= \sup_{(a,b,x)\in\R^{\dimm+2}}\left\{-\lambda^\top x - \zeta a
        -\eta b - h(a,b,x)\right\} \\
        &=
        \sup_{\substack{(\nu,x,a,b)\in\\\calM_{+,m}(\calZ)\times\R^{\dimm+2}}}
        \bigg\{-\lambda^\top x
        - \zeta a -\eta b + \int g(z)\de \nu(z)\:: \\
        &\hphantom{={}}\qquad \int m(z)\de \nu(z) =
        x,\: \int \de \nu(z) = a,\: D_\phi(\nu\,\|\,\Pbase{})\leq b\bigg\} \\
        &=
        \sup_{\substack{(\nu,b)\in\\\calM_{+,m}(\calZ)\times\R}}\left\{\int(g(z)
        -\lambda^\top m(z) - \zeta )\de \nu(z) -\eta b \::\:
        D_\phi(\nu\,\|\,\Pbase{})\leq b\right\}.
    \end{align*}
    First note that $D_\phi(\nu\,\|\,\Pbase{})\geq 0$ for every
    $\nu\in\calM_{+,m}(\calZ)$, as the integrand $\phi^\pi$ in
    Definition~\ref{defi: generalized phi divergence} is nonnegative
    because $\phi\geq0$. It must thus be that $b\geq 0$.
    If $\eta<0$, the supremum is $+\infty$, as letting $b\to+\infty$ at
    $\nu=\Pbase{}$, for which $D_\phi(\Pbase{}\,\|\,\Pbase{})=0\leq b$ and the
    integrals of $g$ and $m$ are finite, sends the objective to $+\infty$. If
    $\eta\geq0$, the objective is nonincreasing in $b$, so that, for each $\nu$ with
    $D_\phi(\nu\,\|\,\Pbase{})<\infty$, it suffices to take
    $b=D_\phi(\nu\,\|\,\Pbase{})$.
    By plugging the constraint into the objective, the optimization problem
    becomes
    \begin{align}
        \sup_{\nu\in\calM_{+,m}(\calZ)}&\int(g(z) -\lambda^\top m(z) - \zeta
        )\de \nu(z) -\eta D_\phi(\nu\,\|\,\Pbase{}) & \nonumber\\
        &= \sup_{\nu\in\calM_{+,m}(\calZ)}\int\left\{(g(z) -\lambda^\top
        m(z) - \zeta )\frac{\de \nu}{\de \varrho}(z) -\eta \phi^\pi\left(\frac{\de
        \nu}{\de \varrho}(z), \frac{\de \Pbase{}}{\de \varrho}(z)\right)\right\}\de
        \varrho(z)
        \label{aux eq: opt problem for each b case 1}
    \end{align}
    where, for a given $\nu$, $\varrho\in\calM_{+,m}(\calZ)$ is any
    measure with $\nu\ll\varrho$ and $\Pbase{}\ll\varrho$, and the equality
    exploits the definition of (generalized) $\phi$-divergence and of perspective
    function (see Definition \ref{defi: perspective function} and
    Definition \ref{defi: generalized phi divergence}). The supremum may be
    restricted to the $\nu$ with $D_\phi(\nu\,\|\,\Pbase{})<\infty$, which we call
    feasible, since for every other $\nu$ the constraint
    $D_\phi(\nu\,\|\,\Pbase{})\leq b$ fails for all real $b$. For each feasible
    $\nu$, the function $g$ is $\nu$-integrable by assumption. In particular,
    $\Pbase{}$ is feasible, as $\Pbase{}\in\calM_{+,m}(\calZ)$ and
    $D_\phi(\Pbase{}\,\|\,\Pbase{})=0$, so that $g(Z)-\lambda^\top m(Z)-\zeta$
    is $\Pbase{}$-integrable.

    Every feasible $\nu$ is
    dominated by $\nu+\Pbase{}$, which lies in $\calM_{+,m}(\calZ)$, and the
    objective in \eqref{aux eq: opt problem for each b case 1} does not depend on
    the choice of $\varrho$ by
    Remark~\ref{remark: unique representation of generalized phi divergence}.
    Hence \eqref{aux eq: opt problem for each b case 1} is the supremum, over the
    $\varrho\in\calM_{+,m}(\calZ)$ with $\Pbase{}\ll\varrho$, of the supremum of
    the same objective over the feasible $\nu\ll\varrho$. Fix such a $\varrho$
    and let $\beta\in L_1(\varrho)$ be a version of $\de\Pbase{}/\de\varrho$.
    For $\alpha\in L_1(\varrho)$, define
    \[
    u_\alpha(z):=(g(z)-\lambda^\top m(z)-\zeta)\,\alpha(z)
    -\eta\,\phi^\pi\big(\alpha(z),\beta(z)\big)
    \quad\text{and}\quad
    I(\alpha):=\int u_\alpha(z)\,\de\varrho(z),
    \]
    where $I(\alpha)$ is assigned the value $-\infty$ whenever the negative part
    of $u_\alpha$ fails to be $\varrho$-integrable. We claim that the supremum
    over the feasible $\nu\ll\varrho$ equals
    $\sup_{\alpha\in L_1(\varrho)}I(\alpha)$. One inequality is immediate, as the
    density $\de\nu/\de\varrho$ of a feasible $\nu\ll\varrho$ lies in
    $L_1(\varrho)$ and $I(\de\nu/\de\varrho)$ equals the objective at $\nu$.
    The reverse inequality is not immediate, because for
    $\alpha\in L_1(\varrho)$ the measure $\alpha\varrho$ need not have
    finite moment $\int\|m\|\alpha \de\varrho$ and finite divergence from $\Pbase{}$. 
    Fix $\alpha\in L_1(\varrho)$ with $I(\alpha)>-\infty$.
    Then $u_\alpha>-\infty$ holds $\varrho$-almost everywhere, and since
    $(g-\lambda^\top m-\zeta)\,\alpha$ is finite $\varrho$-almost everywhere,
    so is $\phi^\pi(\alpha,\beta)$. As $\phi(s)=+\infty$ for $s<0$,
    Definition~\ref{defi: perspective function} gives $\phi^\pi(s,t)=+\infty$
    whenever $s<0$ and $t\geq0$, so that $\alpha\geq0$ holds $\varrho$-almost
    everywhere. For $k\in\N$, let
    \[
    A_k:=\big\{z\in\calZ:\ \|m(z)\|\leq k,\
    \phi^\pi\big(\alpha(z),\beta(z)\big)\leq k\big\}
    \quad\text{and}\quad
    \alpha_k:=\alpha\,\indic_{A_k}+\beta\,\indic_{\calZ\setminus A_k}.
    \]
    The measure $\nu_k:=\alpha_k\varrho$ is nonnegative and finite. It has
    finite moment, as
    \[
    \int\|m(z)\|\,\de\nu_k(z)\leq k\int\alpha(z)\,\de\varrho(z)
    +\E_{\Pbase{}}[\|m(Z)\|]<\infty,
    \]
    and it has finite divergence, as
    \[
    D_\phi(\nu_k\,\|\,\Pbase{})=\int_{A_k}\phi^\pi\big(\alpha(z),\beta(z)\big)\,
    \de\varrho(z)\leq k\,\varrho(\calZ)<\infty,
    \]
    where the equality holds because $\phi^\pi(t,t)=0$ for every $t\geq0$, by
    $\phi(1)=0$ when $t>0$ and by $\phi^\infty(0)=0$ when $t=0$. Hence $\nu_k$
    is feasible, and, by the same property and $\beta\,\de\varrho=\de\Pbase{}$,
    its objective value is
    \[
    I(\alpha_k)=\int_{A_k}u_\alpha(z)\,\de\varrho(z)
    +\E_{\Pbase{}}\big[(g(Z)-\lambda^\top m(Z)-\zeta)\,
    \indic\{Z\notin A_k\}\big].
    \]
    The sets $A_k$ increase to $\{z\in\calZ:\phi^\pi(\alpha(z),\beta(z))<\infty\}$,
    which has full $\varrho$-measure, and hence full $\Pbase{}$-measure as
    $\Pbase{}\ll\varrho$. The second term thus vanishes as $k\to\infty$ by
    dominated convergence, since $g(Z)-\lambda^\top m(Z)-\zeta$ is
    $\Pbase{}$-integrable, while monotone convergence applied separately to
    $u_\alpha^+$ and $u_\alpha^-$ gives
    $\int_{A_k}u_\alpha\,\de\varrho\to I(\alpha)$. Therefore $I(\alpha_k)\to I(\alpha)$,
    which proves the claim.
    Consequently, for each such $\varrho$ and $\beta$, the supremum of the
    objective over the feasible $\nu\ll\varrho$ equals the left-hand side of
    \begin{align}
        \sup_{\alpha\in L_1(\varrho)}&\int\left\{(g(z) -\lambda^\top m(z)
        - \zeta )\alpha(z) -\eta \phi^\pi\left(\alpha(z),
        \beta(z)\right)\right\}\de
        \varrho(z) \nonumber \\
        &=\int\sup_{\alpha\in\R}\left\{(g(z) -\lambda^\top m(z) - \zeta
        )\alpha -\eta \phi^\pi\left(\alpha, \beta(z)\right)\right\}\de \varrho(z),
        \label{aux eq: opt problem for each b 2 case 1}
    \end{align}
    where the equality is the maximization form of Theorem~14.60 in
    \citet{rockafellar2009VariationalAnalysis}, which applies because for each
    $z$ the integrand
    is upper semicontinuous in $\alpha$ (the affine term is continuous and
    $-\eta\,\phi^\pi(\cdot,\beta(z))$ is upper semicontinuous), so its negative
    is a normal
    integrand on the decomposable space $L_1(\varrho)$. Then, recalling that the
    integral runs over $\calZ$, it is possible to
    split $\calZ$ into the set where $\beta$ vanishes and the set where it
    is positive. Formally:
    \[\calZ_+(\beta):=\{z\in\calZ:\beta(z)>0\}\quad\text{and}\quad\calZ_0(\beta):=\{z\in%
    \calZ:\beta(z)=0\}.
    \]
    Using Lemma \ref{lemma: lemma 4.11 kuhn}, \eqref{aux eq: opt problem for
    each b 2 case
    1} can be rewritten as
    \begin{align}\label{eq: average + supremum}
        \int_{\calZ_+(\beta)}\eta\phi^*&\left(\frac{g(z) -\lambda^\top m(z)
        -\zeta}{\eta}\right)\beta(z)\de\varrho(z) + \nonumber\\&\qquad
        \int_{\calZ_0(\beta)}\eta \iota_{\cl(\dom(\phi^*))}\left(\frac{g(z)
        -\lambda^\top m(z) -\zeta}{\eta}\right)\de\varrho(z).
    \end{align}
    The first integral is an expectation over $\Pbase{}$ because $\Pbase{}(\calZ_+(\beta))
    =
    \int_{\calZ_+(\beta)}\de \Pbase{}(z) = \int_{\calZ_+(\beta)}\beta(z)\de\varrho(z) =
    1$, where
    the last equality follows from the definition of the partition
    $\calZ=\calZ_+(\beta)
    \cup \calZ_0(\beta)$.

    Putting everything together, the conjugate of $h$ can be rewritten as
    \begin{align}
        h^*(-\zeta,-\eta,-\lambda) = &\E_{\Pbase{}}\bigg[\eta\phi^*\bigg(\frac{
        g(Z) -\lambda^\top m(Z) -\zeta}{\eta}\bigg)\bigg] \nonumber \\
        &+\sup_{\substack{\varrho \in \calM_{+,m}(\calZ):\, \Pbase{}\ll\varrho, \\
        \beta=\de \Pbase{}/\de
        \varrho\:\text{a.s.}}}\bigg\{\int_{\calZ_0(\beta)}\eta
        \iota_{\cl(\dom(\phi^*))}\left(\frac{g(z) -\lambda^\top m(z)
        -\zeta}{\eta}\right)\de\varrho(z)\bigg\}.
        \label{aux eq: opt problem for each b 3 case 1}
    \end{align}
    If there exists $z_0\in\calZ$ such that $\frac{g(z_0) -\lambda^\top
    m(z_0) -\zeta}{\eta}\not\in\cl(\dom(\phi^*)),$ then the objective evaluates
    to
    $+\infty$. To see this, there are two possibilities:
    \begin{itemize}
        \item if $z_0$ is an atom of $\Pbase{}$, then the first expectation
        evaluates to $+\infty$, since $\phi^*=+\infty$ strictly outside
        $\cl(\dom(\phi^*))$ and the integrand is bounded below by the
        $\Pbase{}$-integrable function $g(Z)-\lambda^\top m(Z)-\zeta$;
        \item if $z_0$ is not an atom of $\Pbase{}$, then one can always choose $\varrho
        = \Pbase{} +
        \delta^{\mathtt{Dirac}}_{z_0}$ and set $\beta(z)=\indic\{z\neq z_0\}$ so that
        the second
        integral picks up the Dirac mass at $z_0$, where the integrand evaluates
        to
        $+\infty$ by definition of the indicator function
        $\iota_{\cl(\dom(\phi^*))}$.
    \end{itemize}
    Lemma \ref{lemma: lemma 4.12 kuhn} shows that $\cl(\dom(\phi^*)) = (-\infty,
    \phi^\infty(1)]$ if $\phi^\infty(1)<\infty$ and $\cl(\dom(\phi^*))=\R$
    otherwise. It
    follows that $\frac{g(z) -\lambda^\top m(z)
    -\zeta}{\eta}\in\cl(\dom(\phi^*))$ is equivalent to
    \begin{align}
        \zeta + \eta \phi^\infty(1) \geq \sup_{z\in\calZ}\{ g(z) - \lambda^\top
        m(z)\}.
        \label{aux eq: non-infinite objective case 1}
    \end{align}
    Whenever the above condition is satisfied, the second integral in \eqref{aux
    eq: opt
    problem for each b 3 case 1} is identically 0 by definition of the indicator
    function
    $\iota_{\cl(\dom(\phi^*))}$. Thus, the convex conjugate of $h$ is
    \begin{align*}
        h^*(-\zeta,-\eta,-\lambda) & = 
        \begin{cases}
            \E_{\Pbase{}}\bigg[\eta\phi^*\bigg(\frac{g(Z) -\lambda^\top m(Z)
            -\zeta}{\eta}\bigg)\bigg] \quad &\text{if \eqref{aux eq:
            non-infinite
            objective case 1} holds},\\
            +\infty\quad &\text{otherwise.} 
        \end{cases}
    \end{align*}
    \uline{\textbf{Extension to $\eta=0$.}} Recall that we derived the above
    expression under
    the assumption that $\eta>0$. We now
    claim that
    \begin{align*}
        h^*(-\zeta,-\eta,-\lambda) & = 
        \begin{cases}
            \E_{\Pbase{}}\Big[(\phi^*)^\pi(g(Z) -\lambda^\top m(Z) -\zeta,
            \eta)\Big] \quad &\text{if $\eta\geq 0$ and \eqref{aux eq:
            non-infinite
            objective case 1} holds},\\
            +\infty\quad &\text{otherwise,} 
        \end{cases}
    \end{align*}
    holds. To see why, note that $h^*$ is convex and closed
    \citep[Theorem~12.2]{rockafellar1970ConvexAnalysis}. Furthermore,
    Lemma \ref{lemma: entropy
    function
    proper cvx lsc} gives that $\phi^*$ and $(\phi^*)^\pi$ are both convex,
    proper, and
    lower semicontinuous. By Lemma \ref{lemma: perspective larger than
    argument},
    $(\phi^*)^\pi(z,\eta)\geq z$ for every $\eta\geq0$. Along any convergent
    sequence of
    multipliers $(\zeta_n,\eta_n,\lambda_n)\to(\zeta,\eta,\lambda)$ the
    $(\zeta_n,\lambda_n)$ are bounded, so $g(z)-\lambda_n^\top
    m(z)-\zeta_n\geq g(z)-C(1+\|m(z)\|)$ for a finite constant $C$
    independent of $n$, and this lower bound is $\Pbase{}$-integrable because
    $\E_{\Pbase{}}[g(Z)]$ and $\E_{\Pbase{}}[\|m(Z)\|]$ are finite (as
    $\Pbase{}\in\calM_{+,m}(\calZ)$ with $D_\phi(\Pbase{}\,\|\,\Pbase{})=0$). Hence the
    integrands $(\phi^*)^\pi(g(Z)-\lambda_n^\top m(Z)-\zeta_n,\eta_n)$ are bounded
    below by this single integrable minorant, and the lower semicontinuity of
    $(\phi^*)^\pi$ together with Fatou's lemma gives that
    $(\zeta,\eta,\lambda)\mapsto\E_{\Pbase{}}[(\phi^*)^\pi( g(Z) -\lambda^\top
    m(Z) -\zeta, \eta)]$ is lower semicontinuous, while joint convexity of
    $(\phi^*)^\pi$ makes it convex, hence it is closed and convex.

    Denote by $F(\zeta,\eta,\lambda)$ the right-hand side of the claimed
    display. We have just shown that its expectation component is closed
    and convex. The constraint \eqref{aux eq: non-infinite objective case
    1} defines, in turn, a closed convex subset of
    $\R\times\R_+\times\R^{\dimm}$, because
    $\lambda\mapsto\sup_{z\in\calZ}\{g(z)-\lambda^\top m(z)\}$ is a
    supremum of affine functions of $\lambda$, hence convex and lower
    semicontinuous, while $(\zeta,\eta)$ enter the constraint linearly.
    When $\phi^\infty(1)=+\infty$, the convention of the statement renders
    the constraint vacuous and the set is the entire space. The convention
    is essential here: under the alternative reading $0\cdot(+\infty)=0$,
    the constraint would be vacuous for $\eta>0$ yet active on the slice
    $\{\eta=0\}$, and the resulting set would fail to be closed. As the
    sum of a closed convex function and the indicator of a closed convex
    set, $F$ is closed and convex, and it coincides with $h^*$ on
    $\{\eta>0\}$ by the previous step.

    Agreement on $\{\eta>0\}$ alone does not determine the values of two
    closed convex functions on the boundary slice $\{\eta=0\}$, so we
    further verify that their domains share a relative interior point with
    $\eta>0$. Both functions are nonincreasing in $\eta$: for $h^*$, this
    follows from its representation as a supremum over pairs $(\nu,b)$
    with $b\geq D_\phi(\nu\,\|\,\Pbase{})\geq0$ obtained at the beginning of the
    previous step, whose objective is nonincreasing in $\eta$ pointwise in
    $(\nu,b)$. For $F$, it follows from the identity
    $(\phi^*)^\pi(x,\eta)=\sup_{u\in\dom(\phi)}\{ux-\eta\phi(u)\}$ for
    $\eta\geq0$ together with $\phi\geq0$, and from the fact that the
    left-hand side of \eqref{aux eq: non-infinite objective case 1} is
    nondecreasing in $\eta$, so that feasibility at $(\zeta,0,\lambda)$
    implies feasibility at $(\zeta,\eta,\lambda)$ for every $\eta>0$.
    Neither function takes the value $-\infty$: $F$ is bounded below by
    the integrable minorant of the previous paragraph, and
    $h^*(-\zeta,-\eta,-\lambda)\geq\E_{\Pbase{}}[g(Z)-\lambda^\top
    m(Z)]-\zeta>-\infty$, as follows from evaluating the supremum defining
    it at $(\nu,b)=(\Pbase{},0)$. Consequently, finiteness of either function
    at a point of the slice propagates to every point vertically above it.

    If $h^*=+\infty$ throughout $\{\eta>0\}$, then so is $F$, and the
    propagation property forces both functions to equal $+\infty$ on the
    slice as well, so that they agree everywhere. Otherwise, the set
    $D:=\dom(h^*)\cap\{\eta>0\}=\dom(F)\cap\{\eta>0\}$ is nonempty and
    convex, and the propagation property yields
    $D\subseteq\dom(h^*)\subseteq\cl(D)$, and likewise for $F$. Since a
    convex set and its closure have the same relative interior
    \citep[Theorem~6.3]{rockafellar1970ConvexAnalysis}, it follows
    that $\rint(\dom(h^*))=\rint(\dom(F))=\rint(D)\subseteq\{\eta>0\}$.
    Fix any $y\in\rint(D)$ and any point $x$ on the slice $\{\eta=0\}$.
    For a closed proper convex function $f$, the value at $x$ is recovered
    as the limit along the segment joining $x$ to any point of
    $\rint(\dom(f))$, that is,
    $f(x)=\lim_{t\to0}f\big((1-t)x+ty\big)$
    \citep[Corollary~7.5.1]{rockafellar1970ConvexAnalysis}. Every point
    of the open
    segment has a strictly positive $\eta$-coordinate, where $h^*$ and $F$
    coincide, so the two limits coincide, and hence so do the values at
    $x$. This proves the claim.

    \uline{\textbf{The biconjugate and the bound.}} Using the definition, the
    biconjugate
    of $h$ is then immediate to construct:
    \begin{align}
        h^{**}(a,b,x)
        &=
        \sup_{\substack{(\zeta,\eta,\lambda)\in\\\R\times\R_+\times\R^{\dimm}}}
        -\lambda^\top x - \zeta a
        - \eta b - h^*(-\zeta,-\eta,-\lambda)\nonumber \\
        &= \begin{cases}
            \sup\limits_{\substack{(\zeta,\eta,\lambda)\in\\\R\times\R_+\times\R^{\dimm}}}
            -\lambda^\top x -
            \zeta a - \eta b - \E_{\Pbase{}}\Big[(\phi^*)^\pi(g(Z)
            -\lambda^\top m(Z) -\zeta, \eta)\Big] \\
            \text{s.t. }\quad \zeta + \eta \phi^\infty(1) \geq
            \sup\limits_{z\in\calZ}\{g(z) - \lambda^\top m(z)\}.
        \end{cases}
        \label{eq: bi-conjugate linear loss}
    \end{align}
    Let $\calF$ denote the set of $(\zeta,\eta,\lambda)$ satisfying
    $\eta\geq0$ and the pointwise constraint in
    \eqref{eq: bi-conjugate linear loss}.
    Therefore, by \eqref{aux eq: rewrite primal} and the
    biconjugate inequality $h^{**}\leq h$,
    \begin{align*}
    \sup_{P\in\calP_\phi({\rho},\delta)}\E_P[g(Z)]
    &\leq
    \sup_{\om\in\calM(\rho)}
    \inf_{(\zeta,\eta,\lambda)\in\calF}
    \left\{\lambda^\top\om+\zeta+\eta\delta+
    \E_{\Pbase{}}\!\left[
    (\phi^*)^\pi(g(Z)-\lambda^\top m(Z)-\zeta,\eta)
    \right]\right\}\\
    &\leq
    \inf_{(\zeta,\eta,\lambda)\in\calF}
    \left\{\iota_{\calM(\rho)}^*(\lambda)+\zeta+\eta\delta+
    \E_{\Pbase{}}\!\left[
    (\phi^*)^\pi(g(Z)-\lambda^\top m(Z)-\zeta,\eta)
    \right]\right\},
    \end{align*}
    where the second inequality is the minimax inequality. The support function
    identity established in Section~\ref{subsec: setup} restricts $\lambda$ to $\Rlam$ and
    replaces $\iota_{\calM(\rho)}^*(\lambda)$ with
    $\lambda_{12}^\top\rho$, which gives the stated weak dual bound.
\end{proof}

\subsection{Proof of Lemma~\ref{lemma: fenchel requirements}}
\begin{proof}\label{proof: fenchel requirements}
    \uline{\textbf{Relative-interior qualification.}}
    Write $c:=\E_{\Pbase{}}[m(Z)]$. Recall that $(a,b,x)\in\dom(h)$ exactly when
    some $\nu\in\calM_{+,m}(\calZ)$ has mass $\int\de\nu=a$, moment $\int
    m\,\de\nu=x$ and $D_\phi(\nu\,\|\,\Pbase{})\leq b$, in which case the value of
    \eqref{aux eq: function domain} is smaller than $+\infty$, since
    $-\int g(z)\,\de\nu(z)<\infty$ by hypothesis.

    \uline{\textbf{Uniform strict feasibility near $\om$.}} Since
    $c\in\Gamma(\delta)$,
    because $D_\phi(\Pbase{}\,\|\,\Pbase{})=0\leq\delta$, and
    $\om\in\rint(\Gamma(\delta))$,
    Theorem~6.4 in \citet{rockafellar1970ConvexAnalysis} yields a $\mu_0>1$ such
    that
    $c+\mu_0(\om-c)\in\Gamma(\delta)$. Fixing any $\mu\in(1,\mu_0)$, the point
    \[c+\mu(\om-c)=\big(1-\tfrac{\mu-1}{\mu_0-1}\big)\om+\tfrac{\mu-1}{\mu_0-1}\big(c
    +\mu_%
    0(\om-c)\big)\]
    is a convex combination of $\om\in\rint(\Gamma(\delta))$ and
    $c+\mu_0(\om-c)\in\Gamma(\delta)$, with weight
    $1-\tfrac{\mu-1}{\mu_0-1}>0$ on $\om$. Therefore,
    $c+\mu(\om-c)\in\rint(\Gamma(\delta))$ by Theorem~6.1 in
    \citet{rockafellar1970ConvexAnalysis}. By continuity of $y\mapsto c+\mu(y-c)$, there
    is $r>0$ such that the relative neighborhood
    $V:=\calB_r(\om)\cap\operatorname{aff}(\Gamma(\delta))$ of $\om$ satisfies
    $V\subseteq\Gamma(\delta)$ and
    $c+\mu(y-c)\in\Gamma(\delta)$
    for all $y\in V$. For $y\in V$, fix $P_y\in{\calP(\calZ)}$ so that
    $\E_{P_y}[m(Z)]=c+\mu(y-c)$ and $D_\phi(P_y\,\|\,\Pbase{})\leq\delta$. Set
    $Q_y:=(1-\tfrac1\mu)\Pbase{}+\tfrac1\mu P_y$. Then $\E_{Q_y}[m(Z)]=y$ and, by
    convexity of
    $D_\phi(\cdot\,\|\,\Pbase{})$ with $D_\phi(\Pbase{}\,\|\,\Pbase{})=0$, we get
    \[D_\phi(Q_y\,\|\,\Pbase{})\leq \tfrac1\mu\,D_\phi(P_y\,\|\,\Pbase{})\leq
    \tfrac\delta\mu<\delta .\]

    \uline{\textbf{A relative neighborhood in $\dom(h)$.}} Consider the affine set
    \[
    \mathbb{A}:=\{(a,b,x)\in\R^{\dimm+2}:
    x-(a-1)c\in\operatorname{aff}(\Gamma(\delta))\}.
    \]
    For a target $(a,b,x)\in\mathbb{A}$ in a small neighborhood of $(1,\delta,\om)$,
    let $y:=x-(a-1)c$, which lies in $\operatorname{aff}(\Gamma(\delta))$ and hence in
    $V$ for the neighborhood small enough. Writing
    the Lebesgue decomposition of $P_y$ with respect to $\Pbase{}$ as
    $P_y=w'_y \Pbase{}+P_y^\perp$, where $w'_y:=\de (P_y)_{\mathtt{ac}}/\de \Pbase{}$ is
    the density
    of its absolutely continuous part and $P_y^\perp\perp \Pbase{}$ its singular
    part, set
    \[\nu:=(a-1)\,\Pbase{}+Q_y=\Big(a-\tfrac1\mu+\tfrac1\mu w'_y\Big)\Pbase{}+\tfrac1\mu
    P_y^\perp\in\calM_{+,m}(\calZ).\]
    This $\nu$ is nonnegative if $a\geq\tfrac1\mu$, and has mass
    $\int\de\nu=a$ and moment $\int m\,\de\nu=(a-1)c+y=x$. Once the divergence
    bound below is established, the hypothesis also gives
    $\int g(z)\,\de\nu(z)>-\infty$. Set
    $w_a:=\tfrac{a-1/\mu}{1-1/\mu}$, so the absolutely continuous density
    of
    $\nu$ equals $(1-\tfrac1\mu)w_a+\tfrac1\mu w'_y$. By convexity of
    $\phi$,
    $\Pbase{}$-a.e.,
    \[\phi\Big((1-\tfrac1\mu)w_a+\tfrac1\mu
    w'_y\Big)\leq\big(1-\tfrac1\mu\big)\phi(w_a)+\tfrac1\mu\phi(w'_y),\]
    so, adding the singular contribution
    $\tfrac1\mu\phi^\infty(1)P_y^\perp(\calZ)$ and applying Lemma~\ref{lemma:
    divergence lebesgue decomposition} twice,
    \begin{align*}
    D_\phi(\nu\,\|\,\Pbase{})&\leq\big(1-\tfrac1\mu\big)\phi(w_a)
    +\tfrac1\mu\Big[\textstyle
    \int\phi(w'_y)\,\de
    \Pbase{}+\phi^\infty(1)P_y^\perp(\calZ)\Big]\\
    &=\big(1-\tfrac1\mu\big)\phi(w_a)+\tfrac1\mu
    D_\phi(P_y\,\|\,\Pbase{})\leq\big(1-\tfrac1\mu\big)\phi(w_a)+\tfrac\delta\mu,
    \end{align*}
    which holds uniformly in $y\in V$, the right-hand side depending on the
    target only
    through $a$. As $a\to1$ we have $w_a\to1$ and, since $\phi$ is
    continuous at $1$
    with $\phi(1)=0$, $\phi(w_a)\to0$. Therefore, the upper bound above
    tends to $\tfrac\delta\mu<\delta$,
    because $\mu>1$. Hence, for $(a,b)$ close enough to $(1,\delta)$,
    the upper bound is smaller than $b$, so that
    $D_\phi(\nu\,\|\,\Pbase{})\leq b$.
    This fact, together with $\int\de\nu=a$ and $\int m\,\de\nu=x$, gives
    $(a,b,x)\in\dom(h)$ throughout a relative neighborhood of
    $(1,\delta,\om)$ within $\mathbb{A}$.

    \uline{\textbf{The affine hull of $\dom(h)$.}} To conclude that
    $(1,\delta,\om)\in\rint(\dom(h))$, it remains to verify that
    $\dom(h)\subseteq\mathbb{A}$, so that the neighborhood just constructed is a
    neighborhood relative to $\operatorname{aff}(\dom(h))$. Take
    $(a,b,x)\in\dom(h)$ and let $\nu\in\calM_{+,m}(\calZ)$ have mass $a$,
    moment $x$, and $D_\phi(\nu\,\|\,\Pbase{})\leq b<\infty$. For $s\in(0,1)$ small
    enough so that $1-sa>0$, the probability measure $Q_s:=s\nu+(1-sa)\Pbase{}$ can
    be written as the convex combination $Q_s=s\,\nu+(1-s)\,t_s\Pbase{}$ with
    $t_s:=(1-sa)/(1-s)$, so that, by convexity of $D_\phi(\cdot\,\|\,\Pbase{})$,
    \[
    D_\phi(Q_s\,\|\,\Pbase{})\leq s\,D_\phi(\nu\,\|\,\Pbase{})+(1-s)\,\phi(t_s)\leq
    s\,b+(1-s)\,\phi(t_s),
    \]
    where $D_\phi(t_s\Pbase{}\,\|\,\Pbase{})=\phi(t_s)$ by
    Definition~\ref{defi: generalized
    phi divergence} with $\varrho=\Pbase{}$. Since $t_s\to1$ as $s\downarrow0$ and
    $\phi$ is continuous at $1$ with $\phi(1)=0$, the right-hand side falls
    below $\delta$ for $s$ small enough and
    $\E_{Q_s}[m(Z)]=s\,x+(1-sa)\,c\in\Gamma(\delta)$. Rearranging,
    $x-(a-1)c=c+s^{-1}\big(\E_{Q_s}[m(Z)]-c\big)$, and since
    $\E_{Q_s}[m(Z)]-c$ belongs to the subspace
    $\operatorname{aff}(\Gamma(\delta))-c$, which is closed under scaling, we
    conclude $x-(a-1)c\in\operatorname{aff}(\Gamma(\delta))$, i.e.\
    $(a,b,x)\in\mathbb{A}$. Hence $\dom(h)\subseteq\mathbb{A}$, while the neighborhood
    constructed above is open relative to $\mathbb{A}$, so
    $\operatorname{aff}(\dom(h))=\mathbb{A}$ and
    $(1,\delta,\om)\in\rint(\dom(h))$.

    \uline{\textbf{Properness.}}
    We argue by contradiction. Suppose that $h(a,b,x)=-\infty$ for some
    $(a,b,x)\in\R^{\dimm+2}$. Any $\nu\in\calM_{+,m}(\calZ)$ that is
    feasible for \eqref{aux eq: function domain} at $(a,b,x)$ satisfies
    $D_\phi(\nu\,\|\,\Pbase{})\leq b<\infty$, so that
    $\int|g|\,\de\nu<\infty$ by hypothesis and $\int g\,\de\nu$ is well
    defined and finite. Since $h(a,b,x)=-\infty$, the supremum of
    $\int g\,\de\nu$ over the feasible measures equals $+\infty$, so we may
    select a sequence $\{\nu_k\}_{k\geq1}$ of feasible measures with
    $\int g\,\de\nu_k\geq 4^k$ for every $k\geq1$. In particular,
    $\int[g]_+\,\de\nu_k\geq 4^k$, because $\int g\,\de\nu_k\leq
    \int[g]_+\,\de\nu_k$.

    Let $\nu^\dagger:=\sum_{k\geq1}2^{-k}\nu_k$, a nonnegative measure of
    mass $a$, where here and below every interchange of a sum and an integral follows
    from Tonelli's theorem. Write $\nu_k=w_k\Pbase{}+\nu_k^\perp$ for the Lebesgue
    decomposition of $\nu_k$ with respect to $\Pbase{}$, as in
    Lemma~\ref{lemma: divergence lebesgue decomposition}, where $\nu_k^\perp$ is
    concentrated on a $\Pbase{}$-null set $N_k$. The function
    $w^\dagger:=\sum_{k\geq1}2^{-k}w_k$ is finite $\Pbase{}$-almost everywhere, as
    $\int w^\dagger\,\de\Pbase{}\leq\sum_{k\geq1}2^{-k}\nu_k(\calZ)=a$, and
    $\nu^{\dagger\perp}:=\sum_{k\geq1}2^{-k}\nu_k^\perp$ is concentrated on the
    $\Pbase{}$-null set $\cup_{k\geq1}N_k$. By uniqueness of the Lebesgue
    decomposition, $\nu^\dagger=w^\dagger\Pbase{}+\nu^{\dagger\perp}$ is thus the
    Lebesgue decomposition of $\nu^\dagger$ with respect to $\Pbase{}$, and
    Lemma~\ref{lemma: divergence lebesgue decomposition} gives
    \[D_\phi(\nu^\dagger\,\|\,\Pbase{})=\int_\calZ\phi(w^\dagger(z))\,\de
    \Pbase{}(z)+\phi^\infty(1)\,\nu^{\dagger\perp}(\calZ).\]
    We bound the two terms in turn. As the entropy function $\phi$ is convex,
    lower semicontinuous, and proper, it is the pointwise supremum of the affine
    functions it
    majorizes \citep[Theorem~12.1]{rockafellar1970ConvexAnalysis}. For
    any affine $\ell(t)=\alpha t+\beta$ with $\ell\leq\phi$ and any
    $z\in\calZ$ with $w^\dagger(z)<\infty$,
    \[\ell(w^\dagger(z))
    =\alpha\sum_{k\geq1}2^{-k}w_k(z)+\beta\sum_{k\geq1}2^{-k}
    =\sum_{k\geq1}2^{-k}\ell(w_k(z))
    \leq\sum_{k\geq1}2^{-k}\phi(w_k(z)),\]
    where the first equality uses $\sum_{k\geq1}2^{-k}=1$ and the
    convergence of $\sum_{k\geq1}2^{-k}w_k(z)$. Taking the supremum over
    such $\ell$ yields 
    $\phi(w^\dagger)\leq\sum_{k\geq1}2^{-k}\phi(w_k)$ so that
    \[\int_\calZ\phi(w^\dagger(z))\,\de\Pbase{}(z)
    \leq\sum_{k\geq1}2^{-k}\int_\calZ\phi(w_k(z))\,\de\Pbase{}(z).\]
    For the singular term,
    $\nu^{\dagger\perp}(\calZ)=\sum_{k\geq1}2^{-k}\nu_k^\perp(\calZ)$, so
    that
    $\phi^\infty(1)\,\nu^{\dagger\perp}(\calZ)
    =\sum_{k\geq1}2^{-k}\phi^\infty(1)\,\nu_k^\perp(\calZ)$
    under the convention that the product evaluates to $0$ when the
    singular mass vanishes. Combining the two bounds and applying
    Lemma~\ref{lemma: divergence lebesgue decomposition} to each $\nu_k$, gives
    \[D_\phi(\nu^\dagger\,\|\,\Pbase{})
    \leq\sum_{k\geq1}2^{-k}\Big[\int_\calZ\phi(w_k(z))\,\de\Pbase{}(z)
    +\phi^\infty(1)\,\nu_k^\perp(\calZ)\Big]
    =\sum_{k\geq1}2^{-k}D_\phi(\nu_k\,\|\,\Pbase{})\leq b.\]
    Finally, we get
    \[\int_\calZ[g(z)]_+\,\de\nu^\dagger(z)
    =\sum_{k\geq1}2^{-k}\int_\calZ[g(z)]_+\,\de\nu_k(z)
    \geq\sum_{k\geq1}2^{-k}4^k=+\infty.\]
    The measure $\nu^\dagger$ thus belongs to $\calM_+(\calZ)$ and
    satisfies $D_\phi(\nu^\dagger\,\|\,\Pbase{})\leq b<\infty$, yet
    $\int_{\calZ}|g|\,\de\nu^\dagger\geq\int_{\calZ}[g]_+\,\de\nu^\dagger=+\infty$,
    contradicting the hypothesis.
    {Therefore, $h(a,b,x)>-\infty$ for every
    $(a,b,x)\in\R^{\dimm+2}$. Moreover,
    $\Pbase{}$ is feasible in the definition of
    $h(1,0,\E_{\Pbase{}}[m(Z)])$, and
    $\E_{\Pbase{}}[|g(Z)|]<\infty$ by hypothesis. Hence this value of $h$
    is finite, so $h$ is proper.}
\end{proof}

\subsection{Proof of Theorem~\ref{thm: strong duality}}
\begin{proof}\label{proof: strong duality}
Let $K_\rho$ be the set defined before
Lemma~\ref{lemma: fenchel requirements}, and let $q:=\iota_{K_\rho}$. The set $K_\rho$ is
nonempty and polyhedral,
so $q$ is a proper polyhedral convex function. By the fiber decomposition
\eqref{aux eq: rewrite primal},
\[
-\sup_{P\in\calP_\phi({\rho},\delta)}\E_P[g(Z)]
=
\inf_{y\in\R^{\dimm+2}}\{h(y)+q(y)\}.
\]
The function $h$ is proper and convex, where we established convexity in the proof
of Theorem~\ref{theorem: weak duality}, and {properness follows from
Lemma~\ref{lemma: fenchel requirements}.}

Choose
$\om_0\in\calM(\rho)\cap\rint(\Gamma(\delta))$. By
{Lemma~\ref{lemma: fenchel requirements}},
$(1,\delta,\om_0)\in\rint(\dom(h))$, while
$(1,\delta,\om_0)\in\dom(q)=K_\rho$. Lemma~\ref{lemma: fenchel duality
polyhedral} therefore gives
\[
\inf_{y\in\R^{\dimm+2}}\{h(y)+q(y)\}
=
\sup_{y^*\in\R^{\dimm+2}}\{-h^*(y^*)-q^*(-y^*)\}.
\]

Write $y^*=(-\zeta,-\eta,-\lambda)$. The conjugate of $q$ is
\begin{align*}
q^*(\zeta,\eta,\lambda)
&=\sup_{(1,\delta,\om)\in K_\rho}
\{\zeta+\eta\delta+\lambda^\top\om\}\\
&=\zeta+\eta\delta+\iota_{\calM(\rho)}^*(\lambda).
\end{align*}
The conjugate calculation in the proof of
Theorem~\ref{theorem: weak duality} gives
\[
h^*(-\zeta,-\eta,-\lambda)
=
\E_{\Pbase{}}\!\left[
(\phi^*)^\pi\big(g(Z)-\lambda^\top m(Z)-\zeta,\eta\big)
\right]
\]
when $\eta\geq0$ and
\[
\zeta+\eta\phi^\infty(1)
\geq\sup_{z\in\calZ}\{g(z)-\lambda^\top m(z)\},
\]
and gives $+\infty$ otherwise. Negating the Fenchel identity and using
\[
\iota_{\calM(\rho)}^*(\lambda)
=
\begin{cases}
\lambda_{12}^\top\rho,&\lambda\in\Rlam,\\
+\infty,&\lambda\notin\Rlam,
\end{cases}
\]
yields exactly the dual program in the statement.
\end{proof}

\subsection{Proof of Lemma~\ref{lemma: mixture density bounds}}
\begin{proof}\label{proof: mixture density bounds}
We prove each part separately.

\uline{\textbf{Part (i): baseline as the mixture.}}
The implication $(1)\Rightarrow(2)$ is immediate. For $(2)\Rightarrow(1)$,
define
\[
R(A):=\frac{\Pbase{}(A)-\kappa P(A)}{1-\kappa}.
\]
Then $R$ is a probability measure and $\Pbase{}=\kappa P+(1-\kappa)R$.

Next, $(2)\Rightarrow(3)$: if $\Pbase{}(A)=0$, then $\kappa P(A)\leq \Pbase{}(A)=0$, hence
$P(A)=0$, so $P\ll \Pbase{}$. Let $u:=\de P/\de \Pbase{}$. Then for every
$A\in\mathscr{F}$,
\[
\int_{A} \kappa u\,\de \Pbase{}=\kappa P(A)\leq \Pbase{}(A)=\int_{A} 1\,\de \Pbase{},
\]
so $\kappa u\leq 1$ $\Pbase{}$-a.s., that is $u\leq 1/\kappa$.

Finally, $(3)\Rightarrow(2)$ follows by integrating the bound $0\leq \de P/\de
\Pbase{}\leq 1/\kappa$.
Indeed, for every $A\in\mathscr{F}$,
\[
\kappa P(A)=\int_{A} \kappa \frac{\de P}{\de \Pbase{}}\,\de \Pbase{}\leq \int_{A}\de
\Pbase{}=
\Pbase{}(A),
\]
which is precisely $(2)$.

\uline{\textbf{Part (ii): baseline as a mixing component.}}
The implication $(1)\Rightarrow(2)$ is immediate. For $(2)\Rightarrow(1)$,
define
\[
R(A):=\frac{P(A)-\kappa \Pbase{}(A)}{1-\kappa}.
\]
Then $R$ is a probability measure and $P=\kappa \Pbase{}+(1-\kappa)R$.

Next, $(2)\Rightarrow(3)$. Consider the Lebesgue decomposition of $P$ with
respect to $\Pbase{}$,
\[
P=u\Pbase{}+P^\perp,
\qquad
P^\perp\perp \Pbase{},
\]
where $u:=\de P_{\mathtt{ac}}/\de \Pbase{}$ is the Radon--Nikodym derivative of the
absolutely
continuous part $P_{\mathtt{ac}}\ll \Pbase{}$ of $P$ with respect to $\Pbase{}$, and
$P^\perp$ is its
singular part.
Choose a measurable set $N$ such that $\Pbase{}(N)=0$ and $P^\perp(N^c)=0$. 
Note that such $N$ exists by the Lebesgue decomposition. Define
\[
B:=\{z\in\calZ: u(z)<\kappa\}.
\]
Then
\[
P(B\cap N^c)
=
\int_{B\cap N^c} u\, \de \Pbase{}
<
\kappa \Pbase{}(B\cap N^c),
\]
unless $\Pbase{}(B\cap N^c)=0$. But by $(2)$ we must have
\[
P(B\cap N^c)\geq \kappa \Pbase{}(B\cap N^c).
\]
Hence $\Pbase{}(B\cap N^c)=0$, and since $\Pbase{}(N)=0$, it follows that $\Pbase{}(B)=0.$
Therefore, $u\geq \kappa$ $\Pbase{}$-a.s.

Finally, $(3)\Rightarrow(2)$ follows by integrating:
\[
P(A)
=
\int_A u\, \de \Pbase{} + P^\perp(A)
\geq
\int_A \kappa\, \de \Pbase{}
=
\kappa \Pbase{}(A).
\]

When $P\ll\Pbase{}$, we have $P^\perp=0$ and $u=\de P/\de\Pbase{}$, which
gives the final claim of part~(ii).
\end{proof}

\subsection{Proof of Proposition~\ref{prop: induced entropy}}
\begin{proof}\label{proof: induced entropy}
We prove each part separately, writing throughout the Lebesgue decomposition
$P=w\Pbase{}+P^\perp$
of a candidate $P\in\Delta(\calZ)$ with respect to $\Pbase{}$, with $P^\perp\perp
\Pbase{}$ and
$w:=\de P_{\mathtt{ac}}/\de \Pbase{}$, where $w$ is the Radon--Nikodym derivative of
the absolutely
continuous part $P_{\mathtt{ac}}\ll \Pbase{}$ of $P$ with respect to $\Pbase{}$.

\uline{\textbf{Part (i): baseline as the mixture.}} The function
$\overline{\phi}=\phi+\iota_{[0,1/\kappa]}$ is convex and lower semicontinuous
as the sum of
a convex lower semicontinuous function and the indicator of the closed convex
set
$[0,1/\kappa]$. Moreover, $\overline{\phi}(1)=\phi(1)+\iota_{[0,1/\kappa]}(1)=0$
because
$1\in[0,1/\kappa]$ (as $\kappa<1$), and $\overline{\phi}(s)=+\infty$ for
$s<0$. Hence
$\overline{\phi}$ is an entropy function in the sense of
Definition~\ref{defi: entropy function}. Its effective domain
$\dom(\overline{\phi})=\dom(\phi)\cap[0,1/\kappa]$ is bounded, so
$\overline{\phi}^\infty(1)=+\infty$. By Lemma~\ref{lemma: divergence lebesgue
decomposition}, applied to the entropy function $\overline{\phi}$,
\[
D_{\overline{\phi}}(P\,\|\,\Pbase{})=\int\overline{\phi}(w)\,\de
\Pbase{}+\overline{\phi}^\infty(1)\,P^\perp(\calZ).
\]
By Lemma~\ref{lemma: mixture density bounds-i}, the mixture restriction
\eqref{eq: P mix baseline mixture} is equivalent to $P\ll \Pbase{}$ and $0\leq w\leq
1/\kappa$
$\Pbase{}$-a.s. If it holds, then $P^\perp(\calZ)=0$ and $\iota_{[0,1/\kappa]}(w)=0$
$\Pbase{}$-a.s.,
whence
$D_{\overline{\phi}}(P\,\|\,\Pbase{})=\int\phi(w)\,\de \Pbase{}=D_\phi(P\,\|\,\Pbase{})$.
If it fails, then
either $P^\perp(\calZ)>0$, so the second term is $+\infty$ since
$\overline{\phi}^\infty(1)=+\infty$, or $w>1/\kappa$ on a set of positive
$\Pbase{}$-measure, so
$\iota_{[0,1/\kappa]}(w)=+\infty$ there and the first term is $+\infty$. Either
way,
$D_{\overline{\phi}}(P\,\|\,\Pbase{})=+\infty$. As the mixture
restriction is equivalent to $\kappa P\leq\Pbase{}$ by the same lemma, this proves the
stated identity for $\overline{\phi}$. Therefore
$D_{\overline{\phi}}(P\,\|\,\Pbase{})\leq\delta$ holds
if and only if the mixture restriction holds and $D_\phi(P\,\|\,\Pbase{})\leq\delta$,
which by
\eqref{eq: P mix baseline mixture} is exactly membership in
{$\calP^{\ll}_{\mathsf{mix}}(\rho,\delta)$}. This proves
{$\calP^{\ll}_{\mathsf{mix}}(\rho,\delta)
=\calP_{\overline{\phi}}({\rho},\delta)$}.

\uline{\textbf{Part (ii): baseline as a mixing component.}} The function
$\underline{\phi}=\phi+\iota_{[\kappa,\infty)}$ is convex and lower
semicontinuous as the
sum of a convex lower semicontinuous function and the indicator of the closed
convex set
$[\kappa,\infty)$. Moreover,
$\underline{\phi}(1)=\phi(1)+\iota_{[\kappa,\infty)}(1)=0$
because $1\in[\kappa,\infty)$ (as $\kappa<1$), and $\underline{\phi}(s)=+\infty$
for $s<0$.
Hence $\underline{\phi}$ is an entropy function in the sense of
Definition~\ref{defi: entropy function}. Since
$\dom(\underline{\phi})=\dom(\phi)\cap[\kappa,\infty)$ coincides with
$\dom(\phi)$ for large
arguments, the truncation leaves the upper-tail recession unchanged, so
$\underline{\phi}^\infty(1)=\phi^\infty(1)$. By
Lemma~\ref{lemma: divergence lebesgue decomposition}, applied to the entropy
function $\underline{\phi}$, together with
$\underline{\phi}^\infty(1)=\phi^\infty(1)$,
\[
D_{\underline{\phi}}(P\,\|\,\Pbase{})
=\int\underline{\phi}(w)\,\de \Pbase{}+\underline{\phi}^\infty(1)\,P^\perp(\calZ)
=
\begin{cases}
D_\phi(P\,\|\,\Pbase{}), & w\geq\kappa\ \ \Pbase{}\text{-a.s.},\\
+\infty, & \text{otherwise,}
\end{cases}
\]
since when $w\geq\kappa$ $\Pbase{}$-a.s.\ we have $\iota_{[\kappa,\infty)}(w)=0$
$\Pbase{}$-a.s., so
$\int\underline{\phi}(w)\,\de \Pbase{}=\int\phi(w)\,\de \Pbase{}$ and the singular term
equals
$\phi^\infty(1)\,P^\perp(\calZ)$, whereas otherwise
$\iota_{[\kappa,\infty)}(w)=+\infty$ on
a set of positive $\Pbase{}$-measure. By Lemma~\ref{lemma: mixture density
bounds-ii}, the
condition $w\geq\kappa$ $\Pbase{}$-a.s.\ is exactly the mixture restriction
\eqref{eq: P mix baseline component}, as well as the dominance
$\kappa\Pbase{}\leq P$, which proves the stated identity for $\underline{\phi}$.
Hence $D_{\underline{\phi}}(P\,\|\,\Pbase{})\leq\delta$
holds if and
only if the mixture restriction holds and $D_\phi(P\,\|\,\Pbase{})\leq\delta$, which is
exactly
membership in {$\calP_{\mathsf{mix}}({\rho},\delta)$}. This proves
{$\calP_{\mathsf{mix}}({\rho},\delta)
=\calP_{\underline{\phi}}({\rho},\delta)$}.
\end{proof}

\subsection{Proof of Theorem~\ref{theorem: strong duality with mixture
constraint}}
\begin{proof}\label{proof: strong duality with mixture constraint}
By Proposition~\ref{prop: induced entropy}, each mixture-constrained
robustness set is an ordinary robustness set generated by an induced entropy.
Both parts therefore follow from Theorem~\ref{thm: strong duality} once its
hypotheses are verified for the corresponding entropy.

\uline{\textbf{Part (i): baseline as the mixture.}}
Proposition~\ref{prop: induced entropy-i} gives
\[
\calP^{\ll}_{\mathsf{mix}}(\rho,\delta)
=
\calP_{\overline{\phi}}({\rho},\delta),
\qquad
\overline{\phi}
=
\phi+\iota_{[0,1/\kappa]}.
\]
The induced entropy is continuous at $1$ because it coincides with $\phi$ on
a neighborhood of $1$, and
$\overline{\phi}^\infty(1)=+\infty$. Moreover, every
$\nu\in\calM_+(\calZ)$ with
$D_{\overline{\phi}}(\nu\,\|\,\Pbase{})<\infty$ satisfies
$\nu\ll\Pbase{}$ and
$\de\nu/\de\Pbase{}\leq1/\kappa$ $\Pbase{}$-almost surely. Hence
\[
\int_{\calZ}|g(z)|\,\de\nu(z)
\leq
\frac{1}{\kappa}\E_{\Pbase{}}[|g(Z)|]
<\infty,
\]
where the last inequality holds as $\Pbase{}\in\calP(\calZ)$.
The compatibility condition is imposed in the statement, so
Theorem~\ref{thm: strong duality} applies with
$\overline{\phi}$ in place of $\phi$. Its pointwise constraint is vacuous because the
recession is infinite, and the resulting dual is the one displayed in
part~\textup{(i)}.

\uline{\textbf{Part (ii): baseline as a mixing component.}}
Proposition~\ref{prop: induced entropy-ii} gives
\[
\calP_{\mathsf{mix}}({\rho},\delta)
=
\calP_{\underline{\phi}}({\rho},\delta),
\qquad
\underline{\phi}
=
\phi+\iota_{[\kappa,\infty)}.
\]
Again, the induced entropy is continuous at $1$ because it coincides with
$\phi$ on a neighborhood of $1$, while
$\underline{\phi}^\infty(1)=\phi^\infty(1)$. The integrability and
compatibility conditions required by Theorem~\ref{thm: strong duality} are
imposed directly in the statement. Applying that theorem with
$\underline{\phi}$ in place of $\phi$ therefore gives the dual in part~\textup{(ii)}.
\end{proof}

\subsection{Proofs of Theorem~\ref{thm: ignoring dual constraint} and
Proposition~\ref{cor: ac unconstrained dual}}
\begin{proof}\label{proof: ignoring dual constraint}
We first establish a preliminary fact about the deletion of the constraint, then prove the
proposition, and finally prove the theorem. Throughout, we write
$y_\lambda(z):=g(z)-\lambda^\top m(z)$,
$s(\lambda):=\sup_{z\in\calZ}y_\lambda(z)$, and
$e(\lambda):=\Pbase{}\text{-}\esssup_{z\in\calZ}\,y_\lambda(z)$.
As $\Pbase{}\in\calP(\calZ)$,
$y_\lambda\in L_1(\Pbase{})$, so $e(\lambda)>-\infty$. Recall from
\eqref{eq: dual objective ac} the dual objective $f$ of
Theorem~\ref{theorem: weak duality}, whose expectation term has the integrand
$(\phi^*)^\pi(y_\lambda(Z)-\zeta,\eta)$.
Lemma~\ref{lemma: perspective larger than argument} bounds the
integrand from below by $y_\lambda-\zeta$, so this expectation is well defined,
possibly with value $+\infty$.
Recall also from
\eqref{eq: dual feasible set ac} the dual feasibility set
$\calF=\{(\zeta,\eta,\lambda)\in\R\times\R_+\times\Rlam:\;
\zeta+\eta\,\phi^\infty(1)\geq s(\lambda)\}$, and define its
essential-supremum counterpart
\begin{equation}\label{eq: F esssup}
    \calF^{\ll} :=
    \Big\{(\zeta,\eta,\lambda)\in\R\times\R_+\times\Rlam:\;
    \zeta+\eta\,\phi^\infty(1) \geq e(\lambda)\Big\},
\end{equation}
noting that
$\calF\subseteq\calF^{\ll}\subseteq\R\times\R_+\times\Rlam$, as the
essential supremum never exceeds the supremum. Throughout this
proof the budget $\delta$ is held fixed, and we suppress it from the notation,
writing $f$ for the dual objective at this budget and
$V_{\mathsf{P}}$ for $V_{\mathsf{P}}(\delta)$, and likewise for the three remaining
values of \eqref{eq: primal values ac} and \eqref{eq: dual values ac}.
To prove the theorem we need to show
that $V_{\mathsf{P}}=V^{\ll}_{\mathsf{P}}$, whereas to prove
the proposition we need to show that
$V^{\ll}_{\mathsf{P}}=V^{\circ}_{\mathsf{D}}$.

\uline{\textbf{The essential-supremum constraint.}} We first show that
$f(\zeta,\eta,\lambda)=+\infty$ whenever
$(\zeta,\eta,\lambda)\in\R\times\R_+\times\Rlam$ lies outside
$\calF^{\ll}$. Consequently,
\begin{equation}\label{aux eq: Vcirc over Fll}
V^{\circ}_{\mathsf{D}}
=\inf_{(\zeta,\eta,\lambda)\in\calF^{\ll}}
f(\zeta,\eta,\lambda).
\end{equation}
In this sense, deleting the constraint \eqref{eq: pointwise dual constraint} amounts to
replacing the supremum $s(\lambda)$ in \eqref{eq: dual feasible set ac} with the
essential supremum $e(\lambda)$ in \eqref{eq: F esssup}. To accomplish this, we prove two
claims. Claim~1: if
$\phi^\infty(1)<\infty$, then $[1,+\infty)\subseteq\dom(\phi)$. Claim~2: the
recession function of the conjugate satisfies $(\phi^*)^\infty(x)=+\infty$
for every $x>0$. Claim~1 follows because, by
Definition~\ref{defi: recession function} with base point $z_0=1$,
the quantity
$\alpha^{-1}\{\phi(1+\alpha)-\phi(1)\}$ is nondecreasing in $\alpha>0$ with
limit $\phi^\infty(1)$, so that
$\phi(1+\alpha)\leq\alpha\,\phi^\infty(1)<\infty$ for every $\alpha>0$. 
For Claim~2, $\phi$ is closed, proper,
and convex, so $\phi^{**}=\phi$, and $\phi^*$ is closed, proper, and convex
by Lemma~\ref{lemma: entropy function proper cvx lsc}. Then
Theorem~13.3 in \citet{rockafellar1970ConvexAnalysis}, applied with
$f=\phi^*$, identifies $(\phi^*)^\infty$ with the support function of
$\dom(\phi^{**})=\dom(\phi)$, hence
$(\phi^*)^\infty(x)=\sup_{s\in\dom(\phi)}sx\geq\sup_{s\geq1}sx=+\infty$ for
every $x>0$, in view of Claim~1.

Fix
$(\zeta,\eta,\lambda)\in(\R\times\R_+\times\Rlam)\setminus\calF^{\ll}$, so that
$\zeta+\eta\,\phi^\infty(1)<e(\lambda)$. This forces
$\phi^\infty(1)<\infty$, because if $\phi^\infty(1)=+\infty$, the left-hand side
would equal $+\infty$ for $\eta>0$ and, under the convention stated
after \eqref{eq: pointwise dual constraint}, for $\eta=0$ as well, a contradiction.
Write $c:=\zeta+\eta\,\phi^\infty(1)$ and note that $\Pbase{}\{y_\lambda>c\}>0$, because 
if this probability were zero, $c$
would be an essential upper bound for $y_\lambda$ and $e(\lambda)\leq c$, which is again a
contradiction. Consider two cases. If $\eta>0$, then on the event
$\{y_\lambda>c\}$ we have $(y_\lambda-\zeta)/\eta>\phi^\infty(1)$, which
lies strictly above $\sup\dom(\phi^*)$ in view of 
{Corollary}~\ref{lemma: implicit regularization of entropy}, and so
$\phi^*((y_\lambda-\zeta)/\eta)=+\infty$ there and
$(\phi^*)^\pi(y_\lambda-\zeta,\eta)=\eta\,\phi^*((y_\lambda-\zeta)/\eta)
=+\infty$ on a set of positive $\Pbase{}$-probability. If $\eta=0$, the
constraint reads $\zeta<e(\lambda)$, so
$\Pbase{}\{y_\lambda-\zeta>0\}>0$ by the same argument, and on that event
$(\phi^*)^\pi(y_\lambda-\zeta,0)=(\phi^*)^\infty(y_\lambda-\zeta)=+\infty$
by the consequence recorded above. In either
case the integrand equals $+\infty$ on a set of positive
$\Pbase{}$-probability, so the expectation in \eqref{eq: dual objective ac}
equals $+\infty$. Hence
$f(\zeta,\eta,\lambda)=+\infty$. The infimum of $f$ over
$\R\times\R_+\times\Rlam$ therefore coincides with its infimum over
$\calF^{\ll}$, which proves \eqref{aux eq: Vcirc over Fll}.

\phantomsection\label{proof: ac unconstrained dual}%
\uline{\textbf{Proof of Proposition~\ref{cor: ac unconstrained dual}.}}
Define the convex cone
$\calM^{\ll}_{+,m}(\calZ):=\{\nu\in\calM_{+,m}(\calZ):\nu\ll\Pbase{}\}$
and the function
\[
h^{\ll}(a,b,x):=\inf_{\nu\in\calM^{\ll}_{+,m}(\calZ)}
\Big\{-\int g(z)\,\de\nu(z):\ \int m(z)\,\de\nu(z)=x,\ \int\de\nu(z)=a,\
D_\phi(\nu\,\|\,\Pbase{})\leq b\Big\},
\]
so that, by \eqref{aux eq: rewrite primal},
$V^{\ll}_{\mathsf{P}}
=\sup_{\om\in\calM(\rho)}-h^{\ll}(1,\delta,\om)$.
We compute its conjugate and then apply
Lemma~\ref{lemma: fenchel duality polyhedral}, as in
Theorem~\ref{thm: strong duality}.

\uline{\textbf{Convexity and the conjugate on $\{\eta>0\}$.}} Joint convexity of
$h^{\ll}$ follows exactly as in the proof of Theorem~\ref{theorem: weak
duality}. For the
conjugate, fix $\eta>0$. The computation leading to
\eqref{aux eq: opt problem for each b case 1} is unchanged, with the supremum
now running over the feasible $\nu\ll\Pbase{}$, that is, the
$\nu\in\calM^{\ll}_{+,m}(\calZ)$ with $D_\phi(\nu\,\|\,\Pbase{})<\infty$. We
may thus take $\varrho=\Pbase{}$, so that $\beta\equiv1$ holds
$\Pbase{}$-almost everywhere and $\phi^\pi(\alpha,\beta)=\phi(\alpha)$. By the
claim established in the proof of Theorem~\ref{theorem: weak duality}, the
supremum over the feasible $\nu\ll\Pbase{}$ then equals, writing
$y:=y_\lambda$,
\begin{align}
\sup_{\alpha\in L_1(\Pbase{})}\int\big[(y(z)-\zeta)\alpha(z)
-\eta\,\phi(\alpha(z))\big]\de\Pbase{}(z),
\label{aux eq: ac conjugate program}
\end{align}
where the integral is assigned the value $-\infty$ whenever the negative part
of its integrand fails to be integrable. In particular, neither the moment
restriction nor the finiteness of the divergence needs to be imposed on
$\alpha$. The
interchange of supremum and integral is
granted by Theorem~14.60 in \citet{rockafellar2009VariationalAnalysis}, 
and the definition of the conjugate,
$\sup_{u\in\R}\{tu-\eta\phi(u)\}=\eta\phi^*(t/\eta)$ for $t\in\R$ and $\eta>0$, yields
\[
h^{\ll,*}(-\zeta,-\eta,-\lambda)
=\E_{\Pbase{}}\Big[\eta\,\phi^*\Big(\frac{y(Z)-\zeta}{\eta}\Big)\Big],
\qquad\eta>0.
\]

\uline{\textbf{Extension to $\eta=0$.}} Let
$F^{\circ}(\zeta,\eta,\lambda):=
\E_{\Pbase{}}[(\phi^*)^\pi(y(Z)-\zeta,\eta)]$. The corresponding step of
the proof of Theorem~\ref{theorem: weak duality} shows that this
expectation is closed and convex in $(\zeta,\eta,\lambda)$,
while the conjugate $h^{\ll,*}$ is closed and convex. Both functions 
are nonincreasing in $\eta$: $h^{\ll,*}$
through its representation as a supremum over pairs $(\nu,b)$ with
$\nu\in\calM^{\ll}_{+,m}(\calZ)$ and $b\geq D_\phi(\nu\,\|\,\Pbase{})\geq0$,
whose objective is nonincreasing in $\eta$ pointwise in $(\nu,b)$, and
$F^{\circ}$ through the identity
$(\phi^*)^\pi(x,\eta)=\sup_{u\in\dom(\phi)}\{ux-\eta\,\phi(u)\}$ for
$\eta\geq0$ together with $\phi\geq0$. Neither takes the value $-\infty$:
$F^{\circ}\geq\E_{\Pbase{}}[y(Z)]-\zeta>-\infty$ by Lemma~\ref{lemma:
perspective larger than argument}, and
$h^{\ll,*}(-\zeta,-\eta,-\lambda)\geq\E_{\Pbase{}}[y(Z)]-\zeta>-\infty$ by
evaluating its defining supremum at $(\nu,b)=(\Pbase{},0)$, which is
admissible because $\Pbase{}\in\calM^{\ll}_{+,m}(\calZ)$. Since the two
functions agree on $\{\eta>0\}$ by the previous step, an argument identical
to the one in the proof of Theorem~\ref{theorem: weak duality} 
yields $h^{\ll,*}(-\zeta,-\eta,-\lambda)=F^{\circ}(\zeta,\eta,\lambda)$
for every $(\zeta,\eta,\lambda)\in\R\times\R_+\times\R^{\dimm}$.

\uline{\textbf{The domain of $h^{\ll}$ and the assembly.}} The proof of
Lemma~\ref{lemma: fenchel requirements} applies step by step upon
replacing $\Gamma(\delta)$ with $\Gamma^{\ll}(\delta)$ and choosing every
auxiliary measure to be absolutely continuous. Hence
$(1,\delta,\omega)\in\rint(\dom(h^{\ll}))$ for every
$\omega\in\rint(\Gamma^{\ll}(\delta))$ and $h^{\ll}$ is proper. The set
$K_\rho$ defined before Lemma~\ref{lemma: fenchel requirements} is nonempty and
polyhedral. By the compatibility condition,
there is an $\omega_0\in\calM(\rho)\cap\rint(\Gamma^{\ll}(\delta))$, and
therefore $K_\rho\cap\rint(\dom(h^{\ll}))\neq\emptyset$. Applying
Lemma~\ref{lemma: fenchel duality polyhedral} to
$h^{\ll}$ and $\iota_{K_\rho}$, and using the conjugate computed above together with
the support function of $\calM(\rho)$ recorded in
Section~\ref{subsec: setup} gives
\[
V^{\ll}_{\mathsf{P}}
=\inf_{(\zeta,\eta,\lambda)\in\R\times\R_+\times\Rlam}
f(\zeta,\eta,\lambda)
=V^{\circ}_{\mathsf{D}},
\]
which proves the proposition.

\uline{\textbf{Proof of Theorem~\ref{thm: ignoring dual constraint}.}} Suppose condition
(a) holds, that is, $\phi^\infty(1)=+\infty$. Then, by
Lemma~\ref{lemma: divergence lebesgue decomposition} and $\phi\geq0$, every $P$ with
$D_\phi(P\,\|\,\Pbase{})<\infty$ has a singular part of zero mass, that is,
$P\ll\Pbase{}$. Hence the two robustness sets coincide and
$V_{\mathsf{P}}=V^{\ll}_{\mathsf{P}}$ follows immediately. Suppose instead
that condition (b) holds. Then the equality
$e(\lambda)=s(\lambda)$ for every $\lambda\in\Rlam$, assumed in
condition (b),
makes the defining inequalities of \eqref{eq: dual feasible set ac} and
\eqref{eq: F esssup} identical, so $\calF=\calF^{\ll}$, and
\eqref{aux eq: Vcirc over Fll} gives
$V^{\circ}_{\mathsf{D}}
=\inf_{\calF^{\ll}}f
=\inf_{\calF}f=V_{\mathsf{D}}$. Because 
Proposition~\ref{cor: ac unconstrained dual} holds under the conditions
of the theorem,
we get $V^{\ll}_{\mathsf{P}}=V^{\circ}_{\mathsf{D}}$. Moreover,
$V_{\mathsf{P}}\leq V_{\mathsf{D}}$ by Theorem~\ref{theorem: weak
duality}, whose integrability hypothesis is implied by condition (b), as
$\calM_{+,m}(\calZ)\subseteq\calM_{+}(\calZ)$, while
$V^{\ll}_{\mathsf{P}}\leq V_{\mathsf{P}}$, as the supremum defining
$V_{\mathsf{P}}$ runs over a larger robustness set.
Using all these facts, we obtain
\[
V^{\ll}_{\mathsf{P}}\leq V_{\mathsf{P}}\leq V_{\mathsf{D}}
=V^{\circ}_{\mathsf{D}}=V^{\ll}_{\mathsf{P}},
\]
so that the four values coincide. In particular
$V_{\mathsf{P}}=V^{\ll}_{\mathsf{P}}$, which was to be shown.
\end{proof}

\subsection{Proof of Lemma~\ref{lemma: contamination as constrained tv}}
\begin{proof}\label{proof: contamination as constrained tv}
Recall that
$\calC_\kappa(\Pbase{})=\{P\in\Delta(\calZ):P=\kappa\Pbase{}+(1-\kappa)R,\ R\in\Delta(\calZ)\}$
and
$\calC^{\ll}_\kappa(\Pbase{})=\{P\in\Delta(\calZ):\Pbase{}=\kappa P+(1-\kappa)R,\ R\in\Delta(\calZ)\}$,
and that $\calC_{\kappa,\delta}(\Pbase{})$ and
$\calC^{\ll}_{\kappa,\delta}(\Pbase{})$ are defined in
\eqref{eq: two parameter contamination} by intersecting these two sets with
$\calT_{\delta}(\Pbase{})$. Fix $\kappa\in(0,1)$ and $\delta\in[0,1]$. The endpoints
$\kappa\in\{0,1\}$ are treated at the end.

\uline{\textbf{Step 1: domination constraints and uniqueness of the mixing
distribution.}}
By construction, $P\in\calC_\kappa(\Pbase{})$ if and only if there exists
$R\in\Delta(\calZ)$ with $P=\kappa\Pbase{}+(1-\kappa)R$,
which is condition~(1) of Lemma~\ref{lemma: mixture density bounds-ii}.
By the equivalence (1)$\Leftrightarrow$(2)
of that lemma this holds if and only if $P\geq\kappa\Pbase{}$ as measures, thus
\begin{equation}\label{aux eq: C kappa as domination}
\calC_\kappa(\Pbase{})=\bigl\{P\in\Delta(\calZ):\ P\geq\kappa\Pbase{}\bigr\}.
\end{equation}
Symmetrically, $P\in\calC^{\ll}_\kappa(\Pbase{})$ if and only if there exists
$R\in\Delta(\calZ)$ with $\Pbase{}=\kappa P+(1-\kappa)R$, which is
condition~(1) of Lemma~\ref{lemma: mixture density bounds-i}. By the equivalence
(1)$\Leftrightarrow$(2) of that lemma this holds
if and only if $\kappa P\leq\Pbase{}$ as measures, so that
\begin{equation}\label{aux eq: C ll kappa as domination}
\calC^{\ll}_\kappa(\Pbase{})=\bigl\{P\in\Delta(\calZ):\ \kappa P\leq\Pbase{}\bigr\}.
\end{equation}
Moreover, since $\kappa<1$, the two mixture equations can be solved for
the mixing distribution, giving $R=(P-\kappa\Pbase{})/(1-\kappa)$ and
$R=(\Pbase{}-\kappa P)/(1-\kappa)$, respectively, so that in both cases the mixing
distribution is uniquely determined by $P$ and $\Pbase{}$, as claimed in the statement.

\uline{\textbf{Step 2: the rescaling identities.}} Let
$P\in\calC_\kappa(\Pbase{})$ with mixing distribution $R$. Substituting the
mixture equation gives
$P(A)-\Pbase{}(A)=(1-\kappa)\bigl(R(A)-\Pbase{}(A)\bigr)$ for every
$A\in\mathscr{F}$, and taking absolute values and suprema over $A\in\mathscr{F}$ yields
$\TV(P\,\|\,\Pbase{})=(1-\kappa)\TV(R\,\|\,\Pbase{})$. Let instead
$P\in\calC^{\ll}_\kappa(\Pbase{})$ with mixing distribution $R$. The same
substitution gives
$\Pbase{}(A)-P(A)=(1-\kappa)\bigl(R(A)-P(A)\bigr)$ and
$\Pbase{}(A)-R(A)=\kappa\bigl(P(A)-R(A)\bigr)$ for every $A\in\mathscr{F}$, whence
$\TV(P\,\|\,\Pbase{})=(1-\kappa)\TV(P\,\|\,R)$ and
$\TV(R\,\|\,\Pbase{})=\kappa\,\TV(P\,\|\,R)$. Since the total variation distance
never exceeds one, these identities imply in particular that every
distribution in $\calC_\kappa(\Pbase{})$ or in $\calC^{\ll}_\kappa(\Pbase{})$
lies within total variation distance $1-\kappa$ of the baseline.

\uline{\textbf{Step 3: the two characterizations.}} Consider part (ii)
first. By definition, $\calC_{\kappa,\delta}(\Pbase{})$ collects the
distributions $P=\kappa\Pbase{}+(1-\kappa)R$ with $R\in\Delta(\calZ)$ and
$\TV(P\,\|\,\Pbase{})\leq\delta$. By Step~2, the latter constraint holds if and
only if $\TV(R\,\|\,\Pbase{})\leq\delta/(1-\kappa)$, which proves the first
equality in \eqref{eq:contam-as-tv}, while the second follows upon
intersecting both sides of \eqref{aux eq: C kappa as domination} with
$\calT_{\delta}(\Pbase{})$. The argument for part (i) is identical, with
\eqref{aux eq: C ll kappa as domination} in place of
\eqref{aux eq: C kappa as domination} and with the identity
$\TV(P\,\|\,\Pbase{})=(1-\kappa)\TV(P\,\|\,R)$ turning the constraint
$\TV(P\,\|\,\Pbase{})\leq\delta$ into $\TV(P\,\|\,R)\leq\delta/(1-\kappa)$, which
proves \eqref{eq:reverse-contam-as-tv}. Next, the containment
$\calC_{\kappa,\delta}(\Pbase{})\cup\calC^{\ll}_{\kappa,\delta}(\Pbase{})
\subseteq\calT_{\min\{\delta,1-\kappa\}}(\Pbase{})$ follows by combining
the constraint $\TV(P\,\|\,\Pbase{})\leq\delta$, valid on both sets by
definition, with the bound $\TV(P\,\|\,\Pbase{})\leq1-\kappa$ obtained at the end
of Step~2. Finally, if $\delta\geq1-\kappa$, the same bound gives
$\calC_\kappa(\Pbase{})\subseteq\calT_{1-\kappa}(\Pbase{})\subseteq
\calT_{\delta}(\Pbase{})$ and
$\calC^{\ll}_\kappa(\Pbase{})\subseteq\calT_{\delta}(\Pbase{})$, so
intersecting either contamination neighborhood with
$\calT_{\delta}(\Pbase{})$ leaves it unchanged, that is,
$\calC_{\kappa,\delta}(\Pbase{})=\calC_{\kappa}(\Pbase{})$ and
$\calC^{\ll}_{\kappa,\delta}(\Pbase{})=\calC^{\ll}_{\kappa}(\Pbase{})$.

\uline{\textbf{Endpoints.}} If $\kappa=0$, both mixture equations hold with
the unique choices $R=P$ and $R=\Pbase{}$, respectively, so
$\calC_0(\Pbase{})=\calC^{\ll}_0(\Pbase{})=\Delta(\calZ)$, matching the
domination sets $\{P\geq0\}$ and $\{0\leq\Pbase{}\}$, and the rescaling
identities hold by direct substitution. Steps~2 and~3 then apply verbatim,
and both characterizations reduce to
$\calC_{0,\delta}(\Pbase{})=\calC^{\ll}_{0,\delta}(\Pbase{})=\calT_{\delta}(\Pbase{})$.
If $\kappa=1$, both mixture equations force $P=\Pbase{}$, so
$\calC_1(\Pbase{})=\calC^{\ll}_1(\Pbase{})=\{\Pbase{}\}$. Since
$\TV(\Pbase{}\,\|\,\Pbase{})=0\leq\delta$, intersecting with
$\calT_{\delta}(\Pbase{})$ gives
$\calC_{1,\delta}(\Pbase{})=\calC^{\ll}_{1,\delta}(\Pbase{})=\{\Pbase{}\}$
for every $\delta\in[0,1]$.
\end{proof}

\subsection{Proof of Lemma~\ref{lemma: slater dual form}}
\begin{proof}\label{proof: slater dual form}
    We prove each part separately. Throughout, we fix $\theta\in\Theta$ and recall
    the image $\calN(\theta)$ of the moment function from
    Section~\ref{subsec:existence-dual}.
    For this proof it is convenient to permute the coordinates of $\calM(\rho)$ so that
    the $\mineq$ inequality constraints precede the $\meq$ equality
    constraints, where $\mineq:=\mathsf d_1+\mathsf d_3$ and
    $\meq:=\mathsf d_2+\mathsf d_4$, and to write $\lambda_{\leq}\in\R^{\mineq}$ for the
    inequality multipliers and $\lambda_{=}\in\R^{\meq}$ for the equality multipliers, so
    that
    $\lambda=(\lambda_{\leq},\lambda_{=})$. The mapping with the previous notation is thus
    $\lambda_{\leq}=(\lambda_1,\lambda_3)$ and $\lambda_{=}=(\lambda_2,\lambda_4)$, and
    the vector $\rhopad$ is
    permuted accordingly, so that its first $\mineq$ components are
    $(\rho_1^\top,0_{\mathsf{d}_3}^\top)^\top$ and its last $\meq$
    components are $(\rho_2^\top,0_{\mathsf{d}_4}^\top)^\top$. Finally,
    we write $\lambda_{12}=(\lambda_1^\top,\lambda_2^\top)^\top\in\R^{\dime}$ and use
    the notation $\xi=(\zeta,\eta,\lambda^\top)^\top$ to gather all multipliers together.

    \uline{\textbf{Part (i).}} Under
    Assumption~\ref{assump: dual primitives}, and under
    Assumption~\ref{assump: dual primitives superlinear} in case \textup{(a)} of
    Assumption~\ref{assump: dps-entropy}, the Slater condition of
    Assumption~\ref{assump: dual slater} or of Assumption~\ref{assump: dps-ceiling}
    places the origin in the interior of $\mathrm{co}(\calN(\theta))-\rhopad$. In case
    \textup{(b)}, every distribution $Q$ in the definition of $\calN_{\bar{x}}(\theta)$
    satisfies $\E_Q[\|m(Z;\theta)\|]\leq\bar{x}\,\E_{\Pbase{}}[\|m(Z;\theta)\|]<\infty$,
    as $\Pbase{}\in\calP(\calZ)$, so that the separation argument in the proof of
    Lemma~\ref{lemma: constraint qualification-i} gives
    $\calN_{\bar{x}}(\theta)\subseteq\cl(\mathrm{co}(\calN(\theta)))$. The Slater
    condition of Assumption~\ref{assump: dps-ceiling} then places the origin in the
    interior of $\cl(\mathrm{co}(\calN(\theta)))-\rhopad$, and hence in the interior
    of $\mathrm{co}(\calN(\theta))-\rhopad$, as a convex set whose closure has
    nonempty interior has the same interior as its closure
    \citep[Theorem~6.3]{rockafellar1970ConvexAnalysis}. In either case, the
    origin $0$ lies in the interior of
    $\mathrm{co}(\calN(\theta))-\rhopad=\mathrm{co}(\calN(\theta)-\rhopad)\subseteq\R^{\dimm}$,
    where the equality follows as the convex hull is translation invariant.
    By definition of the interior in $\R^{\dimm}$, there exists $r>0$
    such that $\calB_r(0)\subseteq\mathrm{co}(\calN(\theta)-\rhopad)$. Set
    $\varepsilon:=r/(2\sqrt{\dimm})$ and consider the point $v_+\in\R^{\dimm}$
    with components
    $v_{+,j}:=-\varepsilon$ for $j=1,\ldots,\mineq$ and $v_{+,j}:=+\varepsilon$ for
    $j=\mineq+1,\ldots,\dimm$. Since $\|v_+\|=\varepsilon\sqrt{\dimm}=r/2<r$, we
    have $v_+\in
    \calB_r(0)\subseteq\mathrm{co}(\calN(\theta)-\rhopad)$. By
    Carath\'eodory's
    theorem in $\R^{\dimm}$, there exist points $z_1,\ldots,z_{\dimm+1}\in\calZ$
    and weights
    $\omega_1,\ldots,\omega_{\dimm+1}\geq 0$ with
    $\sum_{i=1}^{\dimm+1}\omega_i=1$ such that
    \[\sum_{i=1}^{\dimm+1}\omega_i\bigl(m(z_i;\theta)-\rhopad\bigr)=v_+.\]
    Use these weights to construct the finitely supported distribution
    \[Q_+:=\sum_{i=1}^{\dimm+1}\omega_i\delta^{\mathtt{Dirac}}_{z_i}\in\Delta(\calZ).\]
    Then,
    $\E_{Q_+}[m_j(Z;\theta)]=\rhopad_j-\varepsilon<\rhopad_j$ for $j\leq\mineq$ and
    $\E_{Q_+}[m_j(Z;\theta)]=\rhopad_j+\varepsilon$ for $j>\mineq$, as required. The
    distribution
    $Q_-$ is constructed symmetrically by replacing $v_+$ with
    $v_-:=-\varepsilon\,1_{\dimm}\in\R^{\dimm}$.

    \uline{\textbf{Part (ii).}} Under
    Assumption~\ref{assump: dual primitives superlinear} the constraint is vacuous,
    hence $\calF(\theta)=\R\times\R_+\times\Rlam$ and
    there is nothing more to prove. Consider now
    Assumption~\ref{assump: dual primitives}, so that
    $\phi^\infty(1)\in(0,+\infty)$ by
    Assumption~\ref{assump: dp-entropy}. Set $\lambda^\circ:=0\in\Rlam$,
    $\eta^\circ:=1$, and $\zeta^\circ:=\bar{g}$, where
    $\sigma(0,\theta)=\bar{g}<\infty$ by Assumption~\ref{assump: dp-reg}.
    Then $\xi^\circ\in\R\times\R_{++}\times\Rlam$ and
    $\zeta^\circ+\eta^\circ\phi^\infty(1)=\bar{g}+\phi^\infty(1)>\bar{g}
    =\sigma(\lambda^\circ,\theta)$, where the inequality follows from
    $\phi^\infty(1)>0$, so $\xi^\circ$ is a strictly feasible point of $\calF(\theta)$,
    which was to be shown.

    \uline{\textbf{Part (iii).}} For a set $S$ and a level $\alpha\in\R$, we write
    $\lev_{\leq\alpha}(f,S):=\{x\in S:f(x)\leq\alpha\}$ for the sublevel set of
    $f$ relative to $S$. The argument rests on the following recession
    criterion for a closed proper convex function on $\R^n$: it has a nonempty bounded
    sublevel set if and only if its recession function is strictly positive at every
    nonzero direction, in which case all its nonempty sublevel sets are bounded. This
    follows from Theorems~8.4 and~8.7 and Corollary~8.7.1 in
    \citet{rockafellar1970ConvexAnalysis}, since all nonempty sublevel sets of such a
    function share the recession cone formed by the directions at which its recession
    function is nonpositive, a nonempty closed convex set is bounded if and only if its
    recession cone is $\{0\}$, and the recession function vanishes at the origin. We
    apply this criterion to the extension of $f(\cdot,\theta)$ by $+\infty$ outside
    $\calF(\theta)$, which we now show is closed, proper, and convex on
    $\R^{\dimm+2}$.

    \uline{\textbf{Properties of $f$.}} By Lemma
    \ref{lemma:
    entropy function proper cvx lsc}, $\phi^*$ and its perspective
    $(\phi^*)^\pi$ are both
    convex, proper, and lower semicontinuous. The term within the expectation
    admits an integrable lower bound that is uniform along any convergent
    sequence of multipliers $\xi_s\to\xi$. Indeed, under
    Assumption~\ref{assump: dual primitives},
    Lemma~\ref{lemma: perspective larger than argument} at $\check{x}=0$
    gives $(\phi^*)^\pi(x,\eta)\geq-\eta\,\phi(0)$ with $\phi(0)<\infty$ by
    Assumption~\ref{assump: dp-entropy}, whereas under
    Assumption~\ref{assump: dual primitives superlinear}, the same lemma at
    $\check{x}=1$ gives
    \[(\phi^*)^\pi(g(z;\theta)-\lambda^\top m(z;\theta)-\zeta,\eta)\geq
    g(z;\theta)-\lambda^\top m(z;\theta)-\zeta\geq
    -\big(|g(z;\theta)|+\sup_s\|\lambda_s\|\,\|m(z;\theta)\|+\sup_s|\zeta_s|\big),\]
    which is a $\Pbase{}$-integrable bound as
    $\Pbase{}\in\calP(\calZ)$. Therefore, the expectation
    $\E_{\Pbase{}}[(\phi^*)^\pi(g(Z;\theta)-\lambda^\top m(Z;\theta)-\zeta,\eta)]$ is
    convex, because integration preserves convexity, and lower semicontinuous.
    Adding the affine continuous term $\lambda_{12}^\top\rho+\zeta+\eta\delta$ preserves
    both properties, so $f(\cdot,\theta)$ is convex and lower
    semicontinuous on $\R^{\dimm+2}$, hence closed.

    \uline{\textbf{Properties of the extension of $f$.}} If
    $\phi^\infty(1)=+\infty$, then the dual constraint is vacuous for every $\eta\geq0$ --
    its
    left-hand side $\zeta+\eta\phi^\infty(1)$ equals $+\infty$, for $\eta>0$ because
    $\eta\,\phi^\infty(1)=+\infty$ and for $\eta=0$ under the convention
    $0\cdot\phi^\infty(1)=+\infty$ -- so that
    $\calF(\theta)$ coincides with the ambient cone $\Lbb$ of
    Section~\ref{subsec:existence-dual}, which after the permutation reads
    $\R\times\R_+\times\R^{\mineq}_+\times\R^{\meq}$ and is closed
    and convex. The closedness and convexity of the extension of
    $f(\cdot,\theta)$ by $+\infty$ outside $\calF(\theta)$ then follow
    directly from the lower semicontinuity and convexity of $f$
    established above. If instead
    $\phi^\infty(1)<\infty$, rewrite
    \[\calF(\theta)=\{\xi\in\R^{\dimm+2}:\lambda_{\leq}\geq 0,\ \eta\geq 0,\
    \sigma(\lambda,\theta)\leq
    \zeta+\eta\phi^\infty(1)\},\]
    with $\sigma(\lambda,\theta)$ the dual support function of
    Section~\ref{subsec:existence-dual},
    so that $\calF(\theta)$ is the intersection of the polyhedral cone
    $\{\lambda_{\leq}\geq
    0,\eta\geq 0\}$ with the set
    $\{\xi:\sigma(\lambda,\theta)\leq\zeta+\eta\phi^\infty(1)\}$, the
    preimage of the epigraph of $\sigma$ under the linear map
    $\xi\mapsto(\lambda,\zeta+\eta\phi^\infty(1))$. Note that since $\sigma$ is a supremum
    of affine functions in $\lambda$, it is convex and its epigraph is convex. Moreover,
    it is the pointwise supremum of continuous functions of $\lambda$, thus it is lower
    semicontinuous and its epigraph is closed. As the preimage of a closed convex set
    under a linear map,
    $\{\xi:\sigma(\lambda,\theta)\leq\zeta+\eta\phi^\infty(1)\}$ is
    closed and convex, and therefore $\calF(\theta)$ is the intersection of two closed
    convex sets, hence itself closed and convex. The extension is then the sum of $f$ and
    the indicator function of the closed convex set $\calF(\theta)$, so its closedness and
    convexity again follow from the corresponding properties of $f$ established above.

    \uline{\textbf{Properness of $f$.}} The function $f(\cdot,\theta)$ never
    takes the value $-\infty$ because at each $\xi$ the term within the expectation
    admits an integrable lower bound, as shown in the previous paragraph. Moreover, $f$ is
    not identically $+\infty$. Under
    Assumption~\ref{assump: dual primitives}, it is finite at the point
    $\xi_0=(\bar{g},1,0)$ with $\bar{g}=\sup_{z\in\calZ}g(z;\theta)<\infty$, where
    the finiteness follows by Assumption~\ref{assump: dp-reg} and which is feasible since
    $\bar{g}+\phi^\infty(1)>\bar{g}=\sigma(0,\theta)$ in view of
    $\phi^\infty(1)>0$. At this value the integrand equals $\phi^*(g(Z;\theta)-\bar{g})$,
    because the perspective at $\eta=1$ reduces to $\phi^*$ itself, and
    $g(z;\theta)-\bar{g}\leq 0$ on $\calZ$. Since $\phi^*$ is nondecreasing with
    $\phi^*(0)=-\inf\phi=0$ and $\phi^*(x)\geq x$ by
    Lemma~\ref{lemma: perspective larger than argument}, the integrand lies between
    $g(Z;\theta)-\bar{g}$ and $0$. Both bounds are $\Pbase{}$-integrable
    as $\Pbase{}\in\calP(\calZ)$, implying that
    $f(\xi_0,\theta)<\infty$. Under
    Assumption~\ref{assump: dual primitives superlinear}, it is finite at the point
    $\xi_0:=\xi^{*}$ of Assumption~\ref{assump: dps-anchor}, which is feasible since
    $\calF(\theta)$ is unrestricted. We conclude that $f$ is proper.

    \uline{\textbf{Weak duality with slack.}} We next record an inequality that is
    used in several steps below and in the proof of
    Lemma~\ref{lemma: outer value regularity}. Fix
    $\xi=(\zeta,\eta,\lambda)\in\calF(\theta)$ and $P\in\calP(\calZ)$ with
    $D_\phi(P\,\|\,\Pbase{})<\infty$. Write
    $w(z):=g(z;\theta)-\lambda^\top m(z;\theta)$ and let $P=r\Pbase{}+P_\perp$ be the
    Lebesgue decomposition of $P$ with respect to $\Pbase{}$, so that
    $D_\phi(P\,\|\,\Pbase{})=\E_{\Pbase{}}[\phi(r(Z))]+\phi^\infty(1)P_\perp(\calZ)$ by
    Lemma~\ref{lemma: divergence lebesgue decomposition}. In particular,
    $0\leq\E_{\Pbase{}}[\phi(r(Z))]\leq D_\phi(P\,\|\,\Pbase{})<\infty$, so that
    $r(z)\in\dom(\phi)$ for $\Pbase{}$-almost every $z\in\calZ$, and
    Lemma~\ref{lemma: perspective larger than argument} at $\check{x}=r(z)$ gives
    \[(\phi^*)^\pi(w(z)-\zeta,\eta)\geq (w(z)-\zeta)\,r(z)-\eta\,\phi(r(z))\]
    for $\Pbase{}$-almost every $z\in\calZ$. As $P\in\calP(\calZ)$, the function
    $(w-\zeta)r$ belongs to $L_1(\Pbase{})$, and so does $\phi(r)$, so integration
    yields
    \begin{align}\label{aux eq: decomposition 1}
        \E_{\Pbase{}}[(\phi^*)^\pi(w(Z)-\zeta,\eta)]\geq
        \int_\calZ (w(z)-\zeta)\,r(z)\,\de \Pbase{}(z) - \eta\,\E_{\Pbase{}}[\phi(r(Z))].
    \end{align}
    As for the singular part, if $\phi^\infty(1)=+\infty$, then
    $D_\phi(P\,\|\,\Pbase{})<\infty$ forces $P_\perp(\calZ)=0$, whereas if
    $\phi^\infty(1)<\infty$, the constraint defining $\calF(\theta)$ gives
    $w(z)-\zeta\leq\eta\,\phi^\infty(1)$ for every $z\in\calZ$. In either case
    \begin{align}\label{aux eq: decomposition 2}
        \int_\calZ (w(z)-\zeta)\,\de P_\perp(z) \leq
    \eta\,\phi^\infty(1)\,P_\perp(\calZ),
    \end{align}
    under the convention that the right-hand side evaluates to $0$ when
    $P_\perp(\calZ)=0$. It then follows that
    \begin{align}\label{eq: slater dual slack}
    \begin{split}
    f(\xi,\theta)
    &=\lambda_{12}^\top\rho+\zeta+\eta\delta
    +\E_{\Pbase{}}[(\phi^*)^\pi(w(Z)-\zeta,\eta)]\\
    &\geq\lambda_{12}^\top\rho+\zeta+\eta\delta+\E_{P}[w(Z)-\zeta]
    -\int_\calZ(w(z)-\zeta)\,\de P_\perp(z)-\eta\,\E_{\Pbase{}}[\phi(r(Z))]\\
    &\geq\lambda_{12}^\top\rho+\zeta+\eta\delta+\E_{P}[w(Z)-\zeta]
    -\eta\,D_\phi(P\,\|\,\Pbase{})\\
    &=\E_{P}[g(Z;\theta)]+\eta\bigl(\delta-D_\phi(P\,\|\,\Pbase{})\bigr)
    +\bigl(\lambda_{12}^\top\rho-\lambda^\top\E_{P}[m(Z;\theta)]\bigr),
    \end{split}
    \end{align}
    where the equality in the first line is the definition of $f$, the first inequality
    combines \eqref{aux eq: decomposition 1} with the decomposition
    $P=r\Pbase{}+P_\perp$, the second inequality follows from
    \eqref{aux eq: decomposition 2} and the decomposition of
    $D_\phi(P\,\|\,\Pbase{})$ displayed above, and the last equality holds because $P$
    is a probability measure. If, in addition, $P$ is primal feasible, that is,
    $D_\phi(P\,\|\,\Pbase{})\leq\delta$ and $\E_P[m(Z;\theta)]\in\calM(\rho)$, then both
    terms after $\E_P[g(Z;\theta)]$ in \eqref{eq: slater dual slack} are nonnegative,
    the second because $\lambda\in\Rlam$ and, by the support function of $\calM(\rho)$
    recorded in Section~\ref{subsec: setup}, $\lambda^\top x\leq\lambda_{12}^\top\rho$
    for every $x\in\calM(\rho)$. In this case $f(\xi,\theta)\geq\E_P[g(Z;\theta)]$.

    \uline{\textbf{Choice of the candidate sublevel set.}} By the recession criterion
    above, applied to the extension of $f(\cdot,\theta)$ by $+\infty$ outside
    $\calF(\theta)$, it suffices to exhibit one nonempty bounded sublevel set
    $\lev_{\leq\alpha}(f,\calF(\theta))$. Indeed, the recession function of the extension
    is then strictly positive at every $d\neq0$. At every
    $d\in\mathrm{rec}(\calF(\theta))$ it coincides with $f^\infty(d,\theta)$, because, by
    Definition~\ref{defi: recession function}, both recession functions can be computed
    along the ray $\xi_0+\tau d$, which stays in $\calF(\theta)$, where the two functions
    agree. This is the claim of part (iii). The rest of the proof does so
    using the
    distributions $Q_\pm$ from part (i), the distribution $P^\circ$ of
    Assumption~\ref{assump: dp-feas} or Assumption~\ref{assump: dps-feas}, and, when
    $\phi^\infty(1)=+\infty$, the full-support condition
    imposed in case (a) of Assumption~\ref{assump: dps-support}.

    To avoid clutter, we
    present the argument for $\meq=\mineq=1$. The result
    readily generalizes to higher (but finite) dimensions: for every unit vector
    $u=(u_{\leq},u_{=})$ with $u_{\leq}\geq0$, Carath\'eodory's theorem
    furnishes, exactly as in the proof of part~(\textit{i}), a distribution
    $Q_u\in\Delta(\calZ)$ with $\E_{Q_u}[m(Z;\theta)]=\rhopad-\varepsilon u$, hence with
    $\E_{Q_u}[u^\top m(Z;\theta)]=-\varepsilon$ after the recentering below, and
    the diverging-multiplier arguments then apply with $Q_u$, for $u$ the
    limiting direction of the multipliers, in place of $Q_+$ or $Q_-$. Without loss of
    generality, let
    $\rhopad=0$. Otherwise, redefine the moment function by setting $\tilde
    m_j(Z;\theta):=m_j(Z;\theta)-\rhopad_j$ for each $j=1,\ldots,\dimm$, so that the
    inequality
    constraints $\E_P[m_j(Z;\theta)]\leq \rhopad_j$ become
    $\E_P[\tilde m_j(Z;\theta)]\leq 0$
    and the equality constraints $\E_P[m_j(Z;\theta)]=\rhopad_j$ become $\E_P[\tilde
    m_j(Z;\theta)]=0$, and replace $m_j$ with $\tilde m_j$ throughout.
    Since $\lambda^\top\rhopad=\lambda_{12}^\top\rho$, the invertible
    affine change of variables $\tilde\zeta:=\zeta+\lambda_{12}^\top\rho$ maps the dual
    objective and the pointwise constraint onto those of the recentered problem, and it
    carries bounded sublevel sets onto bounded sublevel sets. As
    shown in the properness paragraph, $f(\cdot,\theta)$ is finite at
    the feasible point $\xi_0$ identified there so we may pick
    $\alpha^\star:=f(\xi_0,\theta)+1$
    which makes $\lev_{\leq\alpha^\star}(f,\calF(\theta))$ nonempty.

    \uline{\textbf{Case I: $\eta=0$.}} Define
    \[\Lbb_{|0} :=
    \Big\{(\zeta,\eta,\lambda_{\leq},\lambda_{=})\in\R\times\{0\}\times\R_+\times\R:\zeta
    \geq%
    \sup_{z\in\calZ}\{g(z;\theta)-\lambda^\top
    m(z;\theta)\}\Big\}.\]
    We now show that (\textit{a}) the portion of
    $\lev_{\leq\alpha^\star}(f,\calF(\theta))$ on the slice
    $\{\eta=0\}$ is contained in $\Lbb_{|0}$ and (\textit{b}) that
    $f(\xi,\theta)\geq\zeta$ on this slice.

    We start with (\textit{a}). Indeed, note that when $\phi^\infty(1)<\infty$,
    the slice of $\calF(\theta)$ is $\Lbb_{|0}$ itself. When instead
    $\phi^\infty(1)=+\infty$, the slice of $\calF(\theta)=\Lbb$ is
    unconstrained, but every point of it outside $\Lbb_{|0}$ has
    $f(\xi,\theta)=+\infty$ and so belongs to no sublevel set. To see
    this, consider any $\xi=(\zeta,0,\lambda)\in\Lbb$ with
    $\zeta<\sup_{z\in\calZ}\{g(z;\theta)-\lambda^\top m(z;\theta)\}$. The set
    $\{z\in\calZ:g(z;\theta)-\lambda^\top m(z;\theta)>\zeta\}$ is nonempty and open, by
    continuity of $g$ and $m$, and hence has positive $\Pbase{}$-measure, since under
    Assumption~\ref{assump: dps-support} every nonempty open subset of
    $\calZ=\supp(\Pbase{})$ has positive mass. At each such $z$ the term within the
    expectation in $f$ equals
    \[(\phi^*)^\pi(g(z;\theta)-\lambda^\top m(z;\theta)-\zeta,0)
    =(\phi^*)^\infty(g(z;\theta)-\lambda^\top m(z;\theta)-\zeta)=+\infty.\]
    To see this, first note that Definition~\ref{defi: entropy function} and
    the domain condition $[0,\infty)\subseteq \dom(\phi)$ 
    (imposed in case~\textup{(a)} of Assumption~\ref{assump: dps-entropy}, or implied by 
    Assumption~\ref{assump: dp-entropy}) imply that $\dom(\phi)=[0,\infty)$, and then
    \begin{align}\label{eq: recession function on slice eta=0}
    (\phi^*)^\infty(x)=\iota^*_{\dom(\phi)}(x)=\sup_{y\in\dom(\phi)}\{xy\}
    =\sup_{y\geq 0}\{xy\}=
    \begin{cases}
        0,&x\leq 0,\\
        +\infty,&x>0,
    \end{cases}
    \end{align}
    where the first equality follows from Theorem~13.3 in
    \citet{rockafellar1970ConvexAnalysis}. It follows that
    $(\phi^*)^\infty$ equals $+\infty$ whenever its argument 
    is positive. This happens whenever $g(z;\theta)-\lambda^\top m(z;\theta)>\zeta$, which
    is satisfied
    for the particular choice of $z$ above. Since the set of $z$ for which this happens
    has positive $\Pbase{}$-measure, the expectation in $f$ is $+\infty$ as claimed.
        
    We now turn to (\textit{b}). Since $(\phi^*)^\pi(x,0)=(\phi^*)^\infty(x)\geq 0$
    for every $x$ by \eqref{eq: recession function on slice eta=0}, the expectation in $f$
    is nonnegative on the slice $\{\eta=0\}$, hence
    $f(\xi,\theta)\geq\zeta$. The same conclusion holds under option (b)
    of Assumption~\ref{assump: dps-entropy}, where $\dom(\phi)=[0,\bar{x}]$ with
    $\bar{x}=\sup\dom(\phi)\in(1,\infty)$: there
    $(\phi^*)^\infty(x)=\sup_{y\in[0,\bar{x}]}xy=\bar{x}\,[x]_+\geq0$, so
    $f(\xi,\theta)\geq\zeta$ on the slice as well.

    Option (b) of Assumption~\ref{assump: dps-entropy} is treated at the
    end of this proof. The
    relaxed-set analysis that follows therefore concerns
    Assumption~\ref{assump: dual primitives} and option (a) of
    Assumption~\ref{assump: dual primitives superlinear} (and applies
    verbatim, with the $\lambda$-coordinates deleted, when those assumptions are
    invoked with $\dimm=0$).

    We now construct two relaxed constraint sets containing $\Lbb_{|0}$,
    whose intersection with the sublevel set we then show to be bounded. Because
    $\sup_{z\in\calZ}h(z)\geq\E_Q[h(Z)]$ for any $h$
    whenever $\supp(Q)\subseteq\calZ$, replacing the supremum in the
    constraint of $\Lbb_{|0}$ with an expectation under $Q_-$, respectively $Q_+$,
    can only enlarge the set. In particular, we use $\E_{Q_-}[m_2(Z;\theta)]=-\varepsilon$
    and $\E_{Q_+}[m_2(Z;\theta)]=+\varepsilon$ from Part~(i).
    This yields the following two sets
    \[\overline{\Lbb}_{|0,-} :=
    \{(\zeta,\eta,\lambda_{\leq},\lambda_{=})\in\R\times\{0\}\times\R_+\times\R:\zeta\geq
    \E_{Q_-}[g(Z;\theta)]-\lambda_{\leq} \E_{Q_-}[m_1(Z;\theta)] +
    \lambda_{=}\varepsilon\}\supseteq
    \Lbb_{|0}\]
    and
    \[\overline{\Lbb}_{|0,+} :=
    \{(\zeta,\eta,\lambda_{\leq},\lambda_{=})\in\R\times\{0\}\times\R_+\times\R:\zeta\geq
    \E_{Q_+}[g(Z;\theta)]-\lambda_{\leq} \E_{Q_+}[m_1(Z;\theta)] -
    \lambda_{=}\varepsilon\}\supseteq
    \Lbb_{|0}.\]
    We build two such sets because the equality multiplier $\lambda_{=}$ is
    unconstrained in sign: the term $+\lambda_{=}\varepsilon$ in
    $\overline{\Lbb}_{|0,-}$ makes its constraint fail as $\lambda_{=}\to+\infty$,
    while the term $-\lambda_{=}\varepsilon$ in $\overline{\Lbb}_{|0,+}$ makes its
    constraint fail as $\lambda_{=}\to-\infty$.
    The inequality multiplier $\lambda_{\leq}$ is nonnegative and is handled by
    either set, since $\E_{Q_\pm}[m_1(Z;\theta)]<0$ makes
    $-\lambda_{\leq}\E_{Q_\pm}[m_1(Z;\theta)]\to+\infty$ as $\lambda_{\leq}\to+\infty$.

    Consider any sequence with $\|(\zeta,\lambda_{\leq},\lambda_{=})\|\to+\infty$
    lying in
    $\lev_{\leq\alpha^\star}(f,\overline{\Lbb}_{|0,-}\cap\overline{\Lbb}_{|0,+})$. By
    definition of sublevel set, it satisfies
    $f(\xi,\theta)\leq\alpha^\star$ together with the defining
    constraints of $\overline{\Lbb}_{|0,-}$ and $\overline{\Lbb}_{|0,+}$. We show that
    each such sequence must eventually violate one of these requirements thus
    contradicting its membership in the sublevel set. Thus, no diverging sequence can lie
    in $\lev_{\leq\alpha^\star}(f,\overline{\Lbb}_{|0,-}\cap\overline{\Lbb}_{|0,+})$. This
    implies that the set is bounded, and, a fortiori, that
    $\lev_{\leq\alpha^\star}(f,\Lbb_{|0})$ is bounded as well.
    
    The different diverging sequences we must consider are:
    \begin{itemize}
        \item if $|\zeta|\to\infty$, then either $f$ diverges ($\zeta\to\infty$)
        or the
        constraint is violated ($\zeta\to-\infty$);
        \item if $\lambda_{\leq}\to\infty$, then the constraint is violated because
        $\E_{Q_-}[m_1(Z;\theta)]<0$ and $\zeta$ is bounded along this
        sequence;
        \item if $|\lambda_{=}|\to\infty$, then either the constraint in
        $\overline{\Lbb}_{|0,-}$ ($\lambda_{=}\to\infty$) or the one in
        $\overline{\Lbb}_{|0,+}$ ($\lambda_{=}\to-\infty$) is violated;
        \item if $|\zeta|\to\infty$ and $\lambda_{\leq}\to\infty$, then either $f$
        diverges
        ($\zeta\to\infty$) or the constraints in $\overline{\Lbb}_{|0,-}$ and
        $\overline{\Lbb}_{|0,+}$ are violated ($\zeta\to-\infty$);
        \item if $|\zeta|\to\infty$ and $|\lambda_{=}|\to\infty$, then when
        $\zeta\to\infty$
        the function diverges, whereas when $\zeta\to-\infty$, either the constraint in
        $\overline{\Lbb}_{|0,-}$ is violated ($\lambda_{=}\to\infty$) or the one
        in
        $\overline{\Lbb}_{|0,+}$ is violated ($\lambda_{=}\to-\infty$);
        \item if $\lambda_{\leq}\to\infty$ and $|\lambda_{=}|\to\infty$, then either
        the
        constraint in $\overline{\Lbb}_{|0,-}$ is violated
        ($\lambda_{=}\to\infty$) or the
        one in $\overline{\Lbb}_{|0,+}$ is violated ($\lambda_{=}\to-\infty$)
        because
        $\zeta$ is bounded along this sequence;
        \item if $|\zeta|\to\infty, \lambda_{\leq}\to\infty,$ and
        $|\lambda_{=}|\to\infty$, then
        whenever $\zeta\to\infty$ the function diverges, whereas if
        $\zeta\to-\infty$
        either the constraint in $\overline{\Lbb}_{|0,-}$ is violated
        ($\lambda_{=}\to\infty$) or the one in $\overline{\Lbb}_{|0,+}$ is
        violated
        ($\lambda_{=}\to-\infty$).
    \end{itemize}
    Therefore
    $\lev_{\leq\alpha^\star}(f,\overline{\Lbb}_{|0,-}\cap\overline{\Lbb}_{|0,+})$
    is
    bounded and so is $\lev_{\leq\alpha^\star}(f,\Lbb_{|0})$.

    \uline{\textbf{Case II: $\eta>0$.}} Denote by
    \[\Lbb_{>0} :=
    \Big\{(\zeta,\eta,\lambda_{\leq},\lambda_{=})\in\R\times\R_{++}\times\R_+\times\R:
    \zeta +
    \eta \phi^\infty(1) \geq \sup\limits_{z\in \calZ}\{ g(z;\theta) -
    \lambda^\top m(z;\theta)\}\Big\}\]
    the feasible set at $\eta>0$. On $\Lbb_{>0}$ the distributions $Q_+,Q_-$ from
    part~(i) induce the two relaxed inequalities
    \begin{align*}
        \zeta + \eta \phi^\infty(1) &\geq \E_{Q_-}[g(Z;\theta)]
        - \lambda_{\leq}\E_{Q_-}[m_1(Z;\theta)] + \lambda_{=}\varepsilon,\\
        \zeta + \eta \phi^\infty(1) &\geq \E_{Q_+}[g(Z;\theta)]
        - \lambda_{\leq}\E_{Q_+}[m_1(Z;\theta)] - \lambda_{=}\varepsilon.
    \end{align*}
    We first derive two lower bounds on $f$ over $\Lbb_{>0}$.
    
    \uline{\textbf{First bound.}} The first is obtained via the distribution $P^\circ$
    provided
    by Assumption~\ref{assump: dp-feas} or
    Assumption~\ref{assump: dps-feas}, which satisfies the moment restrictions
    together with $\delta_\circ:=D_\phi(P^\circ\,\|\,\Pbase{})<\delta$. The
    distribution $P^\circ$ belongs to $\calP(\calZ)$ by
    Assumption~\ref{assump: dp-feas}, respectively by
    Assumption~\ref{assump: dps-feas}. Fix $\xi\in\Lbb_{>0}\subseteq\calF(\theta)$.
    Inequality
    \eqref{eq: slater dual slack} at $P=P^\circ$, in the recentered problem, gives
    \[f(\xi,\theta)\geq\E_{P^\circ}[g(Z;\theta)]+\eta(\delta-\delta_\circ)
    -\lambda^\top\E_{P^\circ}[m(Z;\theta)]
    \geq\E_{P^\circ}[g(Z;\theta)]+\eta(\delta-\delta_\circ),\]
    where the second inequality holds because $\lambda^\top\E_{P^\circ}[m(Z;\theta)]$
    equals $\lambda_{\leq}\E_{P^\circ}[m_1(Z;\theta)]\leq 0$, in view of
    $\lambda_{\leq}\geq 0$, $\E_{P^\circ}[m_1(Z;\theta)]\leq 0$, and
    $\E_{P^\circ}[m_2(Z;\theta)]=0$. Because we assumed that $\delta>\delta_\circ$,
    the bound above shows that $f$ diverges along any sequence in $\Lbb_{>0}$
    with $\eta\to\infty$, regardless of the behavior of
    $(\zeta,\lambda_{\leq},\lambda_{=})$.
    
    \uline{\textbf{Second bound.}} The second bound handles diverging $\zeta$:
    since $0\in\dom(\phi)$, Lemma~\ref{lemma: perspective larger than
    argument} at $\check{x}=0$ gives
    $(\phi^*)^\pi(w(z)-\zeta,\eta)\geq-\eta\,\phi(0)$, implying that
    $f(\xi,\theta)\geq\zeta+\eta(\delta-\phi(0))$ on $\Lbb_{>0}$.
    We argue by contradiction. Consider any diverging sequence in
    $\lev_{\leq\alpha^\star}(f,\Lbb_{>0})$
    such that $\|(\zeta,\eta,\lambda_{\leq},\lambda_{=})\|\to+\infty$.
    If $\eta\to\infty$ along a subsequence, the first bound gives
    $f(\xi,\theta)\geq\E_{P^\circ}[g(Z;\theta)]+\eta(\delta-\delta_\circ)
    \to+\infty$, already contradicting $f\leq\alpha^\star$. It therefore
    suffices to treat sequences along which $\eta$ remains in a bounded
    subset of $\R_{++}$, so that only $(\zeta,\lambda_{\leq},\lambda_{=})$ 
    diverges.

    Suppose first that $\phi^\infty(1)<\infty$. These cases reduce to the
    $\eta=0$ analysis up to bounded scalars: (a) if
    $\zeta\to+\infty$, the second bound gives
    $f(\xi,\theta)\geq\zeta+\eta(\delta-\phi(0))$, and the right-hand
    side diverges since $\eta(\delta-\phi(0))$ is bounded along the sequence; (b) if
    $\zeta\to-\infty$, the pointwise constraint is eventually violated
    uniformly in $\lambda$. To see this, the two relaxed inequalities read
    \[\zeta+\eta\phi^\infty(1)\geq\E_{Q_-}[g(Z;\theta)]+\varepsilon\lambda_{\leq}
    +\varepsilon\lambda_{=}\]
    and
    \[\zeta+\eta\phi^\infty(1)\geq\E_{Q_+}[g(Z;\theta)]+\varepsilon\lambda_{\leq}
    -\varepsilon\lambda_{=},\]
    since $\E_{Q_\pm}[m_1(Z;\theta)]=-\varepsilon$ after the recentering, and
    taking the larger of the two right-hand sides gives
    \[\zeta+\eta\phi^\infty(1)\geq
    \min_{\pm}\E_{Q_\pm}[g(Z;\theta)]+\varepsilon\lambda_{\leq}+\varepsilon|\lambda_{=}|
    \geq\min_{\pm}\E_{Q_\pm}[g(Z;\theta)],\]
    a finite lower bound that holds for every $\lambda\in\R_+\times\R$.
    Along the sequence the left-hand side diverges to $-\infty$, so 
    the constraint eventually fails regardless of the behavior of $\lambda$;
    (c) if $\lambda_{\leq}\to+\infty$ or
    $|\lambda_{=}|\to\infty$ with $\zeta$ bounded, the same arguments based on
    the $Q_\pm$ distributions used for $\eta=0$ apply. These cases are sufficient
    because the case for $\zeta$ does not depend on $\lambda$.

    Suppose instead that $\phi^\infty(1)=+\infty$, so that, by the
    restriction of this analysis to option (a) of Assumption~\ref{assump: dps-entropy},
    $[0,\infty)\subseteq\dom(\phi)$ and $g,m$ are continuous. The constraint defining
    $\Lbb_{>0}$ and the two relaxed inequalities are then vacuous.
    Consider the following cases. If $\zeta\to+\infty$, the
    argument in (a) applies verbatim, as it does not use the constraint. If
     $\zeta\to-\infty$ with $\lambda$ bounded, then Assumption~\ref{assump: dps-entropy}
     allows us to use Lemma~\ref{lemma: perspective larger than argument} at
     some $\check{x}>1$ in $\dom(\phi)$
     to get
    $(\phi^*)^\pi(w(z)-\zeta,\eta)\geq \check{x}(w(z)-\zeta)-\eta\,\phi(\check{x})$,
    implying that
    \[f(\xi,\theta)\geq \zeta(1-\check{x})
    + \check{x}\,\E_{\Pbase{}}[w(Z)] + \eta(\delta
    -\phi(\check{x}))
    \to+\infty.\]
    If instead $s:=\|\lambda\|\to\infty$ with $\zeta$ bounded above, write
    $\lambda=s\,u_s$ and extract a further subsequence along which
    $u_s\to u=(u_{\leq},u_{=})\in\R_+\times\R$ with $\|u\|=1$, using compactness of
    the unit sphere and closedness of the constraint $\lambda_{\leq}\geq 0$. We
    claim that $\Pbase{}(u^\top m(Z;\theta)<0)>0$. 
    To see this, let $Q=Q_-$ if $u_{=}\geq 0$ and $Q=Q_+$ if
    $u_{=}<0$, with $Q_\pm$ and $\varepsilon$ as in part (i). Then
    \[\E_{Q}[u^\top m(Z;\theta)]
    =u_{\leq}\E_{Q}[m_1(Z;\theta)]+u_{=}\E_{Q}[m_2(Z;\theta)]<0,\]
    where the first term is nonpositive, since $u_{\leq}\geq 0$ and
    $\E_{Q_\pm}[m_1(Z;\theta)]<0$, and strictly negative when $u_{=}=0$, as
    then $u_{\leq}=1$, while the second term equals $-|u_{=}|\varepsilon\leq 0$ by
    the choice of $Q$, with strict inequality when $u_{=}\neq 0$. A random
    variable with strictly negative mean is strictly negative with positive
    probability, so $Q(u^\top m(Z;\theta)<0)>0$ and the set
    $\{z\in\calZ:u^\top m(z;\theta)<0\}$ is nonempty. Moreover, this set is open by
    continuity of $m$, and therefore has positive $\Pbase{}$-measure by
    the full-support condition of
    Assumption~\ref{assump: dps-support}. By
    continuity of $\Pbase{}$ from below there then exists $c>0$ such that the
    event $\calA:=\{z\in\calZ:u^\top m(z;\theta)<-2c\}$ has
    $p:=\Pbase{}(\calA)>0$.
    {As $\Pbase{}\in\calP(\calZ)$ and $u_s\to u$, for all
    sufficiently large $s$}
    \[\int_{\calA}\big|(u_s-u)^\top m(z;\theta)\big|\,\de\Pbase{}(z)\leq
    \|u_s-u\|\,\E_{\Pbase{}}\|m(Z;\theta)\|\leq cp,\]
    so that, using $u^\top m(z;\theta)<-2c$ on $\calA$,
    \begin{align}\label{aux eq: decomposition 3}
        \int_{\calA}u_s^\top m(z;\theta)\,\de\Pbase{}(z)
        \leq\int_{\calA}u^\top m(z;\theta)\,\de\Pbase{}(z)+cp
        \leq-2cp+cp=-cp.
    \end{align}
    Set $\check{K}:=2/p\in\dom(\phi)$. It then follows that
    \begin{align*}
    f(\xi,\theta)
    &\overset{(i)}{=} \zeta+\eta\delta+\E_{\Pbase{}}[(\phi^*)^\pi(w(Z)-\zeta,\eta)]\\
    &\overset{(ii)}{\geq} \zeta+\eta\delta
    +\check{K}\int_{\calA}(w(z)-\zeta)\,\de \Pbase{}(z)
    -\eta\big(p\,\phi(\check{K})+(1-p)\,\phi(0)\big)\\
    &\overset{(iii)}{\geq} \zeta(1-\check{K}p)
    -\check{K}\,\E_{\Pbase{}}[[g(Z;\theta)]_-]+\check{K}cps
    +\eta\big(\delta-p\,\phi(\check{K})-(1-p)\,\phi(0)\big)\\
    &\overset{(iv)}{=} -\zeta-\frac{2}{p}\,\E_{\Pbase{}}[[g(Z;\theta)]_-]+2cs
    +\eta\big(\delta-p\,\phi(2/p)-(1-p)\,\phi(0)\big)\to+\infty,
    \end{align*}
    where $(\textit{i})$ is the definition of $f$ under the centering $\rho=0$;
    $(\textit{ii})$ applies Lemma~\ref{lemma: perspective larger than argument}
    pointwise at $\check{x}=\check{K}$ on $\calA$ and at $\check{x}=0$ on
    $\calZ\setminus\calA$ and integrates against $\Pbase{}$, using $\Pbase{}(\calA)=p$ and
    $\Pbase{}(\calZ\setminus\calA)=1-p$; $(\textit{iii})$ expands
    $w=g-s\,u_s^\top m$, bounds
    $\int_{\calA}g\,\de\Pbase{}\geq-\E_{\Pbase{}}[[g(Z;\theta)]_-]$, which is
    finite by Assumption~\ref{assump: dps-int}, and uses
    \eqref{aux eq: decomposition 3} multiplied by $-s$;
    and $(\textit{iv})$ substitutes $\check{K}=2/p$, so that $\check{K}p=2$ and hence
    $\zeta(1-\check{K}p)=-\zeta$. The limit is $+\infty$ because $-\zeta$ is bounded
    below along this subsequence (recall $\zeta$ is bounded above), $\eta$ is
    bounded, and $cs\to+\infty$. In every case $f\to+\infty$ along a
    subsequence, contradicting $f\leq\alpha^\star$.

    Therefore, we can conclude that $\lev_{\leq\alpha^\star}(f,\Lbb_{>0})$ is bounded.

    \uline{\textbf{Combining the two cases.}} Under
    Assumption~\ref{assump: dual primitives} or option (a) of
    Assumption~\ref{assump: dps-entropy}, we now show that
    $\lev_{\leq\alpha^\star}(f,\calF(\theta))=\lev_{\leq\alpha^\star}(f,\Lbb_{|0})\cup%
    \lev_{\leq\alpha^\star}(f,\Lbb_{>0})$.
    For the inclusion $\subseteq$, note that every point of
    $\lev_{\leq\alpha^\star}(f,\calF(\theta))$ has either $\eta=0$, and then it lies
    in $\Lbb_{|0}$ -- since any point of the $\eta=0$ slice outside $\Lbb_{|0}$ has
    $f=+\infty$, as shown at the start of the $\eta=0$ case -- or $\eta>0$, in which
    case it lies in $\Lbb_{>0}$ by construction. For the reverse inclusion $\supseteq$,
    both $\Lbb_{|0}\subseteq\calF(\theta)$ and
    $\Lbb_{>0}\subseteq\calF(\theta)$ by construction, so the two sublevel sets on
    the right are contained in $\lev_{\leq\alpha^\star}(f,\calF(\theta))$. As the
    union of two bounded sets, $\lev_{\leq\alpha^\star}(f,\calF(\theta))$ is
    bounded. By the reduction in the choice of the candidate sublevel set,
    this proves part (iii) under Assumption~\ref{assump: dual primitives} and under
    option (a) of Assumption~\ref{assump: dps-entropy}.

    \uline{\textbf{Option (b) of Assumption~\ref{assump: dps-entropy}
    for arbitrary $\dimm\geq0$.}} It remains to
    establish part (iii) under option (b) of
    Assumption~\ref{assump: dps-entropy}. The
    relaxed-set analysis above does not apply in this case because 
    it relies on $\dom(\phi)=[0,\infty)$ and on the continuity and full-support
    condition of Assumption~\ref{assump: dps-support}, both available
    only under option (a). We instead verify the claim of part (iii)
    directly. Since $\phi^\infty(1)=+\infty$ under option (b), the
    feasible set is the cone $\calF(\theta)=\Lbb$, so that
    $\mathrm{rec}(\calF(\theta))=\Lbb$ and the claim reads:
    $f^\infty(d,\theta)>0$ for every nonzero
    $d=(d_\zeta,d_\eta,d_\lambda)\in\Lbb$. As before, we work in the
    recentered problem, $\rhopad=0$. The verification has two
    steps. We first exploit the boundedness of $\dom(\phi)$ to show that
    $f(\cdot,\theta)$ is finite on $\Lbb$ and to obtain
    $f^\infty(\cdot,\theta)$ in closed form. We then show that the
    closed form is strictly positive along each applicable class of
    directions. When $\dimm=0$, the $\lambda$-block is absent, we set
    $d_\lambda=0$, and only the first two classes below arise. When
    $\dimm\geq1$, all three classes arise: those with $d_\eta>0$, where the distribution
    $P^\circ$ of Assumption~\ref{assump: dps-feas} provides the required slack; those
    with $d_\eta=0$ and $d_\lambda=0$, where positivity is a
    consequence of $\bar{x}>1$; and those with $d_\eta=0$ and
    $d_\lambda\neq0$, where the condition of
    Assumption~\ref{assump: dps-ceiling} takes over the role that the
    full-support condition played in the analysis above.

    We begin with the closed form. Note first that
    $\dom(\phi)=[0,\bar{x}]$. Write $h:=(\phi^*)^\pi$ and
    $C_1:=\max\{\phi(0),\phi(\bar{x})\}$, which bounds $\phi$ on
    $[0,\bar{x}]$ by convexity and the nonnegativity of $\phi$.
    Evaluating the supremum defining $\phi^*$ at the points $0$ and
    $\bar{x}$ of $\dom(\phi)$ gives the lower bound in
    \[\max\{-\phi(0),\,\bar{x}y-\phi(\bar{x})\}\leq\phi^*(y)\leq
    \bar{x}\,[y]_+,\]
    while the upper bound uses $\phi\geq0$ together with
    $xy\leq\bar{x}\,[y]_+$ for $x\in[0,\bar{x}]$ and every $y\in\R$. Consequently
    $|h(x,\eta)|\leq\bar{x}|x|+C_1\eta$ for every $\eta\geq0$, and when
     $\eta=0$ we have $h(x,0)=(\phi^*)^\infty(x)=\bar{x}\,[x]_+$, as
    computed in Case I. In particular, $\dom(h)=\R\times\R_+$.
    {The displayed linear growth bound, together with
    $\Pbase{}\in\calP(\calZ)$, shows
    that $f(\cdot,\theta)$ is finite on $\Lbb$. The arguments at the
    beginning of this part show that it is closed, proper, and convex.}
    Moreover, $h$ is closed, proper, and convex by
    Lemma~\ref{lemma: entropy function proper cvx lsc}, positively
    homogeneous by construction, and satisfies $h(0,0)=0$, so
    Theorem~8.5 in \citet{rockafellar1970ConvexAnalysis} gives
    $h^\infty=h$. By the same theorem, for every $x_0\in\dom(h)$ and
    every direction $v$, the difference quotients
    $[h(x_0+\tau v)-h(x_0)]/\tau$ are nondecreasing in $\tau>0$ and
    increase to $h^\infty(v)=h(v)$. Fix any
    $\xi_0=(\zeta_0,\eta_0,\lambda_0)\in\Lbb$. Since $\Lbb$ is a convex
    cone containing $d$, the ray $\xi_0+\tau d$ belongs to
    $\Lbb$. At each $z\in\calZ$ the term within the expectation in $f$ is
    $h$ composed with an affine map of $\xi$, whose value at $\xi_0$ lies
    in $\dom(h)$, so the difference quotients of this term along $d$
    increase pointwise to
    $h(-d_\lambda^\top m(z;\theta)-d_\zeta,\,d_\eta)$. 
    Monotone convergence therefore yields
    \[f^\infty(d,\theta)=d_\zeta+d_\eta\delta
    +\E_{\Pbase{}}\big[h\big(-d_\lambda^\top m(Z;\theta)-d_\zeta,\,
    d_\eta\big)\big].\]
     
    We now show that the right-hand side above is strictly positive in each of
    the applicable cases described at the beginning of this part. 
    (1) Directions with $d_\eta>0$. Let
    $\delta_\circ:=D_\phi(P^\circ\,\|\,\Pbase{})<\delta$, with $P^\circ$ the distribution
    of Assumption~\ref{assump: dps-feas}, which belongs to $\calP(\calZ)$ by
    Assumption~\ref{assump: dps-feas}. For every $\tau>0$ the
    point $\xi_0+\tau d$ lies in $\Lbb=\calF(\theta)$, so
    \eqref{eq: slater dual slack} at $P=P^\circ$, in the recentered problem, gives
    \[f(\xi_0+\tau d,\theta)\geq\E_{P^\circ}[g(Z;\theta)]
    +(\eta_0+\tau d_\eta)(\delta-\delta_\circ)
    -(\lambda_0+\tau d_\lambda)^\top\E_{P^\circ}[m(Z;\theta)].\]
    Subtracting the finite value $f(\xi_0,\theta)$, dividing by $\tau$, and letting
    $\tau\to\infty$, Definition~\ref{defi: recession function} yields
    \[f^\infty(d,\theta)\geq d_\eta(\delta-\delta_\circ)
    -d_\lambda^\top\E_{P^\circ}[m(Z;\theta)]\geq d_\eta(\delta-\delta_\circ)>0,\]
    where the second inequality holds because $d_\lambda\in\Rlam$ and
    $\E_{P^\circ}[m(Z;\theta)]$ lies in the recentered moment constraint set, so that
    $d_\lambda^\top\E_{P^\circ}[m(Z;\theta)]\leq0$ by the support function of
    $\calM(\rho)$ recorded in Section~\ref{subsec: setup}, evaluated at $\rho=0$.
    (2) Directions with $d_\eta=0$ and
    $d_\lambda=0$, so that $d_\zeta\neq0$. In this case
     $f^\infty(d,\theta)=d_\zeta+\bar{x}\,[-d_\zeta]_+$,
    which equals $d_\zeta>0$ when $d_\zeta>0$ and $(1-\bar{x})d_\zeta>0$
    when $d_\zeta<0$, in view of $\bar{x}>1$.
    (3) When $\dimm\geq1$, directions with $d_\eta=0$ and
    $d_\lambda\neq0$. Write $Y:=-d_\lambda^\top m(Z;\theta)$, so that 
    $f^\infty(d,\theta)=d_\zeta+\bar{x}\,\E_{\Pbase{}}[[Y-d_\zeta]_+]$.
    We say that a distribution $Q\in\Delta(\calZ)$ is ``ceiling-feasible'' if
    $Q\ll\Pbase{}$ with $\de Q/\de\Pbase{}\leq\bar{x}$
    $\Pbase{}$-almost surely, as in the definition of
    $\calN_{\bar{x}}(\theta)$ in Assumption~\ref{assump: dps-ceiling}.
    For every such $Q$,
    \[\E_Q[Y]\leq d_\zeta+\E_Q[[Y-d_\zeta]_+]\leq
    d_\zeta+\bar{x}\,\E_{\Pbase{}}[[Y-d_\zeta]_+]=f^\infty(d,\theta),\]
    where the first inequality applies $y\leq t+[y-t]_+$ pointwise at
    $t=d_\zeta$ and the second bounds the density of $Q$ by $\bar{x}$. 
    Since Assumption~\ref{assump: dps-ceiling} places
    $0$ in the interior of $\calN_{\bar{x}}(\theta)-\rhopad$,
    there exists $r_0>0$ with
    $-r_0\,d_\lambda/\|d_\lambda\|\in\calN_{\bar{x}}(\theta)-\rhopad$.
    In the recentered problem this says that some ceiling-feasible $Q$
    satisfies $\E_Q[m(Z;\theta)]=-r_0\,d_\lambda/\|d_\lambda\|$, and
    hence $\E_Q[Y]=r_0\|d_\lambda\|$. Therefore
    $f^\infty(d,\theta)\geq r_0\|d_\lambda\|>0$, for any value of
    $d_\zeta$.

    In every applicable case $f^\infty(d,\theta)>0$, which establishes
    part (iii) under option (b) for every $\dimm\geq0$. Parts (i)--(ii) require no
    separate argument: part (ii) is the vacuity statement already proven, and part (i)
    was proven under both options above.
\end{proof}

\subsection{Proof of Theorem~\ref{thm: existence of dual solutions}}
\begin{proof}\label{proof:thm: existence of dual solutions}
Fix $\theta\in\Theta$, and let $\calF(\theta)$ and $f(\cdot,\theta)$ be
the dual feasible set and the dual objective of Section~\ref{subsec:existence-dual}.

\uline{\textbf{Properties of the feasible set and of the objective.}} By the
properties established in the proof of Lemma~\ref{lemma: slater dual form-iii},
$\calF(\theta)$ is closed and convex, $f(\cdot,\theta)$ is a closed proper convex
function, and $f(\xi_0,\theta)<\infty$ at the feasible point $\xi_0$ exhibited there, so
that $\calF(\theta)\cap\dom(f(\cdot,\theta))\neq\emptyset$.

\uline{\textbf{Reduction to a recession condition.}} Since $\calF(\theta)$
is not compact, the Weierstrass theorem does not apply directly, and we rely instead on
Proposition~3.2.2 in \citet{bertsekas2009ConvexOptimizationTheory}. This result
concerns the minimization of a closed convex function over a closed convex set. It states
that the set of minimizers is nonempty and compact if and only if the
set and the function have no nonzero direction of recession in common. Now, a
direction $d$ is a direction of recession of $f(\cdot,\theta)$ exactly
when $f^\infty(d,\theta)\leq0$, whereas
Lemma~\ref{lemma: slater dual form-iii} shows that $f^\infty(d,\theta)>0$
for every nonzero direction of recession $d$ of $\calF(\theta)$. Hence $\calF(\theta)$ and
$f(\cdot,\theta)$ have no nonzero direction of recession in common, and
the set of minimizers of \eqref{eq: dual with moments} is nonempty and compact. Convexity
of the set of minimizers follows from the convexity of $f(\cdot,\theta)$
and $\calF(\theta)$, since the minimizers of a convex function over a convex set form a
convex set.
\end{proof}

\subsection{Proof of Lemma~\ref{lemma: constraint qualification}}
\begin{proof}\label{proof: constraint qualification}
Throughout, $\Gamma(\delta)$ is convex and contains
$c:=\E_{\Pbase{}}[m(Z;\theta)]$, as recalled in
Section~\ref{subsec:strong-duality}.

\uline{\textbf{Part (i).}} For \textup{(a)}, if $\rhopad=c$ we take
$P^{\circ}:=\Pbase{}$. Otherwise, since $\rhopad\in\interior(\Gamma(\delta))$
and $c\in\Gamma(\delta)$, Theorem~6.4 in
\citet{rockafellar1970ConvexAnalysis} yields $\mu_0>1$ such that
$c+\mu_0(\rhopad-c)\in\Gamma(\delta)$, that is, some $P_1\in\calP(\calZ)$ with
$\E_{P_1}[m(Z;\theta)]=c+\mu_0(\rhopad-c)$ and $D_\phi(P_1\,\|\,\Pbase{})\leq\delta$.
Set $P^{\circ}:=(1-\mu_0^{-1})\Pbase{}+\mu_0^{-1}P_1$, which belongs to
$\calP(\calZ)$. Then
$\E_{P^{\circ}}[m(Z;\theta)]=(1-\mu_0^{-1})c+\mu_0^{-1}\big(c+\mu_0(\rhopad-c)\big)=\rhopad$,
and, since $D_\phi(\cdot\,\|\,\Pbase{})$ is convex, as recalled in
Section~\ref{subsec:strong-duality},
$D_\phi(P^{\circ}\,\|\,\Pbase{})\leq(1-\mu_0^{-1})\cdot0+\mu_0^{-1}\delta<\delta$.
For \textup{(b)},
let $P\in\calP(\calZ)$ and suppose that $\E_P[m(Z;\theta)]$ does not belong to
the closed convex set $\mathrm{cl}(\mathrm{co}(\calN(\theta)))$. By the
separating hyperplane theorem there exist $a\in\R^{\dimm}$ and $b\in\R$
such that $a^{\top}\E_P[m(Z;\theta)]>b\geq a^{\top}m(z;\theta)$ for every
$z\in\calZ$. Integrating the second inequality under $P$ gives
$a^{\top}\E_P[m(Z;\theta)]\leq b$, a contradiction. Hence
$\Gamma(\delta)\subseteq\mathrm{cl}(\mathrm{co}(\calN(\theta)))$, so that
$\rhopad$ lies in the interior of $\mathrm{cl}(\mathrm{co}(\calN(\theta)))$.
As this interior is nonempty, the affine hull of
$\mathrm{co}(\calN(\theta))$ is the whole of $\R^{\dimm}$, and, since a convex
set and its closure have the same relative interior
\citep[Theorem~6.3]{rockafellar1970ConvexAnalysis},
$\rhopad\in\interior(\mathrm{co}(\calN(\theta)))$, which is \textup{(b)}.
For \textup{(c)}, let $P\in\calP(\calZ)$ with
$D_\phi(P\,\|\,\Pbase{})\leq\delta$. As
$\bar{x}<\infty$ forces $\phi^\infty(1)=+\infty$,
Lemma~\ref{lemma: divergence lebesgue decomposition} gives $P\ll\Pbase{}$ and
$\int\phi(\de P/\de\Pbase{})\de\Pbase{}<\infty$, so that
$\de P/\de\Pbase{}\in\dom(\phi)\subseteq[0,\bar{x}]$ $\Pbase{}$-almost surely.
Thus $\E_P[m(Z;\theta)]\in\calN_{\bar{x}}(\theta)$, whence
$\Gamma(\delta)\subseteq\calN_{\bar{x}}(\theta)$ and
$\rhopad\in\interior(\calN_{\bar{x}}(\theta))$, which is \textup{(c)}.

\uline{\textbf{Part (ii).}} For $K>0$ let
$m_K(z):=m(z;\theta)\indic\{\|m(z;\theta)\|\leq K\}$, let
$\mu_K:=\E_{P^{\circ}}[m_K(Z)]$, and let
$\Sigma_K:=\E_{P^{\circ}}\big[(m(Z;\theta)-\rhopad)(m_K(Z)-\mu_K)^{\top}\big]$,
which is finite because $\|m_K\|\leq K$ and $m(\cdot;\theta)$ is
$P^{\circ}$-integrable. As $K\to\infty$, $m_K\to m(\cdot;\theta)$ pointwise
with $\|m_K\|\leq\|m(\cdot;\theta)\|$ and $\E_{P^{\circ}}\|m(Z;\theta)\|^2<\infty$,
so the dominated convergence theorem gives
$\Sigma_K\to\V_{P^{\circ}}[m(Z;\theta)]$. As the latter matrix is
nonsingular and the nonsingular matrices form an open set, $\Sigma_K$ is
nonsingular for some $K$, which we fix from now on, and we write
$\sigma_{\min}(\Sigma_K)>0$ for its smallest singular value. Let $B:=2K$,
so that $\|m_K(z)-\mu_K\|\leq B$ for every $z$, fix $v\in\R^{\dimm}$ with
$\|v\|\leq1$ and $t\in\R$ with $|t|\leq1/(2B)$, and define
$r_t(z):=1+t\,v^{\top}(m_K(z)-\mu_K)$, which takes values in
$[\tfrac12,\tfrac32]$, and $P_t:=r_tP^{\circ}$. Since
$\int r_t\,\de P^{\circ}=1+t\,v^{\top}(\mu_K-\mu_K)=1$, the measure $P_t$ is a
probability distribution. It belongs to $\calP(\calZ)$ because
$r_t\leq\tfrac32$ and $P^{\circ}\in\calP(\calZ)$, and
\[
\E_{P_t}[m(Z;\theta)]=\E_{P^{\circ}}\big[m(Z;\theta)\,r_t(Z)\big]
=\rhopad+t\,\E_{P^{\circ}}\big[m(Z;\theta)(m_K(Z)-\mu_K)^{\top}\big]v
=\rhopad+t\,\Sigma_Kv,
\]
where the last equality uses $\E_{P^{\circ}}[m_K(Z)-\mu_K]=0$ and
$\E_{P^{\circ}}[m(Z;\theta)]=\rhopad$. As $\Sigma_K$ is nonsingular, the set
$\{\rhopad+t\Sigma_Kv:\|v\|\leq1\}$ contains the closed ball
$\overline{\calB}_{|t|\sigma_{\min}(\Sigma_K)}(\rhopad)$. It remains
to show that
$D_\phi(P_t\,\|\,\Pbase{})<\delta$ whenever $|t|$ is small enough. Write
$P^{\circ}=w^{\circ}\Pbase{}+P^{\circ\perp}$ for the Lebesgue decomposition
of $P^{\circ}$ with respect to $\Pbase{}$. Then
$P_t=r_tw^{\circ}\Pbase{}+r_tP^{\circ\perp}$ is that of $P_t$, and
Lemma~\ref{lemma: divergence lebesgue decomposition} gives
\[
D_\phi(P_t\,\|\,\Pbase{})=\int\phi(r_tw^{\circ})\,\de\Pbase{}
+\phi^\infty(1)\int r_t\,\de P^{\circ\perp}.
\]
For $r\geq1$ and $w\geq0$, Theorem~8.5 in
\citet{rockafellar1970ConvexAnalysis} gives
$\phi(rw)\leq\phi(w)+(r-1)\,w\,\phi^\infty(1)$, as the recession function of
$\phi$ is $y\mapsto y\,\phi^\infty(1)$ on $[0,\infty)$. For
$r\in[\tfrac12,1)$, convexity of $\phi$ and $\phi\geq0$ give
$\phi(rw)\leq r\phi(w)+(1-r)\phi(0)\leq\phi(w)+(1-r)\phi(0)$. Since
$|r_t-1|\leq|t|B$, in both cases
$\phi(r_tw^{\circ})\leq\phi(w^{\circ})+|t|B\,\big(w^{\circ}\phi^\infty(1)+\phi(0)\big)$
pointwise, and integrating under $\Pbase{}$, using
$\int w^{\circ}\,\de\Pbase{}\leq1$ and
$\int r_t\,\de P^{\circ\perp}\leq(1+|t|B)P^{\circ\perp}(\calZ)$, yields
\[
D_\phi(P_t\,\|\,\Pbase{})\leq D_\phi(P^{\circ}\,\|\,\Pbase{})
+|t|B\,\big(\phi(0)+2\phi^\infty(1)\big).
\]
The right-hand side is smaller than $\delta$ for $|t|$ small enough, because
$D_\phi(P^{\circ}\,\|\,\Pbase{})<\delta$, and for every such $t$ the closed ball
$\overline{\calB}_{|t|\sigma_{\min}(\Sigma_K)}(\rhopad)$ is contained in
$\Gamma(\delta)$, which proves the first claim of \textup{(ii)}. If
$P^{\circ}\ll\Pbase{}$, then $P_t=r_tP^{\circ}\ll\Pbase{}$ for every $t$, so the
same ball is contained in the attainable set of the distributions dominated
by $\Pbase{}$, which proves the second claim.
\end{proof}

\subsection{Proof of Lemma~\ref{lemma: outer value regularity}}
\begin{proof}\label{proof:lemma: outer value regularity}
    Throughout the proof, we suppress the arguments
    $\rho$ and
    $\delta$ and thus write $\calP_\phi(\theta)$ for the primal feasible set and
    $V_{\mathsf{P}}(\theta)$ and $V_{\mathsf{D}}(\theta)$ for the value
    functions. We write $P^\circ$ for the strictly feasible distribution
    of Assumption~\ref{assump: dp-feas} or Assumption~\ref{assump: dps-feas}.

    \uline{\textbf{Part (i): Finiteness.}} We first bound the two value functions from
    below. {By either Assumption~\ref{assump: dp-feas} or
    Assumption~\ref{assump: dps-feas},
    $P^\circ\in\calP_\phi(\theta)$. Therefore}
    $V_{\mathsf{P}}(\theta)\geq\E_{P^\circ}[g(Z;\theta)]>-\infty$. Moreover,
    for every primal-feasible $P$ and dual-feasible $\xi$, inequality
    \eqref{eq: slater dual slack} in the proof of
    Lemma~\ref{lemma: slater dual form-iii}, whose two slack terms are nonnegative at
    primal-feasible $P$, gives $\E_P[g(Z;\theta)]\leq f(\xi,\theta)$.
    Taking the supremum over $P$
    and the infimum over $\xi$ therefore gives
    $V_{\mathsf{D}}(\theta)\geq V_{\mathsf{P}}(\theta)>-\infty$.

    We next bound the two value functions from above. Since
    $V_{\mathsf{P}}(\theta)\leq V_{\mathsf{D}}(\theta)$, it suffices to
    find a feasible point at which the dual objective is finite. Such a
    point is provided by the properness paragraph in the proof of
    Lemma~\ref{lemma: slater dual form-iii}, which shows that $f(\xi_0,\theta)<\infty$
    at the feasible point $\xi_0=(\bar{g},1,0)$, with
    $\bar{g}=\sup_{z\in\calZ}g(z;\theta)$, under
    Assumption~\ref{assump: dual primitives}, and at $\xi_0=\xi^{*}$ of
    Assumption~\ref{assump: dps-anchor} under
    Assumption~\ref{assump: dual primitives superlinear}. Under
    Assumption~\ref{assump: dual primitives}, the point $\xi_0$ is also the strictly
    feasible point constructed in the proof of Lemma~\ref{lemma: slater dual form-ii}.
    Therefore $V_{\mathsf{P}}(\theta)\leq V_{\mathsf{D}}(\theta)\leq
    f(\xi_0,\theta)<\infty$, which together with the lower bound establishes part
    (i).

    \uline{\textbf{Part (ii): Upper semicontinuity.}}
    We introduce the notation
    \[\Psi(\xi,\theta):=\E_{\Pbase{}}\big[(\phi^*)^\pi\big(g(Z;\theta)
    -\lambda^\top
    m(Z;\theta)-\zeta,\,\eta\big)\big],\]
    so that
    $f(\xi,\theta)=\lambda_{12}^\top\rho+\zeta+\eta\delta+\Psi(\xi,\theta)$.
    Since $V_{\mathsf{D}}$ is real valued on $\Theta$ follows from
    part~\ref{lemma: outer value regularity-i}, whose hypotheses hold at every
    $\theta\in\Theta$ under Assumption~\ref{assump: outer regularity}. Fix
    $\theta_0\in\Theta$ and let $(\theta_n)_{n\geq1}\subseteq\Theta$ be an arbitrary
    sequence with $\theta_n\to\theta_0$. We now show that
    $\limsup_{n\to\infty}V_{\mathsf{D}}(\theta_n)\leq
    V_{\mathsf{D}}(\theta_0)$.

    \uline{\textbf{Step 1: strategy and two preliminary facts.}} 
    We start showing two facts that are used repeatedly below. First, for every
    $\xi=(\zeta,\eta,\lambda)\in\R\times\R_+\times\Rlam$, every
    $\theta\in\Theta$, and every $z\in\calZ$,
    \begin{align}\label{aux eq: usc dual minorant}
    (\phi^*)^\pi\big(g(z;\theta)-\lambda^\top
    m(z;\theta)-\zeta,\,\eta\big)\geq
    g(z;\theta)-\lambda^\top m(z;\theta)-\zeta.
    \end{align}
    Indeed, this is Lemma~\ref{lemma: perspective larger than argument}
    with $\check{x}=1$.
    Second, the conjugate $\phi^*$ is
    nondecreasing and finite on $(-\infty,\phi^\infty(1))$, hence
    continuous there, as a finite convex function on an open interval.

    The remainder of the proof implements the following strategy in the finite-recession 
    and superlinear cases. Since
    $V_{\mathsf{D}}(\theta_n)\leq f(\xi,\theta_n)$ for every
    $\xi\in\calF(\theta_n)$, it suffices to construct, for every
    $\varepsilon>0$, a multiplier $\xi_\varepsilon$ such that (a)
    $f(\xi_\varepsilon,\theta_0)\leq
    V_{\mathsf{D}}(\theta_0)+2\varepsilon$, (b)
    $\xi_\varepsilon\in\calF(\theta_n)$ for all $n$ large, and (c)
    $\limsup_{n\to\infty}f(\xi_\varepsilon,\theta_n)\leq
    f(\xi_\varepsilon,\theta_0)$. Indeed, (a)--(c) give
    $\limsup_{n\to\infty}V_{\mathsf{D}}(\theta_n)\leq
    V_{\mathsf{D}}(\theta_0)+2\varepsilon$, and letting
    $\varepsilon\to 0$ concludes.

    \uline{\textbf{Step 2: the finite-recession case.}} Suppose
    Assumption~\ref{assump: dual primitives} holds at every
    $\theta\in\Theta$ and fix $\varepsilon>0$.

    Since $V_{\mathsf{D}}(\theta_0)\in\R$, there exists
    $\xi=(\zeta,\eta,\lambda)\in\calF(\theta_0)$ with
    $f(\xi,\theta_0)\leq V_{\mathsf{D}}(\theta_0)+\varepsilon$. By Step
    1, $f(\xi,\theta_0)$ is finite. Feasibility of $\xi$ and
    $\phi^\infty(1)<\infty$ imply
    $\sigma(\lambda,\theta_0)\leq\zeta+\eta\phi^\infty(1)<\infty$, while
    $\sigma(\lambda,\theta_0)>-\infty$ because $\calZ$ is nonempty. Let
    $\xi^\circ=(\zeta^\circ,\eta^\circ,\lambda^\circ)$ be the strictly feasible point of
    Lemma~\ref{lemma: slater dual form-ii}, with slack
    $s^\circ:=\zeta^\circ+\eta^\circ\phi^\infty(1)-\sigma(\lambda^\circ,\theta_0)>0$
    and $\eta^\circ>0$. We take for $\xi^\circ$ the point constructed in
    the proof of that lemma, which is the point $\xi_0$ of part~(i) at $\theta_0$, so
    that $f(\xi^\circ,\theta_0)<\infty$, while $f(\xi^\circ,\theta_0)>-\infty$ by
    Step~1. For $t\in(0,1]$, define
    $\xi_t:=(1-t)\xi+t\xi^\circ=(\zeta_t,\eta_t,\lambda_t)$, which belongs to the convex
    set $\R\times\R_+\times\Rlam$. Since $\sigma(\cdot,\theta_0)$ is
    convex on $\Rlam$, as a pointwise supremum of functions affine in $\lambda$, the slack
    at $\xi_t$ satisfies
    \[\zeta_t+\eta_t\phi^\infty(1)-\sigma(\lambda_t,\theta_0)
    \geq(1-t)\big\{\zeta+\eta\phi^\infty(1)
    -\sigma(\lambda,\theta_0)\big\}
    +t\,s^\circ\geq t\,s^\circ>0,\]
    so $\xi_t$ is strictly feasible at $\theta_0$, with $\eta_t\geq
    t\eta^\circ>0$. Moreover, $f(\cdot,\theta_0)$ is convex on
    $\R\times\R_+\times\Rlam$, hence
    \[f(\xi_t,\theta_0)\leq(1-t)f(\xi,\theta_0)
    +t\,f(\xi^\circ,\theta_0).\]
    Since both values on the right-hand side are finite, choosing $t$
    small enough yields
    $f(\xi_t,\theta_0)\leq V_{\mathsf{D}}(\theta_0)+2\varepsilon$. We fix
    such a $t$ for the remainder of this step, abbreviate
    $\xi':=\xi_t=(\zeta',\eta',\lambda')$, and write
    $s:=\zeta'+\eta'\phi^\infty(1)-\sigma(\lambda',\theta_0)\geq
    t\,s^\circ>0$ for the associated slack.

    Since $\sigma(\lambda',\theta_0)<\infty$,
    Assumption~\ref{assump: or-usc sigma} applies at $\theta_0$ and yields a radius $r>0$
    such that
    $\sigma(\lambda',\theta)\leq\sigma(\lambda',\theta_0)+s/2$
    for every $\theta\in\calB_r(\theta_0)\cap\Theta$. Then, for every
    $\theta\in\calB_r(\theta_0)\cap\Theta$ we have:
    \[\zeta'+\eta'\phi^\infty(1)=\sigma(\lambda',\theta_0)+s\geq
    \sigma(\lambda',\theta)+s/2,\] 
    so that $\xi'\in\calF(\theta)$ and
    \[\frac{g(z;\theta)-\lambda'^\top m(z;\theta)-\zeta'}{\eta'}
    \leq\frac{\sigma(\lambda',\theta_0)+s/2-\zeta'}{\eta'}
    =\phi^\infty(1)-\frac{s}{2\eta'}<\phi^\infty(1),\]
    for every $z\in\calZ$. It follows that, since $\phi^*$ is nondecreasing and finite on
    $(-\infty,\phi^\infty(1))$ by Step 1, the integrand of
    $\Psi(\xi',\theta)$ is bounded above, uniformly over $z\in\calZ$ and
    $\theta\in\calB_r(\theta_0)\cap\Theta$, by the finite constant
    $C:=\eta'\phi^*\big(\phi^\infty(1)-s/(2\eta')\big)$.

    By Assumption~\ref{assump: or-cont} applied at
    $\theta_0$, the maps $\theta'\mapsto g(z;\theta')$ and $\theta'\mapsto m(z;\theta')$
    are continuous at $\theta_0$ for $\Pbase{}$-almost every $z$, the exceptional null set
    possibly depending on $\theta_0$. 
    Therefore, writing
    $y_n(z):=g(z;\theta_n)-\lambda'^\top m(z;\theta_n)-\zeta'$ and
    $y_0(z):=g(z;\theta_0)-\lambda'^\top m(z;\theta_0)-\zeta'$, we have
    $y_n(z)\to y_0(z)$. Moreover,
    $y_0(z)\leq\sigma(\lambda',\theta_0)-\zeta'=\eta'\phi^\infty(1)-s$,
    so that $y_0(z)/\eta'$ lies in the open interval $(-\infty,\phi^\infty(1))$, on which
    $\phi^*$ is continuous by Step 1. Hence
    $\eta'\phi^*\big(y_n(z)/\eta'\big)\to\eta'\phi^*\big(y_0(z)/\eta'\big)$ for
    $\Pbase{}$-almost every $z$. For all $n$ large enough that
    $\theta_n\in\calB_r(\theta_0)$, the functions
    $z\mapsto C-\eta'\phi^*\big(y_n(z)/\eta'\big)$ are nonnegative, so Fatou's lemma gives
    \[\liminf_{n\to\infty}\E_{\Pbase{}}\Big[C-\eta'\phi^*\big(y_n(Z)/\eta'\big)\Big]
    \geq\E_{\Pbase{}}\Big[C-\eta'\phi^*\big(y_0(Z)/\eta'\big)\Big].\]
    Since
    $\Psi(\xi',\theta_0)>-\infty$ by Step 1, we get
    $\limsup_{n\to\infty}\Psi(\xi',\theta_n)\leq\Psi(\xi',\theta_0)$.
    As the remaining terms of $f(\xi',\cdot)$ do not depend on $\theta$,
    we conclude that
    $\limsup_{n\to\infty}f(\xi',\theta_n)\leq f(\xi',\theta_0)\leq
    V_{\mathsf{D}}(\theta_0)+2\varepsilon$, so that requirements (a)--(c) of Step 1 are
    met by $\xi_\varepsilon=\xi'$, which shows upper semicontinuity.

   \uline{\textbf{Step 3: the superlinear case.}} Suppose
    Assumption~\ref{assump: dual primitives superlinear} holds at every
    $\theta\in\Theta$, so that $\calF(\theta)=\R\times\R_+\times\Rlam$ for every
    $\theta\in\Theta$ by
    Lemma~\ref{lemma: slater dual form-ii}. Requirement (b) of Step 1 
    therefore holds for every
    multiplier, and it remains to show (a) and (c).

    We claim that
    \begin{align}\label{aux eq: usc dual eta reduction}
    V_{\mathsf{D}}(\theta_0)
    =\inf\big\{f(\xi,\theta_0):\xi\in
    \R\times\R_{++}\times\Rlam\big\},
    \end{align}
    which excludes the set of multipliers with $\eta=0$. To see this,
    fix $(\zeta,\lambda)\in\R\times\Rlam$ and write
    $y(z):=g(z;\theta_0)-\lambda^\top m(z;\theta_0)-\zeta$. For every
    $y\in\R$, the map $\eta\mapsto\eta\phi^*(y/\eta)$ is nonincreasing on
    $(0,\infty)$ and increases to $(\phi^*)^\pi(y,0)=(\phi^*)^\infty(y)$
    as $\eta\to0$. Indeed, with $s=1/\eta$ and $\phi^*(0)=-\inf\phi=0$, it equals the
    difference quotient $\{\phi^*(sy)-\phi^*(0)\}/s$, which is nondecreasing in $s$ by
    convexity of $\phi^*$ and converges to $(\phi^*)^\infty(y)$ as $s\to\infty$
    \citep[Theorem~8.5]{rockafellar1970ConvexAnalysis}. Hence, as
    $\eta\to0$, the integrand of $\Psi(\zeta,\eta,\lambda,\theta_0)$
    converges pointwise to the integrand of
    $\Psi(\zeta,0,\lambda,\theta_0)$. Moreover,
    the minorant in \eqref{aux eq: usc dual minorant} implies that
    the shifted integrands
    \[
    \eta\phi^*\big(y(Z)/\eta\big)-y(Z)
    \]
    are nonnegative and increase pointwise to
    $(\phi^*)^\infty(y(Z))-y(Z)$ as $\eta\downarrow0$. Since $y(Z)$ is
    integrable under the maintained conditions, the monotone convergence
    theorem applied to these shifted integrands gives
    $\Psi(\zeta,\eta,\lambda,\theta_0)\to\Psi(\zeta,0,\lambda,\theta_0)$
    as $\eta\to0$. Since also $\eta\delta\to0$, we conclude that
    $f(\zeta,\eta,\lambda,\theta_0)\to f(\zeta,0,\lambda,\theta_0)$
    as $\eta\to0$, so the infimum over multipliers with $\eta>0$ is not larger than the
    value of $f(\cdot,\theta_0)$ at any multiplier with $\eta=0$, which
    proves \eqref{aux eq: usc dual eta reduction}.

    Fix $\varepsilon>0$ and pick
    $\xi=(\zeta,\eta,\lambda)$ with $\eta>0$ and
    $f(\xi,\theta_0)\leq V_{\mathsf{D}}(\theta_0)+2\varepsilon$, which is
    possible in view of \eqref{aux eq: usc dual eta reduction}, so that
    requirement (a) of Step 1 holds.
    It remains to verify requirement (c).
    By Assumption~\ref{assump: or-cont}, for $\Pbase{}$-almost every $z$,
    $y_n(z):=g(z;\theta_n)-\lambda^\top m(z;\theta_n)-\zeta\to
    y_0(z):=g(z;\theta_0)-\lambda^\top m(z;\theta_0)-\zeta$ along the
    given sequence. We now distinguish the two cases of
    Assumption~\ref{assump: dps-entropy}.

    In case (a), the conjugate $\phi^*$ is finite and convex, thus continuous, so
    $\eta\phi^*\big(y_n(z)/\eta\big)\to\eta\phi^*\big(y_0(z)/\eta\big)$
    for $\Pbase{}$-almost every $z$. Assumption~\ref{assump: or-dom} provides
    a radius $r>0$ and a function $G\in L_1(\Pbase{})$
    dominating the integrand of $\Psi(\xi,\cdot)$ on
    $\calB_r(\theta_0)\cap\Theta$.
    Applying Fatou's lemma to the nonnegative functions
    $z\mapsto G(z)-\eta\phi^*\big(y_n(z)/\eta\big)$ yields
    $\limsup_{n\to\infty}\Psi(\xi,\theta_n)\leq\Psi(\xi,\theta_0)$,
    and requirement (c) follows because the remaining terms of
    $f(\xi,\cdot)$ do not depend on $\theta$.

    In case (b), every subgradient of $\phi^*$ lies in
    $\dom(\phi)\subseteq[0,\bar{x}]$, because $\phi^*$ is a pointwise
    supremum of the affine maps $y\mapsto ty-\phi(t)$ indexed by
    $t\in\dom(\phi)$. Hence $\phi^*$ is finite, nondecreasing, and
    Lipschitz with constant $\bar{x}$. In particular, because
    $\phi^*(0)=0$ and the Lipschitz property,
    \[\eta\,\phi^*\big(y_n(z)/\eta\big)\leq\bar{x}\,|y_n(z)|
    \leq\bar{x}\,\big(G_g(z)+\|\lambda\|\,G_m(z)+|\zeta|\big)\]
    for all $n$ large and $\Pbase{}$-almost every $z$, by
    Assumption~\ref{assump: or-dom}. As the integrand converges $\Pbase{}$-almost
    surely by continuity of $\phi^*$, requirement (c) follows as in case
    (a). 

    The requirements (a) and (c) of Step 1 are thus satisfied in both cases
    of Assumption~\ref{assump: dps-entropy}, which shows upper semicontinuity.
\end{proof}

\subsection{Proof of Theorem~\ref{thm: existence full problem}}
\begin{proof}\label{proof:thm: existence full problem}
By Lemma~\ref{lemma: outer value regularity},
$V_{\mathsf{D}}(\cdot\,;\delta)$ is real valued and upper semicontinuous on
$\Theta$. Since $\Theta$ is nonempty and compact, the extreme value theorem
for real valued upper semicontinuous functions on compact sets yields a
point $\theta^\star\in\Theta$ such that
\[
V_{\mathsf{D}}(\theta^\star;\delta)
=\sup_{\theta\in\Theta}V_{\mathsf{D}}(\theta;\delta).
\]
Writing $V_{\mathsf{D}}^\star:=V_{\mathsf{D}}(\theta^\star;\delta)$, the
set of maximizers is
\[
\argsup_{\theta\in\Theta}V_{\mathsf{D}}(\theta;\delta)
=\bigl\{\theta\in\Theta:
V_{\mathsf{D}}(\theta;\delta)\geq V_{\mathsf{D}}^\star\bigr\}.
\]
It is nonempty and closed by upper semicontinuity, and therefore compact as
a closed subset of $\Theta$. Moreover, Assumption~\ref{assump: outer
regularity} supplies, at every $\theta\in\Theta$, the hypotheses of
Theorem~\ref{thm: existence of dual solutions}. Hence, at every outer
maximizer, the set of minimizers of \eqref{eq: dual with moments} is
nonempty, compact, and convex. This proves
part~\ref{thm: existence full problem-i}.

\uline{\textbf{Compatibility for strong duality.}}
Fix an arbitrary $\theta\in\Theta$, let $P^\circ$ denote the strictly feasible
distribution of Assumption~\ref{assump: dp-feas} or
Assumption~\ref{assump: dps-feas}, and let $\Gamma(\theta;b)$ be the attainable moment set
of \eqref{eq: attainable moment set}. The entropy function is continuous at $1$: under
Assumption~\ref{assump: dp-entropy}, $\phi(0)<\infty$ and
$\phi^\infty(1)<\infty$ imply $\dom(\phi)=[0,\infty)$, whereas under
Assumption~\ref{assump: dps-entropy}, convexity and the stated domain
conditions place $1$ in the interior of $\dom(\phi)$. The feasibility
margin also implies $\delta>0$. It remains only to verify
$\calM(\rho)\cap\rint(\Gamma(\theta;\delta))\neq\emptyset$.

When $\dimm=0$ there is no moment constraint and nothing to verify. Suppose
therefore that $\dimm\geq1$, write
$\om^\circ:=\E_{P^\circ}[m(Z;\theta)]\in\calM(\rho)$ and
$\delta_\circ:=D_\phi(P^\circ\,\|\,\Pbase{})<\delta$.
We first construct a constant $K<\infty$, a radius $r>0$, and
distributions $\{Q_v:\|v\|<r\}\subseteq\Delta(\calZ)$ with
$\E_{Q_v}[\|m(Z;\theta)\|]<\infty$, $D_\phi(Q_v\,\|\,\Pbase{})\leq K$, and
$\E_{Q_v}[m(Z;\theta)]=\rhopad+v$.

Under Assumption~\ref{assump: dual primitives}, every
$Q\in\Delta(\calZ)$ satisfies
$D_\phi(Q\,\|\,\Pbase{})\leq\phi(0)+\phi^\infty(1)=:K$, which is finite by
Assumption~\ref{assump: dp-entropy}. Indeed, writing $Q=w\Pbase{}+Q_\perp$
for the Lebesgue decomposition of $Q$ with respect to $\Pbase{}$,
convexity of $\phi$ gives
$\phi(w)\leq(1-w/T)\,\phi(0)+(w/T)\,\phi(T)$ for every $T\geq w$, and
letting $T\to\infty$ yields $\phi(w)\leq\phi(0)+\phi^\infty(1)\,w$.
Integrating against $\Pbase{}$ and invoking
Lemma~\ref{lemma: divergence lebesgue decomposition} then delivers
$D_\phi(Q\,\|\,\Pbase{})\leq\phi(0)+\phi^\infty(1)\,
\{\E_{\Pbase{}}[w(Z)]+Q_\perp(\calZ)\}=\phi(0)+\phi^\infty(1)$.
The Slater condition of Assumption~\ref{assump: dual slater} yields a radius
$r>0$ such that
$\rhopad+\calB_r(0)\subseteq\mathrm{co}(\calN(\theta))$, and
Carath\'eodory's theorem, exactly as in the proof of
Lemma~\ref{lemma: slater dual form-i}, implies that, for every $\|v\|<r$,
there exists a finitely supported $Q_v\in\Delta(\calZ)$ with
$\E_{Q_v}[m(Z;\theta)]=\rhopad+v$.

Under Assumption~\ref{assump: dual primitives superlinear}, for
$C>0$ with $[0,C]\subseteq\dom(\phi)$ let
$\calQ_C:=\{Q\in\Delta(\calZ):Q\ll\Pbase{},\ \de Q/\de\Pbase{}\leq C\}$. 
Every $Q\in\calQ_C$ satisfies
$\E_Q[\|m(Z;\theta)\|]\leq C\,\E_{\Pbase{}}[\|m(Z;\theta)\|]<\infty$,
as $\Pbase{}\in\calP(\calZ)$, and
\[
D_\phi(Q\,\|\,\Pbase{})
=\E_{\Pbase{}}[\phi(\tfrac{\de Q}{\de\Pbase{}}(Z))]
\leq\max\{\phi(0),\phi(C)\}<\infty,
\]
because a convex function on an interval is bounded by its values at the
endpoints. In case (b) of Assumption~\ref{assump: dps-entropy},
Assumption~\ref{assump: dps-ceiling} gives $r>0$ with
$\rhopad+\calB_r(0)\subseteq\calN_{\bar{x}}(\theta)$, so each
point $\rhopad+v$ with $\|v\|<r$ is the moment under some
$Q_v\in\calQ_{\bar{x}}$, and we may take
$K:=\max\{\phi(0),\phi(\bar{x})\}$, finite by
Assumption~\ref{assump: dps-entropy}. In case (a) of
Assumption~\ref{assump: dps-entropy}, let
$\calN_C(\theta):=\{\E_Q[m(Z;\theta)]:Q\in\calQ_C\}$ and
$\calN_\infty(\theta):=\bigcup_{C\geq1}\calN_C(\theta)$, which is
convex. We claim that
$\mathrm{co}(\calN(\theta))\subseteq\cl(\calN_\infty(\theta))$. Fix
$z_0\in\calZ$ and $\varepsilon>0$. The conditional distribution
$\Pbase{}(\cdot\mid \calB_\varepsilon(z_0))$ is well defined because
$\Pbase{}(\calB_\varepsilon(z_0))>0$ in view of the full-support condition
of Assumption~\ref{assump: dps-support}, belongs to $\calQ_C$ for
$C=1/\Pbase{}(\calB_\varepsilon(z_0))$, and its moment vector
$\E_{\Pbase{}}[m(Z;\theta)\mid Z\in\calB_\varepsilon(z_0)]$ converges to
$m(z_0;\theta)$ as $\varepsilon\to0$, by continuity of
$m(\cdot;\theta)$. Finite mixtures of such conditional distributions,
which again belong to some $\calQ_C$, therefore approximate every
convex combination of points of $\calN(\theta)$, proving the claim.
The Slater condition of Assumption~\ref{assump: dps-ceiling} then gives
$r_1>0$ with
$\rhopad+\calB_{r_1}(0)\subseteq\mathrm{co}(\calN(\theta))
\subseteq\cl(\calN_\infty(\theta))$, so that
$\rhopad\in\interior(\cl(\calN_\infty(\theta)))
=\interior(\calN_\infty(\theta))$, where the equality holds because a
convex set and its closure share the same relative interior
\citep[Theorem~6.3]{rockafellar1970ConvexAnalysis} and both sets are of
full dimension. Since $\calN_\infty(\theta)$ thus contains a neighborhood
of $\rhopad$, we may pick $\dimm+1$ points of $\calN_\infty(\theta)$
spanning a simplex whose interior contains $\rhopad$, where
the $i$-th point belongs to $\calN_{C_i}(\theta)$ for some $C_i\geq1$. Letting
$C^\star:=\max_i C_i$, convexity of $\calN_{C^\star}(\theta)$
yields $r>0$ with
$\rhopad+\calB_r(0)\subseteq\calN_{C^\star}(\theta)$, and we take
$K:=\max\{\phi(0),\phi(C^\star)\}$, finite as
$[0,\infty)\subseteq\dom(\phi)$.

With the family $\{Q_v\}$ at hand, fix $t\in(0,1)$ small enough that
$(1-t)\delta_\circ+tK<\delta$ and define
$P_v:=(1-t)P^\circ+tQ_v$ for $\|v\|<r$. The distribution $P_v$ belongs
to $\calP(\calZ)$, since $P^\circ\in\calP(\calZ)$ and $Q_v$ is finitely supported
under Assumption~\ref{assump: dual primitives}, whereas under
Assumption~\ref{assump: dual primitives superlinear} the density
$\de Q_v/\de\Pbase{}$ is bounded and $\Pbase{}\in\calP(\calZ)$. Moreover,
$D_\phi(P_v\,\|\,\Pbase{})\leq(1-t)\delta_\circ+tK<\delta$, by convexity
of $D_\phi(\cdot\,\|\,\Pbase{})$, and
$\E_{P_v}[m(Z;\theta)]=(1-t)\om^\circ+t(\rhopad+v)$. Hence
$\Gamma(\theta;\delta)$ contains $\calB_{tr}(\om_c)$, where
$\om_c:=(1-t)\om^\circ+t\rhopad$, so that
$\om_c\in\interior(\Gamma(\theta;\delta)) \subseteq\rint(\Gamma(\theta;\delta))$, whereas
$\om_c\in\calM(\rho)$ as a convex combination of the points
$\om^\circ$ and $\rhopad$ of the convex set $\calM(\rho)$. Since $\theta$
was arbitrary, the compatibility condition holds at every
$\theta\in\Theta$.

Under the additional integrability condition in
part~\ref{thm: existence full problem-ii},
all the conditions of Theorem~\ref{thm: strong duality} therefore hold at
every $\theta\in\Theta$. Applying that theorem pointwise gives
\[
V_{\mathsf{D}}(\theta;\delta)=V_{\mathsf{P}}(\theta;\delta)
\qquad\text{for every }\theta\in\Theta.
\]
The two value functions therefore have exactly the same maximizers, which
proves part~\ref{thm: existence full problem-ii}.
\end{proof}

\subsection{Proof of Lemma~\ref{lemma: family entropy}}
\begin{proof}\label{proof:lemma: family entropy}
\uline{\textbf{Entropy function and divergence.}} We apply
Proposition~\ref{prop: induced entropy} twice, with $P$ in place of $\Pbase{}$. First,
let $\phi_{\underline{u}}:=\phi_{\TV}+\iota_{[\underline{u},\infty)}$. If
$\underline{u}>0$, part~(ii) of that proposition with $\kappa=\underline{u}\in(0,1)$
shows that $\phi_{\underline{u}}$ is an entropy function with
$\phi_{\underline{u}}^{\infty}(1)=\phi_{\TV}^{\infty}(1)=1/2$, and that
$D_{\phi_{\underline{u}}}(Q\,\|\,P)$ equals $D_{\phi_{\TV}}(Q\,\|\,P)$ if
$\underline{u}P\leq Q$ and $+\infty$ otherwise. If $\underline{u}=0$, then
$\phi_{\underline{u}}=\phi_{\TV}$, as $\phi_{\TV}=+\infty$ on $(-\infty,0)$, and the
same conclusions hold trivially. Second, if $\overline{u}<\infty$, part~(i) with
$\kappa=1/\overline{u}\in(0,1)$, applied to $\phi_{\underline{u}}$, shows that
$\phi_{\mathbb U}=\phi_{\underline{u}}+\iota_{[0,\overline{u}]}$ is an entropy function
with $\phi_{\mathbb U}^{\infty}(1)=+\infty$, and that $D_{\phi_{\mathbb U}}(Q\,\|\,P)$
equals $D_{\phi_{\underline{u}}}(Q\,\|\,P)$ if $Q\leq\overline{u}P$ and $+\infty$
otherwise. If $\overline{u}=+\infty$, then $\phi_{\mathbb U}=\phi_{\underline{u}}$ and
there is nothing to add. Combining the two steps with the definition of
$\calP_{\mathbb U}(P)$, we obtain $D_{\phi_{\mathbb U}}(Q\,\|\,P)=+\infty$ if
$Q\notin\calP_{\mathbb U}(P)$ and
$D_{\phi_{\mathbb U}}(Q\,\|\,P)=D_{\phi_{\TV}}(Q\,\|\,P)=\TV(Q\,\|\,P)$ otherwise, where
the second equality is Lemma~\ref{lem:tv_posneg}. Finally, $\phi_{\mathbb U}$ is
continuous at $1$, as it coincides on $\mathbb{U}$ with the continuous function
$\phi_{\TV}$ and $1$ is an interior point of $\mathbb{U}$. The identity
between \eqref{eq:robustness family} and the divergence neighborhood
follows, since every $Q\in\calP_{\mathbb U}(P)$ belongs to $\calP(\calZ)$
under either condition of the lemma. This is immediate when $g(\cdot;\theta)$ and
$m(\cdot;\theta)$ are bounded. When $\overline{u}<\infty$ and $P\in\calP(\calZ)$, the
inequality $Q\leq\overline{u}P$ gives
$\E_{Q}[|g(Z;\theta)|]\leq\overline{u}\,\E_{P}[|g(Z;\theta)|]<\infty$, and likewise for
$\|m(Z;\theta)\|$.

\uline{\textbf{Conjugate and perspective.}} For $y\in\R$,
$\phi_{\mathbb U}^{*}(y)=\sup_{u\in\mathbb U}\{uy-\tfrac12|u-1|\}$. The
function $u\mapsto uy-\tfrac12|u-1|$ is concave and piecewise linear on
$\mathbb{U}$, with slope $y+\tfrac12$ on $u<1$ and slope $y-\tfrac12$ on
$u>1$, so its supremum over the interval is attained at an endpoint of
$\mathbb{U}$ or at the kink $u=1$. When $\overline{u}<\infty$ this gives
$\phi_{\mathbb U}^{*}(y)=\max_{u\in\{\underline{u},1,\overline{u}\}}
\{uy-\tfrac12|u-1|\}$, a finite function of $y$. When
$\overline{u}=+\infty$, the supremum is $+\infty$ for $y>\tfrac12$, as the
slope on $u>1$ is then positive, and equals
$\max\{\underline{u}y-\tfrac12(1-\underline{u}),\,y\}$ for $y\leq\tfrac12$,
as that slope is then nonpositive. For $\eta>0$, the perspective
is $(\phi_{\mathbb U}^{*})^{\pi}(x,\eta)=\eta\,\phi_{\mathbb U}^{*}(x/\eta)$,
and multiplying each branch through by $\eta$ yields
\eqref{eq:integrand family}, the domain restriction $x/\eta\leq\tfrac12$
becoming $x\leq\eta/2$. At $\eta=0$ the perspective is the recession
function of $\phi_{\mathbb U}^{*}$, that is,
$\lim_{t\to\infty}\phi_{\mathbb U}^{*}(y_{0}+tx)/t$ for any $y_{0}$ in the
domain of $\phi_{\mathbb U}^{*}$. When $\overline{u}<\infty$ this limit is
$\max_{u\in\{\underline{u},1,\overline{u}\}}ux$, which coincides with the
first branch of \eqref{eq:integrand family} at $\eta=0$. When
$\overline{u}=+\infty$ and $x>0$ the argument $y_{0}+tx$ exceeds $\tfrac12$
for $t$ large and the limit is $+\infty$, in agreement with the third
branch at $\eta=0$, whereas for $x\leq0$ the limit is
$\max\{\underline{u}x,x\}$, in agreement with the second branch. Hence
$(\phi_{\mathbb U}^{*})^{\pi}=\mathsf{r}$ on $\R\times\R_{+}$.
\end{proof}

\subsection{Proof of Lemma~\ref{lemma: family duality}}
\begin{proof}\label{proof:lemma: family duality}
Throughout the proof we fix $\theta$ and abbreviate $g(\cdot):=g(\cdot;\theta)$,
$m(\cdot):=m(\cdot;\theta)$, and $\calF:=\calF(\theta)$. Both $g$ and $m$
are bounded, as they are continuous on the compact $\calZ$.

\uline{\textbf{Part (i).}} Let $Q\in\calP_{\mathbb U}(P)$ have Lebesgue
decomposition $Q=wP+Q^{\perp}$ with respect to $P$, so that
$w\in\mathbb{U}$ $P$-almost surely and $Q^{\perp}=0$ when
$\overline{u}<\infty$ (Lemma~\ref{lemma: family entropy}), and let
$\xi=(\zeta,\eta,\lambda)\in\calF$. Put $x(z):=g(z)-\lambda^{\top}m(z)-\zeta$.
By \ref{prop:r-sup}, $w(z)x(z)\leq\mathsf{r}(x(z),\eta)+\eta|w(z)-1|/2$ for
$P$-almost every $z$, whence, integrating against $P$,
\[
\int x\,\de Q=\E_{P}[wx]+\int x\,\de Q^{\perp}
\leq\E_{P}[\mathsf{r}(x,\eta)]+\frac{\eta}{2}\E_{P}|w-1|
+\int x\,\de Q^{\perp}.
\]
When $\overline{u}<\infty$ the last integral vanishes. When
$\overline{u}=+\infty$, dual feasibility gives $x\leq\eta/2$ on $\calZ$, so
that $\int x\,\de Q^{\perp}\leq\tfrac{\eta}{2}Q^{\perp}(\calZ)$. In both
cases, since $\TV(Q\,\|\,P)=\tfrac12\E_{P}|w-1|+\tfrac12Q^{\perp}(\calZ)$ by
Lemmas~\ref{lemma: divergence lebesgue decomposition}
and~\ref{lem:tv_posneg},
\[
\E_{Q}[g]-\lambda^{\top}\E_{Q}[m]-\zeta=\int x\,\de Q
\leq P\ell_{\xi,\theta}+\eta\,\TV(Q\,\|\,P).
\]
Rearranging and using the definition \eqref{eq:Phi-def} of $\Jdual$ gives
the displayed inequality. If $Q\in\calP_{\mathbb U}(\theta;\rho,P,\delta)$,
then $\TV(Q\,\|\,P)\leq\delta$ and, since $\lambda\in\Rlam$ and
$\E_{Q}[m]\in\calM(\rho)$, the support function identity of
Section~\ref{subsec: setup} gives
$\lambda^{\top}\E_{Q}[m]\leq\iota^{*}_{\calM(\rho)}(\lambda)
=\lambda_{12}^{\top}\rho$, so both terms on the right are nonnegative.
Taking the infimum over $\xi\in\calF$ and the supremum over $Q$ shows that
$\Kmax(\theta;\rho,P)$ dominates the upper bound. Moreover, when the robustness
set is nonempty this bound exceeds $-\infty$, as $g$ is bounded, while
$\Kmax(\theta;\rho,P)\leq\sup_{z\in\calZ}g(z)<\infty$ by evaluating $\Jdual$
at the feasible point $(\sup_{z}g(z),0,0)$, at which the integrand equals
$\underline{u}\,(g-\sup_{z}g)\leq0$ by \eqref{eq:integrand family}.

\uline{\textbf{Part (ii).}} We apply Theorem~\ref{thm: strong duality}
with the entropy $\phi_{\mathbb U}$ of Lemma~\ref{lemma: family entropy}
and the baseline $P$, which is legitimate because that theorem is stated
for an arbitrary baseline distribution and an arbitrary entropy function
continuous at $1$. Its hypotheses hold: $\delta>0$;
$\int_{\calZ}|g|\,\de\nu<\infty$ for every
finite measure $\nu$; and the compatibility condition
$\calM(\rho)\cap\rint(\Gamma(\delta))\neq\emptyset$ holds because, by
Lemma~\ref{lemma: family entropy}, the attainable set $\Gamma(\delta)$ of
that theorem coincides with $\Gamma_{\mathbb U}(\theta;P,\delta)$, and
the constraint qualification supplies a point of $\calM(\rho)$ in its interior,
hence in its relative interior. The theorem states that the upper bound
equals the infimum over $(\zeta,\eta,\lambda)\in\R\times\R_{+}\times\Rlam$
of $\lambda_{12}^{\top}\rho+\zeta+\eta\delta+\E_{P}[(\phi_{\mathbb U}^{*})^{\pi}
(g-\lambda^{\top}m-\zeta,\eta)]$ subject to
$\zeta+\eta\phi_{\mathbb U}^{\infty}(1)\geq\sigma(\lambda,\theta)$. By
Lemma~\ref{lemma: family entropy}, the perspective is $\mathsf{r}$ and the
recession constant is $1/2$ when $\overline{u}=+\infty$, so that the
program is $\inf_{\xi\in\calF}\Jdual(\xi,\theta;\rho,P)$, and $+\infty$
when $\overline{u}<\infty$, so that the constraint is vacuous and the
program is again $\inf_{\xi\in\calF}\Jdual(\xi,\theta;\rho,P)$ with
$\calF=\R\times\R_{+}\times\Rlam$. This is \eqref{eq:dual}.

\uline{\textbf{Part (iii).}}
\uline{\textbf{Recession cone.}} When $\overline{u}<\infty$, $\calF$ is the closed
convex cone $\R\times\R_{+}\times\Rlam$ and coincides with its recession
cone. When $\overline{u}=+\infty$, $\calF$ is the sublevel set
$\{\xi:f(\xi)\leq0\}$ of the closed proper convex function
$f(\xi):=\sigma(\lambda,\theta)-\zeta-\eta/2+\iota_{\R_{+}}(\eta)
+\iota_{\Rlam}(\lambda)$, in which $\lambda\mapsto\sigma(\lambda,\theta)$
is finite, continuous, and convex on $\R^{\dimm}$, as a supremum over the
compact $\calZ$ of functions affine in $\lambda$ with bounded intercepts.
Its recession function is
$f^{\infty}(d)=\sigma^{\infty}(\lambda_{d},\theta)-\zeta_{d}-\eta_{d}/2
+\iota_{\R_{+}}(\eta_{d})+\iota_{\Rlam}(\lambda_{d})$: indeed, for any
$\lambda^{\circ}$, $\sigma(\lambda^{\circ}+t\lambda_{d},\theta)/t
=\sup_{z\in\calZ}\{(g(z)-\lambda^{\circ\top}m(z))/t-\lambda_{d}^{\top}m(z)\}
\to\sigma^{\infty}(\lambda_{d},\theta)$ as $t\to\infty$, the first term
in the braces converging to zero uniformly in $z$, while the recession function of the
indicator of a closed convex cone is the indicator itself. The sublevel-set description of
recession cones \citep[Theorem~8.7]{rockafellar1970ConvexAnalysis} then gives
$0^{+}\calF=\{d:f^{\infty}(d)\leq0\}$, which is the set displayed in the statement.

\uline{\textbf{Recession function of the objective.}} Fix $\xi^{\circ}\in\calF$ and
$d\in0^{+}\calF$, and write $x^{\circ}:=g-\lambda^{\circ\top}m-\zeta^{\circ}$
and $x_{d}:=-\lambda_{d}^{\top}m-\zeta_{d}$. For $t>0$, positive
homogeneity \ref{prop:r-hom} gives
$\mathsf{r}(x^{\circ}+tx_{d},\eta^{\circ}+t\eta_{d})/t
=\mathsf{r}(x^{\circ}/t+x_{d},\eta^{\circ}/t+\eta_{d})$, which converges to
$\mathsf{r}(x_{d},\eta_{d})$ pointwise on $\calZ$ as $t\to\infty$: when
$\overline{u}<\infty$ by continuity of $\mathsf{r}$ on
$\R\times\R_{+}$, and when $\overline{u}=+\infty$ by continuity of
$\mathsf{r}$ on its closed domain $\{x\leq\eta/2\}$, in which the points
$(x^{\circ}/t+x_{d},\eta^{\circ}/t+\eta_{d})$ lie for every $t>0$ because
$\xi^{\circ}+td\in\calF$, and to which the limit $(x_{d},\eta_{d})$ belongs
because $d\in0^{+}\calF$. The convergence is dominated, by
\ref{prop:r-lip} and the boundedness of $g$ and $m$, so the dominated
convergence theorem yields
$\lim_{t\to\infty}\{\Jdual(\xi^{\circ}+td,\theta;\rho,P)
-\Jdual(\xi^{\circ},\theta;\rho,P)\}/t=\zeta_{d}+\eta_{d}\delta
+\lambda_{d,12}^{\top}\rho+P\,\mathsf{r}(x_{d},\eta_{d})$, which is the
recession function of the closed proper convex function
$\Jdual(\cdot,\theta;\rho,P)+\iota_{\calF}$ in the direction $d$
(Definition~\ref{defi: recession function}). The function is convex because
$\ell_{\xi,\theta}(z)$ is convex in $\xi$ for every $z$ by
\ref{prop:r-hom}, closed because $\calF$ is closed and
$\xi\mapsto\Jdual(\xi,\theta;\rho,P)$ is continuous on $\calF$ by the
dominated convergence theorem and \ref{prop:r-lip}, and proper because it
is finite on $\calF$.

\uline{\textbf{Positivity.}} Suppose the constraint qualification holds at
$(\theta,\rho,P)$ with radius $r$ and point $w_{0}\in\calM(\rho)$, and let
$d\in0^{+}\calF$ be nonzero. Applying part~(i) to the program with $g$
replaced by $0$, for which $d$ is a feasible point of its dual feasible set
by the description of $0^{+}\calF$ just obtained, we get, for every
$Q\in\calP_{\mathbb U}(P)$,
\begin{equation}\label{eq:recession slack}
\Jdual^{\infty}(d,\theta;\rho,P)
\;\geq\;\eta_{d}\bigl(\delta-\TV(Q\,\|\,P)\bigr)
+\bigl(\lambda_{d,12}^{\top}\rho-\lambda_{d}^{\top}\E_{Q}[m]\bigr).
\end{equation}
We distinguish three cases. If $\eta_{d}>0$, take $Q=Q^{\circ}$, a
distribution of the family with $\E_{Q^{\circ}}[m]\in\calM(\rho)$ and
$\TV(Q^{\circ}\,\|\,P)<\delta$, which exists by part~(iv) below, whose proof
does not use the present part. Then the first term in
\eqref{eq:recession slack} is strictly positive and the second is
nonnegative, as in part~(i). If $\eta_{d}=0$ and $\lambda_{d}\neq0$, the
point $w:=w_{0}-\tfrac{r}{2}\lambda_{d}/\|\lambda_{d}\|$ lies in
$\calB_{r}(w_{0})\subseteq\Gamma_{\mathbb U}(\theta;P,\delta)$, so there
is $Q\in\calP_{\mathbb U}(P)$ with $\E_{Q}[m]=w$, and \eqref{eq:recession
slack} gives
$\Jdual^{\infty}(d,\theta;\rho,P)\geq\lambda_{d,12}^{\top}\rho
-\lambda_{d}^{\top}w_{0}+\tfrac{r}{2}\|\lambda_{d}\|\geq\tfrac{r}{2}
\|\lambda_{d}\|>0$, the second inequality by the support function identity
since $w_{0}\in\calM(\rho)$ and $\lambda_{d}\in\Rlam$. If $\eta_{d}=0$ and
$\lambda_{d}=0$, then $\zeta_{d}\neq0$ and, by
\eqref{eq:integrand family} at $\eta=0$,
$\Jdual^{\infty}(d,\theta;\rho,P)=\zeta_{d}+\mathsf{r}(-\zeta_{d},0)$.
When $\overline{u}=+\infty$ the recession-cone description forces
$\zeta_{d}\geq\sigma^{\infty}(0,\theta)=0$, hence $\zeta_{d}>0$ and
$\mathsf{r}(-\zeta_{d},0)=-\underline{u}\zeta_{d}$, so that the recession
function equals $(1-\underline{u})\zeta_{d}>0$. When $\overline{u}<\infty$
the same computation applies if $\zeta_{d}>0$, whereas if $\zeta_{d}<0$
then $\mathsf{r}(-\zeta_{d},0)=-\overline{u}\zeta_{d}$ and the recession
function equals $(\overline{u}-1)|\zeta_{d}|>0$, where the
inequality follows from $\overline{u}>1$.

\uline{\textbf{Existence.}} The closed proper convex function
$\Jdual(\cdot,\theta;\rho,P)+\iota_{\calF}$ has no nonzero direction of
recession, since its recession function is $+\infty$ off $0^{+}\calF$ and
strictly positive on $0^{+}\calF\setminus\{0\}$. By
Proposition~3.2.2 in \citet{bertsekas2009ConvexOptimizationTheory}, its
set of
minimizers is nonempty and compact, and it is convex as the set of
minimizers of a convex function. This set is $\Xmax(\theta;\rho,P)$.

\uline{\textbf{Part (iv).}} Suppose first that the constraint
qualification holds with radius $r$ and point $w_{0}\in\calM(\rho)$. Since
$\Gamma_{\mathbb U}(\theta;P,\delta)\subseteq\Gamma_{\mathbb U}(\theta;P,1)$,
the second requirement holds. For the first, let $b_{P}:=\E_{P}[m]$, which
belongs to $\Gamma_{\mathbb U}(\theta;P,\delta)$ because
$P\in\calP_{\mathbb U}(P)$ and $\TV(P\,\|\,P)=0$. If $b_{P}=w_{0}$, then $Q^{\circ}:=P$
satisfies the first requirement with $\TV(Q^{\circ}\,\|\,P)=0<\delta$. Otherwise, choose
$\varepsilon>0$ so small that $w_{1}:=w_{0}+\varepsilon(w_{0}-b_{P})\in
\calB_{r}(w_{0})$, let $Q_{1}\in\calP_{\mathbb U}(P)$ attain $w_{1}$ with
$\TV(Q_{1}\,\|\,P)\leq\delta$, and set
$Q^{\circ}:=(1+\varepsilon)^{-1}Q_{1}+\varepsilon(1+\varepsilon)^{-1}P$,
which belongs to the convex set $\calP_{\mathbb U}(P)$ and satisfies
$\E_{Q^{\circ}}[m]=(1+\varepsilon)^{-1}(w_{1}+\varepsilon b_{P})=w_{0}
\in\calM(\rho)$ and, by convexity of $\TV(\cdot\,\|\,P)$ and $\TV(P\,\|\,P)=0$,
$\TV(Q^{\circ}\,\|\,P)\leq(1+\varepsilon)^{-1}\TV(Q_{1}\,\|\,P)\leq\delta/(1+\varepsilon)
<\delta$.

Conversely, let $Q^{\circ}$ satisfy the first requirement, with
$\delta^{\circ}:=\TV(Q^{\circ}\,\|\,P)<\delta$ and $w^{\circ}:=
\E_{Q^{\circ}}[m]\in\calM(\rho)$, and let $w_{1}\in\calM(\rho)$ and $r_{1}>0$
be such that
$\calB_{r_{1}}(w_{1})\subseteq\Gamma_{\mathbb U}(\theta;P,1)$.
As $\delta^{\circ}<\delta\leq1$, we may put
$t:=(\delta-\delta^{\circ})/(1-\delta^{\circ})\in(0,1]$. For any
$w\in\calB_{r_{1}}(w_{1})$, attained by some $Q_{w}\in\calP_{\mathbb U}(P)$,
the mixture $Q_{t}:=(1-t)Q^{\circ}+tQ_{w}$ belongs to
$\calP_{\mathbb U}(P)$, has moment vector $(1-t)w^{\circ}+tw$, and satisfies
$\TV(Q_{t}\,\|\,P)\leq(1-t)\delta^{\circ}+t\,\TV(Q_{w}\,\|\,P)\leq(1-t)\delta^{\circ}+t
=\delta$, using $\TV\leq1$. Hence
$\Gamma_{\mathbb U}(\theta;P,\delta)$ contains $\calB_{tr_{1}}((1-t)w^{\circ}+tw_{1})$,
whose center is a point of the convex set $\calM(\rho)$, which is the constraint
qualification with radius $tr_{1}$. Finally, the first requirement says precisely that the
infimum \eqref{eq:mindiv-primal} is strictly below $\delta$.
\end{proof}

\subsection{Proof of Lemma~\ref{lemma: mindiv dual}}
\begin{proof}\label{proof:lemma: mindiv dual}
Throughout the proof we fix $\theta$ and abbreviate $m(\cdot):=m(\cdot;\theta)$,
$\calF_{\mathsf D}:=\calF_{\mathsf D}(\theta)$, and
$\sigma^{\infty}(\lambda):=\sigma^{\infty}(\lambda,\theta)$. We recall
that $\lambda^{\top}\rhopad=\lambda_{12}^{\top}\rho$ for every
$\lambda\in\R^{\dimm}$, and set $\overline{M}:=\sup_{z\in\calZ}\|m(z)\|$,
which is finite because $m$ is continuous on the compact $\calZ$. For
$\lambda\in\R^{\dimm}$ we write $h_{\lambda}:=-\lambda^{\top}m(\cdot)$, a
continuous function on $\calZ$ with $\sup_{z}h_{\lambda}(z)
=\sigma^{\infty}(\lambda)\leq\|\lambda\|\overline M$, and we call
$\zeta$ admissible for $\lambda$ if $(\zeta,\lambda)\in\calF_{\mathsf D}$,
that is, if $\zeta\geq\sigma^{\infty}(\lambda)-\tfrac12$ when
$\overline{u}=+\infty$ and always when $\overline{u}<\infty$.

\uline{\textbf{A variational inequality.}} We first show that for every
$(\zeta,\lambda)\in\calF_{\mathsf D}$ and every
$Q\in\calP_{\mathbb U}(P)$,
\begin{equation}\label{eq:varrep}
  \E_{Q}[h_{\lambda}]-\zeta\;\leq\;
  \E_{P}\bigl[\mathsf{r}(h_{\lambda}-\zeta,1)\bigr]+\TV(Q\,\|\,P).
\end{equation}
This is the inequality of Lemma~\ref{lemma: family duality-i}, applied with
$0$ in place of $g(\cdot;\theta)$ and with $\eta=1$, once the terms $\delta$ and
$\lambda_{12}^{\top}\rho$ common to both sides cancel. Indeed, for this choice the dual
support function reduces to $\sigma^{\infty}(\lambda)$, so that $(\zeta,1,\lambda)$
belongs to the dual feasible set \eqref{eq:F-def} if and only if
$(\zeta,\lambda)\in\calF_{\mathsf D}$.

\uline{\textbf{Attainment of the primal projection.}} Let
$\mathcal K_{\theta,\rho}:=\{Q\in\calP_{\mathbb U}(P):\E_{Q}[m(Z;\theta)]
\in\calM(\rho)\}$. Because $\calZ$ is compact, $\Delta(\calZ)$ is weakly
compact. The set $\calP_{\mathbb U}(P)$ is weakly closed: if $Q_{n}\Rightarrow
Q$ with $Q_{n}\in\calP_{\mathbb U}(P)$, the portmanteau theorem gives
$Q(U)\leq\liminf_{n}Q_{n}(U)\leq\overline{u}P(U)$ for every open $U$ and
$Q(C)\geq\limsup_{n}Q_{n}(C)\geq\underline{u}P(C)$ for every closed $C$. 
Continuity of $m$ and closedness of $\calM(\rho)$ imply that
$\mathcal K_{\theta,\rho}$ is weakly closed, hence weakly compact. Moreover,
on the compact metric space $\calZ$,
$2\TV(Q\,\|\,P)=\sup\{|\E_{Q}[a]-\E_{P}[a]|:a\in C(\calZ),\|a\|_{\infty}\leq1\}$,
so $Q\mapsto\TV(Q\,\|\,P)$ is weakly lower semicontinuous, as a supremum of
weakly continuous functions. It follows that the infimum in
\eqref{eq:mindiv-primal} is attained whenever $\mathcal K_{\theta,\rho}$ is
nonempty. The same argument shows that
$\{Q\in\calP_{\mathbb U}(P):\TV(Q\,\|\,P)\leq\delta'\}$ is weakly compact for
every $\delta'\geq0$, so that, as $Q\mapsto\E_{Q}[m]$ is weakly continuous
and linear, the attainable sets $\Gamma_{\mathbb U}(\theta;P,\delta')$
are compact and convex.

\uline{\textbf{Part (i).}} Suppose $\calB_{r}(w_{0})\subseteq
\Gamma_{\mathbb U}(\theta;P,1)$ with $w_{0}\in\calM(\rho)$. We prove the
weak direction first, then the multiplier and intercept bounds, then the
strong direction, and finally attainment.

\uline{\textbf{Weak direction.}} Fix $(\zeta,\lambda)\in\calF_{\mathsf D}$ and let
$Q$ be feasible for the projection problem, so that
$w:=\E_{Q}[m]\in\calM(\rho)$. Since $\lambda\in\Rlam$, the support function
of $\calM(\rho)$ gives $\lambda^{\top}w\leq\lambda_{12}^{\top}\rho$, whence
$\E_{Q}[h_{\lambda}]=-\lambda^{\top}w\geq-\lambda_{12}^{\top}\rho$. Combining
with \eqref{eq:varrep}, $\TV(Q\,\|\,P)\geq v_{P}(\zeta,\lambda)$. Taking the
infimum over feasible $Q$ and the supremum over
$(\zeta,\lambda)\in\calF_{\mathsf D}$ yields
$\mathsf D(\theta;\rho,P)\geq\sup_{\calF_{\mathsf D}}v_{P}$.

\uline{\textbf{Multiplier bound.}} For $(\zeta,\lambda)\in\calF_{\mathsf D}$ with
$\lambda\neq0$ and any $r'<r$, the point
$w:=w_{0}-r'\lambda/\|\lambda\|$ lies in $\calB_{r}(w_{0})$ and is therefore
attained by some $Q_{w}\in\calP_{\mathbb U}(P)$. Applying \eqref{eq:varrep}
to $Q_{w}$ and using $\TV\leq1$,
\[
  1\;\geq\;\TV(Q_{w}\,\|\,P)\;\geq\;-\zeta-\lambda^{\top}w
  -\E_{P}\bigl[\mathsf{r}(h_{\lambda}-\zeta,1)\bigr]
  \;=\;v_{P}(\zeta,\lambda)+\lambda_{12}^{\top}\rho-\lambda^{\top}w
  \;\geq\;v_{P}(\zeta,\lambda)+r'\|\lambda\|,
\]
the last inequality because $\lambda_{12}^{\top}\rho\geq\lambda^{\top}w_{0}$
and $\lambda^{\top}(w_{0}-w)=r'\|\lambda\|$. The point
$(\zeta,\lambda)=(-\tfrac12,0)$ is admissible and, by
\eqref{eq:integrand family}, $v_{P}(-\tfrac12,0)=-\mathsf{r}(\tfrac12,1)
+\tfrac12=0$, so $\sup_{\calF_{\mathsf D}}v_{P}\geq0$, and every
$(\zeta,\lambda)$ with $v_{P}(\zeta,\lambda)\geq0$, in particular every
maximizer, satisfies $\|\lambda\|\leq1/r'$ for every $r'<r$, hence
$\|\lambda\|\leq1/r$.

\uline{\textbf{The intercept.}} Fix $\lambda$ and consider
$F_{\lambda}(\zeta):=\zeta+\E_{P}[\mathsf{r}(h_{\lambda}-\zeta,1)]$ on the
set of admissible $\zeta$, so that $v_{P}(\zeta,\lambda)=-F_{\lambda}(\zeta)
-\lambda_{12}^{\top}\rho$. The function $F_{\lambda}$ is finite and convex,
as $\mathsf{r}(\cdot,1)$ is convex, and its one-sided derivatives are
\[
F'_{\lambda,+}(\zeta)=1-\E_{P}\bigl[\mathsf{r}'_{-}(h_{\lambda}-\zeta)\bigr],
\qquad
F'_{\lambda,-}(\zeta)=1-\E_{P}\bigl[\mathsf{r}'_{+}(h_{\lambda}-\zeta)\bigr],
\]
where $\mathsf{r}'_{\pm}$ denote the one-sided derivatives of
$x\mapsto\mathsf{r}(x,1)$ on its domain, by the dominated convergence
theorem applied to difference quotients bounded by $u^{\dagger}$
(\ref{prop:r-lip}). From \eqref{eq:integrand family}, $\mathsf{r}'_{-}(x)$
equals $\underline{u}$ for $x\leq-\tfrac12$, $1$ for
$-\tfrac12<x\leq\tfrac12$, and $\overline{u}$ for $x>\tfrac12$, while
$\mathsf{r}'_{+}(x)$ equals $\underline{u}$ for $x<-\tfrac12$, $1$ for
$-\tfrac12\leq x<\tfrac12$, and $\overline{u}$ for $x\geq\tfrac12$, where these
last two values are relevant only when $\overline{u}<\infty$. In every
case the one-sided derivatives take values in $\mathbb{U}$ and the
subdifferential of $\mathsf{r}(\cdot,1)$ at $x$ is the interval
$[\mathsf{r}'_{-}(x),\mathsf{r}'_{+}(x)]\subseteq\mathbb{U}$. We claim that
$F_{\lambda}$ attains its minimum over the admissible set and that every
minimizer $\zeta^{*}$ satisfies $|\zeta^{*}|\leq\|\lambda\|\overline M
+\tfrac12$. When $\overline{u}<\infty$ every $\zeta\in\R$ is admissible. For
$\zeta>\sigma^{\infty}(\lambda)+\tfrac12$ we have $h_{\lambda}-\zeta<-\tfrac12$
on $\calZ$, so $F'_{\lambda,+}(\zeta)=1-\underline{u}>0$, and for
$\zeta<\inf_{z}h_{\lambda}(z)-\tfrac12$ we have $h_{\lambda}-\zeta>\tfrac12$,
so $F'_{\lambda,-}(\zeta)=1-\overline{u}<0$, as
$\overline{u}>1$. The convex function
$F_{\lambda}$ is therefore increasing to the right of
$\sigma^{\infty}(\lambda)+\tfrac12$ and decreasing to the left of
$\inf_{z}h_{\lambda}-\tfrac12$, so it attains its minimum, and only there,
on the compact interval between these two points, whose endpoints are
bounded by $\|\lambda\|\overline M+\tfrac12$ in absolute value. When
$\overline{u}=+\infty$, the admissible set is the closed half line
$[\zeta_{\lambda},\infty)$ with $\zeta_{\lambda}:=\sigma^{\infty}(\lambda)
-\tfrac12$, on which $h_{\lambda}-\zeta\leq\tfrac12$ and hence
$\mathsf{r}'_{-}(h_{\lambda}-\zeta)\leq1$, so that $F'_{\lambda,+}\geq0$
throughout: $F_{\lambda}$ is nondecreasing on the half line and the
endpoint $\zeta_{\lambda}$ is a minimizer. Any other minimizer $\zeta^{*}$
makes $F_{\lambda}$ constant on $[\zeta_{\lambda},\zeta^{*}]$, hence
$\E_{P}[\mathsf{r}'_{-}(h_{\lambda}-\zeta')]=1$, that is
$P(h_{\lambda}-\zeta'\leq-\tfrac12)=0$, for every
$\zeta'\in(\zeta_{\lambda},\zeta^{*})$, which forces $\zeta^{*}\leq
\inf_{z}h_{\lambda}(z)+\tfrac12$ up to $P$-null sets and in particular
$|\zeta^{*}|\leq\|\lambda\|\overline M+\tfrac12$. Combined with the
multiplier bound, every maximizer of $v_{P}$ over $\calF_{\mathsf D}$
satisfies $|\zeta^{*}|\leq\overline M/r+\tfrac12$, as claimed.

\uline{\textbf{Strong direction.}} Define $T:\R^{\dimm}\to[0,+\infty]$ by
\[
  T(w):=\inf\bigl\{\TV(Q\,\|\,P):Q\in\calP_{\mathbb U}(P),\ \E_{Q}[m]=w\bigr\},
\]
with $T(w)=+\infty$ when no such $Q$ exists, so that
$\mathsf D(\theta;\rho,P)=\inf_{w\in\R^{\dimm}}\{T(w)+\iota_{\calM(\rho)}(w)\}$
by definition. The function $T$ is proper, since $T\geq0$ and
$T(\E_{P}[m])=0$, its effective domain is
$\Gamma_{\mathbb U}(\theta;P,1)$, and it is convex: given $w,w'$ with
near-optimal $Q,Q'$ and $t\in(0,1)$, the mixture $tQ+(1-t)Q'$ belongs to
the convex set $\calP_{\mathbb U}(P)$, attains the moment vector
$tw+(1-t)w'$, and, by convexity of $\TV(\cdot\,\|\,P)$, satisfies
$\TV(tQ+(1-t)Q'\,\|\,P)\leq t\,\TV(Q\,\|\,P)+(1-t)\,\TV(Q'\,\|\,P)$. We apply
Lemma~\ref{lemma: fenchel duality polyhedral} to the proper convex
functions $T$ and $\iota_{\calM(\rho)}$, where the latter is polyhedral. The
hypothesis of the lemma is that $\calM(\rho)$ meets the relative interior
of $\dom(T)$, and $w_{0}\in\calM(\rho)$ is an interior point of $\dom(T)$
by assumption. The lemma yields, with the supremum attained because the common
value is finite,
\[
  \mathsf D(\theta;\rho,P)
  =\sup_{x^{*}\in\R^{\dimm}}
  \{-T^{*}(x^{*})-\iota_{\calM(\rho)}^{*}(-x^{*})\}
  =\max_{\lambda\in\Rlam}\bigl\{-\lambda_{12}^{\top}\rho
  -T^{*}(-\lambda)\bigr\},
\]
where, at $\lambda:=-x^{*}$, the support function
$\iota^{*}_{\calM(\rho)}(\lambda)$ equals $\lambda_{12}^{\top}\rho$ when
$\lambda\in\Rlam$ and $+\infty$ otherwise. It remains to compute, for
$\lambda\in\Rlam$,
\[
  T^{*}(-\lambda)=\sup_{w\in\R^{\dimm}}\{-\lambda^{\top}w-T(w)\}
  =\sup_{Q\in\calP_{\mathbb U}(P)}\bigl\{\E_{Q}[h_{\lambda}]-\TV(Q\,\|\,P)\bigr\},
\]
where the second equality collapses the partial minimization inside $T$.
We claim that
\begin{equation}\label{eq:scalar-conjugate}
  \sup_{Q\in\calP_{\mathbb U}(P)}\bigl\{\E_{Q}[h_{\lambda}]-\TV(Q\,\|\,P)\bigr\}
  =\min_{\zeta\ \text{admissible for}\ \lambda}F_{\lambda}(\zeta).
\end{equation}
The inequality ``$\leq$'' is \eqref{eq:varrep} rearranged, valid for every
admissible $\zeta$. For ``$\geq$'' we exhibit, for a minimizer $\zeta^{*}$
of $F_{\lambda}$ over the admissible set, a distribution $Q^{*}\in
\calP_{\mathbb U}(P)$ with $\E_{Q^{*}}[h_{\lambda}]-\TV(Q^{*}\,\|\,P)
=F_{\lambda}(\zeta^{*})$, by matching the mass of the pointwise maximizers
in \ref{prop:r-sup}. Recall that $u\in[\mathsf{r}'_{-}(x),\mathsf{r}'_{+}(x)]$
if and only if $u\in\mathbb{U}$ attains the supremum in \ref{prop:r-sup}
at $(x,1)$, that is, $ux=\mathsf{r}(x,1)+\tfrac12|u-1|$, the standard
characterization of the subdifferential of a conjugate function. Suppose
first that $\overline{u}<\infty$, so that every $\zeta\in\R$ is
admissible and any minimizer $\zeta^{*}$ satisfies
$F'_{\lambda,-}(\zeta^{*})\leq0\leq F'_{\lambda,+}(\zeta^{*})$. Then
$\E_{P}[\mathsf{r}'_{-}(h_{\lambda}-\zeta^{*})]\leq1\leq
\E_{P}[\mathsf{r}'_{+}(h_{\lambda}-\zeta^{*})]$, and there is $t\in[0,1]$
such that $w^{*}:=(1-t)\mathsf{r}'_{-}(h_{\lambda}-\zeta^{*})
+t\,\mathsf{r}'_{+}(h_{\lambda}-\zeta^{*})$ satisfies $\E_{P}[w^{*}]=1$.
As $\mathsf{r}'_{-}$ and $\mathsf{r}'_{+}$ take values in
$\mathbb{U}$, so does $w^{*}$. Set $Q^{*}:=w^{*}P\in\calP_{\mathbb
U}(P)$. Since $w^{*}(z)$ lies in the subdifferential of $\mathsf{r}(\cdot,1)$
at $h_{\lambda}(z)-\zeta^{*}$,
$w^{*}(h_{\lambda}-\zeta^{*})=\mathsf{r}(h_{\lambda}-\zeta^{*},1)
+\tfrac12|w^{*}-1|$ pointwise, and integrating against $P$ gives
$\E_{Q^{*}}[h_{\lambda}]-\zeta^{*}=\E_{P}[\mathsf{r}(h_{\lambda}-\zeta^{*},1)]
+\TV(Q^{*}\,\|\,P)$, that is, $\E_{Q^{*}}[h_{\lambda}]-\TV(Q^{*}\,\|\,P)
=F_{\lambda}(\zeta^{*})$. Suppose next that
$\overline{u}=+\infty$, and take $\zeta^{*}=\zeta_{\lambda}$, which is a
minimizer by the intercept step and at which
$F'_{\lambda,+}(\zeta_{\lambda})\geq0$, that is,
$\E_{P}[w^{*}]\leq1$ for $w^{*}:=\mathsf{r}'_{-}(h_{\lambda}-\zeta_{\lambda})
\in\{\underline{u},1\}$. Let $z^{*}\in\calZ$ attain the supremum of the
continuous $h_{\lambda}$ on the compact $\calZ$, set
$s:=1-\E_{P}[w^{*}]\geq0$, and $Q^{*}:=w^{*}P+s\,\delta^{\mathtt{Dirac}}_{z^{*}}$, a
distribution in $\calP_{\mathbb U}(P)$ because $Q^{*}\geq w^{*}P\geq
\underline{u}P$ and no cap is imposed. Then, using
$h_{\lambda}(z^{*})-\zeta_{\lambda}=\tfrac12$ and the subdifferential
identity for $w^{*}$,
\[
  \E_{Q^{*}}[h_{\lambda}]-\zeta_{\lambda}
  =\E_{P}[w^{*}(h_{\lambda}-\zeta_{\lambda})]+\tfrac{s}{2}
  =\E_{P}[\mathsf{r}(h_{\lambda}-\zeta_{\lambda},1)]
  +\tfrac12\E_{P}|w^{*}-1|+\tfrac{s}{2}
  \geq\E_{P}[\mathsf{r}(h_{\lambda}-\zeta_{\lambda},1)]+\TV(Q^{*}\,\|\,P),
\]
the last inequality because $Q^{*}-P=(w^{*}-1)P+s\,\delta^{\mathtt{Dirac}}_{z^{*}}$ and
the total variation norm is subadditive, so that
$\TV(Q^{*}\,\|\,P)\leq\tfrac12\E_{P}|w^{*}-1|+\tfrac{s}{2}$ by
Lemma~\ref{lem:tv_posneg}.
Hence $\E_{Q^{*}}[h_{\lambda}]-\TV(Q^{*}\,\|\,P)\geq F_{\lambda}(\zeta_{\lambda})$,
which proves \eqref{eq:scalar-conjugate}. Combining
\eqref{eq:scalar-conjugate} with the Fenchel display,
\[
  \mathsf D(\theta;\rho,P)
  =\max_{\lambda\in\Rlam}\Bigl\{-\lambda_{12}^{\top}\rho
  -\min_{\zeta\ \text{admissible}}F_{\lambda}(\zeta)\Bigr\}
  =\sup_{(\zeta,\lambda)\in\calF_{\mathsf D}}v_{P}(\zeta,\lambda),
\]
which, together with the weak direction, proves the equality
in~\eqref{eq:mindiv}.

\uline{\textbf{Attainment.}} By the multiplier bound the supremum in
\eqref{eq:mindiv} may be restricted to $\|\lambda\|\leq1/r$, and by the
intercept analysis the supremum over $\zeta$ at each such $\lambda$ is
attained on the compact interval
$I_{\lambda}:=[-\overline M/r-\tfrac12,\overline M/r+\tfrac12]$
intersected with the admissible set, which is nonempty. The map
$(\zeta,\lambda)\mapsto v_{P}(\zeta,\lambda)$ is jointly continuous on
$\{(\zeta,\lambda)\in\calF_{\mathsf D}:\|\lambda\|\leq1/r\}$ by the
dominated convergence theorem and \ref{prop:r-lip}, and the correspondence
$\lambda\mapsto I_{\lambda}\cap\{\zeta:(\zeta,\lambda)\in\calF_{\mathsf D}\}$
is compact-valued and continuous on the compact set
$\{\lambda\in\Rlam:\|\lambda\|\leq1/r\}$, since $\sigma^{\infty}$ is
continuous as the support function of the compact set
$\{-m(z):z\in\calZ\}$. Berge's maximum theorem then makes
$\lambda\mapsto\max_{\zeta}v_{P}(\zeta,\lambda)$ continuous on that compact
set, so the supremum over $\lambda$ is attained as well.

\uline{\textbf{Part (ii).}} \uline{\textbf{Finite side.}} If
$\mathsf D(\theta;\rho,P)\leq\delta$, then by the attainment of the primal
projection there is $Q^{\circ}\in\calP_{\mathbb U}(P)$ with
$\E_{Q^{\circ}}[m]\in\calM(\rho)$ and $\TV(Q^{\circ}\,\|\,P)\leq\delta$, so the
robustness set \eqref{eq:robustness family} is nonempty and
$\Kmax(\theta;\rho,P)$ is finite by Lemma~\ref{lemma: family duality-i}.

\uline{\textbf{Infinite side.}} Suppose the hypothesis of part~(i) holds and
$\mathsf D(\theta;\rho,P)>\delta$. By part~(i) there is
$(\tilde\zeta,\tilde\lambda)\in\calF_{\mathsf D}$ with
$v_{P}(\tilde\zeta,\tilde\lambda)>\delta$. Fix any $\xi^{0}\in\calF(\theta)$,
which exists as the set is nonempty, and consider the ray $\xi^{0}+t\,d$, $t\geq0$, in
the direction $d:=(\tilde\zeta,1,\tilde\lambda)$. Each point of the ray is
dual feasible: $\eta^{0}+t\geq0$, $\lambda^{0}+t\tilde\lambda\in\Rlam$
because $\Rlam$ is a convex cone, and, when $\overline{u}=+\infty$, by
subadditivity of the supremum together with the two feasibility
constraints,
\[
  \sigma(\lambda^{0}+t\tilde\lambda,\theta)
  \leq\Bigl(\zeta^{0}+\frac{\eta^{0}}{2}\Bigr)
  +t\Bigl(\tilde\zeta+\frac12\Bigr)
  =(\zeta^{0}+t\tilde\zeta)+\frac{\eta^{0}+t}{2}.
\]
Along the ray, by positive homogeneity \ref{prop:r-hom} and the dominated
convergence theorem, exactly as in the computation of the recession
function in the proof of Lemma~\ref{lemma: family duality-iii},
\[
  \lim_{t\to\infty}\frac{\Jdual(\xi^{0}+t\,d,\theta;\rho,P)}{t}
  =\tilde\zeta+\delta+\tilde\lambda_{12}^{\top}\rho
  +\E_{P}\bigl[\mathsf{r}(-\tilde\zeta-\tilde\lambda^{\top}m(Z),1)\bigr]
  =\delta-v_{P}(\tilde\zeta,\tilde\lambda)<0 .
\]
Hence $\Jdual(\xi^{0}+t\,d,\theta;\rho,P)\to-\infty$ and
$\Kmax(\theta;\rho,P)=-\infty$.

\uline{\textbf{Part (iii).}} Write
$\overline{M}_{\Theta}:=\sup_{(z,\theta')\in\calZ\times\Theta}\|m(z;\theta')\|$
and $d_{m}(\theta_{1},\theta_{2}):=\sup_{z\in\calZ}\|m(z;\theta_{1})
-m(z;\theta_{2})\|$, which tends to zero as $\theta_{2}\to\theta_{1}$ by the
uniform continuity of $m$ on the compact $\calZ\times\Theta$.

\uline{\textbf{Stability of the attainable set.}} For a unit vector $u\in\R^{\dimm}$,
$\theta'\in\Theta$, and $P'\in\Delta(\calZ)$, let
$h_{\Gamma}(u,\theta';P'):=\sup\{u^{\top}w:w\in\Gamma_{\mathbb U} (\theta';P',\delta)\}$
be the support function of the attainable set, a compact convex set by the attainment
step. This is the upper bound over the family with integrand $u^{\top}m(\cdot;\theta')$
and no moment restrictions, so by Lemma~\ref{lemma: family duality-ii}, whose constraint
qualification is vacuous in the absence of moment restrictions,
\[
  h_{\Gamma}(u,\theta';P')=\inf\bigl\{\zeta+\eta\delta
  +P'\mathsf{r}\bigl(u^{\top}m(\cdot;\theta')-\zeta,\eta\bigr):
  (\zeta,\eta)\in\calF_{0}(u,\theta')\bigr\},
\]
where $\calF_{0}(u,\theta'):=\{(\zeta,\eta)\in\R\times\R_{+}:\zeta+\eta/2
\geq\sup_{z}u^{\top}m(z;\theta')\}$ when $\overline{u}=+\infty$ and
$\calF_{0}(u,\theta'):=\R\times\R_{+}$ otherwise. The near-minimizers of
this program lie in a fixed compact set. Indeed, the point
$(\overline{M}_{\Theta},0)$ is feasible with objective at most
$\overline{M}_{\Theta}$, as the integrand is $\underline{u}\,(u^{\top}m
-\overline{M}_{\Theta})\leq0$ there by \eqref{eq:integrand family}, so
$h_{\Gamma}\leq\overline{M}_{\Theta}$. On the other hand, for any feasible
$(\zeta,\eta)$, Lemma~\ref{lemma: family duality-i} with $Q=P'$ gives an
objective at least $-\overline{M}_{\Theta}+\eta\delta$, and
\ref{prop:r-lb} gives an objective at least
$(1-\underline{u})\zeta-\underline{u}\,\overline{M}_{\Theta}
-\tfrac{1-\underline{u}}{2}\eta$ and, when $\overline{u}<\infty$, at least
$(1-\overline{u})\zeta-\overline{u}\,\overline{M}_{\Theta}
-\tfrac{\overline{u}-1}{2}\eta$, whereas when $\overline{u}=+\infty$
feasibility gives $\zeta\geq-\overline{M}_{\Theta}-\eta/2$. Hence every
feasible $(\zeta,\eta)$ whose objective is at most $h_{\Gamma}+1$ satisfies
$0\leq\eta\leq C_{\eta}:=(2\overline{M}_{\Theta}+1)/\delta$ and
$|\zeta|\leq C_{\zeta}$ for a constant $C_{\zeta}$ depending only on
$\overline{M}_{\Theta}$, $\delta$, and $\mathbb{U}$. Let
$\calA_{\Gamma}:=\{(\zeta,\eta,\lambda):|\zeta|\leq C_{\zeta},\ 0\leq\eta\leq
C_{\eta},\ \|\lambda\|\leq1\}$, a compact set, and note that the integrand
$\mathsf{r}(u^{\top}m(\cdot;\theta')-\zeta,\eta)$ of a feasible
$(\zeta,\eta)$ with $(\zeta,\eta,-u)\in\calA_{\Gamma}$ is the member of
$\calH^{\infty}_{\calA_{\Gamma}}$ with index $(\zeta,\eta,-u)$, as,
when $\overline{u}=+\infty$, feasibility
is exactly the requirement $\zeta+\eta/2\geq\sigma^{\infty}(-u,\theta')$
of \eqref{eq:recession class}, while this requirement is
absent when $\overline{u}<\infty$. Now let $(\theta',P')$ and $(\theta'',P'')$ be given,
fix $\varepsilon>0$, and pick $(\zeta,\eta)\in\calF_{0}(u,\theta')$
$\varepsilon$-optimal at $(\theta',P')$ with $\varepsilon\leq1$, so that
$(\zeta,\eta,-u)\in \calA_{\Gamma}$. When $\overline{u}=+\infty$ the shift
$s:=[\sup_{z}u^{\top}m(z;\theta'')-\sup_{z}u^{\top}m(z;\theta')]_{+}\leq
d_{m}(\theta',\theta'')$ makes $(\zeta+s,\eta)$ feasible at $\theta''$, and we set $s:=0$
when $\overline{u}<\infty$. Then, by \ref{prop:r-lip},
\begin{align*}
  h_{\Gamma}(u,\theta'';P'')
  &\leq\zeta+s+\eta\delta+P''\mathsf{r}\bigl(u^{\top}m(\cdot;\theta'')
  -\zeta-s,\eta\bigr)\\
  &\leq\zeta+\eta\delta+P'\mathsf{r}\bigl(u^{\top}m(\cdot;\theta')-\zeta,
  \eta\bigr)+\|P'-P''\|_{\calH^{\infty}_{\calA_{\Gamma}}}
  +s+u^{\dagger}\bigl(d_{m}(\theta',\theta'')+s\bigr)\\
  &\leq h_{\Gamma}(u,\theta';P')+\varepsilon
  +\|P'-P''\|_{\calH^{\infty}_{\calA_{\Gamma}}}
  +(1+2u^{\dagger})\,d_{m}(\theta',\theta''),
\end{align*}
where the middle inequality splits the $P''$-integral into the
$P'$-integral of the same function, whose index lies in $\calA_{\Gamma}$, plus
the difference in baselines, plus the $P''$-integral of the pointwise
change in the integrand, which the Lipschitz property bounds. When
$\overline{u}=+\infty$ both arguments of $\mathsf{r}$ lie in its domain by
feasibility. Letting $\varepsilon\to0$ and exchanging the roles of
the two pairs,
\[
  \sup_{\|u\|=1}\bigl|h_{\Gamma}(u,\theta';P')
  -h_{\Gamma}(u,\theta'';P'')
  \bigr|\leq(1+2u^{\dagger})\,d_{m}(\theta',\theta'')
  +\|P'-P''\|_{\calH^{\infty}_{\calA_{\Gamma}}}.
\]

\uline{\textbf{Persistence.}} Let $w_{0}\in\calM(\rho)$ with $\calB_{r}(w_{0})
\subseteq\Gamma_{\mathbb U}(\theta;P,\delta)$, and let
$(\theta',\rho',P')$ satisfy $(1+2u^{\dagger})d_{m}(\theta,\theta')
+\|P-P'\|_{\calH^{\infty}_{\calA_{\Gamma}}}<r/3$ and $\|\rho'-\rho\|<r/3$.
Writing $h'$ for the support function of
$\Gamma':=\Gamma_{\mathbb U} (\theta';P',\delta)$ and $h$ for that of
$\Gamma:=\Gamma_{\mathbb U} (\theta;P,\delta)$, the inclusion
$\calB_{r}(w_{0})\subseteq\Gamma$ reads $h(u)\geq u^{\top}w_{0}+r$ for every unit $u$, so
the display gives $h'(u)\geq u^{\top}w_{0}+2r/3$ for every unit $u$, that is,
$\calB_{2r/3}(w_{0})\subseteq\Gamma'$, as the set $\Gamma'$ is closed and
convex. Let $w_{0}':=w_{0}+\rhopad'-\rhopad$, where $\rhopad'$ is the padded
target of $\rho'$. Then $w_{0}'\in\calM(\rho')$, the shift acting on the
first two blocks only and preserving the inequalities and equalities
defining $\calM(\rho)$, and $\|w_{0}'-w_{0}\|=\|\rho'-\rho\|<r/3$, so that
$\calB_{r/3}(w_{0}')\subseteq\calB_{2r/3}(w_{0})\subseteq\Gamma'$. This is
the constraint qualification at $(\theta',\rho',P')$ with radius $r/3$.
Choosing $\varepsilon>0$ such that $\|\theta'-\theta\|<\varepsilon$ implies
$(1+2u^{\dagger})d_{m}(\theta,\theta')<r/6$, which is possible by uniform
continuity, and $\varepsilon\leq r/6$, the persistence claim holds on the
neighborhood of the statement for any compact $\calA_{\mathsf D}\supseteq
\calA_{\Gamma}$, since $\|\cdot\|_{\calH^{\infty}_{\calA_{\Gamma}}}\leq
\|\cdot\|_{\calH^{\infty}_{\calA_{\mathsf D}}}$.

\uline{\textbf{Lipschitz continuity.}} Let $(\theta_{1},\rho_{1},P_{1})$ and
$(\theta_{2},\rho_{2},P_{2})$ lie in the neighborhood just constructed, so
that the constraint qualification holds at both with radius $r/3$, and
write $\mathsf D_{i}:=\mathsf D(\theta_{i};\rho_{i},P_{i})$,
$d_{m}:=d_{m}(\theta_{1},\theta_{2})$, and
$\sigma^{\infty}_{i}(\lambda):=\sigma^{\infty}(\lambda,\theta_{i})$. By
part~(i) applied at $(\theta_{1},\rho_{1},P_{1})$ with radius $r/3$, there
is a maximizer $(\zeta_{1},\lambda_{1})$ of $v_{P_{1}}$ with
$\|\lambda_{1}\|\leq3/r$ and $|\zeta_{1}|\leq\tfrac12+3\overline{M}_{\Theta}/r$.
When $\overline{u}=+\infty$ the shift
$s:=[\sigma^{\infty}_{2}(\lambda_{1})-\sigma^{\infty}_{1}(\lambda_{1})]_{+}
\leq\|\lambda_{1}\|d_{m}$ makes $(\zeta_{1}+s,\lambda_{1})$ admissible at
$\theta_{2}$, and we set $s:=0$ when $\overline{u}<\infty$. Set
$C_{2}:=\tfrac12+3\overline{M}_{\Theta}/r+6\overline{M}_{\Theta}/r$ and let
$\calA_{\mathsf D}$ be a compact set containing $\calA_{\Gamma}$ and
$\{(\zeta,1,\lambda):|\zeta|\leq C_{2},\ \|\lambda\|\leq3/r\}$, so that the
integrands $f_{1}:=\mathsf{r}(-\zeta_{1}-\lambda_{1}^{\top}m(\cdot;
\theta_{1}),1)$ and $\mathsf{r}(-\zeta_{1}-s-\lambda_{1}^{\top}m(\cdot;
\theta_{2}),1)$ belong to $\calH^{\infty}_{\calA_{\mathsf D}}$, their indices
satisfying the recession-cone requirement of \eqref{eq:recession class} by
admissibility when $\overline{u}=+\infty$, while this
requirement is absent when $\overline{u}<\infty$. Evaluating the $\theta_{2}$-problem at
$(\zeta_{1}+s,\lambda_{1})$, which bounds $\mathsf D_{2}$ from below by the
weak direction of part~(i), and using that $\mathsf{r}(\cdot,1)$ is
$u^{\dagger}$-Lipschitz on its domain,
\begin{align*}
  \mathsf D_{2}
  &\geq -\E_{P_{2}}\bigl[\mathsf{r}\bigl(-\zeta_{1}-s
  -\lambda_{1}^{\top}m(Z;\theta_{2}),1\bigr)\bigr]-\zeta_{1}-s
  -\lambda_{1,12}^{\top}\rho_{2}\\
  &\geq -\E_{P_{1}}[f_{1}]-\zeta_{1}-\lambda_{1,12}^{\top}\rho_{1}
  -\|P_{1}-P_{2}\|_{\calH^{\infty}_{\calA_{\mathsf D}}}
  -u^{\dagger}\bigl(s+\|\lambda_{1}\|d_{m}\bigr)-s
  -\|\lambda_{1}\|\|\rho_{1}-\rho_{2}\|\\
  &\geq \mathsf D_{1}
  -C\Bigl(\|\rho_{1}-\rho_{2}\|+d_{m}
  +\|P_{1}-P_{2}\|_{\calH^{\infty}_{\calA_{\mathsf D}}}\Bigr),
  \qquad C:=\max\Bigl\{1,\ \frac{3(2u^{\dagger}+1)}{r}\Bigr\},
\end{align*}
where the middle inequality splits
$|\E_{P_{2}}[f_{2}]-\E_{P_{1}}[f_{1}]|\leq\|P_{1}-P_{2}\|_{\calH^{\infty}
_{\calA_{\mathsf D}}}+\E_{P_{2}}[|f_{2}-f_{1}|]$ with $f_{2}$ the first
integrand displayed, both of whose arguments lie in the domain of
$\mathsf{r}(\cdot,1)$, and the last inequality uses
$v_{P_{1}}(\zeta_{1},\lambda_{1})=\mathsf D_{1}$, the strong direction at
$(\theta_{1},\rho_{1},P_{1})$, together with $s\leq(3/r)d_{m}$ and
$\|\lambda_{1}\|\leq3/r$. Exchanging the roles of the two problems gives
the reverse bound, whence the asserted Lipschitz inequality. The constants
$C$, $C_{2}$, and the set $\calA_{\mathsf D}$ depend only on $r$, $\delta$,
$\mathbb{U}$, and $\overline{M}_{\Theta}$, as claimed.
\end{proof}

\subsection{Proof of Lemma~\ref{lem:donsker-EB}}
\begin{proof}\label{proof:lem:donsker-EB}
\uline{\textbf{Envelope.}} Recall $a$ from
Section~\ref{subsec:ls-auxiliary},
so that $\ell_{\xi,\theta}=\mathsf{r}(a(\cdot;\theta,\zeta,\lambda),\eta)$.
Under Assumption~\ref{assump: reg-EB}, $g$ and $m$ are continuous on the
compact $\calZ\times\Theta$, hence bounded, $|g|\leq\overline G$ and
$\|m\|\leq\overline M$. Over the compact index $\calA$ one has
$\|\lambda\|\leq L_{0}$, $|\zeta|\leq C_{0}$, and $\eta\leq\eta_{0}$, so
$|a|\leq\overline G+L_{0}\overline M+C_{0}=:R_{0}$ and therefore, by
\ref{prop:r-lip}, $|\ell_{\xi,\theta}|\leq F_{0}:=u^{\dagger}R_{0}
+\tfrac{u^{\dagger}}{2}\eta_{0}<\infty$ on the index set of
$\calH^{\Jdual}_{\calA}$. As a constant, $F_{0}$ is a
$\Pbase{}$-square-integrable envelope with $\|F_{0}\|_{Q,2}=F_{0}$ for every
$Q$. 

\uline{\textbf{A larger class.}} By \eqref{eq:integrand family},
on its domain $\mathsf{r}(x,\eta)$ is the pointwise maximum of the affine
functions $\underline{u}x-\tfrac{1-\underline{u}}{2}\eta$ and $x$, and,
when $\overline{u}<\infty$, of $\overline{u}x-\tfrac{\overline{u}-1}{2}\eta$
as well. Let $\widetilde{\mathsf{r}}:\R\times\R_{+}\to\R$ denote this
maximum of two or three affine functions on all of $\R\times\R_{+}$, which
coincides with $\mathsf{r}$ on the domain of the latter and is finite
everywhere, and let $\widetilde{\calH}^{\Jdual}_{\calA}:=
\{\widetilde{\mathsf{r}}(a(\cdot;\theta,\zeta,\lambda),\eta):\theta\in\Theta,
\ \xi\in \calA\}$. Then
$\calH^{\Jdual}_{\calA}\subseteq\widetilde{\calH}^{\Jdual}_{\calA}$,
because the restriction to $\xi\in\calF(\theta)$ places the argument of
$\mathsf{r}$ in its domain at every $z$, and it suffices to show that
$\widetilde{\calH}^{\Jdual}_{\calA}$ is $\Pbase{}$-Donsker, as a subclass of a
Donsker class is Donsker.

\uline{\textbf{VC-subgraph in the multipliers.}} Fix $\theta$. As
$(\zeta,\eta,\lambda)$ ranges over $\calA$, each affine branch
$u\,a(\cdot;\theta,\zeta,\lambda)-\tfrac{|u-1|}{2}\eta$, for
$u\in\{\underline{u},1,\overline{u}\}$, varies over the finite-dimensional
vector space
$V_{\theta}:=\operatorname{span}\{g(\cdot;\theta),m_{1}(\cdot;\theta),\dots,
m_{\dimm}(\cdot;\theta),1\}$, of dimension at most $\dimm+2$, which is
VC-subgraph
\citep[Lemma~2.6.15]{vandervaart1996WeakConvergenceEmpirical}.
Since the pairwise maximum of two VC-subgraph classes is again VC-subgraph
\citep[Lemma~2.6.18]{vandervaart1996WeakConvergenceEmpirical}, applied
once or twice, the class
$\{\widetilde{\mathsf{r}}(a(\cdot;\theta,\zeta,\lambda),\eta):\xi\in \calA\}$
is VC-subgraph with index $V$ bounded by a constant depending only on
$\dimm$, uniformly in $\theta$. By
Theorem~2.6.7 in \citet{vandervaart1996WeakConvergenceEmpirical} its
uniform covering numbers obey, with envelope $F_{0}$,
\[
  \sup_{Q}N\bigl(\varepsilon\|F_{0}\|_{Q,2},\{\widetilde{\mathsf{r}}
  (a(\cdot;\theta,\zeta,\lambda),\eta):\xi\in \calA\},
  L_{2}(Q)\bigr)\leq C\,\varepsilon^{-2(V-1)},\qquad 0<\varepsilon<1,
\]
with $C$ not depending on $\theta$.

\uline{\textbf{Lipschitz control in $\theta$.}} The $\theta$-direction is
handled by Assumption~\ref{assump: reg-EB}: since $\widetilde{\mathsf{r}}
(\cdot,\eta)$ is $u^{\dagger}$-Lipschitz,
\begin{align*}
  \|\widetilde{\mathsf{r}}(a(\cdot;\theta,\zeta,\lambda),\eta)
  &-\widetilde{\mathsf{r}}(a(\cdot;\theta',\zeta,\lambda),\eta)\|_{\infty} \\
  &\leq u^{\dagger}\Bigl(\sup_{z}|g(z;\theta)-g(z;\theta')|
   +\|\lambda\|\,\sup_{z}\|m(z;\theta)-m(z;\theta')\|\Bigr)
  \leq u^{\dagger}(L+L_{0}L)\,\|\theta-\theta'\|
\end{align*}
uniformly over $\xi\in \calA$, where we use $\|\lambda\|\leq L_{0}$ and the
uniform Lipschitz-in-$\theta$ moduli of \ref{assump: reg-EB}. Covering the
compact $\Theta$ by $N(\varepsilon,\Theta,\|\cdot\|)\leq
C'(\operatorname{diam}\Theta/\varepsilon)^{\dimt}$ balls and pairing the VC
bound at each center with this modulus, which is legitimate because the
envelope is the single constant $F_{0}$, yields, for the full class,
\[
  \sup_{Q}N\bigl(\varepsilon\|F_{0}\|_{Q,2},\widetilde{\calH}^{\Jdual}_{\calA},
  L_{2}(Q)\bigr)
  \leq C''\,\varepsilon^{-(\dimt+2(V-1))},\qquad 0<\varepsilon<1,
\]
so the uniform entropy integral
$\int_{0}^{1}\sup_{Q}\sqrt{\log N(\varepsilon\|F_{0}\|_{Q,2},
\widetilde{\calH}^{\Jdual}_{\calA},L_{2}(Q))}\,\de\varepsilon$ is finite.

\uline{\textbf{Conclusion.}} Separability of $\Theta\times \calA$ and continuity
of $(\theta,\xi)\mapsto\widetilde{\mathsf{r}}(a(z;\theta,\zeta,\lambda),\eta)$
make the class pointwise measurable. A pointwise-measurable class with
finite uniform entropy integral and a $\Pbase{}$-square-integrable envelope
is $\Pbase{}$-Donsker
\citep[Theorem~2.5.2]{vandervaart1996WeakConvergenceEmpirical}, so
$\widetilde{\calH}^{\Jdual}_{\calA}$, and hence $\calH^{\Jdual}_{\calA}$, is
$\Pbase{}$-Donsker. Let $\calA':=\calA\cup\{(\zeta,1,\lambda):
(\zeta,\lambda)\in\Pi_{\zeta\lambda}(\calA)\}$, which is compact as the union of
$\calA$ and a continuous image of the compact set $\Pi_{\zeta\lambda}(\calA)$.
The identical argument with $g\equiv0$ and $\calA'$ in place of
$\calA$ shows that the
class $\{\widetilde{\mathsf{r}}(-\lambda^{\top}m(\cdot;\theta)-\zeta,\eta):
\theta\in\Theta,\ \xi\in \calA'\}$ is $\Pbase{}$-Donsker, and both
$\calH^{\infty}_{\calA}$ and $\calH^{\mathsf D}_{\calA}$ are subclasses of
it, as the restrictions on their indices place the argument of
$\mathsf{r}$ in its domain, where $\widetilde{\mathsf{r}}$ and $\mathsf{r}$
coincide. 
Finally, a $\Pbase{}$-Donsker class with bounded envelope is
$\Pbase{}$-Glivenko--Cantelli.
\end{proof}

\subsection{Details on the Construction of \texorpdfstring{$\Xi$ and
$\calH$}{the Class H}}\label{subsec:ls-construction}

This section verifies the properties claimed in
Section~\ref{subsec:ls-auxiliary} for the radii $\varepsilon_{\theta}$ and
for the sets $\calN_{0}$ and $\Xi$ constructed there. 

\uline{\textbf{The radii $\varepsilon_{\theta}$.}}
Fix $\theta\in\Tmax(\rho_{0},\Pbase{})$, and let $\theta_{j}$ be a center
of the subcover with $\|\theta-\theta_{j}\|<\varepsilon^{\mathsf D}_{\theta_{j}}/2$.
On the set of triples $(\theta',\rho',P')$ with
$\|\theta'-\theta\|+\vertiii{(\rho',P')-(\rho_{0},\Pbase{})}_{\calH_{0}}
<\varepsilon^{\mathsf D}_{\theta_{j}}/2$, the constraint qualification holds
with radius $r_{\theta_{j}}/3$ and $\mathsf D(\theta';\rho',P')$ is within
$C_{\theta_{j}}$ times the displayed distance of
$\mathsf D(\theta;\rho_{0},\Pbase{})<\delta$ (Assumption~\ref{assump:
feas-EB}), so that, after shrinking the radius to some
$\varepsilon_{\theta}>0$, the strict feasibility
$\mathsf D(\theta';\rho',P')<\delta$ persists with a uniform margin. On
such a neighborhood Lemma~\ref{lemma: family duality-iii} makes every set
$\Xmax(\theta';\rho',P')$ nonempty and compact, and the proof of
part~(ii)
of Lemma~\ref{lem:aux-EB} below, which uses only the slack inequality of
Lemma~\ref{lemma: family duality-i}, the persisting constraint
qualification, and the uniform feasibility margin, bounds every minimizer
realized on the neighborhood explicitly, so that their union is bounded.
This delivers the radius $\varepsilon_{\theta}$ of
Section~\ref{subsec:ls-auxiliary}, and the set $\Xi$ defined there is
bounded as a finite union of bounded sets.

\uline{\textbf{The set $\calN_{0}$.}}
Since
$\vertiii{(\widehat\rho_{n},\widehat P_{n})-(\rho_{0},\Pbase{})}_{\calH_{0}}
\to0$ in probability by Assumption~\ref{assump: clt-EB} and the
Glivenko--Cantelli property of Lemma~\ref{lem:donsker-EB}, with probability
approaching one this distance is smaller than $\min_{k}\varepsilon_{k}$.
On that
event, the triple $(\theta',\widehat\rho_{n},\widehat P_{n})$ lies in $\calN_{0}$ for
every
$\theta'$ with $\|\theta'-\theta_{k}\|$ smaller than $\varepsilon_{k}$ minus this distance
for some $k$, and every dual optimizer realized at such a $\theta'$ has its index in
$\Xi$.

\uline{\textbf{The class $\calH$.}}
With $\calA$
and $\calH$ as fixed in Section~\ref{subsec:ls-auxiliary},
$\calA_{0}\subseteq \calA$ gives
$\calH_{0}\subseteq\calH$, so that
$\|\cdot\|_{\calH_{0}}\leq
\|\cdot\|_{\calH}$ and hence $\vertiii{\cdot}_{\calH_{0}}\leq
\vertiii{\cdot}_{\calH}$. The bound obtained above for
$\vertiii{\cdot}_{\calH_{0}}$-neighborhoods therefore holds
a fortiori on the smaller
$\vertiii{\cdot}_{\calH}$-neighborhoods, and the set $\Xi$ remains valid
when the baseline is measured in $\vertiii{\cdot}_{\calH}$.

\subsection{Proof of Lemma~\ref{lem:aux-EB}}
\begin{proof}\label{proof:lem:aux-EB}
Throughout, $\overline G:=\sup_{(z,\theta')\in\calZ\times\Theta}|g(z;\theta')|$
and $\overline M:=\sup_{(z,\theta')\in\calZ\times\Theta}\|m(z;\theta')\|$,
both finite by Assumption~\ref{assump: reg-EB}, and we use repeatedly that
$|P\ell-P'\ell|\leq\|P-P'\|_{\calH}$ for every $\ell\in\calH$ and all
$P,P'\in\Delta(\calZ)$.

\uline{\textbf{Part (i).}} Fix any $\theta\in\Tmax(\rho_{0},\Pbase{})$.
By
Assumption~\ref{assump: feas-EB},
\begin{equation}\label{eq:strict-feas}
  \mathsf{D}(\theta;\rho_{0},\Pbase{})<\delta .
\end{equation}
Let $\theta_{j}$ be a center of the finite subcover fixed in
Section~\ref{subsec:ls-auxiliary} with
$\|\theta-\theta_{j}\|<\varepsilon^{\mathsf D}_{\theta_{j}}/2$.
Lemma~\ref{lemma: mindiv dual-iii} at $(\theta_{j},\rho_{0},\Pbase{})$,
where the constraint qualification holds by
Assumption~\ref{assump: slater-EB}, gives
\[
  \bigl|\mathsf D(\theta;\rho,P)-\mathsf D(\theta;\rho_{0},\Pbase{})\bigr|
  \;\leq\;C_{\theta_{j}}\bigl(\|\rho-\rho_{0}\|
  +\|P-\Pbase{}\|_{\calH^{\infty}_{\calA_{\mathsf D,\theta_{j}}}}\bigr)
  \;\leq\;C_{\theta_{j}}\,\vertiii{(\rho,P)-(\rho_{0},\Pbase{})}_{\calH}
\]
whenever $\|\rho-\rho_{0}\|+\|P-\Pbase{}\|_{\calH^{\infty}
_{\calA_{\mathsf D,\theta_{j}}}}<\varepsilon^{\mathsf D}_{\theta_{j}}/2$, where
the second inequality follows from
$\calH^{\infty}_{\calA_{\mathsf D,\theta_{j}}}\subseteq\calH_{0}\subseteq\calH$.
Set
$\overline C:=\max_{1\leq j\leq J}C_{\theta_{j}}$ and
$\overline\varepsilon:=\min_{1\leq j\leq J}\varepsilon^{\mathsf D}_{\theta_{j}}/2$,
so that the display above holds with $\overline C$ in place of $C_{\theta_{j}}$ at
every $\theta\in\Tmax(\rho_{0},\Pbase{})$ whenever
$\vertiii{(\rho,P)-(\rho_{0},\Pbase{})}_{\calH}<\overline\varepsilon$. The
same Lipschitz estimate, read in its $\theta$-argument together with the
Lipschitz continuity of $m$ in $\theta$ (Assumption~\ref{assump: reg-EB}),
shows that $\theta\mapsto\mathsf D(\theta;\rho_{0},\Pbase{})$ is continuous on
the compact set $\Tmax(\rho_{0},\Pbase{})$, so that, by
\eqref{eq:strict-feas},
$\overline{\mathsf D}:=\sup_{\theta\in\Tmax(\rho_{0},\Pbase{})}
\mathsf D(\theta;\rho_{0},\Pbase{})<\delta$. With
$r:=\min\{\overline\varepsilon,(\delta-\overline{\mathsf D})/(2\overline C)\}$
we obtain $\mathsf D(\theta;\rho,P)<\overline{\mathsf D}+\overline C\,r\leq
\tfrac12(\delta+\overline{\mathsf D})<\delta$, that is,
$\theta\in\Theta_{\delta}(\rho,P)$, for every $\theta\in\Tmax(\rho_{0},\Pbase{})$
and every $(\rho,P)$ in the $\vertiii{\cdot}_{\calH}$-ball
$\calB_{r}(\rho_{0},\Pbase{})$, which eventually contains any sequence
$(\rho_{n},P_{n})\to(\rho_{0},\Pbase{})$. This proves part~(i).

\uline{\textbf{Part (ii).}} Fix any $\theta\in\Tmax(\rho_{0},\Pbase{})$
and let $\theta_{j}$ be as in part~(i), with constraint-qualification
radius $r_{\mathrm{cq}}:=r_{\theta_{j}}$. By Lemma~\ref{lemma: mindiv dual-iii} at
$(\theta_{j},\rho_{0},\Pbase{})$, together with \eqref{eq:strict-feas} and
the Lipschitz estimate of part~(i), there is $\varepsilon>0$ such that,
on
the ball $\calB_{\varepsilon}(\theta,\rho_{0},\Pbase{})$ of the statement,
the constraint qualification holds with radius $r_{\mathrm{cq}}/3$ and
$\mathsf D(\theta';\rho',P')\leq\delta-c$, where
$c:=\tfrac12(\delta-\mathsf D(\theta;\rho_{0},\Pbase{}))>0$. By
Lemma~\ref{lemma: family duality-iii}, the set $\Xmax(\theta';\rho',P')$
is then nonempty, compact, and convex at every
$(\theta',\rho',P')\in\calB_{\varepsilon}(\theta,\rho_{0},\Pbase{})$, and it
remains to bound these sets uniformly, which we do through the slack
inequality of Lemma~\ref{lemma: family duality-i}. Fix
$(\theta',\rho',P')\in\calB_{\varepsilon}(\theta,\rho_{0},\Pbase{})$
and $\xi^{*}=(\zeta^{*},\eta^{*},\lambda^{*})\in\Xmax(\theta';\rho',P')$.
Since the point $(\sup_{z}g(z;\theta'),0,0)$ is feasible with objective at
most $\overline G$,
$\Jdual(\xi^{*},\theta';\rho',P')=\Kmax(\theta';\rho',P')\leq\overline G$.
By the attainment part of Lemma~\ref{lemma: mindiv dual}, there is
$Q^{\circ}\in\calP_{\mathbb U}(P')$ with
$\E_{Q^{\circ}}[m(Z;\theta')]\in\calM(\rho')$ and
$\TV(Q^{\circ}\,\|\,P')=\mathsf D(\theta';\rho',P')\leq\delta-c$, so
Lemma~\ref{lemma: family duality-i} gives $\overline G\geq
\Jdual(\xi^{*},\theta';\rho',P')\geq-\overline G+\eta^{*}c$, that is,
$\eta^{*}\leq2\overline G/c$. Next, let $w_{0}'\in\calM(\rho')$ be the point
of the constraint qualification at $(\theta',\rho',P')$, so that
$\calB_{r_{\mathrm{cq}}/3}(w_{0}')\subseteq\Gamma_{\mathbb U}(\theta';P',\delta)$. If
$\lambda^{*}\neq0$, the point
$w:=w_{0}'-\tfrac{r_{\mathrm{cq}}}{6}\lambda^{*}/\|\lambda^{*}\|$
lies in that ball and is attained by some $Q_{w}\in\calP_{\mathbb U}(P')$
with $\TV(Q_{w}\,\|\,P')\leq\delta$, and Lemma~\ref{lemma: family duality-i}
gives
\[
\overline G\;\geq\;\Jdual(\xi^{*},\theta';\rho',P')
\;\geq\;-\overline G+\eta^{*}\bigl(\delta-\TV(Q_{w}\,\|\,P')\bigr)
+\bigl(\lambda^{*\top}_{12}\rho'-\lambda^{*\top}w_{0}'\bigr)
+\tfrac{r_{\mathrm{cq}}}{6}\|\lambda^{*}\|
\;\geq\;-\overline G+\tfrac{r_{\mathrm{cq}}}{6}\|\lambda^{*}\|,
\]
using $\lambda^{*\top}_{12}\rho'\geq\lambda^{*\top}w_{0}'$, whence
$\|\lambda^{*}\|\leq12\overline G/r_{\mathrm{cq}}$.
Finally, \ref{prop:r-lb} and the bounds just obtained give
\[
\overline G\geq\Jdual(\xi^{*},\theta';\rho',P')
\geq(1-\underline{u})\zeta^{*}-\|\lambda^{*}\|\,\|\rho'\|
-\underline{u}\bigl(\overline G+\|\lambda^{*}\|\overline M\bigr)
-\tfrac{1-\underline{u}}{2}\,\eta^{*},
\]
which bounds $\zeta^{*}$ from above by a constant depending only on
$\overline G$, $\overline M$, $r_{\mathrm{cq}}$, $c$, $\mathbb{U}$, and
$\sup\{\|\rho'\|:\|\rho'-\rho_{0}\|<\varepsilon\}$. From below, $\zeta^{*}$
is bounded by $\sigma(\lambda^{*},\theta')-\eta^{*}/2\geq
-\overline G-\|\lambda^{*}\|\overline M-\eta^{*}/2$ when
$\overline{u}=+\infty$, by dual feasibility, and by the third inequality
of \ref{prop:r-lb}, which gives $\overline G\geq(1-\overline{u})\zeta^{*}
-\|\lambda^{*}\|\|\rho'\|-\overline{u}(\overline G+\|\lambda^{*}\|\overline M)
-\tfrac{\overline{u}-1}{2}\eta^{*}$, when $\overline{u}<\infty$. Hence every
minimizer realized on $\calB_{\varepsilon}(\theta,\rho_{0},\Pbase{})$
lies in a single bounded set, which proves part~(ii).

\uline{\textbf{Properties of the dual.}} Before turning to
parts~(iii)--(vi), we record four properties of the dual program that
are
used repeatedly below. \uline{\textbf{First.}} The objective
$\Jdual(\xi,\theta';\rho',P')$ is jointly continuous in
$(\xi,\theta',\rho',P')$ along every sequence
$(\xi_{n},\theta_{n},\rho_{n},P_{n})\to(\xi,\theta',\rho',P')$ with
$\xi_{n}\in\calF(\theta_{n})$ for every $n$, $\xi\in\calF(\theta')$, and
$\|P_{n}-P'\|_{\calH^{\Jdual}_{\calK}}\to0$ for some compact $\calK$
containing $\{\xi_{n}\}_{n\geq1}$.
Indeed, $\Jdual$ is affine in $\rho'$ and satisfies
\[
  |P_{n}\ell_{\xi_{n},\theta_{n}}-P'\ell_{\xi,\theta'}|
  \leq\|P_{n}-P'\|_{\calH^{\Jdual}_{\calK}}
  +|P'\ell_{\xi_{n},\theta_{n}}-P'\ell_{\xi,\theta'}|\to0,
\]
where $\calK$ is the compact set of the hypothesis, the first term
vanishes by hypothesis, and the second by the dominated convergence theorem,
because the integrand is continuous in $(\xi,\theta)$ at every $z$ on the domain
of $\mathsf r$, by continuity of $g(\cdot;\theta)$ and $m(\cdot;\theta)$ on
$\calZ\times\Theta$ (Assumption~\ref{assump: reg-EB}), and bounded by
\ref{prop:r-lip}. \uline{\textbf{Second.}} The correspondence
$\theta'\mapsto\calF(\theta')$ has closed graph and is lower
hemicontinuous. It is constant when $\overline{u}<\infty$. When
$\overline{u}=+\infty$, the right-hand side of the constraint
$\zeta+\eta/2\geq\sigma(\lambda,\theta')$ is jointly continuous in
$(\theta',\lambda)$ by Berge's maximum theorem on the compact $\calZ$, which
gives the closed graph. For lower hemicontinuity at $\theta$, take a
strictly feasible point $\xi^{\circ}$ of $\calF(\theta)$, that is, one with
$\zeta^{\circ}+\eta^{\circ}/2>\sigma(\lambda^{\circ},\theta)$, which
exists because $\sigma(\cdot,\theta)$ is finite. For
$\xi\in\calF(\theta)$ and $t\in(0,1]$, convexity of the constraint function in $\xi$ makes
$(1-t)\xi+t\xi^{\circ}$ strictly feasible at $\theta$, hence feasible at every $\theta'$
near $\theta$ by joint continuity, and letting $t\to0$ along a diagonal produces the
required approximating sequence. \uline{\textbf{Third.}} As a consequence, the dual value
$(\theta',\rho',P')\mapsto\Kmax(\theta';\rho',P')$ is jointly upper semicontinuous along
every sequence $(\theta_{n},\rho_{n},P_{n})\to(\theta,\rho,P)$ with
$\|P_{n}-P\|_{\calH^{\Jdual}_{K}}\to0$ for every compact $K$: given $\xi\in\calF(\theta)$,
lower hemicontinuity supplies $\xi_{n}\in\calF(\theta_{n})$ with $\xi_{n}\to\xi$, whence
$\limsup_{n}\Kmax(\theta_{n};\rho_{n},P_{n})
\leq\lim_{n}\Jdual(\xi_{n},\theta_{n};\rho_{n},P_{n})
=\Jdual(\xi,\theta;\rho,P)$ by the first property, and taking the infimum
over $\xi\in\calF(\theta)$ gives
$\limsup_{n}\Kmax(\theta_{n};\rho_{n},P_{n})\leq\Kmax(\theta;\rho,P)$. In particular,
$\Kmax(\cdot;\rho,P)$ is upper semicontinuous on $\Theta$ for every fixed $(\rho,P)$.
Finally, at every $(\theta,\rho_{0},\Pbase{})$ with $\theta\in\Tmax(\rho_{0},\Pbase{})$
the dual value is jointly continuous along sequences $(\theta_{n},\rho_{n},P_{n})\to
(\theta,\rho_{0},\Pbase{})$ converging in $\vertiii{\cdot}_{\calH}$. Indeed, writing
\[
  \Kmax(\theta_{n};\rho_{n},P_{n})-\Kmax(\theta;\rho_{0},\Pbase{})
  =\bigl\{\Kmax(\theta_{n};\rho_{n},P_{n})-\Kmax(\theta_{n};\rho_{0},\Pbase{})\bigr\}
  +\bigl\{\Kmax(\theta_{n};\rho_{0},\Pbase{})-\Kmax(\theta;\rho_{0},\Pbase{})\bigr\},
\]
the first term is bounded by
$M\vertiii{(\rho_{n},P_{n})-(\rho_{0},\Pbase{})}_{\calH}\to0$ by the
Lipschitz estimate of part~(v) below, whose proof rests on
part~(ii) and
\hyperref[prop:linearity-EB]{(L)} alone, and the second term vanishes
because $\theta'\mapsto\Kmax(\theta';\rho_{0},\Pbase{})$ is continuous at
$\theta$. The latter follows from Berge's maximum theorem: by
part~(ii)
the optimizer sets $\Xmax(\theta';\rho_{0},\Pbase{})$ for $\theta'$ near
$\theta$ are nonempty and contained in a compact ball $\overline{\calB}^{*}$, which we
enlarge to contain the strictly feasible point $\xi^{\circ}$, so that the
infimum over $\calF(\theta')$ coincides with the minimum over the compact
set $\calF(\theta')\cap \overline{\calB}^{*}$. The correspondence
$\theta'\mapsto\calF(\theta')\cap \overline{\calB}^{*}$ is compact-valued and continuous.
Upper
hemicontinuity follows from the closed graph, and lower hemicontinuity from the argument
just given, as the points $(1-t)\xi+t\xi^{\circ}$ lie in the convex set
$\overline{\calB}^{*}$. At the
fixed baseline $\Pbase{}$, the objective is jointly continuous in $(\xi,\theta')$ by the
first property. \uline{\textbf{Fourth.}} For every $h\in\R^{\dime}$ and every
$\gamma\in UC(\calH,e_{\Pbase{}})$, the map
$(\theta',\xi)\mapsto\lambda_{12}^{\top}h+\gamma\ell_{\xi,\theta'}$ is continuous, and
hence uniformly continuous, on the set of pairs $(\theta',\xi)\in\Theta\times\calA$
with $\xi\in\calF(\theta')$, on which $\ell_{\xi,\theta'}\in\calH$. Indeed, this set is
compact by the closed graph of $\calF(\cdot)$ and the compactness of $\Theta$ and
$\calA$, the linear part of the map is continuous in $\xi$, and
$e_{\Pbase{}}(\ell_{\xi_{n},\theta_{n}},\ell_{\xi,\theta'})\to0$ along every sequence
$(\theta_{n},\xi_{n})\to(\theta',\xi)$ in this set, by the dominated convergence
theorem applied as in the first property, while $\gamma$ is uniformly continuous with
respect to $e_{\Pbase{}}$.

\uline{\textbf{Part (iii).}} Let $(\theta_{n},\rho_{n},P_{n})\to
(\theta,\rho_{0},\Pbase{})$ with $\theta\in\Tmax(\rho_{0},\Pbase{})$. By
part~(ii) and the construction of Section~\ref{subsec:ls-auxiliary}, for
$n$ large the sets $\Xmax(\theta_{n};\rho_{n},P_{n})$ are nonempty and
contained in $\Xi\subseteq\calA$, so any selection
$\xi_{n}\in\Xmax(\theta_{n};\rho_{n},P_{n})$ admits a subsequence
$\xi_{n_{l}}\to\xi^{\infty}$, with $\xi^{\infty}\in\calF(\theta)$ by the
closed graph of $\calF(\cdot)$. Since $\xi_{n_{l}}\in\calA$, the first
property above gives
$\Jdual(\xi_{n_{l}},\theta_{n_{l}};\rho_{n_{l}},P_{n_{l}})\to
\Jdual(\xi^{\infty},\theta;\rho_{0},\Pbase{})$, while
$\Jdual(\xi_{n_{l}},\theta_{n_{l}};\rho_{n_{l}},P_{n_{l}})
=\Kmax(\theta_{n_{l}};\rho_{n_{l}},P_{n_{l}})\to\Kmax(\theta;\rho_{0},\Pbase{})$
by the joint continuity of the dual value at $(\theta,\rho_{0},\Pbase{})$.
Hence $\Jdual(\xi^{\infty},\theta;\rho_{0},\Pbase{})$
equals $\Kmax(\theta;\rho_{0},\Pbase{})$,
that is, $\xi^{\infty}\in\Xmax(\theta;\rho_{0},\Pbase{})$, and thus
$\lambda_{12,n_{l}}\to\lambda^{\infty}_{12}\in\Lmax(\theta;\rho_{0},\Pbase{})$.
This proves part~(iii).

\uline{\textbf{Part (iv).}} We apply Berge's maximum theorem to
$v(\theta;h,\gamma)=\min_{\xi\in\Xmax(\theta;\rho_{0},\Pbase{})}
\{\lambda_{12}^{\top}h+\gamma\ell_{\xi,\theta}\}$ at any
$\theta\in\Tmax(\rho_{0},\Pbase{})$. By part~(v), $\calN$ is a
neighborhood of $\theta$ and, for every $\theta'\in\calN\cap\Theta$, the set
$\Xmax(\theta';\rho_{0},\Pbase{})$ is contained in $\Xi\subseteq\calA$, so that
$\ell_{\xi,\theta'}\in\calH$ and $\gamma\ell_{\xi,\theta'}$ is defined for every $\xi$
in it. The inner objective is continuous in
$(\theta,\xi)$ by the fourth property above. The correspondence
$\theta'\mapsto\Xmax(\theta';\rho_{0},\Pbase{})$ is nonempty- and
compact-valued for $\theta'$ near $\theta$ by part~(ii), lower
hemicontinuous at $\theta$ by Assumption~\ref{assump: mult-EB}, and upper
hemicontinuous at $\theta$ by part~(iii) applied with
$(\rho_{n},P_{n})\equiv(\rho_{0},\Pbase{})$. Berge's theorem therefore gives
continuity of $v(\cdot;h,\gamma)$ at $\theta$, which proves part~(iv).

\uline{\textbf{Part (v).}} We first construct a compact neighborhood of
$\Tmax(\rho_{0},\Pbase{})$ on which the optimizer sets are uniformly
controlled. For each $\theta\in\Tmax(\rho_{0},\Pbase{})$, let
$\varepsilon_{\theta}>0$ be the radius supplied by part~(ii). The balls
$\calB_{\varepsilon_{\theta}/2}(\theta)$ cover the compact set
$\Tmax(\rho_{0},\Pbase{})$, so we can extract a finite subcover with centers
$\theta_{1},\dots,\theta_{K}$ and set
\[
  \calN:=\bigcup_{k=1}^{K}\overline{\calB}_{\varepsilon_{\theta_{k}}/2}(\theta_{k}),
  \qquad
  r:=\min_{1\leq k\leq K}\tfrac{\varepsilon_{\theta_{k}}}{2}>0 .
\]
Then $\calN$ is compact and contains $\Tmax(\rho_{0},\Pbase{})$ in its
interior. Any $\theta\in\calN\cap\Theta$ lies within
$\varepsilon_{\theta_{k}}/2$ of
some center $\theta_{k}$, so that, for any $(\rho,P)$ with
$\vertiii{(\rho,P)-(\rho_{0},\Pbase{})}_{\calH}<r$,
\[
  \|\theta-\theta_{k}\|+\vertiii{(\rho,P)-(\rho_{0},\Pbase{})}_{\calH}
  \;<\;\tfrac{\varepsilon_{\theta_{k}}}{2}+r
  \;\leq\;\varepsilon_{\theta_{k}},
\]
and part~(ii) applies at the center $\theta_{k}$. Hence
$\Xmax(\theta;\rho,P)$ is nonempty and compact, and these sets are bounded
uniformly over all such $(\theta,\rho,P)$. Since
$\vertiii{\cdot}_{\calH_{0}}\leq\vertiii{\cdot}_{\calH}$, we may shrink
$\calN$ and $r$ so that every such triple lies in the cover
$\calN_{0}$ from which the set $\Xi$ was formed, whence
$\Xmax(\theta;\rho,P)\subseteq\Xi\subseteq \calA$ throughout. Let $M\in[1,\infty)$ bound
$\|\lambda_{12}\|$ over $\Pi_{12}(\calA)$. This establishes
every claim of the second assertion of part~(v) except the Lipschitz
estimate, which we prove next.

\uline{\textbf{Lipschitz condition.}} Fix any $\theta\in\calN\cap\Theta$ and two
pairs
$(\rho,P),(\rho',P')$ in the $\vertiii{\cdot}_{\calH}$-ball
$\calB_{r}(\rho_{0},\Pbase{})$. By
\hyperref[prop:linearity-EB]{(L)}, $(\rho,P)\mapsto\Kmax(\theta;\rho,P)$ is
concave, as a pointwise infimum of the jointly affine maps
$(\rho,P)\mapsto\Jdual(\xi,\theta;\rho,P)=\zeta+\eta\delta
+\lambda_{12}^{\top}\rho+P\ell_{\xi,\theta}$, and at any optimizer
$\xi\in\Xmax(\theta;\rho,P)\subseteq\calA$ it admits the supergradient
$(\lambda_{12},\ell_{\xi,\theta})$. Concavity then gives
\[
  \lambda_{12}^{\top}(\rho'-\rho)+(P'-P)\ell_{\xi,\theta}
  \;\geq\;\Kmax(\theta;\rho',P')-\Kmax(\theta;\rho,P)
  \;\geq\;\lambda_{12}'^{\top}(\rho'-\rho)+(P'-P)\ell_{\xi',\theta},
\]
with $\xi'\in\Xmax(\theta;\rho',P')\subseteq \calA$. Since
$\|\lambda_{12}\|,\|\lambda_{12}'\|\leq M$ and
$\ell_{\xi,\theta},\ell_{\xi',\theta}\in\calH$, each side is bounded in
absolute value by $M\|\rho-\rho'\|+\|P-P'\|_{\calH}\leq
M\,\vertiii{(\rho,P)-(\rho',P')}_{\calH}$, using $M\geq1$. Hence
$|\Kmax(\theta;\rho,P)-\Kmax(\theta;\rho',P')|\leq
M\,\vertiii{(\rho,P)-(\rho',P')}_{\calH}$, uniformly over
$\theta\in\calN\cap\Theta$.

\uline{\textbf{Continuity of the bound.}} We prove lower and upper
semicontinuity separately. For lower semicontinuity, fix $\varepsilon'>0$
and any $\theta^{*}\in\Tmax(\rho_{0},\Pbase{})$, which attains
$\bmax(\delta,\rho_{0},\Pbase{})$. By part~(i),
$\theta^{*}\in\Theta_{\delta}(\rho,P)$ on a neighborhood of
$(\rho_{0},\Pbase{})$, and by the Lipschitz condition just established
$\Kmax(\theta^{*};\rho,P)>\Kmax(\theta^{*};\rho_{0},\Pbase{})-\varepsilon'$
for $(\rho,P)$ close enough to $(\rho_{0},\Pbase{})$. Therefore, on that
neighborhood,
\[
  \bmax(\delta,\rho,P)\;\geq\;\Kmax(\theta^{*};\rho,P)
   \;>\;\bmax(\delta,\rho_{0},\Pbase{})-\varepsilon',
\]
which proves lower semicontinuity. For upper semicontinuity, suppose, towards
contradiction, that there is a sequence $(\rho_{n},P_{n})\to(\rho_{0},
\Pbase{})$, with $\|P_{n}-\Pbase{}\|_{\calH^{\Jdual}_{K}}\to0$ for every
compact $K$, along which $\bmax(\delta,\rho_{n},P_{n})\geq\bmax(\delta,\rho_{0},\Pbase{})
+2\varepsilon'$ for some $\varepsilon'>0$. Pick
$\theta_{n}\in\Theta_{\delta}(\rho_{n},P_{n})$ with
$\Kmax(\theta_{n};\rho_{n},P_{n})>\bmax(\delta,\rho_{n},P_{n})-\varepsilon'$, which
exists because $\bmax(\delta,\rho_{n},P_{n})$ is the supremum of
$\Kmax(\cdot;\rho_{n},P_{n})$ over $\Theta_{\delta}(\rho_{n},P_{n})$,
nonempty for $n$ large by part~(i). Compactness of $\Theta$
(Assumption~\ref{assump: reg-EB}) gives a subsequence with
$\theta_{n_{l}}\to\overline\theta\in\Theta$. By the joint upper
semicontinuity of the dual value along such sequences, recorded above,
\[
  \Kmax(\overline\theta;\rho_{0},\Pbase{})
  \;\geq\;\limsup_{l}\Kmax(\theta_{n_{l}};\rho_{n_{l}},P_{n_{l}})
  \;\geq\;\limsup_{l}\,\bigl(\bmax(\delta,\rho_{n_{l}},P_{n_{l}})-\varepsilon'\bigr)
  \;\geq\;\bmax(\delta,\rho_{0},\Pbase{})+\varepsilon' .
\]
On the other hand, by Assumption~\ref{assump: feas-EB} the argmax of
$\Kmax(\cdot;\rho_{0},\Pbase{})$ over all of $\Theta$ lies inside
$\Theta_{\delta}(\rho_{0},\Pbase{})$, so that
$\bmax(\delta,\rho_{0},\Pbase{})=\sup_{\theta\in\Theta}\Kmax(\theta;\rho_{0},\Pbase{})
\geq\Kmax(\overline\theta;\rho_{0},\Pbase{})$, a contradiction. This proves
part~(v).

\uline{\textbf{Part (vi).}} Fix an open set
$\calU\supseteq\Tmax(\rho_{0},\Pbase{})$ and set $\calK:=\Theta\setminus\calU$,
a compact set disjoint from $\Tmax(\rho_{0},\Pbase{})$. The map
$\Kmax(\cdot;\rho_{0},\Pbase{})$ is upper semicontinuous on $\calK$, by the
third property above, and bounded above by $\overline G$, so
$\beta_{\calK}:=\sup_{\theta\in\calK}\Kmax(\theta;\rho_{0},\Pbase{})
\in[-\infty,\infty)$ is attained on $\calK$ whenever finite. Since
$\bmax(\delta,\rho_{0},\Pbase{})$ is the supremum of
$\Kmax(\cdot;\rho_{0},\Pbase{})$ over all of $\Theta$ and is attained only on
$\Tmax(\rho_{0},\Pbase{})$ (Assumption~\ref{assump: feas-EB}), we have
$\beta_{\calK}<\bmax(\delta,\rho_{0},\Pbase{})$. Let
$\varepsilon_{\calK}:=\min\{1,(\bmax(\delta,\rho_{0},\Pbase{})-\beta_{\calK})/2\}>0$,
with $\varepsilon_{\calK}:=1$ when $\beta_{\calK}=-\infty$. We claim that
$\sup_{\theta\in\calK}\Kmax(\theta;\rho_{n},P_{n})
\leq\bmax(\delta,\rho_{0},\Pbase{})-\varepsilon_{\calK}$ for all $n$ large.
Otherwise there
are infinitely many $n$ and points $\theta_{n}\in\calK$ with
$\Kmax(\theta_{n};\rho_{n},P_{n})>\bmax(\delta,\rho_{0},\Pbase{})-\varepsilon_{\calK}$,
and by compactness a subsequence has $\theta_{n}\to\theta_{\star}\in\calK$, so
that the joint upper semicontinuity of the dual value along
$(\theta_{n},\rho_{n},P_{n})$ yields
\[
  \bmax(\delta,\rho_{0},\Pbase{})-\varepsilon_{\calK}
   \;\leq\;\limsup_{n}\Kmax(\theta_{n};\rho_{n},P_{n})
   \;\leq\;\Kmax(\theta_{\star};\rho_{0},\Pbase{})
   \;\leq\;\beta_{\calK}
   \;\leq\;\bmax(\delta,\rho_{0},\Pbase{})-2\varepsilon_{\calK},
\]
a contradiction in view of $\varepsilon_{\calK}>0$, the last inequality holding by the
definition of $\varepsilon_{\calK}$ and trivially when $\beta_{\calK}=-\infty$. On the
other hand, $\bmax(\delta,\rho_{n},P_{n})\to\bmax(\delta,\rho_{0},\Pbase{})$ by
the first assertion of part~(v) and $\nu_{n}\to0$, so that
$\bmax(\delta,\rho_{n},P_{n})-\nu_{n}>\bmax(\delta,\rho_{0},\Pbase{})-\varepsilon_{\calK}$
for all $n$ large. For such $n$ no $\theta\in\calK$ satisfies
$\Kmax(\theta;\rho_{n},P_{n})\geq\bmax(\delta,\rho_{n},P_{n})-\nu_{n}$, which
proves part~(vi).
\end{proof}

\subsection{Proof of Theorem~\ref{thm:cons-EB}}
\begin{proof}\label{proof:thm:cons-EB}
For $j\geq1$, let
$K_j:=\{\xi\in\R\times\R_{+}\times\R^{\dimm}:\|\xi\|\leq j\}$.
By Assumption~\ref{assump: clt-EB}, $\widehat\rho_{n}\cprob\rho_{0}$, whereas
Lemma~\ref{lem:donsker-EB} gives
$\|\widehat P_n-\Pbase{}\|_{\calH}\cprob0$ and
$\|\widehat P_n-\Pbase{}\|_{\calH^{\Jdual}_{K_j}}\cprob0$ for every $j$.
By Borel-Cantelli, from any subsequence we may therefore extract a further subsequence
along which all these convergences hold almost surely. We do so to be able to
apply Lemma~\ref{lem:aux-EB-vi}, which holds for deterministic sequences. 
Every compact
$K\subseteq\R\times\R_{+}\times\R^{\dimm}$ is contained in some $K_j$, so
Lemma~\ref{lem:aux-EB-vi} gives
$\bmax(\delta,\widehat\rho_n,\widehat P_n)\to
\bmax(\delta,\rho_0,\Pbase{})$ almost surely along that further subsequence. The
subsequence characterization of convergence in probability then yields
$\bmaxhat(\delta)\cprob\bmax(\delta,\rho_0,\Pbase{})$.
The argument for $\bminhat(\delta)$ is identical with $g\mapsto-g$.
\end{proof}

\subsection{Proof of Theorem~\ref{thm:HD-EB}}
\begin{proof}\label{proof:thm:HD-EB}
We prove the result for $\bmax$. That for $\bmin$ is symmetric, and the joint
statement then follows because both coordinates are functionals of the same
pair $(\widehat\rho_{n},\widehat P_{n})$. Recall that
$\mathsf b_{0}=(\Pbase{}\ell)_{\ell\in\calH}$ is the element of
$\ell^{\infty}(\calH)$ induced by $\Pbase{}$, as defined in
Section~\ref{subsec:ls-hd}, and fix sequences
$t_{n}\downarrow 0$, $h_{n}\to h$ in $\R^{\dime}$, and $\gamma_{n}\to\gamma$
in $\ell^{\infty}(\calH)$ with $\gamma\in UC(\calH,e_{\Pbase{}})$. Write
$\rho_{n}:=\rho_{0}+t_{n}h_{n}$ and $\mathsf b_{n}:=\mathsf b_{0}
+t_{n}\gamma_{n}$. By \hyperref[prop:independence-EB]{(I)} the feasible set
$\calF(\theta)$ does not depend on $(\rho,\mathsf b)$, so a dual point
feasible at one $(\rho,\mathsf b)$ is feasible at every other. By
\hyperref[prop:linearity-EB]{(L)} the objective
$\Jdual(\xi,\theta;\cdot)$ is jointly affine in $(\rho,\mathsf b)$,
so that, for every $\theta\in\Theta$ and every
$\xi\in\calF(\theta)\cap\calA$,
\begin{equation*}
  \Jdual(\xi,\theta;\rho_{n},\mathsf b_{n})
  -\Jdual(\xi,\theta;\rho_{0},\mathsf b_{0})
  =t_{n}\bigl\{\lambda_{12}^{\top}h_{n}+\gamma_{n}\ell_{\xi,\theta}\bigr\}.
\end{equation*}

\uline{\textbf{Step 1: directional derivative.}} The argument consists of
a lower bound obtained by holding a population
optimizer fixed and an upper bound obtained along near-optimizers of the
perturbed program.

\uline{\textbf{(a) Lower bound.}} Fix $\theta\in\Tmax(\rho_{0},\Pbase{})$.
Then $\theta\in\calN_{\Theta}$, and
$\mathsf D(\theta;\rho_{n},\mathsf b_{n})<\delta$ for all $n$ large. Indeed,
$\mathsf D(\theta;\rho_{0},\mathsf b_{0})=\mathsf D(\theta;\rho_{0},\Pbase{})
<\delta$ by Assumption~\ref{assump: feas-EB}, and, by its dual definition
restricted to $\calA$, the map $\mathsf b\mapsto\mathsf D(\theta;\rho,\mathsf
b)$ is $1$-Lipschitz for the $\ell^{\infty}(\calH)$ norm and the map
$\rho\mapsto\mathsf D(\theta;\rho,\mathsf b)$ is Lipschitz, in both cases
uniformly in the remaining argument. This is the counterpart of
Lemma~\ref{lem:aux-EB-i} with $P$ replaced by $\mathsf b$. Hence $\theta$
belongs to the constraint set of the supremum defining
$\bmax(\delta,\rho_{n},\mathsf b_{n})$ for all $n$ large.
Note that $\Kmax(\theta;\rho_{n},\mathsf b_{n})$ and
$\Kmax(\theta;\rho_{n},\Pbase{})$ are infima of objectives that differ by
$(\mathsf b_{n}-\mathsf b_{0})(\ell_{\xi,\theta})
=t_{n}\gamma_{n}\ell_{\xi,\theta}$, where
$\ell_{\xi,\theta}\in\calH^{\Jdual}_{\calA}\subseteq\calH$ for every
$\xi\in\calF(\theta)\cap \calA$, so that the two values differ by at most
$t_{n}\|\gamma_{n}\|_{\ell^{\infty}(\calH)}$. Moreover,
$\Kmax(\theta;\rho_{n},\Pbase{})$ is finite for all $n$ large. We may
thus choose $\xi^{*}_{n}=(\zeta^{*}_{n},\eta^{*}_{n},\lambda^{*}_{n})\in
\calF(\theta)\cap \calA$ with $\Jdual(\xi^{*}_{n},\theta;\rho_{n},\mathsf
b_{n})\leq\Kmax(\theta;\rho_{n},\mathsf b_{n})+t_{n}^{2}$. Since
$\xi^{*}_{n}$ is $t_{n}^{2}$-optimal at $(\rho_{n},\mathsf b_{n})$ and
feasible at $(\rho_{0},\mathsf b_{0})$, so that
$\Kmax(\theta;\rho_{0},\mathsf b_{0})
\leq\Jdual(\xi^{*}_{n},\theta;\rho_{0},\mathsf b_{0})$, subtracting the
two inequalities and using \hyperref[prop:linearity-EB]{(L)} yields
\begin{equation*}
  \Kmax(\theta;\rho_{n},\mathsf b_{n})-\Kmax(\theta;\rho_{0},\mathsf b_{0})
   \;\geq\;
   t_{n}\bigl\{(\lambda^{*}_{12,n})^{\top}h_{n}
   +\gamma_{n}\ell_{\xi^{*}_{n},\theta}\bigr\}-t_{n}^{2}.
\end{equation*}
The sequence $\{\xi^{*}_{n}\}$ lies in the compact set
$\calF(\theta)\cap \calA$. Let
$\xi^{*}_{n_{l}}\to\xi^{\infty}\in\calF(\theta)\cap \calA$ along a
subsequence. By \hyperref[prop:linearity-EB]{(L)}
and boundedness of $\calA$,
$|\Jdual(\xi,\theta;\rho_{n_{l}},\mathsf b_{n_{l}})
-\Jdual(\xi,\theta;\rho_{0},\mathsf b_{0})|\to0$ uniformly over
$\xi\in\calF(\theta)\cap \calA$, so that
$\Kmax(\theta;\rho_{n_{l}},\mathsf b_{n_{l}})\to
\Kmax(\theta;\rho_{0},\mathsf b_{0})$ and, by the near-optimality of
$\xi^{*}_{n_{l}}$,
$\Jdual(\xi^{*}_{n_{l}},\theta;\rho_{0},\mathsf b_{0})
\to\Kmax(\theta;\rho_{0},\mathsf b_{0})$, whereas
$\Jdual(\xi^{*}_{n_{l}},\theta;\rho_{0},\mathsf b_{0})\to
\Jdual(\xi^{\infty},\theta;\rho_{0},\mathsf b_{0})$ by the dominated
convergence theorem, as $\mathsf b_{0}$ is induced by $\Pbase{}$. Hence
$\Jdual(\xi^{\infty},\theta;\rho_{0},\mathsf b_{0})
=\Kmax(\theta;\rho_{0},\mathsf b_{0})$, that is,
$\xi^{\infty}\in\Xmax(\theta;\rho_{0},\mathsf b_{0})$. Since, moreover,
$|\gamma_{n}\ell_{\xi,\theta}-\gamma\ell_{\xi,\theta}|
\leq\|\gamma_{n}-\gamma\|_{\ell^{\infty}(\calH)}\to0$ uniformly in $\xi$ and
$\xi\mapsto\gamma\ell_{\xi,\theta}$ is continuous on $\calF(\theta)\cap \calA$
by the fourth property of the dual in the proof of
Lemma~\ref{lem:aux-EB}, passing to a subsequence
along which the $\liminf$ is attained yields
\[
  \liminf_{n}\,
   \bigl\{(\lambda^{*}_{12,n})^{\top}h_{n}+\gamma_{n}\ell_{\xi^{*}_{n},\theta}\bigr\}
  \;\geq\;\min_{\xi^{*}\in\Xmax(\theta;\rho_{0},\mathsf b_{0})}
   \{(\lambda_{12}^{*})^{\top}h+\gamma\ell_{\xi^{*},\theta}\},
\]
where the minimum on the right-hand side is attained because
the map $\xi\mapsto\lambda_{12}^{\top}h+\gamma\ell_{\xi,\theta}$ is
continuous on the nonempty compact set $\Xmax(\theta;\rho_{0},\mathsf b_{0})
=\Xmax(\theta;\rho_{0},\Pbase{})$ and where the equality 
follows from $\Xmax(\theta;\rho_{0},\Pbase{})\subseteq\Xi\subseteq\calA$.
Next, $\bmax(\delta,\rho_{n},\mathsf b_{n})\geq
\Kmax(\theta;\rho_{n},\mathsf b_{n})$ for all $n$ large, because $\theta$
belongs to the constraint set of the supremum, and
$\bmax(\delta,\rho_{0},\mathsf b_{0})=\Kmax(\theta;\rho_{0},\mathsf b_{0})$,
by the identity $\bmax(\delta,\rho_{0},\mathsf b_{0})
=\bmax(\delta,\rho_{0},\Pbase{})$ and
$\theta\in\Tmax(\rho_{0},\Pbase{})$. Dividing by $t_{n}$ and taking the supremum over
$\theta\in\Tmax(\rho_{0},\Pbase{})$, we obtain
\begin{equation}\label{eq:lower-EB}
  \liminf_{n}\frac{\bmax(\delta,\rho_{n},\mathsf b_{n})-\bmax(\delta,\rho_{0},\mathsf
  b_{0})}{t_{n}}
   \;\geq\;\max_{\theta^{*}}\min_{\xi^{*}}
   \{(\lambda_{12}^{*})^{\top}h+\gamma\ell_{\xi^{*},\theta^{*}}\}.
\end{equation}

\uline{\textbf{(b) Upper bound.}} Fix an arbitrary subsequence of
$\{(t_{n},h_{n},\gamma_{n})\}_{n\geq1}$, which we continue to index by $n$.
Along it, choose $\theta_{n}\in\calN_{\Theta}$ with $\mathsf D(\theta_{n};\rho_{n},\mathsf
b_{n})<\delta$ and
$\Kmax(\theta_{n};\rho_{n},\mathsf b_{n})\geq\bmax(\delta,\rho_{n},\mathsf
b_{n})-t_{n}^{2}$, and, by compactness of $\calN_{\Theta}$, extract a
further subsequence along which $\theta_{n}\to\overline\theta\in
\calN_{\Theta}$. We first show that
$\overline\theta\in\Tmax(\rho_{0},\Pbase{})$. Since the difference quotient
in \eqref{eq:lower-EB} has a finite limit inferior and $t_{n}\to0$, part~(a)
gives $\liminf_{n}\bmax(\delta,\rho_{n},\mathsf b_{n})
\geq\bmax(\delta,\rho_{0},\mathsf b_{0})$. Conversely, the extended criterion
is jointly upper semicontinuous along $(\theta_{n},\rho_{n},\mathsf b_{n})
\to(\overline\theta,\rho_{0},\mathsf b_{0})$, as it is $1$-Lipschitz in
$\mathsf b$ uniformly in the remaining arguments and equal, at
$\mathsf b_{0}$ and on $\calN_{\Theta}$, to the jointly upper semicontinuous
map $(\theta,\rho)\mapsto\Kmax(\theta;\rho,\Pbase{})$ for $\rho$ in a
neighborhood of $\rho_{0}$. Hence
\[
  \bmax(\delta,\rho_{0},\mathsf b_{0})
  \;\geq\;\Kmax(\overline\theta;\rho_{0},\mathsf b_{0})
  \;\geq\;\limsup_{n}\Kmax(\theta_{n};\rho_{n},\mathsf b_{n})
  \;\geq\;\limsup_{n}\bmax(\delta,\rho_{n},\mathsf b_{n}),
\]
where the first inequality holds because $\bmax(\delta,\rho_{0},\Pbase{})$
is the supremum of $\Kmax(\cdot;\rho_{0},\Pbase{})$ over all of $\Theta$
(Assumption~\ref{assump: feas-EB}), and the last inequality
follows from the choice of $\theta_{n}$. Together with the limit inferior bound,
this shows that $\bmax(\delta,\rho_{n},\mathsf b_{n})
\to\bmax(\delta,\rho_{0},\mathsf b_{0})$ and
$\Kmax(\overline\theta;\rho_{0},\Pbase{})=\bmax(\delta,\rho_{0},\Pbase{})$,
that is, $\overline\theta\in\Tmax(\rho_{0},\Pbase{})$.

Next, we argue as in part~(a) with the roles of $(\rho_{0},\mathsf b_{0})$
and $(\rho_{n},\mathsf b_{n})$ interchanged. Let $\xi^{*}_{n}$ minimize
$\xi\mapsto\lambda_{12}^{\top}h_{n}+\gamma\ell_{\xi,\theta_{n}}$ over the
compact set $\Xmax(\theta_{n};\rho_{0},\mathsf b_{0})$, where the minimum is
attained by the continuity of $\xi\mapsto\gamma\ell_{\xi,\theta_{n}}$ noted
in part~(a). Since $\xi^{*}_{n}$ is optimal at $(\rho_{0},\mathsf b_{0})$
and feasible at $(\rho_{n},\mathsf b_{n})$,
\hyperref[prop:linearity-EB]{(L)} bounds $\Kmax(\theta_{n};\rho_{n},\mathsf
b_{n})-\Kmax(\theta_{n};\rho_{0},\mathsf b_{0})$ from above by
$t_{n}\{(\lambda^{*}_{12,n})^{\top}h_{n}
+\gamma_{n}\ell_{\xi^{*}_{n},\theta_{n}}\}$. Combining this bound with the
near-optimality of $\theta_{n}$, with
$\Kmax(\theta_{n};\rho_{0},\mathsf b_{0})\leq\bmax(\delta,\rho_{0},\mathsf
b_{0})$, which holds as in the first inequality of the preceding display,
and with the replacement of $(h_{n},\gamma_{n})$ by $(h,\gamma)$ at a cost
that vanishes uniformly in $\xi$, exactly as in part~(a), we obtain, along
the subsequence,
\[
  \limsup_{n}\frac{\bmax(\delta,\rho_{n},\mathsf b_{n})-\bmax(\delta,\rho_{0},\mathsf
  b_{0})}{t_{n}}
   \;\leq\;\limsup_{n}\,\min_{\xi\in\Xmax(\theta_{n};\rho_{0},\mathsf b_{0})}
    \{\lambda_{12}^{\top}h+\gamma\ell_{\xi,\theta_{n}}\}.
\]
It remains to bound the right-hand side. Fix
$\xi^{*}\in\Xmax(\overline\theta;\rho_{0},\mathsf b_{0})$. By the lower
hemicontinuity of $\theta\mapsto\Xmax(\theta;\rho_{0},\Pbase{})$ at
$\overline\theta\in\Tmax(\rho_{0},\Pbase{})$
(Assumption~\ref{assump: mult-EB}) and the equality
$\Xmax(\theta_{n};\rho_{0},\mathsf b_{0})=\Xmax(\theta_{n};\rho_{0},\Pbase{})$
on $\calN_{\Theta}$, there exist
$\xi_{n}\in\Xmax(\theta_{n};\rho_{0},\mathsf b_{0})$ with
$\xi_{n}\to\xi^{*}$. The map $(\theta,\xi)\mapsto\lambda_{12}^{\top}h
+\gamma\ell_{\xi,\theta}$ is continuous on the pairs $(\theta,\xi)$
with $\xi\in\calF(\theta)\cap\calA$, by the fourth property of the dual in the proof of
Lemma~\ref{lem:aux-EB}, so that the right-hand side of the preceding
display is at most $(\lambda_{12}^{*})^{\top}h
+\gamma\ell_{\xi^{*},\overline\theta}$. Minimizing over $\xi^{*}$ and using
$\overline\theta\in\Tmax(\rho_{0},\Pbase{})$, we obtain, along the
subsequence,
\[
  \limsup_{n}\frac{\bmax(\delta,\rho_{n},\mathsf b_{n})-\bmax(\delta,\rho_{0},\mathsf
  b_{0})}{t_{n}}
   \;\leq\;\min_{\xi^{*}\in\Xmax(\overline\theta;\rho_{0},\mathsf b_{0})}
    \{(\lambda_{12}^{*})^{\top}h+\gamma\ell_{\xi^{*},\overline\theta}\}
   \;\leq\;\max_{\theta^{*}}\min_{\xi^{*}}
    \{(\lambda_{12}^{*})^{\top}h+\gamma\ell_{\xi^{*},\theta^{*}}\}.
\]
Since the initial subsequence was arbitrary and the bound does not depend
on it, the same bound holds for the limit superior along the full sequence.
Together with~\eqref{eq:lower-EB}, this proves the expression for
$d\bmaxAt{(\rho_{0},\Pbase{})}[h,\gamma]$ in~\eqref{eq:HD-EB-formula}.
Finally, the map $(h,\gamma)\mapsto d\bmaxAt{(\rho_{0},\Pbase{})}[h,\gamma]$
is continuous on $\R^{\dime}\times UC(\calH,e_{\Pbase{}})$. Indeed, it is
Lipschitz,
\[
  \bigl|d\bmaxAt{(\rho_{0},\Pbase{})}[h,\gamma]
  -d\bmaxAt{(\rho_{0},\Pbase{})}[h',\gamma']\bigr|
  \;\leq\;M_{\calA}\|h-h'\|+\|\gamma-\gamma'\|_{\ell^{\infty}(\calH)},
  \qquad M_{\calA}:=\sup\{\|\lambda_{12}\|:\xi\in\calA\},
\]
where $M_{\calA}<\infty$ as $\calA$ is compact. The bound follows from
$|\max_{S}f-\max_{S}f'|\leq\sup_{S}|f-f'|$ and the same inequality for minima,
because, for every $\theta^{*}\in\Tmax(\rho_{0},\Pbase{})$ and
$\xi^{*}\in\Xmax(\theta^{*};\rho_{0},\Pbase{})\subseteq\Xi\subseteq\calA$, we have
$\|\lambda_{12}^{*}\|\leq M_{\calA}$ and $\ell_{\xi^{*},\theta^{*}}\in\calH$, so that
$|(\gamma-\gamma')\ell_{\xi^*,\theta^*}|\leq\|\gamma-\gamma'\|_{\ell^{\infty}(\calH)}$.

\uline{\textbf{Step 2: asymptotic distribution.}} By
Lemma~\ref{lem:donsker-EB} and Assumption~\ref{assump: clt-EB}, the class
$\calH$ augmented by the $\dime$ coordinates of the influence function
$\psi$ is $\Pbase{}$-Donsker, so that
\[
  \sqrt n\bigl((\widehat\rho_{n}-\rho_{0}),\,(\widehat P_{n}-\Pbase{})\bigr)
   \;\rightsquigarrow\;(\mathbb Z,\mathbb G)
   \quad\text{in }\R^{\dime}\times\ell^{\infty}(\calH),
\]
where the limit $(\mathbb Z,\mathbb G)$ is a tight Gaussian process whose support is
contained in $\R^{\dime}\times\allowbreak UC(\calH,e_{\Pbase{}})$. This is
the tangent set relative to which Step~1 established the Hadamard
directional differentiability of
$(\rho,\mathsf b)\mapsto(\bmin,\bmax)(\delta,\rho,\mathsf b)$.
The hypotheses of the directional functional delta method
\citep*[][Theorem~2.1]{fang2019InferenceDirectionallyDifferentiable} are
therefore met, and, in view of the two identities established below, it
yields
\[
  \sqrt n\!\begin{pmatrix}\bminhat(\delta)-\bmin(\delta,\rho_{0},\Pbase{})\\
   \bmaxhat(\delta)-\bmax(\delta,\rho_{0},\Pbase{})\end{pmatrix}
   \;\rightsquigarrow\;
   \begin{pmatrix}d\bminAt{(\rho_{0},\Pbase{})}[\mathbb Z,\mathbb G]\\
   d\bmaxAt{(\rho_{0},\Pbase{})}[\mathbb Z,\mathbb G]\end{pmatrix}.
\]

Two identities underlie the passage from the extended functional to the
estimators and population bounds appearing in that display. First, the
centering constants equal the population bounds, by the identity
$\bmax(\delta,\rho_{0},\mathsf b_{0})=\bmax(\delta,\rho_{0},\Pbase{})$. 
Second, $\bmaxhat(\delta)=\bmax(\delta,\widehat\rho_{n},\widehat{\mathsf b}_{n})$,
with $\widehat{\mathsf b}_{n}:=(\widehat P_{n}\ell)_{\ell\in\calH}$, with
probability approaching one, so that the conclusion of the delta method for
the extended functional transfers to $\bmaxhat(\delta)$. We prove the
second identity by showing that, with probability approaching one, the
supremum over $\Theta$ defining $\bmaxhat(\delta)$ is unaffected by
restricting it to $\calN_{\Theta}$. On the event
$\vertiii{(\widehat\rho_{n},\widehat P_{n})-(\rho_{0},\Pbase{})}_{\calH_{0}}
<\varepsilon_{*}$, whose probability tends to one by
Assumption~\ref{assump: clt-EB} and the Glivenko--Cantelli property of
Lemma~\ref{lem:donsker-EB}, the extended program coincides with the sample
program on $\calN_{\Theta}$, so that
$\bmax(\delta,\widehat\rho_{n},\widehat{\mathsf b}_{n})
=\sup\{\Kmax(\theta;\widehat\rho_{n},\widehat P_{n}):
\theta\in\calN_{\Theta},\ \mathsf D(\theta;\widehat\rho_{n},\widehat
P_{n})<\delta\}\leq\bmaxhat(\delta)$.
For the reverse inequality, we pass, as in
the proof of Theorem~\ref{thm:cons-EB}, to a subsequence along which
$\vertiii{(\widehat\rho_{n},\widehat P_{n})-(\rho_{0},\Pbase{})}_{\calH}\to0$
and $\|\widehat P_{n}-\Pbase{}\|_{\calH^{\Jdual}_{K}}\to0$ for every compact
$K$ almost surely, and apply Lemma~\ref{lem:aux-EB-vii} with the open
set $\calU=\{\theta:d(\theta,\Tmax(\rho_{0},\Pbase{}))<\varepsilon_{\Theta}\}$ and
$\nu_{n}=1/n$. For all $n$ large,
every $\theta\in\Theta_{\delta}(\widehat\rho_{n},\widehat P_{n})$ with
$\Kmax(\theta;\widehat\rho_{n},\widehat P_{n})\geq\bmaxhat(\delta)-1/n$ then
lies in $\calU\cap\Theta\subseteq\calN_{\Theta}$, so that the supremum
of
$\Kmax(\cdot;\widehat\rho_{n},\widehat P_{n})$ over
$\Theta_{\delta}(\widehat\rho_{n},\widehat P_{n})\setminus\calN_{\Theta}$ is
at most $\bmaxhat(\delta)-1/n<\bmaxhat(\delta)$, where $\bmaxhat(\delta)$ is
finite for $n$ large by Lemmas~\ref{lem:aux-EB-i} and~\ref{lem:aux-EB-vi}.
Hence the supremum over $\Theta_{\delta}(\widehat\rho_{n},\widehat P_{n})$
defining $\bmaxhat(\delta)$ equals the supremum over
$\Theta_{\delta}(\widehat\rho_{n},\widehat P_{n})\cap\calN_{\Theta}$, and,
as the subsequence is arbitrary, this holds with probability approaching
one, that is,
$\bmaxhat(\delta)=\bmax(\delta,\widehat\rho_{n},\widehat{\mathsf b}_{n})$.
The same argument applies to $\bminhat(\delta)$ with $g\mapsto-g$, which
completes the proof.
\end{proof}

\subsection{Proof of Theorem~\ref{thm:boot-EB}}
\begin{proof}\label{proof:thm:boot-EB}
Throughout, we let $\calN$, $r$, and $M$ be the neighborhood, radius, and constant of
Lemma~\ref{lem:aux-EB-vi}, and we use the Hausdorff distance $d_{\mathrm H}$.
To prove the main result, by Theorem~3.2 in
\citet{fang2019InferenceDirectionallyDifferentiable},
it suffices to verify the four conditions \textbf{(FS1)}--\textbf{(FS4)} below.

\uline{\textbf{(FS1).}} Hadamard directional differentiability of
$(\rho,P)\mapsto(\bmin,\bmax)(\delta,\rho,P)$ at $(\rho_{0},\Pbase{})$ tangentially
to $\R^{\dime}\times\allowbreak UC(\calH,e_{\Pbase{}})$, with continuous
derivative, is Theorem~\ref{thm:HD-EB}.     

\uline{\textbf{(FS2).}} $\sqrt n((\widehat\rho_{n},\widehat P_{n})-
(\rho_{0},\Pbase{}))\rightsquigarrow(\mathbb Z,\mathbb G)$ is established in Step~2
of the proof of Theorem~\ref{thm:HD-EB}. 

\uline{\textbf{(FS3).}} The conditional bootstrap law of
$\sqrt n((\widehat\rho_{n}^{*},\widehat P_{n}^{*})-(\widehat\rho_{n},\widehat
P_{n}))$ converges to the same $(\mathbb Z,\mathbb G)$ in probability. This is the
consistency of the nonparametric bootstrap for the $\Pbase{}$-Donsker class
$\calH$ adjoined with $\psi$
\citep[Theorems~3.6.1--3.6.2]{vandervaart1996WeakConvergenceEmpirical},
which gives the claim for the pair $(\mathbb G_{n}^{*}\psi,\mathbb G_{n}^{*})$, in the
bounded-Lipschitz metric and in outer probability. The bootstrap statistic differs from
this pair only in its first coordinate, by
$R_{n}^{*}:=\sqrt n(\widehat\rho_{n}^{*}-\widehat\rho_{n})-\mathbb G_{n}^{*}\psi$,
which is $\op(1)$ by the linearization assumed in Theorem~\ref{thm:boot-EB}. As every
function $f$ bounded by one and $1$-Lipschitz satisfies
$|f(a)-f(a')|\leq\min\{\|a-a'\|,2\}$, the bounded-Lipschitz distances of the two
conditional laws from the law of $(\mathbb Z,\mathbb G)$ differ by at most
$\E[\min\{\|R_{n}^{*}\|,2\}\mid Z_{1},\ldots,Z_{n}]$. This term converges to zero in
outer probability by Markov's inequality, as its expectation
$\E[\min\{\|R_{n}^{*}\|,2\}]$ vanishes by bounded convergence.

\uline{\textbf{(FS4).}} As in the extension of Section~\ref{subsec:ls-hd},
we consider the localized derivative, defined as $\widehat{d\bmax}_{n}$
but with the outer supremum restricted to
$\Tmaxhat\cap\calN_{\Theta}$ and the inner infimum to
$\Xmax(\theta;\widehat\rho_{n},\widehat P_{n})\cap\calA$, which is what places
$\ell_{\xi,\theta}$ in $\calH$.
Neither restriction binds with probability approaching one. Indeed, by
Assumption~\ref{assump: clt-EB} and Lemma~\ref{lem:donsker-EB},
$\vertiii{(\widehat\rho_{n},\widehat P_{n})-(\rho_{0},\Pbase{})}_{\calH}\cprob0$
and $\|\widehat P_{n}-\Pbase{}\|_{\calH^{\Jdual}_{K}}\cprob0$ for every
compact $K$, while $\widehat\nu_{n}\sqrt{\log n/n}\cprob0$, as
$\widehat\nu_{n}=\Op(1)$, so that, exactly
as in the proof of Theorem~\ref{thm:cons-EB}, from any subsequence we may
extract a further subsequence along which these convergences hold almost
surely. Along it, Lemma~\ref{lem:aux-EB-vii}, applied with $\calU$ the
intersection of $\interior(\calN)$ and the open set
$\{\theta:d(\theta,\Tmax(\rho_{0},\Pbase{}))<\varepsilon_{\Theta}\}$, which contains
$\Tmax(\rho_{0},\Pbase{})$, shows that $\Tmaxhat\subseteq\calU$, and
hence $\Tmaxhat\subseteq\calU\cap\Theta\subseteq\calN_{\Theta}\cap\calN$, for all $n$
large. Since the subsequence was arbitrary, the probability of the event
\begin{equation}\label{eq:boot-localization}
  \Tmaxhat\subseteq\calN_{\Theta}\cap\calN
  \ \text{and}\
  \vertiii{(\widehat\rho_{n},\widehat P_{n})-(\rho_{0},\Pbase{})}_{\calH}<r
\end{equation}
approaches 1.
On the event in \eqref{eq:boot-localization}, Lemma~\ref{lem:aux-EB-vi}
makes the sets $\Xmax(\theta;\widehat\rho_{n},\widehat P_{n})$,
$\theta\in\Tmaxhat$, nonempty, compact, and contained in $\Xi\subseteq\calA$.
Both restrictions are then vacuous: $\Tmaxhat\subseteq\calN_{\Theta}$ gives
$\Tmaxhat\cap\calN_{\Theta}=\Tmaxhat$, and
$\Xmax(\theta;\widehat\rho_{n},\widehat P_{n})\subseteq\calA$ gives
$\Xmax(\theta;\widehat\rho_{n},\widehat P_{n})\cap\calA
=\Xmax(\theta;\widehat\rho_{n},\widehat P_{n})$ for every $\theta\in\Tmaxhat$,
so that on this event the localized derivative coincides with the
expression of Section~\ref{subsec:ls-bootstrap},
\[
  \sup_{\theta\in\Tmaxhat\cap\calN_{\Theta}}\,
  \inf_{\xi\in\Xmax(\theta;\widehat\rho_{n},\widehat P_{n})\cap\calA}
  \bigl\{\lambda_{12}^{\top}h+\gamma\,\ell_{\xi,\theta}\bigr\}
  \;=\;\sup_{\theta\in\Tmaxhat}\,
  \inf_{\xi\in\Xmax(\theta;\widehat\rho_{n},\widehat P_{n})}
  \bigl\{\lambda_{12}^{\top}h+\gamma\,\ell_{\xi,\theta}\bigr\}
  \;=\;\widehat{d\bmax}_{n}[h,\gamma].
\]
Let $\widetilde{d\bmax}_{n}$ be equal to the localized derivative, and
hence to $\widehat{d\bmax}_{n}$, on the intersection of the event in
\eqref{eq:boot-localization} with $\{\Tmaxhat\neq\emptyset\}$, an event whose
probability also tends to one by part~(i) of the first fact below, and set
$\widetilde{d\bmax}_{n}:=0$ off it. On that event the outer supremum is over a
nonempty set and each inner infimum over a nonempty subset of the compact set
$\calA$, so that $\widetilde{d\bmax}_{n}$ is Lipschitz on all of
$\R^{\dime}\times\ell^{\infty}(\calH)$,
\[
  \bigl|\widetilde{d\bmax}_{n}[h,\gamma]
  -\widetilde{d\bmax}_{n}[h',\gamma']\bigr|
  \;\leq\;M_{\calA}\|h-h'\|+\|\gamma-\gamma'\|_{\ell^{\infty}(\calH)},
\]
with $M_{\calA}$ as in the proof of Theorem~\ref{thm:HD-EB},
by $|\sup_{S}f-\sup_{S}f'|\leq\sup_{S}|f-f'|$, the same inequality for infima,
and $|(\gamma-\gamma')\ell_{\xi,\theta}|\leq\|\gamma-\gamma'\|_{\ell^{\infty}(\calH)}$
for $\ell_{\xi,\theta}\in\calH$. By
\citet[Remark~3.4 and Lemma~S.3.6]{fang2019InferenceDirectionallyDifferentiable},
their Assumption~4 therefore holds for $\widetilde{d\bmax}_{n}$ once we
establish pointwise convergence on the tangent set, which, as
$\widetilde{d\bmax}_{n}$ and $\widehat{d\bmax}_{n}$ coincide with probability
approaching one, amounts to
\begin{equation}\label{eq:boot-pointwise}
  \widehat{d\bmax}_{n}[h,\gamma]\;\cprob\;d\bmaxAt{(\rho_{0},\Pbase{})}[h,\gamma]
  \qquad\text{for every }(h,\gamma)\in\R^{\dime}\times UC(\calH,e_{\Pbase{}}).
\end{equation}
Fix $(h,\gamma)\in\R^{\dime}\times UC(\calH,e_{\Pbase{}})$ and write
\[
  \Phi(\theta,\xi):=\lambda_{12}^{\top}h+\gamma\ell_{\xi,\theta},\qquad
  v(\theta):=v(\theta;h,\gamma),\qquad
  \widehat v_{n}(\theta):=\min_{\xi\in\Xmax(\theta;\widehat\rho_{n},\widehat P_{n})}
  \Phi(\theta,\xi),
\]
where
$v(\theta;h,\gamma)=\min_{\xi\in\Xmax(\theta;\rho_{0},\Pbase{})}\Phi(\theta,\xi)$ is
the inner value of Lemma~\ref{lem:aux-EB-v}, so that
$d\bmaxAt{(\rho_{0},\Pbase{})}[h,\gamma]
=\max_{\theta\in\Tmax(\rho_{0},\Pbase{})}v(\theta)$ and, on the event in
\eqref{eq:boot-localization},
$\widehat{d\bmax}_{n}[h,\gamma]=\sup_{\theta\in\Tmaxhat}\widehat
v_{n}(\theta)$,
with $v$ well defined on $\calN\cap\Theta$ by
Lemma~\ref{lem:aux-EB-vi}. By the fourth property of the dual in the
proof of Lemma~\ref{lem:aux-EB}, $\Phi$ is uniformly continuous on the compact set of
pairs $(\theta,\xi)\in\Theta\times\calA$ with $\xi\in\calF(\theta)$, with a
nondecreasing modulus $\omega$
satisfying $\omega(t)\to0$ as $t\downarrow0$. The minima above, taken
over nonempty compact sets, are therefore attained, and so is the maximum of $v$ over
the compact set $\Tmax(\rho_{0},\Pbase{})$, on which $v$ is continuous by
Lemma~\ref{lem:aux-EB-v}. The set $\Tmaxhat$ need not be closed, as
$\Theta_{\delta}(\widehat\rho_{n},\widehat P_{n})$ is defined by a strict inequality,
so we keep the supremum over it. For any $\theta\in\Theta$ and
nonempty compact $B,B'\subseteq\calF(\theta)\cap\calA$, choosing for each $\xi\in B$ a
point $\xi'\in B'$ with
$\|\xi-\xi'\|\leq d_{\mathrm H}(B,B')$ and conversely gives
\begin{equation}\label{eq:boot-modulus}
  \Bigl|\min_{\xi\in B}\Phi(\theta,\xi)-\min_{\xi\in B'}\Phi(\theta,\xi)\Bigr|
  \;\leq\;\omega\bigl(d_{\mathrm H}(B,B')\bigr).
\end{equation}
The convergence \eqref{eq:boot-pointwise} rests on two facts.

\uline{\textbf{First fact.}} We show that
$d_{\mathrm H}(\Tmaxhat,\Tmax(\rho_{0},\Pbase{}))\cprob0$.
We show the two inclusions $\Tmax(\rho_{0},\Pbase{})\subseteq\Tmaxhat$ and
$\Tmaxhat\subseteq\calU_{\varepsilon}:=\{\theta:d(\theta,\Tmax(\rho_{0},\Pbase{}))
<\varepsilon\}$, for arbitrary $\varepsilon>0$, each with probability
approaching one, which bound the two directed distances by $0$ and by
$\varepsilon$, respectively.

(i) $\Tmax(\rho_{0},\Pbase{})\subseteq\Tmaxhat$. By
Lemma~\ref{lem:aux-EB-i}, $\Tmax(\rho_{0},\Pbase{})\subseteq
\Theta_{\delta}(\widehat\rho_{n},\widehat P_{n})$ on the event
$\{\vertiii{(\widehat\rho_{n},\widehat P_{n})-(\rho_{0},\Pbase{})}_{\calH}<r\}$,
whose probability tends to one, and on that event, for every
$\theta^{*}\in\Tmax(\rho_{0},\Pbase{})\subseteq\calN$,
\[
  \Kmax(\theta^{*};\widehat\rho_{n},\widehat P_{n})-\bmaxhat(\delta)
  =\bigl\{\Kmax(\theta^{*};\widehat\rho_{n},\widehat P_{n})
   -\Kmax(\theta^{*};\rho_{0},\Pbase{})\bigr\}
  +\bigl\{\bmax(\delta,\rho_{0},\Pbase{})-\bmaxhat(\delta)\bigr\}
  \;\geq\;-n^{-1/2}R_{n},
\]
where we used $\Kmax(\theta^{*};\rho_{0},\Pbase{})=\bmax(\delta,\rho_{0},\Pbase{})$
and set $R_{n}:=\sqrt n\bigl(M\vertiii{(\widehat\rho_{n},\widehat P_{n})
-(\rho_{0},\Pbase{})}_{\calH}+|\bmaxhat(\delta)-\bmax(\delta,\rho_{0},\Pbase{})|\bigr)$%
. The first brace is bounded by the Lipschitz estimate of
Lemma~\ref{lem:aux-EB-vi}. Now $R_{n}=\Op(1)$, by the root-$n$ rates of
Assumption~\ref{assump: clt-EB} and Lemma~\ref{lem:donsker-EB} for the
first term and by the weak limit of Theorem~\ref{thm:HD-EB} for the second.
Since $\widehat\nu_{n}^{-1}=\Op(1)$ and $\sqrt{\log n}\to\infty$, we have
$\widehat\nu_{n}\sqrt{\log n}\to\infty$ in probability, so that
$\PP(\widehat\nu_{n}\sqrt{\log n}\geq R_{n})\to1$ by the tightness of
$R_{n}$. On the intersection of these events every
$\theta^{*}\in\Tmax(\rho_{0},\Pbase{})$ satisfies the inequality
$\Kmax(\theta^{*};\widehat\rho_{n},\widehat P_{n})\geq\bmaxhat(\delta)
-\widehat\nu_{n}\sqrt{\log n/n}$ defining $\Tmaxhat$, which proves~(i).

(ii) $\Tmaxhat\subseteq\calU_{\varepsilon}$. This is the argument
that gave \eqref{eq:boot-localization}, with $\calU_{\varepsilon}$ in place of
$\calU$: along any almost-sure subsequence, Lemma~\ref{lem:aux-EB-vii}
places $\Tmaxhat$ inside $\calU_{\varepsilon}$ for all $n$ large, and hence
$\PP(\Tmaxhat\subseteq\calU_{\varepsilon})\to1$. In particular, along any
almost-sure subsequence,
$\sup_{\theta\in\Tmaxhat}d(\theta,\Tmax(\rho_{0},\Pbase{}))\to0$ almost
surely, as $\varepsilon$ is arbitrary.

\uline{\textbf{Second fact.}} For every $\theta\in\Tmax(\rho_{0},\Pbase{})$ and
every sequence $\theta_{n}\to\theta$, along any almost-sure subsequence,
\[
  d_{\mathrm H}\bigl(\Xmax(\theta_{n};\widehat\rho_{n},\widehat P_{n}),
   \Xmax(\theta;\rho_{0},\Pbase{})\bigr)\to0 .
\]
Both sets are nonempty and compact for $n$ large, by
Lemma~\ref{lem:aux-EB-iii}. The directed distance from
$\Xmax(\theta_{n};\widehat\rho_{n},\widehat P_{n})$ to
$\Xmax(\theta;\rho_{0},\Pbase{})$ vanishes by the joint upper
hemicontinuity of Lemma~\ref{lem:aux-EB-iv}. To see this, were it bounded below by some
$\varepsilon>0$ along a subsequence, selections $\xi_{n}$ with
$d(\xi_{n},\Xmax(\theta;\rho_{0},\Pbase{}))\geq\varepsilon$ would admit a further
subsequence converging to a point of $\Xmax(\theta;\rho_{0},\Pbase{})$, a
contradiction. The directed distance from $\Xmax(\theta;\rho_{0},\Pbase{})$
to $\Xmax(\theta_{n};\widehat\rho_{n},\widehat P_{n})$ vanishes by
Assumption~\ref{assump: M-joint-EB}, which gives
$d(\xi^{*},\Xmax(\theta_{n};\widehat\rho_{n},\widehat P_{n}))\to0$ for each
$\xi^{*}\in\Xmax(\theta;\rho_{0},\Pbase{})$. As these maps are $1$-Lipschitz in
$\xi^{*}$, the pointwise convergence is uniform on the compact set
$\Xmax(\theta;\rho_{0},\Pbase{})$. Applied with
$(\widehat\rho_{n},\widehat P_{n})$ replaced by the constant sequence
$(\rho_{0},\Pbase{})$, the same argument shows that
$\theta'\mapsto\Xmax(\theta';\rho_{0},\Pbase{})$ is Hausdorff continuous at
$\theta$.

\uline{\textbf{Conclusion.}} On the event in \eqref{eq:boot-localization},
using $|\sup_{S}f-\sup_{S}f'|\leq\sup_{S}|f-f'|$ for bounded $f,f'$,
\[
  \bigl|\widehat{d\bmax}_{n}[h,\gamma]-d\bmaxAt{(\rho_{0},\Pbase{})}[h,\gamma]\bigr|
  \;\leq\;
  \underbrace{\sup_{\theta\in\Tmaxhat}
    \bigl|\widehat v_{n}(\theta)-v(\theta)\bigr|}_{(a)}
  +\underbrace{\Bigl|\sup_{\theta\in\Tmaxhat}v(\theta)
    -\max_{\theta\in\Tmax(\rho_{0},\Pbase{})}v(\theta)\Bigr|}_{(b)} ,
\]
and it suffices to show that $(a)\cprob0$ and $(b)\cprob0$. For $(a)$,
\eqref{eq:boot-modulus} gives $(a)\leq\omega(S_{n})$, where
\[
  S_{n}:=\sup_{\theta\in\Tmaxhat}
  d_{\mathrm H}\bigl(\Xmax(\theta;\widehat\rho_{n},\widehat P_{n}),
  \Xmax(\theta;\rho_{0},\Pbase{})\bigr),
\]
so it suffices that $S_{n}\cprob0$, which we verify along an arbitrary
almost-sure subsequence. Were $S_{n}\geq\varepsilon$ for infinitely many $n$
on an event of positive probability, we could select
$\theta_{n}\in\Tmaxhat$ at which the Hausdorff distance in the definition
of $S_{n}$ is at least $\varepsilon/2$ and, by part~(ii) of the first fact
and the compactness of $\Theta$, a further subsequence with
$\theta_{n}\to\theta\in\Tmax(\rho_{0},\Pbase{})$. The triangle inequality
\[
  d_{\mathrm H}\bigl(\Xmax(\theta_{n};\widehat\rho_{n},\widehat P_{n}),
   \Xmax(\theta_{n};\rho_{0},\Pbase{})\bigr)
  \leq d_{\mathrm H}\bigl(\Xmax(\theta_{n};\widehat\rho_{n},\widehat P_{n}),
   \Xmax(\theta;\rho_{0},\Pbase{})\bigr)
  +d_{\mathrm H}\bigl(\Xmax(\theta;\rho_{0},\Pbase{}),
   \Xmax(\theta_{n};\rho_{0},\Pbase{})\bigr)
\]
and the second fact, applied to each term on the right, then give a
contradiction. Hence $(a)\cprob0$. For $(b)$, fix $\varepsilon>0$. By
Lemma~\ref{lem:aux-EB-v}, $v$ is continuous at every
$\theta^{*}\in\Tmax(\rho_{0},\Pbase{})$, so each $\theta^{*}$ has a radius
$s_{\theta^{*}}>0$ with $|v(\theta)-v(\theta^{*})|<\varepsilon$ whenever
$\|\theta-\theta^{*}\|<s_{\theta^{*}}$. Extracting a finite subcover of
$\Tmax(\rho_{0},\Pbase{})$ by the balls of half these radii and letting
$\varsigma>0$ be the smallest half radius, every $\theta$ with
$d(\theta,\Tmax(\rho_{0},\Pbase{}))<\varsigma$ lies within $s_{\theta_{k}}$ of
some center $\theta_{k}$ and so satisfies
$v(\theta)<v(\theta_{k})+\varepsilon\leq\max_{\Tmax(\rho_{0},\Pbase{})}v+\varepsilon$.
On the event $\{\Tmax(\rho_{0},\Pbase{})\subseteq\Tmaxhat\subseteq
\calU_{\varsigma}\}$, whose probability tends to one by the first fact, we
therefore have $0\leq(b)\leq\varepsilon$. As $\varepsilon$ is arbitrary,
$(b)\cprob0$, which proves \eqref{eq:boot-pointwise} and hence~(FS4).

With (FS1)--(FS4) verified,
\citet[][Theorem~3.2]{fang2019InferenceDirectionallyDifferentiable} delivers
the conditional weak convergence of $\widetilde{d\bmax}_{n}$, evaluated at
the bootstrap directions, to
$d\bmaxAt{(\rho_{0},\Pbase{})}[\mathbb Z,\mathbb G]$. The event on which
$\widetilde{d\bmax}_{n}=\widehat{d\bmax}_{n}$ depends on the data alone and has
probability approaching one. 
The same conditional weak convergence therefore holds for the bootstrap
statistic of Algorithm~\ref{alg:FS-EB}.

\uline{\textbf{Coverage.}} Coverage follows by inverting the conditional
convergence in distribution just established. Write
\[
  F^{\mathrm{up}}(t):=\PP\bigl(d\bmaxAt{(\rho_{0},\Pbase{})}[\mathbb Z,\mathbb G]
  \leq t\bigr),
  \qquad
  F^{\mathrm{low}}(t):=\PP\bigl(d\bminAt{(\rho_{0},\Pbase{})}[\mathbb Z,\mathbb G]
  \leq t\bigr)
\]
for the limiting distribution functions and let
$q^{\mathrm{up}}(\tau):=\inf\{t\in\R:F^{\mathrm{up}}(t)\geq\tau\}$ and
$q^{\mathrm{low}}(\tau):=\inf\{t\in\R:F^{\mathrm{low}}(t)\geq\tau\}$ denote
the associated quantile functions. Let
$\widehat{\mathfrak{c}}^{\mathrm{up}}_{\tau}$ and
$\widehat{\mathfrak{c}}^{\mathrm{low}}_{\tau}$
denote the $\tau$-quantiles of the conditional laws of the two bootstrap
statistics, which the empirical quantiles over the $B$ bootstrap draws of
Algorithm~\ref{alg:FS-EB} approximate. Fix
$\tau\in\{\alpha/2,\alpha\}$ and $\tau'\in\{1-\alpha,1-\alpha/2\}$. By
hypothesis, $F^{\mathrm{up}}$ is continuous and strictly increasing at
$q^{\mathrm{up}}(\tau)$, so that $q^{\mathrm{up}}$ is continuous at $\tau$, and likewise
$q^{\mathrm{low}}$ is continuous at $\tau'$. The conditional weak
convergence holds in the bounded-Lipschitz metric in outer probability, so
that every subsequence admits a further subsequence along which the
conditional laws converge weakly almost surely. Along such a subsequence
Lemma~21.2 in
\citet{vandervaart1998AsymptoticStatistics}, which gives the
convergence of quantiles at every continuity point of the limiting quantile
function,
yields $\widehat{\mathfrak{c}}^{\mathrm{up}}_{\tau}\to q^{\mathrm{up}}(\tau)$
and
$\widehat{\mathfrak{c}}^{\mathrm{low}}_{\tau'}\to q^{\mathrm{low}}(\tau')$
almost surely. As the initial subsequence was arbitrary, this gives
\begin{equation}\label{eq:quantile-consistency}
  \widehat{\mathfrak{c}}^{\mathrm{up}}_{\tau}\;\cprob\;q^{\mathrm{up}}(\tau),
  \qquad
  \widehat{\mathfrak{c}}^{\mathrm{low}}_{\tau'}\;\cprob\;
  q^{\mathrm{low}}(\tau').
\end{equation}
Finally, combining \eqref{eq:quantile-consistency} with the marginal limit
$\sqrt n\bigl(\bmaxhat(\delta)-\bmax(\delta,\rho_{0},\Pbase{})\bigr)\rightsquigarrow
d\bmaxAt{(\rho_{0},\Pbase{})}[\mathbb Z,\mathbb G]$ of Theorem~\ref{thm:HD-EB}, the
continuity of $F^{\mathrm{up}}$ at $q^{\mathrm{up}}(\alpha)$, and Slutsky's
lemma, we obtain
\[
  \liminf_{n\to\infty}
  \PP\Bigl(\bmax(\delta,\rho_{0},\Pbase{})\in
   \bigl(-\infty,\,\bmaxhat(\delta)
   -\widehat{\mathfrak{c}}^{\mathrm{up}}_{\alpha}/\sqrt n\,\bigr]
  \Bigr)
  \;=\;1-F^{\mathrm{up}}\!\bigl(q^{\mathrm{up}}(\alpha)^{-}\bigr)
  \;\geq\;1-\alpha.
\]
The analogous identities for the lower one-sided and two-sided sets
complete the proof.
\end{proof}

\subsection{Proof of Proposition~\ref{thm: shadow price budget}}
\begin{proof}\label{proof: shadow price budget}
We prove the results for $\overline{\tau}$. Those for $\underline{\tau}$ follow by
substituting $-g$ for $g$, which leaves the integrability condition
$\E_{\Pbase{}}[|g(Z)|]<\infty$
unaffected. Write $b:=(1-\kappa)\delta$ and $\overline{v}(b):=\overline{\tau}(\delta)$, so
that by \eqref{eq: tv set summary}
\begin{align}
\overline{v}(b)=\sup\bigl\{\E_P[g(Z)]:\ P\in\calP(\calZ),\ \kappa P\leq\Pbase{},\
\TV(P\,\|\,\Pbase{})\leq b\bigr\},
\qquad b\in\R_+.
\label{eq: v budget reparametrization}
\end{align}
The reparametrization is affine with slope $1-\kappa$, and the dual problem
\eqref{eq: dual summary tv} depends on the budget only through
$\eta(1-\kappa)\delta=\eta b$, so that $\Xmax$ is its solution set at $b$ and
$\partial_{\pm}\overline{\tau}(\delta)=(1-\kappa)\,\partial_{\pm}\overline{v}(b)$. It
therefore suffices to prove the claims for $\overline{v}$ on $(0,\infty)$, with $\eta$ in
place of $(1-\kappa)\eta$ and $[1-\kappa,\infty)$ in place of $[1,\infty)$. Let
\[
f_{b}(\xi):=\zeta+\eta b
+\E_{\Pbase{}}\bigl[\mathsf{r}_{\TV}\bigl(g(Z)-\zeta,\eta\bigr)\bigr],
\qquad \xi=(\zeta,\eta)\in\calF:=\R\times\R_+,
\]
be the dual objective at budget $b$, so that
\begin{align}
f_{b}(\xi)=f_{b_0}(\xi)+\eta\,(b-b_0)
\qquad\text{for all }\xi\in\calF\text{ and }b,b_0\in\R_+.
\label{eq: budget affine shift}
\end{align}
By properties \ref{prop:r-hom} and \ref{prop:r-lip} of the integrand
\eqref{eq:integrand family} with $\mathbb{U}=[0,1/\kappa]$, $\mathsf{r}_{\TV}$ is jointly
convex with $|\mathsf{r}_{\TV}(x,\eta)|\leq\kappa^{-1}|x|+(2\kappa)^{-1}\eta$ and
$|\mathsf{r}_{\TV}(x,\eta)-\mathsf{r}_{\TV}(x',\eta')|\leq\kappa^{-1}|x-x'|
+(2\kappa)^{-1}|\eta-\eta'|$, so that, as $\E_{\Pbase{}}[|g(Z)|]<\infty$, $f_b$ is
finite, convex, and Lipschitz continuous on the closed convex set $\calF$.

\uline{\textbf{Step 1: regularity of $\overline{v}$.}} Fix $b\in\R_+$. The baseline is
feasible in \eqref{eq: v budget reparametrization}, so
$\overline{v}(b)\geq\E_{\Pbase{}}[g(Z)]>-\infty$, and every feasible $P$ satisfies
$P\ll\Pbase{}$ with density $w\leq1/\kappa$ by
Lemma~\ref{lemma: mixture density bounds-i}, so
$\E_P[g(Z)]\leq\kappa^{-1}\E_{\Pbase{}}[|g(Z)|]<\infty$. Monotonicity holds because the
feasible set grows with $b$, and $\overline{v}$ is constant on $[1-\kappa,\infty)$
because $\kappa P\leq\Pbase{}$ already implies $\TV(P\,\|\,\Pbase{})\leq1-\kappa$
(Lemma~\ref{lemma: contamination as constrained tv}). For concavity, let $P_1$ and $P_2$
be feasible at $b_1$ and $b_2$, let $t\in(0,1)$, and set $P_t:=tP_1+(1-t)P_2$. Then
$\kappa P_t\leq\Pbase{}$ and
$\TV(P_t\,\|\,\Pbase{})\leq t\,\TV(P_1\,\|\,\Pbase{})+(1-t)\,\TV(P_2\,\|\,\Pbase{})
\leq tb_1+(1-t)b_2$, by the convexity of
$P\mapsto\TV(P\,\|\,\Pbase{})=\sup_{A\in\mathscr{F}}|P(A)-\Pbase{}(A)|$
(Lemma~\ref{lem:tv_posneg}), a supremum of convex functions. Hence
$\overline{v}(tb_1+(1-t)b_2)\geq\E_{P_t}[g(Z)]=t\,\E_{P_1}[g(Z)]+(1-t)\,\E_{P_2}[g(Z)]$,
and taking suprema over $P_1$ and $P_2$ gives concavity on $\R_+$. Consequently,
$\overline{v}$ is continuous on $(0,\infty)$, and its one-sided derivatives
exist and are finite at every $b>0$ and are nonincreasing in $b$, with
$\partial_{+}\overline{v}(b)\leq\partial_{-}\overline{v}(b)$
\citep[Theorem~10.1 and Section~24]{rockafellar1970ConvexAnalysis}. The map
$\partial_{+}\overline{v}$ is nonnegative by monotonicity, and, as a
concave function on an open interval, $\overline{v}$ is differentiable at all but
countably many $b>0$. Finally, for continuity at $0$, a feasible $P$ at
$b$ has $|w-1|\leq1/\kappa$ and $\E_{\Pbase{}}[|w-1|]=2\,\TV(P\,\|\,\Pbase{})\leq2b$
(Lemma~\ref{lem:tv_posneg}), so that for every $M>0$
\[
|\E_P[g(Z)]-\E_{\Pbase{}}[g(Z)]|
\leq\E_{\Pbase{}}\bigl[|w-1|\,|g(Z)|\bigr]
\leq 2bM+\kappa^{-1}\,\E_{\Pbase{}}\bigl[|g(Z)|\,\indic\{|g(Z)|>M\}\bigr].
\]
Taking the supremum over feasible $P$, letting $b\downarrow0$, and then $M\to\infty$ by
dominated convergence gives $\overline{v}(b)\to\overline{v}(0)$. This proves
part~\ref{thm: shadow price budget-i}.

\uline{\textbf{Step 2: strong duality and dual attainment.}} Fix $b>0$. We verify
Assumption~\ref{assump: dual primitives superlinear} for the entropy
$\overline{\phi}_{\TV}$ at budget $b$ with $\dimm=0$. As $0$ and
$1/\kappa$ lie in the domain of $\phi_{\TV}$, Remark~\ref{remark: feasibility mixture}
shows that $\overline{\phi}_{\TV}$ has infinite recession and falls under case~(b) of
Assumption~\ref{assump: dps-entropy} with $\bar{x}=1/\kappa$, and that the
integrability condition in Assumption~\ref{assump: dps-int}, which is the one of
Theorem~\ref{thm: existence full problem-ii}, follows from
$\E_{\Pbase{}}[|g(Z)|]<\infty$.
Since $(\overline{\phi}_{\TV}^{*})^{\pi}=\mathsf{r}_{\TV}$ by
Lemma~\ref{lemma: family entropy}, the point $(\zeta^{*},\eta^{*})=(0,1)$ satisfies
Assumption~\ref{assump: dps-anchor} by the bound on $\mathsf{r}_{\TV}$ recorded above.
Assumption~\ref{assump: dps-feas} holds with $P^\circ=\Pbase{}$, as
$D_{\overline{\phi}_{\TV}}(\Pbase{}\,\|\,\Pbase{})=0<b$.
Assumption~\ref{assump: dps-support} concerns case~(a) only, and
Assumption~\ref{assump: dps-ceiling} is vacuous because $\dimm=0$. Hence
Theorem~\ref{thm: existence of dual solutions} shows that $\Xmax$, the set of minimizers
of $f_b$ over $\calF$, is nonempty, compact, and convex, where the constraint in
\eqref{eq: dual with moments} is vacuous since
$\overline{\phi}_{\TV}^\infty(1)=+\infty$, and Lemma~\ref{lemma: slater dual form-iii}
shows that the recession function of $f_b$ is strictly positive on every nonzero
recession direction of $\calF$. Moreover,
Theorem~\ref{theorem: strong duality with mixture constraint-i} applies with
$\phi=\phi_{\TV}$ and budget $b$, its compatibility condition holding under the
conventions $\calM(\rho)=\Gamma_{\overline{\phi}_{\TV}}(b)=\{0\}$ of
Section~\ref{subsec: setup}, and its robustness set is the feasible set of
\eqref{eq: v budget reparametrization} by \eqref{eq: mixture sets as contamination} and
Lemma~\ref{lemma: contamination as constrained tv-i}. Hence
\begin{align}
\overline{v}(b)=\inf_{\xi\in\calF}f_{b}(\xi)=f_b(\xi^{*})
\qquad\text{for every }\xi^{*}\in\Xmax.
\label{eq: v strong duality}
\end{align}

\uline{\textbf{Step 3: optimal multipliers are supergradients.}} Fix $b_0>0$ and
$\xi^{\star}=(\zeta^{\star},\eta^{\star})\in\overline{\Xi}_{b_0}$, indexing solution sets
by the budget $b$. For every $b>0$,
$\overline{v}(b)\leq f_{b}(\xi^{\star})=f_{b_0}(\xi^{\star})+\eta^{\star}(b-b_0)
=\overline{v}(b_0)+\eta^{\star}(b-b_0)$, by \eqref{eq: v strong duality},
\eqref{eq: budget affine shift}, and the optimality of $\xi^{\star}$ at $b_0$. Hence
$\eta^{\star}$ is a supergradient of $\overline{v}$ at $b_0$, and
\begin{align}
\partial_{+}\overline{v}(b_0)\;\leq\;\overline{\eta}^{\min}_{b_0}
\;\leq\;\overline{\eta}^{\max}_{b_0}\;\leq\;\partial_{-}\overline{v}(b_0).
\label{eq: supergradient sandwich}
\end{align}

\uline{\textbf{Step 4: compactness of nearby solutions.}} Fix
$0<b_1\leq b_0\leq b_2<\infty$ and $b_a\in(0,b_1)$. For $b\in[b_1,b_2]$ and
$\xi^{\star}\in\overline{\Xi}_{b}$ with multiplier $\eta^{\star}$,
\eqref{eq: supergradient sandwich} and concavity give
$\eta^{\star}\leq\partial_{-}\overline{v}(b)\leq\partial_{+}\overline{v}(b_a)=:K<\infty$,
and then \eqref{eq: budget affine shift}, optimality at $b$, and monotonicity give
$f_{b_0}(\xi^{\star})=\overline{v}(b)+\eta^{\star}(b_0-b)
\leq\overline{v}(b_2)+K(b_2-b_1)=:c$. Hence
$\bigcup_{b\in[b_1,b_2]}\overline{\Xi}_{b}\subseteq
L_c:=\{\xi\in\calF: f_{b_0}(\xi)\leq c\}$, a nonempty closed convex set, since
$f_{b_0}$ is continuous and convex on the closed convex set $\calF$. Its recession cone
satisfies $0^{+}(L_c)\subseteq 0^{+}(\calF)\cap\{d: f^{\infty}_{b_0}(d)\leq0\}=\{0\}$,
by Corollary~8.3.3 and Theorem~8.7 in
\citet{rockafellar1970ConvexAnalysis} and Step~2,
so that $L_c$ is compact \citep[Theorem~8.4]{rockafellar1970ConvexAnalysis}.

\uline{\textbf{Step 5: the one-sided derivatives.}} Fix $b_0>0$, let $b_k\downarrow b_0$
with $b_k\leq b_2$ for some $b_2>b_0$, and pick
$\xi^{\star}_k=(\zeta_k,\eta_k)\in\overline{\Xi}_{b_k}$, which is possible by Step~2.
Step~3 applied at $b_0$ and evaluated at $b_k$, and applied at $b_k$ and evaluated at
$b_0$, gives
\[
\eta_k\;\leq\;\frac{\overline{v}(b_k)-\overline{v}(b_0)}{b_k-b_0}\;
\leq\;\overline{\eta}^{\min}_{b_0},
\]
and the difference quotients converge to $\partial_{+}\overline{v}(b_0)$ by Step~1. By
Step~4, along a subsequence $\xi^{\star}_k\to\xi^{\circ}=(\zeta^{\circ},\eta^{\circ})$,
and $\xi^{\circ}\in\calF$ since $\calF$ is closed. Moreover,
$f_{b_0}(\xi^{\circ})=\lim_k f_{b_0}(\xi^{\star}_k)
=\lim_k\{\overline{v}(b_k)+\eta_k(b_0-b_k)\}=\overline{v}(b_0)$, by the continuity of
$f_{b_0}$, by \eqref{eq: budget affine shift} with the optimality of $\xi^{\star}_k$ at
$b_k$, and by the continuity of $\overline{v}$ together with the boundedness of
$(\eta_k)$. Hence $\xi^{\circ}\in\overline{\Xi}_{b_0}$ and
$\eta^{\circ}\geq\overline{\eta}^{\min}_{b_0}$, so that along the subsequence
$\overline{\eta}^{\min}_{b_0}\geq\partial_{+}\overline{v}(b_0)\geq\lim_k\eta_k
=\eta^{\circ}\geq\overline{\eta}^{\min}_{b_0}$, that is,
$\partial_{+}\overline{v}(b_0)=\overline{\eta}^{\min}_{b_0}$. The symmetric argument
with $b_k\uparrow b_0$ gives $\partial_{-}\overline{v}(b_0)=\overline{\eta}^{\max}_{b_0}$.
Translating back through
$\partial_{\pm}\overline{\tau}(\delta)=(1-\kappa)\,\partial_{\pm}\overline{v}(b)$ proves
part~\ref{thm: shadow price budget-ii}.

\uline{\textbf{Step 6: parts~\ref{thm: shadow price budget-iii}
and~\ref{thm: shadow price budget-iv}.}} For $0<b_1<b_2$, Step~5 and the monotonicity
of the difference quotients of the concave $\overline{v}$ give
$\overline{\eta}^{\min}_{b_1}=\partial_{+}\overline{v}(b_1)
\geq\{\overline{v}(b_2)-\overline{v}(b_1)\}/(b_2-b_1)
\geq\partial_{-}\overline{v}(b_2)=\overline{\eta}^{\max}_{b_2}$, which is
part~\ref{thm: shadow price budget-iii}, as $\delta\mapsto b$ is increasing. By Step~5,
$\overline{v}$ is differentiable at $b$ if and only if
$\overline{\eta}^{\min}_{b}=\overline{\eta}^{\max}_{b}$, and since
$\{\eta:(\zeta,\eta)\in\overline{\Xi}_{b}\}$ is the compact interval
$[\overline{\eta}^{\min}_{b},\overline{\eta}^{\max}_{b}]$, as the image of the compact
convex set $\overline{\Xi}_{b}$ under a linear map, this is the uniqueness of the
optimal budget multiplier. The same holds for $\overline{\tau}$ at
$\delta=b/(1-\kappa)$, with $\overline{\tau}'(\delta)=(1-\kappa)\overline{v}'(b)$.
Nonnegativity of $\overline{\eta}$ holds because $\eta\geq0$ on $\calF$, and its
monotonicity follows from part~\ref{thm: shadow price budget-iii}.
\end{proof}

\subsection{Proof of Proposition~\ref{prop: certified breakdown}}
\begin{proof}\label{proof: certified breakdown}
Write $\widehat{\delta}:=\widehat{\delta}_{\mathsf{cb}}(\alpha)$,
$\mathfrak{c}:=\mathfrak{c}_{1-\alpha/2}$, $\|\cdot\|_{\infty}$ for the supremum norm on
$[0,1]$, $B:=\underline{\tau}-\tau_\star$,
$G_n:=\sqrt{n}\,(\widehat{\underline{\tau}}_n-\underline{\tau})$, and
$c_n:=\widehat{\underline{\tau}}_n-n^{-1/2}\mathfrak{c}\,\widehat{\sigma}_n
=\underline{\tau}+n^{-1/2}(G_n-\mathfrak{c}\,\widehat{\sigma}_n)$, so that
$\widehat{\delta}=\inf\{\delta\in[0,1]:c_n(\delta)\leq\tau_\star\}$. By
Assumption~\ref{assump: breakdown regularity-i}, $B$ is continuous and nonincreasing,
$B(0)>0$, and $\delta_{\mathsf{b}}<1$. Moreover,
$\overline{\tau}(\delta)\geq\overline{\tau}(0)\geq\underline{\tau}(0)>\tau_\star$ for
every $\delta\in[0,1]$ by Proposition~\ref{thm: shadow price budget-i}, so that the upper
inequality in \eqref{eq: breakdown point} holds throughout,
$\delta_{\mathsf{b}}=\inf\{\delta\in[0,1]:B(\delta)\leq0\}$, and $B>0$ on
$[0,\delta_{\mathsf{b}})$. Hence the closed
nonempty set $\{\delta\in[0,1]:B(\delta)\leq0\}$ contains its infimum
$\delta_{\mathsf{b}}$, whence $B(\delta_{\mathsf{b}})=0$ by continuity from the left, and
differentiability at $\delta_{\mathsf{b}}$ gives
\begin{align}
B(\delta)=-\underline{\eta}(\delta_{\mathsf{b}})\,(\delta-\delta_{\mathsf{b}})
+R(\delta),
\qquad R(\delta)=o(|\delta-\delta_{\mathsf{b}}|).
\label{eq: breakdown expansion}
\end{align}
Since $\underline{\eta}(\delta_{\mathsf{b}})>0$, there exists $\varepsilon_0>0$ with
$[\delta_{\mathsf{b}}-\varepsilon_0,\delta_{\mathsf{b}}+\varepsilon_0]\subseteq(0,1)$ and
$B(\delta)\leq-\{\underline{\eta}(\delta_{\mathsf{b}})/2\}(\delta-\delta_{\mathsf{b}})<0$
on $(\delta_{\mathsf{b}},\delta_{\mathsf{b}}+\varepsilon_0]$.

\uline{\textbf{Step 1: consistency.}} By
Assumptions~\ref{assump: breakdown regularity-ii}--\ref{assump: breakdown regularity-iii}
and the continuous mapping theorem, $\|G_n\|_{\infty}=\Op(1)$ and
$\|\widehat{\sigma}_n\|_{\infty}=\Op(1)$, so that
$\|c_n-\underline{\tau}\|_{\infty}=\Op(n^{-1/2})$. Fix $\varepsilon\in(0,\varepsilon_0]$
and set $\gamma_{\varepsilon}:=\min[B(\delta_{\mathsf{b}}-\varepsilon),
\{\underline{\eta}(\delta_{\mathsf{b}})/2\}\varepsilon]>0$. On the event
$\{\|c_n-\underline{\tau}\|_{\infty}<\gamma_{\varepsilon}\}$, whose probability tends to
one, $c_n-\tau_\star>B-\gamma_{\varepsilon}\geq
B(\delta_{\mathsf{b}}-\varepsilon)-\gamma_{\varepsilon}\geq0$ on
$[0,\delta_{\mathsf{b}}-\varepsilon]$, since $B$ is nonincreasing, whereas
$c_n(\delta_{\mathsf{b}}+\varepsilon)-\tau_\star<B(\delta_{\mathsf{b}}+\varepsilon)
+\gamma_{\varepsilon}\leq0$. Hence
$|\widehat{\delta}-\delta_{\mathsf{b}}|\leq\varepsilon$ on that event, which proves
part~\ref{prop: certified breakdown-i}.

\uline{\textbf{Step 2: the crossing identity.}} On the event
$\{|\widehat{\delta}-\delta_{\mathsf{b}}|\leq\varepsilon_0/2\}$, whose probability tends
to one by Step~1, $0<\widehat{\delta}<1$. The set
$\{\delta\in[0,1]:c_n(\delta)\leq\tau_\star\}$ is nonempty and closed, because $c_n$ has
continuous paths by
Assumptions~\ref{assump: breakdown regularity-ii}--\ref{assump: breakdown regularity-iii},
so it contains $\widehat{\delta}$, whereas $c_n>\tau_\star$ on $[0,\widehat{\delta})$ and
continuity give $c_n(\widehat{\delta})\geq\tau_\star$. Hence
$c_n(\widehat{\delta})=\tau_\star$ with probability approaching one, that is,
\begin{align}
B(\widehat{\delta})=n^{-1/2}A_n,
\qquad
A_n:=\mathfrak{c}\,\widehat{\sigma}_n(\widehat{\delta})-G_n(\widehat{\delta}).
\label{eq: crossing identity}
\end{align}
Moreover, $A_n\rightsquigarrow\mathfrak{c}\,\sigma(\delta_{\mathsf{b}})
-\mathbb{G}(\delta_{\mathsf{b}})$. Indeed,
$\widehat{\sigma}_n(\widehat{\delta})\cprob\sigma(\delta_{\mathsf{b}})$ by
Assumption~\ref{assump: breakdown regularity-iii}, the continuity of $\sigma$, and
Step~1, and $G_n(\widehat{\delta})-G_n(\delta_{\mathsf{b}})=\op(1)$ because
$(G_n,\widehat{\delta})\rightsquigarrow(\mathbb{G},\delta_{\mathsf{b}})$ by Slutsky's
lemma, $\mathbb{G}$ has continuous paths, and the extended continuous mapping theorem
\citep[Theorem~1.11.1]{vandervaart1996WeakConvergenceEmpirical} applies
to
$(x,t)\mapsto x(t)-x(\delta_{\mathsf{b}})$, while
$G_n(\delta_{\mathsf{b}})\rightsquigarrow\mathbb{G}(\delta_{\mathsf{b}})
\sim\normal(0,\sigma^2(\delta_{\mathsf{b}}))$.

\uline{\textbf{Step 3: asymptotic distribution and coverage.}} Fix
$\varepsilon\in(0,\underline{\eta}(\delta_{\mathsf{b}}))$ and $\nu\in(0,\varepsilon_0]$
with $|R(\delta)|\leq\varepsilon|\delta-\delta_{\mathsf{b}}|$ whenever
$|\delta-\delta_{\mathsf{b}}|\leq\nu$. On the intersection of
$\{|\widehat{\delta}-\delta_{\mathsf{b}}|\leq\nu\}$ with the event of Step~2, whose
probability tends to one, \eqref{eq: breakdown expansion} and
\eqref{eq: crossing identity} give
$\{\underline{\eta}(\delta_{\mathsf{b}})-\varepsilon\}
|\widehat{\delta}-\delta_{\mathsf{b}}|\leq n^{-1/2}|A_n|$, so that
$\sqrt{n}\,(\widehat{\delta}-\delta_{\mathsf{b}})=\Op(1)$ by Step~2, and hence
$\sqrt{n}R(\widehat{\delta})=\op(1)$. Substituting \eqref{eq: breakdown expansion} into
\eqref{eq: crossing identity},
\[
\sqrt{n}\,(\widehat{\delta}-\delta_{\mathsf{b}})
=\frac{\sqrt{n}\,R(\widehat{\delta})-A_n}{\underline{\eta}(\delta_{\mathsf{b}})}
\;\rightsquigarrow\;
\frac{\mathbb{G}(\delta_{\mathsf{b}})-\mathfrak{c}\,\sigma(\delta_{\mathsf{b}})}
{\underline{\eta}(\delta_{\mathsf{b}})}
\;\sim\;\normal\bigl(-\mathfrak{c}\,\varsigma_{\mathsf{b}},\
\varsigma_{\mathsf{b}}^{2}\bigr),
\]
by Step~2, Slutsky's lemma, and \eqref{eq: robustness se}, which is
part~\ref{prop: certified breakdown-ii}. Since the limit is continuous,
\[
\PP\bigl(\widehat{\delta}\leq\delta_{\mathsf{b}}\bigr)
=\PP\bigl(\sqrt{n}(\widehat{\delta}-\delta_{\mathsf{b}})\leq0\bigr)
\;\to\;\PP\bigl\{\normal(-\mathfrak{c}\varsigma_{\mathsf{b}},
\varsigma_{\mathsf{b}}^{2})\leq0\bigr\}
=\PP\bigl\{\normal(0,1)\leq \mathfrak{c}\bigr\}=1-\alpha/2,
\]
which is part~\ref{prop: certified breakdown-iii}.
\end{proof}

\subsection{Proof of Proposition~\ref{prop: exchange rate}}
\begin{proof}\label{proof: exchange rate}
Write $B:=\underline{\tau}-\tau_\star$ and $\mathfrak{c}:=\mathfrak{c}_{1-\alpha/2}$. As
shown in the proof of Proposition~\ref{prop: certified breakdown}, $B$ is continuous,
$B>0$ on $[0,\delta_{\mathsf{b}})$, $B(\delta_{\mathsf{b}})=0$, and $B$ is differentiable
at $\delta_{\mathsf{b}}$ with slope
$-\underline{\eta}(\delta_{\mathsf{b}})\in(-\infty,0)$, so that
\begin{align}
B(\delta)=\underline{\eta}(\delta_{\mathsf{b}})\,(\delta_{\mathsf{b}}-\delta)\,\{1+o(1)\}
\qquad\text{as }\delta\uparrow\delta_{\mathsf{b}}.
\label{eq: frontier expansion}
\end{align}
At every $\delta\in(0,\delta_{\mathsf{b}})$ at which $\underline{\tau}$ is
differentiable, $B'(\delta)=-\underline{\eta}(\delta)$ and $\sigma$ is differentiable, so
$\log n^{\star}(\delta;\alpha)=2\log\mathfrak{c}+2\log\sigma(\delta)-2\log B(\delta)$ is
differentiable with derivative
$2\{\sigma'(\delta)/\sigma(\delta)+\underline{\eta}(\delta)/B(\delta)\}$, which is
\eqref{eq: exchange rate formula}. For the limit, the derivative of the finite convex
function $\underline{\tau}$ is nondecreasing where it exists and converges from the left
to
$\partial_{-}\underline{\tau}(\delta_{\mathsf{b}})=\underline{\tau}'(\delta_{\mathsf{b}})$
\citep[Section~24]{rockafellar1970ConvexAnalysis}, so that
$\underline{\eta}(\delta)\to\underline{\eta}(\delta_{\mathsf{b}})$ as
$\delta\uparrow\delta_{\mathsf{b}}$ and, by \eqref{eq: frontier expansion},
$\underline{\eta}(\delta)/B(\delta)=\{1+o(1)\}/(\delta_{\mathsf{b}}-\delta)$. Since
$\sigma$ is continuously differentiable and strictly positive on $[0,1]$,
$\sigma'/\sigma$ is bounded there, so the second term in
\eqref{eq: exchange rate formula} is of relative order $O(\delta_{\mathsf{b}}-\delta)$,
and the claim follows.
\end{proof}

\subsection{Proof of Theorem~\ref{theorem: tv dual recovers lee bound}}
\begin{proof}\label{proof:theorem: tv dual recovers lee bound}
We derive in full the upper bound $\overline{\mu}^{\mathsf{C}}(1;\delta)$ on the complier
mean, which delivers the lower bound $\underline{\tau}(\delta)$ on $\tau_0$, and
indicate at the end the symmetric argument for the remaining bound.
Write $\phi_\TV(u)=\tfrac12|u-1|$ for the total variation entropy, so that
$D_{\phi_\TV}(Q\,\|\,\PT)=\TV(Q\,\|\,\PT)$. Recall from
Section~\ref{subsec: rct attrition} that $\calP_{\mathsf{RCT}}(\delta)$ is the set
$\calP_{\TV}(\delta)$ of \eqref{eq: tv set summary} with $(\kappa,\Pbase{})=(\pi,\PT)$,
which Proposition~\ref{prop: induced entropy-i} identifies with the neighborhood generated
by the induced entropy $\overline{\phi}_{\TV}=\phi_\TV+\iota_{[0,1/\pi]}$ at budget
$(1-\pi)\delta$. Since
$\pi\PT\leq \PT$ and $\TV(\PT\,\|\,\PT)=0$, we have
$\PT\in\calP_{\mathsf{RCT}}(\delta)$, so the set is nonempty.

\uline{\textbf{Dual reduction.}} We apply
Theorem~\ref{theorem: weak duality} to the induced entropy $\overline{\phi}_{\TV}$,
which is an entropy function by Proposition~\ref{prop: induced entropy-i}, with
$g(y)=y$, no moment function, and divergence budget $(1-\pi)\delta$. Its integrability
hypothesis holds. Indeed, if $\nu\in\calM_+(\calY)$ satisfies
$D_{\overline{\phi}_{\TV}}(\nu\,\|\,\PT)<\infty$, then
Lemma~\ref{lemma: divergence lebesgue decomposition}, together with
$\overline{\phi}_{\TV}^\infty(1)=+\infty$ and $\overline{\phi}_{\TV}=+\infty$ off
$[0,1/\pi]$, gives $\nu\ll\PT$ with $\de\nu/\de\PT\leq1/\pi$ $\PT$-almost surely, so
that $\int|y|\,\de\nu(y)\leq\pi^{-1}\E_{\PT}[|Y|]<\infty$. The pointwise constraint of
the theorem is vacuous because $\overline{\phi}_{\TV}^\infty(1)=+\infty$, and the
moment multiplier is absent, so that weak duality reads
\begin{equation}\label{eq: lee dual reduced}
\overline{\mu}^{\mathsf{C}}(1;\delta)
\leq
\inf_{(\zeta,\eta)\in\R\times\R_+}
\Big\{
\zeta+\eta(1-\pi)\delta
+\E_{\PT}\big[(\overline{\phi}_{\TV}^*)^\pi(Y-\zeta,\eta)\big]
\Big\}.
\end{equation}
The conjugate $\overline{\phi}_{\TV}^*$ and its perspective
$(\overline{\phi}_{\TV}^*)^\pi$ are recorded in closed form in
Example~\ref{ex: tv truncated duals}. In particular, for $\eta\geq 0$ the pointwise
supremum defining
$(\overline{\phi}_{\TV}^*)^\pi(s,\eta)=\sup_{0\leq u\leq 1/\pi}\{su-\tfrac{\eta}{2}|u-1|\}$
is attained at
\[
u^\star(s,\eta)
=
\begin{cases}
0,& s\leq-\eta/2,\\
1,& -\eta/2\leq s\leq\eta/2,\\
1/\pi,& s\geq\eta/2.
\end{cases}
\]

Since $\delta\leq 1$ we have
$(1-\pi)\delta\leq 1-\pi\delta$, hence $t_\mathsf{L}\leq t_\mathsf{U}$. Set
\[
\zeta^\dag:=\tfrac12(t_\mathsf{L}+t_\mathsf{U}),
\qquad
\eta^\dag:=t_\mathsf{U}-t_\mathsf{L}\geq 0,
\]
so that $Y-\zeta^\dag\leq-\eta^\dag/2\Leftrightarrow Y\leq t_\mathsf{L}$ and
$Y-\zeta^\dag\geq\eta^\dag/2\Leftrightarrow Y\geq t_\mathsf{U}$. The corresponding
maximizer $u^\star(Y-\zeta^\dag,\eta^\dag)$ is
\[
w^\dag(y):=
\begin{cases}
0,& y\leq t_\mathsf{L},\\
1,& t_\mathsf{L}<y<t_\mathsf{U},\\
1/\pi,& y\geq t_\mathsf{U}.
\end{cases}
\]
Because $\PT$ has no atoms at $t_\mathsf{L}$ or $t_\mathsf{U}$,
\[
\PT(Y\leq t_\mathsf{L})=(1-\pi)\delta,
\qquad
\PT(Y\geq t_\mathsf{U})=\pi\delta,
\]
hence
\[
\E_{\PT}[w^\dag(Y)]
=\big(1-(1-\pi)\delta-\pi\delta\big)+\tfrac{1}{\pi}\,\pi\delta=1,
\quad
\tfrac12\E_{\PT}\big[|w^\dag(Y)-1|\big]
=\tfrac12(1-\pi)\delta+\tfrac12\Big(\tfrac{1}{\pi}-1\Big)\pi\delta
=(1-\pi)\delta.
\]
Thus the measure $P^\dag$ with $\de P^\dag/\de\PT=w^\dag$ is a probability
measure with $\TV(P^\dag\,\|\,\PT)=(1-\pi)\delta$. Moreover, by
Lemma~\ref{lemma: mixture density bounds-i}
$0\leq w^\dag\leq 1/\pi$ gives $\pi P^\dag\leq\PT$. Hence,
$P^\dag\in\calP_{\mathsf{RCT}}(\delta)$ and, by primal feasibility,
\begin{equation}\label{eq: lee primal lower}
\overline{\mu}^{\mathsf{C}}(1;\delta)\geq\E_{P^\dag}[Y]=\E_{\PT}[Y\,w^\dag].
\end{equation}
By the rescaling identity of
Lemma~\ref{lemma: contamination as constrained tv-i}, the mixing distribution
$R^\dag=(\PT-\pi P^\dag)/(1-\pi)$ satisfies $\TV(P^\dag\,\|\,R^\dag)=\delta$: the
attaining decomposition places the complier and always observed outcome
distributions at total variation distance exactly $\delta$, saturating the
sensitivity restriction.

Since $w^\dag$ is a pointwise
maximizer, Example~\ref{ex: tv truncated duals} gives, $\PT$-almost surely,
\[
(\overline{\phi}_{\TV}^*)^\pi(Y-\zeta^\dag,\eta^\dag)
=(Y-\zeta^\dag)w^\dag-\tfrac{\eta^\dag}{2}|w^\dag-1|.
\]
Evaluating the objective in~\eqref{eq: lee dual reduced} at
$(\zeta^\dag,\eta^\dag)$ and using $\E_{\PT}[w^\dag]=1$ together with
$\tfrac12\E_{\PT}[|w^\dag-1|]=(1-\pi)\delta$,
\begin{align*}
&\zeta^\dag+\eta^\dag(1-\pi)\delta+\E_{\PT}\big[(\overline{\phi}_{\TV}^*)^\pi
  (Y-\zeta^\dag,\eta^\dag)\big]\\
&\qquad=\E_{\PT}[Y\,w^\dag] +\zeta^\dag\big(1-\E_{\PT}[w^\dag]\big) +\eta^\dag\Big((1
-\pi)\delta-\tfrac12\E_{\PT}[|w^\dag-1|]\Big)\\
&\qquad=\E_{\PT}[Y\,w^\dag].
\end{align*}
Therefore the infimum in~\eqref{eq: lee dual reduced} is at most
$\E_{\PT}[Y\,w^\dag]$. Combining this with~\eqref{eq: lee primal lower} and
the weak duality inequality~\eqref{eq: lee dual reduced} gives
\[
\overline{\mu}^{\mathsf{C}}(1;\delta)
=\E_{\PT}[Y\,w^\dag]
=\E_{\PT}\big[Y\indic\{t_\mathsf{L}<Y<t_\mathsf{U}\}\big]
+\frac{1}{\pi}\E_{\PT}\big[Y\indic\{Y\geq t_\mathsf{U}\}\big],
\]
where the last equality is the definition of $w^\dag$. This is the sharp upper bound
on the complier mean.

\uline{\textbf{From the complier mean to $\tau_0$.}} By the mixture
\eqref{eq: mixture AT and C}, the definition of the always observed bounds in
\eqref{eq: at bounds def}, and $\mu_{\mathsf{T}}=\E_{\PT}[Y]$,
\[
\underline{\mu}^{\mathsf{AO}}(1;\delta)
=\frac{\E_{\PT}[Y]-\pi\,\overline{\mu}^{\mathsf{C}}(1;\delta)}{1-\pi}.
\]
Splitting $\E_{\PT}[Y]=\E_{\PT}[Y\indic\{Y\leq t_\mathsf{L}\}]
+\E_{\PT}[Y\indic\{t_\mathsf{L}<Y<t_\mathsf{U}\}]
+\E_{\PT}[Y\indic\{Y\geq t_\mathsf{U}\}]$ and inserting the closed form for
$\overline{\mu}^{\mathsf{C}}(1;\delta)$ just obtained, the upper-tail terms cancel and
\[
\E_{\PT}[Y]-\pi\,\overline{\mu}^{\mathsf{C}}(1;\delta)
=\E_{\PT}\big[Y\indic\{Y\leq t_\mathsf{L}\}\big]
+(1-\pi)\,\E_{\PT}\big[Y\indic\{t_\mathsf{L}<Y<t_\mathsf{U}\}\big].
\]
Dividing by $1-\pi$ gives the closed form for $\underline{\mu}^{\mathsf{AO}}(1;\delta)$.
Under \ref{assump: Mono} a control unit is observed only if $S(0)=1$, in which case
$S(1)\geq S(0)$ forces $S(1)=1$, so that the observed control units are exactly the
always observed ones and $\mu^{\mathsf{AO}}(0)=\E_{P_0}[Y\mid D=0,S=1]$
does not depend on the distribution of the complier outcome. Subtracting it as
in \eqref{eq: tau bounds def} yields the stated closed form for
$\underline{\tau}(\delta)$. The closed form for $\overline{\tau}(\delta)$ follows
from the symmetric argument, upon replacing $w^\dag$ with the reweighting equal to $1/\pi$
on $\{Y\leq s_\mathsf{L}\}$, to $1$ on $(s_\mathsf{L},s_\mathsf{U})$, and to $0$ on
$\{Y\geq s_\mathsf{U}\}$, which is feasible and attains
$\underline{\mu}^{\mathsf{C}}(1;\delta)$ by the same computation.

\uline{\textbf{The bounds are attained and the value set is an interval.}} Both bounds are
attained, the reweightings just displayed exhibiting distributions in
$\calP_{\mathsf{RCT}}(\delta)$ at which the complier mean equals
$\overline{\mu}^{\mathsf{C}}(1;\delta)$ and $\underline{\mu}^{\mathsf{C}}(1;\delta)$. It
remains to show that every intermediate value is attained as well. The set
$\calP_{\mathsf{RCT}}(\delta)$ is convex, as the intersection of the $\TV$ ball of
radius $(1-\pi)\delta$ around $\PT$ with the set $\{Q\in\Delta(\calY):\pi Q\leq\PT\}$,
both of which are convex. Moreover, $Q\mapsto\E_Q[Y]$ is affine, hence so is the map
\eqref{eq: tau map attrition},
\[
Q\longmapsto\tau(Q)=
\frac{\mu_{\mathsf{T}}-\pi\,\E_Q[Y]}{1-\pi}-\mu^{\mathsf{AO}}(0),
\]
which sends a distribution of the complier outcome to the implied value of
$\tau_0$. The image of a convex set under an affine map into $\R$ is an interval,
and its endpoints are attained, so that image is exactly
$[\underline{\tau}(\delta),\overline{\tau}(\delta)]$.

Finally, every $Q\in\calP_{\mathsf{RCT}}(\delta)$ is consistent with the observed
distribution and the sensitivity restriction. Indeed, by
Lemma~\ref{lemma: contamination as constrained tv-i}, every such $Q$ admits the
unique mixing distribution $R=(\PT-\pi Q)/(1-\pi)$, and $\PT=\pi Q+(1-\pi)R$ reproduces
the mixture \eqref{eq: mixture AT and C} with $Q$ in the role of the complier outcome
distribution and $R$ in that of the always observed one. By the rescaling identity
of the same lemma, $\TV(Q\,\|\,\PT)=(1-\pi)\TV(Q\,\|\,R)$, so the budget
$\TV(Q\,\|\,\PT)\leq(1-\pi)\delta$ is equivalent to $\TV(Q\,\|\,R)\leq\delta$, which is
the maintained restriction.
Moreover, the pair $(Q,R)$ is rationalized by an actual data-generating
process: taking $Q$ as the complier and $R$ as the always observed law of
$Y(1)$, the identified law of $Y(0)$ on the always observed, namely that of
$Y\mid D=0,S=1$, and any coupling of the remaining coordinates, one obtains a
joint law of $(Y(0),Y(1),S(0),S(1),D)$ that satisfies random assignment and
\ref{assump: Mono} and reproduces the observed law of $(YS,S,D)$. No feasible
$Q$ is therefore excluded by the data. Together with the
previous paragraph, the set of values of $\tau_0$ consistent with $\PT$,
$\pi$, $\mu^{\mathsf{AO}}(0)$, and the restriction is the closed interval
$[\underline{\tau}(\delta),\overline{\tau}(\delta)]$.
\end{proof}

\subsection{Proof of Proposition~\ref{prop: attrition inference}}
\begin{proof}\label{proof:prop: attrition inference}
We work directly with the closed form bound of
Theorem~\ref{theorem: tv dual recovers lee bound}, and we view the plug-in estimator
as a functional of the
empirical measure $\widehat P_n$ of $W_1,\ldots,W_n$, where $W=(YS,S,D)$, acting on the
class of functions of $W$
\[
\calH_{\mathsf{RCT}}:=\{f_t,\ g_t:\ t\in\R\}\cup\{D,\ 1-D,\ DS,\ (1-D)S,\ (1-D)YS\},
\]
where $f_t:=DS\,\indic\{YS\leq t\}$ and $g_t:=DS\,YS\,\indic\{YS>t\}$.
Each of the families $\{f_t\}$ and $\{g_t\}$ multiplies a fixed function by the
indicators of the half lines $\{YS\leq t\}$ or $\{YS>t\}$, so that it is VC-subgraph
\citep[Lemma~2.6.18(vi)]{vandervaart1996WeakConvergenceEmpirical}. 
Hence $\calH_{\mathsf{RCT}}$ has a finite uniform
entropy integral
\citep[Theorem~2.6.7]{vandervaart1996WeakConvergenceEmpirical}, and its
envelope $1+|YS|$ is square integrable by
Assumption~\ref{assump: attrition inference}. The class is also measurable, 
thus it is $P_0$-Donsker
\citep[Theorem~19.14]{vandervaart1998AsymptoticStatistics}, that is,
$\sqrt n(\widehat P_n-P_0)\rightsquigarrow\mathbb{G}$ in
$\ell^\infty(\calH_{\mathsf{RCT}})$, where $\mathbb G$ is a tight Gaussian process whose
sample paths lie in the closed subspace $\mathbb{D}_0$ of the
$h\in\ell^\infty(\calH_{\mathsf{RCT}})$ that are uniformly continuous with respect to the
$L_2(P_0)$ semimetric.

For a probability measure $P$ on the range of $W$ with $P(D)\in(0,1)$, $P(DS)>0$, and
$P((1-D)S)>0$, let $F_P(t):=P(f_t)/P(DS)$ and $M_P(t):=P(g_t)/P(DS)$ be the distribution
function and the upper-tail moment of the treated observed outcome, so that
$F_{P_0}=F_{\PT}$, let
$\pi_P:=1-\{P((1-D)S)/P(1-D)\}/\{P(DS)/P(D)\}$ be the complier share, let
$c_P:=F_P^{-1}(\pi_P\delta)$ and $d_P:=F_P^{-1}(1-(1-\pi_P)\delta)$ be the trimming
points, and let
\[
\Phi(P):=M_P(c_P)+\frac{\pi_P}{1-\pi_P}\,M_P(d_P)-\frac{P((1-D)YS)}{P((1-D)S)}.
\]
Write $c:=s_\mathsf{L}(\delta)$, $d:=s_\mathsf{U}(\delta)$, and $M:=M_{P_0}$. Since
$\vartheta_+(y;\delta)=y\,\indic\{y>c\}+\tfrac{\pi}{1-\pi}\,y\,\indic\{y\geq d\}$ and
$\PT$ has no atoms, $\Phi(P_0)=\overline\tau(\delta)$. Moreover, $\Phi(\widehat P_n)$
differs from $\widehat{\overline\tau}(\delta)$ only through the atom of the empirical
distribution of the treated observed outcomes at its own quantile
$\widehat s_\mathsf{U}(\delta)$, a difference of order $O_\PP(n^{-1})$ because these
outcomes have no ties almost surely.

The map $\Phi$ is Hadamard differentiable at $P_0$ tangentially to $\mathbb{D}_0$. Indeed,
it is a composition of the following maps, each Hadamard differentiable at the relevant
point, so that the chain rule
\citep[Theorem~20.9]{vandervaart1998AsymptoticStatistics}
applies.
\begin{itemize}
    \item The evaluations $P\mapsto P(h)$ at the five functions of the finite part of
    $\calH_{\mathsf{RCT}}$, and smooth functions of them on the stated range, which
    deliver $\pi_P$, $P(DS)$, equal to $p\,r_1$ at $P_0$, and the control mean.
    \item The maps $P\mapsto F_P$ and $P\mapsto M_P$ into $\ell^\infty(\R)$, whose
    derivatives at $P_0$ in a direction $h$ are
    $h_F:=\{h(f_{\,\cdot\,})-F_{\PT}\,h(DS)\}/(p\,r_1)$ and
    $h_M:=\{h(g_{\,\cdot\,})-M\,h(DS)\}/(p\,r_1)$. For $h\in\mathbb{D}_0$, both are
    continuous, because for $t<t'$ the squared $L_2(P_0)$ distances between $f_t$ and
    $f_{t'}$ and between $g_t$ and $g_{t'}$ equal $p\,r_1\,\PT((t,t'])$ and
    $p\,r_1\,\E_{\PT}[Y^2\indic\{t<Y\leq t'\}]$, respectively, which vanish as
    $t'-t\to0$ because $\PT$ has no atoms.
    \item The quantile maps $(F,u)\mapsto F^{-1}(u)$ at $(F_{\PT},\pi\delta)$ and at
    $(F_{\PT},1-(1-\pi)\delta)$. Writing $F^{-1}(u)$ as the quantile of $F+u_0-u$ at the
    fixed level $u_0$, Lemma~21.3 in \citet{vandervaart1998AsymptoticStatistics} and the
    chain rule show that each map is Hadamard differentiable tangentially to the
    directions $(h_F,\dot u)$ with $h_F$ continuous at $t_0:=F_{\PT}^{-1}(u_0)$, with
    derivative $(\dot u-h_F(t_0))/f_{\PT}(t_0)$, as $f_{\PT}$ is continuous and positive
    at $t_0\in\{c,d\}$.
    \item The evaluation maps $(M',t)\mapsto M'(t)$ at $(M,c)$ and at $(M,d)$. For
    $t_0\in\{c,d\}$, $h_s\to h_M$ uniformly with $h_M$ continuous at $t_0$, and
    $\dot t_s\to\dot t$,
    \[
    \frac{(M+s\,h_s)(t_0+s\,\dot t_s)-M(t_0)}{s}
    =\frac{M(t_0+s\,\dot t_s)-M(t_0)}{s}+h_s(t_0+s\,\dot t_s)
    \;\to\;-t_0f_{\PT}(t_0)\,\dot t+h_M(t_0)
    \]
    as $s\to0$, where the limit of the first term follows from
    $M(t)=\E_{\PT}[Y\indic\{Y>t\}]$ and the continuity of $f_{\PT}$ at $t_0$.
\end{itemize}

\uline{\textbf{(i) Asymptotic normality.}}
The remaining dependence of the bound on $\pi$ runs through the budget
$(1-\pi)\delta$ (hence the cut-point levels), the cap $1/\pi$, and the
reconstruction weight $1/(1-\pi)$, all smooth on $\pi\in(0,1)$. It
remains to compute the derivative $\Phi'_{P_0}$. Let $h\in\mathbb{D}_0$ induce the
directions $(h_F,h_M,\dot\pi)$, and let $\dot c,\dot d$ be the
induced threshold movements. Differentiating $F_{\PT}(c)=\pi\delta$ and
$F_{\PT}(d)=1-(1-\pi)\delta$ gives
\[
  \dot c=\frac{\delta\,\dot\pi-h_F(c)}{f_{\PT}(c)},\qquad
  \dot d=\frac{\delta\,\dot\pi-h_F(d)}{f_{\PT}(d)}.
\]
Differentiating
$\overline\mu(\delta)=M(c)+\tfrac{\pi}{1-\pi}\,M(d)$ with the last two maps above and
substituting, the density factors $f_{\PT}(c),f_{\PT}(d)$
cancel and the derivative is linear in $(h_F,h_M,\dot\pi)$: the part linear
in $(h_F,h_M)$ is
$h_M(c)+\tfrac{\pi}{1-\pi}\,h_M(d)+c\,h_F(c)+\tfrac{\pi}{1-\pi}\,d\,h_F(d)$,
while the coefficient on $\dot\pi$ is the envelope derivative $\varkappa(\delta)$ of the
statement. As $\Phi'_{P_0}$ depends on $h$ only through finitely many
evaluations, it is linear and continuous on all of $\ell^\infty(\calH_{\mathsf{RCT}})$,
and the functional delta method
\citep[Theorem~20.8]{vandervaart1998AsymptoticStatistics} gives
\[
\sqrt n\bigl(\widehat{\overline\tau}(\delta)-\overline\tau(\delta)\bigr)
=\Phi'_{P_0}\bigl(\sqrt n(\widehat P_n-P_0)\bigr)+\op(1)
=\frac{1}{\sqrt n}\sum_{i=1}^n\Phi'_{P_0}(h_{W_i})+\op(1),
\]
where $h_W$ denotes the direction $f\mapsto f(W)-P_0(f)$. At $h=h_W$, the directions are
$h_F=\tfrac{DS}{p\,r_1}(\indic\{Y\leq\,\cdot\,\}-F_{\PT})$ and
$h_M=\tfrac{DS}{p\,r_1}(Y\indic\{Y>\,\cdot\,\}-M)$, the direction of $\pi_P$ is
$\dot\pi=\psi_{\pi}(W)$ by the delta method on the two selection frequencies, and the
control mean contributes $-\psi_0(W)$, so that $\Phi'_{P_0}(h_W)$ equals the influence
function
$\psi_{\mathsf{full}}(\,\cdot\,;\delta)=\psi_{\mathsf{T}}(\,\cdot\,;\delta)+\varkappa(\delta)\,\psi_{\pi}-\psi_0$,
whose first and
third terms reproduce the fixed-$\pi$ influence
$\psi_2(\,\cdot\,;\delta)=\psi_{\mathsf{T}}(\,\cdot\,;\delta)-\psi_0$ of the
reduction. Each component is mean zero and square
integrable by the second-moment condition, so
$\sqrt n(\widehat{\overline\tau}(\delta)-\overline\tau(\delta))
=n^{-1/2}\sum_i\psi_{\mathsf{full}}(W_i;\delta)+\op(1)\rightsquigarrow\normal(0,\sigma_{\mathsf{full}}^2(\delta))$.

\uline{\textbf{(ii) Consistency of the variance estimators.}}
On $\{D=1,S=1\}$, the support of $\psi_{\mathsf{T}}$, and on $\{D=0,S=1\}$, the support of
$\psi_0$, the selection-frequency influence $\psi_{\pi}$ is constant.
Since $\psi_{\mathsf{T}}$ and $\psi_0$ are mean zero this gives
$\E[\psi_{\mathsf{T}}\psi_{\pi}]=\E[\psi_0\psi_{\pi}]=0$, hence
$\E[\psi_2\psi_{\pi}]=0$ and
$\sigma_{\mathsf{full}}^2(\delta)=\E[(\psi_2(W;\delta)+\varkappa(\delta)\psi_{\pi})^2]=\sigma_2^2(\delta)+\varkappa(\delta)^2
V_{\pi}$.
The estimated influence functions $\widehat\psi_{\mathsf{full}}$
and $\widehat\psi_2$ are obtained from $\psi_{\mathsf{full}}$ and $\psi_2$ by
replacing the vector of population inputs
$\vartheta_0:=(s_\mathsf{L}(\delta),s_\mathsf{U}(\delta),\pi,r_0,r_1,p,
\mu^{\mathsf{AO}}(0),\overline\mu(\delta),\varkappa(\delta))$ with its plug-in
counterpart $\widehat\vartheta_n$, which is consistent by the law of large numbers
together with the continuous mapping theorem. Write $\psi_{\bullet}(\cdot\,;\vartheta)$,
$\bullet\in\{\mathsf{full},2\}$, for the influence function evaluated at a
generic input vector $\vartheta$, and consider the classes
$\{w\mapsto\psi_{\bullet}(w;\vartheta)^2:\vartheta\in K\}$, where $K$ is a
compact neighborhood of $\vartheta_0$. Each member is a fixed linear
combination, with bounded coefficients, of products of indicators of
half lines in $y$, the functions $y$ and $y\,\indic\{y\geq t\}$, and
indicators of the events $\{D=1,S=1\}$ and $\{D=0,S=1\}$, so the
classes are Glivenko--Cantelli with integrable
envelope proportional to$(1+Y^2)$. The uniform law
of large numbers over $K$, combined with the continuity of
$\vartheta\mapsto\E[\psi_{\bullet}(W;\vartheta)^2]$ at $\vartheta_0$ yields
$\widehat V_{\mathsf{full}}(\delta)\cprob\sigma_{\mathsf{full}}^2(\delta)$ and
$\widehat V_2(\delta)\cprob\sigma_2^2(\delta)$, that is
$\widehat\sigma_{\mathsf{NO}}^2(\delta)\cprob\sigma_{\mathsf{full}}^2(\delta)$
and $\widehat\sigma_{\mathsf{naive}}^2(\delta)\cprob\sigma_2^2(\delta)$.

Turning to the bootstrap, the class $\calH_{\mathsf{RCT}}$ is $P_0$-Donsker
with a square-integrable envelope, so that the bootstrap empirical process converges
conditionally to $\mathbb{G}$ in $\ell^\infty(\calH_{\mathsf{RCT}})$
\citep[Theorem~23.7]{vandervaart1998AsymptoticStatistics}. As $\Phi$ is
Hadamard
differentiable at $P_0$ tangentially to $\mathbb{D}_0$, in which $\mathbb{G}$ takes its
values, and the Hadamard derivative is linear, the delta
method for the bootstrap
\citep[Theorem~23.9]{vandervaart1998AsymptoticStatistics} gives
conditional
weak convergence of
$\sqrt n\,(\widehat{\overline\tau}{}^{*}(\delta)-\widehat{\overline\tau}(\delta))$
to a $\normal(0,\sigma_{\mathsf{full}}^2(\delta))$, 
where $\widehat{\overline\tau}{}^{*}(\delta)$ denotes the plug-in estimator
computed on a bootstrap resample. Consistency of
$\widehat\sigma_{\mathsf{boot}}^2(\delta)$ as $B\to\infty$ follows because
the uniform integrability condition (3.21) in Section~3.1.6 of
\citet{shao1995JackknifeBootstrap} is satisfied.
Therefore,
$\widehat\sigma_{\mathsf{boot}}^2(\delta)\cprob\sigma_{\mathsf{full}}^2(\delta)$.

We now establish the identity
$\sigma_{\mathsf{full}}^2(1)=V_{\mathsf{Lee},\mathsf{U}}+V_{\mathsf{Lee},\mathsf{C}}$.
At $\delta=1$, the two trimming quantiles are identical, that is
$s_\mathsf{L}(1)=s_\mathsf{U}(1)=F_{\PT}^{-1}(\pi)=:y_{\pi}$, so that
$\vartheta_+(y;1)=\tfrac{1}{1-\pi}\,y\,\indic\{y\geq y_{\pi}\}$, the bound becomes
$\overline{\mu}(1)=\E_{\PT}[Y\mid Y\geq y_{\pi}]$, and
$\varkappa(1)=(\overline{\mu}(1)-y_{\pi})/(1-\pi)$ is the derivative of the
population trimmed mean with respect to the trimming proportion. Define
$\psi_{\mathsf{U}}:=\psi_{\mathsf{T}}(\,\cdot\,;1)+\varkappa(1)\,\psi_{\pi}$,
so that $\psi_{\mathsf{full}}(\,\cdot\,;1)=\psi_{\mathsf{U}}-\psi_0$. By the
derivation of part~(i) at $\delta=1$, which applies as $f_{\PT}$ is
continuous and strictly positive at $y_{\pi}$, $\psi_{\mathsf{U}}$ is the
influence function of the estimated treated trimmed mean, accounting for the
estimation of both the trimming threshold and the trimming proportion, so
that $\E[\psi_{\mathsf{U}}^2]=V_{\mathsf{Lee},\mathsf{U}}$. Likewise
$\psi_0$ is the influence function of the estimated control mean, so that
$\E[\psi_0^2]=V_{\mathsf{Lee},\mathsf{C}}$. Since $\psi_{\mathsf{T}}$ and
$\psi_0$ have disjoint supports and $\E[\psi_0\psi_{\pi}]=0$ as established
above, we get $\E[\psi_{\mathsf{U}}\psi_0]=0$, and therefore
\[
\sigma_{\mathsf{full}}^2(1)
=\E\big[(\psi_{\mathsf{U}}-\psi_0)^2\big]
=\E[\psi_{\mathsf{U}}^2]+\E[\psi_0^2]
=V_{\mathsf{Lee},\mathsf{U}}+V_{\mathsf{Lee},\mathsf{C}}.
\]
Finally, when $r_1<1$,
$\widehat\sigma_{\mathsf{Lee}}^2=\widehat V_{\mathsf{Lee},\mathsf{U}}+\widehat
V_{\mathsf{Lee},\mathsf{C}}$ is the GMM variance estimator of
\citet[Proposition~3]{lee2009TrainingWagesSample}, hence
$\widehat\sigma_{\mathsf{Lee}}^2\cprob\sigma_{\mathsf{full}}^2(1)$, which was to be shown.
\end{proof}

\subsection{Proof of Lemma~\ref{lem:gumbel-coercive}}
\begin{proof}\label{proof:lem:gumbel-coercive}

We check the requirements of condition~(b) of
Theorem~\ref{thm: ignoring dual constraint} in turn, and only the
essential-supremum identity requires an argument. Because the same shock vector
enters the two maxima,
\[
\min_j(t_{jb}-t_{jb'})
\leq g(u;\theta)\leq
\max_j(t_{jb}-t_{jb'}),
\]
so $g$ is bounded. The entropy $\underline{\phi}_{\TV}$ is continuous at $1$ and
boundedness of $g$ implies
$\int|g(u;\theta)|\,\de\nu(u)<\infty$ for every finite measure $\nu$. The singleton
moment set $\{\rhopad_0\}$ is nonempty, convex, and closed, and
$\rint(\{\rhopad_0\})=\{\rhopad_0\}\subseteq
\rint(\Gamma^{\ll}_{\underline{\phi}_{\TV}}(\theta;\delta))$ by the
maintained relative interior condition.

It remains to compare the supremum and essential supremum. Fix $\lambda$ and
write $h_\lambda:=g-\lambda^\top m$. At fixed $\theta$, the ties 
partition $\R^{J+1}$ into finitely many open sets. On each set,
$s_1$ is constant and $h_\lambda$ is continuous. At a boundary point $u$,
consider the perturbed vector $u^r$, whose $j$-th entry is $u_j-rj$. For all
sufficiently small $r>0$, this perturbation preserves every strict
comparison and selects, in both choice problems, the lowest-indexed
alternative among the alternatives tied at $u$. It therefore approaches $u$
through an open set on which $s_1$ is constant and agrees with its value at
$u$. Since $g$ and $s_2$ are continuous, we get $h_\lambda(u^r)\to h_\lambda(u)$.
Continuity of $h_\lambda$ on that set therefore implies that values
arbitrarily close to $h_\lambda(u)$ occur on a nonempty open set. The Gumbel
baseline has full support, as its coordinates are independent with
strictly positive densities on $\R$, so every such open set has positive
$\Pbase{}$-probability. It follows that
$h_\lambda(u)\leq\Pbase{}\text{-}\esssup h_\lambda$ for every $u$.
Taking the supremum over $u$ gives
$\sup_{u}h_\lambda(u)\leq\Pbase{}\text{-}\esssup h_\lambda$, whereas the
reverse inequality holds because the supremum is an essential upper bound.
Hence
\[
\sup_{u\in\R^{J+1}}h_\lambda(u)
=
\Pbase{}\text{-}\esssup_{u\in\R^{J+1}}h_\lambda(u).
\]
The argument uses $g$ only through its continuity, so it applies verbatim to
$-g-\lambda^{\top}m$. Together with the checks recorded at the start of the
proof, this establishes condition~(b) of
Theorem~\ref{thm: ignoring dual constraint} for both signs, and the equality
of the bounds then follows from that theorem.
\end{proof}

\subsection{Proof of Proposition~\ref{prop: formula instruments contamination}}
\begin{proof}\label{proof: formula instruments contamination}
Throughout the proof we use the fact that $g_y$ and $g_x$ are bounded by
\eqref{eq: fi bounded criteria}. By the definition of
$\calP_{\mathsf{FI}}(\delta)$, every $P$ in that set is of the form
$P=(1-\delta)\widetilde{P}+\delta R$ for some $R\in\Delta(\calV)$, so that
\begin{align}
\E_P[h]=(1-\delta)\,\E_{\widetilde{P}}[h]+\delta\,\E_R[h]
\label{eq: mixture expectation}
\end{align}
for every bounded $h$ on $\calV$. Combining \eqref{eq: mixture expectation} with the
definitions of $G_y(\widetilde{P})$, $G_x(\widetilde{P})$, $G_y(v';\delta)$ and
$G_x(v';\delta)$, and using that $R$ is a probability measure, gives
\begin{align}
G_y(P)=\int_{\calV}G_y(v';\delta)\,\de R(v'),
\qquad
G_x(P)=\int_{\calV}G_x(v';\delta)\,\de R(v').
\label{eq: mixture numerator denominator}
\end{align}
\uline{\textbf{Part (i).}} Write $\underline{G}_x:=\inf_{v'\in\calV}G_x(v';\delta)$
and note that, by the definition of $G_x(v';\delta)$,
\[
\underline{G}_x
=G_x(\widetilde{P})-\delta\sup_{v'\in\calV}\bigl\{g_x(v')-\overline{g}_x^*\bigr\},
\]
so that $\underline{G}_x>0$ precisely when
 $\delta\in\calI_{\mathsf{FS}}$. That set is an interval containing
the origin, since the displayed expression is affine and nonincreasing in $\delta$
and equals $G_x(\widetilde{P})>0$ at $\delta=0$, and its right endpoint is
$\delta_{\mathsf{FS}}$, as the ratio appearing in that definition
is the budget at which the displayed expression vanishes. Suppose
$\underline{G}_x>0$. Then \eqref{eq: mixture numerator denominator} gives
$G_x(P)\geq\underline{G}_x>0$ for every $P\in\calP_{\mathsf{FI}}(\delta)$, so the
estimator is well defined there. Moreover,
$|G_y(P)|\leq\sup_{v'\in\calV}|G_y(v';\delta)|<\infty$ by
\eqref{eq: mixture numerator denominator} and the boundedness of $g_y$, so that
$|\widehat\beta(P)|$ is bounded by
$\sup_{v'}|G_y(v';\delta)|/\underline{G}_x$ and both bounds in
\eqref{eq: fi cont bounds} are finite. Suppose instead $\underline{G}_x<0$. Since
$\sup_{v'\in\calV}G_x(v';\delta)\geq G_x(\widetilde{P})>0$, the interval of values
of $G_x(P)$, which by \eqref{eq: mixture numerator denominator} has endpoints
$\underline{G}_x$ and $\sup_{v'}G_x(v';\delta)$, contains zero in its interior,
so the neighborhood contains assignment processes at which the recentered first stage
vanishes.

\uline{\textbf{Part (ii).}} Fix $\delta\in\calI_{\mathsf{FS}}$, so
that $\underline{G}_x>0$ by part (i), and set
$\overline{c}:=\sup_{v'\in\calV}G_y(v';\delta)/G_x(v';\delta)$, which is finite
because the numerators are bounded and the denominators are bounded below by
$\underline{G}_x>0$. For every $v'\in\calV$ we have
$G_y(v';\delta)\leq\overline{c}\,G_x(v';\delta)$, and integrating this
inequality against any $R\in\Delta(\calV)$ gives $G_y(P)\leq\overline{c}\,G_x(P)$ by
\eqref{eq: mixture numerator denominator}. Since $G_x(P)>0$, this reads
$\widehat\beta(P)\leq\overline{c}$, so that
$\overline{\beta}(\delta)\leq\overline{c}$. Conversely, taking $R$ to
be the point mass at $v'$ gives $\widehat\beta(P)=G_y(v';\delta)/G_x(v';\delta)$,
so that the supremum defining $\overline{c}$ is approached within
$\calP_{\mathsf{FI}}(\delta)$ and
$\overline{\beta}(\delta)=\overline{c}$, with attainment whenever the
supremum over $\calV$ is attained. The argument for
$\underline{\beta}(\delta)$ is symmetric. For the intermediate
values, let $v_1,v_2\in\calV$ and let
$R_t:=(1-t)\,\delta^{\mathtt{Dirac}}_{v_1}+t\,\delta^{\mathtt{Dirac}}_{v_2}$
for $t\in[0,1]$. The corresponding $\widehat\beta$ is a continuous function of
$t$, so by the intermediate value theorem it takes every value between the two
ratios attached to $v_1$ and $v_2$, and choosing $v_1$ and $v_2$ so that those
two ratios are arbitrarily close to $\underline{\beta}(\delta)$
and to $\overline{\beta}(\delta)$ shows that every value strictly
between the two bounds is taken.
\end{proof}

\subsection{Proof of Lemma~\ref{lemma: expectation decomposition in a TV neighborhood}}
\begin{proof}\label{proof:lemma: expectation decomposition in a TV neighborhood}
We prove the three parts in turn. The Hahn--Jordan construction behind
part~(i) also yields an expectation-decomposition identity, which we
record
along the way and then use to prove part~(ii).
Part~(iii) follows instead from the density
characterization of the decomposition neighborhood in
Lemma~\ref{lemma: mixture density bounds-i}, combined with the quantile
change of variables already employed in the proof of part~(ii).

\uline{\textbf{Part (i)}.} Define the signed measure $\nu:=P-Q$. By the Hahn
decomposition theorem, there exist sets $\calY^+,\calY^-\in\mathscr{F}$ such that
$\calY^-\cup\calY^+=\calY$, $\calY^-\cap \calY^+ = \emptyset$, $\nu(E)\geq 0$ for every
measurable $E\subseteq\calY^+$, and $\nu(E)\leq 0$ for every measurable
$E\subseteq\calY^-$. Define the nonnegative measures $\nu^-(E):=-\nu(E\cap\calY^-)$ and
$\nu^+(E):=\nu(E\cap\calY^+)$ for each $E\in\mathscr{F}$. By
construction they have disjoint supports. Since $P$ and $Q$ are probability
measures, $\nu(\calY)=P(\calY)-Q(\calY)=0$, and therefore $\nu^+(\calY) =
\nu^-(\calY)$ because $\nu^+(\calY) - \nu^-(\calY)= P(\calY) - Q(\calY) = 0.$

It remains to identify this common mass with $\TV(P\,\|\,Q)$. For every $E\in\mathscr{F}$,
$\nu(E)=\nu^+(E)-\nu^-(E)\leq \nu^+(E)\leq \nu^+(\calY)$, so
$\sup_{A\in\mathscr{F}}\nu(A)\leq \nu^+(\calY)$. Taking $A=\calY^+$ gives
$\nu(\calY^+)=\nu^+(\calY^+)-\nu^-(\calY^+)=\nu^+(\calY^+)-0=\nu^+(\calY)$, so the
supremum is achieved and $\sup_{A\in\mathscr{F}}\nu(A)=\nu^+(\calY)$. Applying the same
argument to $-\nu$, whose positive part is $\nu^-$, gives
$-\inf_{A\in\mathscr{F}}\nu(A)=\nu^-(\calY)$. Hence
\[
\TV(P\,\|\,Q)
= \sup_{A\in\mathscr{F}}|\nu(A)|
= \max\left\{\sup_{A\in\mathscr{F}}\nu(A),\,
-\inf_{A\in\mathscr{F}}\nu(A)\right\}
= \max\{\nu^+(\calY),\nu^-(\calY)\}
=\nu^+(\calY)=\nu^-(\calY),
\]
which together with $\TV(P\,\|\,Q)\leq\delta$ gives the asserted mass identity.

It remains to record the attainment claim. By construction
$\nu^+(\calY\setminus\calY^+)=\nu\big((\calY\setminus\calY^+)\cap\calY^+\big)=0$
and $\nu^-(\calY^+)=-\nu(\calY^+\cap\calY^-)=0$, so $\calY^+$ carries $\nu^+$ in
the stated sense. Moreover $\nu(\calY^+)=\nu^+(\calY)=\TV(P\,\|\,Q)\geq 0$ by the
previous display, whence
\[
|P(\calY^+)-Q(\calY^+)|=|\nu(\calY^+)|=\nu^+(\calY)=\TV(P\,\|\,Q),
\]
so the supremum defining the total variation is attained at $\calY^+$. This
proves fact (\textit{i}).

Finally, we record the decomposition identity used below. Since $\nu=P-Q$, we
have $P=Q+\nu=Q+\nu^+-\nu^-$. Moreover, $\nu^+\leq P$ and $\nu^-\leq Q$ as
measures, since for every $E\in\mathscr{F}$ we have
$\nu^+(E)=P(E\cap\calY^+)-Q(E\cap\calY^+)\leq P(E)$ and
$\nu^-(E)=Q(E\cap\calY^-)-P(E\cap\calY^-)\leq Q(E)$, so every $f\in L_1(P)\cap L_1(Q)$ is
integrable with respect to both $\nu^+$ and $\nu^-$, and
\begin{equation}\label{eq:tv-expectation-decomposition}
\int f\,\de P
=
\int f\,\de Q+\int f\,\de \nu^+-\int f\,\de \nu^-.
\end{equation}

\uline{\textbf{Part (ii)}.} Let $q=F_Q^{-1}$. Since $Q$ is the image of the uniform
distribution on $(0,1)$ under $q$, we have $\int h\,\de Q=\int_0^1 h(q(u))\,\de u$ for
every $h\in L_1(Q)$, and in particular $\E_Q[Y] = \int_0^1 q(u)\,\de u.$ By
fact (\textit{i}),
$\nu^+(\calY)=\nu^-(\calY)=\TV(P\,\|\,Q)\leq\delta$. Write
$t:=\nu^+(\calY)=\nu^-(\calY)=\TV(P\,\|\,Q)\leq\delta$ for the common mass of the two
parts. Taking $f(y)=y$ in~\eqref{eq:tv-expectation-decomposition},
\[
\E_P[Y] = \E_{Q}[Y] +\int y\,\de \nu^+(y) -\int y\,\de \nu^-(y).
\]
Because
$y\in[\underline{Y},\overline{Y}]$ on
the support,
\[
\int y\,\de\nu^+(y)\leq \overline{Y}\,\nu^+(\calY)= t\,\overline{Y},
\]
and so
\[
\E_P[Y] \leq \E_{Q}[Y]+t\,\overline{Y}-\int y\,\de\nu^-(y).
\]
We now bound $\int y\,\de\nu^-(y)$ from below. Since $\nu^-\leq Q$, we have
$\nu^-\ll Q$, so the Radon--Nikodym theorem yields a density
$r:=\de\nu^-/\de Q$, which satisfies $0\leq r\leq 1$ $Q$-a.e.\ in view of
$\nu^-\leq Q$, together with $\int r(y)\,\de Q(y)=\nu^-(\calY)=t$. Therefore,
by the
change of variables $\int h(x)\,\de Q(x)=\int_0^1 h(q(u))\,\de u$ applied with
$h(x)=x\,r(x)$,
\[
\int y\,\de\nu^-(y)=\int y\,r(y)\,\de Q(y) =\int_0^1 q(u)\,w(u)\,\de u,
\]
where $w(u):=r(q(u))$ satisfies $0\leq w\leq 1$ and $\int_0^1 w(u)\,\de u=t$.
Minimizing
$\int_0^1 q(u)\,w(u)\,\de u$ over all $w$ with $0\leq w\leq1$ and
$\int_0^1 w(u)\,\de u=t$ -- a class at least as large as the one induced by
the measures $\nu^-\leq Q$ of mass $t$ -- and using that $q$ is weakly
increasing,
the minimum over this class is attained at $w^\star(u)=\indic\{u\leq t\}$, whence
\[
\int y\,\de\nu^-(y)\geq\int_0^t q(u)\,\de u
\qquad\text{for every }\nu^-\leq Q\text{ with }\nu^-(\calY)=t.
\]
Substituting this lower bound, every $P$ with $\TV(P\,\|\,Q)=t\leq\delta$ obeys
\[
\E_P[Y]\leq \E_{Q}[Y]+G(t),\qquad G(t):=t\,\overline{Y}-\int_0^t q(u)\,\de u .
\]
The map $G$ is nondecreasing on $[0,1]$, since
$G(t)=\int_0^t(\overline{Y}-q(u))\,\de u$ and the integrand is nonnegative
because $q(u)\leq\overline{Y}$ for all $u$. Hence the bound is largest at the
maximal
admissible
mass $t=\delta$, giving the closed form upper bound
\[
\overline{\mu}_{Q}(\delta) = \E_{Q}[Y]+\delta\,\overline{Y}-\int_0^\delta q(u)\,\de
u ,
\]
attained: let $\nu^\star$ be the sub-measure of $Q$ of total mass $\delta$
carried by the lowest quantiles, that is, $\de\nu^\star/\de
Q=\indic\{y<q(\delta)\}+\gamma\,\indic\{y=q(\delta)\}$ with $\gamma\in[0,1]$
chosen so that $\nu^\star(\calY)=\delta$ -- such a $\gamma$ exists since
$F(q(\delta)-)\leq\delta\leq F(q(\delta))$ by definition of the quantile
function -- and set
$P^\star:=Q-\nu^\star+\delta\,\delta^{\mathtt{Dirac}}_{\overline{Y}}$. Then
$P^\star$ is a nonnegative measure in view of $\nu^\star\leq Q$, its total
mass equals $1-\delta+\delta=1$, and $\TV(P^\star\,\|\,Q)\leq\delta$ because
$P^\star-Q=\delta\,\delta^{\mathtt{Dirac}}_{\overline{Y}}-\nu^\star$ implies
$|P^\star(A)-Q(A)|\leq\max\{\delta,\nu^\star(\calY)\}=\delta$ for every
$A\in\mathscr{F}$, and
\[
\int y\,\de\nu^\star = \int_0^{F(q(\delta)-)} q(u)\,\de u
+ q(\delta)\bigl(\delta-F(q(\delta)-)\bigr) = \int_0^\delta q(u)\,\de u,
\]
where the last equality uses that $q(u)=q(\delta)$ for
$u\in(F(q(\delta)-),\delta]$. Hence both one-sided bounds hold with equality
and $\E_{P^\star}[Y]=\overline{\mu}_{Q}(\delta)$.
A symmetric argument gives the lower bound.

Finally, because $\E_Q[Y]=\int_0^1q(u)\de u$, we can also rewrite the bounds as
\[
\overline{\mu}_{Q}(\delta) = \delta\,\overline{Y}+\int_\delta^1 q(u)\,\de u,
\qquad
\underline{\mu}_{Q}(\delta) = \delta\,\underline{Y}+\int_0^{1-\delta} q(u)\,\de u.
\]

\uline{\textbf{Part (iii)}.} Suppose first that $\kappa=1$. In
this case, $\calC^{\ll}_{1}(Q)=\{Q\}$ and both displayed formulas reduce to the
identity $\E_Q[Y]=\int_0^1 q(u)\,\de u$. Fix then $\kappa\in(0,1)$. By
Lemma~\ref{lemma: mixture density bounds-i}, a probability measure
$\tilde P$ belongs to $\calC^{\ll}_{\kappa}(Q)$ if and only if
$\tilde P\ll Q$ and its density $\tilde r:=\de\tilde P/\de Q$ satisfies
$\tilde r\leq\kappa^{-1}$ $Q$-a.e. In particular, every
$\tilde P\in\calC^{\ll}_{\kappa}(Q)$ satisfies
$\E_{\tilde P}[|Y|]\leq\kappa^{-1}\E_Q[|Y|]<\infty$.

By the change of variables $\int h(y)\,\de Q(y)=\int_0^1 h(q(u))\,\de u$
applied with $h(y)=y\,\tilde r(y)$,
\[
\E_{\tilde P}[Y]=\int y\,\tilde r(y)\,\de Q(y)=\int_0^1 q(u)\,w(u)\,\de u,
\]
where $w(u):=\tilde r(q(u))$ satisfies $0\leq w\leq \kappa^{-1}$ and
$\int_0^1 w(u)\,\de u=\int \tilde r(y)\,\de Q(y)=1$. We now maximize
$\int_0^1 q(u)\,w(u)\,\de u$ over all $w$ with $0\leq w\leq\kappa^{-1}$ and
$\int_0^1 w(u)\,\de u=1$ and claim that the maximum over this class is attained at
$w^\star(u):=\kappa^{-1}\indic\{u>1-\kappa\}$. Indeed, for any feasible $w$,
\[
\int_0^1 q(u)\,\bigl(w^\star(u)-w(u)\bigr)\,\de u
=\int_0^1 \bigl(q(u)-q(1-\kappa)\bigr)\bigl(w^\star(u)-w(u)\bigr)\,\de u
\geq 0,
\]
where the equality holds because $\int_0^1(w^\star(u)-w(u))\,\de u=1-1=0$,
and the inequality follows because the integrand is nonnegative. To see this, 
note that for
$u>1-\kappa$ we have $q(u)\geq q(1-\kappa)$, since $q$ is weakly increasing,
and $w^\star(u)-w(u)=\kappa^{-1}-w(u)\geq0$, whereas for $u\leq1-\kappa$ we
have $q(u)\leq q(1-\kappa)$ and $w^\star(u)-w(u)=-w(u)\leq0$. Hence every
$\tilde P\in\calC^{\ll}_{\kappa}(Q)$ obeys
\[
\E_{\tilde P}[Y]\leq\int_0^1 q(u)\,w^\star(u)\,\de u
=\frac1\kappa\int_{1-\kappa}^1 q(u)\,\de u .
\]

It remains to show that the bound is attained. Let $\nu^\star$ be the
sub-measure of $Q$ of total mass $\kappa$ carried by the largest quantiles,
that is,
$\de\nu^\star/\de Q=\indic\{y>q(1-\kappa)\}+\gamma\,\indic\{y=q(1-\kappa)\}$
with $\gamma\in[0,1]$ chosen so that $\nu^\star(\calY)=\kappa$. Note that such a
$\gamma$ exists because $Q(Y>q(1-\kappa))=1-F(q(1-\kappa))\leq\kappa$ and
$Q(Y\geq q(1-\kappa))=1-F(q(1-\kappa)-)\geq\kappa$ by the definition of the
quantile function. Set $\tilde P^\star:=\kappa^{-1}\nu^\star$. Then
$\tilde P^\star$ is a probability measure by construction, and
$\kappa\tilde P^\star=\nu^\star\leq Q$ as measures, so
$\tilde P^\star\in\calC^{\ll}_{\kappa}(Q)$ in view of
Lemma~\ref{lemma: mixture density bounds-i}. Moreover,
\[
\E_{\tilde P^\star}[Y]=\frac1\kappa\int y\,\de\nu^\star
=\frac1\kappa\left(\int_{F(q(1-\kappa))}^1 q(u)\,\de u
+ q(1-\kappa)\bigl(\kappa-1+F(q(1-\kappa))\bigr)\right)
=\frac1\kappa\int_{1-\kappa}^1 q(u)\,\de u,
\]
where the second equality collects the mass that $\nu^\star$ places on
$\{y>q(1-\kappa)\}$, which contributes $\int_{F(q(1-\kappa))}^1 q(u)\,\de u$
by the change of variables above and the identity
$\{u\in(0,1):q(u)>q(1-\kappa)\}=(F(q(1-\kappa)),1)$ up to null sets, and the
mass $\gamma\,Q(\{q(1-\kappa)\})=\kappa-1+F(q(1-\kappa))$ that $\nu^\star$
places on the atom at $q(1-\kappa)$, while the last equality uses that
$q(u)=q(1-\kappa)$ for $u\in(1-\kappa,F(q(1-\kappa))]$. This proves the
formula for the supremum. A symmetric argument, with the mass carried by the
lowest quantiles and $w^\star(u)=\kappa^{-1}\indic\{u\leq\kappa\}$, yields
the formula for the infimum together with its attainment.
\end{proof}

\subsection{Proof of Theorem~\ref{thm: endpoint sharp bounds}}
\begin{proof}\label{proof:thm: endpoint sharp bounds}
Throughout, for $P\in\Delta(\calZ)$ we write $P=w\Pbase{}+P^\perp$
for its Lebesgue decomposition with respect to $\Pbase{}$, where
$P^\perp\perp\Pbase{}$ and $w$ is the Radon--Nikodym derivative of the
absolutely continuous part of $P$ with respect to $\Pbase{}$, so that
$\E_{\Pbase{}}[w(Z)]+P^\perp(\calZ)=P(\calZ)=1$. By
Lemma~\ref{lemma: divergence lebesgue decomposition},
\begin{align}
D_\phi(P\,\|\,\Pbase{})
=\int_\calZ\phi(w(z))\,\de\Pbase{}(z)+\phi^\infty(1)\,P^\perp(\calZ).
\label{eq: endpoint divergence decomposition}
\end{align}
Since $\phi$ is convex and finite at $\underline{x}$, and at $\bar{x}$ when
the latter is finite, its effective domain is $[\underline{x},\bar{x}]$ if
$\bar{x}<\infty$ and $[\underline{x},\infty)$ if $\bar{x}=\infty$. Moreover,
$\underline{x}\leq1\leq\bar{x}$ because $\phi(1)=0$.

\uline{\textbf{Finite divergence and domination.}}
We show that $\calD_\phi(\Pbase{})$ coincides with the set of
$P\in\Delta(\calZ)$ with $D_\phi(P\,\|\,\Pbase{})<\infty$, and that
$D_\phi(P\,\|\,\Pbase{})\leq\overline{\delta}_\phi$ on this set.

\uline{\textbf{Case $\bar{x}<\infty$.}}
Since $\phi(t)=+\infty$ for $t>\bar{x}$,
Definition~\ref{defi: recession function} with $z_0=1$ gives
$\phi^\infty(1)=+\infty$. Hence, in view of
\eqref{eq: endpoint divergence decomposition} and $\phi\geq0$,
$D_\phi(P\,\|\,\Pbase{})<\infty$ forces $P^\perp(\calZ)=0$ and
$w\in[\underline{x},\bar{x}]$ $\Pbase{}$-almost surely. By the
equivalence of~(2) and~(3) in Lemma~\ref{lemma: mixture density bounds}, which holds for
every $\kappa>0$, applied with $\kappa=1/\bar{x}$ in part~(i) and, if
$\underline{x}>0$, with $\kappa=\underline{x}$ in part~(ii), these two conditions hold
if and only if $\underline{x}\Pbase{}\leq P\leq\bar{x}\Pbase{}$. Conversely, let $P$
satisfy these inequalities, so that $P=w\Pbase{}$ with
$\underline{x}\leq w\leq\bar{x}$ $\Pbase{}$-almost surely. For
every $t\in[\underline{x},\bar{x}]$, convexity of $\phi$ gives
\[
\phi(t)
\leq\frac{(\bar{x}-t)\,\phi(\underline{x})+(t-\underline{x})\,\phi(\bar{x})}{\bar{x}-\underline{x}},
\]
and integrating this bound at $t=w(z)$ against $\Pbase{}$, using
$\E_{\Pbase{}}[w(Z)]=1$, delivers
$D_\phi(P\,\|\,\Pbase{})\leq\{(\bar{x}-1)\phi(\underline{x})+(1-\underline{x})\phi(\bar{x})\}/(\bar{x}-\underline{x})=\overline{\delta}_\phi$,
which is finite by assumption.

\uline{\textbf{Case $\bar{x}=\infty$.}}
By \eqref{eq: endpoint divergence decomposition} and
$\phi\geq0$, $D_\phi(P\,\|\,\Pbase{})<\infty$ forces $w\geq\underline{x}$
$\Pbase{}$-almost surely, and therefore $P\geq w\Pbase{}\geq\underline{x}\Pbase{}$.
Conversely, if $\underline{x}\Pbase{}\leq P$, then
$w\geq\underline{x}$ $\Pbase{}$-almost surely, trivially if $\underline{x}=0$ and by
Lemma~\ref{lemma: mixture density bounds-ii} with $\kappa=\underline{x}$ otherwise, as
the equivalence of~(2) and~(3) there holds for every $\kappa>0$. For every
$t\geq\underline{x}$ and every $T>\max\{t,1\}$, convexity of $\phi$ gives
\[
\phi(t)\leq\frac{(T-t)\,\phi(\underline{x})+(t-\underline{x})\,\phi(T)}{T-\underline{x}},
\]
and, since $\phi(1)=0$, Definition~\ref{defi: recession function} with
$z_0=1$ gives $\phi^\infty(1)=\lim_{T\to\infty}\phi(T)/T$, so that letting
$T\to\infty$ yields $\phi(t)\leq\phi(\underline{x})+(t-\underline{x})\,\phi^\infty(1)$.
Integrating this bound at $t=w(z)$ against $\Pbase{}$ and using
\eqref{eq: endpoint divergence decomposition}, we obtain
\[
D_\phi(P\,\|\,\Pbase{})
\leq\phi(\underline{x})+\phi^\infty(1)\bigl\{\E_{\Pbase{}}[w(Z)]-\underline{x}
+P^\perp(\calZ)\bigr\}
=\phi(\underline{x})+(1-\underline{x})\,\phi^\infty(1)=\overline{\delta}_\phi,
\]
where the equality follows from $\E_{\Pbase{}}[w(Z)]+P^\perp(\calZ)=1$, and
the right-hand side is finite by assumption.

\uline{\textbf{Nonparametric sharp bounds.}}
Fix $\delta\geq\overline{\delta}_\phi$ and $\theta\in\Theta$. If
$P\in\calP_\phi(\theta;\rho,\Pbase{},\delta)$, then
$D_\phi(P\,\|\,\Pbase{})\leq\delta<\infty$, so $P\in\calD_\phi(\Pbase{})$ by
the first part of the proof, and $P\in\calE(\theta;\rho)$ by definition of the robustness
set. Conversely, if $P\in\calD_\phi(\Pbase{})\cap\calE(\theta;\rho)$, then
$D_\phi(P\,\|\,\Pbase{})\leq\overline{\delta}_\phi\leq\delta$ by the first part of the
proof, so $P$ satisfies both constraints defining
$\calP_\phi(\theta;\rho,\Pbase{},\delta)$. The two sets therefore coincide for every
$\theta\in\Theta$, and the displayed identities follow because, at every $\theta$, the two
inner optimization problems share the same feasible set and the same objective, so that
the outer optimizations over $\theta\in\Theta$ coincide as well.
\end{proof}

\begin{spacing}{0.8}
\setlength{\bibsep}{2pt plus .3ex}
\bibliographystyle{aer}
\bibliography{references}
\end{spacing}